\documentclass[defended]{cit_thesis}
\usepackage{geometry}
\usepackage[nohyperlinks,printonlyused]{acronym}

\usepackage[colorlinks,citecolor=blue,linkcolor=blue,urlcolor=blue]{hyperref}

\usepackage[sc]{mathpazo}
\usepackage{inconsolata}

\usepackage{microtype}

\newcommand{\acrolowercase}[1]{\lowercase{\textls[0]{#1}}}

\usepackage{natbib}
\usepackage{amsmath,amsthm,amssymb}
\usepackage{stmaryrd} 
\usepackage{upgreek} 
\usepackage{multirow}
\newtheorem{thm}{Theorem}
\newcommand{\Msun}{\ensuremath{M_{\odot}}}

\newcommand{\tmpsamps}{\ensuremath{N}}
\newcommand{\numtmps}{\ensuremath{M}}
\newcommand{\numslices}{\ensuremath{S}}
\newcommand{\svdtmps}[1]{\ensuremath{L^#1}}
\newcommand{\numsvdtmps}{\svdtmps{s}}
\newcommand{\slicesamps}[1]{\ensuremath{N^#1}}
\newcommand{\slicessamps}{\slicesamps{s}}
\newcommand{\fftblock}{\ensuremath{D}}

\newcommand{\gstlal}{{\tt gstlal}}
\newcommand{\gstreamer}{GStreamer}

\DeclareMathOperator{\expectation}{E}
\DeclareMathOperator{\var}{var}
\DeclareMathOperator{\sinc}{sinc}

\def\clap#1{\hbox to 0pt{\hss#1\hss}}

\def\mathrlap{\mathpalette\mathrlapinternal}
\def\mathclap{\mathpalette\mathclapinternal}
 
\def\mathrlapinternal#1#2{\rlap{$\mathsurround=0pt#1{#2}$}} 
\def\mathclapinternal#1#2{\clap{$\mathsurround=0pt#1{#2}$}}

\usepackage{emulateapj-symbols}

\usepackage{epigraph}
\newif\ifinthesis
\inthesistrue

\usepackage{import}

\usepackage{pdflscape}
\usepackage{deluxetable}
\newenvironment{deluxetable*}[1]{\begin{deluxetable}{#1}\tablewidth{0pt}}{\end{deluxetable}}

\usepackage{listings}
\usepackage{bayestar-localization-paper/bib/aas_macros}
\usepackage{bm}
\usepackage{bigdelim}
\usepackage{mathrsfs}
\usepackage{protosem}
\DeclareMathOperator{\Tr}{Tr}
\DeclareMathOperator{\cov}{cov}
\DeclareMathOperator{\std}{std}

\DeclareMathOperator{\erf}{erf}

\DeclareMathOperator*{\argmax}{\arg\!\max}
\newcommand\transpose{\ensuremath{^{^\mathsf{T}}}}

\usepackage{multirow}
\usepackage{textcomp}

\usepackage[colorlinks]{hyperref}

\copyyear{2015}
\date{November 24, 2014}
\title{The needle in the 100 deg$^2$ haystack: \\
The hunt for binary neutron star mergers with \acs{LIGO} and \acl{PTF}}
\author{Leo P. Singer}

\begin{document}

\maketitle

\begin{dedication}
\vspace{2cm}
\begin{flushright}
To the love of my life, my wife Kristin, and our precious son Isaac.
\end{flushright}
\vspace{2cm}
\epigraph{
Bacon in his instruction tells us that the scientific student ought not to be as the ant, who gathers merely, nor as the spider who spins from her own bowels, but rather as the bee who both gathers and produces. All this is true of the teaching afforded by any part of physical science. Electricity is often called wonderful, beautiful; but it is so only in common with the other forces of nature. The beauty of electricity or of any other force is not that the power is mysterious, and unexpected, touching every sense at unawares in turn, but that it is under law, and that the taught intellect can even now govern it largely. The human mind is placed above, and not beneath it, and it is in such a point of view that the mental education afforded by science is rendered super-eminent in dignity, in practical application and utility; for by enabling the mind to apply the natural power through law, it conveys the gifts of God to man.}{Michael Faraday, Lecture notes of 1858, quoted in \emph{The Life and Letters of Faraday} (1870) by Bence Jones, Vol. 2, p. 404}
\end{dedication}

\begin{acknowledgments}
This is \acs{LIGO} Document Number \acs{LIGO}\nobreakdashes-P1400223\nobreakdashes-v10. I carried out the work presented in this thesis within the \acf{LSC} and the \acf{IPTF} collaboration. The methods and results I present are under review and are potentially subject to change. The opinions expressed here are my own and not necessarily those of the \ac{LSC} or \ac{IPTF}.

\null

I gratefully acknowledge funding from the United States \acf{NSF} for the construction and operation of the \acs{LIGO} Laboratory, which provided support for this work. \acs{LIGO} was constructed by the California Institute of Technology and Massachusetts Institute of Technology with funding from the \ac{NSF} and operates under cooperative agreement PHY\nobreakdashes-0107417. I thank the \ac{NSF} for supporting my research directly through a Graduate Research Fellowship. This work is based on observations obtained with the \acl{P48} and the \acl{P60} at the Palomar Observatory as part of the Intermediate Palomar Transient Factory project, a scientific collaboration among the California Institute of Technology, Los Alamos National Laboratory, the University of Wisconsin, Milwaukee, the Oskar Klein Center, the Weizmann Institute of Science, the TANGO Program of the University System of Taiwan, and the Kavli Institute for the Physics and Mathematics of the Universe. The work in this thesis is partly funded by \emph{Swift} Guest Investigator Program Cycle 9 award 10522 (NASA grant NNX14AC24G) and Cycle 10 award 10553 (NASA grant NNX14AI99G).

\null

Thank you, Mom, thank you Dad, for an upbringing full of love, learning, and love of learning.

Thank you, my wife Kristin, thank you, my son Isaac, for your love and for your patience with me.

Thank you, Susan Bates, for your tutoring in problem solving that has resonated with me from elementary school through every day of my scientific career.

Thank you, John Jacobson, Amanda Vehslage, Tambra Walker, and Dr. Philip Terry\nobreakdashes-Smith, for the most inspiring courses in my high school education, and for molding me into a responsible and well\nobreakdashes-rounded individual.

Thank you, Profs. Luis Orozco and Betsy Beise, for your mentoring and friendship as well as the University of Maryland undergraduate physics courses that I enjoyed so much. Thank you for initiating me into physics research, and for sending me to graduate school so well prepared.

Thank you, Prof. Alan Weinstein, for being an outstanding (and, when necessary, forbearing) thesis advisor, for engineering the many wonderful collaborations that I have been a part of at Caltech, and for showing me how to thrive within a Big Science experiment.

Thank you, Prof. Shri Kulkarni, for recruiting me into \acs{PTF}, for engineering a totally original cross-disciplinary research opportunity in physics and astronomy, and for placing trust in me. I am continually in awe of how that trust has paid off. I thank my colleagues in \acs{PTF} for welcoming me into their highly capable and exciting team.

Thank you, Prof. Christian Ott, for teaching me two formative courses. I was able to write \acs{BAYESTAR}, my greatest contribution so far to Advanced \acs{LIGO}, only because the latter of these courses (Ay\,190: Computational Astrophysics) was fresh in my head.

Thank you, Prof. David Reitze, for making me feel like the success of Advanced \acs{LIGO} depends upon me. (I think that you inspire that same feeling in everyone at \acs{LIGO} Laboratory.)

Thank you, Rory Smith, my officemate, for ducking good-naturedly whenever I wanted to chuck a chair out the window of 257 West Bridge. (Despite many strong oaths, no chairs were actually chucked during the writing of this thesis.)

Thank you, Nick Fotopoulos, Larry Price, Brad Cenko, and Mansi Kasliwal, for your collaboration and friendship throughout my studies, friendships that I hope to keep and to nurture.

\end{acknowledgments}

\begin{abstract}
The Advanced \acs{LIGO} and Virgo experiments are poised to detect \acp{GW} directly for the first time this decade. The ultimate prize will be joint observation of a compact binary merger in both gravitational and electromagnetic channels. However, \ac{GW} sky locations that are uncertain by hundreds of square degrees will pose a challenge. I describe a real-time detection pipeline and a rapid Bayesian parameter estimation code that will make it possible to search promptly for optical counterparts in Advanced \acs{LIGO}. Having analyzed a comprehensive population of simulated \ac{GW} sources, we describe the sky localization accuracy that the \ac{GW} detector network will achieve as each detector comes online and progresses toward design sensitivity. Next, in preparation for the optical search with the \ac{IPTF}, we have developed a unique capability to detect optical afterglows of \acp{GRB} detected by the \emph{Fermi} \ac{GBM}. Its comparable error regions offer a close parallel to the Advanced \acs{LIGO} problem, but \emph{Fermi}'s unique access to MeV\nobreakdashes--GeV photons and its near all\nobreakdashes-sky coverage may allow us to look at optical afterglows in a relatively unexplored part of the \ac{GRB} parameter space. We present the discovery and broadband follow-up observations (X\nobreakdashes-ray, UV, optical, millimeter, and radio) of eight \ac{GBM}\nobreakdashes--\ac{IPTF} afterglows. Two of the bursts (\ac{GRB}~130702A~/~iPTF13bxl and \ac{GRB}~140606B~/~iPTF14bfu) are at low redshift ($z=0.145$ and $z = 0.384$, respectively), are sub\nobreakdashes-luminous with respect to ``standard'' cosmological bursts, and have spectroscopically confirmed broad\nobreakdashes-line type Ic supernovae. These two bursts are possibly consistent with mildly relativistic shocks breaking out from the progenitor envelopes rather than the standard mechanism of internal shocks within an ultra-relativistic jet. On a technical level, the \ac{GBM}--\ac{IPTF} effort is a prototype for locating and observing optical counterparts of \ac{GW} events in Advanced \acs{LIGO} with the \acl{ZTF}.

\end{abstract}

\tableofcontents
\listoffigures
\listoftables

\mainmatter

\renewcommand\author[1]{%
    \begin{center}%
    \setstretch{1.0}%
    \sc%
    \begingroup
    \hyphenpenalty=100000%
    #1%
    \endgroup%
    \end{center}%
    \vspace{6pt}%
}
\newcommand\altaffilmark[1]{$^{#1}$}
\newcommand\altaffiltext[2]{%
    \begin{center}%
    \footnotesize%
    \setlength{\parskip}{-6pt}%
    \setstretch{1.0}%
    \begingroup
    \hyphenpenalty=100000%
    $^{#1}$%
    #2%
    \endgroup%
    \end{center}%
}
\newcommand\attribution[1]{%
    \null%
    \begin{quote}%
    \centering%
    \itshape%
    \footnotesize%
    #1%
    \end{quote}%
    \null%
}

\chapter{The road to Advanced \acs{LIGO}}

Einstein's general theory of relativity holds that the laws of motion play out in a curved space\nobreakdashes-time, with curvature caused by the presence of matter and energy. This strange statement has some even stranger consequences. One of the earliest solutions of Einstein's equation predicted \acp{BH}, stars made of pure space-time curvature, whose gravitational wells are so deep that nothing, not even light, can escape. We now know that when a massive star exhausts the last of its fuel, it can collapse to form a \ac{NS}---the densest possible stable arrangement of matter, something akin to a gigantic nucleus with atomic number $10^{57}$---or a stellar-mass \ac{BH}. This gravitational collapse can be messy and loud. It may produce a relativistic shock wave that powers a long \ac{GRB}, and it may drive a supernova explosion that outshines the late star's host galaxy in visible light for several weeks. Long afterward, the strong gravitational field of a compact object can have other interesting consequences. If the star has a binary companion from which it can accrete matter, it can power a wide range of high-energy transient phenomena. However, all of these processes occur in basically static (but strongly curved) space-time. 

\begin{figure}[t]
    \centering
    \includegraphics{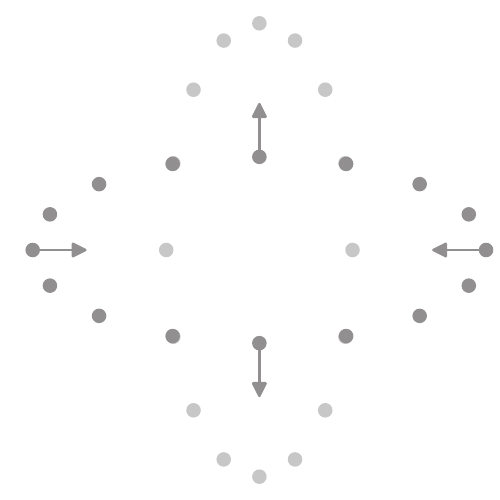}\qquad\includegraphics{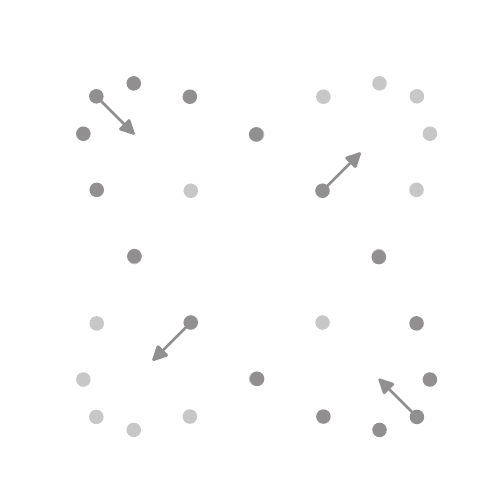}
    \caption[Effect of a \acs{GW} on a ring of free-falling test particles]{\label{fig:polarizations}Effect of a \acs{GW} on a ring of free\nobreakdashes-falling test particles. Left: a `$+$' polarized \ac{GW}, causing the test particles to be alternately squeezed or stretched in two orthogonal directions. Right: a `$\times$' polarized \ac{GW}, causing a stretching and squeezing in a sense that is rotated $45\arcdeg$ relative to the `$+$' polarization.}
\end{figure}

\begin{figure}
    \centering
    \includegraphics[width=0.75\textwidth]{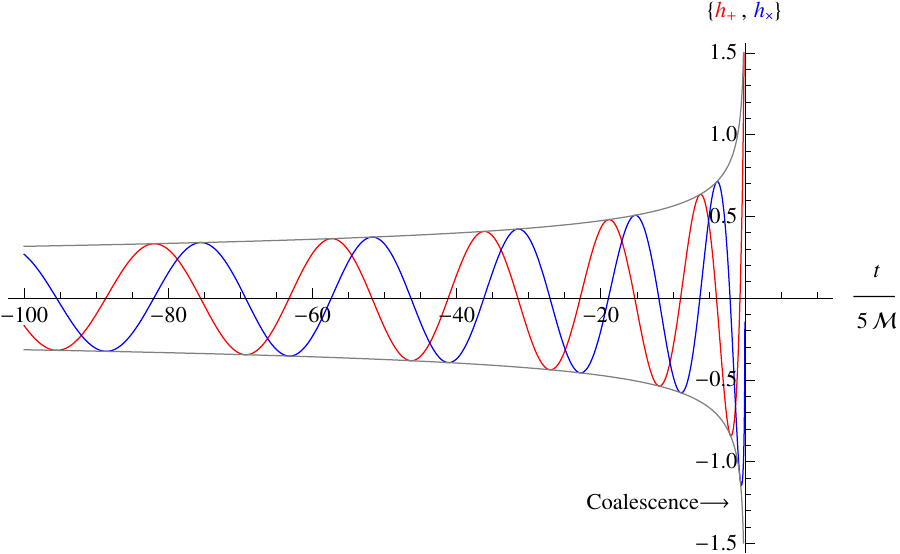}
    \caption[A basic \acs{BNS} inspiral waveform]{\label{fig:inspiral-waveform}A basic \ac{CBC} ``inspiral'' waveform. The red and blue traces correspond to two orthogonal \acs{GW} polarizations (see Chapter~\ref{chap:fisher}). At the lowest post\nobreakdashes-Newtonian order, the signal is shrunk or dilated in time by a single mass parameter: the chirp mass, $\mathcal{M}$, a combination of the component masses defined in Section~\ref{sec:prospects-detection}.}
\end{figure}

In the dynamical regime, Einstein's theory predicts \acp{GW} that transmit energy via propagating disturbances in space\nobreakdashes-time, much as the dynamical solutions of Maxwell's equations carry energy as light. Operationally, the effect of a passing \ac{GW} is to slightly change the separation between free\nobreakdashes-falling objects (see Figure~\ref{fig:polarizations}). The brightest source of gravitational waves that we think nature can make is a binary system of two compact objects (\acp{NS} and/or \acp{BH}). If a compact binary is in a tight enough orbit, gravitational radiation can efficiently carry away energy and angular momentum. This orbital decay was famously observed in the binary pulsar PSR~1913+16 \citep{1975ApJ...195L..51H,1982ApJ...253..908T}, for which Hulse and Taylor received the Nobel Prize in Physics in 1993. The energy loss eventually will become a runaway process, as the orbital separation decreases and the system radiates even more gravitational waves. See Figure~\ref{fig:inspiral-waveform} for an illustration of the basic \ac{GW} ``inspiral'' waveform due to a \ac{CBC}. Ultimately, the two compact objects will coalesce: they will become a single perturbed \ac{BH}, which will ring down as it settles into rotationally symmetric stationary state.

If one or both of the binary companions is a \ac{NS}, the the merger process itself can also be messy and loud. The immense tidal forces can tear apart the \ac{NS} before it takes the final plunge. The resultant hot, highly magnetized accretion flow may create the conditions necessary for a highly relativistic jet \citep{2011ApJ...732L...6R}. This process is thought to power short \acp{GRB} \citep{1986ApJ...308L..43P,1989Natur.340..126E,1992ApJ...395L..83N,2011ApJ...732L...6R}.

The \acf{LIGO} and Virgo have been constructed with the aim of directly detecting \acp{GW} from \acp{CBC} of \acp{BNS}, among other potential sources. This will provide a singularly dramatic confirmation of Einstein's relativity in the otherwise largely untested strong\nobreakdashes-field dynamical regime. \ac{GW} observations could even test alternative theories of gravity~\citep[Section 6,][]{lrr-2006-3,MeasuringNSEquationOfStateCanBeDone} or constrain the NS equation of state~\citep{MeasuringNSEquationOfState}. A temporal coincidence between a \ac{CBC} event and a short \ac{GRB} would also settle the question of the progenitors of at least some of these elusive explosions. No \ac{GW} events were detected in \ac{LIGO}--Virgo observing runs at an initial sensitivity \citep{LowMassSearchS6VSR23}. However, \ac{LIGO} is just now finishing its transformation into Advanced \ac{LIGO}, with Advanced Virgo soon to follow suit. Both designed to be ultimately ten times more sensitive than their predecessors, they will be able to monitor a thousand times more volume within the local Universe. The first detections are expected over the next few years \citep{LIGORates}.

\begin{figure*}
    \centering
    \includegraphics[width=0.32\textwidth]{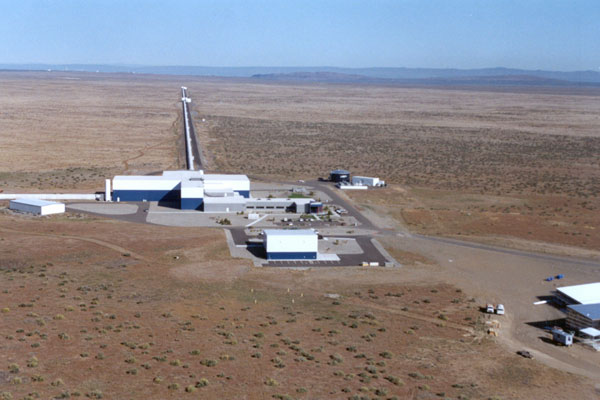}
    \includegraphics[width=0.32\textwidth]{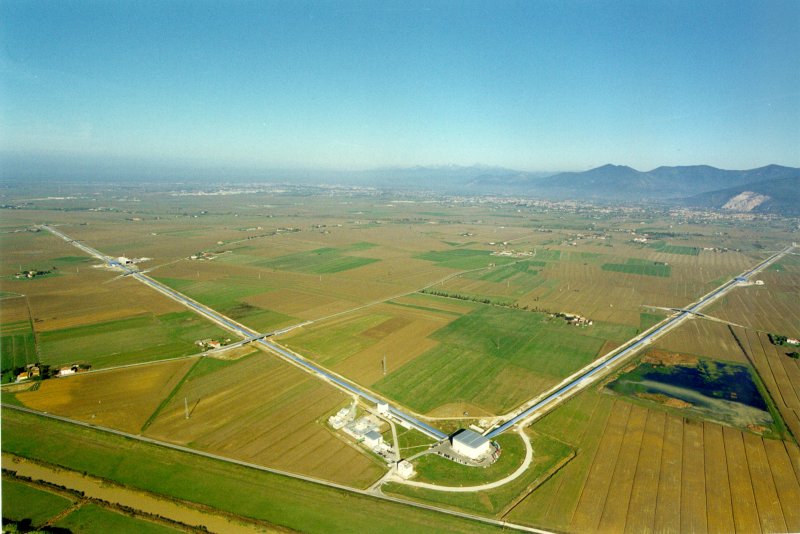}
    \includegraphics[width=0.32\textwidth]{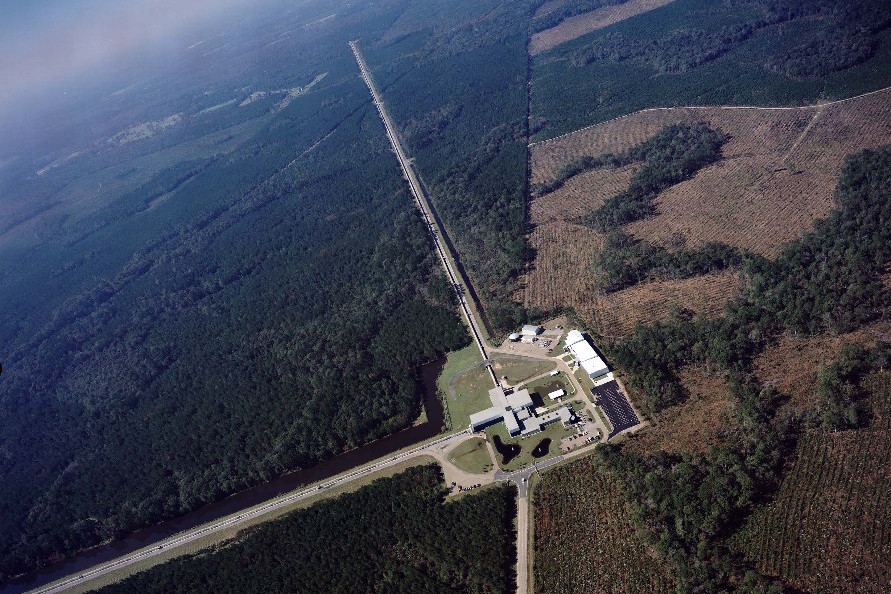}
    \caption[Aerial photographs of ground-based \acs{GW} detectors]{\label{fig:detectors}From left to right: aerial views of \acl{LHO} (reproduced from \url{http://ligo.org}), Virgo (Reproduced from \url{http://virgo.lal.in2p3.fr}), and \acl{LLO} (reproduced from \url{http://ligo.org}).}
\end{figure*}

\begin{figure}
    \centering
    \includegraphics{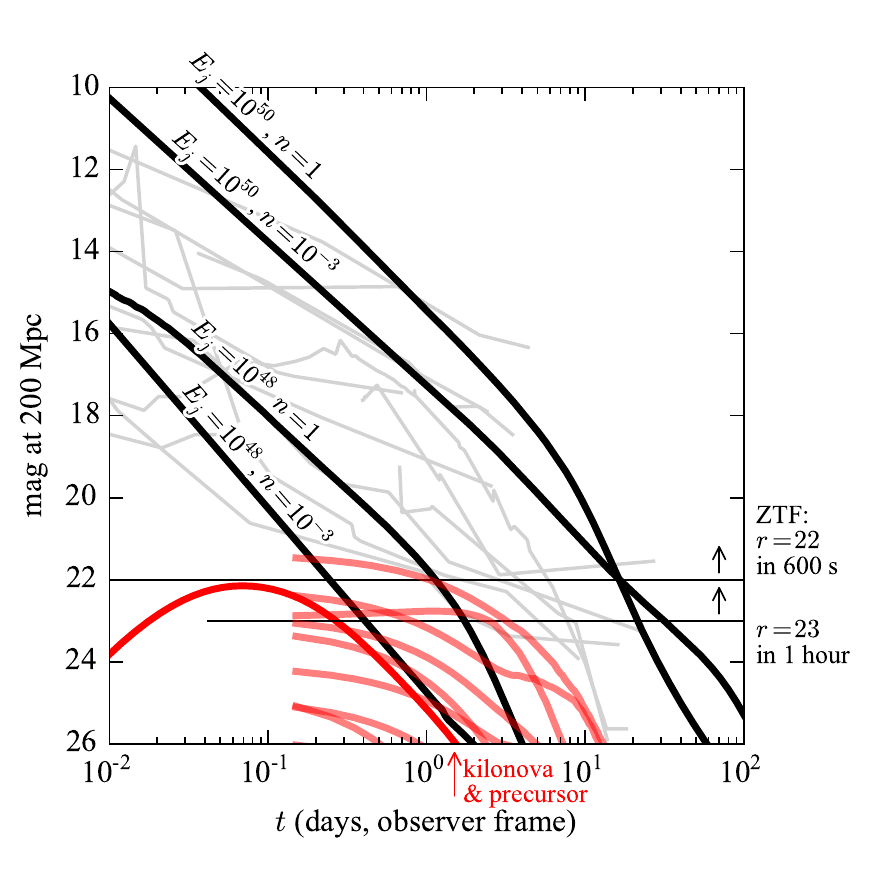}
    \caption[Typical light curves of Advanced \acs{LIGO} optical counterparts]{\label{fig:sgrb-kcorrected}Light curves of short \ac{GRB} afterglows, scaled to an Advanced \ac{LIGO} range of 200~Mpc. Thin gray lines are afterglows of \emph{Swift} short \acp{GRB} that have known redshifts. Thick black lines are synthetic on\nobreakdashes-axis afterglows from \citet{SyntheticSGRBAfterglows} with jet half\nobreakdashes-opening angles of 0.2~rad and observer angles of 0~rad. Jet energies $E_j$, in units of ergs, and circumburst densities $n$, in units of cm$^{-3}$, are labeled on the plot. The solid, deep red line is the $r$\nobreakdashes-band neutron\nobreakdashes-powered kilonova precursor model from \citet{KilonovaPrecursor} with opacity $\kappa_r = 30$~cm$^2$\,g$^{-1}$, free neutron mass $m_n = 10^{-4}$~$M_\sun$, and electron fraction $Y_e = 0.05$. The solid, light red lines represent kilonova models from \citet{KilonovaHighOpacities} with ejected masses of $m_\mathrm{ej} = 10^{-3}$,~$10^{-2}$,~or $10^{-1}~M_\sun$ and characteristic velocities $\beta = v/c = 0.1$,~0.2, or~0.3. The \citet{KilonovaPrecursor} kilonova precursors are blue: they peak at about 0.1~mag brighter in the $g$ band than in $r$. The \citet{KilonovaHighOpacities} kilonova models are red: they are about 1~mag brighter in the $i$ band than in $r$.}
\end{figure}

A perhaps even greater prize would be detecting both the \ac{GW} signal and an optical transient resulting from the same \ac{BNS} merger event. An optical afterglow \citep{SyntheticSGRBAfterglows} would aid in the understanding of the physics of the relativistic jet (for the ``classic'' model, see \citealt{AfterglowSpectra}), and a bright on-axis afterglow would be the most obvious signpost by which to locate the host galaxy. These signatures, however, are expected to be rare because, like the short \ac{GRB} itself, we have to be inside the collimated cone of the jet to see them. Perhaps a more promising optical signature \citep{MostPromisingEMCounterpart} would be that of a roughly omnidirectional ``kilonova'' powered by the radioactive decay of the hot $r$-process ejecta \citep{Kilonova,KilonovaHighOpacities} or a ``kilonova precursor'' powered by free neutrons in the fast-moving outer layers of the ejecta \citep{KilonovaPrecursor}. A kilonova could inform us about the nature and distribution of the ejecta, and tell us whether the merged compact object collapsed directly to a black hole or went through a brief phase as a hyper\nobreakdashes-massive neutron star \citep{KilonovaRedOrBlue}. See Figure~\ref{fig:sgrb-kcorrected} for typical $r$-band light curves of these optical signatures. If we could detect \ac{GW} and \ac{EM} emission from a sufficiently large number of \acp{CBC}, then we could simultaneously measure their luminosity distances and redshifts, thereby adding an almost calibration-free ``standard siren'' to the cosmological distance ladder \citep{SchutzStandardSirens,HolzStandardSirens,DalalStandardSirens,StandardSirens}.\footnote{Third\nobreakdashes-generation detectors such as the proposed Einstein Telescope will be able to simultaneously determine distances and redshifts of \ac{BNS} mergers \emph{through \acp{GW} observations alone} by measuring the orbital frequency at which tidal disruption occurs; see \citet{MessengerStandardSirens}.}

\section{Challenges}

Some first steps toward multimessenger observations were taken in the last \ac{LIGO}--Virgo science run. The first low\nobreakdashes-latency \ac{CBC} search was deployed, including an online matched filter analysis for detection (\acl{MBTA}; \acsu{MBTA}), a fast but ad hoc algorithm for sky localization \citep{CBCLowLatency}, and a system for sending alerts to optical facilities \citep{kanner2008}. The whole process from data acquisition to alerts took about half an hour (dominated by a final human\nobreakdashes-in\nobreakdashes-the\nobreakdashes-loop check; see Chapter~\ref{chap:detection} for a full timing budget). This period also saw the development of the first practical Bayesian parameter estimation codes, which at the time took a few weeks to thoroughly map the parameter space of any detection candidate \citep{S6PE}. A consortium of X\nobreakdashes-ray, optical, and radio telescopes participated in searching for \ac{EM} counterparts \citep{FirstPromptSearchGWTransientsEMCounterparts,FirstSearchesOpticalCounterparts}.

Although these were important proofs of concept, the increased sensitivity of the Advanced \ac{LIGO} detectors will force us toward more sophisticated approaches at each stage of the process (detection, sky localization, and \ac{EM} follow\nobreakdashes-up). The first challenge is longer \ac{GW} signals. A significant part of Advanced \ac{LIGO}'s expanded detection volume comes from better sensitivity at low frequencies, moving the seismic noise cutoff from $\sim$40~Hz to $\sim$10~Hz (see Appendix~\ref{sec:low-frequency-cutoff} for a detailed discussion of the sensitivity as a function of low frequency cutoff). \ac{CBC} signals are chirps, ramping from low to high frequency as $f \propto t^{-3/8}$. Although a typical \ac{BNS} merger signal would remain in band for Initial \ac{LIGO} for about 25~s, it would be detectable by Advanced \ac{LIGO} for as long as 1000~s (see Equation~(\ref{eq:fgw}) in Chapter~\ref{chap:detection}). A second consequence of longer signals is that the signal can accumulate more power and a larger total phase shift while in band, improving the ability to measure the mass of the binary but dramatically increasing the number of \ac{GW} templates required to adequately tile the parameter space. A third problem is that we cannot assume that the detector and the instrument noise are in a stationary state for the durations of these long signals; we must adaptively condition or \emph{whiten} the data as the noise level rises or falls, and we must be able to carry on integrating the signal over gaps or glitches. These are all formidable problems for traditional \ac{FFT}-based matched filter pipelines, which have inflexible data handling, whose latency grows with the length of the signal, and whose computational requirements increase with both the length and number of template signals. To effectively search for these signals in real time we need a detection pipeline whose latency and computational demands do not scale much with the duration of the \ac{GW} signal.

The second challenge is that the sky localization most be both fast \emph{and} accurate. The original rapid sky localization and full Bayesian parameter estimation codes entailed an undesirable tradeoff of response time and accuracy: the former took only minutes, but produced sky areas that were 20 times larges than the latter, which could take days \citep{SiderySkyLocalizationComparison}. This compromise was somewhat acceptable in Initial \ac{LIGO} because, given the small number of galaxies within the detectable volume, one could significantly reduce the area to be searched by selecting fields that contained nearby galaxies \citep{GWHostGalaxyCatalog,GWGC}. With the expanded range of Advanced \ac{LIGO} enclosing many more galaxies, this will still be a valuable strategy, but will be somewhat less effective \citep{NissankeKasliwalEMCounterparts}. Given that the predicted optical signatures of \ac{BNS} mergers are faint (with kilonovae predicted to be fainter than $R > 22$~mag) and may peak in under a day, it is essential that the rapid localization be as accurate as possible. Ideally, it should be just as accurate as the localization from the full Bayesian parameter estimation.

Third and most importantly, we have to build the instruments, software, collaborations, and observational discipline to search through areas of hundreds of deg$^2$ for the faint, rapidly fading optical counterparts. We need deep, wide\nobreakdashes-field optical survey telescopes to scan the \ac{GW} localizations and detect new transient or variable sources, robotic follow-up telescopes to track photometric evolution and obtain color information, a network of 5\nobreakdashes-m class and larger telescopes to secure spectroscopic classifications, as well as X\nobreakdashes-ray and radio telescopes that can act on \acp{TOO}. To identify which among tens or hundreds of thousands of optical transients to follow up we need real-time image subtraction, machine learning, and integration with archival survey data, not to mention a team of human observers in the loop and executing deep spectroscopic observations on large\nobreakdashes-aperture telescopes.

\begin{figure}[t]
    \centering
    \includegraphics[width=0.5\textwidth]{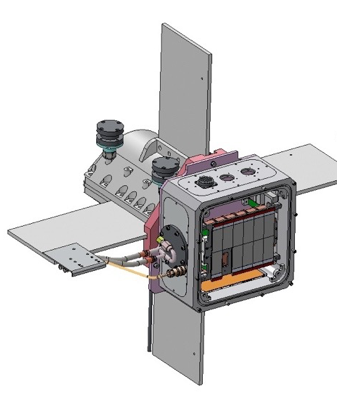}%
    \includegraphics[width=0.5\textwidth]{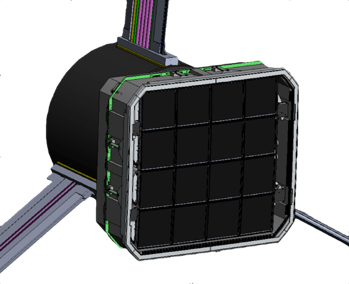}%
    \caption[\acs{PTF} and \acs{ZTF} cameras]{\label{fig:cameras}The \ac{PTF} (left) and \ac{ZTF} (right) cameras. Reproduced from a presentation by E.~Bellm.}
\end{figure}

There is a fortunate convergence between the construction of Advanced \ac{LIGO} and Virgo and the deployment of deep, high\nobreakdashes-cadence, synoptic, optical transient surveys. Experiments like the \acl{PTF} (\acsu{PTF}; \citealt{PTFRau,PTFLaw}) have focused on discovering rare or rapidly rising optical transients, but should also be well suited to searching for optical counterparts of \ac{GW} sources. The key instrument in \ac{PTF} is the \ac{CFH12k} camera~\citep{P48PTF} on the \acf{P48}, capable of reaching limits of $R \approx 20.6$~mag in 60~s over a wide, 7.1~deg$^2$ \ac{FOV}. In its planned successor, the \acf{ZTF}, this will be replaced by a new 47~deg$^2$ camera. With a larger \ac{FOV} and faster readout electronics, \ac{ZTF} will achieve an order of magnitude faster volumetric survey rate (see Figure~\ref{fig:cameras} for an illustration of the \ac{PTF} and \ac{ZTF} cameras and Table~\ref{tab:cameras} for a comparison of survey speeds). A real\nobreakdashes-time image subtraction and machine learning pipeline supplies a stream of new optical transient candidates from which a team of human observers selects the most interesting targets for multicolor photometry on the \acf{P60} and spectroscopic classification on the \acf{P200} and other large telescopes. Lessons learned by \ac{PTF} will inform the planning and operation of future optical transient surveys such as BlackGEM\footnote{\url{https://www.astro.ru.nl/wiki/research/blackgemarray}} (which will be dedicated to following up \ac{GW} sources) and the \acf{LSST}, as relates to both blind transient searches and targeted searches for optical counterparts of \ac{GW} candidates.

\begin{deluxetable*}{lcc}
\tablewidth{0pt}
\tablecaption{\label{tab:cameras}Comparison of the survey speeds of the \acs{PTF} and \acs{ZTF} cameras. Reproduced from a presentation by E.~Bellm.}
\tablehead{\colhead{} & \colhead{\acs{PTF}} & \colhead{\acs{ZTF}}}
\startdata
Active area & 7.26~deg$^2$ & 47~deg$^2$ \\
Overhead time & 46~s & $<15$~s \\
Optimal exposure time & 60~s & 30~s \\
Relative areal survey rate & 1x & 15.0x \\
Relative volumetric survey rate & 1x & 12.3x
\enddata
\end{deluxetable*}

\section{Aims of this thesis}

The aim of my thesis is to deliver the major, fully and realistically characterized and tested, pieces of the search for optical counterparts of \ac{BNS} mergers, including detection and parameter estimation as well as the optical transient search itself. Here is a chapter\nobreakdashes-by\nobreakdashes-chapter summary of the content of this thesis.

\null

\noindent \textbf{Chapter~\ref{chap:fisher}} introduces the basic principles of a matched filter bank \ac{GW} search. We describe the range of a \ac{GW} detector in terms of its directional sensitivity or antenna pattern, its noise \ac{PSD}, and the \ac{SNR}. We then apply the Fisher information matrix formalism to compute the approximate sky resolution of a network of \ac{GW} detectors. There is a great deal of prior literature on this topic that considers \ac{GW} sky localization in terms of timing triangulation (see, for instance, \citealt{FairhurstTriangulation}). Our calculation captures the additional contributions of the phases and amplitudes on arrival at the detectors, which we show to be significant, especially near the plane of the detectors where timing triangulation is formally degenerate. Our derivation is extremely compact, and evaluating it is only marginally more complicated than the timing triangulation approach. We discuss the sky localization accuracy as a function of direction in the sky, and build some intuition that we will rely upon in future chapters. This chapter is in preparation as a separate paper and as a proposed update to a living document describing the \ac{GW} detector commissioning and observing schedule \citep{LIGOObservingScenarios}.

\null

\noindent \textbf{Chapter~\ref{chap:detection}} describes a novel matched filtering algorithm that is capable of detecting a \ac{GW} signal within seconds after the merger, or even seconds \emph{before}. This algorithm, called \acs{LLOID}, uses orthogonal decomposition and multirate signal processing to bring the computational demands of an online \ac{BNS} search within the scope of current resources. My contributions to \acs{LLOID} include: working on the pipeline to drive the latency to $\sim$10~s, improving data handling to be able to skip over glitches in the data efficiently without unduly sacrificing \ac{SNR}, studying the signal processing and computational aspects of the algorithm, improving the time and phase accuracy of triggers, and preparing the first complete description of it for the literature \citep{Cannon:2011vi}. \acs{LLOID} has been extensively tested both offline and in real time with simulated and commissioning data in a series of ``engineering runs,'' and will serve as the flagship low\nobreakdashes-latency \ac{BNS} detection pipeline in Advanced \ac{LIGO}. This chapter is in preparation as a standalone technical paper.

\null

\begin{figure*}
    \centering
    \includegraphics[width=\textwidth]{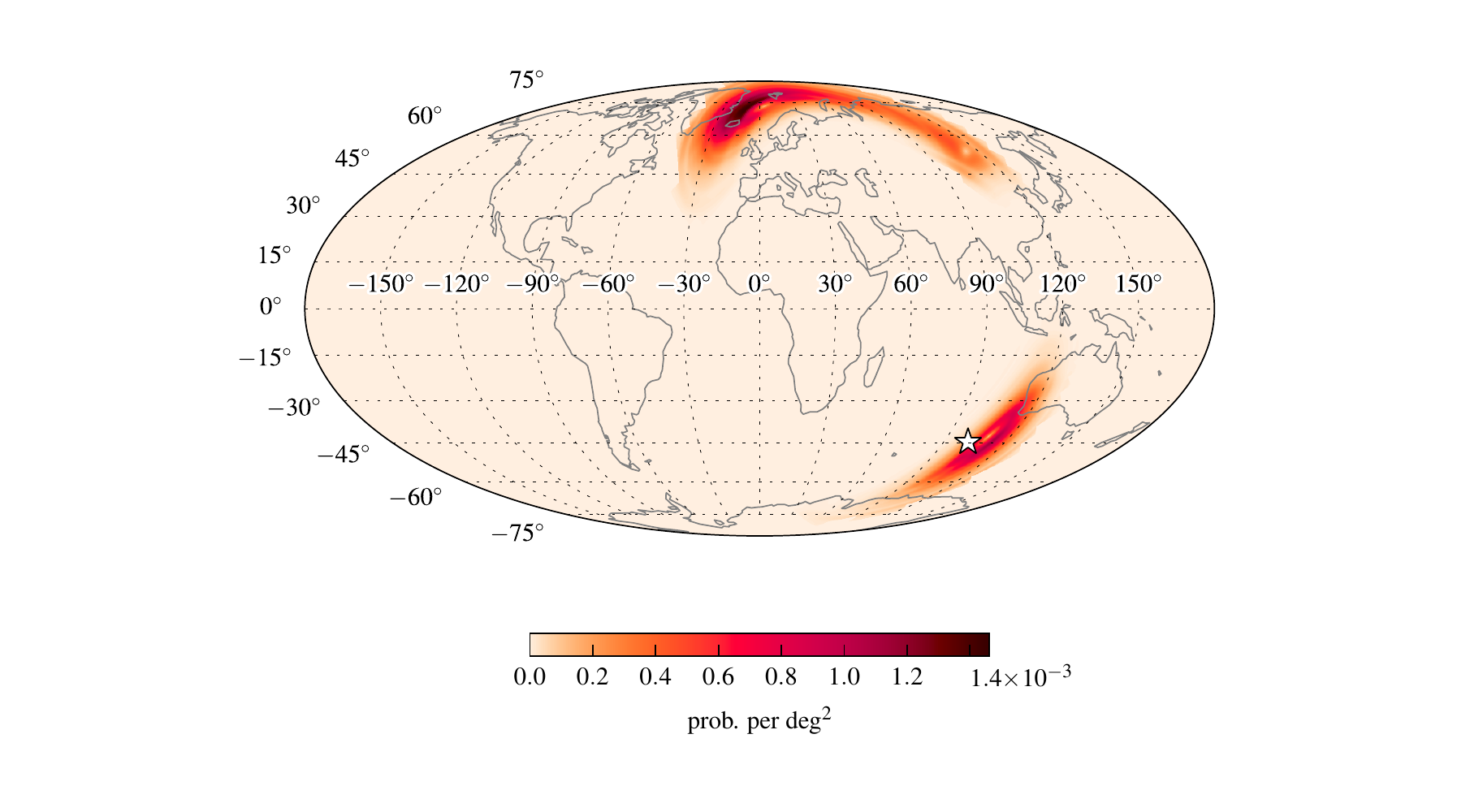}
    \\
    $\downarrow$
    \\
    \includegraphics[width=\textwidth]{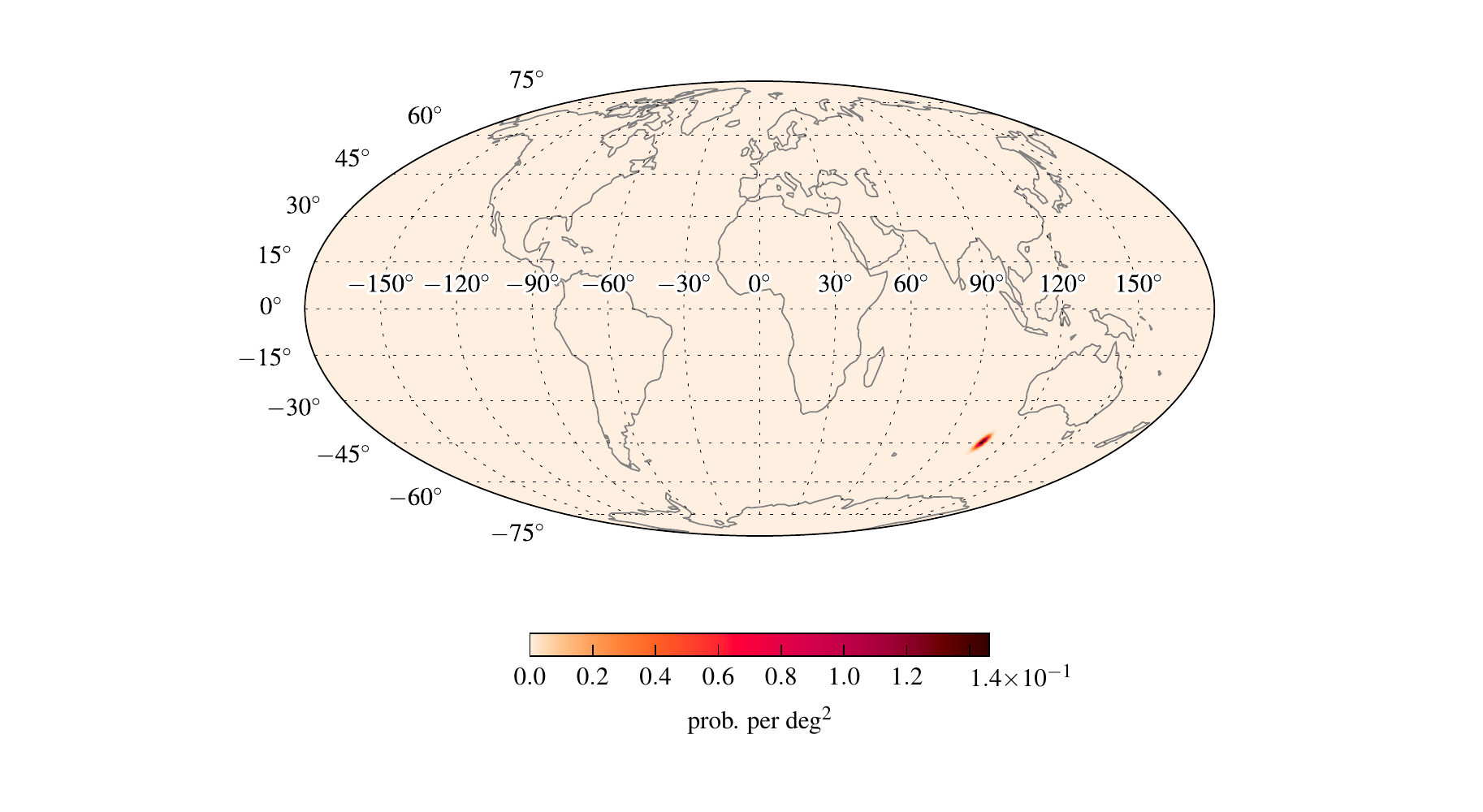}
    \caption[Example \acs{GW} localization, comparing Initial \acs{LIGO} method to \acs{BAYESTAR}]{\label{fig:skymap-improvement-versus-s6}Probability sky maps for a simulated three\nobreakdashes-detector, $(1.26,\,1.49)\,M_\sun$ merger event from the third engineering run. Above is the sky map from the Initial \ac{LIGO} rapid localization code. Below is the sky map from \acs{BAYESTAR}. (This is event G71031.)}
\end{figure*}

\noindent \textbf{Chapter~\ref{chap:bayestar}} develops a new rapid sky localization algorithm, \acs{BAYESTAR}, that takes just tens of seconds, but achieves about the same accuracy (see Figure~\ref{fig:skymap-improvement-versus-s6}) as the full parameter estimation. It owes its speed to three innovations. First, like the ad hoc Initial \ac{LIGO} code, it takes as input the matched filter parameter estimates from the detection pipeline rather than the full \ac{GW} time series. Second, using a result from Chapter~\ref{chap:fisher}, it concerns itself with sky location only and not the masses of the signal, exploiting the fact that the errors in the intrinsic and extrinsic parameters of a \ac{BNS} signal are approximately uncorrelated. Third, though fully Bayesian, unlike the full parameter estimation it does not use \ac{MCMC} sampling; instead it uses an adaptive sampling grid and low-order Gaussian quadrature. The result is both inherently fast and also highly parallelizable. Like \acs{LLOID}, it has been tested with both extensive offline simulations and in online engineering runs.

\null

\noindent Having assembled the Advanced \ac{LIGO} real\nobreakdashes-time \ac{BNS} pipeline in Chapters~\ref{chap:detection}~and~\ref{chap:bayestar}, in \textbf{Chapter~\ref{chap:first2years}} we provide a detailed description of what the first Advanced \ac{LIGO} detections and sky localizations may look like. Because our new rapid sky localization algorithm is orders of magnitude faster than the full parameter estimation, for the first time we can perform end-to-end analyses of thousands of events, thereby providing a statistically meaningful and comprehensive description of the areas and morphologies that will arise in the early Advanced \ac{LIGO} configurations. The first 2015 observing run is expected to involve only the two \ac{LIGO} detectors in Hanford, Washington, and Livingston, Louisiana, and not the Virgo detector in Cascina, Italy. \citet{LIGOObservingScenarios} assumed, based on timing triangulation considerations, that two detector networks would always produce localizations that consist of degenerate annuli spanning many thousands of deg$^2$. We find that the interplay between the phase and amplitude on arrival (i.e., the \ac{GW} polarization) and prior distribution powerfully break this degeneracy (see also \citealt{Raymond:2009,KasliwalTwoDetectors}), limiting \emph{almost all} areas to below 1000~deg$^2$, with a median of about 600~deg$^2$. We elucidate one curious degeneracy that survives, that causes most source localizations to equally favor the true position of the source as well as a position at the polar opposite. We then model the 2016 observing run, which has the \ac{LIGO} detectors operating with somewhat deeper sensitivity and has Advanced Virgo online. Even with Virgo's sensitivity delayed with the staggered commissioning, adding the third detector shrinks areas to a median of $\sim$200~deg$^2$. As the detectors continue approaching final design sensitivity and as more detectors come online, areas will continue to shrink to $\sim$10~deg$^2$ and below. This chapter is published as \citet{FirstTwoYears}. The supplementary data described in Appendix~\ref{chap:first2years-data} contains a browsable catalog of simulated \ac{GW} sky maps, in the format that will be used for sending \ac{GW} alerts (which is described in Appendix~\ref{chap:sky-map-format}).

\null

\begin{figure*}[t]
    \centering
    \includegraphics[width=\textwidth]{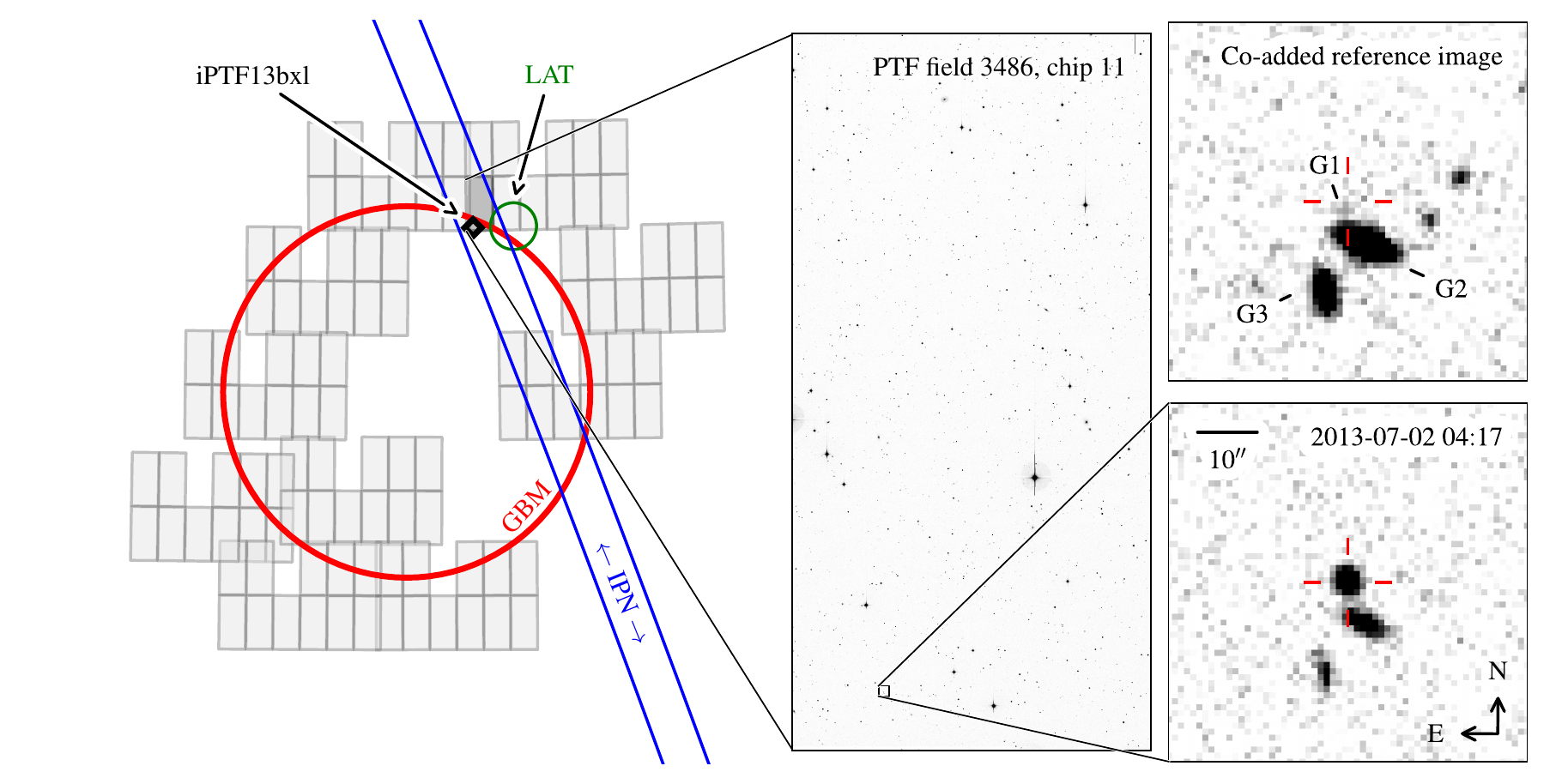}
    \caption[Discovery of GRB~130702A~/~iPTF13bxl]{\label{fig:iPTF13bxl}\ac{P48} imaging of GRB~130702A and discovery of iPTF13bxl. The left panel illustrates the $\gamma$-ray localizations (red circle: 1$\sigma$ \ac{GBM}; green circle: \ac{LAT}; blue lines: 3$\sigma$ \ac{IPN}) and the 10 \ac{P48} reference fields that were imaged (light gray rectangles).  For each \ac{P48} pointing, the locations of the 11 CCD chips are indicated with smaller rectangles (one CCD in the camera is not currently operable). The small black diamond is the location of iPTF13bxl. The right panels show \ac{P48} images of the location of iPTF13bxl, both prior to (top) and immediately following (bottom) discovery. Reproduced from \citet{iPTF13bxl}.}
\end{figure*}

\noindent \textbf{Chapter~\ref{chap:iptf13bxl}} confronts the search for optical transients in large error regions with the \acl{IPTF}; (\acsu{IPTF}; \citealt{iPTF}) and its planned successor, the \acl{ZTF} (\acsu{ZTF}; \citealt{ZTF,ZTFBellm,ZTFSmith}). The surest way to convince ourselves that the search for optical counterparts of \ac{GW} transients will be effective is to try it out and discover something. As a model problem, for the past year we have searched for optical counterparts of \acp{GRB} detected by the \acl{GBM} (\acsu{GBM}; \citealt{GBM}) instrument onboard the \emph{Fermi} satellite. Like \ac{LIGO}, \emph{Fermi} \ac{GBM} produces coarse localizations that are uncertain by $\sim$100~deg$^2$, and though afterglows of long \acp{GRB} are much brighter than anticipated \ac{LIGO} optical counterparts, the important timescales for follow-up observations are similar. \emph{Fermi} \ac{GBM} bursts are also interesting in their own right. \emph{Fermi} \ac{GBM} and the \emph{Swift} \acl{BAT} (\acsu{BAT}; \citealt{BAT}) have highly complementary strengths: fields of view of 70\% and 10\% of the sky respectively, and energy bandpasses of $\sim$few~keV~to~300~GeV (when including the \emph{Fermi} \acl{LAT} or \acsu{LAT}; \citealt{LAT}) and 15\nobreakdashes--150~keV respectively. However, with the \ac{GBM}'s coarse localization, very few \emph{Fermi} bursts have been studied outside the gamma\nobreakdashes-ray band (the exception being bursts that are coincidentally also detected by \emph{Fermi} \ac{LAT} or \emph{Swift} \ac{BAT}). In this chapter, we relate the discovery, redshift, and broadband observations of \ac{GRB}~130702A and its optical afterglow, iPTF13bxl \citet{iPTF13bxl}. This is the first discovery of an optical afterglow based solely on a \emph{Fermi} \ac{GBM} localization (see discovery image in Figure~\ref{fig:iPTF13bxl}). This is a notable event in and of itself for several other reasons. First, its redshift places it among the nearest \acp{GRB} ever recorded. Second, its prompt energy release in gamma rays is intermediate between bright, cosmologically distant, ``standard'' bursts, and nearby \acp{llGRB} which comprise many of the well\nobreakdashes-studied \ac{GRB}--\ac{SN}. Finally, because of its low redshift we were able to spectroscopically detect its associated \acl{SNIcBL}, establishing it as a test for the \ac{GRB}--\ac{SN} connection.

\null

\noindent \textbf{Chapter~\ref{chap:iptf-gbm}} reports on the total of eight \ac{GBM}--\ac{IPTF} afterglows that we have discovered in one year of this experiment. In this chapter, we present our broadband follow\nobreakdashes-up including spectroscopy as well as X\nobreakdashes-ray, UV, optical, sub\nobreakdashes-millimeter, millimeter, and radio observations. We study possible selection effects in the context of the total \emph{Fermi} and \emph{Swift} \ac{GRB} samples. We identify one new outlier on the Amati relation, challenging its application to standardize \ac{GRB} luminosities. We find that two bursts are consistent with a mildly mildly relativistic shock breaking out from the progenitor star, rather than the ultra\nobreakdashes-relativistic internal shock mechanism that powers standard cosmological bursts. Finally, in the context of the \acf{ZTF}, we discuss how we will continue to expand this effort to find optical counterparts of \acl{BNS} mergers that should soon be detected Advanced \acs{LIGO} and Virgo.

\chapter{Range and sky resolution of \acs{GW} detector networks}
\label{chap:fisher}

\attribution{
This chapter is reproduced from a work in preparation, of which I will be the sole author. Section~\ref{sec:detector-sensitivity} is reproduced from \citet{FirstTwoYears}, copyright~\textcopyright{}~2014 The American Astronomical Society.
}

In this chapter, we will use a basic description of the signal and noise received by a \ac{GW} detector network to derive a matched filter bank, the prevailing technique used to search for well\nobreakdashes-modeled \ac{CBC} signals in \ac{LIGO} data. This model will allow us to calculate the range and angular resolution of a network of detectors.

\section{Basic matched filter search}
\label{sec:basic-matched-filter-search}

With interferometric detectors like LIGO and Virgo, the astrophysical signal is embedded in a time series measurement, the strain or the differential change in the lengths of the detectors' two arms. Many noise sources enter the detector in different subsystems, get filtered by the detector's response, and add to the measured strain. There are ``fundamental'' noise sources, such as quantum fluctuations in the laser field that result in shot noise at low frequency and radiation pressure noise at high frequency. Other noise sources are ``technical,'' meaning that they arise from the implementation of the detector as a realizable non-ideal system; examples include glitches due to scattered light, laser frequency fluctuations, cross\nobreakdashes-coupling between length degrees of freedom, coupling between angular and length degrees of freedom, and time\nobreakdashes-varying alignment drifts. Other noise sources are ``environmental,'' such seismic or anthropogenic ground motion noise.

For the purpose of \ac{GW} data analysis, the most important division is between quasi-stationary Gaussian-like noise and transient noise sources (``glitches''). Extracting astrophysical signals from the data requires \ac{FD} techniques (whitening, matched filtering) to suppress the former and \ac{TD} approaches (coincidence, candidate ranking, time slides) to deal with the latter.

\ac{CBC} searches are greatly aided by the fact that their \ac{GW} signals can (at least in principle) be predicted with exquisite precision throughout LIGO's sensitive band. Therefore, a standard approach to \ac{CBC} detection is matched filtering; a representative set of model waveforms is assembled into a template bank with which the data is convolved.

In the \ac{TD}, the strain observed by a single \ac{GW} interferometer is
\begin{equation}
    y_i(t) = x_i (t; \bm\theta) + n_i (t).
\end{equation}
In the \ac{FD},
\begin{equation}\label{eq:signal-model}
    Y_i (\omega) = \int_{-\infty}^\infty y(t) e^{-i \omega t} dt = X_i (\omega; \bm\theta) + N_i (\omega),
\end{equation}
where $X_i (\omega; \bm\theta)$ is the \ac{GW} signal given a parameter vector $\bm\theta$ that describes the \ac{GW} source, and $N_i (\omega)$ is that detector's Gaussian noise with one\nobreakdashes-sided \ac{PSD} $S_i(\omega) = E\left[\left|n_i(\omega)\right|^2\right] + E\left[\left|n_i(-\omega)\right|^2\right] = 2 E\left[\left|n_i(\omega)\right|^2\right]$. We shall denote the combined observation from a network of detectors as $\mathbf Y (\omega) \equiv \{Y_i (\omega)\}_i$.

Under the assumptions that the detector noise is Gaussian and that the noise from different detectors are uncorrelated, the likelihood of the observation, $\mathbf y$, conditioned on the value of $\bm\theta$, is a product of Gaussian distributions:
\begin{equation}\label{eq:gaussian-likelihood}
    \mathcal{L}(\mathbf Y; \bm\theta) = \prod_i p(Y_i | \bm\theta)
        \propto \exp \left[
        - \frac{1}{2} \sum_i \int_0^\infty \frac{\left|Y_i (\omega)
            - X_i(\omega; \bm\theta) \right|^2}{S_i(\omega)} \, d\omega
    \right].
\end{equation}

A \ac{CBC} source is specified by a vector of extrinsic parameters describing its position and orientation, and intrinsic parameters describing the physical properties of the binary components:
\begin{equation}\label{eq:params}
    \bm\theta = \left[
    \begin{array}{cl@{\quad}p{4.25cm}lp{3.75cm}}
        \alpha & \rdelim]{11}{0mm} & right ascension & \rdelim\}{7}{1mm} & \multirow{7}{2cm}{extrinsic parameters, $\bm\theta_\mathrm{ex}$} \\
        \delta && declination & \\
        r && distance & \\
        t_\oplus && arrival time at geocenter & \\
        \iota && inclination angle & \\
        \psi && polarization angle & \\
        \phi_c && coalescence phase & \\
        \cline{1-1}\cline{3-3}
        m_1 && first component's mass & \rdelim\}{4}{1mm} & \multirow{4}{2cm}{intrinsic parameters, $\bm\theta_\mathrm{in}$.}\\
        m_2 && second component's mass & \\
        \mathbf S_1 && first component's spin & \\
        \mathbf S_2 && second component's spin & \\
    \end{array}\right.
\end{equation}
This list of parameters involves some simplifying assumptions. Eccentricity is omitted: although it does play a major role in the evolution and waveforms of \ac{NSBH} and \ac{BBH} sources formed by dynamical capture~\citep{PhysRevD.87.043004}, \ac{BNS} systems formed by binary stellar evolution should almost always circularize due to tidal interaction~\citep{0004-637X-572-1-407} and later \ac{GW} emission~\citep{PhysRev.136.B1224} long before the inspiral enters LIGO's frequency range of $\sim$10\nobreakdashes--1000~kHz. Tidal deformabilities of the \acp{NS} are omitted because the signal imprinted by the companions' material properties is so small that it will only be detectable by an Einstein Telescope\nobreakdashes-class \ac{GW} observatory~\citep{PhysRevD.81.123016}. Furthermore, in \ac{GW} detection efforts, especially those focused on \ac{BNS} systems, the component spins $\mathbf{S}_1$ and $\mathbf{S}_2$ are often assumed to be nonprecessing and aligned with the system's total angular momentum and condensed to a single scalar parameter $\chi$, or even neglected entirely: $\mathbf{S}_1 = \mathbf{S}_2 = 0$.

\begin{figure}
    \centering
    \includegraphics[width=\textwidth]{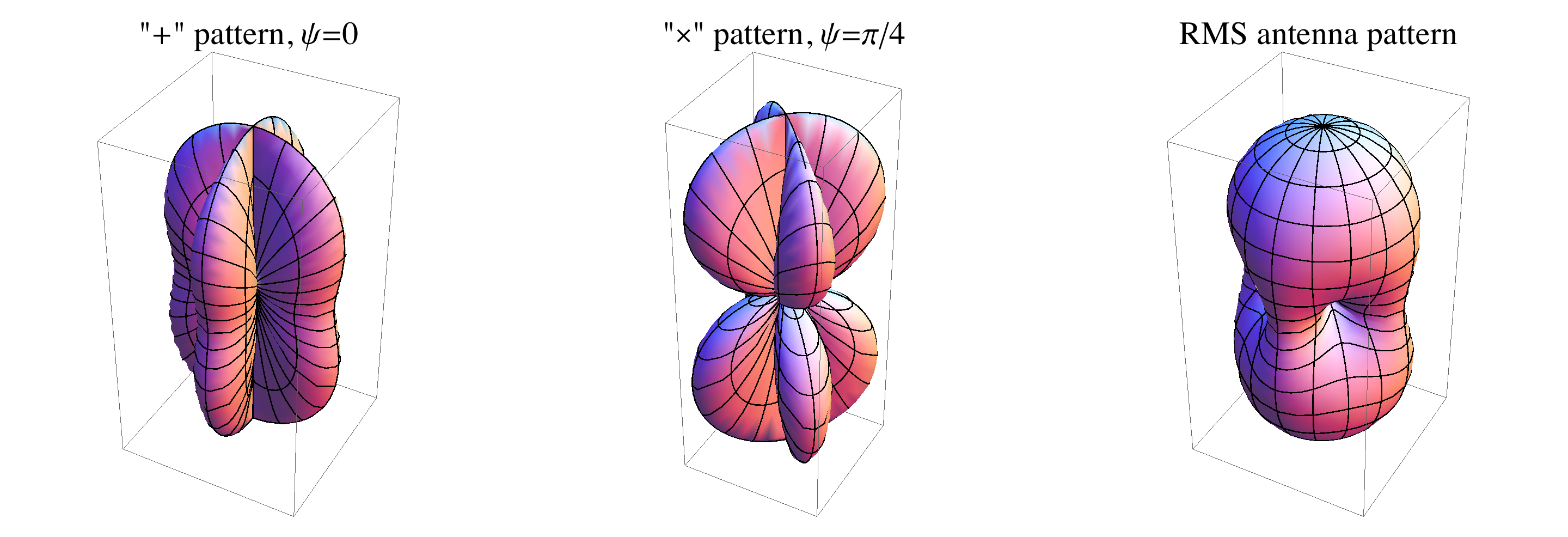}
    \caption[Antenna patterns]{\label{fig:antenna-pattern}The directional dependence of the $+$, $\times$, and \ac{RMS} antenna patterns of a \ac{LIGO}\nobreakdashes-style \ac{GW} detector. The detector is at the center of the light box, with its two arms parallel to the horizontal edges.}
\end{figure}

Assuming circular orbits and no spin precession, we can write the \ac{GW} signal in each detector as a linear combination of two basis waveforms, $H_0$ and $H_{\pi/2}$. For nonprecessing systems, $H_0$ and $H_{\pi/2}$ are approximately ``in quadrature'' in the same sense as the sine and cosine functions, being nearly orthogonal and out of phase by ${\pi/2}$ at all frequencies. If $H_0$ and $H_{\pi/2}$ are Fourier transforms of real functions, then $H_0(\omega) = H_0^*(-\omega)$ and $H_{\pi/2}(\omega) = H_{\pi/2}^*(-\omega)$, and we can write (assuming an arbitrary phase convention)
\begin{equation}
    H_{\pi/2}(\omega) = H_0(\omega) \cdot
    \begin{cases}
        -i & \text{if } \omega \geq 0 \\
        i & \text{if } \omega < 0
    \end{cases}.
\end{equation}
For brevity, we define $H \equiv H_{0}$ and write all subsequent equations in terms of the $H$ basis vector alone. Then, we can write the signal model in a way that isolates all dependence on the extrinsic parameters, $\bm\theta_\mathrm{ex}$, into the coefficients and all dependence on the intrinsic parameters, $\bm\theta_\mathrm{in}$, into the basis waveform, by taking the Fourier transform of Equation~(2.8) of \cite{PhysRevD.83.084002}:
\begin{equation}\label{eq:full-signal-model}
    X_i(\omega; \bm\theta) = e^{-i \omega (t_\oplus - \mathbf{d}_i \cdot \mathbf{n})}
    \frac{r_{1,i}}{r}
    e^{2 i \phi_c}
    \left[
    \frac{1}{2} \left(1 + \cos^2 \iota\right) \Re \left\{\zeta\right\} - i
    \left(\cos\iota\right) \Im \left\{\zeta\right\}
    \right]
    H(\omega; \bm\theta_\mathrm{in})
\end{equation}
for $\omega \geq 0$, where
\begin{equation}
    \zeta = e^{-2 i \psi} \left(
    F_{+,i}(\alpha, \delta, t_\oplus) +
    i F_{\times,i}(\alpha, \delta, t_\oplus)
    \right).
\end{equation}
The quantities $F_{+,i}$ and $F_{\times,i}$ are the dimensionless detector antenna factors, defined such that $0 \leq {F_{+,i}}^2 + {F_{\times,i}}^2 \leq 1$. They depend on the orientation of detector $i$ as well as the sky location (as depicted in Figure~\ref{fig:antenna-pattern}) and sidereal time of the event and are presented in~\cite{ExcessPower}. In a coordinate system with the $x$ and $y$ axes aligned with the arms of a detector, its antenna pattern is given in spherical polar coordinates as
\begin{align}
    F_+ &= -\frac{1}{2}(1 + \cos^2 \theta) \cos{2\phi}, \\
    F_\times &= -\cos\theta \sin{2\phi}.
\end{align}
The unit vector $\mathbf{d}_i$ represents the position of detector $i$ in units of light travel time.\footnote{When considering transient \ac{GW} sources such as those that we are concerned with in this thesis, the origin of the coordinate system is usually taken to be the geocenter. For long\nobreakdashes-duration signals such as those from statically deformed neutron stars, the solar system barycenter is a more natural choice.} The vector $\mathbf{n}$ is the direction of the source. The negative sign in the dot product $-\mathbf{d}_i \cdot \mathbf{n}$ is present because the direction of travel of the \ac{GW} signal is opposite to that of its sky location. The quantity $r_{1,i}$ is a fiducial distance at which detector $i$ would register SNR=1 for an optimally oriented binary (face\nobreakdashes-on, and in a direction perpendicular to the interferometer's arms):
\begin{equation}\label{eq:horizon}
r_{1,i} = 1 / \sigma_i, \qquad {\sigma_i}^2 = \int_0^\infty \frac{\left|H(\omega; \boldsymbol \theta_\mathrm{in})\right|^2}{S_i(\omega)} \,d\omega.
\end{equation}

More succinctly, we can write the signal received by detector $i$ in terms of observable extrinsic parameters $\bm\theta_i = (\rho_i, \gamma_i, \tau_i)$, the amplitude $\rho_i$, phase $\gamma_i$, and time delay $\tau_i$ on arrival at detector $i$:
\begin{equation}\label{eq:signal-model}
    X_i (\omega; \bm\theta_i, \bm\theta_\mathrm{in}) = X_i (\omega; \rho_i, \gamma_i, \tau_i, \bm\theta_\mathrm{in}) = \frac{\rho_i}{\sigma_i} e^{i (\gamma_i - \omega \tau_i)} H(\omega; \bm\theta_\mathrm{in}).
\end{equation}

The prevailing technique for detection of \acp{GW} from \acp{CBC} is to realize a \ac{MLE} from the likelihood in Equation~(\ref{eq:gaussian-likelihood}) and the signal model in Equation~(\ref{eq:signal-model}). Concretely, this results in a bank of matched filters, or the cross-correlation between a template waveform and the incoming data stream,
\begin{equation}
z_i(\tau_i;\bm\theta_\mathrm{in}) = \frac{1}{\sigma_i (\bm\theta_\mathrm{in})} \int_0^\infty \frac{H^*(\omega; \bm\theta_\mathrm{in}) Y_i(\omega) e^{i \omega \tau_i}}{S_i(\omega)} \,d\omega.
\end{equation}
The \ac{ML} point estimates of the signal parameters, $\mathrm{MLE}(\mathbf{y}) = \{\{ \hat{\bm\theta}_i \}_i, \hat{\bm\theta}_\mathrm{in}\} = \{\left\{ \hat\rho_i, \hat\gamma_i, \hat\tau_i \right\}_i, \hat{\bm\theta}_\mathrm{in}\}$, are given by
\begin{eqnarray}
    \label{eq:optimal-tau}
    \hat{\bm\theta}_\mathrm{in}, \{\hat\tau_i\}_i
        &=& \argmax_{\bm\theta_\mathrm{in}, \{\hat\tau_i\}_i}
        \sum_i \left| z_i\left(\tau_i;
        \bm\theta_\mathrm{in}\right) \right|^2, \\
    \label{eq:optimal-rho}
    \hat\rho_i &=& \left| z_i\left(\hat\tau_i;
        \hat{\bm\theta}_\mathrm{in}\right) \right|, \\
    \label{eq:optimal-gamma}
    \hat\gamma_i &=& \arg z_i\left(\hat\tau_i;
        \hat{\bm\theta}_\mathrm{in}\right).
\end{eqnarray}
A detection candidate consists of $\{\left\{ \hat\rho_i, \hat\gamma_i, \hat\tau_i \right\}_i, \hat{\bm\theta}_\mathrm{in}\}$. There are various ways to characterize the significance of a detection candidate. In Gaussian noise, the maximum likelihood for the network is obtained by maximizing the network \ac{SNR}, $\rho_\mathrm{net}$,
\begin{equation}
    \hat\rho_\mathrm{net} = \max_{\bm\theta} \sum_i {|z_i({\bm\theta})|}^2 = \sqrt{\sum_i {\hat\rho_i}^2};
\end{equation}
this, therefore, is the simplest useful candidate ranking statistic.

\section{Measures of detector sensitivity}
\label{sec:detector-sensitivity}

The sensitivity of a single GW detector is customarily described by the horizon distance, or the maximum distance at which a particular source would create a signal with a maximum fiducial single\nobreakdashes-detector \ac{SNR}, $\rho$. It is given by
\begin{equation}
    \label{eq:horizon-distance}
    d_\mathrm{H} \approx \frac{G^{5/6}M^{1/3}\mu^{1/2}}{c^{3/2}\pi^{2/3}\rho}\sqrt{\frac{5}{6} \int_{f_1}^{f_2} \frac{f^{-7/3}}{S(f)}\,\mathrm{d}f},
\end{equation}
where $G$ is Newton's gravitational constant, $c$ is the speed of light, $M$ the sum of the component masses, $\mu$ the reduced mass, $f^{-7/3}$ the approximate power spectral density (PSD) of the inspiral signal, and $S(f)$ the PSD of the detector's noise. The lower integration limit $f_1$ is the low\nobreakdashes-frequency extent of the detector's sensitive band. For the Advanced LIGO and Virgo detectors, ultimately limited at low frequency by ground motion \citep{GWDetectionLaserInterferometry}, we take $f_1 = 10$~Hz. Using a typical value of the detector sensitivity $S(100\,\text{Hz}) = 10^{-46}\,\text{Hz}^{-1}$, we can write Equation~(\ref{eq:horizon-distance}) as a scaling law:
\begin{multline}
    \label{eq:horizon-distance-scaling}
    d_\mathrm{H} \approx
        72.5\,\text{Mpc}
        \left(\frac{M}{M_\odot}\right)^{1/3}
        \left(\frac{\mu}{M_\odot}\right)^{1/2}
        \left(\frac{1}{\rho}\right) \\
    \cdot
        \left[
            \int_{\frac{f_1}{\text{Hz}}}^{\frac{f_2}{\text{Hz}}}
            \left(\frac{f}{100\,\text{Hz}}\right)^{-7/3} 
            \left(\frac{10^{-46}~\text{Hz}^{-1}}{S(f)}\right)
            d\left(\frac{f}{\text{Hz}}\right) \right]^{1/2}.
\end{multline}
For BNS masses, the inspiral ends with a merger and black hole ring down well outside LIGO's most sensitive band. A reasonable approximation is to simply truncate the \ac{SNR} integration at the last stable orbit of a Schwarzschild black hole with the same total mass \citep{maggiore2008gravitational},
\begin{equation}
    \label{eq:f-lso}
    f_2 = \frac{c^3}{6 \sqrt{6} \pi G M} \approx (4400\text{~Hz}) \frac{M_\odot}{M}.
\end{equation}
Usually, $\rho=8$ is assumed because $\rho=8$ signals in two detectors (for a root\nobreakdashes-sum\nobreakdashes-squared network \ac{SNR} of $\rho_\mathrm{net} = 8\sqrt{2} = 11.3$) is nearly adequate for a confident detection (see discussion of detection thresholds in Section~\ref{sec:detection-and-position-reconstruction}). Another measure of sensitivity is the BNS range $d_\mathrm{R}$, the volume-, direction-, and orientation\nobreakdashes-averaged distance of a source with $\rho \geq 8$, drawn from a homogeneous population. Due to the directional sensitivity or antenna pattern of interferometric detectors, the range is a factor of 2.26 smaller than the horizon distance for the same \ac{SNR} threshold. See also \citet{FINDCHIRP,S6InspiralRange}.\footnote{Even at its final design sensitivity, Advanced LIGO's range for BNS mergers is only 200~Mpc or $z = 0.045$ \citep[assuming the \emph{WMAP} nine\nobreakdashes-year $\Lambda$CDM cosmology;][]{WMAP9}. The horizon distance, 452~Mpc or $z = 0.097$, is only modestly cosmic. Because of the small distances considered in this study, we do not distinguish between different distance measures, nor do our gravitational waveforms contain any factors of $(1+z).$}

\section{Fisher information matrix: single detector}

We can predict the uncertainty in the \ac{ML} estimates without working out its full distribution. The \ac{CRLB} gives its covariance in the asymptotic limit of high \ac{SNR}. The \ac{CRLB} has been widely applied in \ac{GW} data analysis to estimate parameter estimation uncertainty (for example, \citealt{1996PhRvD..53.3033B,FairhurstTriangulation,2009PhRvD..79h4032A,WenLocalizationAdvancedLIGO,LIGOObservingScenarios,FairhurstLIGOIndia})\footnote{The Fisher matrix is also used in construction of \ac{CBC} matched filter banks. The common procedure is to place templates uniformly according to the determinant of the signal space metric, which is the Fisher matrix. This is equivalent to uniformly sampling the Jeffreys prior. In practice, this is done either by constructing a hexagonal lattice \citep{PhysRevD.76.102004} or sampling stochastically \citep{2009PhRvD..80j4014H,2009PhRvD..80b4009V,SBank,2010PhRvD..81b4004M,2014PhRvD..89b4003P}}. We will momentarily consider the likelihood for a single detector:
\begin{equation}\label{eq:gaussian-likelihood-spa}
    \mathcal{L}\left(Y_i; \rho_i, \gamma_i, \tau_i,
        \bm\theta_\mathrm{in}\right)
    \propto \exp \left[
        - \frac{1}{2} \int_0^\infty \frac{\left|Y_i (\omega)
            - X_i\left(\omega; \rho_i, \gamma_i, \tau_i,
                \bm\theta_\mathrm{in}\right)
        \right|^2}{S_i(\omega)} \, d\omega
    \right],
\end{equation}
with $X_i(\omega; \rho_i, \gamma_i, \tau_i, \bm\theta_\mathrm{in})$ given by Equation~(\ref{eq:signal-model}).

The Fisher information matrix for a measurement $y$ described by the unknown parameter vector $\bm{\theta}$ is the conditional expectation value
\begin{equation}\label{eq:general-fisher-matrix}
    \mathcal{I}_{jk} = \mathrm{E} \, \left[
        \left(\frac{\partial \log
            \mathcal{L}(Y_i ; \bm\theta)}
            {\partial \theta_j}\right)
        \left(\frac{\partial \log
            \mathcal{L}(Y_i ; \bm\theta)}
            {\partial \theta_k}\right)
    \middle| \bm\theta
    \right].
\end{equation}
Note that if $\log\mathcal{L}$ is twice differentiable in terms of $\bm\theta$, then the Fisher matrix can also be written in terms of second derivatives as
\begin{equation}\label{eq:general-fisher-matrix-second-derivatives}
    \mathcal{I}_{jk} = \mathrm{E} \, \left[
        -\frac{\partial^2 \log
            \mathcal{L}(Y_i ; \bm\theta)}
            {\partial \theta_j \partial \theta_k}
    \middle| \bm\theta
    \right].
\end{equation}
In this form, we can recognize the Fisher matrix as the expectation value, conditioned on the true parameter values, of the Hessian matrix of the log likelihood. It describes how strongly the likelihood depends, on average, on the parameters. If $\hat{\bm\theta}$ is an unbiased estimator of $\bm\theta$, $\tilde{\bm\theta} = \hat{\bm\theta} - \bm\theta$ is the measurement error, and $\Sigma = \mathrm{E} \, [\tilde{\bm\theta}\tilde{\bm\theta}\transpose]$ is the covariance of the measurement error, then the \ac{CRLB} says that $\Sigma \geq \mathcal{I}^{-1}$, in the sense that $\left(\Sigma - \mathcal{I}^{-1}\right)$ is positive semi\nobreakdashes-definite.

When (as in our assumptions) the likelihood is Gaussian, Equation~(\ref{eq:general-fisher-matrix}) simplifies to
\begin{equation}\label{eq:gaussian-fisher-matrix}
    \mathcal{I}_{jk} = \int_0^\infty \Re \left[
        \left(\frac{\partial X_i}{\partial \theta_j}\right)^*
        \left(\frac{\partial X_i}{\partial \theta_k}\right)
    \right] \frac{1}{S_i(\omega)} \, d\omega.
\end{equation}
This form is useful because it involves manipulating the signal $X_i (\omega)$ rather than the entire observation $Y (\omega)$. In terms of the $k$th \ac{SNR}-weighted moment of angular frequency,
\begin{equation}\label{eq:angular-frequency-moments}
    {\overline{\omega^k}}_i =
        \left[ \int_0^\infty \frac{|h (\omega)|^2}{S_i(\omega)} \omega^k \, d\omega \right]
        \left[ \int_0^\infty \frac{|h (\omega)|^2}{S_i(\omega)} \, d\omega \right]^{-1},
\end{equation}
the Fisher matrix for the signal in the $i$th detector is
\begin{equation}
    \mathcal{I}_i = \left(
        \begin{array}{cc}
            \mathcal{I}_{\bm\theta_i,\bm\theta_i} &
            \mathcal{I}_{\bm\theta_i,\bm\theta_\mathrm{in}} \\
            {\mathcal{I}_{\bm\theta_i,\bm\theta_\mathrm{in}}}\transpose &
            {\rho_i}^2 \mathcal{I}_{\bm\theta_\mathrm{in},\bm\theta_\mathrm{in}}
        \end{array}
    \right)
\end{equation}
where
\begin{equation}\label{eq:fisher-matrix}
    \mathcal{I}_{\bm\theta_i,\bm\theta_i} = \bordermatrix{
        ~ & \rho_i & \gamma_i & \tau_i \cr
        \rho_i & 1 & 0 & 0 \cr
        \gamma_i & 0 & {\rho_i}^2 & -{\rho_i}^2 {\overline{\omega}}_i \cr
        \tau_i & 0 & -{\rho_i}^2 {\overline{\omega}}_i & {\rho_i}^2 {\overline{\omega^2}}_i
    }.
\end{equation}
(This is equivalent to an expression given in \citealt{Grover:2013}.) The information matrix elements that relate to the intrinsic parameters can also be expressed as linear combinations of the angular frequency moments. However, as we will see in the next section, we need not compute these matrix elements if we are only interested in sky localization accuracy.

\section{Independence of intrinsic and extrinsic errors}
\label{sec:independence-intrinsic-extrinsic}

If all of the detectors have the same noise \acp{PSD} up to multiplicative factors, $c_1 S_1(\omega) = c_2 S_2(\omega) = \cdots = c_n S_n(\omega) \equiv S(\omega)$, then we can show that the errors in the intrinsic parameters (masses) are not correlated with sky position errors. This is because we can change variables from amplitudes, phases, and times to amplitude ratios, phase differences, and time differences. With $N$ detectors, we can form a single average amplitude, time, and phase, plus $N-1$ linearly independent differences. The averages are correlated with the intrinsic parameters, but neither are correlated with the differences. Since only the differences inform sky location, this gives us license to neglect uncertainty in masses when we are computing sky resolution.

This is easiest to see if we make the temporary change of variables $\rho \rightarrow \varsigma = \log \rho$. This allows us to factor out the \ac{SNR} dependence from the single-detector Fisher matrix. The extrinsic part becomes
\begin{equation}\label{eq:fisher-matrix}
    \mathcal{I}_{\bm\theta_i,\bm\theta_i} = \bordermatrix{
        ~ & \varsigma_i & \gamma_i & \tau_i \cr
        \varsigma_i & {\rho_i}^2 & 0 & 0 \cr
        \gamma_i & 0 & {\rho_i}^2 & -{\rho_i}^2 {\overline{\omega}}_i \cr
        \tau_i & 0 & -{\rho_i}^2 {\overline{\omega}}_i & {\rho_i}^2 {\overline{\omega^2}}_i
    } = {\rho_i}^2 \left( \begin{array}{ccc}
        1 & 0 & 0 \\
        0 & 1 & -{\overline{\omega}}_i \\
        0 & -{\overline{\omega}}_i & {\overline{\omega^2}}_i
    \end{array} \right).
\end{equation}
Due to our assumption that the detectors' \acp{PSD} are proportional to each other, the noise moments are the same for all detectors, ${\overline{\omega^k}}_i \equiv {\overline{\omega^k}}$. Then we can write the single-detector Fisher matrix as
\begin{equation}
    \mathcal{I}_i = {\rho_i}^2 \left(
        \begin{array}{cc}
            A & B \\
            B\transpose & C
        \end{array}
    \right),
\end{equation}
with the top-left block $A$ comprising the extrinsic parameters and the bottom-right block $C$ the intrinsic parameters.

Information is additive, so the Fisher matrix for the whole detector network is
\begin{equation}
    \mathcal{I}_\mathrm{net} = \left(
    \begin{array}{ccccc}
    {\rho_1}^2 A & 0 & \cdots & 0 & {\rho_1}^2 B \\
    0 & {\rho_2}^2 A & & \vdots & {\rho_1}^2 B \\
    \vdots & & \ddots & 0 & \vdots \\
    0 & 0 & \cdots & {\rho_N}^2 A & {\rho_N}^2 B \\
    {\rho_1}^2 B\transpose & {\rho_2}^2 B\transpose & \cdots & {\rho_N}^2 B\transpose & {\rho_\mathrm{net}}^2 C
    \end{array}
    \right).
\end{equation}
Now we introduce the change of variables that sacrifices the $N$th detector's extrinsic parameters for the network averages,
\begin{align}
    &\begin{array}{c@{\quad\rightarrow\quad}l}
    \varsigma_N & \overline{\varsigma} = \left(\sum_i {\rho_i}^2 \varsigma_i\right) / {\rho_\mathrm{net}}^2, \\
    \gamma_N & \overline{\gamma} = \left(\sum_i {\rho_i}^2 \gamma_i\right) / {\rho_\mathrm{net}}^2, \\
    \tau_N & \overline{\tau} = \left(\sum_i {\rho_i}^2 \tau_i\right) / {\rho_\mathrm{net}}^2, \\
    \end{array}
\intertext{and replaces the first $N-1$ detectors' extrinsic parameters with differences,}
    &\left.
    \begin{array}{c@{\quad\rightarrow\quad}l}
    \varsigma_i & \delta\varsigma_i = \varsigma_i - \overline{\varsigma} \\
    \gamma_i & \delta\gamma_i = \gamma_i - \overline{\gamma} \\
    \tau_i & \delta\tau_i = \tau_i - \overline{\tau} \\
    \end{array}
    \right\} \text{ for } i = 1, \dots, N - 1.
\end{align}
The Jacobian matrix that describes this change of variables is
\begin{equation}
    J = \left(
    \begin{array}{cccccc}
    1 & 0 & \cdots & 0 & 1 & 0 \\
    0 & 1 & & 0 & 1 & 0 \\
    \vdots & & \ddots & & \vdots & \vdots \\
    0 & 0 & \cdots & 1 & 1 & 0 \\
    \frac{-{\rho_1}^2}{{\rho_N}^2} & \frac{-{\rho_2}^2}{{\rho_N}^2} & \cdots & \frac{-{\rho_{N-1}}^2}{{\rho_N}^2} & 1 & 0 \\
    0 & 0 & \cdots & 0 & 0 & 1 \\
    \end{array}
    \right).
\end{equation}
The transformed network Fisher matrix is block diagonal,
\begin{equation}
    \mathcal{I}_\mathrm{net} \rightarrow J\transpose \mathcal{I}_\mathrm{net} J = \left(
    \begin{array}{cccccc}
    {\rho_1}^2(1+\frac{1}{{\rho_1}^4}) A & \frac{{\rho_1}^2 {\rho_2}^2}{{\rho_N}^2} A & \cdots & \frac{{\rho_1}^2 {\rho_{N-1}}^2}{{\rho_N}^2} A & 0 & 0 \\
    \frac{{\rho_1}^2 {\rho_2}^2}{{\rho_N}^2} A & {\rho_2}^2(1+\frac{1}{{\rho_1}^4}) A & & \frac{{\rho_2}^2 {\rho_{N-1}}^2}{{\rho_N}^2} A & 0 & 0 \\
    \vdots & & \ddots & \vdots & \vdots & \vdots \\
    \frac{{\rho_1}^2 {\rho_{N-1}}^2}{{\rho_N}^2} A & \frac{{\rho_2}^2 {\rho_{N-1}}^2}{{\rho_N}^2} A & \cdots & {\rho_{N-1}}^2(1+\frac{1}{{\rho_1}^4}) A & 0 & 0 \\
    0 & 0 & \cdots & 0 & {\rho_\mathrm{net}}^2 A & {\rho_\mathrm{net}}^2 B \\
    0 & 0 & \cdots & 0 & {\rho_\mathrm{net}}^2 B\transpose & {\rho_\mathrm{net}}^2 C \\
    \end{array}
    \right).
\end{equation}
The top\nobreakdashes-left block contains $N-1$ relative amplitudes, phases, and times on arrival, all potentially correlated with each other. The bottom\nobreakdashes-right block contains the average amplitudes, phases, and times, as well as the masses. The averages and the masses are correlated with each other, but are not correlated with the differences. Because only the differences are informative for sky localization, we drop the intrinsic parameters from the rest of the Fisher matrix calculations in this chapter.

\section{Interpretation of phase and time errors}

We take a brief digression to discuss the physical interpretation of the time and amplitude errors. For our likelihood, the \ac{CRLB} implies that
\begin{equation}\label{eq:covariance-matrix}
    \cov{
        \left(
        \begin{array}{c}
            \tilde{\rho}_i \\
            \tilde{\gamma}_i \\
            \tilde{\tau}_i
        \end{array}
        \right)
    } \geq \mathcal{I}^{-1} = \left(
        \begin{array}{ccc}
            1 & 0 & 0 \\
            0 & {\rho_i}^2 {\overline{\omega^2}}_i/{\omega_{\mathrm{rms},i}}^2 & {\rho_i}^2 {\overline{\omega}}_i/{\omega_{\mathrm{rms},i}}^2 \\
            0 & {\rho_i}^2 {\overline{\omega}}_i/{\omega_{\mathrm{rms},i}}^2 & {\rho_i}^2/{\omega_{\mathrm{rms},i}}^2
        \end{array}
    \right),
\end{equation}
where ${\omega_{\mathrm{rms},i}}^2 = {\overline{\omega^2}}_i - {{\overline{\omega}}_i}^2$. Reading off the $\tau \tau$ element of the covariance matrix reproduces the timing accuracy in Equation~(24) of \cite{FairhurstTriangulation},
\begin{equation}\label{eq:timing-crlb}
    \std \left(\hat{\tau}_i - \tau_i \right) \geq \sqrt{\left(\mathcal{I}^{-1}\right)_{\tau\tau}} = \frac{\rho_i}{\omega_{\mathrm{rms},i}}.
\end{equation}

The Fisher matrix in Equation~(\ref{eq:fisher-matrix}) is block diagonal, which implies that estimation errors in the signal amplitude $\rho$ are uncorrelated with the phase $\gamma$ and time $\tau$. A sequence of two changes of variables lends some physical interpretation to the nature of the coupled estimation errors in $\gamma$~and~$\tau$.

First, we put the phase and time on the same footing by measuring the time in units of $1 / \sqrt{\overline{\omega^2}}$ with a change of variables from $\tau$ to $\gamma_\tau = \sqrt{\overline{\omega^2}} \tau$:
\begin{equation}
    \mathcal{I}' = \bordermatrix{
        ~ & \rho_i & \gamma_i & \gamma_{\tau,i} \cr
        \rho_i & 1 & 0 & 0 \cr
        \gamma_i & 0 & {\rho_i}^2 & -{\rho_i}^2\frac{{\overline{\omega}}_i}{\sqrt{{\overline{\omega^2}}_i}} \cr
        \gamma_{\tau,i} & 0 & -{\rho_i}^2\frac{{\overline{\omega}}_i}{\sqrt{{\overline{\omega^2}}_i}} & {\rho_i}^2
    }.
\end{equation}
The second change of variables, from $\gamma$ and $\gamma_\tau$ to $\gamma_\pm = \frac{1}{\sqrt{2}}(\gamma \pm \gamma_\tau)$, diagonalizes the Fisher matrix:
\begin{equation}\label{eq:fisher-matrix-extrinsic-one-detector}
    \mathcal{I}'' = \bordermatrix{
        ~ & \rho_i & \gamma_{+,i} & \gamma_{-,i} \cr
        \rho_i & 1 & 0 & 0 \cr
        \gamma_{+,i} & 0 & \left(1 - \frac{\overline{\omega}_i}{\sqrt{\overline{\omega^2}_i}}\right){\rho_i}^2 & 0 \cr
        \gamma_{-,i} & 0 & 0 & \left(1 + \frac{\overline{\omega}_i}{\sqrt{\overline{\omega^2}_i}}\right){\rho_i}^2
    }.
\end{equation}
Thus, in the appopriate time units, the \textit{sum and difference} of the phase and time of the signal are measured independently.

\section{Position resolution}

Finally, we will calculate the position resolution of a network of \ac{GW} detectors. We could launch directly into computing derivatives of the full signal model from Equation~(\ref{eq:full-signal-model}) with respect to all of the parameters, but this would result in a very complicated expression. Fortunately, we can take two shortcuts. First, since we showed in Section~\ref{sec:independence-intrinsic-extrinsic} that the intrinsic parameters are correlated only with an overall nuisance average arrival time, amplitude, and phase, we need not consider the derivatives with respect to mass at all. Second, we can reuse the extrinsic part of the single detector Fisher matrix from Equation~(\ref{eq:fisher-matrix}) by computing the much simpler Jacobian matrix to transform from the time, amplitude, and phase on arrival, to the parameters of interest.

We begin by transforming the single\nobreakdashes-detector Fisher matrix from a polar to a rectangular representation of the complex amplitude given in Equations~(\ref{eq:optimal-rho}, \ref{eq:optimal-tau}), $\rho_i, \gamma_i \rightarrow \Re[z_i] = \rho_i \cos \gamma_i, \Im[z_i] = \rho_i \sin \gamma_i$:
\begin{equation}\label{eq:fisher-matrix-extrinsic-one-detector-cartesian}
    \mathcal{I}_i = \bordermatrix{
        ~ & \Re[z_i] & \Im[z_i] & \tau_i \cr
        \Re[z_i] &
        1 &
        0 &
        {\overline{\omega}}_i b_i \cr
        \Im[z_i] &
        0 &
        1 &
        -{\overline{\omega}}_i b_i \cr
        \tau_i &
        {\overline{\omega}}_i b_i &
        -{\overline{\omega}}_i b_i &
        {\rho_i}^2 {\overline{\omega^2}}_i
    }.
\end{equation}

Consider a source in a ``standard'' orientation with the direction of propagation along the $+z$ axis, such that the \ac{GW} polarization tensor may be written in Cartesian coordinates as
\begin{equation}
    H = \frac{1}{r} e^{2 i \phi_c} \left(
    \begin{array}{ccc}
        \frac{1}{2}(1 + \cos^2 \iota) & i \cos\iota & 0 \\
        i \cos\iota & -\frac{1}{2}(1 + \cos^2 \iota) & 0 \\
        0 & 0 & 0
    \end{array}
    \right).
\end{equation}
Now introduce a rotation matrix $R$ that actively transforms this source to the Earth-relative polar coordinates $\theta, \phi$, and gives the source a polarization angle $\psi$ (adopting temporarily the notation $\mathrm{c}_\theta = \cos\theta,\,\mathrm{s}_\theta = \sin\theta$):
\begin{align}
    R &= R_z(\phi) R_y(\theta) R_z(\psi) R_y(\pi) \\
    &= \left(
        \begin{array}{ccc}
            \mathrm{c}_\phi & -\mathrm{s}_\phi & 0 \\
            \mathrm{s}_\phi & -\mathrm{c}_\phi & 0 \\
            0 & 0 & 1
        \end{array}
    \right)
    \left(
        \begin{array}{ccc}
            \mathrm{c}_\theta & 0 & \mathrm{s}_\theta \\
            0 & 1 & 0 \\
            -\mathrm{s}_\theta & 0 & \mathrm{c}_\theta
        \end{array}
    \right)
    \left(
        \begin{array}{ccc}
            \mathrm{c}_\psi & -\mathrm{s}_\psi & 0 \\
            \mathrm{s}_\psi & -\mathrm{c}_\psi & 0 \\
            0 & 0 & 1
        \end{array}
    \right)
    \left(
        \begin{array}{ccc}
            -1 & 0 & 0 \\
            0 & 1 & 0 \\
            0 & 0 & -1
        \end{array}
    \right).
\end{align}
(The rightmost rotation reverses the propagation direction so that the wave is traveling \emph{from} the sky position $\theta, \phi$.) With the (symmetric) detector response tensor $D_i$, we can write the received amplitude and arrival time as
\begin{align}
    z_i &= r_{1,i} \Tr \left[ D_i R \, H \, R\transpose \right], \\
    \tau_i &= t_\oplus + {\mathbf{d}_i}\transpose R \, \mathbf{k}.
\end{align}
Equivalently, we can absorb the rotation $R$ and the horizon distance $r_{1,i}$ into the polarization tensor, detector response tensors, and positions,
\begin{align}
    H &\rightarrow H^\prime = R_z(\psi) \, R_y(\pi) \, H \, R_y(\pi)\transpose  \, R_z(\psi)\transpose, \\
    D_i &\rightarrow D_i^\prime = r_{1,i} \, R_y(\theta)\transpose \, R_z(\phi)\transpose \, D_i \, R_z(\phi) \, R_y(\theta), \\
    \mathbf{d}_i &\rightarrow \mathbf{d}_i^\prime = R_y(\theta)\transpose \, R_z(\phi)\transpose \, \mathbf{d}_i, \\
    \mathbf{k} &\rightarrow \mathbf{k}^\prime = (0, 0, -1).
\end{align}
Now the model becomes
\begin{align}
    \label{eq:H-hplus-hcross}
    H^\prime &= \left(
        \begin{array}{ccc}
            h_+ & h_\times & 0 \\
            h_\times & -h_+ & 0 \\
            0 & 0 & 0
        \end{array}
    \right), \\
    \label{eq:z-hplus-hcross}
    z_i &= \Tr \left[ D_i^\prime H^\prime \right] = h_+ (D_{00}^\prime - D_{11}^\prime) + 2 h_\times D_{01}^\prime, \\
    \tau_i &= t_\oplus + (\mathbf{d}_i^\prime) \cdot \mathbf{k}, \\
    \intertext{where}
    \label{eq:hplus}
    h_+ &= \frac{1}{r} e^{2 i \phi_c} \left[\frac{1}{2} (1+\cos^2\iota) \cos{2\psi} + i \cos\iota \sin{2\psi}\right], \\
    \label{eq:hcross}
    h_\times &= \frac{1}{r} e^{2 i \phi_c} \left[\frac{1}{2} (1+\cos^2\iota) \sin{2\psi} - i \cos\iota \cos{2\psi}\right].
\end{align}
We insert an infinitesimal rotation $\delta R$ to perturb the source's orientation from the true value:
\begin{align}
    z_i &= \Tr \left[ D_i^\prime (\delta R) H^\prime (\delta R)\transpose \right], \\
    \tau_i &= t_\oplus + (\mathbf{d}_i^\prime)\transpose (\delta R) \mathbf{k}^\prime.
\end{align}
We only need a first order expression for $\delta R$, because we will be taking products of first derivatives of it\footnote{Caution: the angles $\delta\theta$ and $\delta\phi$ represent displacements in two orthogonal directions, but are \emph{not} necessarily simply related to $\theta$ and $\phi$.}:
\begin{equation}
    \delta R = \left(
        \begin{array}{ccc}
            1 & 0 & \delta\theta \\
            0 & 1 & \delta\phi \\
            -\delta\theta & -\delta\phi & 1
        \end{array}
    \right).
\end{equation}
We construct a Jacobian matrix $J_i$ to transform from the single\nobreakdashes-detector observables $(\Re[z_i], \Im[z_i], \tau_i)$ to the position perturbations, polarization components, and geocentered arrival time \\ $(\delta\theta, \delta\phi, \Re[h_+], \Im[h_+], \Re[h_\times], \Im[h_\times], t_\oplus)$:
\begin{equation}\label{eq:jacobian-detector-tensor}
    {J_i}\transpose = \bordermatrix{
        ~ & \Re[z_i] & \Im[z_i] & \tau_i \cr
        \delta\theta &
        -2\Re[h_+]D_{02}^\prime -2\Re[h_\times]D_{12}^\prime &
        -2\Im[h_+]D_{02}^\prime -2\Im[h_\times]D_{12}^\prime &
        -d_0^\prime \cr
        \delta\phi &
        -2\Re[h_\times]D_{02}^\prime +2\Re[h_+]D_{12}^\prime &
        -2\Im[h_\times]D_{02}^\prime +2\Im[h_+]D_{12}^\prime &
        -d_1^\prime \cr
        \Re[h_+] & D_{00}^\prime - D_{11}^\prime & 0 & 0 \cr
        \Im[h_+] & 0 & D_{00}^\prime - D_{11}^\prime & 0 \cr
        \Re[h_\times] & 2D_{01}^\prime & 0 & 0 \cr
        \Im[h_\times] & 0 & 2D_{01}^\prime & 0 \cr
        t_\oplus & 0 & 0 & 1
    }.
\end{equation}
We transform and sum the information from each detector,
\begin{equation}\label{eq:summed-information}
    \mathcal{I}_\mathrm{net} = \sum_i {J_i}\transpose \mathcal{I}_i {J_i}.
\end{equation}

\subsection{Marginalization over nuisance parameters}

To extract an area from the Fisher matrix, we must first marginalize or discard the nuisance parameters. Note that marginalizing parameters of a multivariate Gaussian distribution amounts to simply dropping the relevant entries in the mean vector and covariance matrix. Since the information is the inverse of the covariance matrix, we need to invert the Fisher matrix, drop all but the first two rows and columns, and then invert again.

This procedure has a shortcut called the Schur complement (see, for example, \citealt{numerical-recipes-inversion-partition}). Consider a partitioned square matrix $M$ and its inverse:
\begin{equation}
    M = \left(
        \begin{array}{cc}
            A & B \\
            C & D
        \end{array}
    \right), \qquad
    M^{-1} = \left(
        \begin{array}{cc}
            \tilde{A} & \tilde{B} \\
            \tilde{C} & \tilde{D}
        \end{array}
    \right).
\end{equation}
If $A$ and $B$ are square matrices, then the upper-left block of the inverse can be written as
\begin{equation}
    \tilde{A}^{-1} = A - B D^{-1} C.
\end{equation}
If we partition the $\mathcal{I}_\mathrm{net}$ similarly, the $A$ block consists of the first two rows and columns and $D$ is the lower right block that describes all other parameters. Because the Fisher matrix is symmetric, the off\nobreakdashes-diagonal blocks satisfy $C = B\transpose$. Then the Schur complement
\begin{equation}\label{eq:marginal-fisher-matrix}
    \mathcal{I}_\mathrm{marg} = A - B D^{-1} B\transpose
\end{equation}
gives us the information matrix marginalized over all parameters but $\delta\theta$ and $\delta\phi$.

\subsection{Spatial interpretation}

How do we extract the dimensions of the localization from the Fisher matrix? If there are $N \leq 2$ detectors, then the Fisher matrix must be degenerate, because there are $3N$ measurements and 7 parameters:
\begin{equation*}
    \left\{
    \begin{array}{c}
        \delta\theta \\
        \delta\phi \\
        \Re[h_+] \\
        \Im[h_+] \\
        \Re[h_\times] \\
        \Im[h_\times] \\
        t_\oplus
    \end{array}
    \right\}
    = 7 \text{ parameters}
    \qquad
    \longleftrightarrow
    \qquad
    \left\{
    \begin{array}{c}
        \Re[z_i] \\
        \Im[z_i] \\
        \tau_i
    \end{array}
    \right\}
    \times N = 3N \text{ observables.}
\end{equation*}
Therefore, for $N=2$ detectors, the marginalized Fisher matrix $\mathcal{I}_\mathrm{marg}$ is singular. Its only nonzero eigenvalue $\lambda$ describes the width of an annulus on the sky. The width of the annulus that contains probability $p$ is given by
\begin{equation}\label{eq:l-pth-quantile}
    L_p = 2 \sqrt{2} \erf^{-1}(p) / \sqrt{\lambda}.
\end{equation}
The prefactor $2 \sqrt{2} / \erf^{-1}(p)$ is the central interval of a normal distribution that contains a probability $p$, and is $\approx 3.3$ for $p = 0.9$. \emph{Caution:} for two\nobreakdashes-detector networks, priors play an important role in practical parameter estimation and areas can be much smaller than one would predict from the Fisher matrix (see Chapter~\ref{chap:first2years} for more discussion).

For $N \geq 3$ detectors, the parameters are over\nobreakdashes-constrained by the data and the Fisher matrix describes the dimensions of an ellipse. Within a circle of radius $r$ centered on the origin, the enclosed probability $p$ is
\begin{equation}
    p = \int_0^{2\pi} \int_0^r \frac{1}{2\pi} e^{-s^2/2} s \, ds \, d\phi = 1 - e^{-r^2/2}.
\end{equation}
Therefore the radius $r$ of the circle that contains a probability $p$ is
\begin{equation}
    r = \sqrt{-2\ln (1 - p)}.
\end{equation}
Suppose that the eigenvalues of the Fisher matrix are $\lambda_1$ and $\lambda_2$. This describes a $1\sigma$ uncertainty ellipse that has major and minor radii ${\lambda_1}^{-1/2}$, ${\lambda_2}^{-1/2}$, and area $A_{1\sigma} = \pi / \sqrt{\lambda_1 \lambda_2} = \pi / \sqrt{\det\mathcal{I}}$. Then the area of an ellipse containing probability $p$ is
\begin{equation}\label{eq:a-pth-quantile}
    A_p = -2\pi\ln(1-p) / \sqrt{\det \mathcal{I}},
\end{equation}
or, more memorably for the $90$th percentile, $A_{0.9} = 2\pi\ln(10) / \sqrt{\det{\mathcal{I}}}$.

\subsection{Outline of calculation}
\label{sec:fisher-matrix-area-outline}

Using the above derivation, we arrive at a prediction for the sky resolution of a \ac{GW} detector network. We took some shortcuts that allowed us to avoid directly evaluating the complicated derivatives of the signal itself with respect to sky location. As a result, the expressions involved in each step are simple enough to be manually entered into a computer program. However, because the procedure involves several steps, we outline it once again below.

\begin{enumerate}
    \item Compute, for each detector, the horizon distance $r_{1,i}$, the angular frequency moments ${\overline{\omega}}_i$ and ${\overline{\omega^2}}_i$, and $(h_+, h_\times)$ from Equations~(\ref{eq:hplus},~\ref{eq:hcross}). (These can be reused for multiple source positions as long as the masses and the detector noise \acp{PSD} are the same.)
    \item For a given $\phi, \theta, \psi$, compute the complex received amplitude $z_i$ from Equations~(\ref{eq:H-hplus-hcross},~\ref{eq:z-hplus-hcross}), the extrinsic Fisher matrix from Equation~(\ref{eq:fisher-matrix}), and the Jacobian from Equation~(\ref{eq:jacobian-detector-tensor}).
    \item Sum the information from all detectors using Equation~\ref{eq:summed-information}.
    \item Compute the marginalized Fisher matrix from the Schur complement using Equation~(\ref{eq:marginal-fisher-matrix}).
    \item If there are two detectors, find the width $L_p$ of the ring describing the $p$th quantile using Equation~(\ref{eq:l-pth-quantile}). If there are three or more detectors, find the area $A_p$ of the $p$th quantile using Equation~(\ref{eq:a-pth-quantile}).
    \item (Optionally, convert from (ste)radians to (square) degrees.)
\end{enumerate}

See code listing in Appendix~\ref{sec:fisher-matrix-code-listing}.

\subsection{Example calculation for HLV network}

\begin{figure*}
    (a) Coherent \\
    \includegraphics[width=\textwidth]{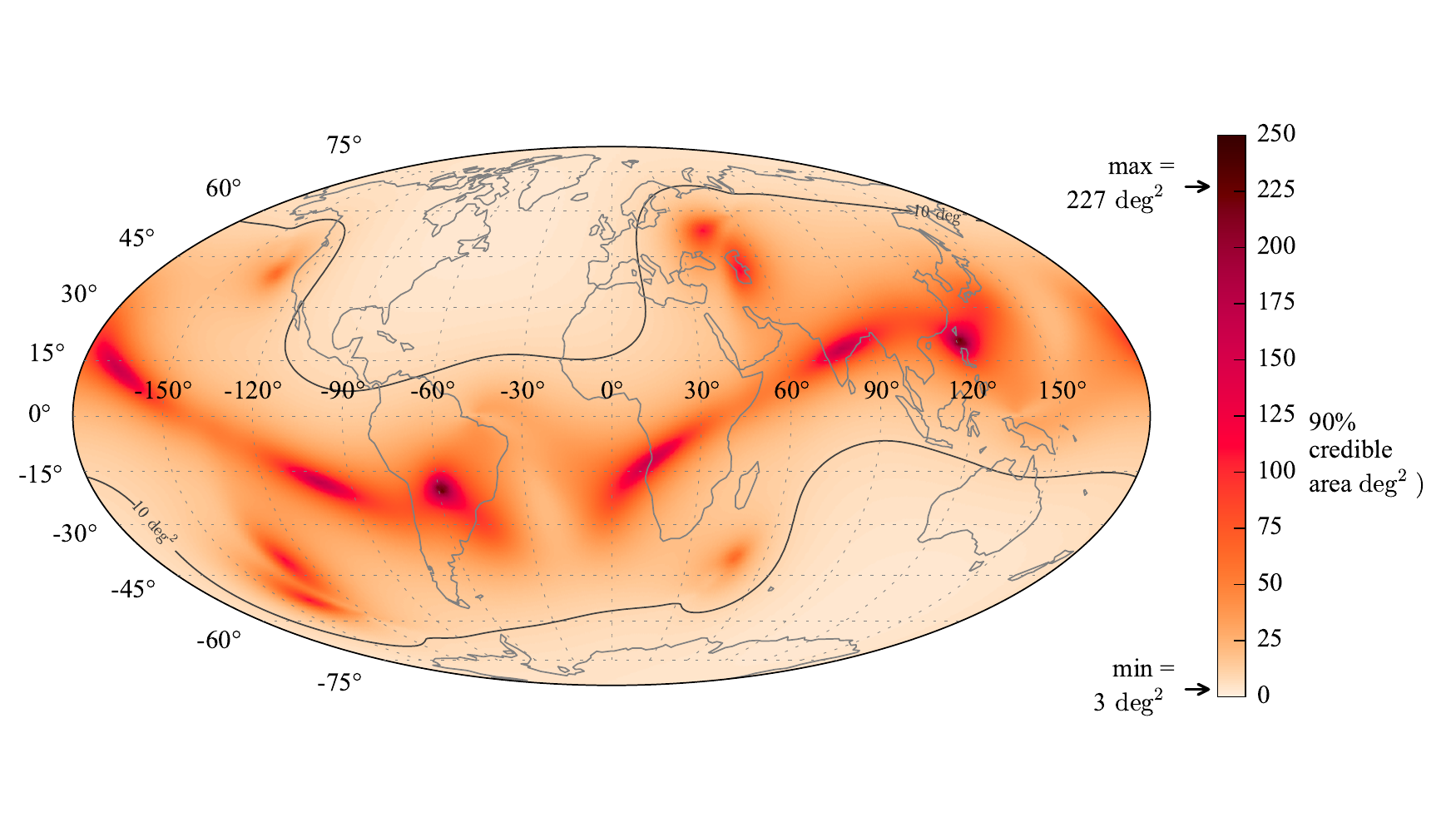} \\
    (b) Time of arrival only \\
    \includegraphics[width=\textwidth]{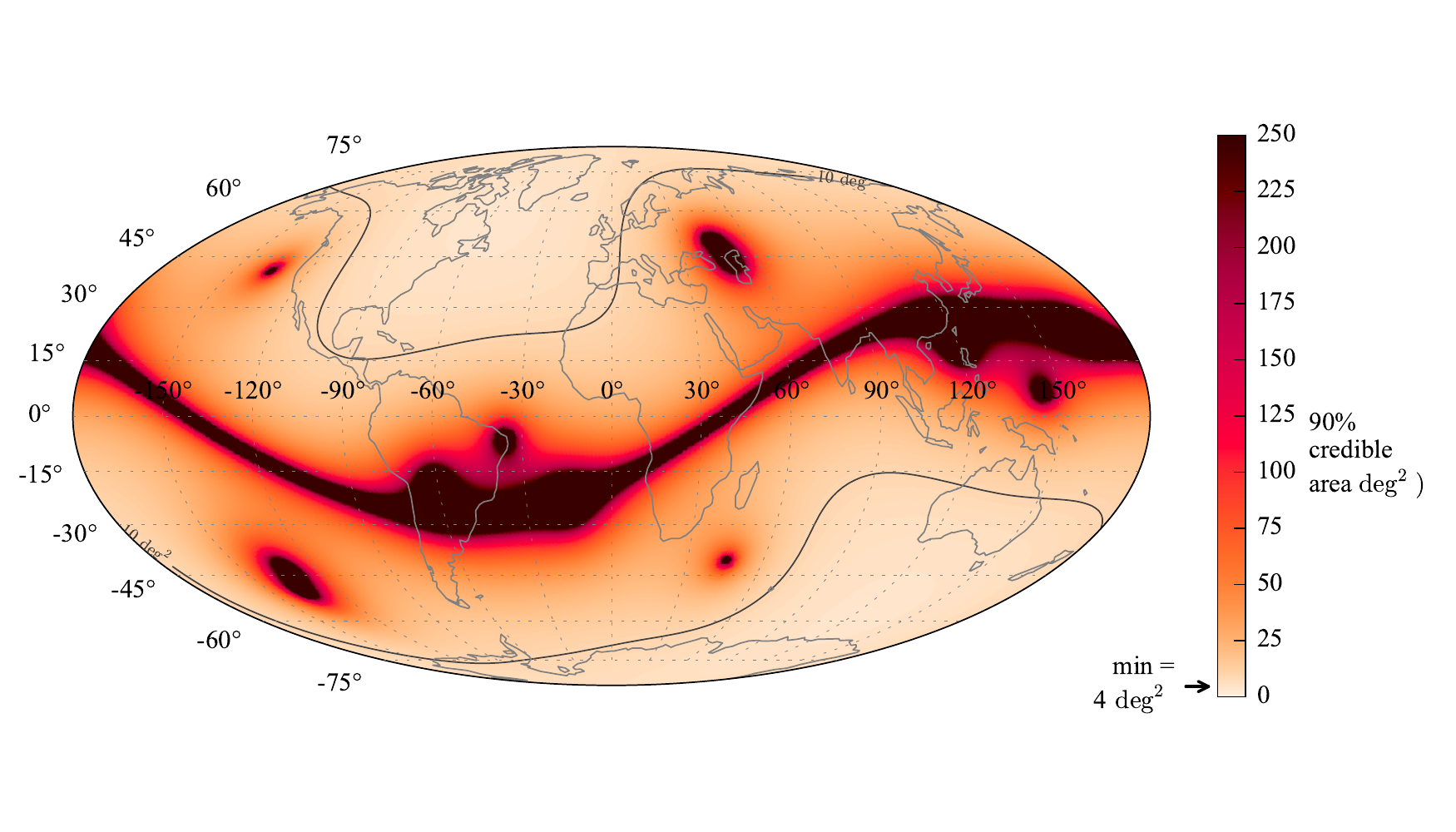}
    \caption[Position resolution as a function of sky position for a three detector network]{\label{fig:fisher-area-map}Position resolution as a function of sky position for a three\nobreakdashes-detector (HLV) network at final design sensitivity. The distance of the source is set to a constant value of 180~Mpc and the inclination angle is fixed to $\iota = 30\arcdeg$. At each sky location, the polarization angle $\psi$ is varied to minimize the position uncertainty. Panel~(a) shows the area of the 90\% credible region as a function of geographic coordinates. The minimum and maximum areas are marked on the color bar. The contour within which sources are localized to 10~deg$^2$ is shown as a black curve. Panel~(b) shows the area that would be found using time of arrival information only (but assuming the same polarization angle). Note that in (b), the maximum area is off the scale; black regions of the plot are localized to areas worse than 250~deg$^2$.}
\end{figure*}

As an example, we calculate the 90\% area as a function of sky position. We consider a three\nobreakdashes-detector, Hanford\nobreakdashes--Livingston\nobreakdashes--Virgo (HLV) network at final design sensitivity. Our source is a 1.4\nobreakdashes--1.4~$M_\sun$ \ac{BNS} merger at a fixed distance of 180~Mpc and a fixed inclination angle of $\iota = 30\arcdeg$. The signal model is a stationary phase approximation waveform accurate to an order of 3.5~PN\footnote{The ``TaylorF2'' waveform.}. Aside from sky location, the polarization angle $\psi$ remains a free parameter; for the purpose of this example calculation, at each sky location we pick the value of $\psi$ that minimizes the area. The result is shown in Figure~\ref{fig:fisher-area-map}(a). The plot is shown in geographic coordinates to preserve spatial relationships to the Earth-fixed detector locations and antenna patterns.

For these sources, the area ranges from about 3~to~200~deg$^2$. It is smaller than 10~deg$^2$ across much of the sky, but much larger than 10~deg$^2$ in a broad ring that is oblique to but near the equator. This is the great circle that is parallel to the plane of the detectors. Here, the times of arrival are more sensitive than anywhere else to the azimuthal angle around the ring, but are constant to first order in the elevation angle relative to this plane. In Figure~\ref{fig:fisher-area-map}(b), we show the sky resolution calculated using times of arrival only. Without the coherence between detectors, the Fisher matrix becomes singular everywhere along this ring.

\begin{figure*}
    \includegraphics[width=\textwidth]{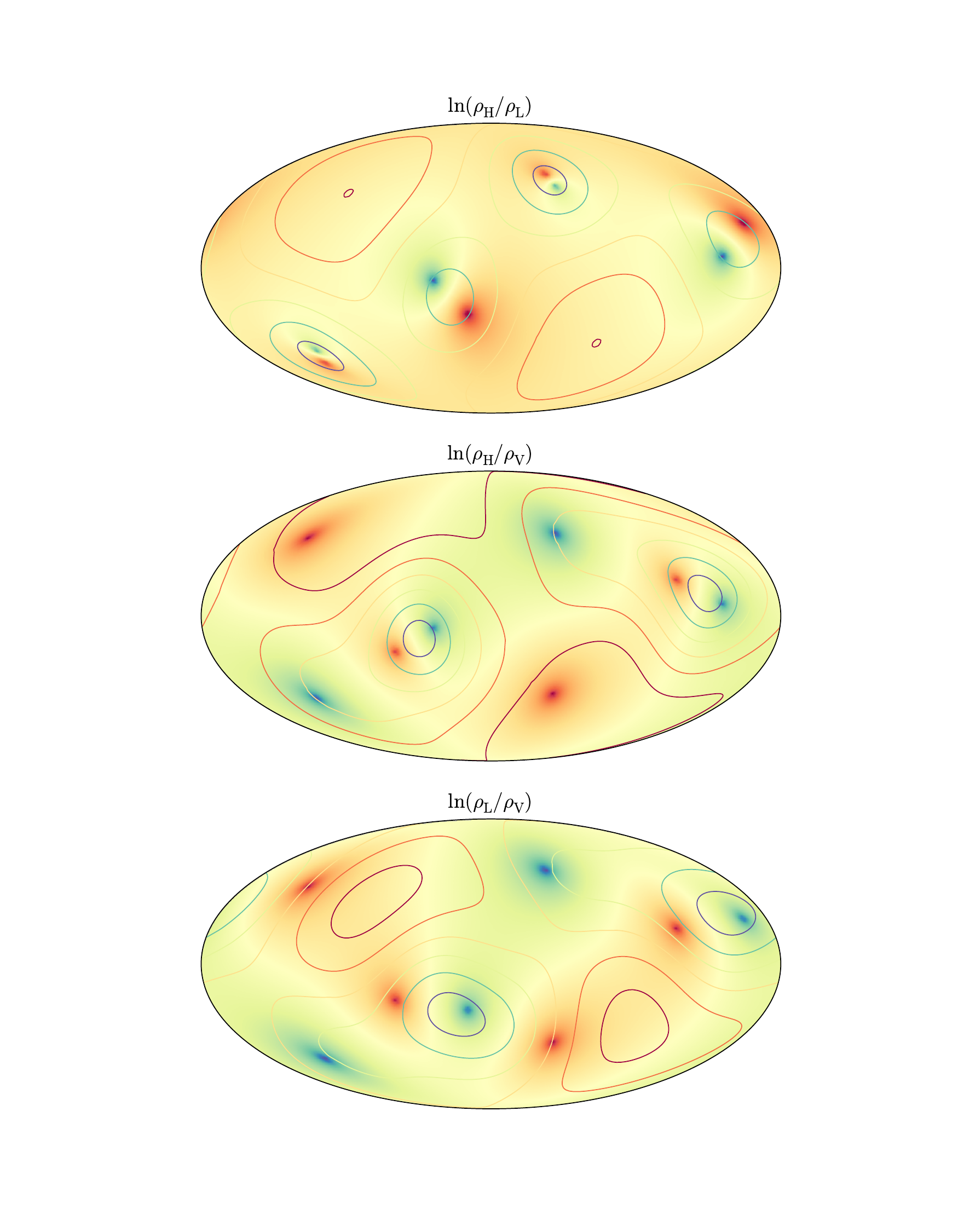}
    \caption[Log ratios of \acsp{SNR} in pairs of detectors]{\label{fig:fisher-snr-ratios}Log ratios of \acp{SNR} in pairs of detectors, $\ln (\rho_i / \rho_j)$ for detectors $i$ and $j$. Contour lines show the network \ac{SNR} for the pair of detectors, $\sqrt{{\rho_i}^2 + {\rho_j}^2}$.}
\end{figure*}

Other prominent features in Figure~\ref{fig:fisher-area-map} include four small knots of coarse resolution at large angles to the great circle described above. These are positions where the \emph{ratio} of the antenna patterns of any two detectors is large---for instance, positions where the Hanford detector is very sensitive but the Livingston detector is not. In Figure~\ref{fig:fisher-snr-ratios}, we have plotted a heatmap of the logarithm of the ratios of the \acp{SNR} in each of the three pairs of detectors; these plots exhibit hotspots at the same locations as those isolated spots of large area uncertainty in Figure~\ref{fig:fisher-area-map}.

\subsection{Improvement in localization due to coherence}

\begin{figure*}
    \includegraphics[width=\textwidth]{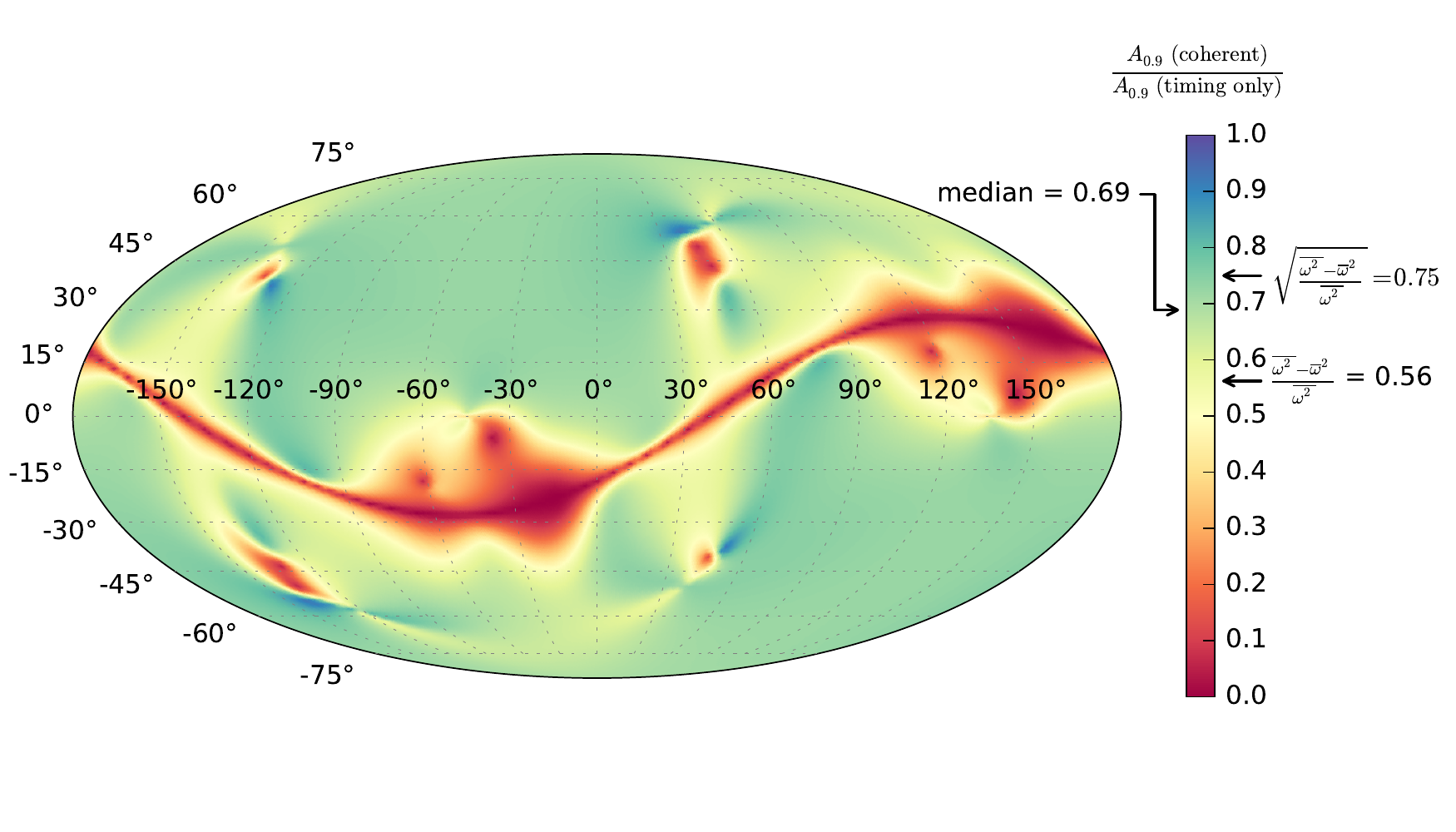}
    \caption[Improvement in sky resolution due to coherence]{\label{fig:fisher-area-improvement}Improvement in localization by including coherence. The color represents the ratio between the area as calculated assuming amplitude and phase consistency between detectors, and the area as calculated from just time delays on arrival. The median value and the two predictions from \citep{Grover:2013} are marked on the color bar.}
\end{figure*}

Much of the current expectations for \ac{GW} sky localization accuracy are based on Fisher matrix calculations that use time of arrival information alone \citep{FairhurstTriangulation,WenLocalizationAdvancedLIGO,LIGOObservingScenarios,FairhurstLIGOIndia}. There is a question as to whether these estimates can be straightforwardly `calibrated' with a scale factor to give realistic areas that account for coherence. \citet{Grover:2013} studied the relationship between the areas predicted this way versus areas found using the full Fisher matrix or by actually performing \ac{MCMC} parameter estimation. They postulated two possible expressions for the ratio between the area computed with coherence versus the area computed from timing only:
\begin{align}
    \label{eq:grover-ratio}
    \frac{A\text{ (coherent)}}{A\text{ (timing only)}} &= \frac{\overline{\omega^2} - {\overline{\omega}}^2}{\overline{\omega^2}} \\
    \intertext{or}
    \label{eq:grover-ratio-sqrt}
    \frac{A\text{ (coherent)}}{A\text{ (timing only)}} &= \sqrt{\frac{\overline{\omega^2} - {\overline{\omega}}^2}{\overline{\omega^2}}}.
\end{align}
This is motivated by the fact that the linear dimensions of the uncertainty ellipse, using only timing, are proportional to $1/\omega_\mathrm{rms} = 1 / \sqrt{\overline{\omega^2} - {\overline{\omega}}^2}$, whereas if the phase was known exactly (and removed altogether from the Fisher matrix, without marginalization), the linear dimensions would be proportional to $1/\sqrt{\overline{\omega^2}}$.

Note that in a two\nobreakdashes-detector network, the \emph{area} scales as $A \propto 1 / \sqrt{\omega_\mathrm{rms}}$, because the localization regions are annuli whose areas are determined by a single linear dimension. For a network of three or more detectors, the area scales as $A \propto 1 / \omega_\mathrm{rms}$, because the area depends on two linear dimensions. (Similar scaling relations would hold for \emph{phase\nobreakdashes-only} localization). Equation~(\ref{eq:grover-ratio}), therefore, may apply to three, but not two, detectors; conversely, we should expect that Equation~(\ref{eq:grover-ratio}) holds for two but not three detectors.

However, there are other problems with Equations~(\ref{eq:grover-ratio})~and~(\ref{eq:grover-ratio-sqrt}). First, for two detector networks there is \emph{no} improvement in sky localization due to adding coherence; the marginal Fisher matrix is \emph{identical} whether computed with timing only or with the full signal model. When combined with a \emph{distance prior}, coherence can partially break degeneracies and improve localization relative to triangulation, but this effect is not represented by the Fisher matrix calculation. Second, they do not account for the fact that in networks of three or more detectors, coherence \emph{itself} can inform sky localization, even \emph{without} time of arrival information; the effect of phase measurement goes beyond the correlations with time of arrival measurement errors. Third, wherever coherence breaks degeneracies in the timing analysis, the ratio in areas can be arbitrarily extreme because the area computed from timing can be almost arbitrarily large. Fourth, in configurations where the posterior probability distributions have multiple modes, coherence and priors can reduce areas significantly (see Chapter~\ref{chap:first2years}) in a way that is not captured by the Fisher matrix. This is a problem with the Fisher matrix itself, inherited by Equations~(\ref{eq:grover-ratio})~and~(\ref{eq:grover-ratio-sqrt}).

In Figure~\ref{fig:fisher-area-improvement} we show the ratio between the area computed using the coherent Fisher matrix versus the area computed using timing alone. The median ratio of 0.69 is marked on the color bar, along with the values of the two \citet{Grover:2013} formulas, Equation~(\ref{eq:grover-ratio})\,$\approx 0.56$ and Equation~(\ref{eq:grover-ratio-sqrt})\,$\approx 0.75$. The median is a little closer to Equation~(\ref{eq:grover-ratio-sqrt}), contrary to what would be expected for a three\nobreakdashes-detector network. There are small pockets where the ratio is almost 1, i.e., no improvement. There is also a thin strip and two broad regions near the detector plane where the ratio is almost arbitrarily small (because the triangulation localization has a degeneracy that is broken by coherence). Neither extreme is well described by the \citet{Grover:2013} formulas, and the median is not particularly close to one formula or the other. The improvement in area seems to be highly sensitive to the detector configuration and the sky position.

\subsection{Revision to \acs{LIGO} observing scenarios document}
\label{sec:revision-observing-scenarios}

Recently, the \acs{LIGO}/Virgo collaboration published a document outlining possible commissioning and observing timetables from 2015 through 2022 \citep{LIGOObservingScenarios}. This document estimates the detection rate and sky localization accuracy as the detector network evolves. Sky areas are estimated using triangulation considerations, as in \citet{FairhurstTriangulation}. As we have now shown, this can modestly overestimate the true uncertainties for three\nobreakdashes-detector networks.

\begin{figure}
    \centering
    \includegraphics[width=\textwidth]{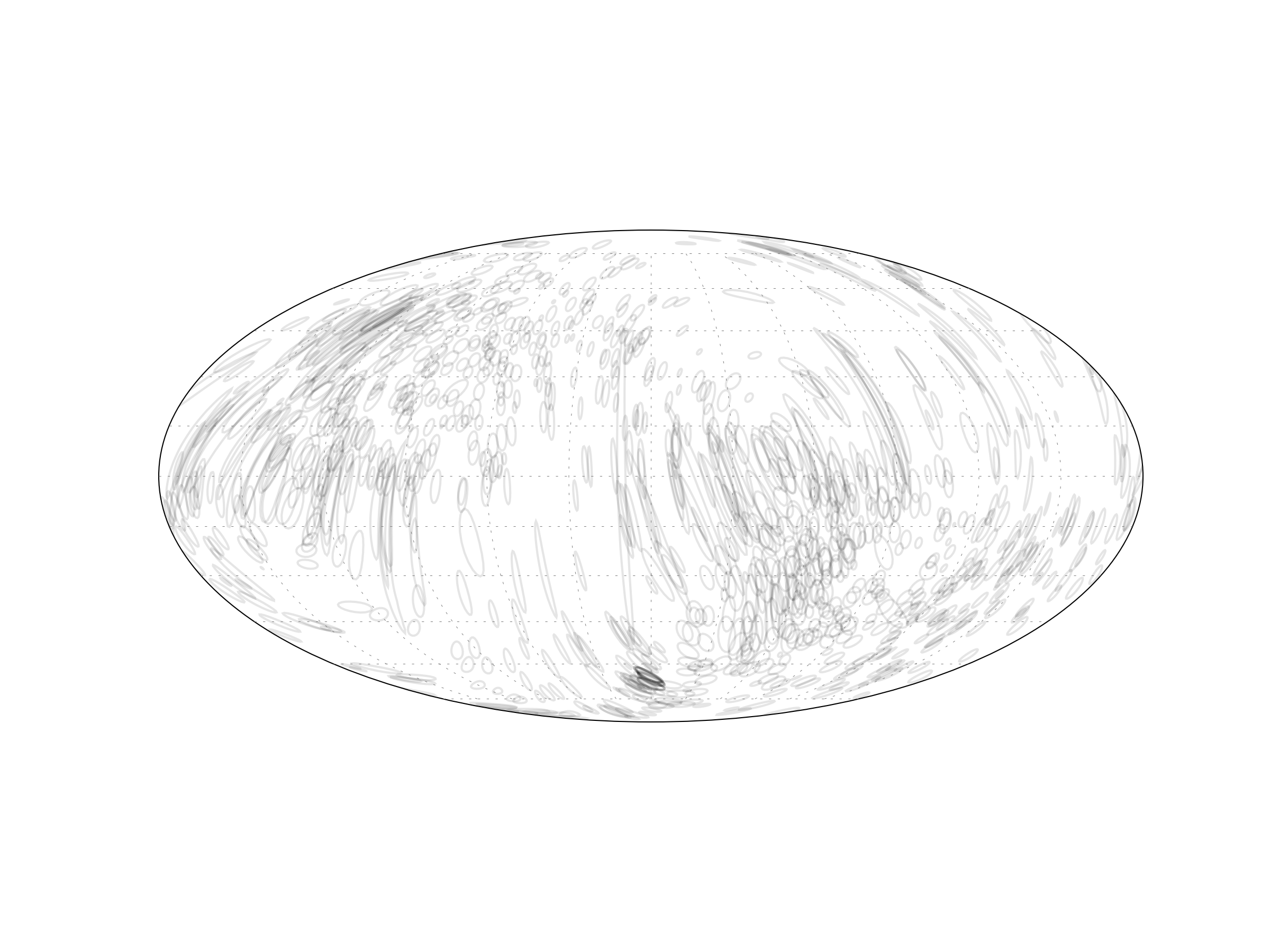}
    \caption[Fisher matrix error ellipses of a random sample of \acs{GW} events]{\label{fig:fisher-ellipses}90\% confidence ellipses of a random sample of sources that would be detectable by a three\nobreakdashes-detector, HLV network at final design sensitivity.}
\end{figure}

Here, we characterize the resolution of an HLV network assuming full coherence. First, we generated a sample of detectable \ac{BNS} signals by drawing samples from a spatially uniform, isotropic distribution and checking whether the \ac{SNR} was $\geq 4$ in at least two detectors and the network \ac{SNR} in those detectors, $\rho_\mathrm{net}$, was at least 12.\footnote{This is not to be confused with the different simulated detection population that we introduce in Chapter~\ref{chap:first2years}, which arises from HL and HLV detector networks at \emph{early} Advanced \ac{LIGO}/Virgo sensitivity.} Each detector has a random and independent duty cycle of 80\%. For all surviving sources, we computed the area from the Fisher matrix, assuming full coherence and also using only timing information. The 90\% confidence ellipses for all of these sources are shown in Figure~\ref{fig:fisher-ellipses}.

These assumptions are similar to, but not exactly the same as, the ``2019'' scenario from \citet{LIGOObservingScenarios}. They assume that Virgo is slightly less sensitive due to differing commissioning timetables. We assume a Virgo \ac{BNS} range of 154~Mpc while they assume 65\nobreakdashes-–130~Mpc. They assumed a single\nobreakdashes-detector threshold of 5, and calculated the network \ac{SNR} from all operating detectors, rather than just those detectors with \acp{SNR} above the single\nobreakdashes-detector threshold. Furthermore, \citet{LIGOObservingScenarios} neglected the contribution to sky localization from detectors with $\rho < 3$.

Summary statistics compared against \citet{LIGOObservingScenarios} are shown in Table~\ref{table:fisher-comparison-observing-scenarios}. Percentiles were computed by dividing the number of three-detector events localized within a given area by the total number of events (i.e., the denominator included both two- and three\nobreakdashes-detector events; equivalent to treating the areas for two\nobreakdashes-detector events as infinite). For timing only, we find that 4.5\% of sources are localized within 5~deg$^2$, consistent with the range of 3\nobreakdashes--8\% in \citet{LIGOObservingScenarios}. However, we find that 30\% of sources are localized within 20~deg$^2$, at the high end of the claimed range of 8\nobreakdashes--28\%. This moderate disagreement is probably due to differing Virgo sensitivities and slightly different detection criteria. Adding full coherence, we find that 7.4\% of sources are found within 5~deg$^2$ and 38\% within 20~deg$^2$. The median decreases from 83~to~46~deg$^2$, shrinking in area by a factor of 0.55.

Assuming statistically independent single\nobreakdashes-detector duty cycles\footnote{In previous \ac{LIGO}/Virgo science runs, the duty cycles of distinct detectors were somewhat correlated. The detectors tended to remain locked longer at night due to reduced local anthropogenic ground motion. Advanced detectors with improved seismic isolation systems may be more resistant to seismically triggered lock loss and have duty cycles that are less correlated.} of 0.8, all three detectors are operational about half the time; most of the rest of the time only a pair of detectors is observing. Including the localization of two-detector events would modestly improve the median area.

\begin{deluxetable*}{c|cc|cc|c}
    \tablecaption{\label{table:fisher-comparison-observing-scenarios}Revised sky resolution predictions for Advanced \acs{LIGO} observing scenarios}
    \tablehead{
        \colhead{} &
        \multicolumn{2}{c}{Range (Mpc)} &
        \multicolumn{2}{c}{\% Localized within} &
        \colhead{Median} \\
        \colhead{Epoch} &
        \colhead{LIGO} &
        \colhead{Virgo} &
        \colhead{5~deg$^2$} &
        \colhead{20~deg$^2$} &
        \colhead{area (deg$^2$)}
    }
    \startdata
    2019+ \citep{LIGOObservingScenarios} & 200 & 65--130 & 3--8 & 8--28 & N/A \\
    HLV (coherent) & 199 & 154 & 5.5 & 36 & 43 \\
    HLV (timing) & 199 & 154 & 3.4 & 27 & 73
    \enddata
    \tablecomments{Revised predictions of sky resolution at final Advanced \ac{LIGO}+Virgo design sensitivity, compared with \citet{LIGOObservingScenarios}. (The median value is omitted from the first row because \citet{LIGOObservingScenarios} does not specify one.)}
\end{deluxetable*}

\section{Summary}

We have presented an idealized model of the noise and signal content in \ac{GW} observations with a network of laser interferometric detectors. This model led us to the matched filter, the maximum likelihood estimator, and the network \ac{SNR} as the simplest possible statistic for discriminating between the presence or absence of an astrophysical signal. Having defined \ac{SNR}, we worked out the horizon distance of a \ac{GW} detector. Next, we used the Fisher matrix formalism to calculate the sky resolution of a \ac{GW} detector network. Previous approaches (\citealt{FairhurstTriangulation}, etc.) have considered only arrival time measurements, but our computation also accounts for measurements of phase and amplitude on arrival. Despite this, our method as outlined in Section~\ref{sec:fisher-matrix-area-outline} is only slightly more complicated than the timing\nobreakdashes-only calculation. Though the Fisher matrix analysis cannot be applied to two\nobreakdashes-detector networks, we do expect it to accurately predict the sky localization accuracy of signals that are confidently detected by networks of three or more detectors of comparable sensitivity. We endorse it as the most sophisticated analysis worth carrying out, short of performing full Bayesian parameter estimation on a population of simulated signals. We advocate using it to revise the overly pessimistic sky resolution predictions for the 2019 and later scenarios in \citet{LIGOObservingScenarios}.

The horizon distance, the observables and parameters in \ac{GW} observations of compact binaries, the Fisher matrix, and estimates of sky resolution will recur in later chapters. In the next chapter, we will use all of these results to study the prospects for detecting and localizing \ac{BNS} signals in near real\nobreakdashes-time, from hundreds of seconds before to seconds after merger.

\chapter{Early warning \acs{GW} detection}
\label{chap:detection}

\author{Kipp~Cannon\altaffilmark{1},
Romain~Cariou\altaffilmark{2},
Adrian~Chapman\altaffilmark{3},
Mireia~Crispin-Ortuzar\altaffilmark{4},
Nickolas~Fotopoulos\altaffilmark{3},
Melissa~Frei\altaffilmark{5,6},
Chad~Hanna\altaffilmark{7},
Erin~Kara\altaffilmark{8},
Drew~Keppel\altaffilmark{9,10},
Laura~Liao\altaffilmark{11},
Stephen~Privitera\altaffilmark{3},
Antony~Searle\altaffilmark{3},
Leo~Singer\altaffilmark{3},
and~Alan~Weinstein\altaffilmark{3}}

\altaffiltext{1}{Canadian Institute for Theoretical Astrophysics, Toronto, ON, Canada}
\altaffiltext{2}{D\'{e}partement de Physique, \'{E}cole Normale Sup\'{e}rieure de Cachan, Cachan, France}
\altaffiltext{3}{LIGO Laboratory, California Institute of Technology, MC 100-36, 1200 E. California Blvd., Pasadena, CA, USA}
\altaffiltext{4}{Facultat de F\'{i}sica, Universitat de Val\`{e}ncia, Burjassot, Spain} 
\altaffiltext{5}{Department of Physics, University of Texas at Austin, Austin, TX, USA}
\altaffiltext{6}{Center for Computational Relativity and Gravitation and School of Mathematical Sciences, Rochester Institute of Technology, Rochester, NY, USA}
\altaffiltext{7}{Perimeter Institute for Theoretical Physics, Waterloo, ON, Canada}
\altaffiltext{8}{Department of Physics and Astronomy, Barnard College, Columbia University, New York, NY, USA}
\altaffiltext{9}{Albert-Einstein-Institut, Max-Planck-Institut f\"{u}r Gravitationphysik, Hannover, Germany}
\altaffiltext{10}{Institut f\"{u}r Gravitationsphysik, Leibniz Universit\"{a}t Hannover, Hannover, Germany}
\altaffiltext{11}{Department of Chemistry and Biology, Ryerson University, Toronto, ON, Canada}

\attribution{This chapter is reproduced in part from \citet{Cannon:2011vi}, which was published in \textnormal{The Astrophysical Journal} as ``Toward early-warning detection of gravitational waves from compact binary coalescence,'' copyright~\textcopyright{}~2012 The American Astronomical Society. Note that the author list is alphabetical because of the large number of contributors to this algorithm and codebase, initiated and organized by K.C., C.H., and D.K. My contributions to that code are related to the interpolation, decimation, and triggering stages. My contributions to the project include all of the calculations related to ``early warning'' detection and localization, all of the accounting of the computational cost, and the measurement of the mismatch of the template bank. I prepared all of the figures and tables in this publication and about 80\% of the text, and was the corresponding author.}

In the first generation of ground-based laser interferometers, the \ac{GW} community initiated a project to send alerts when potential \ac{GW} transients were observed in order to trigger follow-up observations by \ac{EM} telescopes. The typical latencies were 30 minutes \citep{HugheyGWPAW2011}. This was an important achievement, but too late to catch any prompt (i.e., simultaneous with gamma\nobreakdashes-ray emission) optical flash and later than would be desirable to search for an on-axis optical afterglow (which fades rapidly as a power law in time; see for example \citealt{MostPromisingEMCounterpart}). Since the \ac{GW} signal is in principle detectable even before the tidal disruption, one might have the ambition of reporting \ac{GW} candidates not minutes after the merger, but seconds before. We explore one essential ingredient of this problem, a computationally inexpensive real-time filtering algorithm for detecting inspiral signals in \ac{GW} data. We also consider the prospects for advanced \ac{GW} detectors and discuss other areas of work that would be required for rapid analysis.

In October 2010, \ac{LIGO} completed its sixth science run (S6) and Virgo completed its third science run (VSR3). While both \ac{LIGO} detectors and Virgo were operating, several all-sky detection pipelines operated in a low-latency configuration to send astronomical alerts, namely, \ac{cWB}, Omega, and \acl{MBTA} \citep[\acs{MBTA};][]{CBCLowLatency,FirstPromptSearchGWTransientsEMCounterparts}. \ac{cWB} and Omega are both unmodeled searches for bursts based on time-frequency decomposition of the \ac{GW} data. \ac{MBTA} is a novel kind of template-based inspiral search that was purpose-built for low latency operation. \ac{MBTA} achieved the best \ac{GW} trigger-generation latencies, of 2--5 minutes. Alerts were sent with latencies of 30--60 minutes, dominated by human vetting. Candidates were sent for \ac{EM} follow-up to several telescopes; \textit{Swift}, LOFAR, ROTSE, TAROT, QUEST, SkyMapper, Liverpool Telescope, Pi of the Sky, Zadko, and Palomar Transient Factory imaged some of the most likely sky locations \citep{FirstPromptSearchGWTransientsEMCounterparts,SwiftFollowupGWTransients,FirstSearchesOpticalCounterparts}.

There were a number of sources of latency associated with the search for \ac{CBC} signals in S6/VSR3 \citep{HugheyGWPAW2011}, listed here.

\paragraph{Data acquisition and aggregation ($\gtrsim$100~ms)}%
The \ac{LIGO} data distribution system collects data in real time, but distributes it to computers in the control rooms 16 times a second, and archives it for immediate offsite replication in blocks of 16~s~\citep{Bork2001}. Data are copied from all of the \ac{GW} observatories to the analysis clusters over the Internet, which is capable of high bandwidth but only modest latency. Altogether, it takes about 16~s to transmit the data to the analysis clusters, but with moderate changes in infrastructure could be reduced to $\sim 100$~ms if the data were streamed to the computing clusters in real\nobreakdashes-time without blocking it into 16~s chunks.

\paragraph{Data conditioning ($\sim$1~min)}%
Science data must be calibrated using the detector's frequency response to gravitational radiation. Currently, data are calibrated in blocks of 16~s. Within $\sim$1~minute, data quality is assessed in order to create veto flags. These are both technical sources of latency that might be addressed with improved calibration and data quality software for advanced detectors.

\paragraph{Trigger generation (2--5~min)}%
Low-latency data analysis pipelines deployed in S6/VSR3 achieved an impressive latency of minutes. However, second to the human vetting process, this dominated the latency of the entire \ac{EM} follow-up process. Even if no other sources of latency existed, this trigger generation latency is too long to catch prompt or even extended emission. Low-latency trigger generation will become more challenging with advanced detectors because inspiral signals will stay in band up to 10 times longer. In this work, we will focus on reducing this source of latency.

\paragraph{Alert generation (2--3~min)}%
S6/VSR3 saw the introduction of low-latency astronomical alerts, which required gathering event parameters and sky localization from the various online analyses, downselecting the events, and calculating telescope pointings. If other sources of latency improve, the technical latency associated with this infrastructure could dominate, so work should be done to improve it.

\paragraph{Human validation (10--20~min)}%
Because the new alert system was commissioned during S6/VSR3, all alerts were subjected to quality control checks by human operators before they were disseminated. This was by far the largest source of latency during S6/VSR3. Hopefully, confidence in the system will grow to the point where no human intervention is necessary before alerts are sent, so we give it no further consideration here.

~

This chapter will focus on reducing the latency of trigger production. Data analysis strategies for advance detection of \acp{CBC} will have to strike a balance between latency and throughput. \ac{CBC} searches consist of banks of matched filters, or cross-correlations between the data stream and a bank of nominal ``template'' signals. There are many different implementations of matched filters, but most have high throughput at the cost of high latency, or low latency at the cost of low throughput. The former are epitomized by the overlap-save algorithm for \ac{FD} convolution, currently the preferred method in \ac{GW} searches. The most obvious example of the latter is direct \ac{TD} convolution, which can be done in real\nobreakdashes-time. However, its cost in floating point operations per second is linear in the length of the templates, so it is prohibitively expensive for long templates. The computational challenges of low-latency \ac{CBC} searches are still more daunting for advanced detectors for which the inspiral signal remains in band for a large fraction of an hour~(see Appendix~\ref{sec:low-frequency-cutoff}).

Fortunately, the morphology of inspiral signals can be exploited to offset some of the computational complexity of known low-latency algorithms. First, the signals evolve slowly in frequency, so that they can be broken into contiguous band-limited time intervals and processed at possibly lower sample rates. Second, inspiral filter banks consist of highly similar templates, admitting methods such as the \ac{SVD}~\citep{Cannon:2010p10398} or the Gram-Schmidt process~\citep{rbf} to reduce the number of templates.

Several efforts that exploit one or both of these properties are under way to develop low-latency \ac{CBC} search pipelines with tractable computing requirements. One example is \ac{MBTA}~\citep{Marion2004, Buskulic2010}, which was deployed in S6/VSR3. \ac{MBTA} consists of multiple, usually two, template banks for different frequency bands, one which is matched to the early inspiral and the other which is matched to the late inspiral. An excursion in the output of any filter bank triggers coherent reconstruction of the full matched filtered output. Final triggers are built from the reconstructed matched filter output. Another novel approach using networks of parallel, second-order \ac{IIR} filters is being explored by \citet{shaunIIR} and \citet{linqingIIR}.

We will use both properties to demonstrate that a very low latency detection statistic is possible with current computing resources. Assuming the other technical sources of latency can be reduced significantly, this could make it possible to send prompt ($< 1$~minute) alerts to the astronomical community.

The chapter is organized as follows. First, we describe the standard, offline \ac{CBC} detection process. Using a simple model of the detection and sky localization accuracy of this search, we study the prospects for early-warning detection. Then, we provide an overview of our novel method for detecting \ac{CBC} signals near real-time. We then describe a prototype implementation using open source signal processing software. To validate our approach we present a case study focusing on a particular subset of the \ac{NS}--\ac{NS} parameter space. We conclude with some remarks on what remains to be prepared for the advanced detector era.

\section{Prospects for early-warning detection and \acs{EM} follow-up}
\label{sec:prospects-detection}

\begin{figure}[t]
\centering
\includegraphics{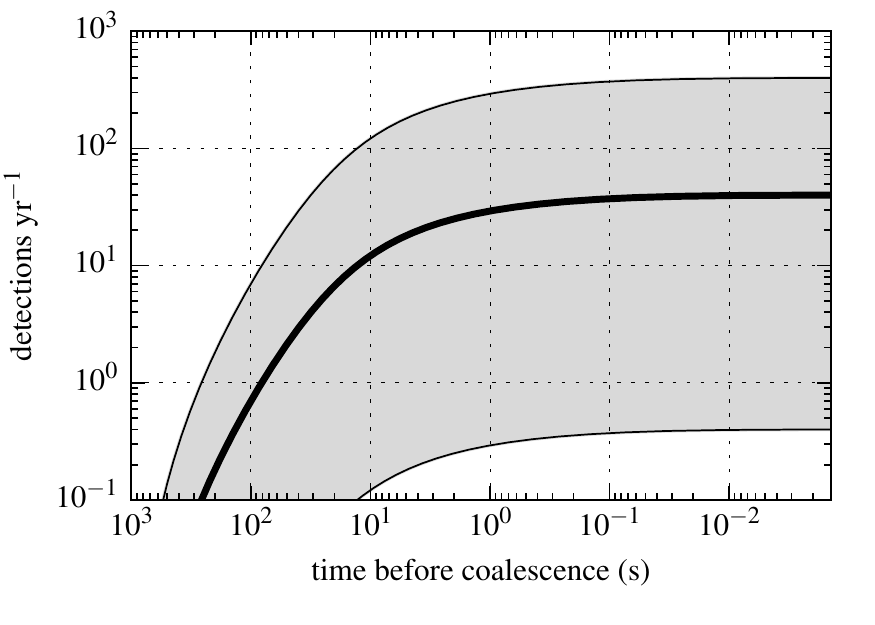}
\caption[Expected number of \acs{BNS} sources detectable before coalescence]{\label{fig:earlywarning}Expected number of \ac{NS}--\ac{NS} sources that could be detectable by Advanced \ac{LIGO} a given number of seconds before coalescence. The heavy solid line corresponds to the most probable yearly rate estimate from~\citet{LIGORates}. The shaded region represents the 5\%--95\% confidence interval arising from substantial uncertainty in predicted event rates.}
\end{figure}
Before the \ac{GW} signal leaves the detection band, we can imagine examining the \ac{SNR} accumulated up to that point and, if it is already significant,
releasing an alert immediately, trading \ac{SNR} and sky localization accuracy for pre-merger detection.

In the quadrupole approximation, the instantaneous frequency of the GW inspiral signal is related to the time $t$ relative to coalescence ~\citep[Section 5.1 of][]{livrev12} through
\begin{equation} \label{eq:fgw}
	f(t) = \frac{1}{\pi \mathcal{M}_\mathrm{t}}
		\left[ \frac{5}{256}\frac{\mathcal{M}_\mathrm{t}}{t} \right]^{3/8},
\end{equation}
where $\mathcal{M}=M^{2/5} \mu^{3/5}$ is the chirp mass of the binary, $\mathcal{M}_\mathrm{t}=G \mathcal{M} / c^3$ is the chirp mass in units of time, $M$ is the total mass, and $\mu$ is the reduced mass. The expected value of the single-detector \ac{SNR} for an optimally oriented (source at detector's zenith or nadir, orbital plane face-on) inspiral source is~\citep{LIGORates}
\begin{equation}
	\label{eq:expected-snr}
	\rho =
		\frac{{\mathcal{M}_\mathrm{t}}^{5/6} c}{\pi^{2/3} D}
		\sqrt{
			\frac{5}{6} \int_{f_\mathrm{low}}^{f_\mathrm{high}}
			\frac{f^{-7/3}}{S(f)} \mathrm{d}f},
\end{equation}
where $D$ is the luminosity distance and $S(f)$ is the one-sided power spectral density of the detector noise.  $f_\mathrm{low}$ and $f_\mathrm{high}$ are low- and high- frequency limits of integration which may be chosen to extend across the entire bandwidth of the detector.  If we want to trigger at a time $t$ before merger, then we must cut off the SNR integration at $f_\mathrm{high} = f(t)$, with $f(t)$ given by Equation~(\ref{eq:fgw}) above.

Figure~\ref{fig:earlywarning} shows projected early detectability rates for \ac{NS}--\ac{NS} binaries in Advanced \ac{LIGO} assuming the anticipated detector sensitivity for the `zero detuning, high power' configuration described in \citet{ALIGONoise} and \ac{NS}--\ac{NS} merger rates estimated in \citet{LIGORates}. The merger rates have substantial measurement uncertainty due to the small sample of known double pulsar systems that will merge within a Hubble time; they also have systematic uncertainty due to sensitive dependence on the pulsar luminosity distribution function~\citep{KalogeraRates}. The most probable estimates indicate that at a single-detector \ac{SNR} threshold of 8 we will observe a total of 40~events~yr$^{-1}$; $\sim$10~yr$^{-1}$ will be detectable within 10~s of merger and $\sim$5~yr$^{-1}$ will be detectable within 25~s of merger if analysis can proceed with near zero latency.

We emphasize that any practical \ac{GW} search will include technical delays due to light travel time between the detectors, detector infrastructure, and the selected data analysis strategy. Figure~\ref{fig:earlywarning} must be understood in the context of all of the potential sources of latency, some of which are avoidable and some of which are not.

\begin{figure}[t]
\centering
\includegraphics{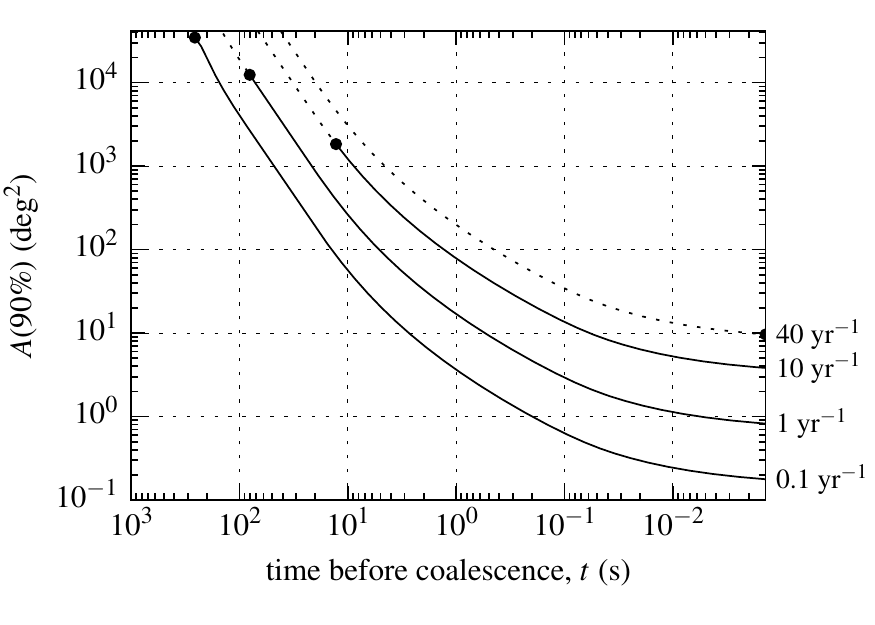}
\caption[Localization area as a function of time before coalescence]{\label{fig:sky-localization-accuracy}Area of the 90\% confidence
region as a function of time before coalescence for sources with anticipated
detectability rates of 40, 10, 1, and 0.1~yr$^{-1}$. The heavy dot indicates
the time at which the accumulated \ac{SNR} exceeds a single-detector threshold of~8.}
\end{figure}
\begin{table}[h]
\caption[Number of sources that are localizable before merger]{\label{table:sky-localization-accuracy}Horizon distance, \ac{SNR} at
merger, and area of 90\% confidence at selected times before merger for sources
with expected detectability rates of 40, 10, 1, and 0.1~yr$^{-1}$.}
\begin{center}
\begin{tabular}{rrrrrrr}
\tableline\tableline
Rate & Horizon & Final & \multicolumn{4}{c}{$A$(90\%) (deg$^2$)} \\
\cline{4-7}
yr$^{-1}$ & (Mpc) & \ac{SNR} & 25 s & 10 s & 1 s & 0 s \\
\tableline
40\phd\phn & 445 & 8.0 & ----- & ----- & ----- & 9.6 \\
10\phd\phn & 280 & 12.7 & ----- & 1200 & 78 & 3.8 \\
1\phd\phn & 130 & 27.4 & 1300 & 260 & 17 & 0.8 \\
0.1 & 60 & 58.9 & 280 & 56 & 3.6 & 0.2 \\
\tableline
\end{tabular}
\tablecomments{A dash (-----) signifies that the confidence area\\is omitted because at the indicated time the SNR would\\not have crossed the detection threshold of 8.}
\end{center}
\end{table}

\ac{EM} follow-up requires estimating the location of the \ac{GW} source. The localization uncertainty can be estimated from the uncertainty in the time of arrival of the \acp{GW}, which is determined by the signal's effective bandwidth and \ac{SNR} \citep{FairhurstTriangulation}. Table~\ref{fig:earlywarning} and Figure~\ref{fig:sky-localization-accuracy} show the estimated 90\% confidence area versus time of the loudest coalescence events detectable by Advanced \ac{LIGO} and Advanced Virgo. This is the \emph{minimum} area; localization is best at high elevation from the plane containing the detectors, and worst at zero elevation. \citeauthor{FairhurstTriangulation} also cautions that his Fisher matrix calculation fails to capture disconnected patches of probability, which occur prominently in networks of three detectors where there are generally two local maxima on opposite sides of the plane of the detectors. Aside from the mirror degeneracy, characterizing the uncertainty region by the Fisher matrix alone tends to overestimate, rather than underestimate, the area for low-\ac{SNR} events, but this effect is generally more than compensated for by the source being in an unfavorable sky location. For these reasons, the localization uncertainty estimated from timing is highly optimistic and will only suffice for an order-of-magnitude estimate. Once per year, we expect to observe an event with a final single-detector \ac{SNR} of $\approx$27 whose location can be constrained to about 1300~deg$^2$ (3.1\% of the sky) within 25~s of merger, 260~deg$^2$ (0.63\% of the sky) within 10~s of merger, and 0.82~deg$^2$ (0.0020\% of the sky) at merger.

\begin{figure*}
    \includegraphics{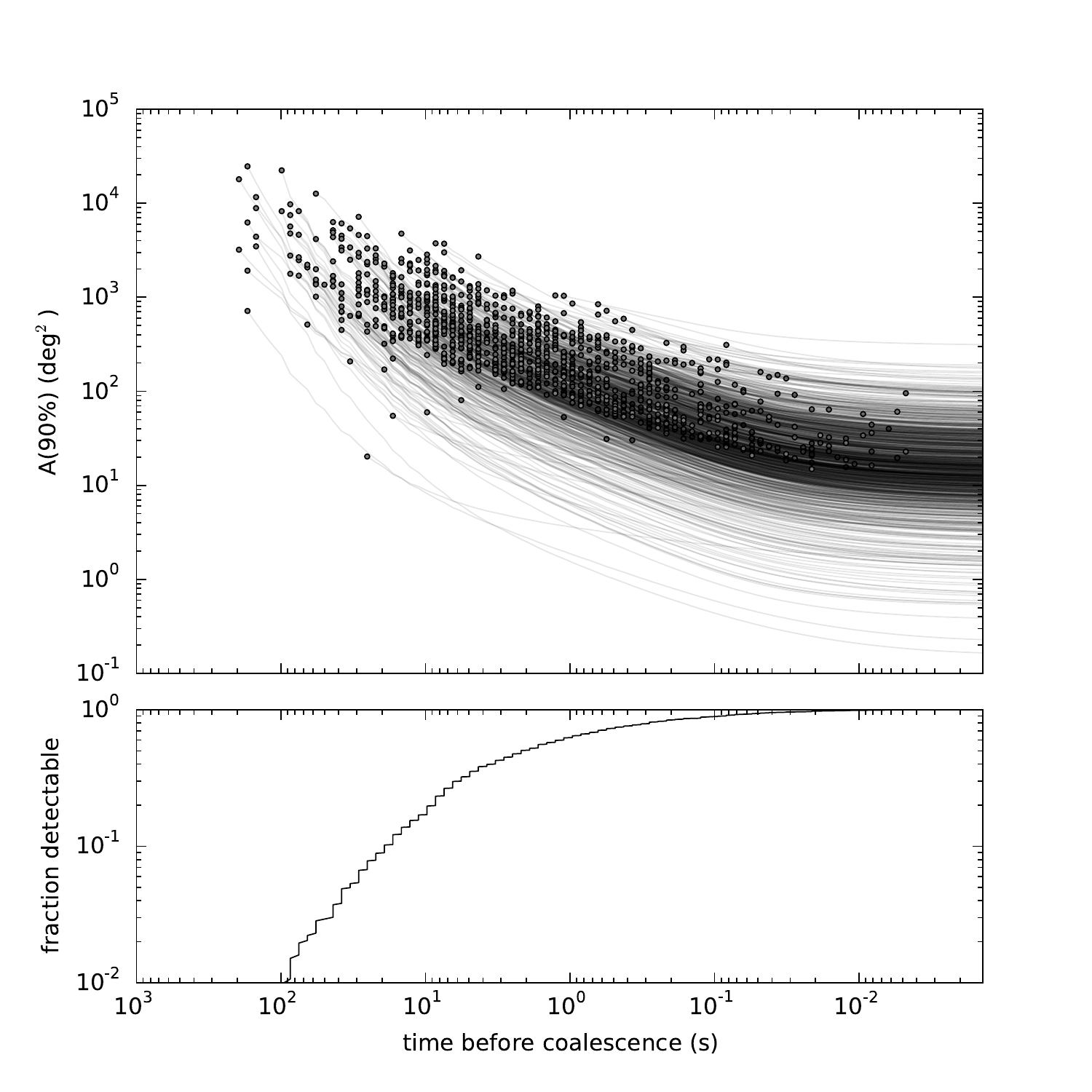}
    \caption[Sky resolution versus time before merger for a random sample of events]{\label{fig:early-warning-localization-sample}Trajectories of 90\% localization area versus time before coalescence for a random sample of \ac{BNS} signals. Each track starts when the signal crosses a single\nobreakdashes-detector \ac{SNR} threshold of $\rho \geq 4$ in at least two detectors, and when all of the detectors that are above threshold together yield a network \ac{SNR} $\rho_\mathrm{net} \geq 12$. The bottom panel shows the cumulative fraction of events that are detectable a given time before coalescence.}
\end{figure*}

The picture is qualitatively similar when we track sky localization area versus time for individual events using the coherent Fisher matrix approach described in Section~\ref{sec:fisher-matrix-area-outline}. In Figure~\ref{fig:early-warning-localization-sample}, we plot the 90\% confidence area as a function of time before coalescence for the event sample described in Section~\ref{sec:revision-observing-scenarios}. The track begins at the earliest time that the source is ``confidently'' detectable (single\nobreakdashes-detector \ac{SNR} threshold of $\rho \geq 4$ in at least two detectors, and the network \ac{SNR} from all of the detectors that are above threshold together yield a network \ac{SNR} $\rho_\mathrm{net} \geq 12$). We find that about 1\% of events are detectable 100~s before merger, with areas of $\sim 10^3$\nobreakdashes--$10^4$~deg$^2$. About 20\% of sources are detectable 10~s before merger, with areas of $\sim 10^2$\nobreakdashes--$10^3$~deg$^2$. By the time that the full signal has been acquired, the areas shrink to $\sim 10$\nobreakdashes--100~deg$^2$.

\emph{After} merger, typical short \ac{GRB} optical afterglows, scaled to Advanced \ac{LIGO} distances, should be within reach of moderately deep optical transient experiments such as the \acl{PTF} (\acsu{PTF}; \citealt{PTFRau,PTFLaw}), its successor \ac{ZTF} \citep{ZTF,ZTFBellm,ZTFSmith}, the eagerly awaited \acl{LSST} (\acsu{LSST}; \citealt{LSST}), and also the BlackGEM array (dedicated to \ac{GW} candidate follow\nobreakdashes-up). However, the exposure and overhead times of these instruments are too long to search the pre\nobreakdashes-merger error boxes. It is possible to reduce the localization uncertainty by only looking at galaxies from a catalog that lie near the sky location and luminosity distance estimate from the \ac{GW} signal~\citep{galaxy-catalog} as was done in S6/VSR3. Within the expected Advanced \ac{LIGO} \ac{NS}--\ac{NS} horizon distance, the number of galaxies that can produce a given signal amplitude is much larger than in Initial \ac{LIGO} and thus the catalog will not be as useful for downselecting pointings for most events. However, exceptional \ac{GW} sources will necessarily be extremely close. Within this reduced volume there will be fewer galaxies to consider for a given candidate and catalog completeness will be less of a concern. For some nearby events, using a galaxy catalog in connection with \emph{pre\nobreakdashes-merger} triggers could make it possible to search for optical flashes with rapidly slewing telescopes such as TAROT. Last, if \emph{Swift} had a fully automated \ac{TOO} mode, then for exceptional sources that are detectable $\sim$100~s before merger, one could slew \emph{Swift} so as to use \ac{BAT} to monitor the area of $\sim 1000$~deg$^2$ for the \ac{GRB} itself, building a virtual all\nobreakdashes-sky \ac{GRB} monitor out of \ac{LIGO} and \emph{Swift} combined (Sathyaprakash et al., in preparation).

\section{Novel real-time algorithm for \acs{CBC} detection}
\label{sec:method}

In this section, we describe a decomposition of the \ac{CBC} signal space that reduces \ac{TD} filtering cost sufficiently to allow for the possibility of early-warning detection with modest computing requirements. We expand on the ideas of \citet{Marion2004} and \citet{Buskulic2010} that describe a multi-band decomposition of the compact binary signal space that resulted in a search with minutes' latency during S6/VSR3~\citep{HugheyGWPAW2011}. We combine this with the \ac{SVD} rank-reduction method of \citet{Cannon:2010p10398} that exploits the redundancy of the template banks.

\subsection{Discrete-time representation of a matched filter}

Here, we translate the basic matched filter bank search described in Section~\ref{sec:basic-matched-filter-search} to a discrete time representation, suitable for studying the filter design. Suppose that the observed data $y[k]$ consists of a known, nominal signal $x[k]$, and additive, zero-mean noise $n[k]$:
$$
	y[k] = x[k] + n[k].
$$
A matched filter is a linear filter, defined as
$$
	z[k] = \sum_{n=0}^{N-1} h[n] \, y[k-n] = z_s[k] + z_n[k],
$$
where $z_s$ is the response of the filter to the signal alone and $z_n$ is the response of the signal to noise alone.  The matched filter's coefficients maximize the ratio of the expectation of the filter's instantaneous response to the variance in the filter's output:
$$
(\textrm{signal to noise})^2 = \frac{\expectation \left[ z[0] \right]^2}{\var \left[ z[k] \right]} = \frac{z_s[0]^2}{\var \left[ z_n[k] \right]}.
$$
It is well known~\citep[see, for example,][]{matched-filter} that if $n[k]$ is Gaussian and wide-sense stationary, then the optimum is obtained when
$$
\tilde{h}[n] = \tilde{x}^*[n] \, \tilde{S}^{-1}[n],
$$
up to an arbitrary multiplicative constant.  Here, $\tilde{h}[n]$, $\tilde{x}[n]$, and $\tilde{y}[n]$ are the discrete Fourier transforms (DFTs) of $h[k]$, $x[k]$, and $y[k]$, respectively; $\tilde{S}[n] = \expectation \left[ \tilde{n}[n] \tilde{n}^* [n] \right]$ is the folded, two-sided, discrete power spectrum of $n[k]$.  It is related to the continuous, one-sided power spectral density $S(f)$ through
$$
	\tilde{S}[n] =
	\begin{cases}
		S(n) & \textrm{if } n = 0 \textrm{ or } n = N / 2 \\
		S(n f^0 / 2 N) / 2 & \textrm{if } 0 < n < N / 2 \\
		\tilde{S}[N - n] & \textrm{otherwise},
	\end{cases}
$$
where $N$ is the length of the filter and $f^0$ is the sample rate.  (In order to satisfy the Nyquist-Shannon sampling criterion, it is assumed that the detector's continuous noise power spectral density $S(f)$ vanishes for all $f > f^0 / 2$, or alternatively, that the data are low-pass filtered prior to matched filtering.)  The DFT of the output is
\begin{equation}
\label{eq:matched-filter-fd}
\tilde{z}[n] = \tilde{x}^*[n] \, \tilde{S}^{-1}[n] \, \tilde{y}[n]
\equiv \left(\tilde{S}^{-1/2}[n] \, \tilde{x}[n]\right)^* \left(\tilde{S}^{-1/2}[n] \, \tilde{y}[n] \right).
\end{equation}
The placement of parentheses in Equation~(\ref{eq:matched-filter-fd}) emphasizes that the matched filter can be thought of as a cross-correlation of a whitened version of the data with a whitened version of the nominal signal.  In this chapter, we shall not describe the exact process by which the detector's noise power spectrum is estimated and deconvolved from the data; for the remainder of this chapter we shall define $y[k]$ as the \emph{whitened} data stream.  Correspondingly, from this point on we shall use $h[k]$ to describe the \emph{whitened} templates, being the inverse DFT of $\left(\tilde{S}^{-1/2}[n] \, \tilde{x}[n]\right)^*$.

Inspiral signals are continuously parameterized by a set of intrinsic source parameters $\theta$ that determine the amplitude and phase evolution of the \ac{GW} strain. For systems in which the effects of spin can be ignored, the intrinsic source parameters are just the component masses of the binary, $\theta = (m_1, m_2)$. For a given source, the strain observed by the detector is a linear combination of two waveforms corresponding to the `$+$' and `$\times$' \ac{GW} polarizations. Thus, we must design two filters for each $\theta$.

The coefficients for the $\numtmps$ filters are known as templates, and are formed by discretizing and time reversing the waveforms and weighting them by the inverse amplitude spectral density of the detector's noise. To construct a template bank, templates are chosen with $\numtmps/2$ discrete signal parameters $\theta_0,\, \theta_1,\, \dots,\, \theta_{\numtmps/2-1}$. These are chosen such that any possible signal will have an inner product $\geq$0.97 with at least one template. Such a template bank is said to have a {\em minimal match} of 0.97~\citep{Owen:1998dk}.

Filtering the detector data involves a convolution of the data with the templates. For a unit-normalized template $h_i[k]$ and whitened detector data $y[k]$, both sampled at a rate $f^0$, the result can be interpreted as the \ac{SNR}, $z_i[k]$, defined as
%
%
\begin{equation}
	\label{eq:SNRTD}
	z_i [k] = \sum_{n=0}^{N-1} h_{i}[n] \, y [k-n].
\end{equation}
This results in $\numtmps$ \ac{SNR} time series. Local peak-finding across time and template indices results in single-detector triggers. Coincidences are sought between triggers in different \ac{GW} detectors in order to form detection candidates.

Equation~(\ref{eq:SNRTD}) can be implemented in the \ac{TD} as a bank of \ac{FIR} filters, requiring $\mathcal O(\numtmps \tmpsamps)$ floating point operations per sample. However, it is typically much more computationally efficient to use the convolution theorem and the \ac{FFT} to implement fast convolution in the \ac{FD}, requiring only $\mathcal O(\numtmps \lg \tmpsamps)$ operations per sample but incurring a latency of $\mathcal O(\tmpsamps)$ samples.

\subsection{The \acs{LLOID} method}

Here we describe a method for reducing the computational cost of a \ac{TD} search for \acp{CBC}. We give a zero latency, real-time algorithm that competes in terms of floating point operations per second with the conventional overlap-save \ac{FD} method, which by contrast requires a significant latency due to the inherent acausality of the Fourier transform. Our method, called \ac{LLOID}, involves two transformations of the templates that produce a network of orthogonal filters that is far more computationally efficient than the original bank of matched filters.

The first transformation is to chop the templates into disjointly supported intervals, or \emph{time slices}. Since the time slices of a given template are disjoint in time, they are orthogonal with respect to time. Given the chirp-like structure of the templates, the ``early'' (lowest frequency) time slices have significantly lower bandwidth and can be safely downsampled. Downsampling reduces the total number of filter coefficients by a factor of $\sim$100 by treating the earliest part of the waveform at $\sim$$1/100$ of the full sample rate. Together, the factor of 100 reduction in the number of filter coefficients and the factor of 100 reduction in the sample rate during the early inspiral save a factor of $\sim$$10^4$ \ac{FLOPS} over the original (full sample rate) templates.

However, the resulting filters are still not orthogonal across the parameter space and are in fact highly redundant. We use the \ac{SVD} to approximate the template bank by a set of orthogonal \emph{basis filters}~\citep{Cannon:2010p10398}. We find that this approximation reduces the number of filters needed by another factor of $\sim$100. These two transformations combined reduce the number of floating point operations to a level that is competitive with the conventional high-latency \ac{FD}-matched filter approach. In the remainder of this section we describe the \ac{LLOID} algorithm in detail and provide some basic computational cost scaling.

\subsubsection{Selectively reducing the sample rate of the data and templates}
\label{sec:time-slices}

The first step of our proposed method is to divide the templates into time slices in a \ac{TD} analog to the \ac{FD} decomposition employed by \ac{MBTA}~\citep{Marion2004,Buskulic2010}. The application to GW data analysis is foreshadowed by an earlier \ac{FD} convolution algorithm, proposed by \citet{gardner1995efficient}, based on splitting the impulse response of a filter into smaller blocks. We decompose each template $h_{i}[k]$ into a sum of $S$ non-overlapping templates,
\begin{equation}
\label{eq:time-slices}
h_{i}[k] = \sum_{s=0}^{S-1}
	\begin{cases}
		h_i^s[k] & \textrm{if } t^s \leq k / f^0 < t^{s+1} \\
		0 & \textrm{otherwise},
	\end{cases}
\end{equation}
for $S$ integers $\{f^0 t^s\}$ such that $0 = f^0 t^0 < f^0 t^1 < \cdots < f^0 t^S = N$. The outputs of these new time-sliced filters form an ensemble of partial \ac{SNR} streams. By linearity of the filtering process, these partial \ac{SNR} streams can be summed to reproduce the \ac{SNR} of the full template.

Since waveforms with neighboring intrinsic source parameters $\theta$ have similar time-frequency evolution, it is possible to design computationally efficient time slices for an extended region of parameter space rather than to design different time slices for each template.

For concreteness and simplicity, consider an inspiral waveform in the quadrupole approximation, for which the time-frequency relation is given by Equation~(\ref{eq:fgw}). This monotonic time-frequency relationship allows us to choose time slice boundaries that require substantially less bandwidth at early times in the inspiral.

An inspiral signal will enter the detection band with some low frequency $f_\mathrm{low}$ at time $t_\mathrm{low}$ before merger. Usually the template is truncated at some prescribed time $t^0$, or equivalent frequency $f_\mathrm{high}$, often chosen to correspond to the \ac{LSO}. The beginning of the template is critically sampled at $2 f_\mathrm{low}$, but the end of the template is critically sampled at a rate of $2 f_\mathrm{high}$. In any time interval smaller than the duration of the template, the bandwidth of the filters across the entire template bank can be significantly less than the full sample rate at which data are acquired.

Our goal is to reduce the filtering cost of a large fraction of the waveform by computing part of the convolution at a lower sample rate. Specifically we consider here time slice boundaries with the smallest power-of-two sample rates that sub-critically sample the time-sliced templates. The time slices consist of the $S$ intervals $\left[t^0, t^1\right),\, \left[t^1, t^2\right),\, \dots,\, \left[t^{S-1}, t^S\right)$, sampled at frequencies $f^0,\, f^1,\, \dots,\, f^\mathrm{S-1}$, where $f^s$ is at least twice the highest nonzero frequency component of any filter in the bank for the $s$th time slice.

The time-sliced templates can then be downsampled in each interval without aliasing, so we define them as
\begin{equation}
\label{eq:time-sliced-templates}
h_{i}^{s}[k] \equiv
	\begin{cases}
		h_{i}\!\left[k\frac{f}{f^s}\right] & \textrm{if } t^s \leq k/f^s < t^{s+1} \\
		0 & \textrm{otherwise.}
	\end{cases}
\end{equation}
We note that the time slice decomposition in Equation~(\ref{eq:time-slices}) is manifestly orthogonal since the time slices are disjoint in time. In the next section, we examine how to reduce the number of filters within each time slice via \ac{SVD} of the time-sliced templates.

\subsubsection{Reducing the number of filters with the \acs{SVD}}
\label{sec:svd}

As noted previously, the template banks used in inspiral searches are by design highly correlated. \citet{Cannon:2010p10398} showed that applying the \ac{SVD} to inspiral template banks greatly reduces the number of filters required to achieve a particular minimal match. A similar technique can be applied to the time-sliced templates as defined in Equation~(\ref{eq:time-sliced-templates}) above. The \ac{SVD} is a matrix factorization that takes the form
\begin{equation}
h_i^s[k] = \sum_{\mathclap{l=0}}^{\mathclap{M-1}} v_{il}^s \sigma_l^s u_l^s[k] \approx \sum_{\mathclap{l=0}}^{\mathclap{L^s-1}} v_{il}^s \sigma_l^s u_l^s[k],
\label{eq:svddecomp}
\end{equation}
where $u_l^s[k]$ are orthonormal \emph{basis templates} related to the original
time-sliced templates through the \emph{reconstruction matrix}, $v_{il}^s\sigma_l^s$. The expectation value of the fractional loss in \ac{SNR} is the \ac{SVD} tolerance, given by
\begin{equation*}
\left[ \sum_{l=0}^{L^s-1} \left( \sigma_l^s \right)^2 \right]\left[ \sum_{l=0}^{M-1} \left( \sigma_l^s \right)^2 \right]^{-1},
\end{equation*}
determined by the number $\numsvdtmps$ of basis templates that are kept in the approximation. \citet{Cannon:2010p10398} showed that highly accurate approximations of inspiral template banks could be achieved with few basis templates. We find that when combined with the time slice decomposition, the number of basis templates \numsvdtmps\ is much smaller than the original number of templates \numtmps\ and improves on the rank reduction demonstrated in \citet{Cannon:2010p10398} by nearly an order of magnitude.

Because the sets of filters from each time slice form orthogonal subspaces, and the basis filters within a given time slice are mutually orthogonal, the set of all basis filters from all time slices forms an orthogonal basis spanning the original templates.

In the next section, we describe how we form our early-warning detection statistic using the time slice decomposition and the \ac{SVD}.

\subsubsection{Early-warning output}

In the previous two sections, we described two transformations that greatly reduce the computational burden of \ac{TD} filtering. We are now prepared to define our detection statistic, the early-warning output, and to comment on the computational cost of evaluating it.

First, the sample rate of the detector data must be decimated to match sample rates with each of the time slices. We will denote the decimated detector data streams using a superscript ``$s$'' to indicate the time slices to which they correspond. The operator $H^\shortdownarrow$ will represent the appropriate decimation filter that converts between the base sample rate $f^0$ and the reduced sample rate $f^s$:
\begin{equation*}
\label{eq:decomp}
	y^{s}[k] = \left( H^\shortdownarrow y^0\right)[k].
\end{equation*}
We shall use the symbol $H^\shortuparrow$ to represent an interpolation filter that converts between sample rates $f^{s+1}$ and $f^s$ of adjacent time slices,
\begin{equation*}
	y^{s}[k] = \left( H^\shortuparrow y^{s+1}\right)[k].
\end{equation*}

From the combination of the time slice decomposition in Equation~(\ref{eq:time-sliced-templates}) and the \ac{SVD} defined in Equation~(\ref{eq:svddecomp}), we define the early-warning output accumulated up to time slice $s$ using the recurrence relation,
%
%
\begin{equation}
	z_i^s [k] =%
		\overbrace{
			\left(H^\uparrow \rho_i^{s+1}\right)[k]
		}^\textrm{\clap{S/N from previous time slices}}
		+
		\underbrace{
			\sum_{\mathclap{l=0}}^{\mathclap{L^s-1}} v_{il}^s \sigma_l^s
		}_\textrm{\clap{reconstruction}}
		\overbrace{
			\sum_{\mathclap{n=0}}^{\mathclap{N^s-1}} u_l^s[n] y^s[k-n]
		}^\textrm{\clap{orthogonal {\sc fir} filters}} .
\end{equation}
Observe that the early-warning output for time slice 0, $z_i^0[k]$, approximates the \ac{SNR} of the original templates.
\begin{figure*}[h!]
	\begin{center}
		\includegraphics[width=\textwidth]{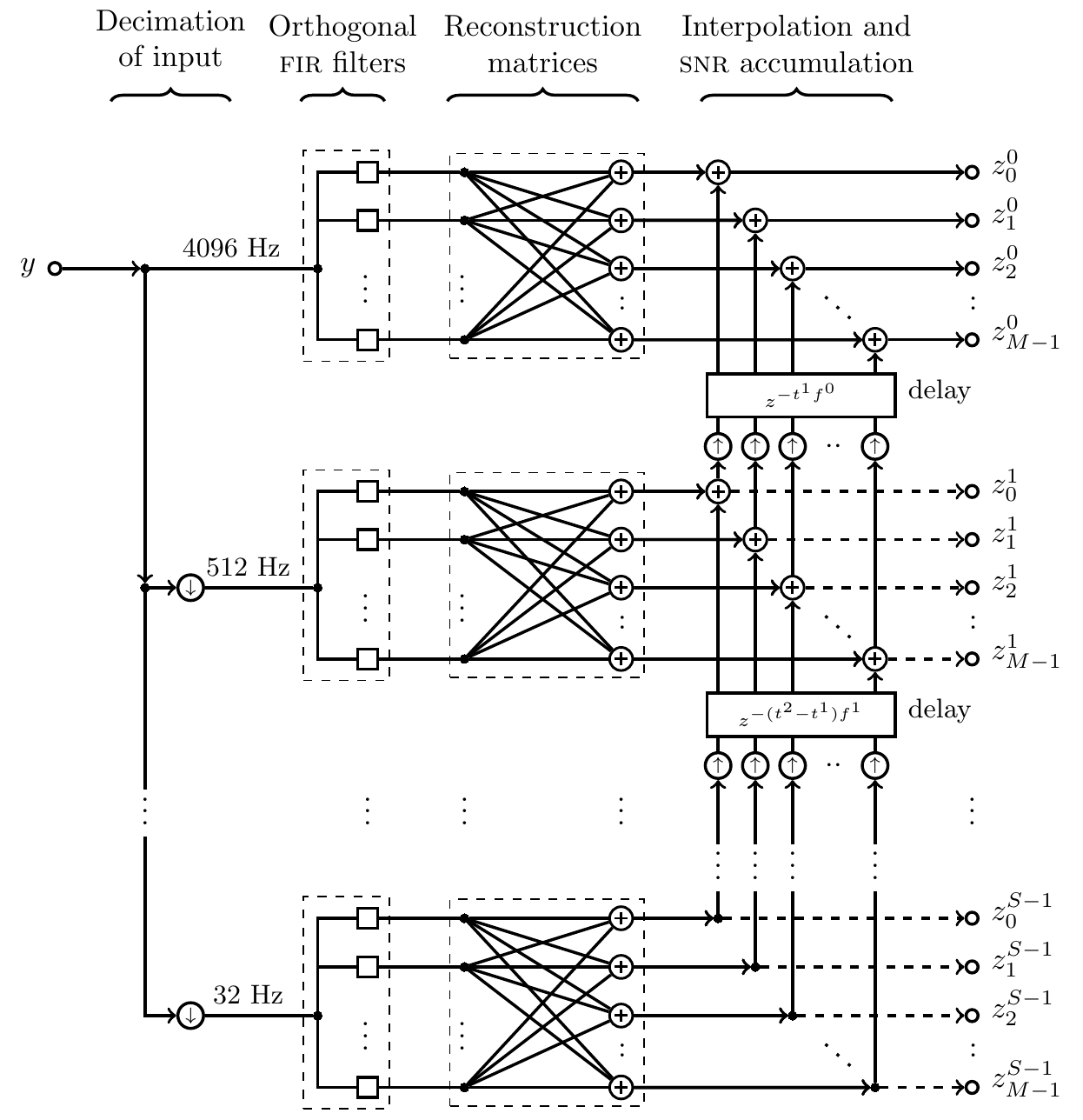}
		\caption[Schematic of \acs{LLOID} pipeline]{\label{fig:pipeline} Schematic of \ac{LLOID} pipeline illustrating signal flow. Circles with arrows represent interpolation \protect\includegraphics{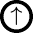} or decimation \protect\includegraphics{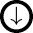}. Circles with plus signs represent summing junctions \protect\includegraphics{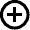}. Squares \protect\includegraphics{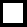} stand for \ac{FIR} filters. Sample rate decreases from the top of the diagram to the bottom. In this diagram, each time slice contains three \ac{FIR} filters that are linearly combined to produce four output channels. In a typical pipeline, the number of \ac{FIR} filters is much less than the number of output channels.}
	\end{center}
\end{figure*}
The signal flow diagram in Figure~\ref{fig:pipeline} illustrates this recursion relation as a multirate filter network with a number of early-warning outputs.

Ultimately, the latency of the entire \ac{LLOID} algorithm is set by the decimation and interpolation filters because they are generally time symmetric and slightly acausal. Fortunately, as long as the latency introduced by the decimation and interpolation filters for any time slice $s$ is less than that time slice's delay $t^s$, the total latency of the \ac{LLOID} algorithm will be zero. To be concrete, suppose that the first time slice, sampled at a rate $f^0 = 4096$~Hz, spans times $[t^0,\,t^1) = [0\,\textrm{s},\,0.5\,\textrm{s})$, and the second time slice, sampled at $f^1 = 512$\,Hz, spans $[t^1,\,t^2) = [0.5\,\textrm{s},\,4.5~\textrm{s})$. Then the second time slice's output, $z_i^1[k]$, will \emph{lead} the first time slice's output, $z_i^0[k]$, by 0.5~s. A decimation filter will be necessary to convert the 4096~Hz input signal $y[k] \equiv y^0[k]$ to the 512~Hz input $y^1[k]$, and an interpolation filter will be necessary to match the sample rates of the two early-warning outputs. In this example, as long as the decimation and interpolation filters are together acausal by less than $t^1 = 0.5$~s, the total \ac{SNR} $z_i^0[k]$ will be available with a latency of zero samples. When zero latency is important, we may take this as a requirement for the decimation and interpolation filter kernels.

In the next section, we compute the expected computational cost scaling of this decomposition and compare it with the direct \ac{TD} implementation of Equation~\eqref{eq:SNRTD} and higher latency blockwise \ac{FD} methods.

\subsection{Comparison of computational costs}

We now examine the computational cost scaling of the conventional \ac{TD} or \ac{FD} matched filter procedure as compared with \ac{LLOID}. For convenience, Table~\ref{tab:recap} provides a review of the notation that we will need in this section.
%
%
\begin{table}
\caption[Notation used to describe filters]{\label{tab:recap}Notation used to describe filters.}
\begin{center}
\begin{tabular}{ll}
\tableline\tableline
& Definition \\
\tableline
$f^s$		& Sample rate in time slice $s$ \\
\numtmps		& Number of templates \\
\tmpsamps	& Number of samples per template \\
\numslices	& Number of time slices \\
\numsvdtmps	& Number of basis templates in time slice $s$ \\
\slicessamps	& Number of samples in decimated time slice $s$\\
$N^\shortdownarrow$ & Length of decimation filter \\
$N^\shortuparrow$ & Length of interpolation filter \\
\tableline
\end{tabular}
\end{center}
\end{table}

\subsubsection{Conventional \acs{TD} method}

The conventional, direct \ac{TD} method consists of a bank of \ac{FIR} filters, or sliding-window dot products. If there are $\numtmps$ templates, each $\tmpsamps$ samples in length, then each filter requires $M N$ multiplications and additions per sample, or, at a sample rate $f^0$,
\begin{equation}
	\label{eq:td-flops}
	2 \numtmps \tmpsamps f^0 \textrm{ \ac{FLOPS}}.
\end{equation}

\subsubsection{Conventional \acs{FD} method}

The most common \ac{FD} method is known as the overlap-save algorithm, described in \citet{numerical-recipes-chapter-13}. It entails splitting the input into blocks of $D$ samples, $D > \tmpsamps$, each block overlapping the previous one by $D - \tmpsamps$ samples. For each block, the algorithm computes the forward \ac{FFT} of the data and each of the templates, multiplies them, and then computes the reverse \ac{FFT}.

Modern implementations of the \ac{FFT}, such as the ubiquitous \texttt{fftw}, require about $2 \fftblock \lg \fftblock$ operations to evaluate a real transform of size $\fftblock$~\citep{Johnson:2007p9654}. Including the forward transform of the data and $M$ reverse transforms for each of the templates, the \ac{FFT} costs $2 (\numtmps + 1) \fftblock \lg \fftblock$ operations per block. The multiplication of the transforms adds a further $2 \numtmps \fftblock$ operations per block. Since each block produces $\fftblock - \tmpsamps$ usable samples of output, the overlap-save method requires
\begin{equation}
	\label{eq:fd-flops}
	f^0 \cdot \frac{2 (\numtmps + 1) \lg \fftblock + 2 \numtmps}{1 - \tmpsamps/\fftblock} \textrm{ \ac{FLOPS}}.
\end{equation}

In the limit of many templates, $M \gg 1$, we can neglect the cost of the forward transform of the data and of the multiplication of the transforms. The computational cost will reach an optimum at some large but finite \ac{FFT} block size $\fftblock \gg \tmpsamps$. In this limit, the \ac{FD} method costs $\approx 2 f^0 \numtmps \lg \fftblock$ \ac{FLOPS}.

By adjusting the \ac{FFT} block size, it is possible to achieve low latency with FD convolution, but the computational cost grows rapidly as the latency in samples $(D-N$) decreases. It is easy to show that in the limit of many templates and long templates, $M, \lg N \gg 1$, the computational cost scales as
$$
\left(1 + \frac{\textrm{template length}}{\textrm{latency}}\right) \left( 2 f^0 M \lg N \right).
$$

\subsubsection{\label{sec:lloid-method}\acs{LLOID} method}

For time slice $s$, the \ac{LLOID} method requires $2 \slicessamps \numsvdtmps f^s$ \ac{FLOPS} to evaluate the orthogonal filters, $2 \numtmps \numsvdtmps f^s$ \ac{FLOPS} to apply the linear transformation from the $\numsvdtmps$ basis templates to the $\numtmps$ time-sliced templates, and $\numtmps f^s$ \ac{FLOPS} to add the resultant partial \ac{SNR} stream.

The computational cost of the decimation of the detector data is a little bit more subtle. Decimation is achieved by applying an \ac{FIR} anti-aliasing filter and then downsampling, or deleting samples in order to reduce the sample rate from $f^{s-1}$ to $f^s$. Naively, an anti-aliasing filter with $(f^{s-1} / f^s) N^\shortdownarrow$ coefficients should demand $2 N^\shortdownarrow (f^{s-1})^2 / f^s$ \ac{FLOPS}. However, it is necessary to evaluate the anti-aliasing filter only for the fraction $f^s / f^{s-1}$ of the samples that will not be deleted. Consequently, an efficient decimator requires only $2 N^\shortdownarrow f^{s-1}$ \ac{FLOPS}. (One common realization is an ingenious structure called a \emph{polyphase decimator}, described in Chapter 1 of \citet{jovanovic2002multirate}.)

The story is similar for the interpolation filters used to match the sample rates of the partial \ac{SNR} streams. Interpolation of a data stream from a sample rate $f^s$ to $f^{s-1}$ consists of inserting zeros between the samples of the original stream, and then applying a low-pass filter with $(f^{s-1} / f^s) N^\shortuparrow$ coefficients. The low-pass filter requires $2 M N^\shortuparrow (f^{s-1})^2 / f^s$ \ac{FLOPS}. However, by taking advantage of the fact that by construction a fraction, $f^s / f^{s-1}$, of the samples are zero, it is possible to build an efficient interpolator that requires only $2 M N^\shortuparrow f^{s-1}$ \ac{FLOPS}. (Again, see \citet{jovanovic2002multirate} for a discussion of \emph{polyphase interpolation}.)

Taking into account the decimation of the detector data, the orthogonal \ac{FIR} filters, the reconstruction of the time-sliced templates, the interpolation of \ac{SNR} from previous time slices, and the accumulation of \ac{SNR}, in total the \ac{LLOID} algorithm requires
\begin{equation}
\label{eq:lloid-flops}
\sum_{\mathclap{s=0}}^{\mathclap{S-1}} \left( 2 \slicessamps \numsvdtmps + 2 \numtmps \numsvdtmps + \numtmps \right) f^s + 2\sum_{\mathclap{f^s \in \{f^k \, : \, 0 < k < S\}}} \left( N^\shortdownarrow f^0 + \numtmps N^\shortuparrow f^{s-1}\right)
\end{equation}
\ac{FLOPS}. The second sum is carried out over the set of distinct sample rates (except for the base sample rate) rather than over the time slices themselves, as we have found that it is sometimes desirable to place multiple adjacent time slices at the same sample rate in order to keep the size of the matrices that enter the \ac{SVD} manageable. Here we have assumed that the decimation filters are connected in parallel, converting from the base sample rate $f^0$ to each of the time slice sample rates $f^1$, $f^2$, $\dots$, and that the interpolation filters are connected in cascade fashion with each interpolation filter stepping from the sample rate of one time slice to the next.

We can simplify this expression quite a bit by taking some limits that arise from sensible filter design. In the limit of many templates, the cost of the decimation filters is negligible as compared to the cost of the interpolation filters. Typically, we will design the interpolation filters with $N^\shortuparrow \lesssim \numsvdtmps$ so that the interpolation cost itself is negligible compared with the reconstruction cost. Finally, if the number of basis templates per time slice $\numsvdtmps$ is not too small, the reconstruction cost dominates over the cost of accumulating the partial \ac{SNR}. In these limits, the cost of \ac{LLOID} is dominated by the basis filters themselves and the reconstruction, totaling $2 \sum_{s=0}^{S-1} f^s \numsvdtmps \left( \slicessamps + \numtmps \right)$ \ac{FLOPS}.

\subsubsection{Speedup of \acs{LLOID} relative to \acs{TD} method}

If the cost of the basis filters dominates, and the frequency of the templates evolves slowly enough in time, then we can use the time-frequency relationship of Equation~(\ref{eq:fgw}) to estimate the speedup relative to the conventional, direct \ac{TD} method. The reduction in \ac{FLOPS} is approximately
\begin{equation}
\label{eq:speedup}
\frac{2 \sum_{s=0}^{S-1} f^s \numsvdtmps \slicessamps}{2 \numtmps \tmpsamps f^0}
\approx \frac{\alpha}{\left(t_\mathrm{low} - t_\mathrm{high}\right) \left(f^0\right)^2} \int_{t_\mathrm{low}}^{t_\mathrm{high}} \left(2 f(t) \right)^2 \, \mathrm{d} t
= \frac{16 \alpha \left(t_\mathrm{low} f^2 (t_\mathrm{low}) - t_\mathrm{high} f^2 (t_\mathrm{high}) \right)}{\left(f^0\right)^2 \left(t_\mathrm{low} - t_\mathrm{high}\right)},
\end{equation}
where $\alpha \approx \numsvdtmps / \numtmps$ is the rank reduction factor, or ratio between the number of basis templates and the number of templates. This approximation assumes that the frequency of the signal is evolving very slowly so that we can approximate the time slice sample rate as twice the instantaneous GW frequency, $f^s \approx 2 f(t)$, and the number of samples in the decimated time slice as the sample rate times an infinitesimally short time interval, $\slicessamps \approx 2 f(t) \, \mathrm{d}t$. The integral is evaluated using the power-law form of $f(t)$ from Equation~(\ref{eq:fgw}). Substituting approximate values for a template bank designed for component masses around (1.4, 1.4) $M_\odot$, $\alpha \approx 10^{-2}$, $t_\mathrm{low} = 10^3$~s, $f_\mathrm{low} = 10^1$~Hz, $f_\mathrm{high} = f_\mathrm{ISCO} \approx 1570$~Hz, $f^0 = 2 f_\mathrm{ISCO}$, and $t_\mathrm{high} = {f_\mathrm{ISCO}}^{-1}$, we find from Equation~(\ref{eq:speedup}) that the \ac{LLOID} method requires only $\sim 10^{-6}$ times as many \ac{FLOPS} as the conventional \ac{TD} method.

\section{Implementation}

In this section we describe an implementation of the \ac{LLOID} method described in Section \ref{sec:method} suitable for rapid \ac{GW} searches for \acp{CBC}. The \ac{LLOID} method requires several computations that can be completed before the analysis is underway. Thus, we divide the procedure into an offline planning stage and an online, low-latency filtering stage. The offline stage can be done before the analysis is started and updated asynchronously, whereas the online stage must keep up with the detector output and produce search results as rapidly as possible. In the next two subsections we describe what these stages entail.

\subsection{Planning stage}

The planning stage begins with choosing templates that cover the space of source parameters with a hexagonal grid~\citep{PhysRevD.76.102004} in order to satisfy a minimal match criterion. This assures a prescribed maximum loss in \ac{SNR} for signals whose parameters do not lie on the hexagonal grid. Next, the grid is partitioned into groups of neighbors called \emph{sub-banks} that are appropriately sized so that each sub-bank can be efficiently handled by a single computer. Each sub-bank contains templates of comparable chirp mass, and therefore similar time-frequency evolution. Dividing the source parameter space into smaller sub-banks also reduces the offline cost of the \ac{SVD} and is the approach considered in \citet{Cannon:2010p10398}. Next, we choose time slice boundaries as in Equation~\eqref{eq:time-sliced-templates} such that all of the templates within a sub-bank are sub-critically sampled at progressively lower sample rates. For each time slice, the templates are downsampled to the appropriate sample rate. Finally, the \ac{SVD} is applied to each time slice in the sub-bank in order to produce a set of orthonormal basis templates and a reconstruction matrix that maps them back to the original templates as described in Equation~\eqref{eq:svddecomp}. The downsampled basis templates, the reconstruction matrix, and the time slice boundaries are all saved to disk.

\subsection{Filtering stage}

The \ac{LLOID} algorithm is amenable to latency-free, real-time implementation. However, a real-time search pipeline would require integration directly into the data acquisition and storage systems of the \ac{LIGO} observatories. A slightly more modest goal is to leverage existing low latency, but not real-time, signal processing software in order to implement the \ac{LLOID} algorithm.

We have implemented a prototype of the low-latency filtering stage using an open-source signal processing environment called \gstreamer\footnote{\url{http://gstreamer.net/}} (version 0.10.33). \gstreamer\ is a vital component of many Linux systems, providing media playback, authoring, and streaming on devices from cell phones to desktop computers to streaming media servers. Given the similarities of \ac{GW} detector data to audio data it is not surprising that \gstreamer\ is useful for our purpose. \gstreamer\ also provides some useful stock signal processing elements such as resamplers and filters. We have extended the \gstreamer\ framework by developing a library called \gstlal\footnote{\url{https://www.lsc-group.phys.uwm.edu/daswg/projects/gstlal.html}} that provides elements for \ac{GW} data analysis.

\gstreamer\ pipelines typically operate with very low (in some consumer applications, imperceptibly low) latency rather than in true real time because signals are partitioned into blocks of samples, or \emph{buffers}. This affords a number of advantages, including amortizing the overhead of passing signals between elements and grouping together sequences of similar operations. However, buffering a signal incurs a latency of up to one buffer length. This latency can be made small at the cost of some additional overhead by making the buffers sufficiently small. In any case, buffering is a reasonable strategy for low-latency \ac{LIGO} data analysis because, as we previously remarked, the \ac{LIGO} data acquisition system has a granularity of $1/16$~s.

\section{Results}

In this section we evaluate the accuracy of the \ac{LLOID} algorithm using our 
\gstreamer{}-based implementation described in the previous section. We calculate the measured \ac{SNR} loss due to the approximations of the \ac{LLOID} method and our implementation of it. Using a configuration that gives acceptable \ac{SNR} loss for our chosen set of source parameters, we then compare the computational cost in \ac{FLOPS} for the direct \ac{TD} method, the overlap-save \ac{FD} method, and \ac{LLOID}.

\subsection{Setup}
\label{sec:bank-setup}

\begin{figure}
    \centering
	\includegraphics[width=0.75\columnwidth]{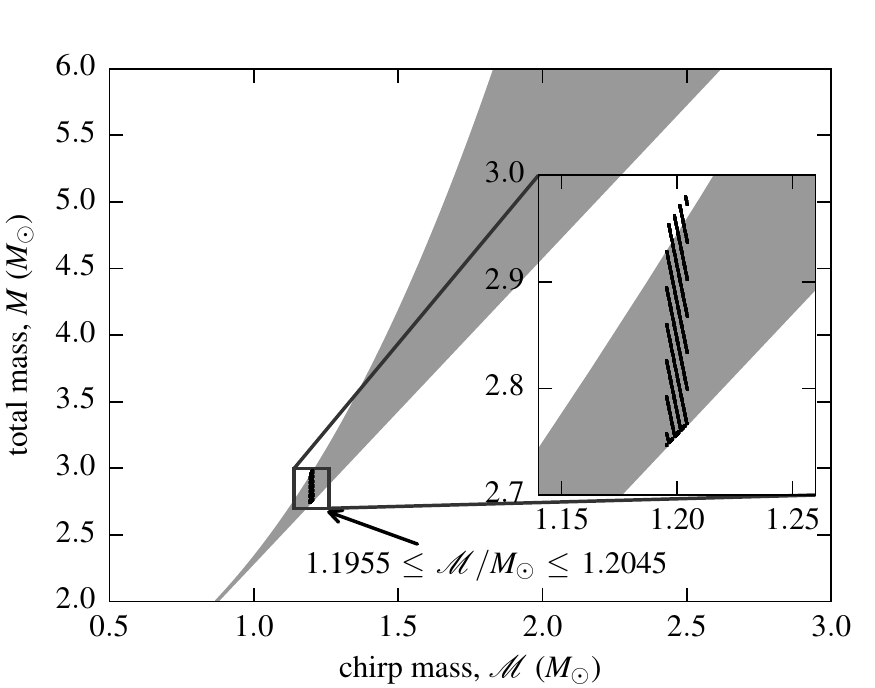}
	\caption[Mass parameters of sub-bank]{\label{fig:tmpltbank}Source parameters selected for sub-bank used in this
case study, consisting of component masses $m_1$ and $m_2$, between 1 and 3~$M_\odot$, and
chirp masses $\mathcal{M}$ between 1.1955 and 1.2045~$M_\odot$.}
\end{figure}

We examine the performance of the \ac{LLOID} algorithm on a small region of compact binary parameter space centered on typical \ac{NS}--\ac{NS} masses.  We begin by constructing a template bank that spans component masses from 1~to~3~$M_\odot$ using a simulated Advanced \ac{LIGO} noise power spectrum~\citep{ALIGONoise}\footnote{\url{http://dcc.ligo.org/cgi-bin/DocDB/ShowDocument?docid=T0900288&version=3}}.  Waveforms are generated in the frequency domain in the stationary phase approximation at (post)$^{3.5}$-Newtonian order in phase and Newtonian order in amplitude \citep[the TaylorF2 waveforms described in][]{TaylorF2}.  Templates are truncated at 10~Hz, where the projected sensitivity of Advanced \ac{LIGO} is interrupted by the ``seismic wall.'' This results in a grid of 98\,544 points, or $2 \times 98\,544 = 197\,088$~templates.  Then we create sub-banks by partitioning the parameter space by chirp mass.  Figure \ref{fig:tmpltbank} illustrates this procedure. We concentrate on a sub-bank with 657 points with chirp masses between 1.1955 and 1.2045~$M_\odot$, or $2 \times 657 = 1314$~templates. With this sub-bank we are able to construct an efficient time slice decomposition that consists of 11 time slices with sample rates of 32\nobreakdashes--4096~Hz, summarized in Table~\ref{tab:time_slices}. We use this sub-bank and decomposition for the remainder of this section.

\begin{landscape}
\begin{table*}
\caption[Filter design for a sub-bank of 1314 templates]{\label{tab:time_slices}Filter design for a sub-bank of 1314 templates.}
\begin{center}
\begin{tabular}{crr@{,\,}lc*{6}{r}}
\tableline\t
ableline
\\ [-1.5ex]
& $f^s$ & $[t^s$&$t^{s+1})$ & &\multicolumn{6}{c}{$-\log_{10}$ (1$-$\ac{SVD} tolerance)} \\
\cline{6-11}
\\[-2ex]
& (Hz) & \multicolumn{2}{c}{(s)} & $N^s$ & $1$ & $2$ & $3$ & $4$ & $5$ & $6$ \\ \tableline\\
\multirow{11}{*}{\includegraphics{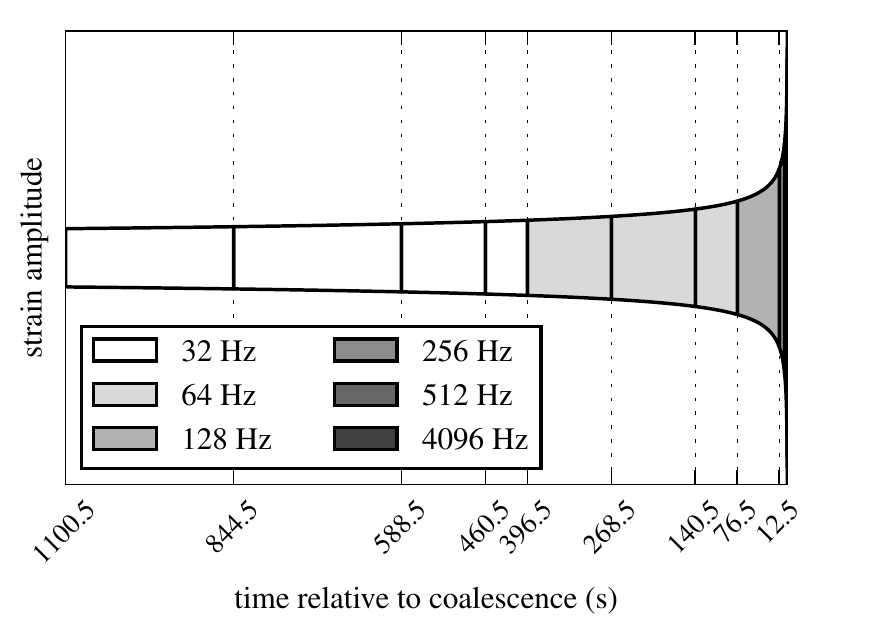}} & 4096 & [0&0.5) & 2048 & 1 & 4 & 6 & 8 & 10 & 14 \\[0.5em]
& 512 & [0.5&4.5) & 2048 & 2 & 6 & 8 & 10 & 12 & 16 \\[0.5em]
& 256 & [4.5&12.5) & 2048 & 2 & 6 & 8 & 10 & 12 & 15 \\[0.5em]
& 128 & [12.5&76.5) & 8192 & 6 & 20 & 25 & 28 & 30 & 32 \\[0.5em]
& 64 & [76.5&140.5) & 4096 & 1 & 8 & 15 & 18 & 20 & 22 \\[0.5em]
& 64 & [140.5&268.5) & 8192 & 1 & 7 & 21 & 25 & 28 & 30 \\[0.5em]
& 64 & [268.5&396.5) & 8192 & 1 & 1 & 15 & 20 & 23 & 25 \\[0.5em]
& 32 & [396.5&460.5) & 2048 & 1 & 1 & 3 & 9 & 12 & 14 \\[0.5em]
& 32 & [460.5&588.5) & 4096 & 1 & 1 & 7 & 16 & 18 & 21 \\[0.5em]
& 32 & [588.5&844.5) & 8192 & 1 & 1 & 8 & 26 & 30 & 33 \\[0.5em]
& 32 & [844.5&1100.5) & 8192 & 1 & 1 & 1 & 12 & 20 & 23 \\[0.5em]
\tableline
\end{tabular}
\tablecomments{From
left to right, this table shows the sample rate, time interval, number of
samples, and number of orthogonal templates for each time slice.  We vary \ac{SVD}
tolerance from $\left(1-10^{-1}\right)$ to $\left(1-10^{-6}\right)$.}
\end{center}
\end{table*}
\end{landscape}

\subsection{Measured \acs{SNR} loss}

The \ac{SNR} loss is to be compared with the mismatch of 0.03 that arises from the discreteness of the template bank designed for a minimal match of 0.97. We will consider an acceptable target \ac{SNR} loss to be a factor of 10 smaller than this, that is, no more than 0.003.

We expect two main contributions to the \ac{SNR} loss to arise in our implementation of the \ac{LLOID} algorithm. The first is the \ac{SNR} loss due to the truncation of the \ac{SVD} at $L^s < M$ basis templates. As remarked upon in \citet{Cannon:2010p10398} and Section~\ref{sec:svd}, this effect is measured by the \ac{SVD} tolerance. The second comes from the limited bandwidth of the interpolation filters used to match the sample rates of the partial \ac{SNR} streams. The maximum possible bandwidth is determined by the length of the filter, $N^\shortuparrow$. \ac{SNR} loss could also arise if the combination of both the decimation filters and the interpolation filters reduces their bandwidth measurably, if the decimation and interpolation filters do not have perfectly uniform phase response, or if there is an unintended subsample time delay at any stage.

To measure the accuracy of our \gstreamer\ implementation of \ac{LLOID} including all of the above potential sources of \ac{SNR} loss, we conducted impulse response tests. The \gstreamer\ pipeline was presented with an input consisting of a unit impulse. By recording the outputs, we can effectively ``play back'' the templates. These impulse responses will be similar, but not identical, to the original, nominal templates. By taking the inner product between the impulse responses for each output channel with the corresponding nominal template, we can gauge exactly how much \ac{SNR} is lost due to the approximations in the \ac{LLOID} algorithm and any of the technical imperfections mentioned above. We call one minus this dot product the \emph{mismatch} relative to the nominal template.

The two adjustable parameters that affect performance and mismatch the most are the \ac{SVD} tolerance and the length of the interpolation filter. The length of the decimation filter affects mismatch as well, but has very little impact on performance.

\paragraph{Effect of \acs{SVD} tolerance}

\begin{figure*}[b]
	\begin{minipage}[t]{0.5\textwidth}
		\begin{center}
			\includegraphics[width=\textwidth]{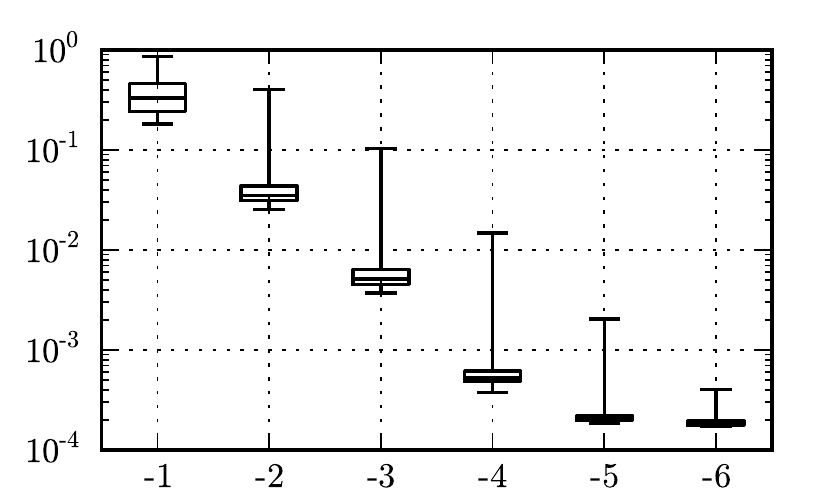}\\
			(a) Mismatch versus \ac{SVD} tolerance
		\end{center}
	\end{minipage}
	\begin{minipage}[t]{0.5\textwidth}
		\begin{center}
			\includegraphics[width=\textwidth]{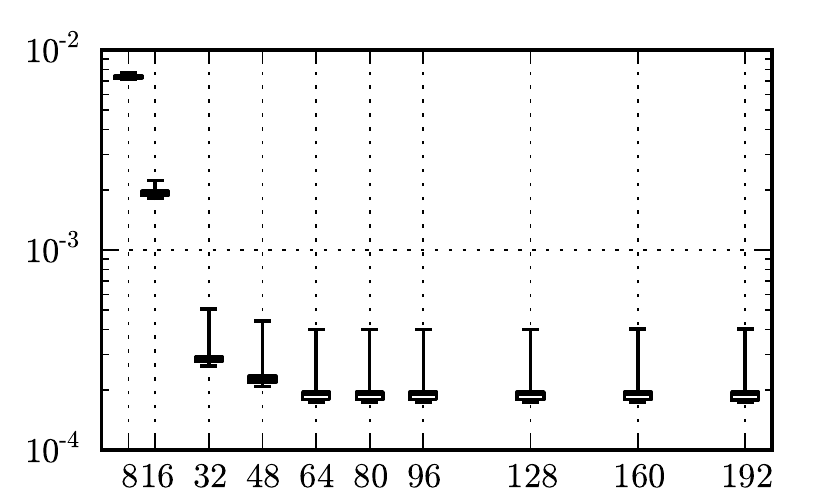}\\
			(b) Mismatch versus $N^\shortuparrow$
		\end{center}
	\end{minipage}
	\caption[Mismatch between \acs{LLOID} filter response and nominal templates]{\label{fig:bw}Box-and-whisker plot of mismatch between nominal template bank and \ac{LLOID} measured impulse responses. The upper and lower boundaries of the boxes show the upper and lower quartiles; the lines in the center denote the medians. The whiskers represent the minimum and maximum mismatches over all templates. In (a) the interpolation filter length is held fixed at $N^\shortuparrow = 192$, while the \ac{SVD} tolerance is varied from $\left(1-10^{-1}\right)$ to $\left(1-10^{-6}\right)$. In (b), the \ac{SVD} tolerance is fixed at $\left(1-10^{-6}\right)$ while $N^\shortuparrow$ is varied from 8 to 192.}
\end{figure*}

We studied how the \ac{SVD} tolerance affected \ac{SNR} loss by holding $N^\shortdownarrow = N^\shortuparrow = 192$ fixed as we varied the \ac{SVD} tolerance from $\left(1-10^{-1}\right)$ to $\left(1-10^{-6}\right)$. The minima, maxima, and median mismatches are shown as functions of \ac{SVD} tolerance in Figure~\ref{fig:bw}(a). As the \ac{SVD} tolerance increases toward 1, the \ac{SVD} becomes an exact matrix factorization, but the computational cost increases as the number of basis filters increases. The conditions presented here are more complicated than in the original work~\citep{Cannon:2010p10398} due to the inclusion of the time-sliced templates and interpolation, though we still see that the average mismatch is approximately proportional to the \ac{SVD} tolerance down to $\left(1-10^{-4}\right)$. However, as the \ac{SVD} tolerance becomes even higher, the median mismatch seems to saturate at around $2 \times 10^{-4}$. This could be the effect of the interpolation, or an unintended technical imperfection that we did not model or expect. However, this is still an order of magnitude below our target mismatch of 0.003. We find that an \ac{SVD} tolerance of $\left(1-10^{-4}\right)$ is adequate to achieve our target \ac{SNR} loss.

\paragraph{Effect of interpolation filter length}

Next, keeping the \ac{SVD} tolerance fixed at $\left(1-10^{-6}\right)$ and the length of the decimation filter fixed at $N^\shortdownarrow = 192$, we studied the impact of the length $N^\shortuparrow$ of the interpolation filter on mismatch. We use GStreamer's stock \texttt{audioresample} element, which provides an FIR decimation filter with a Kaiser-windowed sinc function kernel. The mismatch as a function of $N^\shortuparrow$ is shown in Figure~\ref{fig:bw}(b). The mismatch saturates at $\sim$$2 \times 10^{-4}$ with $N^\shortuparrow = 64$. We find that a filter length of 16 is sufficient to meet our target mismatch of 0.003.

Having selected an \ac{SVD} tolerance of $\left(1-10^{-4}\right)$ and $N^\shortuparrow=16$, we found that we could reduce $N^\shortdownarrow$ to 48 without exceeding a median mismatch of~0.003.

We found that careful design of the decimation and interpolation stages made a crucial difference in terms of computational overhead. Connecting the interpolation filters in cascade fashion rather than in parallel resulted in a significant speedup. Also, only the shortest interpolation filters that met our maximum mismatch constraint resulted in a sub-dominant contribution to the overall cost. There is possibly further room for optimization beyond minimizing $N^\shortuparrow$. We could design custom decimation and interpolation filters, or we could tune these filters separately for each time slice.

\subsection{Other potential sources of \acs{SNR} loss}

One possible source of \ac{SNR} loss for which we have not accounted is the leakage of sharp spectral features in the detector's noise spectrum due to the short durations of the time slices.  In the \ac{LLOID} algorithm, as with many other \ac{GW} search methods, whitening is treated as an entirely separate data conditioning stage.  In this chapter, we assume that the input to the filter bank is already whitened, having been passed through a filter that flattens and normalizes its spectrum.  We elected to omit a detailed description of the whitening procedure since the focus here is on the implementation of a scalable inspiral filter bank.

However, the inspiral templates themselves consist of the GW time series convolved with the impulse response of the whitening filter.  As a consequence, the \ac{LLOID} algorithm must faithfully replicate the effect of the whitening filter.  Since in practice the noise spectra of ground-based \ac{GW} detectors contain both high-$Q$ lines at mechanical, electronic, and control resonances and a very sharp rolloff at the seismic wall, the frequency response of the \ac{LLOID} filter bank must contain both high-$Q$ notches and a very abrupt high-pass filter.  \ac{FIR} filters with rapidly varying frequency responses tend to have long impulse responses and many coefficients.  Since the \ac{LLOID} basis filters have, by design, short impulse responses and very few coefficients, one might be concerned about spectral leakage contaminating the frequency response of the \ac{LLOID} filter bank.

The usual statement of the famous Nyquist--Shannon theorem, stated below as Theorem~\ref{thm:nyquist}, has a natural dual, Theorem~\ref{thm:nyquist-dual}, that addresses the frequency resolution that can be achieved with an \ac{FIR} filter of a given length.

\begin{thm}
	\label{thm:nyquist}
	\citep[After][p. 518]{oppenheim1997signals}
	Let $x(t)$ be a band-limited signal with continuous Fourier transform $\tilde{x}(f)$ such that $\tilde{x}(f') = 0 \; \forall \; f' : |f'| > f_M$.  Then, $x(t)$ is uniquely determined by its discrete samples $x(n/f^0)$, $n = 0, \pm 1, \pm 2, \dots$, if $f^0 > 2 f_M$.
\end{thm}

\begin{thm}
	\label{thm:nyquist-dual}
	Let $x(t)$ be a compactly supported signal such that $x(t') = 0 \; \forall \; t' : |t'| > t_M$.  Then its continuous Fourier transform $\tilde{x}(f)$ is uniquely determined by the discrete frequency components $\tilde{x}(n \, \Delta f)$, $n = 0, \pm 1, \pm 2, \dots$, if $\Delta f < 1 / (2 t_M)$.
\end{thm}

Another way of stating Theorem~\ref{thm:nyquist-dual} is that, provided $x(t)$ is nonzero only for $|t| < 1 / (2 \, \Delta f)$, the continuous Fourier transform can be reconstructed at any frequency $f$ from a weighted sum of $\sinc$ functions centered at each of the discrete frequency components, namely,
$$
	\tilde{x}(f) \propto \sum_{n=-\infty}^{\infty} \tilde{x}\left(n \, \Delta f\right) \sinc \left[ \pi (f - n \, \Delta f) / \Delta f \right].
$$
Failure to meet the conditions of this dual of the sampling theorem results in spectral leakage.  For a \ac{TD} signal to capture spectral features that are the size of the central lobe of the $\sinc$ function, the signal must have a duration greater than $1/\Delta f$.  If the signal $x(t)$ is truncated by sampling it for a shorter duration, then its Fourier transform becomes smeared out; conceptually, power ``leaks'' out into the side lobes of the $\sinc$ functions and washes away sharp spectral features.  In the \ac{GW} data analysis literature, the synthesis of inspiral matched filters involves a step called \emph{inverse spectrum truncation} \citep[see][Section VII]{FairhurstTriangulation} that fixes the number of coefficients based on the desired frequency resolution.

In order to effectively flatten a line in the detector's noise power spectrum, the timescale of the templates must be at least as long as the damping time $\tau$ of the line, $\tau = 2 Q / \omega_0$, where $Q$ is the quality factor of the line and $w_0$ is the central angular frequency. To put this into the context of the sampling theorem, in order to resolve a notch with a particular $Q$ and $f_0$, an \ac{FIR} filter must achieve a frequency resolution of $\Delta f \gtrsim \pi f_0 / Q$ and therefore its impulse response must last for at least a time $1/\Delta f = Q / \pi f_0$. For example, in the S6 detector configuration known as ``Enhanced LIGO,'' the violin modes \citep{ELIGOSusp} had $Q \sim 10^5$ and $\omega_0 \sim (2 \pi) 340$~rad~s$^{-1}$, for a coherence time $\tau \sim 10^2$~s.

In our example template bank, many of the time slices are much shorter than this. However, in summation the time slices have the same duration as the full templates themselves, and the full templates are much longer than many coherence times of the violin mode. For this reason, we speculate that \ac{LLOID} should be just as robust to sharp line features as traditional \ac{FFT}-based searches currently employed in the \ac{GW} field. Future works must verify this reasonable supposition with numerical experiments, including impulse response studies similar to the ones presented here but with detector noise power spectra containing lines with realistically high quality factors.

There could, in principle, be lines with coherence times many times longer than the template duration. For example, the $Q$ of the violin modes may increase by orders of magnitude in Advanced \ac{LIGO}~\citep{ALIGOSusp}. Also, there are certainly narrow lines that are non-stationary. Both of these cases can be dealt with by preprocessing the data with bandstop filters that attenuate the lines themselves and also conservatively large neighborhoods around them. If such bandstops were implemented as an \ac{FIR} filter, they could be built into the time slices without any difficulty.

Another way to deal with line features with coherence times much longer than the templates would be to entirely `factor' the whitening out of the \ac{LLOID} filter bank.  Any line features could be notched out in the whitening stage with \ac{IIR} filters, which can achieve infinitely high $Q$ at just second order.  If the detector data were passed through the whitening filter twice, then time-sliced filters need not depend on the detector's noise power spectral density at all.  In such a variation on the \ac{LLOID} method, the basis filters could be calculated from the \emph{weighted} \ac{SVD}~\citep[Chapter 3.6]{WeightedSVD, jackson2003user} of the time-sliced templates, using the covariance of the detector noise as a weight matrix.

\subsection{Lower bounds on computational cost and latency compared to other methods}

We are now prepared to offer the estimated computational cost of filtering this sub-bank of templates compared to other methods. We used the results of the previous subsections to set the \ac{SVD} tolerance to $\left(1-10^{-4}\right)$, the interpolation filter length to 16, and the decimation filter length to 48. Table~\ref{table:flops} shows the computational cost in \ac{FLOPS} for the sub-bank we described above. For the overlap-save \ac{FD} method, an \ac{FFT} block size of $\fftblock = 2 \tmpsamps$ is assumed, resulting in a latency of $\left(\tmpsamps / f^0\right)$ seconds. Both the \ac{FD} method and \ac{LLOID} are five orders of magnitude faster than the conventional, direct \ac{TD} method. However, the \ac{FD} method has a latency of over half of an hour, whereas the \ac{LLOID} method, with suitable design of the decimation and interpolation filters, has no more latency than the direct \ac{TD} method.
\begin{table}
\caption[Computational cost of \acs{LLOID} versus conventional matched filter methods]{\label{table:flops}Computational cost in \ac{FLOPS} and latency in seconds of the direct \ac{TD} method, the overlap-save \ac{FD} method, and \ac{LLOID}.}
\begin{center}
\begin{tabular}{lllll}
\tableline\tableline
& \ac{FLOPS} & & \ac{FLOPS} & number of \\
Method & (Sub-bank) & Latency (s) & (\ac{NS}--\ac{NS}) & Machines \\[0.1em]
\tableline
Direct (\ac{TD}) & $4.9\times10^{13}$ & 0 & $3.8\times10^{15}$ & $\sim$$3.8\times10^5$ \\
Overlap-save (\ac{FD}) & $5.2\times10^8$ & $2\times10^3$ & $5.9\times10^{10}$ & $\sim$$5.9$ \\
\ac{LLOID} (theory) & $6.6\times10^8$ & 0 & $1.1 \times 10^{11}$ & $\sim$$11$ \\
\ac{LLOID} (prototype) & (0.9 cores) & $0.5$ & ------------ & $\gtrsim$$10$ \\
\tableline
\end{tabular}
\end{center}
\tablecomments{Cost is given for both the sub-bank described in Section~\ref{sec:bank-setup} and a full 1--3~$M_\odot$ \ac{NS}--\ac{NS} search. The last column gives the approximate number of machines per detector required for a full Advanced LIGO \ac{NS}--\ac{NS} search.}
\end{table}

\subsection{Extrapolation of computational cost to an Advanced \acs{LIGO} search}

Table~\ref{table:flops} shows that the \ac{LLOID} method requires $6.6 \times 10^8$ \ac{FLOPS} to cover a sub-bank comprising 657 out of the total 98\,544 mass pairs. Assuming that other regions of the parameter space have similar computational scaling, an entire single-detector search for \ac{NS}--\ac{NS} signals in the 1--3~$M_\odot$ component mass range could be implemented at $(98\,544/657)\approx150$ times the cost, or $9.9 \times 10^{10}$~\ac{FLOPS}.

We computed the computational cost of a full Advanced \ac{LIGO} \ac{NS}--\ac{NS} search a second way by dividing the entire 1--3~$M_\odot$ parameter space into sub-banks of 657 points apiece, performing time slices and \acp{SVD} for each sub-bank, and tabulating the number of floating point operations using Equation~(\ref{eq:lloid-flops}). This should be a much more accurate measure because template length varies over the parameter space. Lower chirp mass templates sweep through frequency more slowly and require more computations while higher chirp mass templates are shorter and require fewer computations. Despite these subtleties, this estimate gave us $1.1 \times 10^{11}$~\ac{FLOPS}, agreeing with the simple scaling argument above.

Modern (circa 2011) workstations can achieve peak computation rates of up to $\sim$$10^{11}$~\ac{FLOPS}. In practice, we expect that a software implementation of \ac{LLOID} will reach average computation rates that are perhaps a factor 10 less than this, $\sim$$10^{10}$~\ac{FLOPS} per machine, due to non-floating point tasks including bookkeeping and thread synchronization. Given these considerations, we estimate that a full Advanced \ac{LIGO}, single-detector, \ac{NS}--\ac{NS} search with \ac{LLOID} will require $\sim$$10$ machines.

By comparison, using the conventional \ac{TD} method to achieve the same latency costs
$4.9 \times 10^{13}$~\ac{FLOPS} for this particular sub-bank, and so simply scaling up by the factor of $150$ suggests that it would require $7.4 \times 10^{15}$~\ac{FLOPS}
to search the full parameter space.  To account for the varying sample rate and template duration across the parameter space, we can also directly calculate the cost for the full \ac{TD} method search using Equation~(\ref{eq:td-flops}), resulting in $3.8 \times 10^{15}$~\ac{FLOPS}, agreeing within an order of magnitude.  This would require~$\gtrsim$$10^5$ current-day machines.  Presently, the \ac{LIGO} Data Grid%
\footnote{\url{https://www.lsc-group.phys.uwm.edu/lscdatagrid/}} consists of
only $\sim$$10^4$ machines, so direct \ac{TD} convolution is clearly impractical.

The overlap-save \ac{FD} method is slightly more efficient than \ac{LLOID} for this particular sub-bank, requiring $5.2 \times 10^8$~\ac{FLOPS}.  The scaling argument projects that a full \ac{FD} search would require $7.8 \times 10^{10}$~\ac{FLOPS}.  The direct calculation from Equation~(\ref{eq:fd-flops}) gives $5.9 \times 10^{10}$~\ac{FLOPS}, in order-of-magnitude agreement.  In this application, the conventional \ac{FD} search is scarcely a factor of two faster than \ac{LLOID} while gaining only $0.3$\% in \ac{SNR}, but only at the price of thousands of seconds of latency.

\subsection{Measured latency and overhead}

Our \gstreamer\ pipeline for measuring impulse responses contained instrumentation that would not be necessary for an actual search, including additional interpolation filters to bring the early-warning outputs back to the full sample rate and additional outputs for recording signals to disk.

We wrote a second, stripped pipeline to evaluate the actual latency and computational overhead. We executed this pipeline on one of the submit machines of the \ac{LIGO}--Caltech cluster, a Sun Microsystems Sun Fire\texttrademark\ X4600~M2 server with eight quad-core 2.7~GHz AMD Opteron\texttrademark\ 8384 processors. This test consumed $\sim$90\% of the capacity of just one out of the 32 cores, maintaining a constant latency of $\sim$0.5~s.

The measured overhead is consistent to within an order of magnitude with the lower bound from the \ac{FLOPS} budget. Additional overhead is possibly dominated by thread synchronization. A carefully optimized \gstreamer\ pipeline or a hand-tuned C implementation of the pipeline might reduce overhead further.

The 0.5~s latency is probably due to buffering and synchronization. The latency might be reduced by carefully tuning buffer lengths at every stage in the pipeline. Even without further refinements, our implementation of the \ac{LLOID} algorithm has achieved latencies comparable to the \ac{LIGO} data acquisition system itself.

\section{Conclusions}

We have demonstrated a computationally feasible filtering algorithm for the rapid and even early-warning detection of \acp{GW} emitted during the coalescence of \acp{NS} and stellar-mass \acp{BH}. It is one part of a complicated analysis and observation strategy that will unfortunately have other sources of latency. However, we hope that it will motivate further work to reduce such technical sources of \ac{GW} observation latency and encourage the possibility of even more rapid \ac{EM} follow-up observations to catch prompt emission in the advanced detector era.

\ac{CBC} events may be the progenitors of some short hard \acp{GRB} and are expected to be accompanied by a broad spectrum of \ac{EM} signals. Rapid alerts to the wider astronomical community will improve the chances of detecting an \ac{EM} counterpart in bands from gamma-rays down to radio. In the Advanced \ac{LIGO} era, it appears possible to usefully localize a few rare events prior to the \ac{GRB}, allowing multi-wavelength observations of prompt emission. More frequently, low-latency alerts will be released after merger but may still yield extended X-ray tails and early on-axis afterglows.

The \ac{LLOID} method is as fast as conventional \ac{FFT}-based, \ac{FD} convolution but allows for latency free, real-time operation. We anticipate requiring $\gtrsim$40 modern multi-core computers to search for binary \acp{NS} using coincident \ac{GW} data from a four-detector network. In the future, additional computational savings could be achieved by conditionally reconstructing the \ac{SNR} time series only during times when a composite detection statistic crosses a threshold~\citep{svd-compdetstat}. However, the anticipated required number of computers is well within the current computing capabilities of the \ac{LIGO} Data Grid.

We have shown a prototype implementation of the \ac{LLOID} algorithm using \gstreamer, an open-source signal processing platform. Although our prototype already achieves latencies of less than one second, further fine tuning may reduce the latency even further. Ultimately the best possible latency would be achieved by tighter integration between data acquisition and analysis with dedicated hardware and software. This could be considered for third-generation detector design. Also possible for third-generation instruments, the \ac{LLOID} method could provide the input for a dynamic tuning of detector response via the signal recycling mirror to match the frequency of maximum sensitivity to the instantaneous frequency of the \ac{GW} waveform. This is a challenging technique, but it has the potential for substantial gains in \ac{SNR} and timing accuracy \citep{PhysRevD.47.2184}.

Although we have demonstrated a computationally feasible statistic for advance detection, we have not yet explored data calibration and whitening, triggering, coincidence, and ranking of \ac{GW} candidates in a framework that supports early \ac{EM} follow-up. One might explore these, and also use the time slice decomposition and the \ac{SVD} to form low-latency signal-based vetoes (e.g.,~$\chi^2$~statistics) that have been essential for glitch rejection used in previous \ac{GW} \ac{CBC} searches. These additional stages may incur some extra overhead, so computing requirements will likely be somewhat higher than our estimates.

Future work must more deeply address sky localization accuracy in a realistic setting as well as observing strategies. Here, we have followed \citet{FairhurstTriangulation} in estimating the area of 90\% localization confidence in terms of timing uncertainties alone, but it would be advantageous to use a galaxy catalog to inform the telescope tiling \citep{galaxy-catalog}. Because early detections will arise from nearby sources, the galaxy catalog technique might be an important ingredient in reducing the fraction of sky that must be imaged. Extensive simulation campaigns incorporating realistic binary merger rates and detector networks will be necessary in order to fully understand the prospects for early-warning detection, localization, and \ac{EM} follow-up using the techniques we have described.

\section*{Acknowledgements}

\ac{LIGO} was constructed by the California Institute of Technology and Massachusetts Institute of Technology with funding from the \ac{NSF} and operates under cooperative agreement PHY-0107417. C.H. thanks Ilya Mandel for many discussions about rate estimates and the prospects of early detection, Patrick Brady for countless fruitful conversations about low-latency analysis methods, and John Zweizig for discussions about \ac{LIGO} data acquisition. N.F. thanks Alessandra Corsi and Larry Price for illuminating discussions on astronomical motivations. L.S. thanks Shaun Hooper for productive conversations on signal processing. This research is supported by the \ac{NSF} through a Graduate Research Fellowship to L.S. and by the Perimeter Institute for Theoretical Physics through a fellowship to C.H. D.K. is supported from the Max Planck Gesellschaft. M.A.F. is supported by \ac{NSF} Grant PHY-0855494.

\chapter{\acs{BAYESTAR}: Rapid Bayesian sky localization of \acs{BNS} mergers}
\label{chap:bayestar}

\attribution{
This chapter is reproduced from a work in preparation for \textnormal{Physical Review D}. The authors will be Leo~P.~Singer and Larry~R.~Price. The introduction of this chapter is reproduced in part from \citet{FirstTwoYears}, copyright~\textcopyright{}~2014 The American Astronomical Society.
}

We expect this decade to bring the first direct detection of \acp{GW} from compact objects. The \ac{LIGO} and Virgo detectors are being rebuilt with redesigned mirror suspensions, bigger optics, novel optical coatings, and higher laser power~\citep{aLIGO, aVirgo}. In their final configuration, Advanced \ac{LIGO} and Virgo are expected to reach $\sim10$ times further into the local universe than their initial configurations did. The best\nobreakdashes-understood sources for \ac{LIGO} and Virgo are \ac{BNS} mergers. They also offer a multitude of plausible \ac{EM} counterparts~\citep{MostPromisingEMCounterpart} including collimated short\nobreakdashes-hard gamma\nobreakdashes-ray bursts (short GRBs; see for example \citealt{1986ApJ...308L..43P,1989Natur.340..126E,1992ApJ...395L..83N,2011ApJ...732L...6R}) and X\nobreakdashes-ray/optical afterglows, near\nobreakdashes-infrared kilonovae \citep[viewable from all angles;][etc.]{Kilonova, BarnesKasenKilonovaOpacities}, and late\nobreakdashes-time radio emission~\citep{NakarPiranRadioFlares,PiranNakarRosswogEMSignals}. Yet, typically poor \ac{GW} localizations of $\gtrsim 100\text{~deg}^2$ will present formidable challenges to observers hunting for their \ac{EM} counterparts.

\begin{figure*}
    \includegraphics[width=1.1\textwidth]{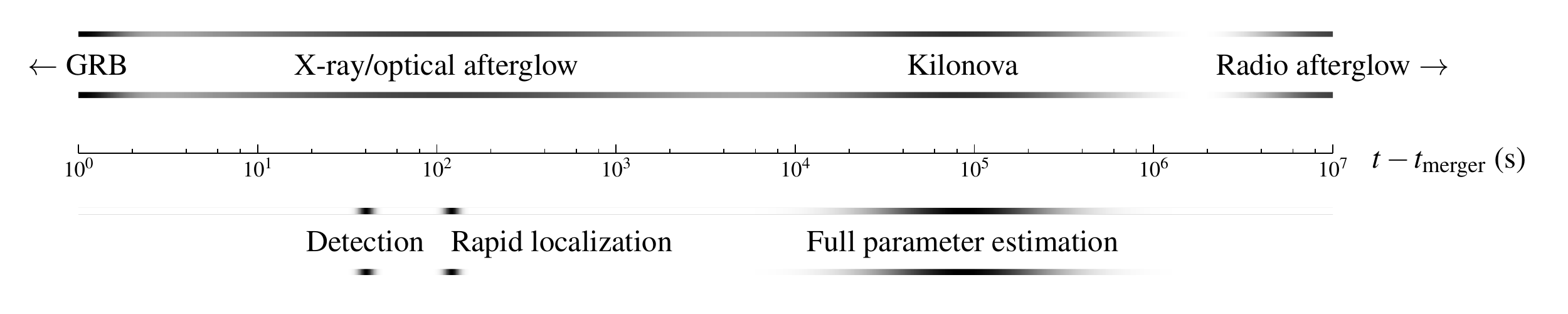}
    \caption[Timeline of \acs{CBC} \acs{EM} counterparts and the Advanced \acs{LIGO} analysis]{\label{fig:timeline}Rough timeline of compact binary merger electromagnetic emissions in relation to the timescale of the Advanced LIGO/Virgo analysis described in this thesis. The time axis measures seconds after the merger.}
\end{figure*}

The final Initial \ac{LIGO}\nobreakdashes--Virgo observing run pioneered the first accurate, practical parameter estimation and position reconstruction methods for \ac{BNS} signals. This included a prompt, semi\nobreakdashes-coherent, ad hoc analysis (Timing++, designed to work with \ac{MBTA}; \citealt{CBCLowLatency}), and the first version of a rigorous Bayesian \ac{MCMC} analysis (\textls{LALINFERENCE}; \citealt{S6PE}). Though important milestones, there was an undesirable compromise made between accuracy and speed: though the former analysis took only minutes, it produced areas that were typically 20 times larges than the latter, which could take days \citep{SiderySkyLocalizationComparison}. To increase the odds of finding a relatively bright but rapidly fading afterglow, one wants localizations that are both prompt \emph{and} accurate, to begin optical searches within minutes to hours of a \ac{GW} detection. To increase the odds of finding a kilonova, one wants localizations that are reliably available in under one day, to allow as much time as possible for multiple deep exposures. See Figure~\ref{fig:timeline} for a timeline of the most promising \ac{EM} counterparts as compared to the response times of the various steps in the \ac{GW} analysis.

To that end, in this chapter we develop a rapid and accurate Bayesian sky localization method that takes mere seconds but achieves approximately the same accuracy as the full \ac{MCMC} analysis. We call this algorithm \ac{BAYESTAR}%
\footnote{A pun on the Cylon battleships in the American television series Battlestar Galactica. The defining characteristic of the Cylons is that they repeatedly defeat humanity by using their superhuman information\nobreakdashes-gathering ability to coordinate overwhelming forces. The name also suggests that, like the Cylons, \ac{GW} detectors may some day rise against us humans.

We do not like to mention the final `L' in the acronym, because then it would be pronounced \textls{BAYESTARL}, which sounds stupid.}%
. It differs from existing techniques in several important ways. In the first place, we treat the matched-filter detection pipeline as a measurement system in and of itself, treating the point\nobreakdashes-parameter estimates that it provides as the experimental input data rather than the full \ac{GW} time series that is used by the \ac{MCMC} analysis. This drastically reduces the dimensionality of both the data and the signal model. It also permits us to avoid directly computing the expensive post-Newtonian model waveforms, making the likelihood itself much faster to evaluate. Finally, instead of of using \ac{MCMC} or some similar method for statistical sampling, we make use of a deterministic quadrature scheme. This algorithm is unique in that it bridges the detection and parameter estimation of \ac{GW} signals, two tasks that have until now involved very different numerical methods and time scales. We expect that \ac{BAYESTAR} will take on a key role in observing \ac{CBC} events in both \ac{GW} and optical channels during the Advanced \ac{LIGO} era.

\section{Bayesian probability and parameter estimation}

In the Bayesian framework, parameters are inferred from the \ac{GW} data by forming the posterior distribution, $p(\bm\theta|\mathbf Y)$, which describes the probability of the parameters given the observations. Bayes' rule relates the likelihood $p(\mathbf Y|\bm\theta)$ to the posterior $p(\bm\theta|\mathbf Y)$,
\begin{equation}\label{bayes}
p(\bm\theta|\mathbf Y) = \frac{p(\mathbf Y|\bm\theta) p(\bm\theta)}{p(\mathbf Y)},
\end{equation}
introducing the prior distribution $p(\bm\theta)$ which encapsulates previous information about the paramters (for example, arising from earlier observations or from known physical bounds on the parameters), and the evidence $p(\mathbf Y)$. In parameter estimation problems such as those we are concerned with in this chapter, the evidence serves as nothing more than a normalization factor. However, when comparing two models with different numbers of parameters, the ratio of the evidences quantifies the relative parsimony of the two models, serving as a precise form of Occam's razor.

The parameters that describe a \ac{BNS} merger signal are listed in Equation~(\ref{eq:params}). The choice of prior is determined by one's astrophysical assumptions, but during \ac{S6} when LIGO's Bayesian \ac{CBC} parameter estimation pipelines were pioneered, the prior was taken to be isotropic in sky location and binary orientation, and uniform in volume, arrival time, and the component masses~\citep{S6PE}.

In Bayesian inference, although it is often easy to write down the likelihood or even the full posterior in closed form, usually one is interested in only a subset $\bm\beta$ of all of the model's parameters, the others, $\bm\lambda$, being nuisance parameters. In this case, we integrate away the nuisance parameters, forming the marginal posterior
\begin{equation}\label{eq:marginal-posterior}
    p(\bm\beta|\mathbf Y) = \int \frac{p(\mathbf Y|\bm\beta,\bm\lambda) p(\bm\beta,\bm\lambda)}{p(\mathbf Y)} \,d\bm\lambda,
\end{equation}
with $\bm\theta = (\bm\beta, \bm\lambda)$. For instance, for the purpose of locating a \ac{GW} source in the sky, all parameters (distance, time, orientation, masses, and spins) except for $(\alpha, \delta)$ are nuisance parameters.

Bayesian parameter estimation has many advantages, including broad generality and the ability to make probabilistically meaningful statements even with very low \ac{SNR} measurements. However, in problems of even modest complexity, the marginalization step involves many-dimensional, ill-behaved integrals. The powerful \ac{MCMC} integration technique has become almost synonymous with Bayesian inference. Though powerful, \ac{MCMC} is inherently non-deterministic and resistant to parallelization, as well as (at least historically) slow. With the most sophisticated \ac{CBC} parameter estimation codes, it still takes days to process a single event. This delay is undesirable for planning targeted \ac{EM} follow\nobreakdashes-up searches of \ac{LIGO} events.

In what follows, we describe a complementary rapid parameter estimation scheme that can produce reliable positions estimates within minutes of a detection. We can even use our scheme to speed up the full \ac{MCMC} analysis and make the refined parameter estimates available more quickly. The key difference is that we start not from the \ac{GW} signal itself, but from the point parameter estimates from the detection. By harnessing the detection pipeline in this manner, we arrive at a simpler Bayesian problem that is amenable to straightforward, deterministic, numerical quadrature.

There are many practical advantages of doing so. For one, there are difficulties in synchronously gathering together the calibrated \ac{GW} strain data, auxiliary instrument channels, and data quality vetoes from all of the sites. The data consumed by the real\nobreakdashes-time detection pipeline are not necessarily final. Longer\nobreakdashes-running follow\nobreakdashes-up analyses can benefit from offline calibration, whereas the rapid sky localization need not re\nobreakdashes-analyze the online data. Moreover, the dimensionality of the problem is greatly reduced, and the problem becomes computationally easier. Finally, by breaking free of the \ac{MCMC} framework, the results are much easier to use for pointing telescopes. With \ac{MCMC} algorithms, it is often desirable to bin or interpolate the cloud of sample points to provide a smooth, high\nobreakdashes-resolution representation of the probability distribution. Reliable post\nobreakdashes-processing of \ac{MCMC} chains often relies on clustering and kernel density estimation, both of which prohibit very large numbers of samples due to rapid growth of computational cost. With \ac{BAYESTAR}, the natural form of the result is an adaptively sampled mesh with high resolution only where it is needed. The output from \ac{BAYESTAR} is therefore extremely convenient for packaging into a \ac{FITS} file for transmission with \ac{GW} alerts (see Appendix~\ref{chap:sky-map-format} for details).

\section{The \acs{BAYESTAR} likelihood}

For the purpose of rapid sky localization, we assume that we do not have access to the \ac{GW} data $\mathbf{Y}$ itself, and that our only contact with it is through the \ac{ML} parameter estimates $\{\left\{ \hat\rho_i, \hat\gamma_i, \hat\tau_i \right\}_i, \hat{\bm\theta}_\mathrm{in}\}$. Although this is a significant departure from conventional \ac{GW} parameter estimation techniques, we can still apply the full Bayesian machinery of Equation~(\ref{eq:marginal-posterior}) to compute a posterior distribution for the sky location.

The relevant likelihood is now the probability of the \ac{ML} estimates, conditioned on the true parameter values, and marginalized over all possible \ac{GW} observations:
\begin{equation}\label{eq:detection-candidate-likelihood}
    p\left(\{\hat{\bm\theta}_i\}_i,
        \hat{\bm\theta}_\mathrm{in}
    \middle| \bm\theta\right)
    \propto \int\limits_{\mathclap{\mathbf{Y} | \mathrm{MLE}(\mathbf{Y}) =
        \{\{\hat{\bm\theta}_i\}_i,
        \hat{\bm\theta}_\mathrm{in}\}}}
    p(\mathbf{Y} | \bm\theta) \, p(\bm\theta)
    \, d\mathbf{Y}.
\end{equation}
Although we may not be able to evaluate this equation directly, with some educated guesses we can create a likelihood that has many properties in common with it. Any valid approximate likelihood must have the same Fisher matrix as shown in Equation~(\ref{eq:fisher-matrix-extrinsic-one-detector}). It must also have the same limiting behavior: it should be periodic in the phase error $\tilde{\gamma}_i$ and go to zero as $\tilde{\tau}_i \rightarrow \pm \infty$, $\hat{\rho}_i \rightarrow 0$, or $\hat{\rho}_i \rightarrow \infty$. Additionally, when $\tilde{\tau}_i = 0$, the distribution of ${\hat{\rho}_i}^2$ should reduce to a noncentral $\chi^2$ distribution with two degrees of freedom, centered about ${\rho_i}^2$, because the complex matched filter time series $z_i(t)$ is Gaussian (under the ideal assumption the \ac{GW} strain time series is Gaussian).

These conditions could be satisfied by realizing a multivariate Gaussian distribution with covariance matrix $\Sigma = \mathcal{I}^\intercal$, and then replacing individual quadratic terms in the exponent of the form $-\tilde{\theta}^2/2$ with $\cos{\tilde{\theta}}$.

Another way is to plug the signal model from Equation~(\ref{eq:signal-model}) \emph{evaluated at the \ac{ML} parameter estimates} into the single-detector likelihood in Equation~(\ref{eq:gaussian-likelihood-spa}):
\begin{align*}
    p\left(\hat{\bm\theta}_i \middle| \bm\theta \right)
    &:=
    p\left(Y_i(\omega) = X_i(\omega; \hat{\bm\theta}_i)
        \middle| \bm\theta \right)
    \\
    &\propto \exp \left[
        - \frac{1}{2} \int_0^\infty \frac{\left|
            \frac{\hat{\rho}_i}{\sigma_i(\hat{\bm\theta}_\mathrm{in})} e^{i (\hat\gamma_i - \omega \hat\tau_i)} H(\omega; \hat{\bm\theta}_\mathrm{in})
            - \frac{\rho_i}{\sigma_i(\bm\theta_\mathrm{in})} e^{i (\gamma_i - \omega \tau_i)} H(\omega; \bm\theta_\mathrm{in})
        \right|^2}{S_i(\omega)} \, d\omega
    \right].
\end{align*}
If we assume that $\hat{\bm\theta}_\mathrm{in} = \bm\theta_\mathrm{in}$, then this reduces to
\begin{equation}\label{eq:autocor-likelihood}
    p\left(\hat\rho_i, \hat\gamma_i, \hat\tau_i
        \middle| \rho_i, \gamma_i, \tau_i \right) \propto
    \exp \left[ - \frac{1}{2}{\hat\rho_i}^2 - \frac{1}{2}{\rho_i}^2
        + \hat\rho_i \rho_i \Re \left\{ e^{i \tilde{\gamma}_i} a_i^*(\tilde{\tau}_i)
        \right\}
    \right],
\end{equation}
with $\tilde{\gamma}_i = \hat\gamma_i - \gamma_i$, $\tilde{\tau}_i = \hat\tau_i - \tau_i$, and the template's autocorrelation function $a_i(t; \bm\theta_\mathrm{in})$ defined as
\begin{equation}\label{eq:autocorrelation-function}
    a_i(t; \bm\theta_\mathrm{in}) := \frac{1}{{{\sigma_i}^2(\bm\theta_\mathrm{in})}} \int_0^\infty \frac{\left| H(\omega; \hat{\bm\theta}_\mathrm{in})\right|^2}{S_i(\omega)} e^{i \omega t} \,d\omega.
\end{equation}

To assemble the joint likelihood for the whole network, we just form the product from the individual detectors:
\begin{equation}
    p\left(\left\{\hat\rho_i, \hat\gamma_i, \hat\tau_i\right\}_i
        \middle| \left\{\rho_i, \gamma_i, \tau_i\right\}_i \right) \propto
    \exp \left[ - \frac{1}{2} \sum_i {\hat\rho_i}^2 - \frac{1}{2} \sum_i {\rho_i}^2
        + \sum_i \hat\rho_i \rho_i \Re \left\{ e^{i \tilde{\gamma}_i} a^*(\tilde{\tau}_i)
        \right\}
    \right].
\end{equation}

\section{Properties}

First, observe that at the true parameter values, $\hat{\bm\theta}_i = \bm\theta_i$, the logarithms of Equation~(\ref{eq:autocor-likelihood}) and Equation~(\ref{eq:gaussian-likelihood-spa}) have the same Jacobian. This is because the derivatives of the autocorrelation function are
\begin{equation*}
    a^{(n)}(t) = i^n \overline{\omega^n},
\end{equation*}
with $\overline{\omega^n}$ defined in Equation~(\ref{eq:angular-frequency-moments}). For example, the first few derivatives are
\begin{equation*}
    a(0) = 1,
    \qquad
    \dot{a}(0) = i \overline{\omega},
    \qquad
    \ddot{a}(0) = -\overline{\omega^2}.
\end{equation*}

Using Equation~(\ref{eq:general-fisher-matrix-second-derivatives}), we can compute the Fisher matrix elements for the autocorrelation likelihood given by Equation~(\ref{eq:autocor-likelihood}), with detector subscript suppressed:
\begin{align}
    \nonumber
    \mathcal{I}_{\rho\rho} &= 1, \\
    \nonumber
    \mathcal{I}_{\rho\gamma} &= 0, \\
    \nonumber
    \mathcal{I}_{\rho\tau} &= 0, \\
    \label{eq:fisher-autocor-gamma-gamma}
    \mathcal{I}_{\gamma\gamma} &= \rho^2
        \int_{-T}^T \left|a(t)\right|^2 w(t; \rho) dt, \\
    \label{eq:fisher-autocor-tau-tau}
    \mathcal{I}_{\tau\tau} &= -\rho^2
        \int_{-T}^T \Re\left[a^*(t) \ddot{a}(t)\right] w(t; \rho) dt, \\
    \label{eq:fisher-autocor-gamma-tau}
    \mathcal{I}_{\gamma\tau} &= -\rho^2
        \int_{-T}^T \Im\left[a^*(t) \dot{a}(t)\right] w(t; \rho) dt,
\intertext{where}
    w(t; \rho) &= \frac{
        \displaystyle
        \exp\left[\frac{\rho^2}{4}\left|a(t)\right|^2\right]
        \left(
        I_0\left[\frac{\rho^2}{4}\left|a(t)\right|^2\right] +
        I_1\left[\frac{\rho^2}{4}\left|a(t)\right|^2\right]
        \right)
    }{
        \displaystyle
        2 \int_{-T}^T
        \exp\left[\frac{\rho^2}{4}\left|a(t')\right|^2\right]
        I_0\left[\frac{\rho^2}{4}\left|a(t')\right|^2\right]
        dt'
    }.
\end{align}
The notation $I_k$ denotes a modified Bessel function of the first kind. Matrix elements that are not listed have values that are implied by the symmetry of the Fisher matrix. Note that the minus signs are correct but a little confusing: despite them, $\mathcal{I}_{\gamma\gamma}, \mathcal{I}_{\tau\tau} \geq 0$ and $\mathcal{I}_{\gamma\tau} \leq 0$. The time integration limits $[-T, T]$ correspond to a flat prior on arrival time, or a time coincidence window between detectors.

We can show that the weighting function $w(t; \rho)$ approaches a Dirac delta function as $\rho \rightarrow \infty$, so that the Fisher matrix for the autocorrelation likelihood approaches the Fisher matrix for the full \ac{GW} data, Equation~(\ref{eq:fisher-matrix}), as $\rho \rightarrow \infty$. The Bessel functions asymptotically approach:
\begin{equation*}
    I_0(x), I_1(x) \rightarrow \frac{e^x}{\sqrt{2 \pi x}}
    \textrm{ as } x \rightarrow \infty.
\end{equation*}
For large $\rho$, the exponential dominates and we can write:
\begin{equation*}
    w(t; \rho) \rightarrow \frac{
        \displaystyle
        \exp\left[\frac{\rho^2}{2}|a(t)|^2\right]
    }{
        \displaystyle
        \int_{-T}^T \exp\left[\frac{\rho^2}{2}|a(t')|^2\right] dt'
    }
    \textrm{ as } \rho \rightarrow \infty.
\end{equation*}
The Taylor expansion of $|a(t)|^2$ is
\begin{equation*}
    |a(t)|^2 = 1 + \frac{1}{2} \left(\frac{\partial^2}{\partial t^2}|a(t)|^2 \Bigg|_{t=0}\right) t^2 + \mathcal{O}(t^4)
    = 1 - {\omega_\mathrm{rms}}^2 t^2 + \mathcal{O}(t^4).
\end{equation*}
Substituting, we find that $w(t; \rho)$ approaches a normalized Gaussian distribution:
\begin{equation*}
    w(t; \rho) \approx \frac{
        \displaystyle
        \exp\left[-\frac{1}{2} \rho^2 {\omega_\mathrm{rms}}^2 t^2\right]
    }{
        \displaystyle
        \int_{-T}^T \exp\left[-\frac{1}{2} \rho^2 {\omega_\mathrm{rms}}^2 (t')^2\right] dt'
    }.
\end{equation*}
And finally, because the Dirac delta function may be defined as the limit of a Gaussian, $w(t; \rho) \rightarrow \delta(t)$ as $\rho \rightarrow \infty$.

\begin{figure*}
    \centering
            \includegraphics[width=0.45\columnwidth]{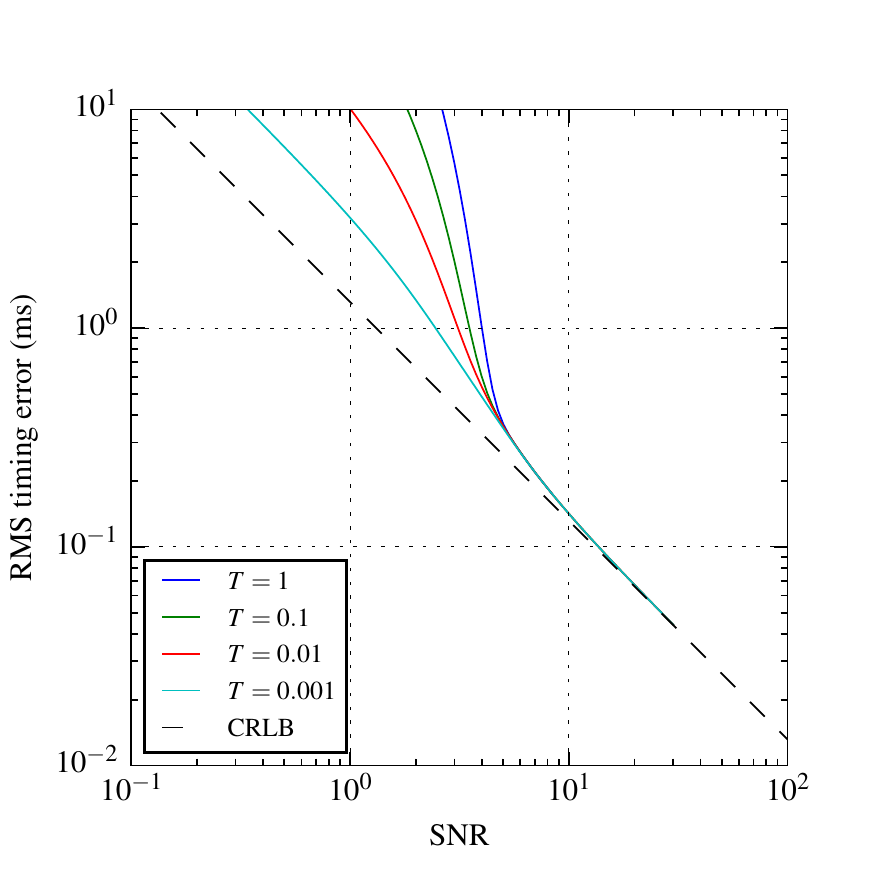}
        \includegraphics[width=0.45\columnwidth]{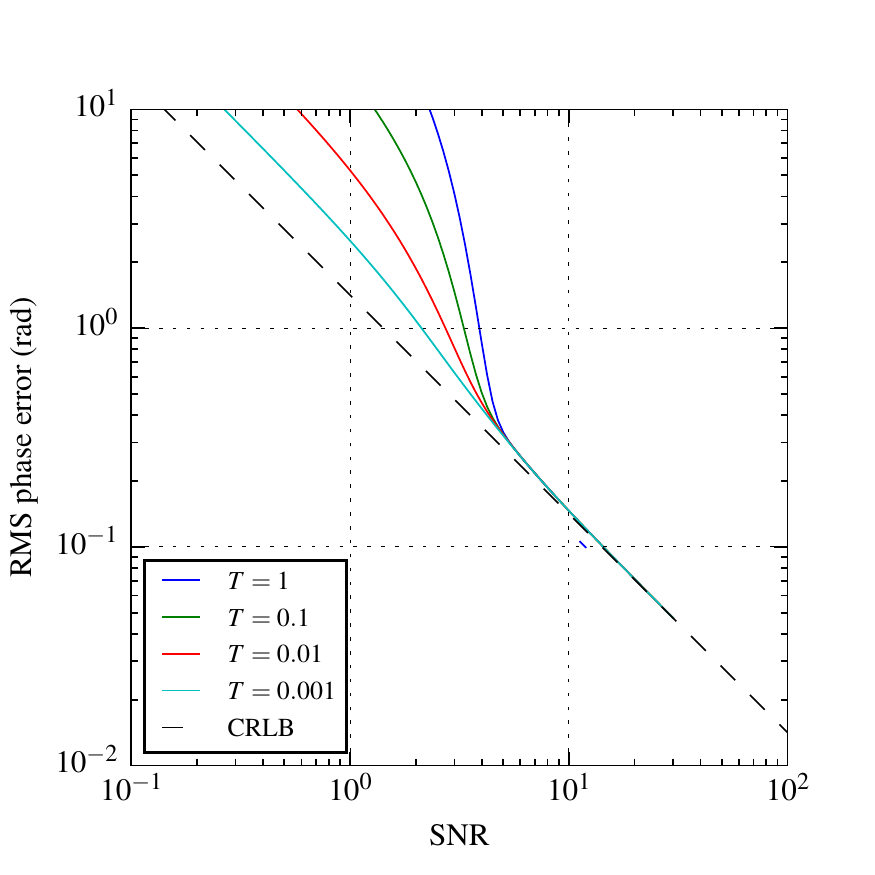}
    \caption[\acl{CRLB} on time and phase accuracy]{\label{fig:crlb-tau}\ac{CRLB} on root\nobreakdashes-mean square timing uncertainty and phase error, using the likelihood for the full \ac{GW} data (Equation~\ref{eq:gaussian-likelihood-spa}; dashed diagonal line) or the autocorrelation likelihood (Equation~\ref{eq:autocor-likelihood}; solid lines) with a selection of arrival time priors.}
\end{figure*}

\begin{figure}
    \centering
    \includegraphics{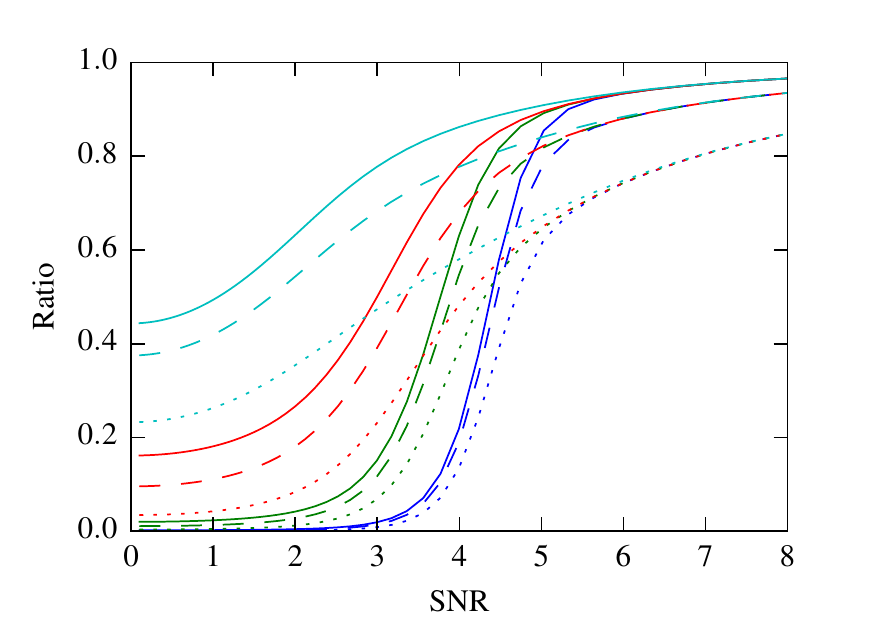}
    \caption[Ratio of Fisher matrix elements between autocorrelation likelihood and full \acs{GW} data]{\label{fig:fishfactor}Ratio between Fisher matrix elements (solid: $\textproto{D}_{\gamma\gamma}$, dashed: $\textproto{D}_{\gamma\tau}$, dotted: $\textproto{D}_{\tau\tau}$) for the autocorrelation likelihood and the full \ac{GW} data. Colors correspond to different arrival time priors as in Figure~\ref{fig:crlb-tau}.}
\end{figure}

We can now write the Fisher matrix for the autocorrelation likelihood in a way that makes a comparison to the full signal model explicit. Define:
\begin{align*}
    \mathcal{I}_{\gamma\gamma} &= \rho^2 \cdot \textproto{D}_{\gamma\gamma}(\rho), \\
    \mathcal{I}_{\tau\tau} &= \rho^2 \overline{\omega^2} \cdot \textproto{D}_{\tau\tau}(\rho), \\
    \mathcal{I}_{\gamma\tau} &= -\rho^2 \overline{\omega} \cdot \textproto{D}_{\gamma\tau}(\rho).
\end{align*}
Now, the $\textproto{D}_{ij}$%
\footnote{The Fish(er) factor.}
contain the integrals from Equations~(\ref{eq:fisher-autocor-gamma-gamma}, \ref{eq:fisher-autocor-tau-tau}, \ref{eq:fisher-autocor-gamma-tau}) and encode the departure of the autocorrelation likelihood from the likelihood of the full data at low \ac{SNR}. All of the $\textproto{D}_{ij}(\rho)$ are sigmoid-type functions that asymptotically approach 1 as $\rho \rightarrow \infty$ (see Figures~\ref{fig:crlb-tau} and~\ref{fig:fishfactor}). The transition \ac{SNR} $\rho_\mathrm{crit}$ is largely the same for all three nontrivial matrix elements, and is determined by the time coincidence window $T$ and the signal bandwidth $\omega_\mathrm{rms}$.

In the limit of large \ac{SNR}, our interpretation is that the point estimates $(\hat\rho, \hat\gamma, \hat\tau)$ contain all of the information about the underlying extrinsic parameters.

On the other hand, in the low \ac{SNR} limit, the diminishing value of $\textproto{D}_{ij}(\rho)$ reflects the fact that some information is lost when the full data $\mathbf{x}$ is discarded. Concretely, as the prior interval $T$ becomes large compared to 1/$\rho\omega_\mathrm{rms}$, the \ac{ML} estimator becomes more and more prone to picking up spurious noise fluctuations far from the true signal. Clearly, when the coincidence window $T$ is kept small as possible, more information is retained in the \ac{ML} point estimates. Put another way, if $T$ is small, then the transition \ac{SNR} $\rho_\mathrm{crit}$ is also small and fainter signals become useful for parameter estimation. In this way, the \ac{BAYESTAR} likelihood exhibits the \emph{threshold effect} that is well\nobreakdashes-known in communication and radar applications \citep{barankin1949locally,mcaulay1969barankin,mcaulay1971barankin}.

In the following sections, we describe the numerical implementation of \ac{BAYESTAR}.

\section{Prior and problem setup}

The detection pipeline supplies a candidate, $\{\left\{ \hat\rho_i, \hat\gamma_i, \hat\tau_i \right\}_i, \hat{\bm\theta}_\mathrm{in}\}$, and discretely sampled noise \acp{PSD}, $S_i(\omega_j)$, for all detectors. We compute the \ac{GW} signal for a source with intrinsic parameters equal to the detection pipeline's estimate, $H(\omega; \hat{\bm\theta}_\mathrm{in})$.
Then we find the \ac{SNR}=1 horizon distance $r_{1,i}$ for each detector by numerically integrating Equation~(\ref{eq:horizon}).

We have no explicit prior on the intrinsic parameters; in our analysis they are fixed at their \ac{ML} estimates, $\hat{\bm\theta}_\mathrm{in}$.

The arrival time prior is connected to the origin of the detector coordinate system. Given the Earth\nobreakdashes-fixed coordinates of the detectors $\mathbf{n}_i$ and the arrival times $\tau_i$, we compute their averages weighted by the timing uncertainty formula:
\begin{equation*}
    \langle \mathbf{n} \rangle = \frac{
        \displaystyle
        \sum_i \frac{\mathbf{n}_i}
            {\left(\hat\rho_i \omega_{\mathrm{rms},i}\right)^2}
    }{
        \displaystyle
        \sum_i \frac{1}{\left(\hat\rho_i \omega_{\mathrm{rms},i}\right)^2}
    },
    \qquad
    \langle \hat\tau \rangle = \frac{
        \displaystyle
        \sum_i \frac{\hat\tau_i}
            {\left(\hat\rho_i \omega_{\mathrm{rms},i}\right)^2}
    }{
        \displaystyle
        \sum_i \frac{1}{\left(\hat\rho_i \omega_{\mathrm{rms},i}\right)^2}
    }.
\end{equation*}
Then, we subtract these means:
\begin{equation*}
    \mathbf{n}_i \leftarrow \mathbf{n}_i - \langle \mathbf{n} \rangle,
    \qquad
    \hat\tau_i \leftarrow \hat\tau_i - \langle \hat\tau \rangle.
\end{equation*}
In these coordinates, now relative to the weighted detector array barycenter, the arrival time prior is uniform in $-T \leq t \leq T$, with $T = \max\limits_i |\mathbf{n}_i| / c + 5~\textrm{ms}$.

The distance prior is a user-selected power of distance,
\begin{equation*}
    p(r) \propto \begin{cases}
        r^m & \text{if } r_\mathrm{min} < r < r_\mathrm{max} \\
        0 & \text{otherwise},
    \end{cases}
\end{equation*}
where $m=2$ for a prior that is uniform in volume, and $m=-1$ for a prior that is uniform in the logarithm of the distance. If a distance prior is not specified, the default is uniform in volume out to the maximum SNR=4 horizon distance:
\begin{equation*}
    m = 2,
    \qquad
    r_\mathrm{min} = 0,
    \qquad
    r_\mathrm{max} = \frac{1}{4} \max_i r_{1,i}.
\end{equation*}

Finally, the prior is uniform in $-1 \leq \cos\iota \leq 1$ and $0 \leq \psi < 2\pi$.

We compute the autocorrelation function for each detector from $t = 0$ to $t = T$ at intervals of $\Delta t = 1/f_s$, where $f_s$ is the smallest power of two that is greater than or equal to the Nyquist rate. Because BNS signals typically terminate at about 1500~Hz, a typical value for $\Delta t$ is $(4096\,\textrm{Hz})^{-1}$. We use a pruned \ac{FFT} because for BNS systems, the \ac{GW} signal remains in LIGO's sensitive band for $\sim$100\nobreakdashes--1000~s, whereas $T \sim 10$~ms.\footnote{See \url{http://www.fftw.org/pruned.html} for a discussion of methods for computing the first $K$ samples of an \ac{FFT} of length $N$.}

\section{Marginal posterior}

The marginal posterior as a function of sky location is
\begin{equation}
    f(\alpha, \delta) \propto
    \int_{0}^{\pi}
    \int_{-1}^{1}
    \int_{-T}^{T}
    \int_{r_\mathrm{min}}^{r_\mathrm{max}}
    \int_{0}^{2\pi}
    \exp \left[ - \frac{1}{2} \sum_i {\rho_i}^2
        + \sum_i \hat\rho_i \rho_i \Re \left\{ e^{i \tilde{\gamma}_i}
        a^*(\tilde{\tau}_i)
        \right\}
    \right] \\
    r^m d\phi_c \, dr \, dt_\oplus \, d\cos{\iota} \, d\psi.
\end{equation}

To marginalize over the coalescence phase, we can write $\tilde{\gamma}_i = \tilde{\gamma}_i^\prime + 2\phi_c$. Then integrating over $\phi_c$ and suppressing normalization factors, we get
\begin{equation}
    f(\alpha, \delta) \rightarrow
    \int_{0}^{\pi}
    \int_{-1}^{1}
    \int_{-T}^{T}
    \int_{r_\mathrm{min}}^{r_\mathrm{max}}
    \exp \left[ - \frac{1}{2} \sum_i {\rho_i}^2 \right] I_0 \left[
            \left| \sum_i \hat\rho_i \rho_i e^{i \tilde{\gamma}_i} a_i^*(\tilde{\tau}_i)
            \right|
    \right] \\
    r^m dr \, dt_\oplus \, d\cos{\iota} \, d\psi.
\end{equation}
In the above equation, we need not distinguish between $\tilde{\gamma}_i$ and $\tilde{\gamma}_i^\prime$ because the likelihood is now invariant under arbitrary phase shifts of all of the detectors' signals.

\subsection{Integral over angles and time}

The integrand is periodic in $\psi$, so simple Newton\nobreakdashes--Cotes quadrature over $\psi$ exhibits extremely rapid convergence. We therefore sample the posterior on a regular grid of 10 points from 0 to $\pi$.

The integral over $\cos\iota$ converges with just as rapidly with Gauss\nobreakdashes--Legendre quadrature, so we use a 10\nobreakdashes-point Gauss\nobreakdashes--Legendre rule for integration over $\cos\iota$.

We sample $t_\oplus$ regularly from $-T$ to $T$ at intervals of $\Delta t$. This is typically $\sim 2 (10\,\mathrm{ms})(4096\,\mathrm{Hz}) \approx 80$~samples. We use Catmull\nobreakdashes--Rom cubic splines to interpolate the real and imaginary parts of the autocorrelation functions between samples.

\subsection{Integral over distance}

\begin{figure}
    \centering
    \caption[Initial subdivisions for radial integral]{\label{fig:radial_integrand}Illustration of initial subdivisions for distance integration scheme. Distance increases from left to right. In the color version, the left\nobreakdashes-hand tail, the left\nobreakdashes- and right\nobreakdashes-hand sides of the maximum likelihood peak, and the right\nobreakdashes-hand tail, are colored cyan, red, green, and blue, respectively.}
    \includegraphics{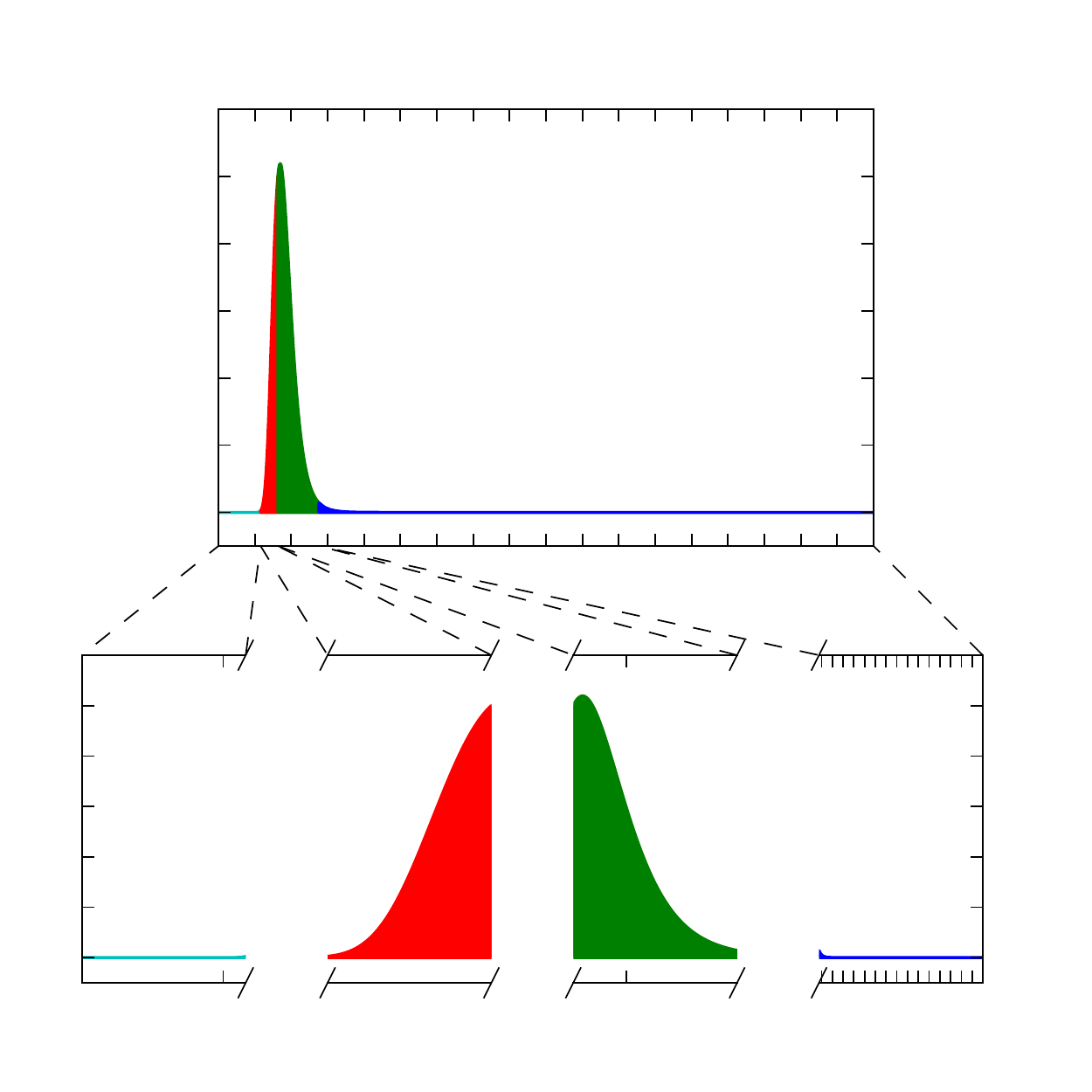}
\end{figure}

The innermost integral over $r$ is a little bit more challenging. The distance integrand has both a sharp maximum\nobreakdashes-likelihood peak and an extended power\nobreakdashes-law prior\nobreakdashes-dominated tail. At any values of the other parameters, either the peak or the tail might dominate. If the peak dominates, it may not be successfully resolved by any general\nobreakdashes-purpose fixed\nobreakdashes-order quadrature scheme. Any general adaptive integrator may take a long time to find the peak.

We tried two ways to evaluate the distance integral, first with adaptive Gaussian quadrature and second with fixed\nobreakdashes-order Gaussian quadrature. The second was ultimately faster and just as accurate for the example events that we tested, but we will describe both methods because they were instructive.

\subsubsection{Adaptive Gaussian quadrature method}

The first method was to use adaptive Gaussian quadrature, but with an educated choice of initial subdivisions. The distance integral can be written as
\begin{align}\label{eq:distance-integral}
    \mathscr{F} &=
        \int_{\mathrlap{r_\mathrm{min}}}^{\mathrlap{r_\mathrm{max}}}
        \exp\left[-\frac{p^2}{r^2}\right]
        I_0\left[\frac{b}{r}\right] r^m dr \nonumber \\
    &=
        \int_{\mathrlap{r_\mathrm{min}}}^{\mathrlap{r_\mathrm{max}}}
        \exp\left[-\frac{p^2}{r^2}+\frac{b}{r}\right]
        \overline{I}_0\left[\frac{b}{r}\right] r^m dr
\end{align}
where
\begin{align}
    p^2 &= \frac{1}{2} r^2 \sum_i {\rho_i}^2 \label{eq:likelihood-p-factor} \\
    b &= r \left| \sum_i \hat\rho_i \rho_i e^{i \tilde{\gamma}_i}
        a_i^*(\tilde{\tau}_i) \right| \label{eq:likelihood-b-factor} \\
    \overline{I}_0(x) &= \exp(-|x|) I_0(x).
\end{align}
The coefficients $p^2$ and $b$ are nonnegative and independent of distance. The symbol $\overline{I}_0$ denotes an exponentially scaled Bessel function. In the limit of large argument, $I_0(|x|) \sim \exp(|x|) / \sqrt{2 \pi |x|}$ \citep{Olver:2010:NHMF,NIST:DLMF}\footnote{\url{http://dlmf.nist.gov/10.40.E1}}. The scaled Bessel is useful for evaluation on a computer because it has a relatively small range $(0, 1]$ and varies slowly in proportion to $x^{1/2}$. If we neglect both the Bessel function and the $r^m$ prior, then the approximate likelihood $\exp(-p^2/r^2 + b/r)$ is maximized when
\begin{equation}\label{eq:ml-distance}
    r = r_0 \equiv 2p^2/b,.
\end{equation}
with $p$ and $b$ defined above in Equations~(\ref{eq:likelihood-p-factor},~\ref{eq:likelihood-b-factor}). The likelihood takes on a factor $\eta$ (say, $\eta=0.01$) of its maximum value when
\begin{equation}
    r = r_\pm = \left(\frac{1}{r_0} \mp \frac{\sqrt{-\log\eta}}{p}\right)^{-1}.
\end{equation}
We have now identified up to five breakpoints that partition the distance integrand into up to four intervals with quantitatively distinct behavior. These intervals are depicted in Figure~\ref{fig:radial_integrand} with distance, $r$, increasing from left to right. There is a left\nobreakdashes-hand or small distance tail in which the integrand is small and monotonically increasing, a left\nobreakdashes- and right\nobreakdashes-hand side of the maximum likelihood peak, and a right\nobreakdashes-hand tail in which the integrand is small and monotonically decreasing. These breakpoints are:
\begin{equation}
    r_\mathrm{break} = \{ r \in
    \left\{
    \begin{array}{c}
    r_\mathrm{min} \\
    r_- \\
    r_0 \\
    r_+ \\
    r_\mathrm{max}
    \end{array}
    \right\} :
    r_\mathrm{min} \leq r
    \leq r_\mathrm{max}\}.
\end{equation}
We used these breakpoints as initial subdivisions in an adaptive Gaussian quadrature algorithm\footnote{for instance, \ac{GSL}'s \texttt{gsl\_integrate\_qagp} function, \url{http://www.gnu.org/software/gsl/manual/html_node/QAGP-adaptive-integration-with-known-singular-points.html}}. This function estimates the integral over each subdivision and each interval's contribution to the total error, then subdivides the interval whose error contribution is largest. Subdivisions continue until a fixed total fractional error is reached. In this way, most integrand evaluations are expended on the most important distance interval, whether that happens to be the tails (when the posterior is dominated by the prior) or the peak (when the posterior is dominated by the observations).

\subsubsection{Fixed order Gaussian quadrature method}

Let's say that we want to evaluate an integral
\begin{equation}\label{eq:importance-sampling-difficult-integral}
    \mathscr{F}(x_1, x_2) = \int_{x_1}^{x_2} f(x') dx'.
\end{equation}
Suppose that this integral is resistant to direct application of standard quadrature techniques, but we have another function $g(x)$ that we can integrate analytically or at least numerically with relative ease,
\begin{equation}
    \mathscr{G}(x) = G(x) - G(x_0) = \int_{x_0}^x g(x') dx'.
\end{equation}
There is an \emph{ultimate}, though tautological, change of variables that makes it trivial to evaluate $\mathscr{G}(x)$,
\begin{equation}
    \mathscr{G}(x) = \int_0^{\mathrlap{\mathscr{G}(x)}} dG.
\end{equation}
If we apply the same change of variables to Equation~(\ref{eq:importance-sampling-difficult-integral}), we get
\begin{equation}\label{eq:importance-sampling}
    \mathscr{F}(x_1, x_2) = \int_{\mathscr{G}(x_1)}^{\mathscr{G}(x_2)} \frac{f(\mathscr{G}^{-1}(G))}{g(\mathscr{G}^{-1}(G))} dG.
\end{equation}
If $f(x)$ and $g(x)$ have sufficiently similar behavior, then their quotient $f/g$ will be better behaved than $f$ itself, and more amenable to any given quadrature technique. When the transformed integral Equation~(\ref{eq:importance-sampling}) is evaluated using Monte Carlo integration, this technique is referred to as importance sampling.

We use the same kind of change of variables with fixed\nobreakdashes-order Gaussian quadrature. We first find the approximate \ac{ML} distance $r_0$ using Equation~(\ref{eq:ml-distance}). If $r_0 < r_\mathrm{min}$ or $r_0 > r_\mathrm{max}$, then the ML peak lies outside of the limits of integration and the original integrand behaves like a low-order power law. In this case, we use 10\nobreakdashes-point Gaussian quadrature to evaluate Equation~(\ref{eq:distance-integral}).

If, on the other hand, $r_\mathrm{min} \leq r_0 \leq r_\mathrm{max}$, then the integrand contains a sharp peak that we can smooth out with importance sampling. We first re-scale the distance integral Equation~(\ref{eq:distance-integral}) so that the peak value of the integrand is $\sim 1$:
\begin{equation}
    \mathscr{F} = \exp\left(\frac{b^2}{4p^2}\right)
        {r_0}^m
        \int_{r_\mathrm{min}}^{r_\mathrm{max}}
        \exp\left[-\left(\frac{p}{r} - \frac{b}{2p}\right)^2\right]
        \left[\frac{r}{r_0}\right]^m
        dr.
\end{equation}
The importance sampling integrand $g(r)$ is
\begin{equation}
    g(r) = \exp\left[-\left(\frac{p}{r} - \frac{b}{2p}\right)^2\right]
        \frac{1}{r^2},
\end{equation}
with definite integral
\begin{equation}
    \mathscr{G}(r) = \frac{\sqrt{\pi}}{p}
        Q\left[\sqrt{2}\left(\frac{p}{r} - \frac{b}{2p}\right)\right],
\end{equation}
where $Q$ is the cumulative distribution function for the upper tail of the standard normal distribution,
\begin{equation}
    Q(x) = \int_x^\infty \frac{1}{\sqrt{2\pi}} \exp \left[-\frac{{x'}^2}{2}\right] dx'.
\end{equation}
Its inverse is
\begin{equation}
    \mathscr{G}^{-1}(G) = \left[\frac{1}{r_0} + \frac{1}{\sqrt{2}p}
    Q^{-1}\left(\frac{p G}{\sqrt{\pi}}\right)\right]^{-1}.
\end{equation}
The transformed integral is
\begin{equation}
    \mathscr{F} = \exp\left(\frac{b^2}{4p^2}\right)
        {r_0}^m
        \int_{\mathscr{G}(r_\mathrm{min})}^{\mathscr{G}(r_\mathrm{max})}
        \overline{I}_0\left[\frac{b}{r}\right]
        \left[\frac{r}{r_0}\right]^m r^2
        \Bigg|_{r = \mathscr{G}^{-1}(G)}
        dG.
\end{equation}
This integral is evaluated numerically with a 10\nobreakdashes-point Gauss\nobreakdashes--Legendre rule.

Last, there are a few special cases where Equation~(\ref{eq:distance-integral}) can be evaluated exactly. If $p^2=0$ (which implies $b = 0$), then the integral may be expressed in terms of logarithms. If $b = 0$ but $p^2 \neq 0$, then the integral may be expressed in terms of the exponential integral function,
\begin{equation}
    E_n(x) = \int_1^\infty \frac{\exp (-xt)}{t^n} dt.
\end{equation}

The distance integral, therefore, takes either one or ten function evaluations.

\section{Adaptive mesh refinement}

We have explained how we evaluate the marginal posterior at a given sky location. Now we must specify where we choose to evaluate it.

Our sampling of the sky is relies completely on the \acl{HEALPix} (\acs{HEALPix}), a special data structure designed for all\nobreakdashes-sky maps. \ac{HEALPix} divides the sky into equal\nobreakdashes-area pixels. There is a hierarchy of \ac{HEALPix} resolutions. A \ac{HEALPix} resolution may be designated by its order $N$. The $N=0$th order or base tiling has 12 pixels. At every successive order, each tile is subdivided into four new tiles. A resolution may also be referred to by the number of subdivisions along each side of the base tiles, $N_\mathrm{side} = 2^N$. There are $N^\mathrm{pix} = 12 {N_\mathrm{side}}^2$ pixels at any given resolution. The \ac{HEALPix} projection uniquely specifies the coordinates of the center of each pixel by providing a mapping from the resolution and pixel index $(N_\mathrm{side}, i_\mathrm{pix})$ to right ascension and declination $(\alpha, \delta)$.

We begin by evaluating the posterior probability density at the center of each of the $N_{\mathrm{pix},0} = 3072$ pixels of an $N_{\mathrm{side},0}=16$ \ac{HEALPix} grid. At this resolution, each pixel has an area of 13.4~deg$^2$. We then rank the pixels by contained probability (assuming constant probability density within a pixel), and subdivide the most probable $N_{\mathrm{pix},0}/4$ pixels into $N_{\mathrm{pix},0}$ new pixels. We then evaluate the posterior again at the pixels that we subdivided, sort again, and repeat seven times, so that we have evaluated the posterior probability density a total of $8 N_{\mathrm{pix},0}$ times. On \emph{most} iterations, we descend one level deeper in \ac{HEALPix} resolution. The resulting map is a tree structure that consists of a mix of several resolutions. We traverse the tree and flatten it into the highest resolution represented. The highest possible resolution is $N_\mathrm{side}=2^{11}$, with an area of $\approx 10^{-3}$~deg$^2$ per pixel.\footnote{Although the resulting sky map contains $N_\mathrm{pix} \approx 5\times10^6$ pixels, at most $\approx 2\times10^4$ pixels have distinct values. For the purpose of delivery to observers, therefore, the output is always \texttt{gzip}\nobreakdashes-compressed with a ratio of $\approx 250:1$.}

Within each iteration, all of the marginalization integrals may be evaluated in parallel. In our C language implementation, the inner loop over pixels is parallelized with OpenMP\footnote{http://openmp.org/}. See Appendix~\ref{sec:code-lalsuite} for source code for \ac{BAYESTAR}.

\section{Run time}

\begin{figure}
    \centering
    \includegraphics{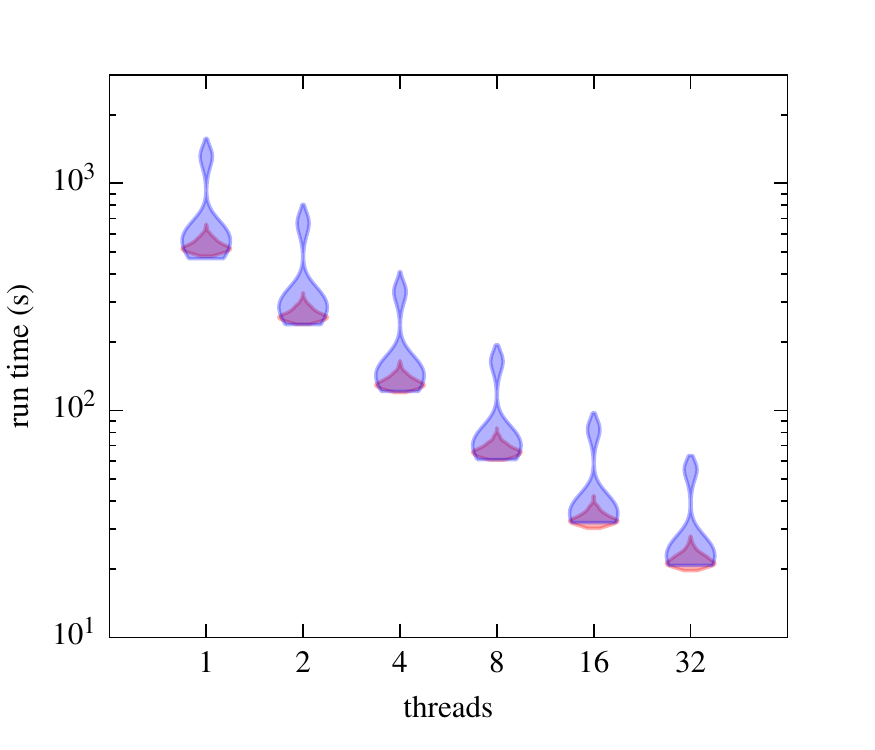}
    \caption[\acs{BAYESTAR} run time]{\label{fig:runtimes}Violin plot of \ac{BAYESTAR} run times as the number of OpenMP threads is varied from 1 to 32. The 2015 scenario is shown in red and the 2016 scenario in blue.}
\end{figure}

Since \ac{BAYESTAR} is designed as one of the final steps in the real\nobreakdashes-time \ac{BNS} search, it is important to characterize how long it takes to calculate a sky map. We compiled \ac{BAYESTAR} with the Intel~C~Compiler~(\texttt{icc}) at the highest architecture\nobreakdashes-specific optimization setting (\texttt{-xhost}). We timed it under Scientific~Linux~6 on a Supermicro~1027GR\nobreakdashes-TRFT system with dual Intel Xeon E5-2670 CPUs clocked at 2.60GHz, providing a total of 16 cores and capable of executing 32 threads simultaneously (with hyperthreading). In Figure~\ref{fig:runtimes}, we show how long it took to calculate a localization with \ac{BAYESTAR} as the number of OpenMP threads was varied from 1 to 32. This is a violin plot, a smoothed vertical histogram. The red surface shows run times for a two\nobreakdashes-detector network (HL) modeled on the first scheduled Advanced \ac{LIGO} observing run in 2015, and the blue surface shows run times for a three\nobreakdashes-detector network (HLV) based on the second planned observing run in 2016. These are the two observing scenarios that will be discussed at length in the following chapter.

Several features are apparent. First, at any number of threads, the two configurations have similar run times, although the 2016 events contain a subpopulation of outliers that take about 2.5 times as long as the 2015 events. These are probably due to taking one of the more expensive code branches in the distance integral. Second, the run times decrease proportionally to the number of threads. A slight flattening at 32 threads may be due to increasing use of hyperthreading. Based on experiences running \ac{BAYESTAR} on the 32\nobreakdashes-core (64 threads with hyperthreading) cluster login machine, we expect the almost ideal parallel speedup to continue on machines with more threads. With just one thread, the \ac{BAYESTAR} analysis takes 500\nobreakdashes--1500~s, already orders of magnitude faster than the full parameter estimation. With 32 threads, \ac{BAYESTAR} takes just 20\nobreakdashes--60~s. This latency is comparable to the other stages (data aggregation, trigger generation, alert distribution) in the real\nobreakdashes-time \ac{BNS} analysis. The 32\nobreakdashes-thread configuration is representative of how \ac{BAYESTAR} might be deployed in early Advanced \ac{LIGO}.\footnote{\ac{BAYESTAR} has been successfully ported to the Intel's Many Integrated Core~(MIC) architecture and has been tested in a 500 thread configuration on a system with dual Intel Xeon Phi 5110P coprocessors.}

\section{Case study}

We have completed our description of the \ac{BAYESTAR} algorithm. In the following chapter, we will describe the first results with \ac{BAYESTAR}, a detailed case study of the sky localization capabilities of the earliest configurations of Advanced \ac{LIGO} and Virgo. Because it is typically three orders of magnitude faster than the conventional \ac{MCMC} analysis, for the first time we are able to generate detailed sky maps for thousands of simulated \ac{GW} sources. With such a large sample, we can for the first time unambiguously describe areas and morphologies that should be typical of networks of just two detectors. This case study will be the subject of the next chapter.

\chapter{The first two years of \acs{EM} follow-up with Advanced \acs{LIGO} and Virgo}
\label{chap:first2years}

\author{Leo~P.~Singer\altaffilmark{1},
Larry~R.~Price\altaffilmark{1},
Ben~Farr\altaffilmark{2,3},
Alex~L.~Urban\altaffilmark{4},
Chris~Pankow\altaffilmark{4},
Salvatore~Vitale\altaffilmark{5},
John~Veitch\altaffilmark{6,3},
Will~M.~Farr\altaffilmark{3},
Chad~Hanna\altaffilmark{7,8},
Kipp~Cannon\altaffilmark{9},
Tom~Downes\altaffilmark{4},
Philip~Graff\altaffilmark{10},
Carl-Johan~Haster\altaffilmark{3},
Ilya~Mandel\altaffilmark{3},
Trevor~Sidery\altaffilmark{3},
and~Alberto~Vecchio\altaffilmark{3}}

\altaffiltext{1}{LIGO Laboratory, California Institute of Technology, Pasadena, CA 91125, USA}
\altaffiltext{2}{Department of Physics and Astronomy \& Center for Interdisciplinary Exploration and Research in Astrophysics (CIERA), Northwestern University, Evanston, IL 60208, USA}
\altaffiltext{3}{School of Physics and Astronomy, University of Birmingham, Birmingham, B15 2TT, UK}
\altaffiltext{4}{Leonard E. Parker Center for Gravitation, Cosmology, and Astrophysics, University of Wisconsin\nobreakdashes--Milwaukee, Milwaukee, WI 53201, USA}
\altaffiltext{5}{Massachusetts Institute of Technology, 185 Albany Street, Cambridge, MA 02139, USA}
\altaffiltext{6}{Nikhef, Science Park 105, Amsterdam 1098XG, The Netherlands}
\altaffiltext{7}{Perimeter Institute for Theoretical Physics, Ontario N2L 2Y5, Canada}
\altaffiltext{8}{The Pennsylvania State University, University Park, PA 16802, USA}
\altaffiltext{9}{Canadian Institute for Theoretical Astrophysics, University of Toronto, Toronto, Ontario, M5S 3H8, Canada}
\altaffiltext{10}{NASA Goddard Space Flight Center, Greenbelt, MD, USA}

\attribution{This chapter is reproduced in part from \citet{FirstTwoYears}, which was published under the same title in \textnormal{The Astrophysical Journal}, copyright~\textcopyright{}~2014 The American Astronomical Society. My contributions included developing the rapid sky localization code, planning the simulation campaign, running the detection pipeline for the 2016 scenario, running the rapid sky localization code for both scenarios, reducing and analyzing the results, generating most of the tables and figures as well as the data release, and writing about 95\% of the text. A.U. developed and ran the pipeline for the 2015 scenario, contributed to the data reduction, and created Figure~\ref{fig:mode_hist}. B.F., S.V., J.V., and P.G. ran the stochastic samplers. The other authors contributed to planning, code development, and editing of the paper.}

Several planned optical astronomy projects with a range of fields of view and apertures are preparing to pursue optical counterparts of \ac{BNS} mergers detected by Advanced \ac{LIGO} and Virgo. These include \ac{ZTF}~\citep{ZTF,ZTFBellm,ZTFSmith}, Pan\nobreakdashes-STARRS\footnote{\url{http://pan-starrs.ifa.hawaii.edu/public/}}, BlackGEM\footnote{\url{https://www.astro.ru.nl/wiki/research/blackgemarray}}, and \ac{LSST}~\citep{LSST}, to name a few. Advanced LIGO is scheduled to start taking data in 2015~\citep{LIGOObservingScenarios}. Preparations for joint \ac{EM} and \ac{GW} observations require a complete understanding of when and how well localized the first \ac{GW} detections will be. Plausible scenarios for the evolution of the configuration and sensitivity of the worldwide \ac{GW} detector network as it evolves from 2015 through 2022, as well as rough estimates of sky localization area, are outlined in \citet{LIGOObservingScenarios}.

To provide a more realistic and complete picture, we have conducted Monte Carlo simulations of the 2015 and 2016 detector network configurations, probing the transition from two to three detectors as Advanced Virgo is scheduled to begin science operation. Prior work has focused on various aspects of position reconstruction with advanced \ac{GW} detectors \citep{FairhurstTriangulation,WenLocalizationAdvancedLIGO,FairhurstLocalizationAdvancedLIGO,2011PhRvD..84j4020V,RodriguezBasicParameterEstimation,NissankeLocalization,NissankeKasliwalEMCounterparts,KasliwalTwoDetectors,Grover:2013,SiderySkyLocalizationComparison}, but ours is the first to bring together a large astrophysically motivated population, an educated guess about the detector commissioning timetable, a realistic \ac{SNR} distribution, and the Advanced \ac{LIGO}/Virgo data analysis pipeline itself.

We have simulated hundreds of \ac{GW} events, recovered them with a real\nobreakdashes-time detection pipeline, and generated sky maps using both real\nobreakdashes-time and thorough off\nobreakdashes-line parameter estimation codes that will be operating in 2015 and beyond. This study contains some of the first results with \ac{BAYESTAR}, a rapid Bayesian position reconstruction code that will produce accurate sky maps less than a minute after any \ac{BNS} merger detection. The \textls{LALINFERENCE\_MCMC} \citep{2008ApJ...688L..61V,Raymond:2009}, \textls{LALINFERENCE\_NEST} \citep{LALINFERENCE_NEST}, and \textls{LALINFERENCE\_BAMBI} \citep{BAMBI,SKYNET} stochastic samplers were also used to follow up a subset of detected \ac{GW} events. Though these analyses are significantly more computationally costly than \ac{BAYESTAR}, taking hours to days, they can provide improved sky location estimates when the \ac{GW} signal is very weak in one detector, and also yield not just sky localization but the full multidimensional probability distribution describing the parameters of a circularized compact binary merger. All four algorithms are part of the \textls{LALINFERENCE} library \citep{S6PE}, developed specifically for estimating the parameters of \ac{GW} sources from ground-based detectors. Together, these analyses will be able to provide sky localizations on time scales that enable searching for all expected electromagnetic counterparts of compact binary mergers (except the \ac{GRB} itself).

With the benefit of a much larger sample size, important features of the 2015 and 2016 configurations come into focus. First, we find that, even in 2015 when only the two LIGO detectors are operating (or in 2016 during periods when the Virgo detector is not in science mode), there is at least a 60\% chance of encountering the source upon searching an area of about 200~deg$^2$. Second, many of these two\nobreakdashes-detector events will not be localized to a single simply connected region in the sky. We elucidate two nearly degenerate sky locations, separated by 180$^\circ$, that arise when only the two \ac{LIGO} detectors are operating. When a \ac{GW} source falls within this degeneracy, its sky map will consist of two diametrically opposed islands of probability. Third, in our simulations, we add a third detector, Advanced Virgo, in 2016. Even though, at that time, Virgo is anticipated to be only one\nobreakdashes-third as sensitive as the other two detectors due to differing \ac{LIGO} and Virgo commissioning timetables, we find that coherence with the signal in Virgo generally breaks the previously mentioned degeneracy and shrinks areas to a third of what they were with two detectors. Fourth and most importantly, a picture of a typical early Advanced \ac{LIGO} event emerges, with most occurring in a limited range of Earth\nobreakdashes-fixed locations, and most sky maps broadly fitting a small number of specific morphologies. 

\section{Sources and sensitivity}
\label{sec:first2years-source-sensitivity}

\ac{BNS} systems are the most promising and best understood targets for joint \ac{GW} and \ac{EM} detection. Though rate estimates remain uncertain, ranging from 0.01 to 10~Mpc$^{-3}$\,Myr$^{-1}$, we choose to work with the ``realistic'' rate obtained from \citet{LIGORates} of 1~Mpc$^{-3}$\,Myr$^{-1}$. This rate leads to a \ac{GW} detection rate of 40~yr$^{-1}$ at final Advanced \ac{LIGO} design sensitivity. Some mergers of \acp{NSBH} are also promising sources of \ac{GW} and \ac{EM} emission. Two Galactic \acp{HMXB} have been identified as possible \ac{NSBH} progenitors \citep{CygX1NSBHBBH,CygX3NSBHBBH}. From these can be extrapolated a lower bound on the \ac{GW} detection rate of at least 0.1~yr$^{-1}$ at Advanced \ac{LIGO}'s final design sensitivity, although rates comparable to \ac{BNS} detections are empirically plausible. Black holes in binaries may possess large spins, causing precession during the inspiral. Precession\nobreakdashes-altered phase evolution can aid in parameter estimation~\citep{2008CQGra..25r4011V,2008ApJ...688L..61V,Harry:2013tca,Nitz:2013mxa,Raymond:2009}, but models of waveforms suitable for rapid detection and parameter estimation are still under active development \citep{Blackman:2014maa,Hannam:2013waveform,Taracchini:2013}. As for the binary black hole mergers detectable by Advanced \ac{LIGO} and Virgo, there are currently no compelling mechanisms for electromagnetic counterparts associated with them. We therefore restrict our attention to \ac{BNS} mergers, because they have the best understood rates, \ac{GW} signal models, and data analysis methods.

\subsection{Observing scenarios}

\begin{figure*}
    \begin{minipage}[b]{0.5\textwidth}
        \begin{center}
            \includegraphics[width=\textwidth]{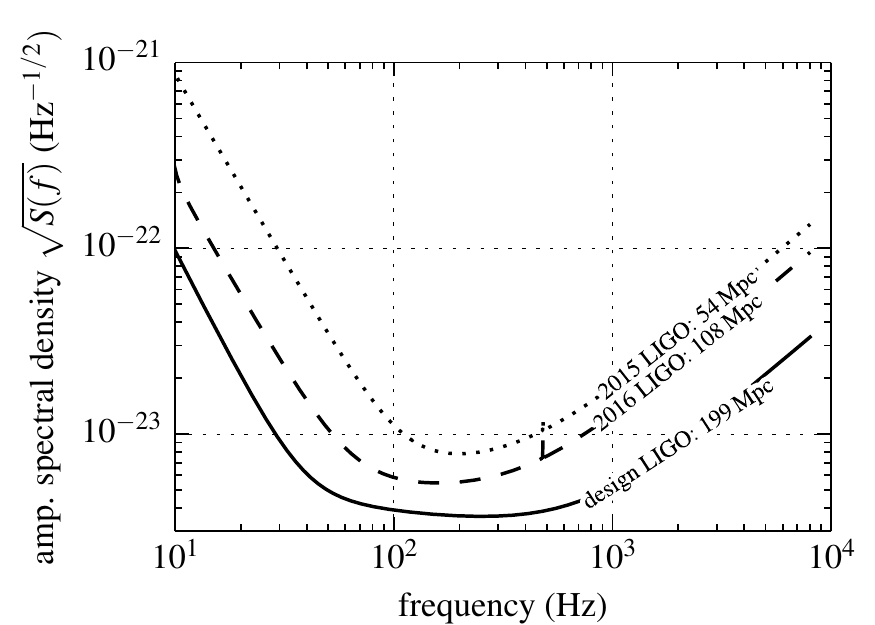}

            (a) \ac{LIGO}
        \end{center}
    \end{minipage}
    \begin{minipage}[b]{0.5\textwidth}
        \begin{center}
            \includegraphics[width=\textwidth]{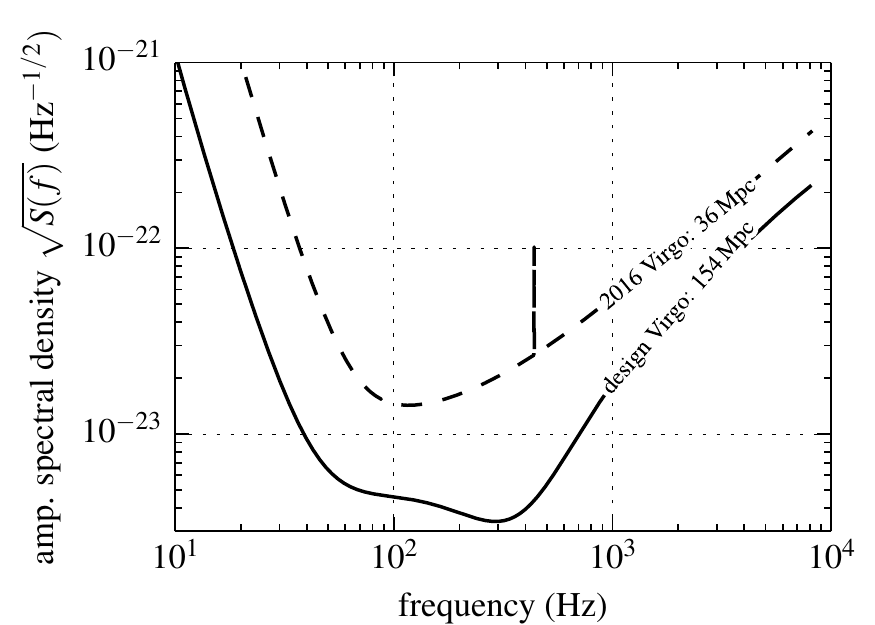}

            (b) Virgo
        \end{center}
    \end{minipage}
    \caption[Amplitude spectral densities for early Advanced \acs{LIGO} configurations]{\label{fig:psds}Model detector noise amplitude spectral density curves. The \ac{LIGO} 2015, 2016, and final design noise curves are shown in the left panel and the Virgo 2016 and final design noise curves in the right panel. The averaged $\rho=8$ range $d_\mathrm{R}$ for $(1.4, 1.4)$~$M_\odot$ \ac{BNS} mergers is given for each detector.}
\end{figure*}

\citet{LIGOObservingScenarios} outline five observing scenarios representing the evolving configuration and capability of the Advanced \ac{GW} detector array, from the first observing run in 2015, to achieving final design sensitivity in 2019, to adding a fourth detector at design sensitivity by 2022. In this study, we focus on the first two epochs. The first, in 2015, is envisioned as a three\nobreakdashes-month science run. \ac{LIGO} Hanford (H) and \ac{LIGO} Livingston (L) Observatories are operating with an averaged $(1.4, 1.4)~M_\odot$ \ac{BNS} range between 40~and~80~Mpc. The second, in 2016\nobreakdashes--2017, is a six\nobreakdashes-month run with H and L operating between 80~and~120~Mpc and the addition of Advanced Virgo (V) with a range between 20~and~60~Mpc. For each configuration, we used model noise \ac{PSD} curves in the middle of the ranges in \citet{LIGOObservingScenarios}, plotted in Figure~\ref{fig:psds}. For H and L, we used the ``early'' and ``mid'' noise curves from \citet{EarlyAdvancedLIGONoiseCurves} for the 2015 and 2016 scenarios respectively. For V in 2016, we used the geometric mean of the high and low curves of \citet{LIGOObservingScenarios}. Final \ac{LIGO} and Virgo design sensitivity is several steps further in the commissioning schedule than that which we consider in this paper.

\subsection{Simulated waveforms}

For each of the two scenarios we made synthetic detector streams by placing post\nobreakdashes-Newtonian inspiral signals into two months of colored Gaussian noise. We used ``SpinTaylorT4'' waveforms, employing the TaylorT4 approximant and accurate to 3.5PN order in phase and 1.5PN order in amplitude \citep{SpinTaylorT4, SpinTaylorT4Erratum, TaylorF2}.\footnote{There is a C language implementation as the function \\ \mbox{\texttt{XLALSimInspiralSpinTaylorT4}} in \textls{lalsimulation}. See \\ Acknowledgments and Appendix~\ref{chap:first2years-data}.} There was an average waiting time of $\approx$100~s between coalescences. At any given time, one \ac{BNS} inspiral signal was entering \ac{LIGO}'s sensitive band while another binary was merging, but both signals were cleanly separated due to their extreme narrowness in time\nobreakdashes-frequency space. The \ac{PSD} estimation used enough averaging that it was unaffected by the overlapping signals. Component masses were distributed uniformly between 1.2~and~1.6~$M_\odot$, bracketing measured masses for components of known \ac{BNS} systems as well as the 1\nobreakdashes-$\sigma$ intervals of the intrinsic mass distributions inferred for a variety of \ac{NS} formation channels~\citep{NeutronStarMass1, NeutronStarMass2}.

We gave each \ac{NS} a randomly oriented spin with a maximum magnitude of $\chi = c\left|\mathbf{S}\right|/Gm^2 \leq 0.05$, where $\mathbf{S}$ is the star's spin angular momentum and $m$ is its mass. This range includes the most rapidly rotating pulsar that has been found in a binary, PSR~J0737\nobreakdashes-3039A \citep{2003Natur.426..531B,DetectingBNSSystemsWithSpin}. However, the fastest\nobreakdashes-spinning millisecond pulsar, PSR~J1748\nobreakdashes-2446ad \citep{FastestSpinningMillisecondPulsar}, has a dimensionless spin parameter of $\sim$0.4, and the theoretical evolutionary limits on \ac{NS} spin\nobreakdashes-up in \ac{BNS} systems are uncertain.

\subsection{Sensitivity to assumptions}

The total detection rate depends on some of these assumptions, and in particular is sensitive to the assumed \ac{NS} mass distribution. As can be seen from Equation~(\ref{eq:horizon-distance}), binaries with the greatest and most symmetric component masses can be detected at the farthest distance. According to Equation~(\ref{eq:f-lso}), for \ac{BNS} systems the merger always occurs at kHz frequencies, on the upward $S(f) \propto f^2$ slope of the noise noise curves in Figure~\ref{fig:psds} in the regime dominated by photon shot noise~\citep{QuantumNoiseSecondGeneration,GWDetectionLaserInterferometry}. As a result, the integral in Equation~(\ref{eq:horizon-distance}) depends only weakly on masses. For equal component masses, the horizon distance scales as $d_\mathrm{H} \propto {m_\mathrm{NS}}^{5/6}$, so the detection rate scales rapidly with mass as $\dot{N} \propto {d_\mathrm{H}}^3 \propto {m_\mathrm{NS}}^{2.5}$.

The normalized distribution of sky localization areas depends only weakly on the distribution of \ac{NS} masses. \citet{FairhurstTriangulation} computes the approximate scaling of sky localization area by considering the Fisher information associated with time of arrival measurement. Valid for moderately high \ac{SNR}, the \ac{RMS} uncertainty in the time of arrival in a given detector is
\begin{equation}
    \label{eq:sigmat}
    \sigma_t = \frac{1}{2 \pi \rho \sqrt{\overline{f^2} - \overline{f}^2}},
\end{equation}
where $\overline{f} = \overline{f^1}$, and $\overline{f^k}$ is the $k$th moment of frequency, weighted by the signal to noise per unit frequency,
\begin{equation}
    \overline{f^k} \approx \left[ \int_{f_1}^{f_2} \frac{|h(f)|^2 f^k}{S(f)}\,\mathrm{d}f \right] \left[ \int_{f_1}^{f_2} \frac{|h(f)|^2}{S(f)}\,\mathrm{d}f \right]^{-1}.
\end{equation}
As in Equation~(\ref{eq:horizon-distance}), we can substitute the approximate inspiral signal spectrum $|h(f)|^2 \propto f^{-7/3}$. The areas then scale as the product of the timing uncertainty in individual detectors, or as simply the square of Equation~(\ref{eq:sigmat}) for a network of detectors with similar (up to proportionality) noise \acp{PSD}. As $m_\mathrm{NS}$ varies from 1 to 2~$M_\odot$, the upper limit of integration $f_2$ given by Equation~(\ref{eq:f-lso}) changes somewhat, but areas change by a factor of less than 1.5. (See also \citealt{Grover:2013} for scaling of sky localization area with mass).

Introducing faster \ac{NS} spins would result in smaller sky localization areas, since orbital precession can aid in breaking \ac{GW} parameter estimation degeneracies \citep{Raymond:2009}. However, rapid spins could require more exotic \ac{BNS} formation channels, and certainly would require using more sophisticated and more computationally expensive \ac{GW} waveforms for parameter estimation.

\subsection{Source locations}

Source locations were random and isotropic, and uniform in distance cubed. The source distribution was cut off at the $\rho=5$, $(1.6, 1.6)~M_\odot$ horizon distance, far enough away that the selection of detected binaries was determined primarily by the sensitivity of the instruments and the detection pipeline, not by the artificial distance boundary.

\subsection{Duty cycle}

Following \citet{LIGOObservingScenarios}, we assumed that each detector had an independent and random $d=80\%$ duty cycle. In the 2015, HL configuration, this implies that both detectors are in operation $d^2=64\%$ of the time. In 2016, there are three detectors operating $d^3 = 51.2\%$ of the time and each of three pairs operating $d^2(1-d)=12.8\%$ of the time. We do not simulate any of the time when there are one or zero detectors operating, but instead fold this into conversion factors from our Monte Carlo counts to detection rates.

\section{Detection and position reconstruction}
\label{sec:detection-and-position-reconstruction}

Searches for \acp{GW} from compact binaries \citep{FINDCHIRP,ihope} employ banks of matched filters, in which the data from all of the detectors are convolved with an array of template waveforms. The output of each filter is the instantaneous \ac{SNR} with respect to that template in that detector. An excursion above a threshold \ac{SNR} in two or more detectors with exactly the same binary parameters and within approximately one light\nobreakdashes-travel time between detectors is considered a coincidence. Coincidences may be accidental, due to chance noise fluctuations or, in real \ac{GW} data streams, environmental disturbances and instrument glitches. Coincidences with sufficiently high $\rho_\mathrm{net}$ (root\nobreakdashes-sum\nobreakdashes-square of the \ac{SNR} in the individual detectors) are considered detection candidates. A $\chi^2$ statistic is used to aid in separating the true, astrophysical signals from accidental coincidences or false positives~\citep{AllenChiSq, HannaThesis, Cannon:2012zt}.

Offline inspiral searches used in past \ac{LIGO}/Virgo science runs will be computationally strained in Advanced \ac{LIGO}/Virgo due to denser template banks and \ac{BNS} signals that remain in band for up to $\sim 10^3$~s. To address these issues and achieve latencies of $\lesssim 1$~minute, a rapid matched\nobreakdashes-filter detection pipeline called \textls{GSTLAL\_INSPIRAL}~\citep{Cannon:2011vi} has been developed. To mimic Advanced \ac{LIGO}/Virgo observations as closely as possible, we used \textls{GSTLAL\_INSPIRAL} to extract simulated detection candidates from our two\nobreakdashes-month data streams.

\subsection{Template waveforms}

The templates were constructed from a frequency domain, post\nobreakdashes-Newtonian model describing the inspiral of two compact objects, accurate to 3.5 post\nobreakdashes-Newtonian order in phase and Newtonian order in amplitude \citep{TaylorF2}.\footnote{These are in \textls{lalsimulation} as the function \\ \mbox{\texttt{XLALSimInspiralTaylorF2}}. See acknowledgements and Appendix~\ref{chap:first2years-data}.} These waveforms neglect spins entirely. This is known to have a minimal impact on detection efficiency for \ac{BNS} sources with low spins \citep{DetectingBNSSystemsWithSpin}. These waveforms are adequate for recovering the weakly spinning simulated signals that we placed into the data stream.

\subsection{Detection threshold}

In our study, we imposed a single-detector threshold \ac{SNR} of 4. A simulated signal is then considered to be detected by \textls{GSTLAL\_INSPIRAL} if it gives rise to a coincidence with sufficiently low false alarm probability as estimated from the \ac{SNR} and $\chi^2$ values. We follow the lead of \citet{LIGOObservingScenarios} in adopting a \acf{FAR} threshold of $\text{\ac{FAR}} \leq 10^{-2}$~yr$^{-1}$. \citet{LIGOObservingScenarios} claim that in data of similar quality to previous \ac{LIGO}/Virgo science runs, this \ac{FAR} threshold corresponds to a network \ac{SNR} threshold of $\rho_\mathrm{net} \geq 12$. Since our data is Gaussian and perfectly free of glitches, to obtain easily reproducible results we imposed a second explicit detection cut of $\rho_\mathrm{net} \geq 12$. We find that our joint threshold on \ac{FAR} and \ac{SNR} differs negligibly from a threshold on \ac{SNR} alone. Because any given simulated signal will cause multiple coincidences at slightly different masses and arrival times, for each simulated signal we keep only the matching candidate with the lowest \ac{SNR}.

\subsection{Sky localization and parameter estimation}

All detection candidates are followed up with rapid sky localization by \ac{BAYESTAR} and a subset were followed up with full parameter estimation by the \textls{LALINFERENCE\_MCMC}/\textls{NEST}/\textls{BAMBI} stochastic samplers. The three different stochastic samplers all use the same likelihood, but serve as useful cross\nobreakdashes-verification. Both \ac{BAYESTAR} and the three stochastic samplers are coherent (exploiting the phase consistency across all detectors) and Bayesian (making use of both the \ac{GW} observations and prior distributions over the source parameters). They differ primarily in their input data.

\ac{BAYESTAR}'s likelihood function depends on only the information associated with the triggers comprising a coincidence: the times, phases, and amplitudes on arrival at each of the detectors. \ac{BAYESTAR} exploits the leading\nobreakdashes-order independence of errors in the extrinsic and intrinsic parameters by holding the masses fixed at the values estimated by the detection pipeline. Marginalized posterior distributions for the sky positions of sources are produced by numerically integrating the posterior in each pixel of the sky map. Because \ac{BAYESTAR}'s analysis explores only a small sector of the full parameter space, never performs costly evaluations of the post-Newtonian \ac{GW} waveforms, and uses highly tuned standard numerical quadrature techniques, it takes well under a minute (see Figure~\ref{fig:timeline}).

On the other hand, the likelihood function used for the stochastic samplers depends on the full \ac{GW} data, and is the combination of independent Gaussian distributions for the residual in each frequency bin after model subtraction. Posterior distributions for the sky position are produced by sampling the full parameter space of the signal model, then marginalizing over all parameters but the sky location. This method requires the generation of a model waveform for each sample in parameter space, making it far more expensive than the \ac{BAYESTAR} approach, but independent of the methods and models used for detection. Most importantly, intrinsic parameters (including spins) can be estimated using these higher-latency methods. For the purposes of this study, parameter estimation used the same frequency\nobreakdashes-domain, non\nobreakdashes-spinning waveform approximant as the detection pipeline. Analyses that account for the spin of the compact objects are more costly, currently taking weeks instead of days to complete, and will be the subject of a future study.

\section{Results}

Of $\sim$100,000 simulated sources spread across the 2015 and 2016 scenarios, $\approx1000$ events survived as confident \ac{GW} detections.\footnote{There were slightly fewer surviving events in the 2016 configuration than in the 2015 configuration. This is because adding a third detector required us to apportion the two months of Gaussian noise to different combinations of detectors. In the 2015 simulation, all two months of data were allocated to the HL network. In 2016 about 43 days were devoted to the HLV and HL configurations, with the remaining 17 days of HV and LV mode contributing few detections.} No false alarms due to chance noise excursions survived our detection threshold; all events which should have been detectable were detected. We constructed probability sky maps using \ac{BAYESTAR} for all events and using \textls{LALINFERENCE\_NEST}/\textls{MCMC} for a randomly selected subsample of 250 events from each scenario.\footnote{Results from \textls{LALINFERENCE\_BAMBI} are not shown in our plots because this sampler was run for only 30 events, and the sampling error bars would overwhelm the plots.}\footnote{The three stochastic samplers \textls{LALINFERENCE\_NEST}/\textls{MCMC}/\textls{BAMBI} were interchangeable to the extent that they used the same likelihood and produced sky maps that agreed with each other.} The top four panels (a, b, c, d) of Figure~\ref{fig:area_hist} show cumulative histograms of the areas in deg$^2$ inside of the smallest 50\% and 90\% confidence regions for each event, for both sky localization methods. These contours were constructed using a `water\nobreakdashes-filling' algorithm: we sampled the sky maps using equal\nobreakdashes-area \acs{HEALPix} \citep[\aclu{Hierarchical Equal Area isoLatitude Pixelization};][]{HEALPix} pixels, ranked the pixels from most probable to least, and finally counted how many pixels summed to a given total probability. In the bottom two panels (e), (f) of Figure~\ref{fig:area_hist}, we also show a histogram of the smallest such constructed region that happened to contain the true location of each simulated source. We call this the searched area.

Panels (a--d) and (e), (f) may be thought of as measuring precision and accuracy respectively. The former measure how dispersed or concentrated each individual sky map is, while the latter describe how far the localization is from the true sky position. The 90\% area histograms and the searched area histograms also answer different but complementary questions that relate to two different strategies for following up \ac{LIGO}/Virgo events. One might decide in 2015 to search for optical counterparts of all \ac{GW} events whose 90\% areas are smaller than, for example, 200~deg$^2$. By finding 200~deg$^2$ on the horizontal axis of the 90\% area histogram, one would find that this corresponds to following up $10\%$ of all \ac{GW} detections. On the other hand, one might decide to always search the most probable 200~deg$^2$ area for every \ac{GW} event, corresponding to a different confidence level for every event. In this case, one would find 200~deg$^2$ on the horizontal axis of the searched area histogram, and find that this strategy would enclose the true location of the \ac{GW} source $64\%$ of the time.

\begin{figure}
    \begin{center}
        \includegraphics{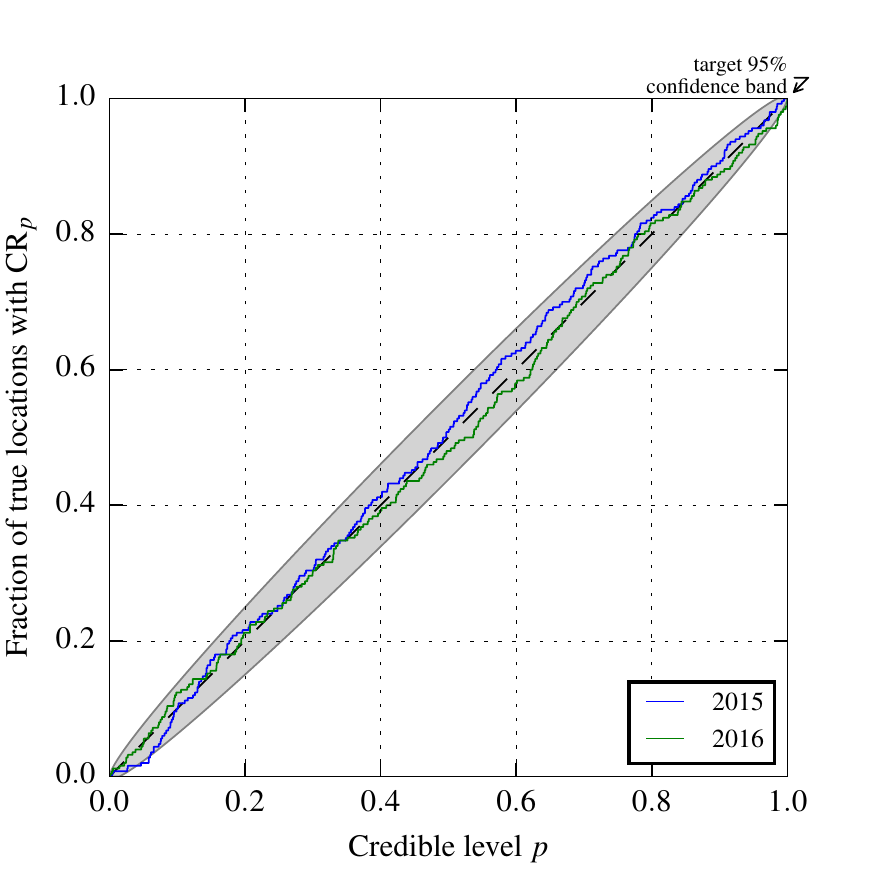}
    \end{center}
    \caption[\acs{BAYESTAR} $P$\nobreakdashes--$P$ plots]{\label{fig:pp}$P$\nobreakdashes--$P$ plots for \ac{BAYESTAR} localizations in the 2015 and 2016 configurations. The gray lozenge around the diagonal is a target 95\% confidence band derived from a binomial distribution.}
\end{figure}

\begin{figure*}
    \begin{minipage}[b]{0.5\textwidth}
        \begin{center}
            \includegraphics[width=\textwidth]{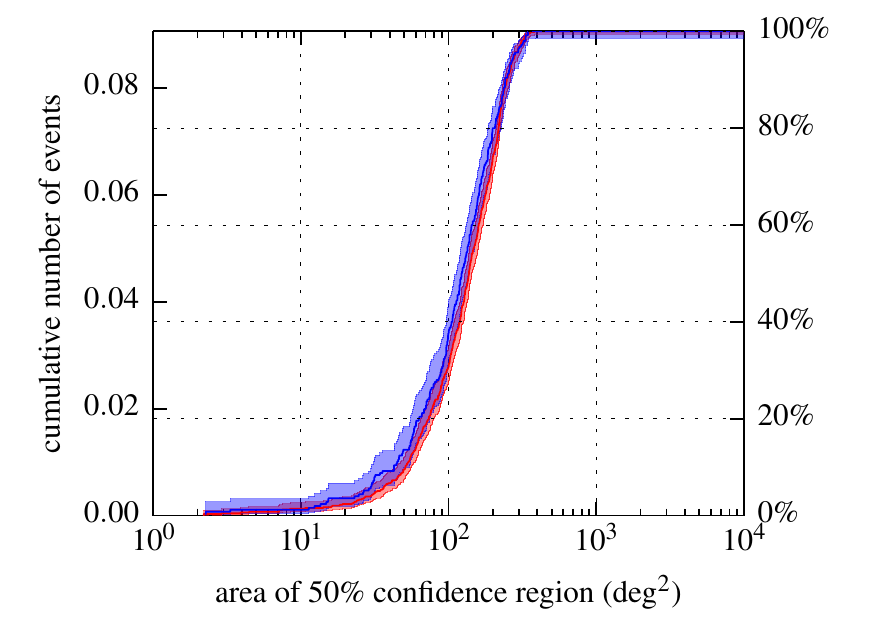}

            (a) 2015, HL
        \end{center}
    \end{minipage}
    \begin{minipage}[b]{0.5\textwidth}
        \begin{center}
            \includegraphics[width=\textwidth]{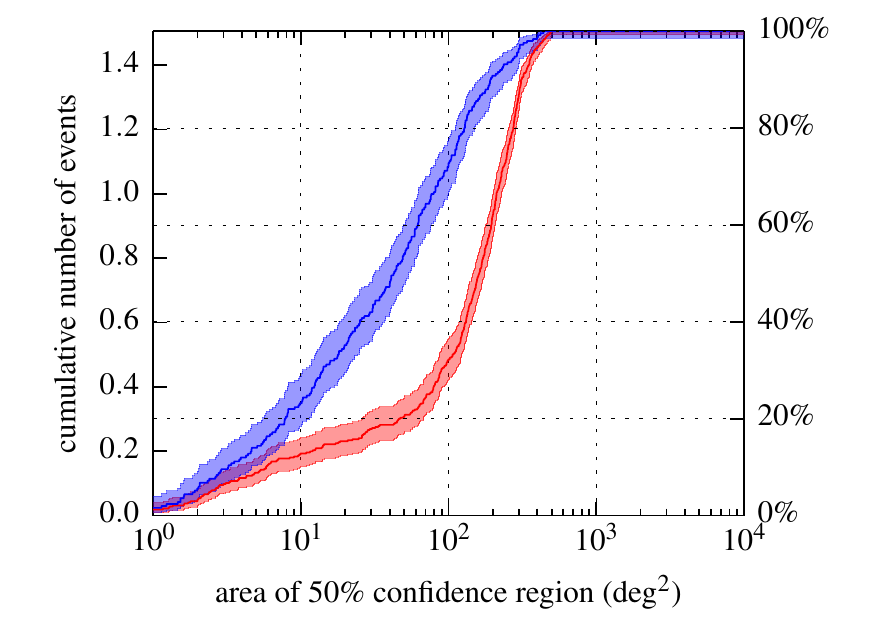}

            (b) 2016, HLV
        \end{center}
    \end{minipage}

    \begin{minipage}[b]{0.5\textwidth}
        \begin{center}
            \includegraphics[width=\textwidth]{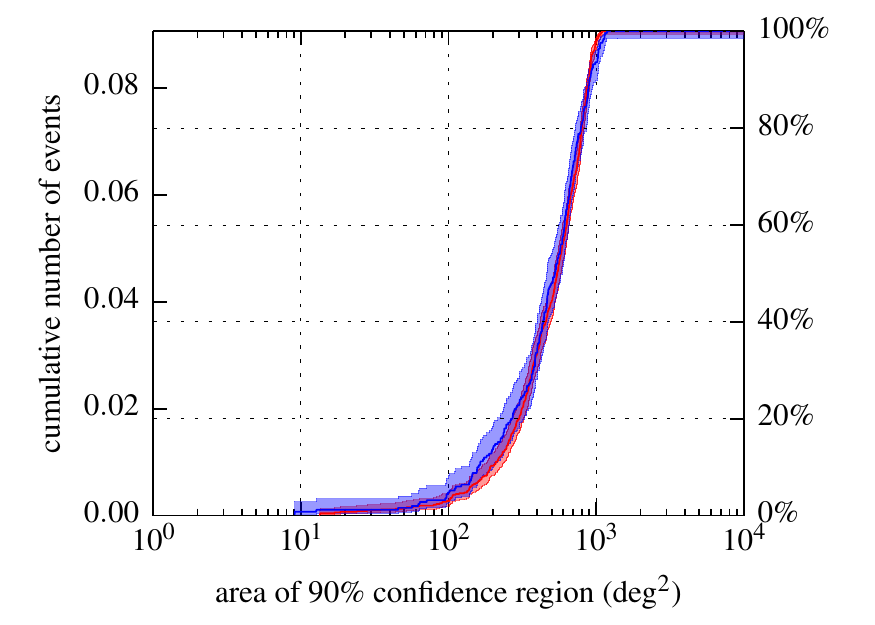}

            (c) 2015, HL
        \end{center}
    \end{minipage}
    \begin{minipage}[b]{0.5\textwidth}
        \begin{center}
            \includegraphics[width=\textwidth]{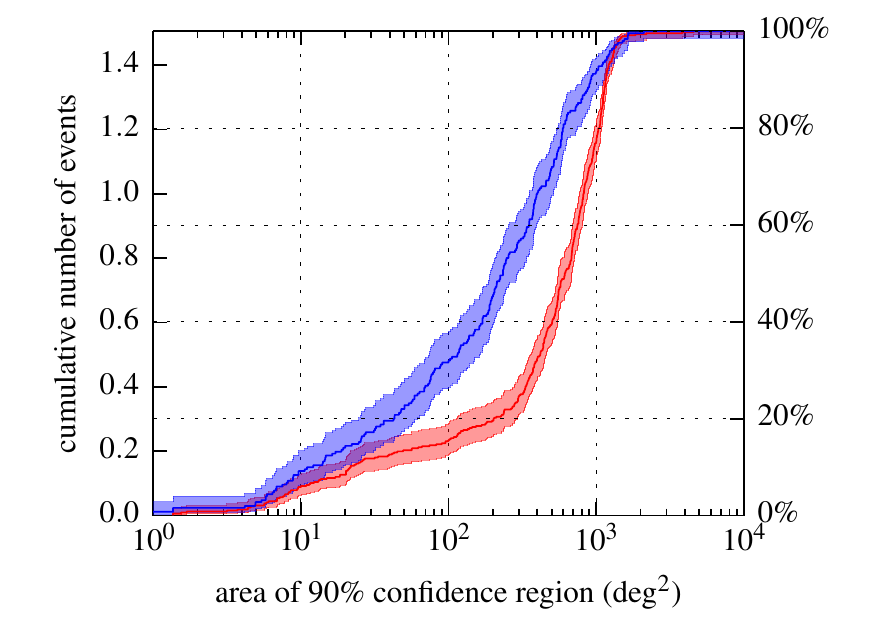}

            (d) 2016, HLV
        \end{center}
    \end{minipage}

    \begin{minipage}[b]{0.5\textwidth}
        \begin{center}
            \includegraphics[width=\textwidth]{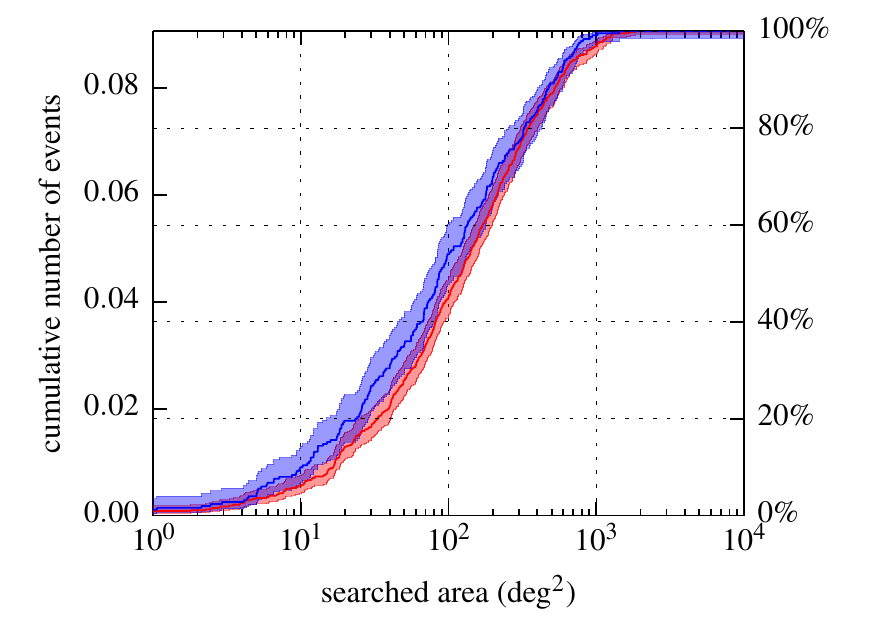}

            (e) 2015, HL
        \end{center}
    \end{minipage}
    \begin{minipage}[b]{0.5\textwidth}
        \begin{center}
            \includegraphics[width=\textwidth]{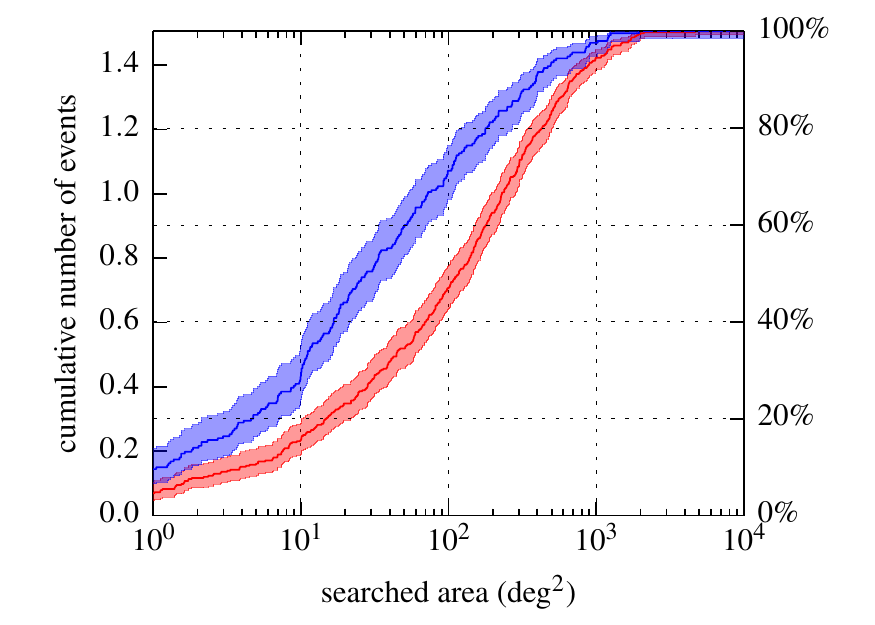}

            (f) 2016, HLV
        \end{center}
    \end{minipage}
    \caption[Cumulative histograms of sky areas]{\label{fig:area_hist}Cumulative histogram of sky localization areas in the 2015 (HL) and 2016 (HLV) scenarios. Plots in the left column (a,~c,~e) refer to the 2015 configuration and in the right column (b,~d,~f) to the 2016 configuration. The first row (a,~b) shows the area of the 50\% confidence region, the second row (c,~d) shows the 90\% confidence region, and the third row (e,~f) shows the ``searched area,'' the area of the smallest confidence region containing the true location of the source. The red lines comprise all detections and their sky maps produced with \ac{BAYESTAR}, and the blue lines represent sky maps for the random subsample of 250 detections analyzed with \textls{LALINFERENCE\_NEST}/\textls{MCMC}. The light shaded region encloses a 95\% confidence interval accounting for sampling errors \citep[computed from the quantiles of a beta distribution;][]{BinomialConfidenceIntervalsAstronomy}. The left axes show the number of detections localized within a given area assuming the ``realistic'' \ac{BNS} merger rates from \citep{LIGORates}. The right axes show the percentage out of all detected events.}
\end{figure*}

One might naively expect that self\nobreakdashes-consistency would require the 90\% confidence area and searched area histograms to intersect at 90\% of detections, but this is not generally required because the posteriors of different events have widely different dimensions. However, it is true that 90\% of sources should be found within their respective 90\% confidence contours. This can be formalized into a graphical self\nobreakdashes-consistency test called a probability\nobreakdashes--probability ($P$\nobreakdashes--$P$) plot (see \citealt{SiderySkyLocalizationComparison} for applications in \ac{GW} parameter estimation). For each event, one follows the water\nobreakdashes-filling procedure to walk outward from the most probable pixel until one reaches the pixel that contains the true sky location, noting the total probability $p$ that has been traversed. A cumulative histogram of $p$ should be diagonal, within a binomial confidence band. It is already well established that \textls{LALINFERENCE\_NEST}/\textls{MCMC}/\textls{BAMBI} satisfy the $P$\nobreakdashes--$P$ plot test when deployed with accurate templates and reasonable priors. We found at first that \ac{BAYESTAR}'s $P$\nobreakdashes--$P$ plots tended to sag below the diagonal, indicating that though the accuracy (i.e., searched area) was comparable to the stochastic samplers, the \emph{precision} was overstated with confidence intervals that were only about 70\% of the correct area. This was rectified by pre\nobreakdashes-scaling the \acp{SNR} from \textls{GSTLAL\_INSPIRAL} by a factor of 0.83 prior to running \ac{BAYESTAR}. This correction factor suggests that, for example, an \ac{SNR} 10 trigger from \textls{GSTLAL\_INSPIRAL} has the \emph{effective information content} of an \ac{SNR} 8.3 signal. The missing information may be due to losses from the discreteness of the template bank, from the \ac{SVD}, from mismatch between the matched\nobreakdashes-filter templates and the simulated signals, from the small but nonzero correlations between masses and intrinsic parameters, or from elsewhere within the detection pipeline. The correction is hard\nobreakdashes-coded into the rapid localization. With it, the $P$\nobreakdashes--$P$ plots are diagonalized without negatively affecting the searched area (see Figure~\ref{fig:pp}).

The left\nobreakdashes-hand axes of all four panels of Figure~\ref{fig:area_hist} show the expected cumulative number of detections, assuming the `realistic' \ac{BNS} merger rates from~\citet{LIGORates}. We stress that the absolute detection rate might be two orders of magnitude smaller or one order of magnitude higher due to the large systematic uncertainty in the volumetric rate of \ac{BNS} mergers, estimated from population synthesis and the small sample of Galactic binary pulsars~\citep{LIGORates}. An additional source of uncertainty in the detection rates is the Advanced \ac{LIGO}/Virgo commissioning schedule given in \citet{LIGOObservingScenarios}. The proposed sensitivity in the 2016 scenario may be considered a plausible upper bound on the performance of the \ac{GW} detector network in 2015, if commissioning occurs faster than anticipated. Likewise, the quoted sensitivity in the 2015 scenario is a plausible lower bound on the performance in 2016. The right\nobreakdashes-hand axes show the cumulative percentage of all detected sources. These percentages depend only on the gross features of the detector configuration and not on the astrophysical rates, so are relatively immune to the systematics described above.

Table~\ref{table:summary} summarizes these results.

\begin{deluxetable*}{rr|rr|rr}
    \tablecaption{\label{table:summary}Detection rate and sky localization accuracy}
    \tablehead{\colhead{} & \colhead{} & \multicolumn{2}{c}{2015} & \multicolumn{2}{c}{2016}}
    \startdata
    \multicolumn{2}{r}{Detectors} & \multicolumn{2}{c}{HL} & \multicolumn{2}{c}{HLV} \\
    \multicolumn{2}{r}{\ac{LIGO} (HL) \ac{BNS} range} & \multicolumn{2}{c}{54 Mpc} & \multicolumn{2}{c}{108 Mpc} \\
    \multicolumn{2}{r}{Run duration} & \multicolumn{2}{c}{3 months} & \multicolumn{2}{c}{6 months} \\
    \multicolumn{2}{r}{No. detections} & \multicolumn{2}{c}{0.091} & \multicolumn{2}{c}{1.5} \\
    \tableline
    \colhead{} & \colhead{} & \colhead{rapid} & \colhead{full PE} & \colhead{rapid} & \colhead{full PE} \\
    \tableline
    
\multirow{5}{*}{\parbox{1.4cm}{\raggedleft Fraction 50\% CR Smaller than}}
& 5\,deg$^2$&---&---&9\%&14\% \\
& 20\,deg$^2$&2\%&3\%&15\%&35\% \\
& 100\,deg$^2$&30\%&37\%&32\%&72\% \\
& 200\,deg$^2$&74\%&80\%&62\%&90\% \\
& 500\,deg$^2$&100\%&100\%&100\%&100\% \\
\tableline
\multirow{5}{*}{\parbox{1.4cm}{\raggedleft Fraction 90\% CR Smaller than}}
& 5\,deg$^2$&---&---&2\%&2\% \\
& 20\,deg$^2$&---&---&8\%&14\% \\
& 100\,deg$^2$&3\%&4\%&15\%&31\% \\
& 200\,deg$^2$&10\%&13\%&19\%&45\% \\
& 500\,deg$^2$&44\%&48\%&39\%&71\% \\
\tableline
\multirow{5}{*}{\parbox{1.4cm}{\raggedleft Fraction Searched Area Smaller than}}
& 5\,deg$^2$&3\%&4\%&11\%&20\% \\
& 20\,deg$^2$&14\%&19\%&23\%&44\% \\
& 100\,deg$^2$&45\%&54\%&47\%&71\% \\
& 200\,deg$^2$&64\%&70\%&62\%&81\% \\
& 500\,deg$^2$&87\%&89\%&83\%&93\% \\
\tableline
\multirow{3}{*}{Median Area $\bigg\{$}
& 50\% CR& 138\,deg$^2$& 124\,deg$^2$& 162\,deg$^2$& 43\,deg$^2$ \\
& 90\% CR& 545\,deg$^2$& 529\,deg$^2$& 621\,deg$^2$& 235\,deg$^2$ \\
& searched& 123\,deg$^2$& 88\,deg$^2$& 118\,deg$^2$& 29\,deg$^2$

    \enddata
    \tablecomments{This table is a summary of the 2015 and 2016 scenarios, listing the participating detectors, \ac{BNS} horizon distance, run duration, and fractions of events localized within 5, 20, 100, 200, or 500~deg$^2$. A dash (---) represents less than 1\% of detections.}
\end{deluxetable*}

\subsection{2015}
\label{sec:2015}

Our 2015 scenario assumes two detectors (HL) operating at an anticipated range of 54~Mpc. About 0.1 detectable \ac{BNS} mergers are expected, though there are nearly two orders of magnitude systematic uncertainty in this number due to the uncertain astrophysical rates. A detection in 2015 is possible, but likely only if the \ac{BNS} merger rates or the detectors' sensitivity are on the modestly optimistic side. A typical or median event (with a localization area in the 50th percentile of all detectable events) would be localized to a 90\% confidence area of $\sim 500$~deg$^2$.

We find that the area histograms arising from the \ac{BAYESTAR} rapid sky localization and the full parameter estimation agree within sampling errors, and that the sky maps resulting from the two analyses are comparable for any individual event. Put differently, the rapid sky localization contains nearly all of the information about sky localization for these events, with the full probability distributions over masses and spins becoming available when the stochastic samplers finish on a timescale of a day.

\begin{figure*}
    \begin{minipage}[b]{0.5\textwidth}
        \begin{center}
            \includegraphics[width=\textwidth]{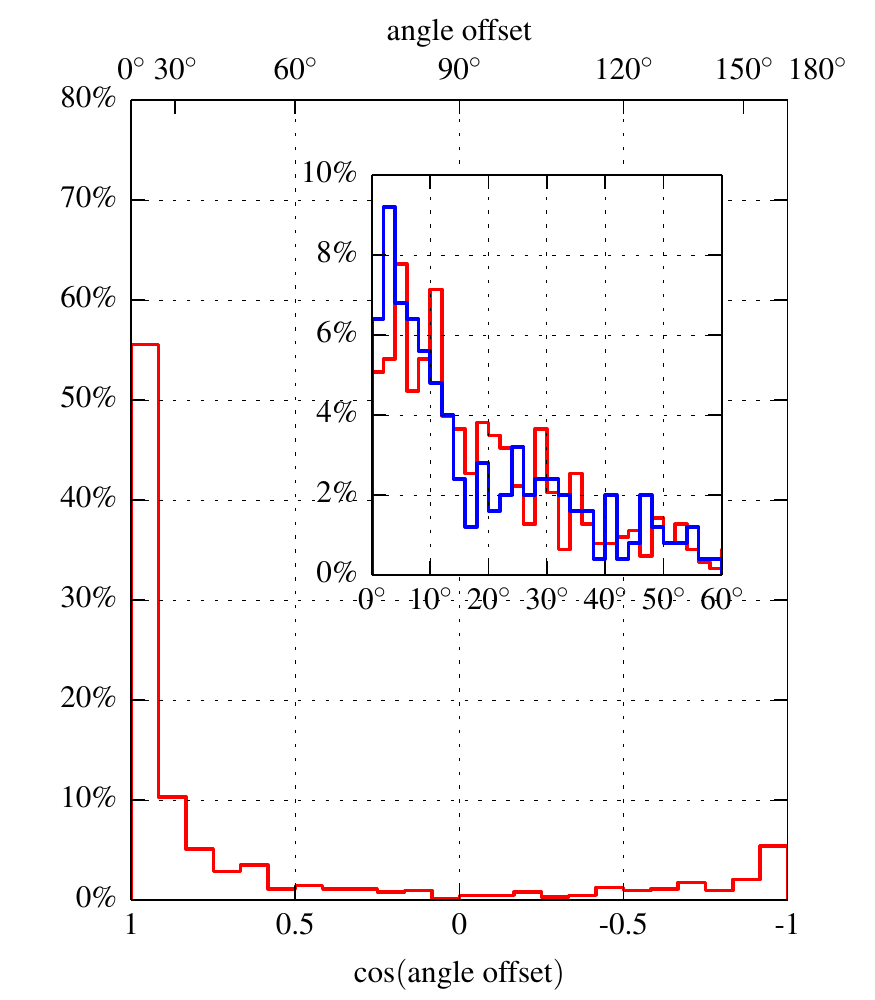}

            (a) 2015, HL
        \end{center}
    \end{minipage}
    \begin{minipage}[b]{0.5\textwidth}
        \begin{center}
            \includegraphics[width=\textwidth]{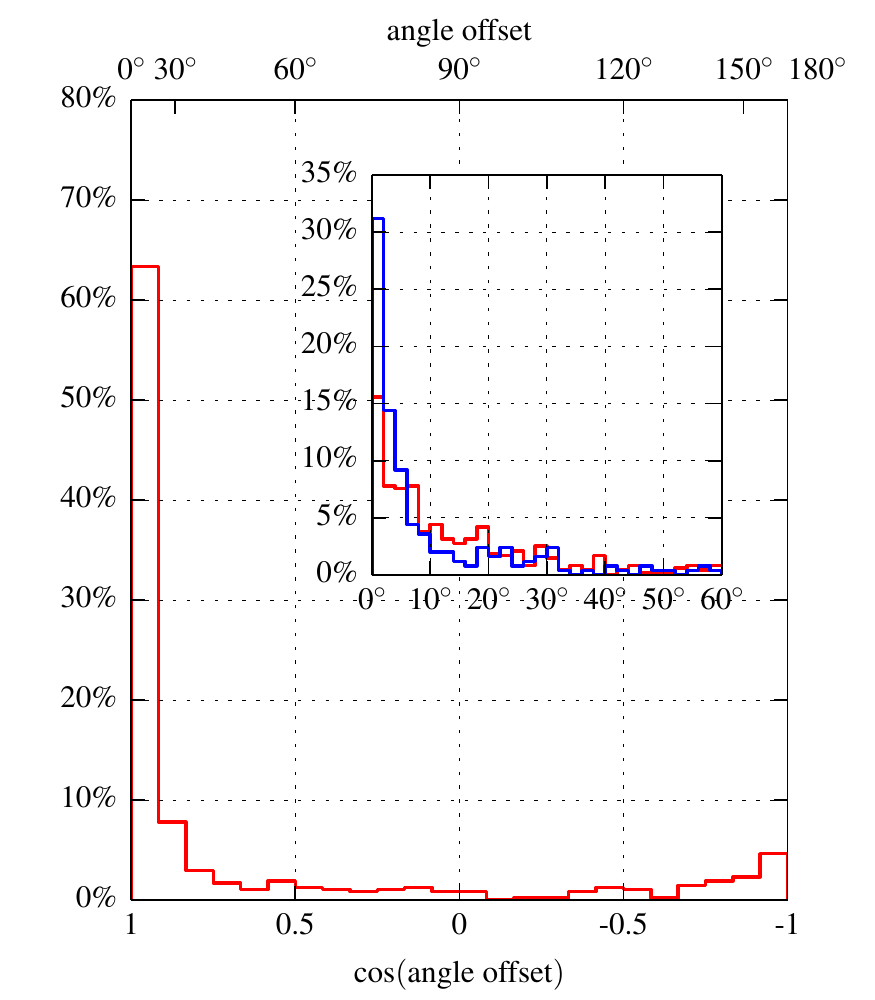}

            (b) 2016, HLV
        \end{center}
    \end{minipage}
    \caption[Cumulative histograms of angle offsets]{\label{fig:offset_hist}Normalized histogram of the cosine angular separation between the location of the simulated \ac{GW} source and the \ac{MAP} location estimate, for (a) the 2015 configuration and (b) the 2016 configuration. The red line encompasses all detections and their \ac{BAYESTAR} localizations, and the blue line the subsample of 250 events analyzed by \textls{LALINFERENCE\_NEST}/\textls{MCMC}. The inset shows the distribution of angle offsets for angles less than 60$^\circ$.}
\end{figure*}

Figure~\ref{fig:offset_hist}(a) shows a histogram of the cosine of the angular separation between the true location of the simulated \ac{GW} source and the \ac{MAP} estimate (the mode of the sky map, or the most probable location). The main feature is a peak at low separation. However, there is a second peak at the polar opposite of the true location, 180$^\circ$ away; about 15\% of events are recovered between 100 and 180$^\circ$ away from the true location.

Correspondingly, for any one event, it is common to find the probability distributed across two antipodal islands on opposite sides of the mean detector plane. We define this plane by finding the average of the two vectors that are normal to the planes of the two detectors' arms, and then taking the plane to which this vector is normal. This plane partitions the sky into two hemispheres. We find that one hemisphere is favored over the other by less than 20\% (which is to say that the odds favoring one hemisphere over the other are as even as 60\%/40\%) for 20\% of events.

The second peak admits a simple explanation as an unavoidable degeneracy due to the relative positions of the H and L interferometers. Before the Hanford and Livingston sites were selected, it was decided that the detectors' arms would be as closely aligned as possible~\citep[Section V\nobreakdashes-C\nobreakdashes-2]{LIGOProposal}. Significant misalignment would have created patches of the sky that were accessible to one detector but in a null of the other detector's antenna pattern, useless for a coincidence test.

\begin{figure*}
    \caption[The HL degeneracy]{\label{fig:degeneracy}HL degeneracy. This, like all sky plots in this paper, is a Mollweide projection in geographic coordinates to emphasize spatial relationships with respect to the Earth\nobreakdashes-fixed \ac{GW} detector network as well as possible ground\nobreakdashes-based telescope sites. Pluses denote the locations of signals whose best-estimate locations are offset by $\geq 100 ^\circ$, comprising the large\nobreakdashes-offset peak that is evident in Figure~\ref{fig:offset_hist}(a). The locations of zero time delay (simultaneous arrival at the H and L detectors) is shown as a thick black line. Shading indicates the \ac{RMS} network antenna pattern, with darker areas corresponding to high sensitivity and white corresponding to null sensitivity.}
    \includegraphics[width=\textwidth]{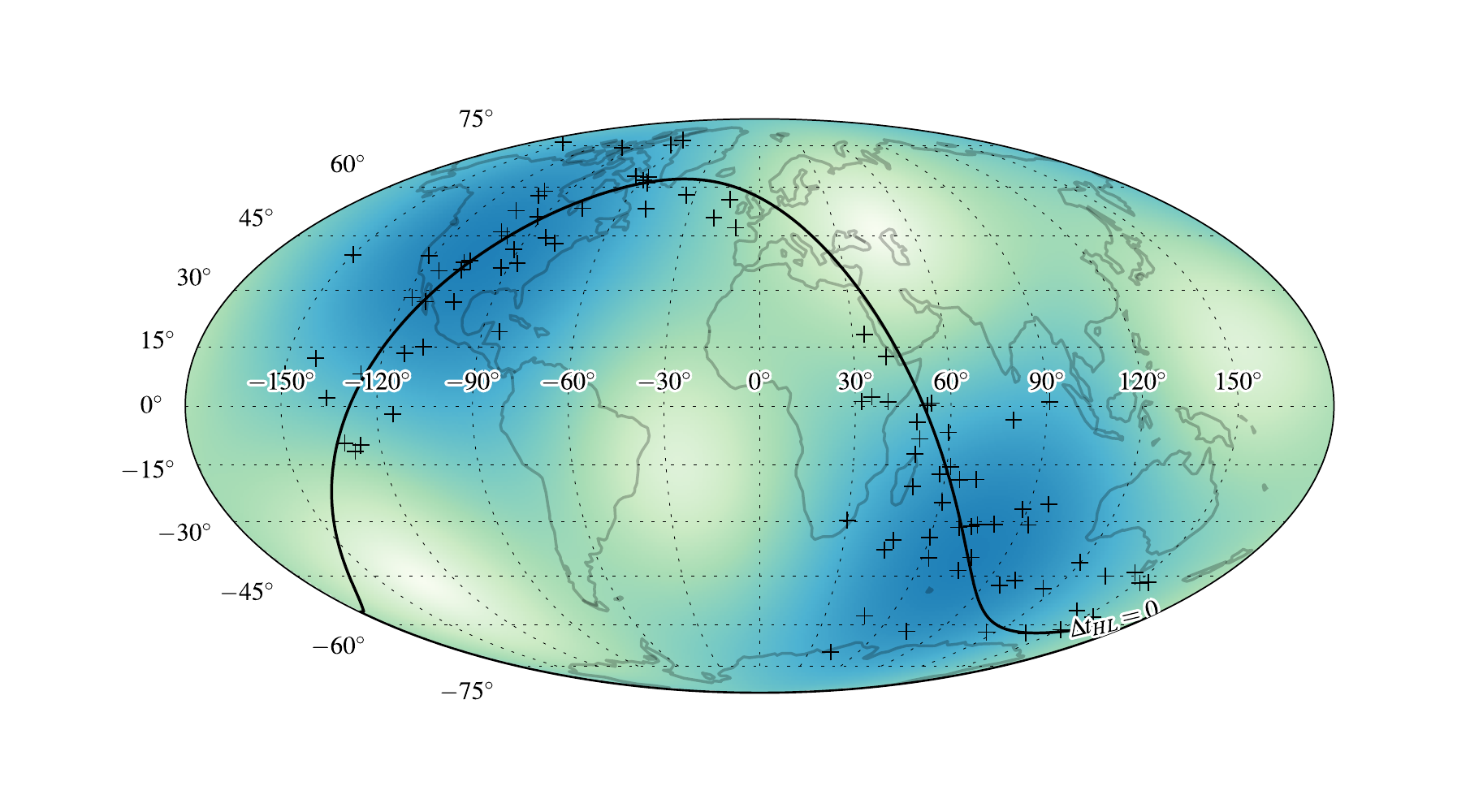}
\end{figure*}

The near alignment maximized the range of the detectors in coincidence, though at a certain expense of parameter estimation. Observe that the sensitivity of an interferometric \ac{GW} detector is identical at antipodal points (i.e., symmetric under all rotations by 180$\arcdeg$). Therefore, any source that lies in the plane of zero time delay between the detectors is always localized to two opposite patches. Because the HL detectors were placed nearby (at continental rather than intercontinental distances) on the surface of the Earth so as to keep their arms nearly coplanar, their combined network antenna pattern has two maxima that lie on opposite sides of that great circle. As a consequence, a large fraction of sources are localized to two islands of probability that cannot be distinguished based on time or amplitude on arrival. See Figure~\ref{fig:degeneracy} for an illustration of these two degenerate patches.

A second undesirable side effect of the aligned antenna patterns is that \ac{GW} polarization, observed as the phase difference on arrival at these two detectors, is of limited help for parameter estimation.

\begin{figure*}
    \caption[A typical bimodal localization, circa 2015]{\label{fig:typical}Localization of a typical circa 2015 \ac{GW} detection. This is a Mollweide projection in geographic coordinates. Shading is proportional to posterior probability per deg$^2$. This is a moderately loud event with $\rho_\mathrm{net}=15.0$, but its 90\% confidence area of 630~deg$^2$ is fairly typical, in the 60th percentile of all detections. The sky map is bimodal with two long, thin islands of probability over the northern and southern antenna pattern maxima. Neither mode is strongly favored over the other. Each island is forked like a snake's tongue, with one fork corresponding to the binary having face\nobreakdashes-on inclination ($\iota \approx 0^\circ$) and the other fork corresponding to face\nobreakdashes-off ($\iota \approx 180^\circ$). \emph{This is event ID 18951 from Tables~\ref{table:2015-sim}~and~\ref{table:2015-coinc} and the online material (see Appendix~\ref{chap:first2years-data} for more details).}}
    \includegraphics[width=\textwidth]{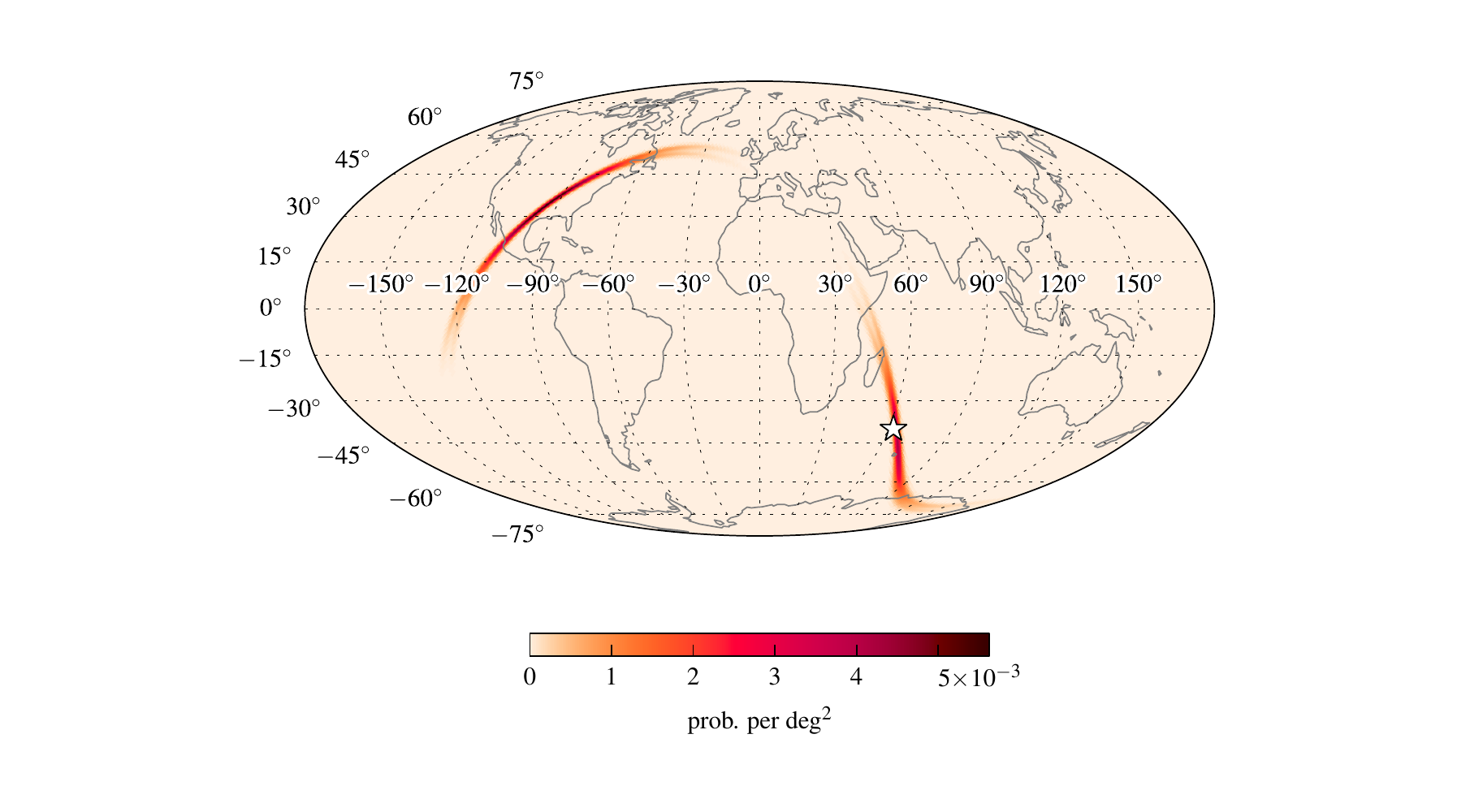}
\end{figure*}

A fairly typical sky map morphology, even at modestly high $\mathrm{SNR}_\mathrm{net}$, will consist of two extended arc-shaped modes, one over North America and a mirror image on the opposite side of the Earth. See Figure~\ref{fig:typical} for a typical event exhibiting this degeneracy. In this example, it is also possible to discern two narrow stripes resembling the forked tongue of a snake. This is a reflection of the HL network's limited polarization sensitivity. It occurs when the \ac{GW} phases on arrival support two different binary inclination angles, with the orbital plane nearly facing the observer but with opposite handedness (usually peaked at $\iota \approx 30\arcdeg$ and $\iota \approx 150\arcdeg$; see \citealt{ShutzThreeFiguresOfMerit}). The two forks cross at a sky location where the \ac{GW} data cannot distinguish between a clockwise or counterclockwise orbit.

The HL degeneracy is even apparent in earlier works on localization of \ac{GW} bursts with networks of four or more detectors: \citet{CWBLocalization} drew a connection between accurate position reconstruction and sensitivity to both the `$+$' and `$\times$' \ac{GW} polarizations, and noted that the close alignment of the HL detector network adversely affects position reconstruction. (They did not, however, point out the common occurrence of nearly $180^\circ$ errors, or note that the worst \ac{GW} localizations paradoxically occur where the HL network's sensitivity is the greatest.)

\begin{figure*}
    \caption[A typical unimodal localization, circa 2015]{\label{fig:typical-unimodal}Localization of a typical circa 2015 \ac{GW} detection. This is a Mollweide projection in geographic coordinates. Shading is proportional to posterior probability per deg$^2$. This event's $\rho_\mathrm{net}=12.7$ is near the threshold, but its 90\% confidence area of 530~deg$^2$ near the median. The sky map consists of a single, long, thin island exhibiting the forked-tongue morphology. \emph{This is event ID 20342 from Tables~\ref{table:2015-sim}~and~\ref{table:2015-coinc} and the online material (see Appendix~\ref{chap:first2years-data} for more details).}}
    \includegraphics[width=\textwidth]{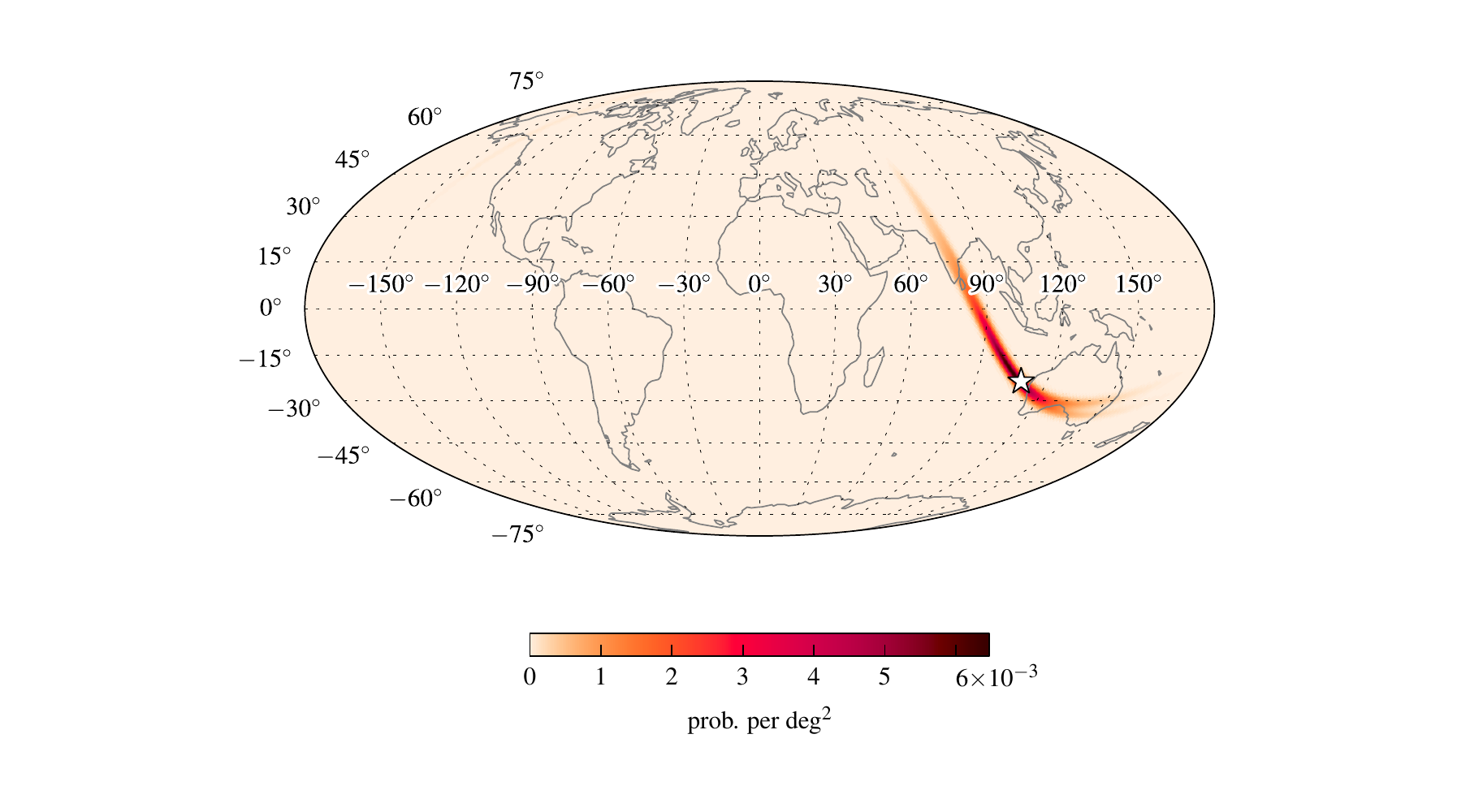}
\end{figure*}

The HL degeneracy affects most events that occur $\lesssim 30\arcdeg$ from one of the antenna pattern maxima. Most events that are $\gtrsim 50\arcdeg$ away have localizations that consist of a single long, thin arc or ring. See Figure~\ref{fig:typical-unimodal} for an example.

\begin{figure*}
    \centering
    \includegraphics{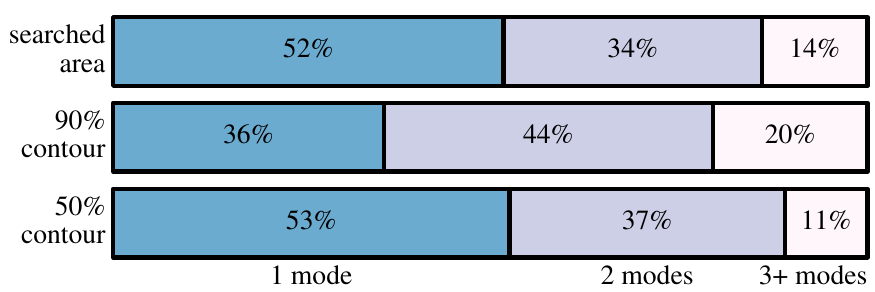}
    \caption[Frequency of localizations with one, two, or more modes]{\label{fig:mode_hist}Frequency with which \ac{GW} sky maps have one, two, or more disconnected modes during 2015. From top to bottom are the number of modes contained within the smallest confidence contour containing each simulated signal, the smallest 90\% contour, and the smallest 50\% contour. In 2015, roughly half of the sky maps will be unimodal, with most of the remainder being bimodal.}
\end{figure*}

In Figure~\ref{fig:mode_hist}, we have plotted a histogram of the number of disconnected modes comprising the 50\% and 90\% confidence regions and the searched area, for the rapid localizations in the 2015 configuration. The ratios of events having one, two, or three or more modes depend weakly on the selected confidence level. In 2015, using either the 50\% contour or the searched area, we find that about half of the events are unimodal and about a third are bimodal, the rest comprising three or more modes. Using the 90\% contour, we find that about a third of the events are unimodal and about half are bimodal.

\subsection{2016}

\begin{figure*}
    \centering
    \includegraphics{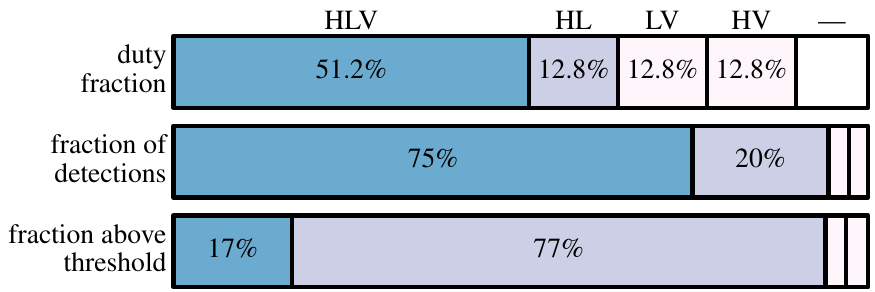}
    \caption[Breakdown of 2016 scenario by detector network]{\label{fig:demographics}Breakdown of 2016 scenario by detector network. The top row shows the duty fraction of each subset of the detector network, the fraction of time when all three detectors~(HLV) are observing, when any pair of detectors are observing~(HL, LV, or HV), or when zero or one detectors are observing~(---). The second row shows the fraction of coincident detections that occur under any given network~(HLV, HL, LV, or HV). The last row shows the fraction of coincident detections for which the given detectors have signals above the single-detector threshold of $\rho=4$.}
\end{figure*}

In our 2016 scenario, the HL detectors double in range to 108~Mpc and the V detector begins observations with a range of 36~Mpc. Over this six\nobreakdashes-month science run we expect $\sim$1.5 detections, assuming a \ac{BNS} merger rate of 1~Mpc$^{-3}$\,Myr$^{-1}$. Figure~\ref{fig:demographics} shows how livetime and duty cycle breaks down according to detector network (HLV, HL, LV, or HV). About half of the time all three detectors are online, with the remaining time divided into four almost equal parts among the three pairs of detectors or $\leq 1$~detector. However, the HLV network accounts for about three\nobreakdashes-quarters of detections and the HL network for most of the rest.

When all three detectors (HLV) are operating, most detections are comprised of H and L triggers, lacking a trigger from V because the signal is below the single\nobreakdashes-detector threshold of $\rho=4$. Slightly more than half (57\%) of all detections have a signal below threshold in one operating detector (almost always~V) while slightly less than half (43\%) consist of triggers from all operating detectors.

The first half consists mainly of HLV events that are detected by HL but not Virgo. For these events, the stochastic samplers provide a marked improvement in sky localization; their 90\% confidence regions have about one\nobreakdashes-third as much area as their rapid localizations. This is because the rapid localization makes use of only the triggers provided by the detection pipeline, lacking information about the signal in Virgo if its \ac{SNR} is $< 4$. The stochastic samplers, on the other hand, can use data from all operating detectors, regardless of \ac{SNR}. Therefore, in the present analysis, an improved sky localization would be available for half of the detections on a timescale of a day. Fortunately, for \ac{BNS} sources, it is immediately known whether an improved localization is possible, since this statement only depends on what detectors were online and which contributed triggers. On the other hand, it may be possible to provide prompt sky localizations for all events by simply lowering the single\nobreakdashes-detector threshold. If the single-detector threshold was dropped to unity, essentially no event would lack a Virgo trigger. There are also efforts to do away with the single\nobreakdashes-detector threshold entirely \citep{2012PhRvD..86l3010K,2013arXiv1307.4158K}. Simultaneously, there is promising work under way in speeding up the full parameter estimation using reduced order quadratures \citep{roq-pe}, interpolation \citep{interpolation-pe}, jump proposals that harness knowledge of the multimodal structure of the posterior \citep{KDEJumpProposal}, hierarchical techniques \citep{HierarchicalParameterEstimation}, and machine learning techniques to accelerate likelihood evaluation \citep{BAMBI,SKYNET}. It seems possible that the the delayed improvement in sky localization may be a temporary limitation that can be overcome well before 2016.

The second half consists of HLV events with triggers from all three detectors and events that occur when only HL, HV, or LV are operating. For these, the \ac{BAYESTAR} analysis and the full stochastic samplers produce comparable sky maps.

\begin{figure*}
    \caption[Typical HLV localization, circa 2016]{\label{fig:typical-hlv}Localization of a typical circa 2016 \ac{GW} detection in the HLV network configuration. This is a Mollweide projection in geographic coordinates. This event consists of triggers in H and L and has $\rho_\mathrm{net} = 13.4$. The rapid sky localization gives a 90\% confidence region with an area of 1100~deg$^2$ and the full stochastic sampler gives 515~deg$^2$. \emph{This is event ID 821759 from Tables~\ref{table:2016-sim}~and~\ref{table:2016-coinc} and the online material (see Appendix~\ref{chap:first2years-data} for more details).}}
    \begin{center}
        \includegraphics[width=\textwidth]{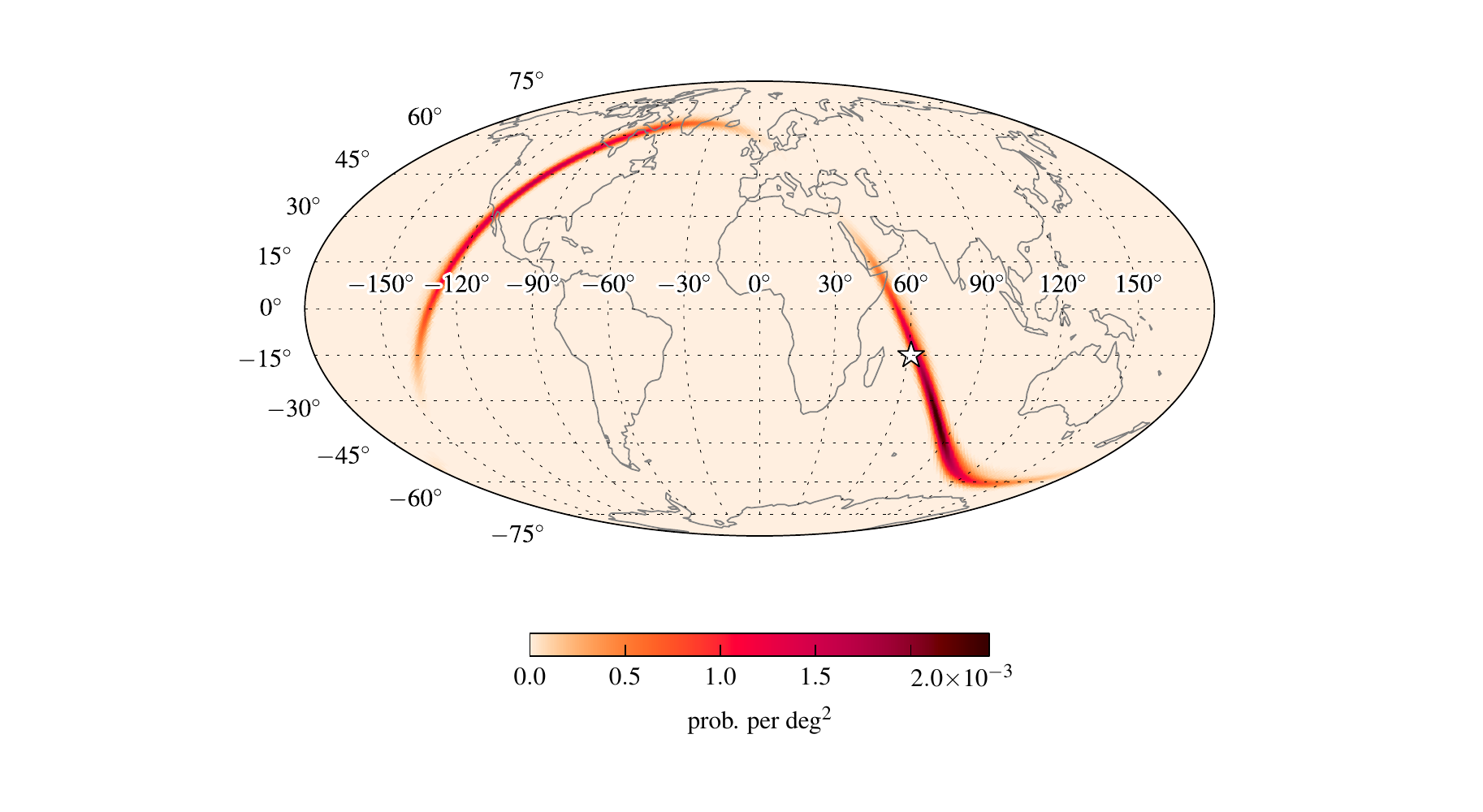}

        (a) \ac{BAYESTAR}

        \includegraphics[width=\textwidth]{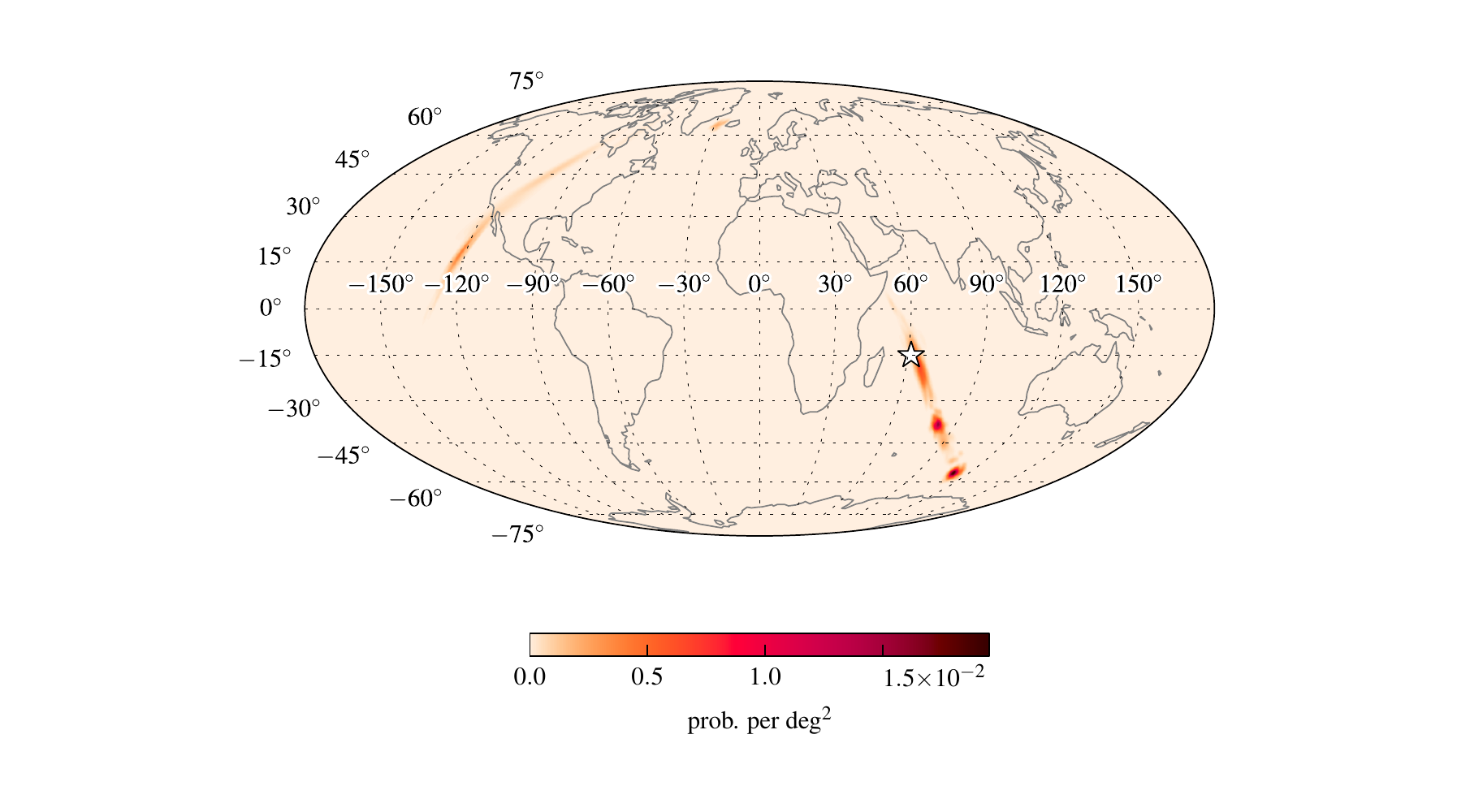}

        (b) \textls{LALINFERENCE\_MCMC}
    \end{center}
\end{figure*}

For nearby loud sources ($\rho_\mathrm{net} \gtrsim 20$), the HLV network frequently produces compact sky localizations concentrated in a single island of probability. However at low SNR ($\rho_\mathrm{net} \lesssim 20$), and especially for the events that are detected as only double coincidence (HL), the refined localizations from the full stochastic samplers often identify many smaller modes. A $\rho_\mathrm{net} =13.4$ example is shown in Figure~\ref{fig:typical-hlv}. In this event, the rapid sky localization has two modes and has a morphology that is well\nobreakdashes-described by the HL degeneracy explained in Section~\ref{sec:2015}. However, the refined, full parameter estimation breaks this into at least four smaller modes.

Of the remaining events, most occur when only the two HL detectors are operating. These look qualitatively the same as those in the 2015 case; their sky maps generally exhibit one or two slender islands of probability. However, percentage\nobreakdashes-wise, two\nobreakdashes-detector events are less well localized in the 2016 scenario than in the 2015 scenario. This unusual result is easily explained. Though the \ac{LIGO} detectors improve in sensitivity at every frequency, with the particular noise curves that we assumed, the signal bandwidth is actually slightly lower with the 2016 sensitivity compared to 2015. This is because of improved sensitivity at low frequency. Applying Equation~(\ref{eq:sigmat}), we find that for a $(1.4, 1.4)~M_\odot$ binary at $\rho=10$, one of the 2015 \ac{LIGO} detectors has an \ac{RMS} timing uncertainty of 131~{\textmu}s, whereas one of the 2016 \ac{LIGO} detectors has a timing uncertainty of 158~{\textmu}s. Clearly, the 2016 detectors will produce more constraining parameter estimates for sources at any fixed distance as the \ac{SNR} improves. However, for constant \ac{SNR} the 2016 \ac{LIGO} detectors should find areas that are $(158/131)^2=1.45$ times larger than events at the same \ac{SNR} in 2015. This is indeed what we find.

\begin{figure*}
    \caption[Typical HV localization, circa 2016]{\label{fig:typical-hv}Rapid localization of a typical circa 2016 \ac{GW} detection in the HV network configuration. This is a Mollweide projection in geographic coordinates. This event's $\rho_\mathrm{net} = 12.2$ is near the detection threshold. Its 90\% confidence area is 4600~deg$^2$, but the true position of the source (marked with the white pentagram) is found after searching 65~deg$^2$. \emph{This is event ID 655803 from Tables~\ref{table:2016-sim}~and~\ref{table:2016-coinc} and the online material (see Appendix~\ref{chap:first2years-data} for more details).}}
    \includegraphics[width=\textwidth]{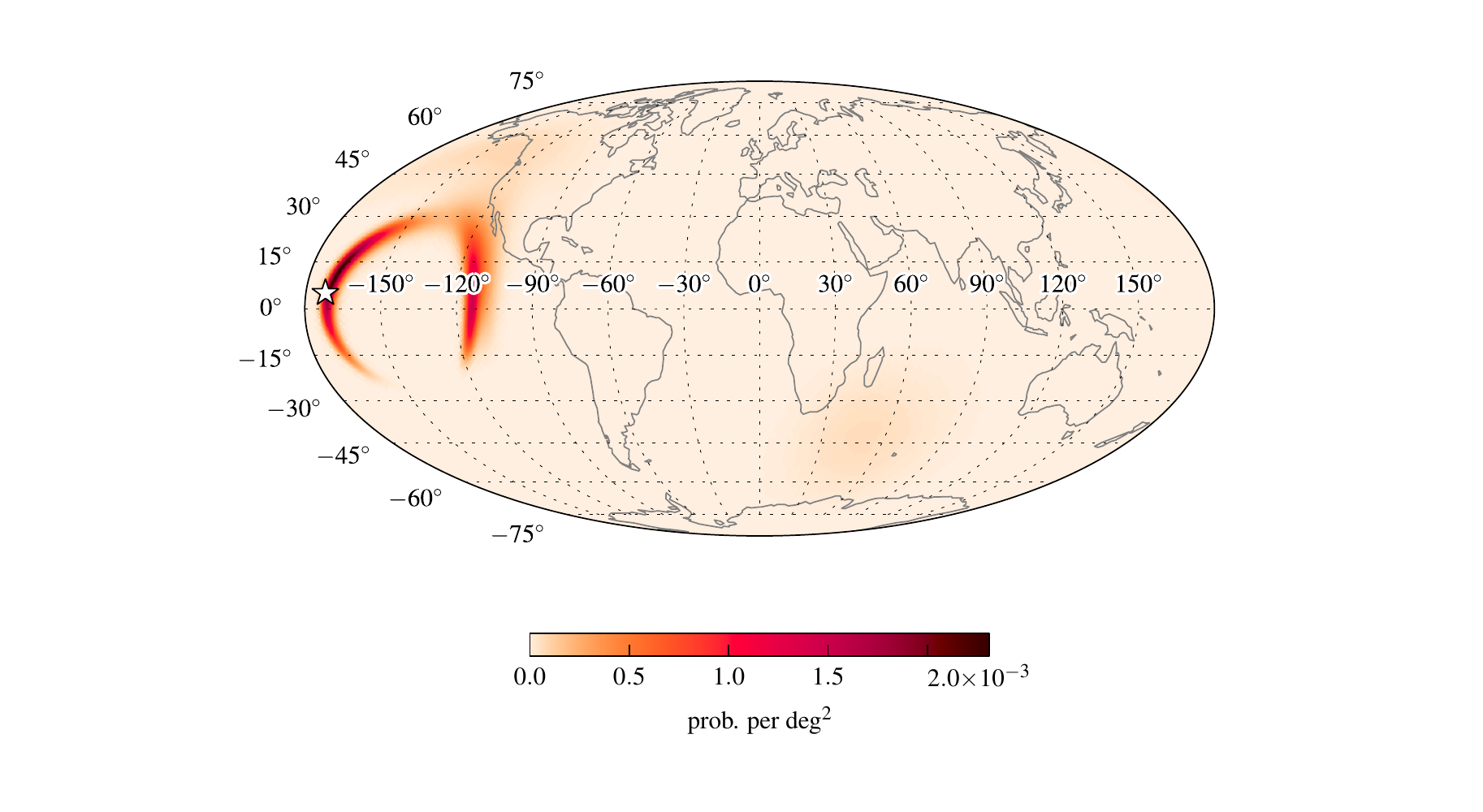}
\end{figure*}

Two\nobreakdashes-detector events involving Virgo (HV and LV) are rare, accounting for only about 6\% of detections. Sky maps for these events sometimes exhibit multiple fringes spread over a quadrant of the sky. These are in part due to the increased importance of phase\nobreakdashes-on\nobreakdashes-arrival due to the oblique alignment of the \ac{LIGO} and Virgo antenna patterns, which gives the network a limited ability to measure \ac{GW} polarization. Occasionally there are also diffuse clouds of probability near the participating \ac{LIGO} detector's two antenna pattern maxima, which may be a vestige of the antenna pattern. A typical HV event that exhibits both features is shown in Figure~\ref{fig:typical-hv}.

\section{Discussion}

\subsection{Caveats}

We reiterate that the scenarios we have described make assumptions about the astrophysical rate of \ac{BNS} mergers and the Advanced \ac{LIGO}/Virgo sensitivity as a function of time. The former is subject to orders of magnitude uncertainty due to the small sample of known galactic binary pulsars as well as model dependence in population synthesis~\citep{LIGORates}. The latter could deviate from \citet{LIGOObservingScenarios} depending on actual Advanced \ac{LIGO}/Virgo commissioning progress. However, the fractions of events localized within a given area are robust with respect to both of these effects.

We have dealt only with \ac{BNS} mergers. \ac{NSBH} mergers are also promising sources for closely related \ac{GW} signals and \ac{EM} transients. A similar, comprehensive investigation of \ac{GW} sky localization accuracy for \ac{NSBH} signals is warranted.

In this simulation, we have used ideal Gaussian noise, but selected a detection threshold that is designed to reproduce expected performance in detectors with realistically wide tails due to instrumental and environmental glitches. If Advanced \ac{LIGO}'s and Virgo's improved seismic isolation and control systems are even more effective at suppressing such glitches than their initial counterparts were, then the $\rho_\mathrm{net}$ threshold for confident detection would decrease, yielding discoveries earlier but with larger typical sky localization areas.

We remind the reader that the events comprising this study would be regarded as confident detections, with $\text{\ac{FAR}} \lesssim 10^{-2}$~yr$^{-1}$, based on \ac{GW} observations alone. In practice, some observers may choose to follow up more marginal detection candidates. For instance, a group with enough resources and telescope time to follow up one candidate per month might filter events with $\text{\ac{FAR}} \leq 12$~yr$^{-1}$. A high false alarm rate threshold will admit correspondingly lower $\rho_\mathrm{net}$ candidates with coarser localizations than those which we have presented here.

Finally, on a positive note, the number of detections is expected to increase considerably as commissioning proceeds toward final design sensitivity. Furthermore, sky localization will improve radically as the HLV detectors approach comparable sensitivity. The addition of two more planned ground\nobreakdashes-based \ac{GW} detectors, \ac{LIGO}\nobreakdashes--India and KAGRA, would likewise increase rates and improve sky localizations dramatically \citep{ShutzThreeFiguresOfMerit,Veitch:2012,FairhurstLIGOIndia,NissankeKasliwalEMCounterparts,LIGOObservingScenarios}.

\subsection{Detection scenarios}

From our representative sample of hundreds of early Advanced \ac{LIGO}/Virgo events emerge a few common morphologies and several possible scenarios for the early detections of \acp{GW} from a \ac{BNS} merger.

We find that in both 2015 and 2016, the detection rate is highly anisotropic and proportional to the cube of the network antenna pattern, with a strong excess above North America and the Indian Ocean and deficits in four spots over the south Pacific, south Atlantic, Caucasus, and north Pacific.

\begin{enumerate}

\item \label{item:hl-unimodal} \textit{HL event, single arc}---This scenario is relevant for the HL network configuration and applies to both 2015 and 2016. Figure~\ref{fig:typical-unimodal} shows a typical sky map for a near\nobreakdashes-threshold detection with $\rho_\mathrm{net} = 12.7$, exhibiting a single long, extended arc spanning $\sim$500~deg$^2$.

\item\label{item:hl-bimodal} \textit{HL event, two degenerate arcs}---This scenario also applies to 2015 or to HL livetime in 2016. Figure~\ref{fig:typical} shows a typical sky map with a moderately high $\rho_\mathrm{net} = 15.0$, localized to $\sim$600~deg$^2$. Its localization embodies the HL degeneracy, with two strong, long, thin modes over North America and the Indian Ocean, separated by nearly $180^\circ$ and therefore 12~hr apart in hour angle. Inevitably, one of these two modes will be nearly Sun\nobreakdashes-facing and inaccessible to optical facilities. Because of the bimodality, these sky maps can span slightly larger areas than case~\ref{item:hl-unimodal}. After taking an inevitable 50\% hit in visibility, such events resemble the single arc scenario.

Whether a given source falls into scenario \ref{item:hl-unimodal} or \ref{item:hl-bimodal} is largely determined by its sky location relative to the network antenna pattern. The transition occurs between $\sim 30\arcdeg$ and $\sim 50\arcdeg$ away from the two points of maximum sensitivity.

\item \textit{HLV event, degeneracy broken by Virgo}---This scenario applies only to the 2016 configuration, while all three instruments are online. The rapid sky localization looks similar to the previous scenario, a pair of long, thin rings over the northern and southern hemispheres, but the full parameter estimation cuts this down to a handful of islands of probability covering as little as half to a third of the area, $\sim 200$~deg$^2$. For such an event, the refined localization could be used to guide $\sim$day\nobreakdashes-cadence kilonova\nobreakdashes-hunting observations or to re\nobreakdashes-target the vetting of afterglow candidates arising from early\nobreakdashes-time observations. Several wide\nobreakdashes-field facilities could be employed to monitor modes that lie in different hemispheres.

\item \textit{HLV event, compactly localized}---This is another 2016, three-detector scenario. It describes many of the events that are detected with triggers in all three instruments. These are many of the best\nobreakdashes-localized events, with 90\% confidence regions only a few times 10~deg$^2$ in area. These events are generally localized to one simply connected region and exhibit a less pronounced preference for particular sky locations. In this scenario, it is most likely that the rapid sky localization and the full parameter estimation will be similar. This is observationally the simplest scenario: just one of the several wide\nobreakdashes-field optical searches (for instance, \ac{ZTF} or BlackGEM) would be able to scan the whole error region at a daily cadence.

\end{enumerate}

\subsection{Comparison with other studies}

This is the first study so far to combine an astrophysically motivated source population, realistic sensitivity and detector network configurations, event selection effects arising from a genuine detection pipeline instead of an ad hoc threshold, and parameter estimation algorithms that will be deployed for \ac{GW} data analysis. This study also has a much larger sample size and lower statistical uncertainty than most of the prior work. It is therefore somewhat difficult to compare results to other studies, which each have some but not all of these virtues.

To the best of the authors' knowledge, \citet{Raymond:2009} were the first to point out the power of Bayesian priors for breaking sky degeneracies in two\nobreakdashes-detector networks, challenging a prevailing assumption at the time that two detectors could only constrain the sky location of a compact binary signal to a degenerate annulus. \citet{LIGOObservingScenarios} speculated that two\nobreakdashes-detector, 2015, HL configuration sky maps would be rings spanning ``hundreds to thousands'' of deg$^2$ and that coherence and amplitude consistency would ``sometimes'' resolve the localizations to shorter arcs. With our simulations, we would only revise that statement to read ``hundreds to \emph{a} thousand'' deg$^2$ and change ``sometimes'' to ``always.'' \citet{KasliwalTwoDetectors} recently argued for the feasibility of optical transient searches (in the context of kilonovae) with two\nobreakdashes-detector \ac{GW} localizations.

\citet{LIGOObservingScenarios} used time\nobreakdashes-of\nobreakdashes-arrival triangulation \citep{FairhurstTriangulation} to estimate the fraction of sources with 90\% confidence regions smaller than 20~deg$^2$, finding a range of 5\nobreakdashes--12\% for 2016. We find 14\%. Our values are more optimistic, but perhaps also more realistic for the assumed detector sensitivity. Our sky localization takes into account phase and amplitude information, which \citet{Grover:2013} points out can produce $\approx$0.4 times smaller areas compared to timing alone. However, it is clear from both \citet{LIGOObservingScenarios} and our study that such well\nobreakdashes-localized events will comprise an exceedingly small fraction of \ac{GW} detections until the end of the decade. We therefore echo \citet{KasliwalTwoDetectors} in stressing the importance of preparing to deal with areas of hundreds of deg$^2$ in the early years of Advanced \ac{LIGO} and Virgo.

\citet{NissankeKasliwalEMCounterparts} used an astrophysical distance distribution, drawing source positions uniformly from comoving volume for distances $d_\mathrm{L} > 200$~Mpc and from a $B$\nobreakdashes-band luminosity\nobreakdashes-weighted galaxy catalog for distances $d_\mathrm{L} \leq 200$~Mpc. They generated sky maps using their own \ac{MCMC} code. In a similar manner to our present study, they imposed a threshold of $\rho_\mathrm{net} > 12$. They explored several different \ac{GW} detector network configurations. The most similar to our 2016 scenario was an HLV network at final design sensitivity. They found a median 95\% confidence region area of $\sim$20~deg$^2$. In comparison, we find a 95~deg$^2$ confidence area of 374~deg$^2$. Our much larger number is explained by several factors. First, we did not draw nearby sources from a galaxy catalog, so we have fewer loud, nearby sources. Second, since we accounted for duty cycle, poorly localized two\nobreakdashes-detector events account for a quarter of our sample. Third, and most important, we assumed Advanced Virgo's anticipated initial sensitivity rather than its final design sensitivity.

\citet{RodriguezBasicParameterEstimation} also studied an HLV network at final design sensitivity. Their simulated signals had identically zero noise content, the average noise contribution among all realizations of zero-mean Gaussian noise. All of their simulated events had a relatively high $\rho_\mathrm{net} = 20$. They found a median $95\%$ confidence area of 11.2~deg$^2$. If we consider all of our 2016 scenario HLV events with $19.5 \leq \rho_\mathrm{net} \leq 20.5$, we find a median area of 126~deg$^2$. Our significantly larger number is once again partly explained by our less sensitive Virgo detector, which introduces several multimodal events even at this high $\rho_\mathrm{net}$.

Similarly, \citet{Grover:2013} and \citet{SiderySkyLocalizationComparison} studied a three\nobreakdashes-detector network, but at Initial \ac{LIGO} design sensitivity. These studies were primarily concerned with evaluating Bayesian parameter estimation techniques with respect to triangulation methods. They found much smaller areas, with a median of about 3~deg$^2$. Both papers used a source population that consisted mainly of very high\nobreakdashes-\ac{SNR} signals with binary black hole masses, with distances distributed logarithmically. All of these effects contribute to unrealistically small areas.

Finally, \citet{KasliwalTwoDetectors} made the first small\nobreakdashes-scale systematic study of localizability with two \ac{LIGO} detectors, albeit at final Advanced \ac{LIGO} design sensitivity. For these noise curves and a $(1.4, 1.4)~M_\odot$ binary with single-detector $\rho=10$, Equation~(\ref{eq:sigmat}) gives a timing uncertainty of 142~{\textmu}s. Their different choice of noise curves should result in areas that are $(131/142)^2 \approx 0.85$ times smaller than ours, at a given $\rho_\mathrm{net}$. As we have, they imposed a network \ac{SNR} threshold of $\rho_\mathrm{net} \geq 12$ on all detections\footnote{\citet{NissankeKasliwalEMCounterparts} and \citet{KasliwalTwoDetectors} present a parallel set of results for a threshold $\rho_\mathrm{net} > 8.5$, relevant for a coherent \ac{GW} search described by \citet{harry-single-spin}. However, the coherent detection statistic described by \citet{harry-single-spin} is designed for targeted searches at a known sky location (for instance, in response to a \ac{GRB}). Thus the $\rho_\mathrm{net} > 8.5$ threshold is not relevant for optical follow\nobreakdashes-up triggered by a detection from an all\nobreakdashes-sky \ac{GW} search. Furthermore, this reduced threshold is not relevant to the HL configuration because the coherent detection statistic reduces to the network \ac{SNR} for networks of two detectors or fewer.}. They found a median 95\% confidence area of $\sim 250$~deg$^2$ from a catalog of 17 events. From our 2015 scenario, we find a median 90\% confidence area that is almost twice as large, $\sim$500~deg$^2$. Though we cannot directly compare our 90\% area to their 95\% area, our 95\% area would be even larger. Several factors could account for this difference, including the smaller sample size in \citet{KasliwalTwoDetectors}. Also, \citet{KasliwalTwoDetectors}, like \citet{NissankeKasliwalEMCounterparts}, drew nearby sources from a galaxy catalog to account for clustering, so their population may contain more nearby, well\nobreakdashes-localized events than ours. Another difference is that \citet{KasliwalTwoDetectors} do not report any multimodal localizations or the 180$^\circ$ degeneracy that we described in Section~\ref{sec:2015}.

\subsection{Conclusion}

Many previous sky localization studies have found that networks of three or more advanced \ac{GW} detectors will localize \ac{BNS} mergers to tens of deg$^2$. 
However, given realistic commissioning schedules, areas of hundreds of deg$^2$ will be typical in the early years of Advanced \ac{LIGO} and Virgo.

We caution that multimodality and long, extended arcs will be a common and persistent feature of Advanced \ac{LIGO}/Virgo detections. We caution that existing robotic follow\nobreakdashes-up infrastructure designed for \acp{GRB}, whose localizations are typically nearly Gaussian and unimodal, will need to be adapted to cope with more complicated geometry. In particular, we advise optical facilities to evaluate the whole \ac{GW} sky map when determining if and when a given event is visible from a particular site.

We have elucidated a degeneracy caused by the relative orientations of the two \ac{LIGO} detectors, such that position reconstructions will often consist of two islands of probability separated by 180$^\circ$. We have shown that this degeneracy is largely broken by adding Virgo as a third detector, even with its significantly lower sensitivity. We have shown that sub\nobreakdashes-threshold \ac{GW} observations are important for sky localization and parameter estimation.

We have demonstrated a real\nobreakdashes-time detection, sky localization, and parameter estimation infrastructure that is ready to deliver Advanced \ac{LIGO}/Virgo science. The current analysis has some limitations for the three\nobreakdashes-detector network, an undesirable trade\nobreakdashes-off of sky localization accuracy and timescale. Work is ongoing to lift these limitations by providing the rapid sky localization with information below the present single\nobreakdashes-detector threshold and by speeding up the full parameter estimation by a variety of methods \citep{roq-pe,interpolation-pe,KDEJumpProposal,HierarchicalParameterEstimation,BAMBI}.

We have exhibited an approach that involves three tiers of analysis, which will likely map onto a sequence of three automated alerts with progressively more information on longer timescales, much as the way in which observers in the \ac{GRB} community are used to receiving a sequence of \ac{GCN} notices about a high\nobreakdashes-energy event.

The maximum timescale of the online \ac{GW} analysis, about a day, is appropriate for searching for kilonova emission. However, the availability of \ac{BAYESTAR}'s rapid localizations within minutes of a merger makes it possible to search for X\nobreakdashes-ray and optical emission. Due to jet collimation, these early\nobreakdashes-time signatures are expected to accompany only a small fraction of \ac{LIGO}/Virgo events. However, the comparative brightness and distinctively short timescale of the optical afterglow makes it an attractive target. \ac{PTF} has recently proved the practicality of wide\nobreakdashes-field afterglow searches through the blind discovery of afterglow\nobreakdashes-like optical transients \citep{PTF11agg,iPTF14yb} and the detection of optical afterglows of \emph{Fermi} \ac{GBM} bursts \citep{iPTF13bxl}. We encourage optical transient experiments such as \ac{ZTF} and BlackGEM to begin searching for \ac{EM} counterparts promptly, based on the rapid \ac{GW} localization. In the most common situation of no afterglow detection, the early observations may be used as reference images for longer\nobreakdashes-cadence kilonova searches.

\section*{Acknowledgements}

L.P.S. and B.F. thank generous support from the \ac{NSF} in the form of Graduate Research Fellowships. B.F. acknowledges support through \ac{NSF} grants DGE\nobreakdashes-0824162 and PHY\nobreakdashes-0969820. A.L.U. and C.P. gratefully acknowledge \ac{NSF} support under grants PHY\nobreakdashes-0970074 and PHY\nobreakdashes-1307429 at the University of Wisconsin\nobreakdashes--Milwaukee (UWM) Research Growth Initiative. J.V. was supported by the research programme of the Foundation for Fundamental Research on Matter, which is partially supported by the Netherlands Organisation for Scientific Research, and by Science and Technology Facilities Council grant ST/K005014/1. P.G. is supported by a NASA Postdoctoral Fellowship administered by the Oak Ridge Associated Universities.

\textls{GSTLAL} analyses were produced on the NEMO computing cluster operated by the Center for Gravitation and Cosmology at UWM under \ac{NSF} Grants PHY\nobreakdashes-0923409 and PHY\nobreakdashes-0600953. \ac{BAYESTAR} analyses were performed on the \ac{LIGO}\nobreakdashes--Caltech computing cluster. The MCMC computations were performed on Northwestern's CIERA High\nobreakdashes-Performance Computing cluster GRAIL\footnote{\url{http://ciera.northwestern.edu/Research/Grail\_Cluster.php}}.

We thank Patrick Brady, Vicky Kalogera, Erik Katsavounidis, Richard O'Shaughnessy, Ruslan Vaulin, and Alan Weinstein for helpful discussions.

This research made use of Astropy\footnote{\url{http://www.astropy.org}} \citep{astropy}, a community-developed core Python package for Astronomy. Some of the results in this paper have been derived using \ac{HEALPix} \citep{HEALPix}. Public-domain cartographic data is courtesy of Natural Earth\footnote{\url{http://www.naturalearthdata.com}} and processed with MapShaper\footnote{\url{http://www.mapshaper.org}}.

\ac{BAYESTAR},
\textls{LALINFERENCE\_NEST}, and
\textls{LALINFERENCE\_MCMC}
are part of the \ac{LIGO} Algorithm Library Suite\footnote{\url{http://www.lsc-group.phys.uwm.edu/cgit/lalsuite/tree}} and the \ac{LIGO} parameter estimation package, \textls{LALINFERENCE}. Source code for
\textls{GSTLAL}%
\footnote{\url{http://www.lsc-group.phys.uwm.edu/cgit/gstlal/tree/}}
and \textls{LALINFERENCE}%
\footnote{\url{http://www.lsc-group.phys.uwm.edu/cgit/lalsuite/tree/lalinference}}
are available online under the terms of the GNU General Public License.

\ac{LIGO} was constructed by the California Institute of Technology and Massachusetts Institute of Technology with funding from the \ac{NSF} and operates under cooperative agreement PHY\nobreakdashes-0757058.

\chapter{Discovery and redshift of an optical afterglow in 71 square degrees: \lowercase{i}PTF13bxl and \acs{GRB}~130702A}
\label{chap:iptf13bxl}

\author{Leo~P.~Singer\altaffilmark{1},
S.~Bradley~Cenko\altaffilmark{2},
Mansi~M.~Kasliwal\altaffilmark{3,13},
Daniel~A.~Perley\altaffilmark{13,4},
Eran~O.~Ofek\altaffilmark{5},
Duncan~A.~Brown\altaffilmark{1,6},
Peter~E.~Nugent\altaffilmark{7,8},
S.~R.~Kulkarni\altaffilmark{4},
Alessandra~Corsi\altaffilmark{9},
Dale~A.~Frail\altaffilmark{10},
Eric~Bellm\altaffilmark{4},
John~Mulchaey\altaffilmark{3},
Iair~Arcavi\altaffilmark{5},
Tom~Barlow\altaffilmark{4},
Joshua~S.~Bloom\altaffilmark{7,8},
Yi~Cao\altaffilmark{4},
Neil~Gehrels\altaffilmark{2},
Assaf~Horesh\altaffilmark{4},
Frank~J.~Masci\altaffilmark{11},
Julie~McEnery\altaffilmark{2},
Arne~Rau\altaffilmark{12},
Jason~A.~Surace\altaffilmark{11},
and~Ofer~Yaron\altaffilmark{5}}

\altaffiltext{1}{LIGO Laboratory, California Institute of Technology, Pasadena, CA 91125, USA}
\altaffiltext{2}{Astrophysics Science Division, NASA Goddard Space Flight Center, Mail Code 661, Greenbelt, MD 20771, USA}
\altaffiltext{3}{Observatories of the Carnegie Institution for Science, 813 Santa Barbara St, Pasadena CA 91101, USA}
\altaffiltext{4}{Cahill Center for Astrophysics, California Institute of Technology, Pasadena, CA 91125, USA}
\altaffiltext{5}{Benoziyo Center for Astrophysics, The Weizmann Institute of Science, Rehovot 76100, Israel}
\altaffiltext{6}{Department of Physics, Syracuse University, Syracuse, NY 13244, USA}
\altaffiltext{7}{Department of Astronomy, University of California Berkeley, B-20 Hearst Field Annex \# 3411, Berkeley, CA, 94720-3411}
\altaffiltext{8}{Physics Division, Lawrence Berkeley National Laboratory, 1 Cyclotron Road MS 50B-4206, Berkeley, CA 94720, USA}
\altaffiltext{9}{George Washington University, Corcoran Hall, Washington, DC 20052, USA}
\altaffiltext{10}{National Radio Astronomy Observatory, P.O. Box O, Socorro, NM 87801, USA}
\altaffiltext{11}{Infrared Processing and Analysis Center, California Institute of Technology, Pasadena, CA 91125, USA}
\altaffiltext{12}{Max-Planck-Institut f\u{u}r extraterrestrische Physik, Giessenbachstrasse 1, 85748 Garching, Germany}
\altaffiltext{13}{Hubble Fellow}

\attribution{This chapter is reproduced from \citet{iPTF13bxl}, which was published under the same title in \textnormal{The Astrophysical Journal Letters}, copyright~\textcopyright{}~2014 The American Astronomical Society. My contributions include developing the \acs{P48} alert and tiling code, reducing the \acs{P48} photometry, assigning candidates for follow\nobreakdashes-up, analyzing the data, preparing all of the figures, and writing about 70\% of the text. S.B.C. and D.A.B. also contributed to candidate vetting. S.B.C. calculated the broadband afterglow models shown in Figure~\ref{fig:sed}. D.A.P., A.C., and A.H. executed and reduced the radio observations. D.A.P. reduced the \acs{P60} photometry and wrote much of Section~\ref{sec:context}. The Magellan Baade spectrum was acquired by M.M.K. and reduced by J.M. The \acs{P200} spectrum was observed and reduced by E.B. Other authors made substantial contributions to the \acs{IPTF} hardware and software.}

\section{Introduction}


Our understanding of \acfp{GRB} has been propelled by our ability to localize these rare and energetic cosmic events precisely. \emph{Compton Gamma\nobreakdashes-ray Observatory}/BATSE's coarse localizations robustly demonstrated that \acp{GRB} were distributed isotropically on the sky and suggested that \acp{GRB} originate at cosmological distances~\citep{GRBsAreExtragalactic}. Prompt arcminute localizations provided by \emph{BeppoSAX} directly enabled the discovery of the first afterglows of long-duration \acp{GRB} \citep{GRBsHaveXrayAfterglows,GRBsHaveOpticalAfterglows,GRBsHaveRadioAfterglows}. Currently, the prompt slewing capabilities of the \textit{Swift} satellite \citep{Swift} enable the on\nobreakdashes-board narrow-field instruments to provide arcsecond localizations for $\approx 90$ \acp{GRB} yr$^{-1}$ within $\approx 100$~s of the burst trigger.

With seven decades of simultaneous energy coverage, \emph{Fermi} has opened a new window into the \ac{GRB} phenomenon, the MeV to GeV regime.  However, \emph{Fermi} remains fundamentally limited by its localization capabilities.  The \aclu{LAT} \citep[\acs{LAT}; 20~MeV\nobreakdashes--300~GeV; 16\% of all-sky;][]{LAT} can localize events with GeV photons to radii as small as $\sim 10$\arcmin.  But the \ac{LAT} only localizes a handful of \acp{GRB} each year.  The \aclu{GBM} (\acs{GBM}; few~keV--30~MeV; 70\% of all-sky; \citealt{FermiGBM}), on the other hand, detects \acp{GRB} at a rate of $\approx 250$~yr$^{-1}$.  However, typical \ac{GBM} \acp{GRB} have localizations of many tens of square degrees (random plus systematic uncertainties).  Consequently, no afterglows have been identified based solely on a \ac{GBM} localization until this work\footnote{The only comparable discovery was the afterglow of \ac{GRB}~120716A in the $\approx 2$~deg$^{2}$ error box from the \ac{IPN} by \citet{GCN13489}.}.

\begin{figure*}
    \ifinthesis
    \centering
    \includegraphics[width=\textwidth]{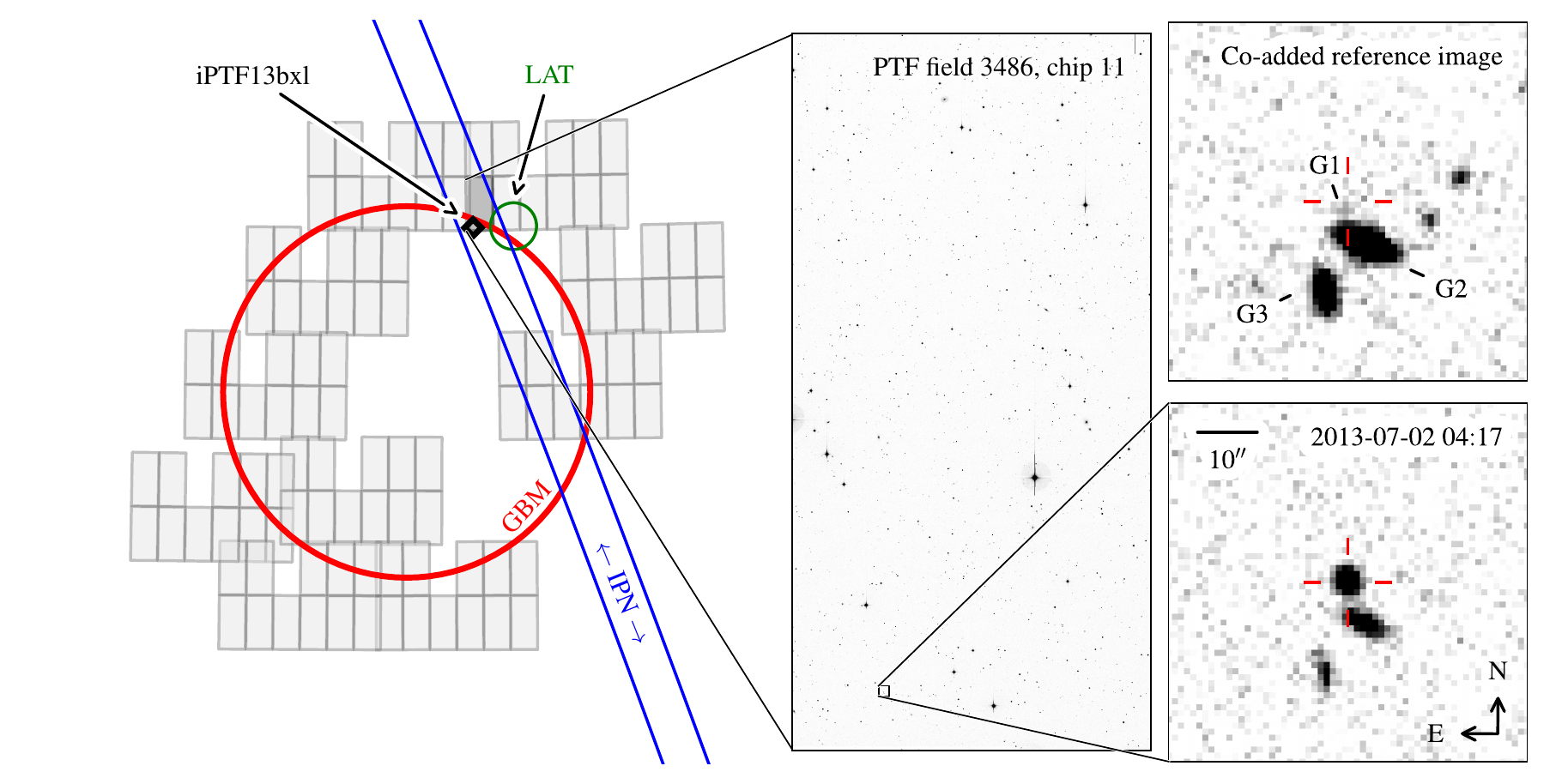}
    \fi
    \caption[Discovery of \acs{GRB}~130702A~/~iPTF13bxl]{\label{fig:discovery}P48 imaging of \ac{GRB}~130702A and discovery of iPTF13bxl. The left panel illustrates the $\gamma$-ray localizations (red circle: 1$\sigma$ \ac{GBM}; green circle: \ac{LAT}; blue lines: 3$\sigma$ \ac{IPN}) and the 10 \ac{P48} reference fields that were imaged (light gray rectangles).  For each P48 pointing, the location of the 11 chips are indicated with smaller rectangles (one CCD in the camera is not currently operable).  Our tiling algorithm places a large weight on the existence of deep P48 pre-explosion imaging (a necessity for high-quality subtraction images); the large gaps inside the \ac{GBM} localization are fields without these reference images.  The small black diamond is the location of iPTF13bxl.  The right panels show \ac{P48} images of the location of iPTF13bxl, both prior to (top) and immediately following (bottom) discovery.  We note that the \ac{LAT} and \ac{IPN} localizations were published \textit{after} our discovery announcement \citep{GCN14967}. }
\end{figure*}

The \acl{PTF} (\acsu{PTF}; \citealt{PTFLaw}) is developing the necessary instrumentation, algorithms, and observational discipline to detect optical counterparts to \ac{GBM} \acp{GRB}. The wide 7.1~deg$^2$ \acl{FOV} and sensitivity ($R \approx 20.6$~mag in 60~s) of the \ac{P48} and \ac{CFH12k} camera~\citep{P48PTF} are well-suited to identifying long-duration \ac{GRB} afterglow candidates. The real\nobreakdashes-time software pipeline (Nugent et al., in prep.) enables rapid panchromatic follow-up with an arsenal of telescopes~(e.g. \citealt{2011ApJ...736..159G}), essential to distinguish the true afterglow from background and foreground contaminants.  Here, we present our discovery of iPTF13bxl, the afterglow of the \emph{Fermi}~\ac{GBM} \ac{GRB}~130702A, found by searching a sky area of 71~deg$^2$ with the \acl{IPTF} (\acsu{IPTF}).

\section{Discovery}
\label{sec:discovery}

On 2013~July~2 at 00:05:23.079~UT, the \emph{Fermi} \ac{GBM} detected trigger 394416326.  The refined human-generated (i.e., ground\nobreakdashes-based) localization, centered on $\alpha = 14^{\mathrm{h}} 35^{\mathrm{m}} 14^{\mathrm{s}}$, $\delta = 12^{\circ} 15\arcmin 00\arcsec$ (J2000.0), with a quoted 68\% containment radius of $4\fdg0$ (statistical uncertainty only), was disseminated less than an hour after the burst (Figure~\ref{fig:discovery}).

\begin{figure*}
    \ifinthesis
    \centering
    \includegraphics[width=\textwidth]{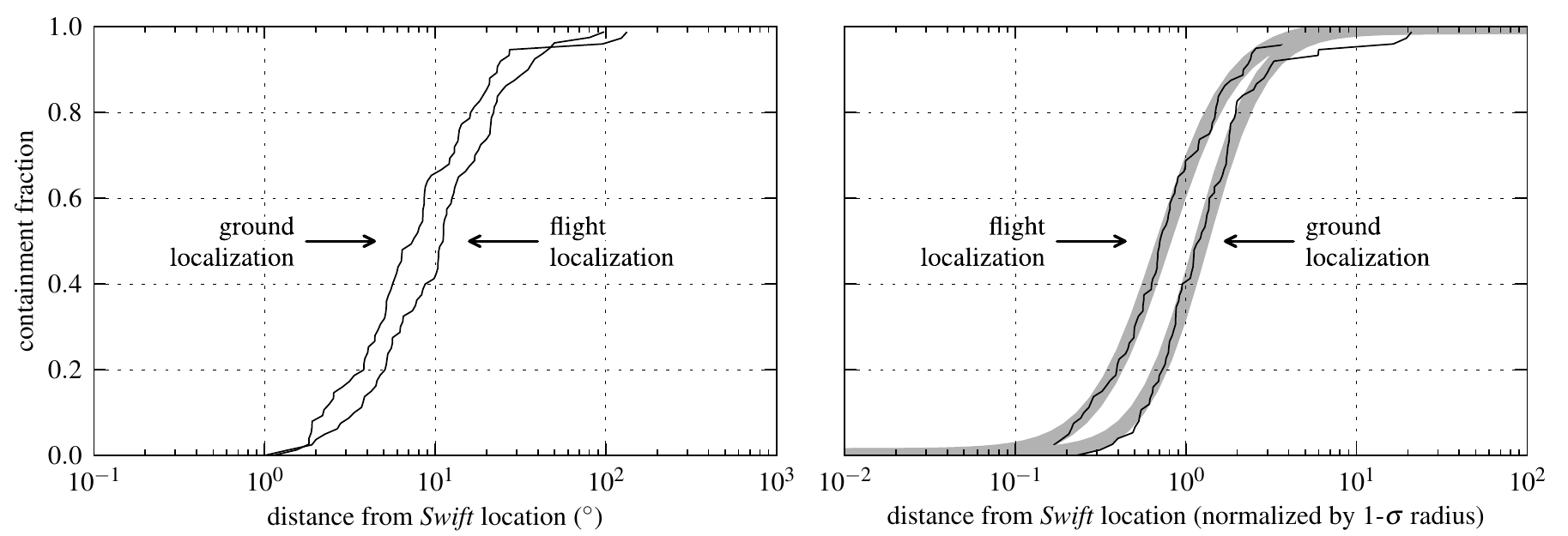}
    \fi
   \caption[Calibration of \emph{Fermi} systematic errors]{\label{fig:fermi-localization}Localization accuracy of \emph{Fermi} \ac{GBM} positions, generated by searching for coincidences with \acp{GRB} detected by the \emph{Swift} satellite.  The left panel shows the fraction of bursts contained within a given distance from the \emph{Swift} position, both for in\nobreakdashes-flight and refined ground\nobreakdashes-based localizations.  Ground-based localizations are on average about half as far from the true \ac{GRB} positions as the in-flight localizations. The right panel shows a cumulative histogram of the \emph{Fermi}\nobreakdashes--\emph{Swift} distance, normalized by each trigger's nominal 1$\sigma$ radius (either ground-based or in-flight).  Although the ground-based localizations are more accurate, the nominal ground\nobreakdashes-based error radii must be interpreted as describing a different confidence level than the  in\nobreakdashes-flight error radius. The thick gray lines are fits to the logistic\nobreakdashes-like function in Equation~\ref{eq:fermi-localization}.}
\end{figure*}

\emph{Fermi}-\ac{GBM} \ac{GRB} positions are known to suffer from significant systematic uncertainties, currently estimated to be $\approx 2^\circ$--$3^\circ$.  To characterize the full radial profile of the localization uncertainty, our \ac{GBM}\nobreakdashes-\ac{IPTF} pipeline automatically computed a probability map for the event, modeled on previous \emph{Fermi}/\emph{Swift} coincidences from 2010~March~30 through 2013~April~4. We fit a sigmoid function:
\begin{equation}
    \label{eq:fermi-localization}
    P(r) = \frac{1}{1 + \left(c_0 r\right)^{c_1}},
\end{equation}
where $r$ is the angular distance to the \emph{Swift} location, normalized by the in\nobreakdashes-flight or ground\nobreakdashes-based error radius for that burst. We find $c_0 = 1.35$, $c_1 = -2.11$ for in\nobreakdashes-flight \ac{GBM} localizations and $c_0 = 0.81$, $c_1 = -2.47$ for ground\nobreakdashes-based \ac{GBM} localizations (Figure~\ref{fig:fermi-localization}). 

Image subtraction within \ac{IPTF} is greatly simplified by observing only pre-defined fields on the plane of the sky; this ensures that sources will fall on approximately the same detector location from night to night, minimizing a possible source of systematic uncertainty.  Using a \acl{HEALPix}~\citep[\acsu{HEALPix};][]{HEALPix} bitmap representation of the probability distribution of the trigger location, we chose ten of these pre-defined fields to maximize the probability of enclosing the true (but still unknown) location of the burst (Figure~\ref{fig:discovery}). In this particular case, the ten selected fields did not include the center of the \ac{GBM} localization because we lacked previous reference images there. Nonetheless, we estimated that these ten fields had a 38\% chance of containing this \ac{GRB}'s location. Given the youth, sky location, and probability of containment, we let our software trigger follow\nobreakdashes-up with the \ac{P48}.

Starting at 04:17:23~UT ($\Delta t \equiv t - t_{\mathrm{GBM}} = 4.2$~hr), we imaged each of these ten fields twice in 60~s exposures with the Mould $R$ filter. These fields were then subjected to the standard \ac{IPTF} transient search: image subtraction, source detection, and ``real/bogus'' machine ranking \citep{BloomMachineLearning,RB2} into likely astrophysical transient sources (``real'', or 1) or likely artifacts (``bogus'', or 0).

The \ac{IPTF} real\nobreakdashes-time analysis found 27,004 transient/variable candidates in these twenty individual subtracted images. Of these, 44 were eliminated because they were determined to be known asteroids in the Minor Planet Checker database\footnote{\url{http://www.minorplanetcenter.org/iau/mpc.html}} using PyMPChecker.\footnote{\url{http://dotastro.org/PyMPC/PyMPC/}} Demanding a real/bogus rank greater than 0.1 reduced the list to 4214.  Rejecting candidates that coincided with point sources in \ac{SDSS} brighter than $r'=21$ narrowed this to 2470.  Further, requiring detection in both \ac{P48} visits and imposing \acs{CCD}\nobreakdashes-wide data quality cuts (e.g., bad pixels) eliminated all but 43 candidates.  Following human inspection, seven sources were saved as promising transients in the \ac{IPTF} database.

Two candidates, iPTF13bxh and iPTF13bxu, were near the cores of bright, nearby galaxies, environments that are inherently variable and also present a challenge to image subtraction. A third, iPTF13bxr, was coincident with a galaxy in \ac{SDSS} with a quasar spectrum (SDSS~J145359.72+091543.3). iPTF13bxt was close to a star in \ac{SDSS}, and so was deemed a likely variable star.  We did not consider these further for the afterglow search. The final three sources, iPTF13bxj (real-bogus score $= 0.86$), iPTF13bxk (real-bogus score $= 0.49$), and iPTF13bxl (real-bogus score $= 0.83$), remained as potential counterparts and were scheduled for $g'r'i'$ photometry with the \acl{P60}~\citep[\acsu{P60};][]{P60Automation} and spectroscopic classification on the \ac{P200}.  iPTF13bxl ($\alpha = 14^\mathrm{h}29^\mathrm{m}14\fs78, \delta = +15\arcdeg46\arcmin, 26\farcs4$) was immediately identified as the most promising candidate because it showed a significant intra\nobreakdashes-night decline. Our panchromatic follow-up (Sections~\ref{sec:followup}~and~\ref{sec:spec})
confirmed iPTF13bxl was indeed the afterglow. Subsequent spectroscopy revealed iPTF13bxj to be a Type II supernova at $z=0.06$ and iPTF13bxk a quasar at $z=2.4$.

Following our discovery announcement \citep{GCN14967}, the \emph{Fermi} \ac{LAT} and \ac{GBM} teams published \ac{GCN} circulars announcing the detection of \ac{GRB}~130702A~\citep{GCN14971,GCN14972}.  As seen by the \ac{GBM}, GRB~130702A had a duration of $t_{90} \approx 59$~s and a 10~keV\nobreakdashes--1~MeV fluence of $f_{\gamma} = (6.3 \pm 2.0) \times 10^{-6}$~erg\,cm$^{-2}$.  The best-fit power-law spectrum may suggest a classification as an X-ray flash.  The \ac{LAT} location was $0\fdg9$ from iPTF13bxl, with a 90\% statistical error radius of $0\fdg5$. An \ac{IPN} triangulation \citep{GCN14974} yielded a 3\nobreakdashes-$\sigma$ annulus that was $0\fdg46$ wide from its center to its edges. iPTF13bxl was $0\fdg16$ from the annulus' centerline (Figure~\ref{fig:discovery}).

\section{Broadband photometric follow-up}
\label{sec:followup}

On 2013~July~3 at 4:10~UT, ($\Delta t = 28.1$~hr), the \ac{P60} obtained two sequences of Sloan $g'r'i'$ observations of the field of iPTF13bxl. \ac{P60} observations were calibrated relative to 20 reference stars in the \ac{SDSS} (AB) system. Final reduction of the \ac{P48} observations was performed automatically at the \acl{IPAC}~\citep{PTFPhotometricCalibration}. We corrected the \ac{P48} and \ac{P60} photometry for Galactic extinction using maps from \citet[][$E(B-V) = 0.024$~mag]{SchlaflyExtinction}.

\begin{figure*}
    \ifinthesis
    \centering
    \includegraphics[width=\textwidth]{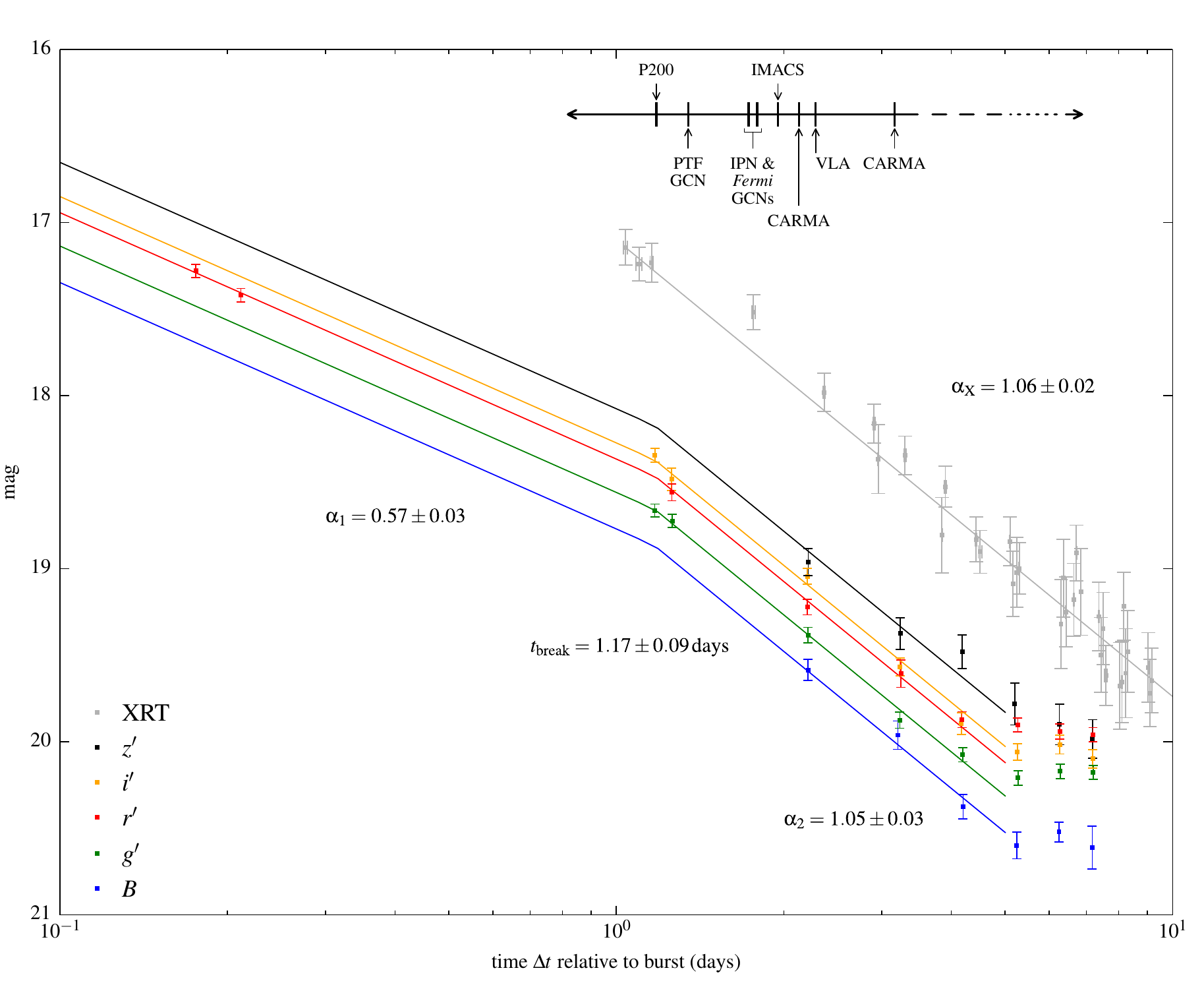}
    \fi
    \caption[Light curve of GRB~130702A~/~iPTF13bxl]{\label{fig:lightcurve}\ac{P48}, \ac{P60}, and \ac{XRT} light curves of iPTF13bxl.  The broken power\nobreakdashes-law fit is shown up to $\Delta t=5$~days. The \ac{XRT} observations, re-binned to improve presentation, are shown in gray as $m(\mathrm{AB})-6.5$ at 1~keV. A timeline in the top right puts the \ac{P48} and \ac{P60} observations in the context of the \ac{XRT} follow\nobreakdashes-up, \ac{PTF}'s discovery \ac{GCN}~\citep{GCN14967}, the announcement of the \ac{LAT}~\citep{GCN14971} and \ac{IPN}~\citep{GCN14974} localizations, and the radio observations. }
\end{figure*}

\begin{figure*}
    \ifinthesis
    \centering
    \includegraphics[width=\textwidth]{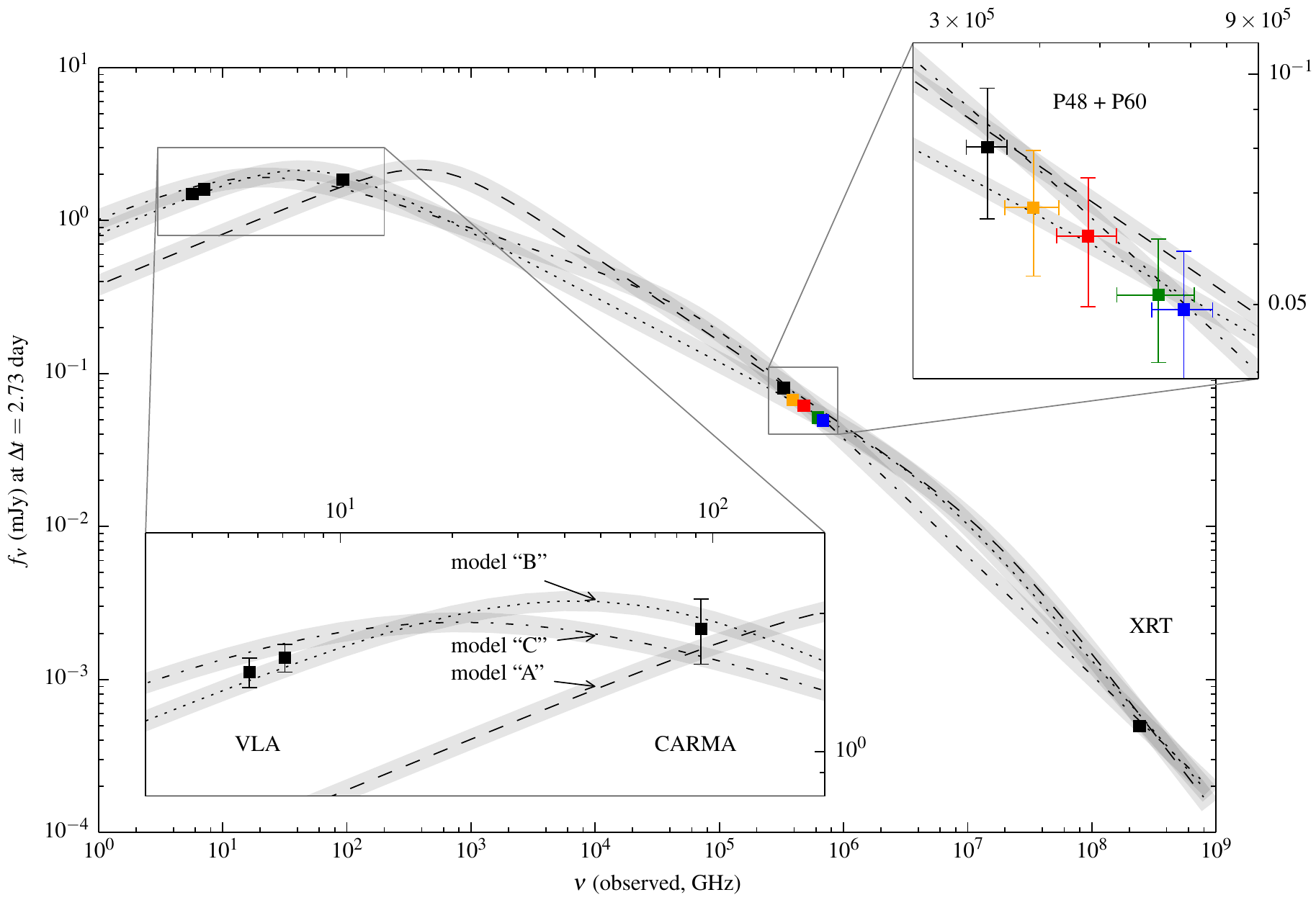}
    \fi
    \caption[Broadband \acs{SED} of GRB~130702A~/~iPTF13bxl]{\label{fig:sed}Broadband \ac{SED} of iPTF13bxl. Two insets show details of the radio and optical observations respectively. The \ac{XRT} and optical observations have been interpolated to the mean time of the radio observations ($\Delta t = 2.6$~days). }
\end{figure*}

The optical light curve is shown in Figure~\ref{fig:lightcurve}. We fit an achromatic broken power law to all filters and all times up to $\Delta t=5$~days after the burst. A spectral slope of $\beta_\mathrm{O}=0.7 \pm 0.1$ is sufficient to characterize the post\nobreakdashes-break color, illustrated in the inset of Figure~\ref{fig:sed}. We note that the optical decay ceased at $r' \approx 20$~mag after $\Delta t \approx~5$~days when the accompanying supernova started to dominate \citep{GCN14994}. This supernova will be the subject of a future work.

Following our discovery of iPTF13bxl, we triggered \ac{TOO} observations with the \emph{Swift} \ac{XRT} \citep{XRT} beginning at 00:50~UT on 2013~July~3 ($\Delta t = 1.03$~days).  We downloaded the data products from the \emph{Swift} \ac{XRT} repository \citep{SwiftXRTRepository}. The resulting 0.3\nobreakdashes--10~keV light curve is plotted in Figure~\ref{fig:lightcurve}. The spectrum is well fit by a power law with photon index $\Gamma = 2.0^{+0.14}_{-0.13}$, while the light curve fades in time with a power-law slope of $\alpha_\mathrm{X} = 1.06 \pm 0.02$, in excellent agreement with the post\nobreakdashes-break optical decay.

After the discovery of the optical counterpart to \ac{GRB}~130702A, we began observations with the \ac{CARMA}. All observations were carried out in single\nobreakdashes-polarization mode with the 3~mm receivers tuned to a frequency of 93~GHz, and were reduced using MIRIAD. We flux\nobreakdashes-calibrated the data using observations of MWC349 and 3C273.  The afterglow was well-detected in both epochs, and we measured flux densities of $1.58 \pm 0.33$~mJy and $1.85 \pm 0.30$~mJy on July 4.13 and 5.17, respectively.

The position of iPTF13bxl was observed with the \ac{VLA} in C\nobreakdashes-band beginning at 6:20~UT on 2013~July~4 ($\Delta t = 2.29$~days).  The observations were conducted using the standard WIDAR correlator setting. Data were reduced using the Astronomical Image Processing System package following standard practice. 3C286 was used for bandpass and flux calibration; J1415+1320 was used for gain and phase calibration. We detected a radio source with flux density of $1.49 \pm 0.08$~mJy at 5.1~GHz at $1.60 \pm 0.08$~mJy at 7.1~GHz. Errors on the measured flux were calculated as the quadrature sum of the map root\nobreakdashes-mean square and a fractional systematic error (of the order of 5\%) to account for uncertainty in the flux density calibration.

The broadband \ac{SED} is shown in Figure~\ref{fig:sed}. We interpolated both the optical and X-ray observations to the mean time of the \ac{VLA} and \ac{CARMA} observations.  In the context of the standard synchrotron afterglow model~\citep{AfterglowSpectra}, the comparable X-ray and optical spectral and temporal indices at this time suggest both bandpasses lie in the same spectral regime, likely with $\nu > \nu_{c}$.  This would imply a relatively hard electron spectral energy index ($N(\gamma_e) \propto \gamma_e^{-p}$) $p \approx 2$, possibly requiring a high-energy cut-off.  

Also in Figure~\ref{fig:sed} we plot three broadband \ac{SED} models synthesized using techniques similar to \citet{PerleyGRB130427A}. Although these models are not formal fits to our highly under-constrained observations, they demonstrate overall consistency with standard synchrotron afterglow theory.  Model ``A'' (dashed line; $\chi^2 = 126$) represents a constant\nobreakdashes-density (\acs{ISM}) circumburst medium with $p = 2.1$, $\epsilon_B = 0.48$, $\epsilon_e = 0.41$, $E = 3\times10^{51}$~erg, $n=1.2 \times 10^{-3}$~cm$^{-3}$.  This model under-predicts the \ac{VLA} bands, but this deviation could be due to scintillation or reverse shock emission at low frequencies.  Model ``B'' (dotted line; $\chi^2 = 7$) is in a wind environment ($\rho \propto r^{-2}$) with $p = 2.1$, $\epsilon_B = 0.32$, $\epsilon_e = 0.43$, $E = 1.4\times10^{51}$~erg, $A* = 4.8\times10^{-3}$~g\,cm$^{-1}$.  This fits the data well except for a small discrepancy with the optical spectral slope.  Model ``C'' (dotted-dashed line; $\chi^2 = 6$) is a similar wind model but with $p = 1.55$.  Of the three, ``C'' fits the data best, but seems non\nobreakdashes-physical (high\nobreakdashes-energy cutoff required).  Accurate determination of the underlying physical parameters would require tracing the evolution of the \ac{SED} with time. 

\section{Optical spectroscopy and host galaxy environment}
\label{sec:spec}

We obtained optical spectra of iPTF13bxl with the \ac{DBSP} mounted on the \ac{P200} on 2013~July~3.17 and the \acl{IMACS}~\citep[\acsu{IMACS};][]{IMACS} mounted on the 6~m Magellan Baade telescope on 2013~July~3.97 ($\Delta t = 1.2$ and 2.0~days, respectively).  The resulting spectra are plotted in Figure~\ref{fig:spectra}.

Our initial \ac{DBSP} spectrum exhibits a largely featureless, blue continuum. The higher \ac{SNR} of the \ac{IMACS} spectrum further reveals faint, narrow emission lines corresponding to [\ion{O}{3}] and H$\alpha$ at a common redshift of $z = 0.145 \pm 0.001$ (luminosity distance $d_{L} = 680$~Mpc), which we adopt as the distance to GRB~130702A. The continuum of both spectra exhibit deviations from a single power-law, with excess flux (when compared with the late-time photometric spectral index of $\beta_\mathrm{O} = 0.7$) visible at shorter wavelengths.  This may suggest some contribution from either shock breakout or the emerging supernova at very early times post-explosion.

Three galaxies are visible in the immediate environment of iPTF13bxl in pre-explosion imaging (labeled ``G1'', ``G2'', and ``G3'' in Figure~\ref{fig:discovery}).  Presumably the emission lines observed from iPTF13bxl arise in G1, given the small spatial offset ($0\farcs6$) and slit orientation (PA $= 90$).  However, our spectra also reveal that galaxies G2 and G3 both lie at redshifts consistent with iPTF13bxl (e.g., Figure~\ref{fig:spectra}).  Observations with DEIMOS on the Keck~II telescope reveal two more galaxies at the same redshift at separations of $1\farcm2$ (SDSS~J142910.29+154552.2) and $2\farcm7$ (SDSS~J142917.67+154352.2) from the transient. The explanation most consistent with past observations of long-duration \ac{GRB} host galaxies (e.g., \citealt{sgl09}) is that GRB~130702A exploded in a dwarf ($M_{r} \approx -16$~mag) member of this association or group, a relatively unusual environment \citep[see also][]{13bxlhost}.

\begin{figure*}
    \ifinthesis
    \centering
    \includegraphics[width=\textwidth]{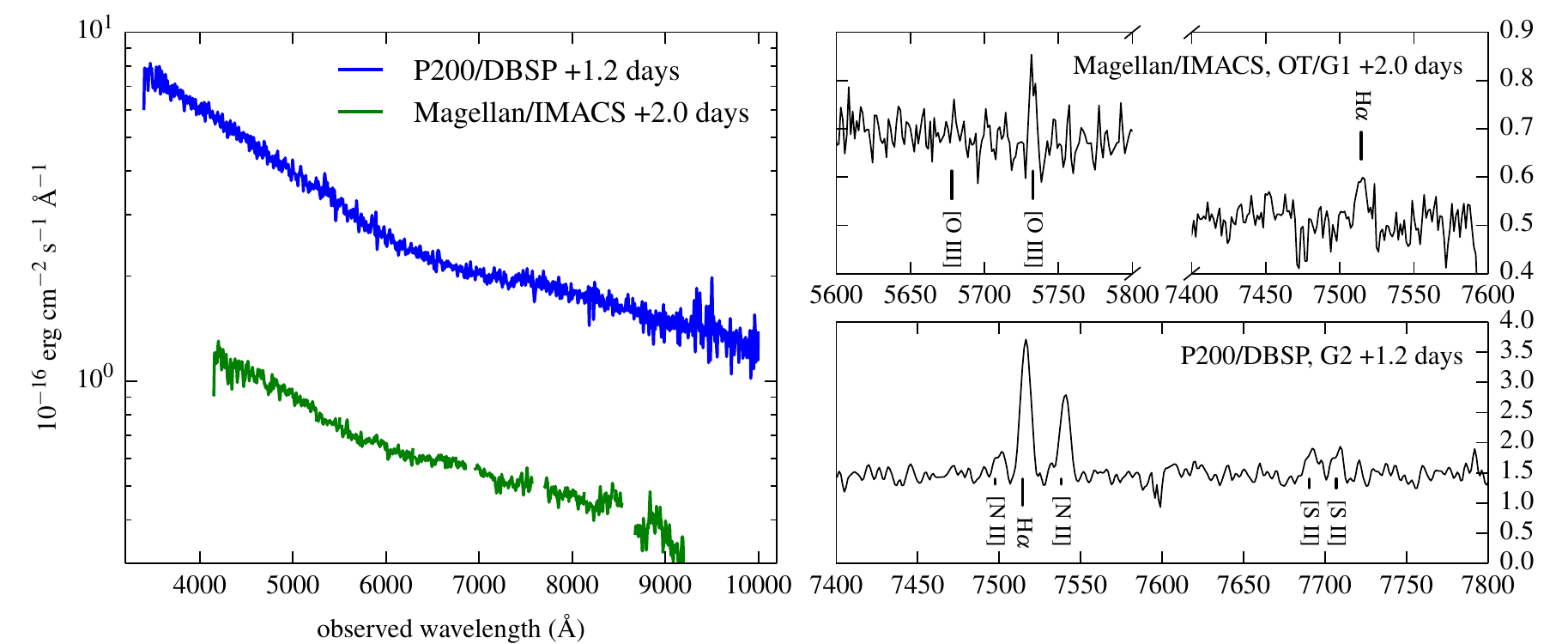}
    \fi
    \caption[Optical spectra]{\label{fig:spectra}Optical spectra of iPTF13bxl and the nearby galaxy SDSS~J142914.57+154619.3 (``G2''). Spectra in the left panel have been smoothed with a Savitzky-Golay filter. Our initial P200 spectrum of the afterglow (left panel, blue) exhibits a largely featureless blue continuum.  A higher \ac{SNR} spectrum taken the following night with \ac{IMACS} (left panel, green) revealed faint emission features corresponding to [\ion{O}{3}] and H$\alpha$ at $z = 0.145$ (top right panel). The bottom right panel shows a spectrum of the nearby galaxy G2, which has the same redshift as iPTF13bxl. }
\end{figure*}

\section{\acs{GRB}~130702A in context}
\label{sec:context}
The prompt $\gamma$-ray isotropic energy release ($E_{\gamma,\mathrm{iso}}$) of \acp{GRB} spans a range of six orders of magnitude, from $\sim 10^{48}$--$10^{54}$~erg.  At $z = 0.145$, the prompt emission from \ac{GRB}~130702A constrains $E_{\gamma,\mathrm{iso}} \lesssim (6.5 \pm 0.1) \times 10^{50}$~erg~\citep[90\% upper limit;][]{GCN15025}. This value is significantly smaller than typical cosmological \acp{GRB} ($E_{\gamma,\mathrm{iso}} \sim 10^{52}$--$10^{54}$~erg; \citealt{2006MNRAS.372..233A,2007ApJ...671..656B}).  Yet \ac{GRB}~130702A greatly outshines the most nearby, sub-luminous events with well-studied supernovae, such as \ac{GRB}~980425 ($E_{\gamma,\mathrm{iso}} = 1.0 \times 10^{48}$~erg; \citealt{paa+00}) and \ac{GRB}~060218 ($E_{\gamma,\mathrm{iso}} = 6.2 \times 10^{49}$~erg; \citealt{GRB060218ShockBreakoutNature}).  

At all wavelengths, the counterpart behaves like a typical \ac{GRB} afterglow scaled down in luminosity by a factor of $\sim$10 compared to a ``typical'' \emph{Swift} burst (or $\sim$100 lower than a luminous pre-\emph{Swift} burst) as observed at the same time.  This is intuitively explained by an equivalent scaling down of the overall energy (per solid angle) of the burst and shockwave relative to more typical, high-luminosity bursts.  It is not yet clear whether this energy difference is due primarily to the release of less relativistic ejecta by the burst overall, a wider jet, or a partially off-axis view of a structured jet.  Late-time radio follow-up should help distinguish these models: an intrinsically low-energy \ac{GRB} should produce a much earlier jet break than a widely-beamed burst, while a structured jet will actually produce an \emph{increase} in flux at late times as the jet core spreads and its radiation enters our sightline.

Events with similar energetics have been found by \emph{Swift}, e.g., \ac{GRB}~050826 at $z=0.30$ and \ac{GRB}~120422A at $z=0.28$ \citep{2007ApJ...661L.127M,zfs+12}.  However, given their low intrinsic luminosities and higher redshift, the afterglows were too faint to identify late-time breaks and establish their shock energies $E_K$, making them difficult to physically interpret.  \ac{GRB}~130702A's proximity avoids both these problems. Our observations suggest---and further observations should confirm---that its $\gamma$-ray and afterglow energetics are intermediate between these two previously quite-disparate classes of \acp{GRB}, helping to fill in the ``gap'' between the well-studied cosmological population and the class of less-luminous local \acp{GRB} and relativistic Type Ic supernovae (e.g., \citealt{2004Natur.430..648S,scp+10}).

\section{Conclusion}
\label{sec:conclusion}
Using the infrastructure outlined above, we estimate that a dedicated \ac{IPTF} program would recover $\sim$10 \ac{GBM} afterglows each year.  The addition of other surveys with comparably wide \aclp{FOV} and apertures (e.g., Pan\nobreakdashes-STARRS, SkyMapper, CRTS) could increase this number, assuming they had similar real\nobreakdashes-time transient detection and follow-up programs in place.  Since \ac{GBM} detects \acp{GRB} in the classical $\gamma$-ray band, their optical counterparts should more closely resemble the pre-\textit{Swift} population ($\approx 1$~mag brighter at a fixed time; \citealt{KanSwiftAfterglowsI}).  Even if only a single event per year as nearby as \ac{GRB}~130702A were uncovered, this would still represent a remarkable advance in our understanding of the \ac{GRB}\nobreakdashes--supernova connection.

Furthermore, this work sets the stage for more discoveries in ongoing and future physics experiments that are limited by similarly coarse position reconstruction. Later this decade, a network of advanced \ac{GW} detectors including the \acl{LIGO} (\acsu{LIGO}) and Virgo is expected to detect $\sim 0.4$\nobreakdashes--400 binary neutron star mergers per year~\citep{LIGORates}, but with positions uncertain to tens to hundreds of deg$^2$~\citep{FairhurstTriangulation,NissankeLocalization,LIGOObservingScenarios}.

Optical counterparts to \ac{GW} sources will rarely (due to jet collimation) include bright, on\nobreakdashes-axis short\nobreakdashes-hard burst afterglows. Fainter $r$\nobreakdashes-process\nobreakdashes-fueled kilonovae \citep{Kilonova} or yet fainter off\nobreakdashes-axis afterglows \citep{offaxis} are expected to accompany binary neutron star mergers. Both of these signatures are predicted to be several magnitudes fainter than iPTF13bxl. Optical searches will be inundated with astrophysical false positives~\citep{NissankeKasliwalEMCounterparts}. This problem will only be exacerbated for future surveys covering larger areas (e.g., \acl{ZTF}; \citealt{ZTF}) and/or with larger apertures (e.g., \acl{LSST}; \citealt{LSST}).  However, a breathtakingly complete astrophysical picture could reward us: masses and spins measured in \acp{GW}; host galaxy and disruption ejecta in optical; circumstellar environment in radio. The case of \ac{GRB}~130702A demonstrates for the first time that optical transients can be recovered from localization areas of $\sim$100~deg$^2$, reaching a crucial milestone on the road to Advanced \ac{LIGO}.

\section*{Acknowledgements}

Optical photometry and spectroscopy referred to in this work will be made available via WISeREP\footnote{\url{http://www.weizmann.ac.il/astrophysics/wiserep/}} \citep{yg12}.

We acknowledge A. Weinstein, A. Gal-Yam, R. Quimby, V. Connaughton, and the \emph{Fermi}-\ac{GBM} team for valuable discussions, S. Caudill, S. Tinyanont, D. Khatami for \ac{P200} observing, and the developers of the COSMOS package for Magellan data reduction.

This research is supported by the \ac{NSF} through a Graduate Research Fellowship for L.P.S., award PHY-0847611 for D.A.B., and NSF-CDI grant 0941742 for J.S.B.  M.M.K. acknowledges generous support from the Carnegie-Princeton Fellowship. M.M.K. and D.A.P. are supported by NASA through the Hubble Fellowship grants HST\nobreakdashes-HF\nobreakdashes-51293.01 and HST\nobreakdashes-HF\nobreakdashes-51296.01\nobreakdashes-A, awarded by the Space Telescope Science Institute, which is operated by the Association of Universities for Research in Astronomy, Inc., for NASA, under contract NAS~5\nobreakdashes-26555. E.O.O. is the incumbent of
the Arye Dissentshik career development chair and is supported by grants from the Israeli Ministry of Science and the I-CORE Program.  D.A.B. is further supported by an RCSA Cottrell Scholar award.

This research made use of Astropy\footnote{\url{http://www.astropy.org}} \citep{astropy}, a community-developed core Python package for Astronomy.  The National Radio Astronomy Observatory is a facility of the \acl{NSF} operated under cooperative agreement by Associated Universities, Inc.
\chapter{\emph{Fermi}, \acs{IPTF}, and the \acs{GRB}--supernova connection}
\label{chap:iptf-gbm}

\author{Leo~P.~Singer\altaffilmark{1},
Mansi~M.~Kasliwal\altaffilmark{2},
S.~Bradley~Cenko\altaffilmark{3,4},
Daniel~A.~Perley\altaffilmark{27,5},
Gemma~E.~Anderson\altaffilmark{6,7},
G.~C.~Anupama\altaffilmark{8},
Iair~Arcavi\altaffilmark{9,10},
Varun~Bhalerao\altaffilmark{11},
Brian~D.~Bue\altaffilmark{12},
Yi~Cao\altaffilmark{5},
Valerie~Connaughton\altaffilmark{13},
Alessandra~Corsi\altaffilmark{14},
Antonino~Cucchiara\altaffilmark{28,3},
Rob~P.~Fender\altaffilmark{6,7},
Neil~Gehrels\altaffilmark{3},
Adam~Goldstein\altaffilmark{28,15},
Assaf~Horesh\altaffilmark{16},
Kevin~Hurley\altaffilmark{17},
Joel~Johansson\altaffilmark{18},
D.~A.~Kann\altaffilmark{19,20},
Chryssa~Kouveliotou\altaffilmark{15},
Kuiyun~Huang\altaffilmark{21},
S.~R.~Kulkarni\altaffilmark{5},
Frank~Masci\altaffilmark{22},
Peter~Nugent\altaffilmark{23,24},
Arne~Rau\altaffilmark{20},
Umaa~D.~Rebbapragada\altaffilmark{12},
Tim~D.~Staley\altaffilmark{6,7},
Dmitry~Svinkin\altaffilmark{25},
Yuji~Urata\altaffilmark{26},
and~Alan~Weinstein\altaffilmark{1}}

\altaffiltext{1}{LIGO Laboratory, California Institute of Technology, Pasadena, CA 91125, USA}
\altaffiltext{2}{Observatories of the Carnegie Institution for Science, 813 Santa Barbara St, Pasadena CA 91101, USA}
\altaffiltext{3}{Astrophysics Science Division, NASA Goddard Space Flight Center, Mail Code 661, Greenbelt, MD 20771, USA}
\altaffiltext{4}{Joint Space-Science Institute, University of Maryland, College Park, MD 20742, USA}
\altaffiltext{5}{Cahill Center for Astrophysics, California Institute of Technology, Pasadena, CA 91125, USA}
\altaffiltext{6}{Astrophysics, Department of Physics, University of Oxford, Keble Road, Oxford OX1 3RH, UK}
\altaffiltext{7}{Physics \& Astronomy, University of Southampton, Southampton SO17 1BJ, UK}
\altaffiltext{8}{Indian Institute of Astrophysics, Koramangala, Bangalore 560 034, India}
\altaffiltext{9}{Las Cumbres Observatory Global Telescope Network, 6740 Cortona Dr., Suite 102, Goleta, CA 93117, USA}
\altaffiltext{10}{Kavli Institute for Theoretical Physics, University of California, Santa Barbara, CA 93106, USA}
\altaffiltext{11}{Inter-University Centre for Astronomy and Astrophysics (IUCAA), Post Bag 4, Ganeshkhind, Pune 411007, India}
\altaffiltext{12}{Jet Propulsion Laboratory, California Institute of Technology, Pasadena, CA 91109, USA}
\altaffiltext{13}{CSPAR and Physics Department, University of Alabama in Huntsville, 320 Sparkman Dr., Huntsville, AL 35899, USA}
\altaffiltext{14}{Texas Tech University, Physics Department, Lubbock, TX 79409-1051}
\altaffiltext{15}{Astrophysics Office, ZP12, NASA Marshall Space Flight Center, Huntsville, AL 35812, USA}
\altaffiltext{16}{Benoziyo Center for Astrophysics, Weizmann Institute of Science, 76100 Rehovot, Israel}
\altaffiltext{17}{Space Sciences Laboratory, University of California-Berkeley, Berkeley, CA 94720, USA}
\altaffiltext{18}{The Oskar Klein Centre, Department of Physics, Stockholm University, SE 106 91 Stockholm, Sweden}
\altaffiltext{19}{Th\"uringer Landessternwarte Tautenburg, Sternwarte 5, 07778 Tautenburg, Germany}
\altaffiltext{20}{Max-Planck Institut f\"ur Extraterrestrische Physik, Giessenbachstr. 1, 85748 Garching, Germany}
\altaffiltext{21}{Department of Mathematics and Science, National Taiwan Normal University, Lin-kou District, New Taipei City 24449, Taiwan}
\altaffiltext{22}{Infrared Processing and Analysis Center, California Institute of Technology, Pasadena, CA 91125, USA}
\altaffiltext{23}{Department of Astronomy, University of California, Berkeley, CA 94720-3411, USA}
\altaffiltext{24}{Physics Division, Lawrence Berkeley National Laboratory, Berkeley, CA 94720, USA}
\altaffiltext{25}{Ioffe Physical-Technical Institute, Politekhnicheskaya 26, St Petersburg 194021, Russia}
\altaffiltext{26}{Institute of Astronomy, National Central University, Chung-Li 32054, Taiwan}
\altaffiltext{27}{Hubble Fellow}
\altaffiltext{28}{NASA Postdoctoral Fellow}

\attribution{This chapter is reproduced from a paper to be titled ``The Needle in the 100~\lowercase{deg}$^2$ Haystack: Uncovering Afterglows of \emph{Fermi} GRBs with the Palomar Transient Factory,'' in preparation for \textnormal{The Astrophysical Journal}.}

\section{Introduction}

Deep synoptic optical surveys including the \acl{PTF} (\acsu{PTF}; \citealt{PTFRau,PTFLaw}) and Pan\nobreakdashes-STARRS \citep{PanSTARRS} have revealed a wealth of new transient and variable phenomena across a wide range of characteristic luminosities and time scales \citep{KasliwalThesis}. With a wide (7~deg$^2$) instantaneous field of view, moderately deep sensitivity (reaching $R = 20.6$~mag in 60~s), a consortium of follow\nobreakdash-up telescopes, sophisticated image subtraction and machine learning pipelines, and an international team of human\nobreakdashes-in\nobreakdashes-the\nobreakdashes-loop observers, \ac{PTF} has been a wellspring of new or rare kinds of explosive transients (for instance, \citealt{CaRichGapTransients,SLSNe}) and early\nobreakdashes-time observations of \acp{SN} or their progenitors (see, for example, \citealt{PTF11fe,PTF10vgv,PTF10tel,iPTF13ast}). \ac{PTF} has even blindly detected the optical emission (\citealt{GCN15883}; Cenko~et~al., in~preparation) from the rarest, brightest, and briefest of all known cosmic explosions, \acp{GRB}, hitherto only discoverable with the aid of precise localizations from space\nobreakdashes-based gamma\nobreakdashes-ray observatories. \ac{PTF} has also detected explosions that optically resemble \ac{GRB} afterglows but may entirely lack gamma\nobreakdashes-ray emission \citep{PTF11agg}.

\acp{GRB} and their broadband afterglows are notoriously challenging to capture. They naturally evolve from bright to faint, and from high (gamma- and hard X\nobreakdashes-ray) to low (optical and radio) photon energies, with information encoded on energy scales from 1~to~$10^{16}$~GHz \citep{PerleyGRB130427A} and time scales from $10^{-3}$~to~10$^7$~s. Only with a rapid sequence of handoffs between facilities graded by energy passband, field of view, and position accuracy, have we been able to find them, pinpoint their host galaxies, and constrain their physics. The \emph{Swift} mission \citep{Swift}, with its 1.4~sr\nobreakdashes-wide (50\% coded) \acl{BAT} (\acsu{BAT}; \citealt{BAT}) and its ability to slew and train its onboard \acl{XRT} (\acsu{XRT}; \citealt{XRT}) and \acl{UVOT} (\acsu{UVOT}; \citealt{UVOT}) on the location of a new burst within 100~s, has triumphed here: in 9~years of operation, it has tracked down $\approx 700$ X\nobreakdashes-ray afterglows and enabled extensive panchromatic observations by a worldwide collaboration of ground\nobreakdashes-based optical and radio facilities.

Meanwhile, the \emph{Fermi} satellite has opened up a new energy regime extending up to 300~GeV, with the \acl{LAT} (\acsu{LAT}; \citep{LAT}) detecting high\nobreakdashes-energy photons for about a dozen bursts per year. The \acl{GBM} (\acsu{GBM}; \citealt{GBM}), an all\nobreakdashes-sky instrument sensitive from 8~keV to 40~MeV, detects \acp{GRB} prolifically at a rate of $\approx 250$~yr$^{-1}$, with a large number (about 44~yr$^{-1}$) belonging to the rarer short, hard bursts \citep{FermiGBMCatalog}. Although \ac{LAT} can provide localizations that are as accurate as $\sim$10\arcmin, \emph{Fermi} \ac{GBM} produces error circles that are several degrees across. Consequently, most \emph{Fermi} bursts do not receive deep, broadband follow\nobreakdashes-up, with the properties of their afterglows largely unknown.

As part of the \ac{IPTF}, over the past year we have developed the ability to rapidly tile these $\sim 100$~deg$^2$ \ac{GBM} error circles and pinpoint the afterglows. This \ac{TOO} capability uses and briefly redirects the infrastructure of the ongoing synoptic survey, notably the machine learning software and the instrumental pipeline composed of the \acl{P48} (\acsu{P48}; \citealt{P48PTF}), the \acl{P60} (\acsu{P60}; \citealt{P60Automation}), and associated spectroscopic resources including the \acf{P200}.

In \citet{iPTF13bxl}, we announced the first discovery of an optical afterglow based solely on a \emph{Fermi} \ac{GBM} localization.\footnote{There are two earlier related cases. The optical afterglow of \ac{GRB}~090902B was detected ex post facto in tiled observations with \ac{ROTSE} about 80~min after the burst, but the afterglow was initially discovered with the help of an X\nobreakdashes-ray detection in \emph{Swift} observations of the \ac{LAT} error circle. \ac{GRB}~120716A was identified by \ac{IPTF} by searching a $\approx 2$~deg$^{2}$ \acs{IPN} error box \citep{GCN13489}.} That explosion, \ac{GRB}~130702A~/~iPTF13bxl, was noteworthy for several reasons. First, it was detected by \emph{Fermi} \ac{LAT}. Second, it was at moderately low redshift, $z = 0.145$, yet had prompt energetics that bridged the gap between ``standard'', bright cosmically distant bursts and nearby sub\nobreakdashes-luminous bursts and \acp{XRF}. Third, due to its low redshift, the accompanying \ac{SN} was spectroscopically detectable.

In this work, we begin with a detailed description of the operation of the \ac{IPTF} \ac{GRB} afterglow search. We then present seven more \ac{GBM}--\ac{IPTF} afterglows from the first 13 months of this project. In all eight cases, the association between the optical transient and the \ac{GRB} was proven by the presence of high\nobreakdashes-redshift absorption lines in the optical spectra and the coincident detection of a rapidly fading X\nobreakdashes-ray source with \emph{Swift} \ac{XRT}. In two cases, the positions were further corroborated by accurate \emph{Fermi} \ac{LAT} error circles, and in four cases by accurate \ac{IPN} triangulations involving distant spacecraft. In one case (\ac{GRB}~140508A), the \ac{IPN} triangulation was performed rapidly and was instrumental in selecting which optical transient candidates to follow up. In six cases, radio afterglows were detected. Our discovery rate of eight out of 35 events is consistent with the ages and searched areas of the \ac{GBM} bursts, combined with the luminosity function of optical afterglows. Consequently, by tiling larger areas and/or stacking exposures, the \ac{IPTF} afterglow search should be able to scale to more coarse localizations, such as those associated with short \acp{GRB}.

Next, we present extensive follow\nobreakdashes-up observations, including $R$\nobreakdashes-band photometry from the \ac{P48}, multicolor photometry from the \ac{P60}, spectroscopy (acquired with the \ac{P200}, Keck, Gemini, APO, Magellan, and VLT), and radio observations with the \acl{VLA}\footnote{\url{http://www.vla.nrao.edu}} (\acsu{VLA}), the \acl{CARMA} (\acsu{CARMA}; \citealt{CARMA1,CARMA2}), the \acl{ATCA} (\acsu{ATCA}; \citealt{ATCA}), and the \acl{AMI} (\acsu{AMI}; \citealt{AMI}). We provide basic physical interpretations of the broadband \acp{SED} of these afterglows. We find that seven of the events are consistent with the standard model of synchrotron cooling of electrons that have been accelerated by a single forward shock encountering either the constant-density circumburst \acl{ISM} (\acsu{ISM}; broadband behavior predicted in \citealt{AfterglowSpectra}) or a stellar (i.e., Wolf\nobreakdashes-Rayet) wind environment \citep{AfterglowSpectraWind}. The exception, \ac{GRB}~140620A~/~iPTF14cva, may be explained by an additional reverse shock or an inverse Compton component.

Two of the afterglows (\ac{GRB}~130702A~/~iPTF13bxl and \ac{GRB}~140606B~/~iPTF14bfu) faded away to reveal spectroscopically detected \acp{SNIcBL}. Despite the abundant photometric evidence for \acp{SN} in afterglow light curves (see \citealt{LightCurvesGRBSNe} and references therein), the distinction of \ac{SN} spectroscopy has been shared by scarcely tens\footnote{Between photometric, late\nobreakdashes-time red bumps and unambiguous spectral identifications, there are also \ac{GRB}\nobreakdashes--\acp{SN} that have some \ac{SN}\nobreakdashes-associated spectral features. The number of \acp{GRB} with spectroscopic \acp{SN} is, therefore, ill defined. See \citet{HjorthGRBSNConnection} and references therein for a more complete census.} out of $\approx$800 long \emph{Swift} bursts in nine years of operation.

We estimate the kinetic energies of the relativistic blast waves for each burst from their X\nobreakdashes-ray afterglows \citep{EnergyOfGammaRayBursts}. We find that although the gamma\nobreakdashes-ray energetics of these eight bursts are broadly similar to the \emph{Swift} sample, two low\nobreakdashes-luminosity bursts (\acp{GRB}~130702A~and~140606B) have significantly lower kinetic energies. We discuss the possibility that these two bursts arise not from a standard ultra\nobreakdashes-relativistic internal shock, but from a mildly relativistic shock as it breaks out from the progenitor star (see, for example, \citealt{RelativisticShockBreakoutRelation}).

We conclude by discussing prospects for targeted optical transient searches in wide areas. This is especially relevant for optical counterparts of gravitational wave events. We illustrate that optical afterglows of short bursts, which are intimately linked to the prime sources for the Advanced \acf{LIGO} and Virgo, should be well within the reach of a similar approach using \ac{ZTF} \citep{ZTF,ZTFBellm,ZTFSmith}.

\section{Search methodology}
\label{sec:afterglow-search-method}

\subsection{Automated \acs{TOO} Marshal: alerts and tiling}

A program called the \ac{IPTF} \ac{TOO} Marshal monitors the stream of \ac{GCN} notices\footnote{\url{http://gcn.gsfc.nasa.gov}} from the three redundant, anonymous NASA/GSFC VOEvent servers. It listens for notices of type \texttt{FERMI\_GBM\_GND\_POS}, sent by \ac{GBM}'s automated on\nobreakdashes-ground localization, or \texttt{FERMI\_GBM\_FIN\_POS}, sent by the \ac{GBM} burst advocate's human\nobreakdashes-in\nobreakdashes-the\nobreakdashes-loop localization.\footnote{Usually, the \emph{Fermi} team suppresses the notices if the burst is detected and localized more accurately by \emph{Swift} \ac{BAT}.}

Upon receiving either kind of notice, the \ac{TOO} Marshal determines if the best-estimate sky position is observable from Palomar at any time within the 24 hours after the trigger. The criterion for observability is that the position is at an altitude $> 23\fdg5$ (i.e. airmass $\lesssim 2.5$), at least $20\arcdeg$ from the center of the moon, at an hour angle between $\pm 6\fh5$, and that the Sun is at least $12\arcdeg$ below the horizon at Palomar.

If the position is observable and the 1\nobreakdashes-$\sigma$ statistical error radius $r_\mathrm{stat}$ reported in the \ac{GCN} notice is less than $10\arcdeg$, the \ac{TOO} Marshal selects a set of ten \ac{P48} fields that optimally cover the error region.\footnote{We followed up but did not detect afterglows of two bursts that were beyond our nominal cutoff error radius of $10\arcdeg$.} It converts the \ac{GBM} position estimate and radius into a probability distribution by applying a well\nobreakdashes-known empirical prescription of the systematic errors of the \ac{GBM} localization. \citet{FermiGBMFirstTwoYears} states the total effective error radius in the \texttt{FERMI\_GBM\_FIN\_POS} localizations is well described by the quadrature sum of the statistical radius and a systematic contribution, where the systematic is $2\fdg6$ for 72\% of bursts and $10\fdg4$ for 28\% of bursts. We use the weighted \ac{RMS} of these two values, $r_\mathrm{sys} = \sqrt{0.72(2\fdg6)^2 + 0.28(10\fdg4)^2} \approx 6\arcdeg$. The total error radius is then $r_\mathrm{eff} = \sqrt{{r_\mathrm{stat}}^2 + {r_\mathrm{sys}}^2}$. We construct a Fisher\nobreakdashes--von~Mises distribution, centered on the best\nobreakdashes-estimate position, with a concentration parameter of
\begin{equation}
    \kappa = \left[1 - \cos \left( \frac{\pi}{180\arcdeg} r_\mathrm{eff} \right)\right]^{-1}.
\end{equation}

With the \texttt{FERMI\_GBM\_FIN\_POS} alert, the \emph{Fermi} \ac{GBM} team also distributes a detailed localization map that accounts for the systematic effects \citep{GBMLocalization}. The \ac{TOO} Marshal retrieves from the \emph{Fermi} data archive a file that describes the 1-, 2-, and 3\nobreakdashes-$\sigma$ significance contours. If the localization has significant asymmetry, we also retrieve a 2D FITS image whose pixel values correspond to the \ac{GBM} localization significance, and use this instead of the Fisher\nobreakdashes--von~Mises distribution.

Giving preference to fields for which deep coadded reference images exist, the \ac{TOO} Marshal selects ten \ac{P48} fields spanning an area of $\approx 72$~deg$^2$ to maximize the probability of enclosing the true (but as yet unknown) location of the source, assuming the above distribution.

The Marshal then immediately contacts a team of humans (the authors) by SMS text message, telephone, and e-mail. The humans are directed to a mobile\nobreakdashes-optimized web application to trigger the \ac{P48} (see Fig.~\ref{fig:screenshot}).

\ifinthesis
\begin{figure*}
\else
\begin{figure}
\fi
    \centering
    \ifinthesis
    \includegraphics[width=\textwidth]{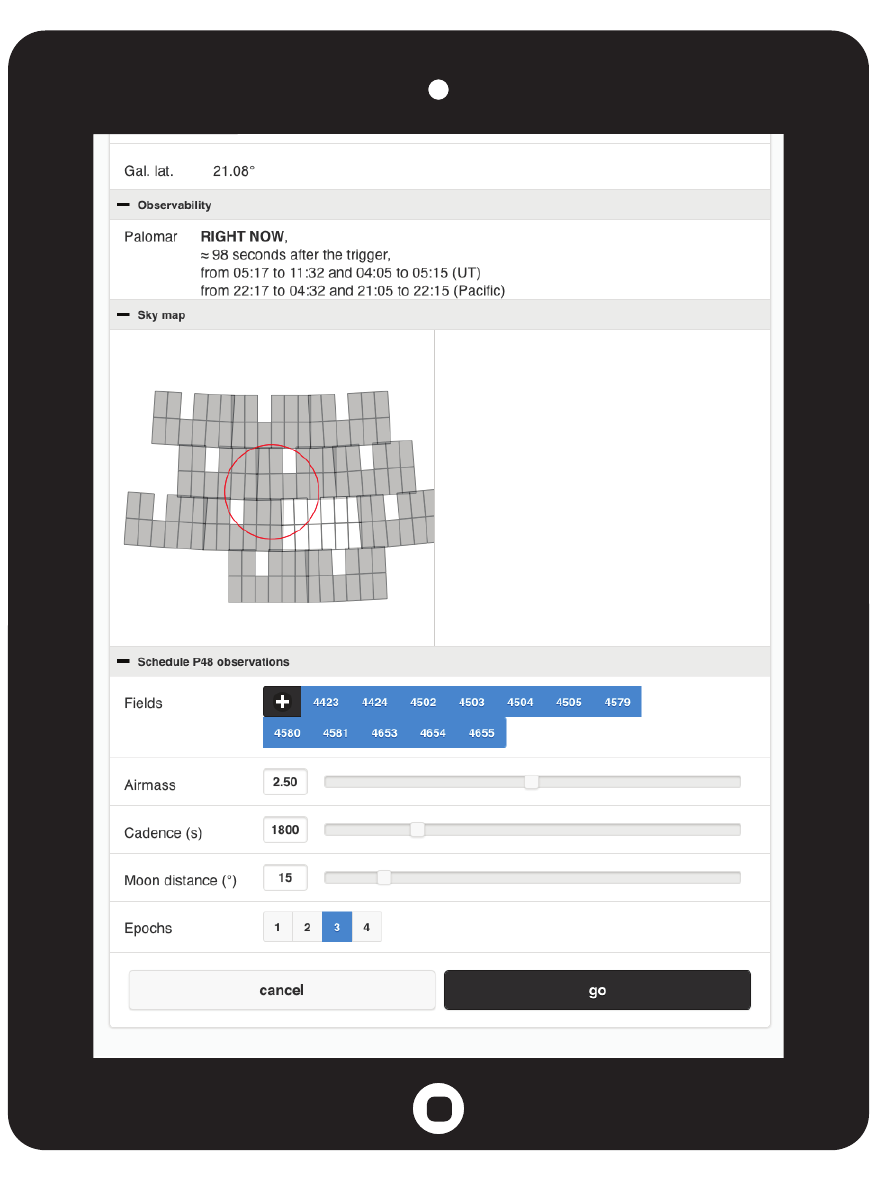}
    \fi
    \caption[Screen shot of \acs{IPTF} \acs{TOO} Marshal]{\label{fig:screenshot}Screen shot of the \ac{IPTF} \ac{TOO} Marshal shortly after a \emph{Fermi} \ac{GBM} detection. At this stage, the application presents the recommended \ac{P48} fields, the time window of observability, and the history of \ac{GCN} notices and circulars related to the trigger. It gives the human participants the option to customize the \ac{P48} sequence by adding or removing \ac{P48} fields and tuning the airmass limit, cadence, or number of images.
}
\ifinthesis
\end{figure*}
\else
\end{figure}
\fi

\subsection{Triggering the \acs{P48}}

Within the above constraints, we decide whether to follow up the burst based on the following criteria. The event must be $\lesssim$12 hours old when it first becomes observable from Palomar and we must cover enough of the error circle to have a $\gtrsim$30\% chance of enclosing the position of the source. We discard any bursts that are detected and accurately localized by \emph{Swift} \ac{BAT}, because these are more efficiently followed up by conventional means. We also give preference to events that are out of the Galactic plane and that are observable for at least 3 hours.

There are some exceptional circumstances that override these considerations. If the burst's position estimate is accessible within an hour after the burst, we may select it even if the observability window is very brief. If the burst is very well localized or has the possibility of a substantially improved localization later due to a \ac{LAT} or \ac{IPN} detection, we may select it even if it is in the Galactic plane.

The default observing program is three epochs of \ac{P48} images at a 30\nobreakdashes-minute cadence. The human may shorten or lengthen the cadence if the burst is very young or old, change the number of epochs, or add and remove \ac{P48} fields. When the human presses the ``Go'' button, the \ac{TOO} Marshal sends a machine-readable e-mail to the \ac{P48} robot. The robot adds the requested fields to the night's schedule with the highest possible priority, ensuring that they are observed as soon as visible.

\subsection{Automated candidate selection}

As the night progresses, the \ac{TOO} Marshal monitors the progress of the observations and the \ac{IPTF} real\nobreakdashes-time image subtraction pipeline (Nugent~et~al., in~preparation). The real\nobreakdashes-time pipeline creates difference images between the new \ac{P48} observations and coadded references composed of observations from months or years earlier. It generates candidates by performing source extraction on the difference images. A machine learning classifier assigns a \emph{real/bogus} score (RB2; \citealt{RB2}) to each candidate that predicts how likely the candidate is to be a genuine astrophysical source (rather than a radiation hit, a ghost, an imperfect image subtraction residual, or any other kind of artifact).

Table~\ref{table:vetting} lists the number of candidates that remain after each stage of candidate selection. First, requiring candidates to have \ac{SNR}$>5$ gives us a median of 35\,000 candidates. This number varies widely with galactic latitude and the area searched (a median of $\sim$500~deg$^{-2}$). Second, we only select candidates that have RB2$>0.1$, reducing the number of candidates to a median of 36\% of the original list.\footnote{This RB2 threshold is somewhat deeper than that which is used in the \ac{IPTF} survey. An improved classifier, RB4 \citep{RB4}, entered evaluation in August 2014 shortly before \ac{GRB}~140808A.} Third, we reject candidates which coincide with known stars in reference catalogs (\ac{SDSS} and the \ac{PTF} reference catalog), cutting the list to 17\%. Fourth, we eliminate asteroids catalogued by the Minor Planet Center, reducing the list to 16\%. Fifth, we demand at least two secure \ac{P48} detections after the \ac{GBM} trigger, reducing the list to a few percent, or $\sim 500$ candidates.

When the image subtraction pipeline has finished analyzing at least two successive epochs of any one field, the \ac{TOO} Marshal contacts the humans again and the surviving candidates are presented to the humans via the Treasures portal.

\begin{deluxetable}{rrrrrrr}
\tablewidth{0pt}
\tablecaption{\label{table:vetting}Number of optical transient candidates surviving each vetting stage}
\tablehead{
    \colhead{} &
    \colhead{\acs{SNR}} &
    \colhead{RB2} &
    \colhead{not} &
    \colhead{not in} &
    \colhead{detected} &
    \colhead{saved for} \\
    \colhead{GRB} &
    \colhead{$>5$} &
    \colhead{$>0.1$} &
    \colhead{stellar} &
    \colhead{MPC\tablenotemark{a}} &
    \colhead{twice} &
    \colhead{follow\nobreakdashes-up}
}
\startdata
130702A  &   14\,629  &   2\,388  &   1\,346  &   1\,323  &     417  &  11  \\
131011A  &   21\,308  &   8\,652  &   4\,344  &   4\,197  &     434  &  23  \\
131231A  &    9\,843  &   2\,503  &   1\,776  &   1\,543  &  1\,265  &  10  \\
140508A  &   48\,747  &  22\,673  &   9\,970  &   9\,969  &     619  &  42  \\
140606B  &   68\,628  &  26\,070  &  11\,063  &  11\,063  &  1\,449  &  28  \\
140620A  &  152\,224  &  50\,930  &  17\,872  &  17\,872  &  1\,904  &  34  \\
140623A  &   71\,219  &  29\,434  &  26\,279  &  26\,279  &     442  &  23  \\
140808A  &   19\,853  &   4\,804  &   2\,349  &   2\,349  &      79  &  12  \\
\tableline
\multicolumn{2}{r}{median reduction} & 36\% & 17\% & 16\% & 1.7\% & 0.068\%
\enddata
\tablenotetext{a}{Not in Minor Planet Center database}
\end{deluxetable}

\subsection{Visual scanning in Treasures Portal}

The remaining candidate vetting steps currently involve human participation, and are informed by the nature of the other transients that \ac{IPTF} commonly detects: foreground \acp{SN} (slowly varying and in low\nobreakdashes-$z$ host galaxies), \acp{AGN}, cataclysmic variables, and M\nobreakdashes-dwarf flares.

\begin{figure}
    \centering
    \includegraphics[width=\columnwidth]{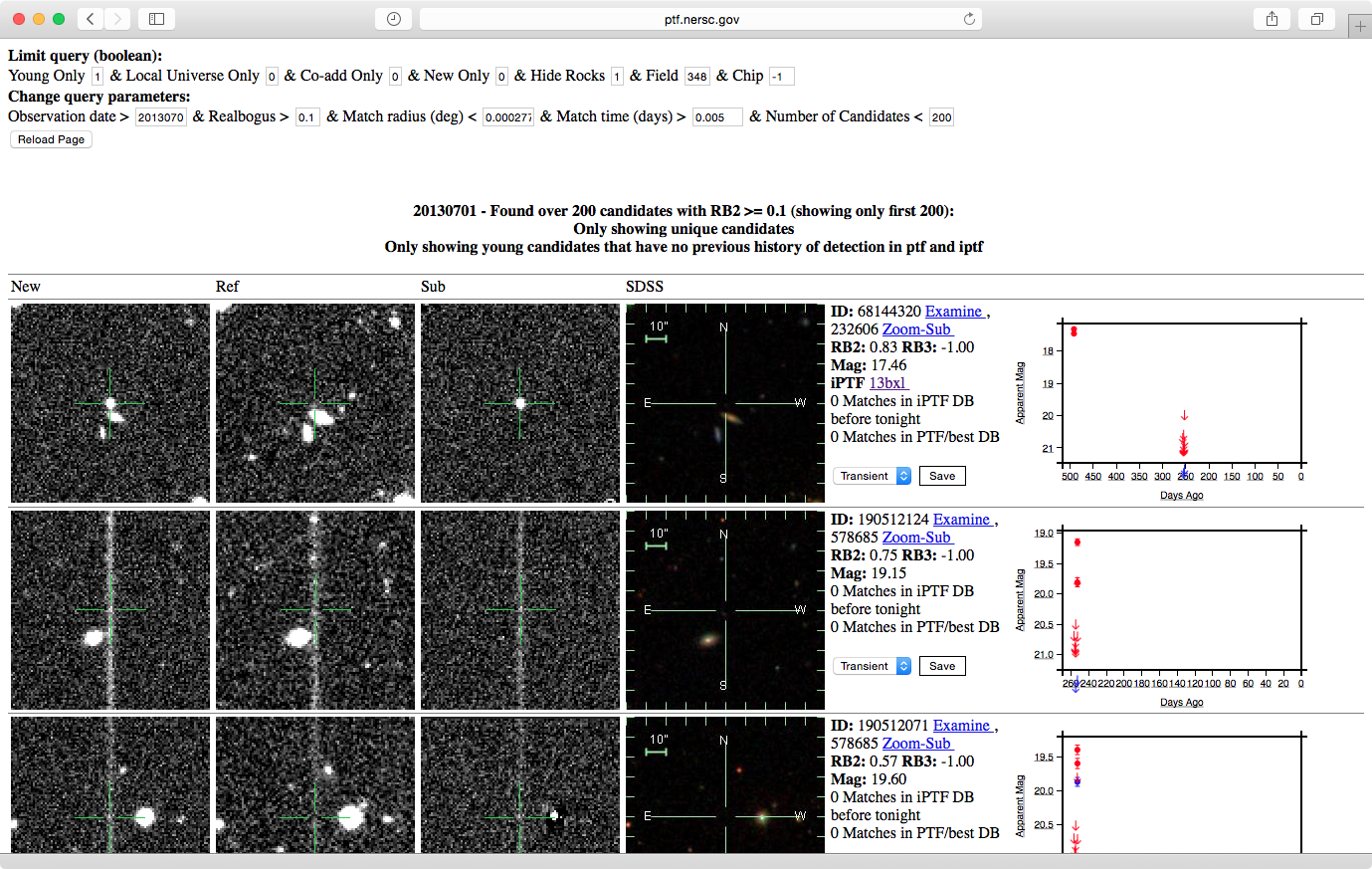}
    \caption[Screen shot of \acs{IPTF} Treasures portal]{\label{fig:treasures}Screen shot of the Treasures portal, showing new, reference, subtraction, and archival \ac{SDSS} images as well as \ac{P48} light curves. This page is for the date and field containing \ac{GRB}~130702A~/~iPTF13bxl.}
\end{figure}

In the Treasures portal, we visually scan through the automatically selected candidates one \ac{P48} field at a time, examining $\sim$10 objects per field (see Figure~\ref{fig:treasures} for a screen shot of the Treasures portal). We visually assess each candidate's image subtraction residual compared to the neighboring stars of similar brightness in the new image. If the residual resembles the new image's PSF, then the candidate is considered likely to be a genuine transient or variable source.

Next, we look at the photometric history of the candidates. Given the time $t$ of the optical observation relative to the burst and the cadence, $\delta t$, we expect that a typical optical afterglow that decays as a power law $F_\nu \propto t^-\alpha$, with $\alpha=1$, would fade by $\delta m = 2.5 \log_{10} (1+\delta t/t)$~mag over the course of our observations. Any source that exhibits statistically significant fading ($\delta m / m \gg 1$) consistent with an afterglow decay becomes a prime target.\footnote{A source that exhibits a statistically significant rise is generally also followed up, but as part of the main \ac{IPTF} transient survey, rather than as a potential optical afterglow.}

Note that a $1\sigma$ decay in brightness requires such a source to be
\begin{equation}
    -2.5\log_{10} \left(\frac{\delta t}{t\sqrt{2}}\right)
\end{equation}
brighter than the $1\sigma$ limiting magnitude of the exposures. For example, given the \ac{P48}'s typical limiting magnitude of $R = 20.6$ and the standard cadence of $\delta t = 0.5$~hour, if a burst is observed $t = 3$~hours after the trigger, its afterglow may be expected to have detectable photometric evolution only if it is brighter than $R = 18.3$. Noting that long \acp{GRB} preferentially occur at high redshifts and in intrinsically small, faint galaxies \citep{GRBSNHostGalaxies}, we consider faint sources that do not display evidence of fading if they are not spatially coincident with any sources in \ac{SDSS} or archival \ac{IPTF} observations.

Therefore, we consider faint sources that do not display evidence of fading if they have no plausibly associated host galaxy in \ac{IPTF} reference images or SDSS (indicating either a faint quiescent stellar source or a distant host galaxy).

If a faint source is near a spatially resolved galaxy, then we compute its distance modulus using the galaxy's redshift or photometric redshift from SDSS. We know that long \ac{GRB} optical afterglows at $t=1$~day typically have absolute magnitudes of $-25 < M_B < -21$ (1$\sigma$ range; see Figure~9 of \citealt{KannTypeITypeIIOpticalAfterglows}). Most \acp{SN} are significantly fainter: type~Ia are typically $M_B \sim -19$ whereas Ibc and II are $M_B \sim -17$, with luminous varieties of both Ibc and II extending to $M_B \sim -19$ \citep{RichardsonComparativeSupernovae,LickSupernovaLuminosityFunction}. Therefore, if the candidate's presumed host galaxy would give it an absolute magnitude $M_R < -20$, it is considered promising. This criterion is only useful for long \acp{GRB} because short \ac{GRB} afterglows are typically $\sim 6$~mag fainter than long \ac{GRB} afterglows \citep{KannTypeITypeIIOpticalAfterglows}.

The human saves all candidates that are considered promising by these measures to the \ac{IPTF} Transient Marshal database. This step baptizes them with an \ac{IPTF} transient name, which consists of the last two digits of the year and a sequential alphabetic designation.

\subsection{Archival vetting in the Transient Marshal}

\begin{figure}
    \centering
    \includegraphics[width=\columnwidth]{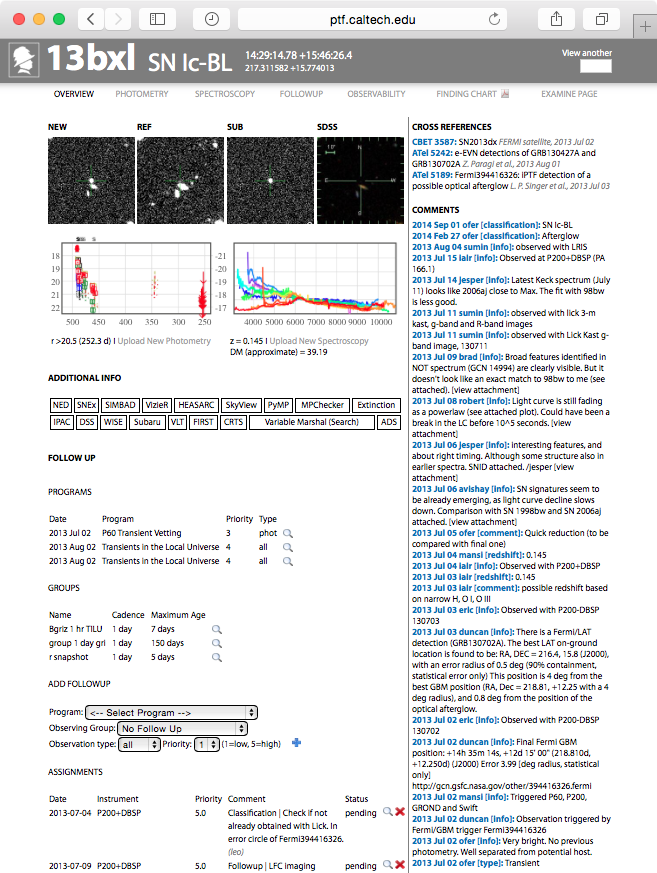}
    \caption[Screen shot of \acs{IPTF} Transient Marshal]{\label{fig:transient-marshal}Screen shot of the \ac{IPTF} Transient Marshal, showing \ac{GRB}~130702A~/~iPTF13bxl.}
\end{figure}

Once named in the Transient Marshal, we perform archival vetting of each candidate using databases including VizieR \citep{VizieR}, NED\footnote{http://ned.ipac.caltech.edu}, the \acf{HEASARC}\footnote{http://heasarc.gsfc.nasa.gov}, and \acl{CRTS} (\acsu{CRTS}; \citealt{CRTS}), in order to check for any past history of variability at that position (see Figure~\ref{fig:transient-marshal} for a screen shot of the Transient Marshal).

We check for associations with known quasars or active galactic nuclei in \citet{QuasarAtlas} or with \ac{AGN} candidates in \citet{ARXA}.

M\nobreakdashes-dwarfs can produce bright, blue, rapidly fading optical flares than can mimic optical afterglows. To filter our M\nobreakdashes-dwarfs, we check for quiescent infrared counterparts in WISE \citep{ALLWISE}. Stars of spectral type L9\nobreakdashes--M0 peak slightly blue\nobreakdashes-ward of the WISE bandpass, with typical colors \citep{WISEOnOrbit}
\begin{eqnarray*}
    3 \lesssim& [R-W1] &\lesssim 12 \\
    0.1 \lesssim& [W1-W2] &\lesssim 0.6 \\
    0.2 \lesssim& [W2-W3] &\lesssim 1 \\
    0 \lesssim& [W3-W4] &\lesssim 0.2.
\end{eqnarray*}
Therefore, a source that is detectable in WISE but that is either absent from or very faint in the \ac{IPTF} reference images suggests a quiescent dwarf star.

\subsection{Photometric, spectroscopic, and broad-band follow-up}

The above stages usually result in $\sim$10 promising optical transient candidates that merit further follow\nobreakdashes-up. If, by this point, data from \emph{Fermi} \ac{LAT} or from \ac{IPN} satellites is available, we can use the improved localization to select an even smaller number of follow\nobreakdashes-up targets.

For sources whose photometric evolution is not clear, we perform photometric follow\nobreakdashes-up. We may schedule additional observations of some of the \ac{P48} fields if a significant number of candidates are in the same field. We may also use the \ac{P48} to gather more photometry for sources that are superimposed on a quiescent source or galaxy, in order to make use of the image subtraction pipeline to automatically obtain host\nobreakdashes-subtracted magnitudes. For isolated sources, we schedule one or more epochs of $r$\nobreakdashes-band photometry with the \ac{P60}. If, by this point, any candidates show strong evidence of fading, we begin multicolor photometric monitoring with the \ac{P60}.

Next, we acquire spectra for 1\nobreakdashes--3 candidates per burst using the \ac{P200}, Gemini, Keck, Magellan, or \ac{HCT}. A spectrum that has a relatively featureless continuum and high redshift absorption lines secures the classification of the candidate as an optical afterglow.

Once any single candidate becomes strongly favored over the others based on photometry or spectroscopy, we trigger X\nobreakdashes-ray and UV observations with \emph{Swift} and radio observations with \ac{VLA}, \ac{CARMA}, and \ac{AMI}. Detection of a radio or X\nobreakdashes-ray afterglow typically confirms the nature of the optical transient, even without spectroscopy.

Finally, we promptly release our candidates, upper limits, and/or confirmed afterglow discovery in \ac{GCN} Circulars.

\subsection{Long-term monitoring and data reduction}

To monitor the optical evolution of afterglows identified by our program, we typically request nightly observations in $ri$ (and occasionally $gz$) filters for as long as the afterglow remained detectable. Bias subtraction, flat-fielding, and other basic reductions are performed automatically at Palomar by the \ac{P60} automated pipeline using standard techniques. Images are then download and stacked as necessary to improve the \ac{SNR}. Photometry of the optical afterglow is then performed in IDL using a custom aperture\nobreakdashes-photometry routine, calibrated relative to \ac{SDSS} secondary standards in the field (when available) or using our own solution for secondary field standards constructed during a photometric night (for fields outside the \ac{SDSS} footprint).

For some bursts (GRB~140606B), we also obtain photometry with the \acf{LMI} mounted on the 4.3\,m \acf{DCT} in Happy Jack, Arizona. Standard CCD reduction techniques (e.g., bias subtraction, flat-fielding, etc.) are applied using a custom IRAF pipeline. Individual exposures are aligned with respect to astrometry from the \acl{2MASS} (\acsu{2MASS}; \citealt{2MASS}) using SCAMP \citep{SCAMP} and stacked with SWarp \citep{SWarp}.

We also usually monitor \ac{GBM}--\ac{IPTF} afterglows with \ac{CARMA}, a millimeter\nobreakdashes-wave interferometer located at Cedar Flat near Big Pine, California. All observations are conducted at 93~GHz in single-polarization mode in the array's C, D, or E configuration. Targets are typically observed once for 1\nobreakdashes--3~hours within a few days after the \ac{GRB}, establishing the phase calibration using periodic observations of a nearby phase calibrator and the bandpass and the flux calibration by observations of a standard source at the start of the track. If detected, we acquire additional observations in approximately logarithmically-spaced time intervals until the afterglow flux falls below detection limits. All observations are reduced using MIRIAD using standard interferometric flagging and cleaning procedures.

\ac{VLA} observations are reduced using the \ac{CASA} package. The calibration is performed using the \ac{VLA} calibration pipeline. After running the pipeline, we inspect the data (calibrators and target source) and apply further flagging when needed. The \ac{VLA} measurement errors are a combination of the \ac{RMS} map error, which measures the contribution of small unresolved fluctuations in the background emission and random map fluctuations due to receiver noise, and a basic fractional error (here estimated to be $\approx 5\%$) which accounts for inaccuracies of the flux density calibration. These errors are added in quadrature and total errors are reported in Table~\ref{tab:radio}.

\ac{AMI} is composed of eight 12.8~m dishes operating in the 13.9\nobreakdashes--17.5~GHz range (central frequency of 15.7~GHz) when using frequency channels 3\nobreakdashes--7 (channels 1, 2, and 8 are disregarded due to their currently susceptibility to radio interference). For further details on the reduction and analysis performed on the AMI observations please see \citep{GRB130427A-AMI}.

\section{The \acs{GBM}--\acs{IPTF} bursts}

\begin{figure*}
    \centering
    \includegraphics[width=\textwidth]{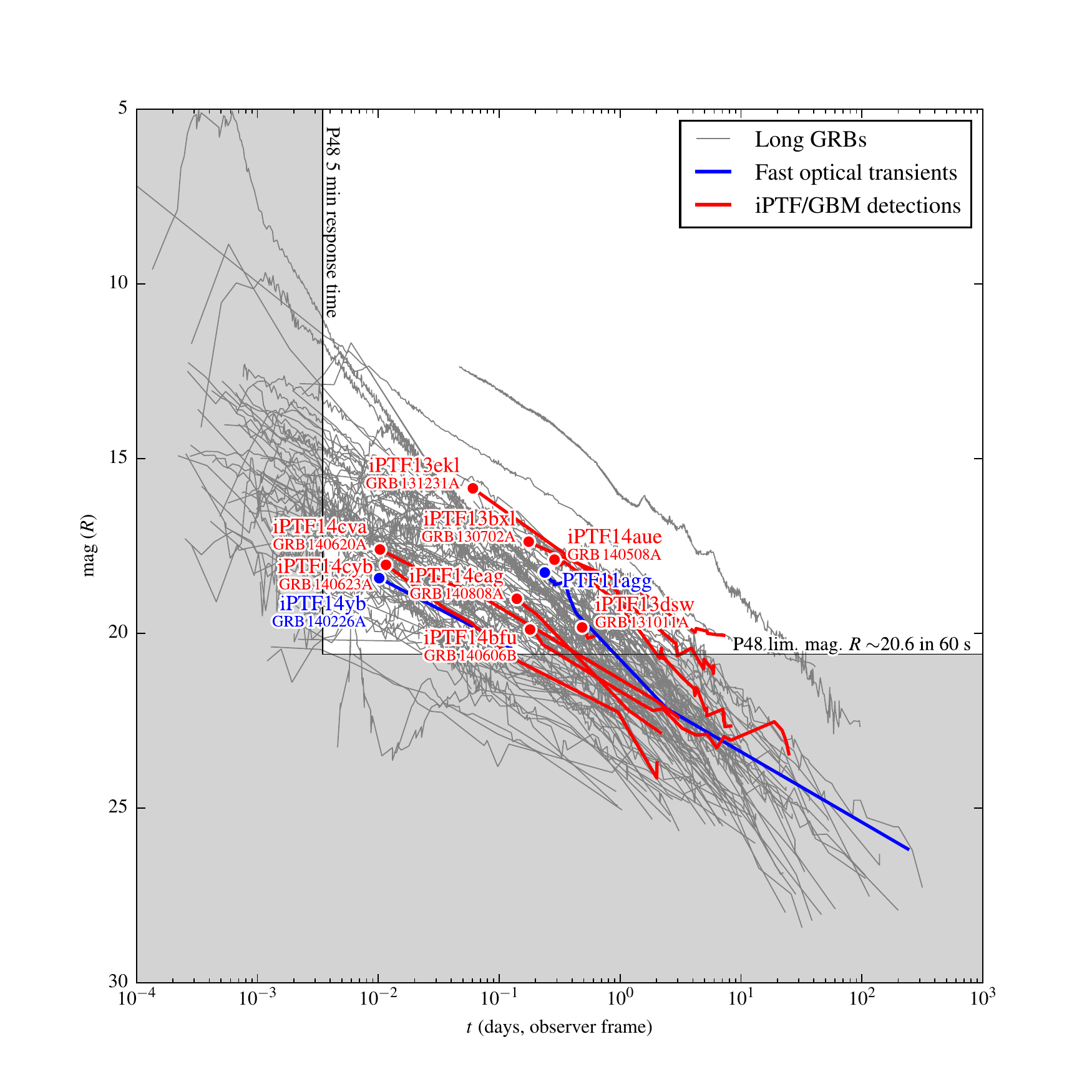}
    \caption[Optical light curves of long \acsp{GRB}]{\label{fig:lightcurve-zoo}Optical light curves of \emph{Fermi}\nobreakdashes--\ac{IPTF} afterglows to date. The light curves of the eight \ac{IPTF}/\ac{GBM} bursts are shown in red. For comparison, the gray lines show a comprehensive sample of long \ac{GRB} optical light curves from \citet{DarkBurstsSwiftEra}, \citet{KanSwiftAfterglowsI}, \citet{PerleyGRB130427A}, and D. A. Kann (private communication). The white area outside of the light gray shading illustrates the range of \ac{GRB} afterglows that are accessible given a half\nobreakdashes-hour cadence and the \ac{P48}'s 60~s limiting magnitude of $R = 20.6$. The two light curves shown in blue are other related \ac{IPTF} transients. The first is PTF11agg, an afterglow\nobreakdashes-like transient with no detected gamma\nobreakdashes-ray emission \citep{PTF11agg}. The second is \ac{GRB}~140226A~/~iPTF14yb, reported initially by \ac{IPTF} from its optical afterglow (\citealt{GCN15883}; Cenko et al., in preparation), and later by \ac{IPN} from its gamma-ray emission \citep{GCN15888}.}
\end{figure*}

To date, we have successfully followed up 35 \emph{Fermi} \ac{GBM} bursts and detected eight optical afterglows. The detections are listed in Table~\ref{table:detections}, and all of the \ac{P48} tilings are listed in  Table~\ref{table:nondetections}. In Figure~\ref{fig:lightcurve-zoo}, the light curves are shown in the context of a comprehensive sample of long and short \ac{GRB} afterglows compiled by D.~A.~Kann~(private communication).

{
\ifinthesis
\tabletypesize{\scriptsize}
\fi

\begin{deluxetable*}{ll|rrr|lr@{$\pm$}lr@{$\pm$}lr@{$\pm$}lr}
\tablecaption{\label{table:detections}\acs{GBM}--\acs{IPTF} detections.}
\tablehead{
    \colhead{} & \colhead{} & \colhead{RA} & \colhead{Dec} & \colhead{Gal.} & \colhead{} & \multicolumn{2}{c}{$E_\mathrm{peak}$} & \multicolumn{2}{c}{$E_{\gamma,\mathrm{iso}}$} & \colhead{} & \colhead{} & \colhead{} \\
    \colhead{GRB} & \colhead{OT} & \colhead{(J2000)} & \colhead{(J2000)} & \colhead{lat.\tablenotemark{a}} & \colhead{$z$} & \multicolumn{2}{c}{(keV, rest)} & \multicolumn{2}{c}{($10^{52}$\,erg, rest)\tablenotemark{b,c}} & \multicolumn{2}{c}{$T_{90}$ (s)} & \colhead{$m_R(t_\mathrm{P48})\tablenotemark{d}$}
}
\startdata
    GRB\,130702A & iPTF13bxl & 14$^\mathrm{h}$29$^\mathrm{m}$15$^\mathrm{s}$ & +15$\arcdeg$46$\arcmin$26$\arcsec$ & 65$\arcdeg$ & 0.145 & 18 & 3 & $<$0.065 & 0.001 & 58.9 & 6.2 & 17.38 \\
    GRB\,131011A & iPTF13dsw & 02$^\mathrm{h}$10$^\mathrm{m}$06$^\mathrm{s}$ & -4$\arcdeg$24$\arcmin$40$\arcsec$ & -61$\arcdeg$ & 1.874 & 625 & 92 & 14.606 & 1.256 & 77.1 & 3 & 19.83 \\
    GRB\,131231A & iPTF13ekl & 00$^\mathrm{h}$42$^\mathrm{m}$22$^\mathrm{s}$ & -1$\arcdeg$39$\arcmin$11$\arcsec$ & -64$\arcdeg$ & 0.6419 & 291 & 6 & 23.015 & 0.278 & 31.2 & 0.6 & 15.85 \\
    GRB\,140508A & iPTF14aue & 17$^\mathrm{h}$01$^\mathrm{m}$52$^\mathrm{s}$ & +46$\arcdeg$46$\arcmin$50$\arcsec$ & 38$\arcdeg$ & 1.03 & 534 & 28 & 24.529 & 0.86 & 44.3 & 0.2 & 17.89 \\
    GRB\,140606B & iPTF14bfu & 21$^\mathrm{h}$52$^\mathrm{m}$30$^\mathrm{s}$ & +32$\arcdeg$00$\arcmin$51$\arcsec$ & -17$\arcdeg$ & 0.384 & 801 & 182 & 0.468 & 0.04 & 22.8 & 2.1 & 19.89 \\
    GRB\,140620A & iPTF14cva & 18$^\mathrm{h}$47$^\mathrm{m}$29$^\mathrm{s}$ & +49$\arcdeg$43$\arcmin$52$\arcsec$ & 21$\arcdeg$ & 2.04 & 387 & 34 & 7.28\phn & 0.372 & 45.8 & 12.1 & 17.60 \\
    GRB\,140623A & iPTF14cyb & 15$^\mathrm{h}$01$^\mathrm{m}$53$^\mathrm{s}$ & +81$\arcdeg$11$\arcmin$29$\arcsec$ & 34$\arcdeg$ & 1.92 & 834 & 317 & 3.58\phn & 0.398 & 114.7 & 9.2 & 18.04 \\
    GRB\,140808A & iPTF14eag & 14$^\mathrm{h}$44$^\mathrm{m}$53$^\mathrm{s}$ & +49$\arcdeg$12$\arcmin$51$\arcsec$ & 59$\arcdeg$ & 3.29 & 503 & 35 & 8.714 & 0.596 & 4.5 & 0.4 & 19.01
\enddata
\tablenotetext{a}{Galactic latitude of optical afterglow. This is one of the main factors that influences the number of optical transient candidates in Table~\ref{table:vetting}.}
\tablenotetext{b}{$E_{\gamma,\mathrm{iso}}$ is given for a 1\,keV\nobreakdashes--10\,MeV rest frame bandpass.}
\tablenotetext{c}{The rest\nobreakdashes-frame spectral properties, $E_\mathrm{peak}$ and $E_{\gamma,\mathrm{iso}}$, for GRB~130702A are reproduced from \citet{GCN15025}. For all other bursts, we calculated these quantities from the spectral fits (the \texttt{scat} files) in the \emph{Fermi} \ac{GBM} catalog \citep{GBMSpectralCatalog} using the $k$\nobreakdashes-correction procedure described by \citet{KCorrectionGRBs}.}
\tablenotetext{d}{$R$\nobreakdashes-band apparent magnitude in initial \ac{P48} detection.}
\end{deluxetable*}

}

{
\ifinthesis
\tabletypesize{\scriptsize}
\fi

\begin{deluxetable}{rr@{$\pm$}lrrrr}
\tablewidth{0pt}
\tablecaption{\label{table:nondetections}Log of \acs{P48} tilings for \emph{Fermi} \acs{GBM} bursts.}
\tablehead{
    \colhead{} & \multicolumn{2}{c}{\acs{GBM}} & \colhead{$t_\mathrm{P48}$} & \colhead{P48} & \colhead{} & \colhead{} \\
    \colhead{GRB time\tablenotemark{a}} & \multicolumn{2}{c}{fluence\tablenotemark{b}} & \colhead{$-t_\mathrm{burst}$\tablenotemark{c}} & \colhead{area\tablenotemark{d}} & \colhead{Prob.\tablenotemark{e}} & \colhead{}
}
\startdata
    
    2013-06-28 20:37:57 & 10\phd\phn & 0.1 & 10.02 & 73 & 32\% \\
    $\rightarrow$\bfseries
    2013-07-02 00:05:20 & 57\phd\phn & 1.2 & 4.20 & 74 & 38\% \\
    
    2013-08-28 07:19:56 & 372\phd\phn & 0.6 & 20.28 & 74 & 64\% \\
    
    2013-09-24 06:06:45 & 37\phd\phn & 0.6 & 23.24 & 74 & 28\% \\
    
    2013-10-06 20:09:48 & 18\phd\phn & 0.6 & 15.26 & 74 & 18\% \\
    $\rightarrow$\bfseries
    2013-10-11 17:47:30 & 89\phd\phn & 0.6 & 11.56 & 73 & 54\% \\
    
    2013-11-08 00:34:39 & 28\phd\phn & 0.5 & 4.69 & 73 & 37\% \\
    
    2013-11-10 08:56:58 & 33\phd\phn & 0.3 & 17.47 & 73 & 44\% \\
    
    2013-11-25 16:32:47 & 5.5 & 0.3 & 11.72 & 95 & 26\% \\
    
    2013-11-26 03:54:06 & 17\phd\phn & 0.3 & 6.94 & 109 & 59\% \\
    
    2013-11-27 14:12:14 & 385\phd\phn & 1.4 & 13.46 & 60 & 50\% \\
    
    2013-12-30 19:24:06 & 41\phd\phn & 0.4 & 7.22 & 80 & 38\% \\
    $\rightarrow$\bfseries
    2013-12-31 04:45:12 & 1519\phd\phn & 1.2 & 1.37 & 30 & 32\% \\
    
    2014-01-04 17:32:00 & 333\phd\phn & 0.6 & 18.57 & 15 & 11\% \\
    
    2014-01-05 01:32:57 & 6.4 & 0.1 & 7.63 & 74 & 22\% \\
    
    2014-01-22 14:19:44 & 9.1 & 0.5 & 11.97 & 75 & 34\% \\
    
    2014-02-11 02:10:41 & 7.4 & 0.3 & 1.77 & 44 & 19\% \\
    
    2014-02-19 19:46:32 & 28\phd\phn & 0.5 & 7.01 & 71 & 14\% \\
    
    2014-02-24 18:55:20 & 24\phd\phn & 0.6 & 7.90 & 72 & 30\% \\
    
    2014-03-11 14:49:13 & 40\phd\phn & 1.2 & 12.18 & 73 & 54\% \\
    
    2014-03-19 23:08:30 & 71\phd\phn & 0.3 & 3.88 & 74 & 48\% \\
    
    2014-04-04 04:06:48 & 82\phd\phn & 0.2 & 0.11 & 109 & 69\% \\
    
    2014-04-29 23:24:42 & 6.2 & 0.2 & 10.99 & 74 & 15\% \\
    $\rightarrow$\bfseries
    2014-05-08 03:03:55 & 614\phd\phn & 1.2 & 6.68 & 73 & 67\% \\
    
    2014-05-17 19:31:18 & 45\phd\phn & 0.4 & 8.60 & 95 & 69\% \\
    
    2014-05-19 01:01:45 & 39\phd\phn & 0.5 & 4.42 & 73 & 41\% \\
    $\rightarrow$\bfseries
    2014-06-06 03:11:52 & 76\phd\phn & 0.4 & 4.08 & 74 & 56\% \\
    
    2014-06-08 17:07:11 & 19\phd\phn & 0.6 & 11.20 & 73 & 49\% \\
    $\rightarrow$\bfseries
    2014-06-20 05:15:28 & 61\phd\phn & 0.6 & 0.17 & 147 & 59\% \\
    $\rightarrow$\bfseries
    2014-06-23 05:22:07 & 61\phd\phn & 0.6 & 0.18 & 74 & 4\% \\
    
    2014-06-28 16:53:19 & 18\phd\phn & 1.0 & 16.16 & 76 & 20\% \\
    
    2014-07-16 07:20:13 & 2.4 & 0.3 & 0.17 & 74 & 28\% \\
    
    2014-07-29 00:36:54 & 81\phd\phn & 0.7 & 3.43 & 73 & 65\% \\
    
    2014-08-07 11:59:33 & 13\phd\phn & 0.1 & 15.88 & 73 & 54\% \\
    $\rightarrow$\bfseries
    2014-08-08 00:54:01 & 32\phd\phn & 0.3 & 3.25 & 95 & 69\%
\enddata
\tablenotetext{a}{Time of \emph{Fermi} \ac{GBM} trigger. $\rightarrow$\textbf{Afterglow detections} are marked with an arrow and set in bold face. The corresponding entries in Table~\ref{table:detections} can be found by matching the date to the \ac{GRB} name (\ac{GRB}~YYMMDDA).}
\tablenotetext{b}{Observed \emph{Fermi} \acs{GBM} fluence in the 10\nobreakdashes--1000~keV band, in units of~$10^{-7}$~erg\,cm$^{-2}$. This quantity is taken from the \texttt{bcat} files from the \emph{Fermi} \ac{GRB} catalog at \ac{HEASARC}.}
\tablenotetext{c}{Age in hours of the burst at the beginning of the \ac{P48} observations.}
\tablenotetext{d}{Area in deg$^2$ spanned by the \ac{P48} fields.}
\tablenotetext{e}{Probability, given the \emph{Fermi} \ac{GBM} localization, that the source is contained within the \ac{P48} fields.}
\end{deluxetable}

}

\begin{figure}
    \centering
    \includegraphics{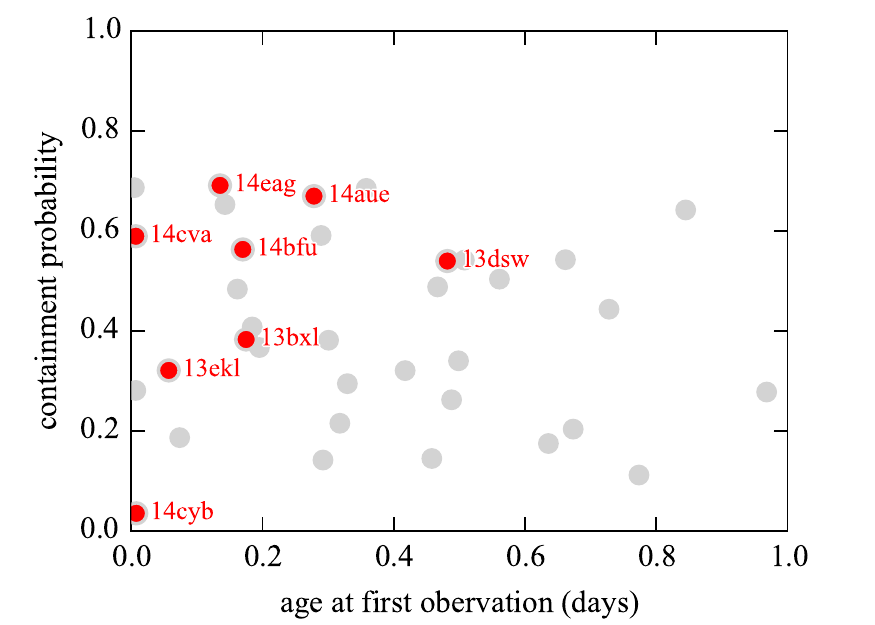}
    \caption[Containment probability and age]{\label{fig:time-prob}
    Prior probability of containing the burst's location within the \ac{P48} fields versus age of the burst at the beginning of \ac{P48} observations. Afterglow detections are shown in red and non\nobreakdashes-detections are shown in gray.}
\end{figure}

The outcome of an individual afterglow search is largely determined by two factors: how much probability is contained within the \ac{P48} footprints, and how bright the afterglow is at the time of the observations (see Figure~\ref{fig:time-prob}). We calculate the expected success rate as follows. For each burst, we find the prior probability that the position is contained within the \ac{P48} fields that we observed. We then compute the fraction of afterglows from Kann's sample (which has a mean and standard deviation of $22 \pm 2$~mag at $t = 1$~day) that are brighter than $R = 20.6$~mag at the same age as when the \ac{P48} observations started. The product of these two numbers is the prior probability of detection for that burst. By summing over all of the \ac{IPTF}/\ac{GBM} bursts, we obtain the expected number of detections. Within 95\% confidence bootstrap error bars, we find an expected 5.5\nobreakdashes--8.5 detections or a success rate of 16\%\nobreakdashes--24\%. This is consistent with the actual success rate of 23\%.

This suggests that the success rate is currently limited by the survey area and the response time (dictated by sky position and weather). We could increase the success rate by decreasing the maximum time since trigger at which we begin follow\nobreakdashes-up. We could increase the success rate without adversely affecting the number of detections by simply searching a greater area for coarsely localized events.

Over the next few sections, we summarize the observations and general physical interpretation of all of the \ac{GBM}\nobreakdashes--\ac{IPTF} afterglows detected to date. Figures~\ref{fig:grb-130702A-sed}\nobreakdashes--\ref{fig:grb-140808A-sed} show light curves and \acp{SED} spanning X\nobreakdashes-ray, UV, optical, IR, and radio frequencies. In 
    \ifinthesis%
    Appendix~\ref{chap:iptf-gbm-observations}%
    \fi%
, Table~\ref{tab:photometry} lists a selection of UV/O/IR observations including all of our \ac{P48} and \ac{P60} observations. Table~\ref{tab:radio} lists all of our radio detections.

The reported \ac{P48} magnitudes are all in the Mould $R$ band and in the AB system \citep{ABMags}, calibrated with respect to either $r^{\prime}$ point sources from \ac{SDSS} or for non\nobreakdashes-\ac{SDSS} fields using the methods described in \citet{PTFPhotometricCalibration}.

\ifinthesis
\begin{deluxetable*}{llllp{2.5cm}l}
\tabletypesize{\footnotesize}
\else
\begin{deluxetable*}{llllll}
\fi
\tablewidth{0pt}
\tablecaption{\label{table:spectra}Log of spectroscopic observations}
\tablehead{
    \colhead{Date} &
    \colhead{Telescope} &
    \colhead{Instrument} &
    \colhead{Wavelengths (\AA)} &
    \colhead{Lines} &
    \colhead{References}
}
\startdata
\cutinhead{GRB~131011A / iPTF13dsw}
2013-10-12~08:56 &
Gemini South &
GMOS &
5100--9300 &
none &
\citet{GCN15324} \\
2013-10-13~03:59 &
ESO/VLT UT3 &
X-shooter &
3100--5560 &
Ly$\alpha$, \ion{Si}{2}, \ion{C}{2}, \ion{C}{4}, \ion{Al}{2} &
\citet{GCN15330} \\
\nodata &
\nodata &
\nodata &
5550--10050 &
\ion{Fe}{2}, \ion{Mg}{2} &
\nodata \\
\cutinhead{GRB~131231A / iPTF13ekl}
2014-01-01~02:15 &
Gemini South &
GMOS &
6000--10000 &
[\ion{O}{2}], [\ion{O}{3}], \ion{Ca}{2} H+K &
\citet{GCN15652} \\
\cutinhead{GRB~140508A / iPTF14aue}
2014-05-08~18:55 &
HCT &
HFOSC &
3800--8400 &
\ion{Fe}{2}, \ion{Mg}{2} &
\citet{GCN16244} \\
2014-05-09 06:33 &
APO &
DIS &
3200--9800 &
none &
none \\
\cutinhead{GRB~140606B / iPTF14bfu}
2014-06-07~19:16 &
Keck~II &
DEIMOS &
4500--9600 &
[\ion{O}{2}], [\ion{O}{3}], H$\alpha$, \ion{Ca}{2} H+K &
\citet{GCN16365} \\
\cutinhead{GRB~140620A / iPTF14cva}
2014-06-20~14:00 &
Gemini North &
GMOS &
5090--9300 &
\ion{Mg}{1}, \ion{Mg}{2}, \ion{Fe}{2}, \ion{Al}{2}, \ion{Si}{2}, \ion{Si}{2}$^*$ &
\citet{GCN16425} \\
\nodata &
\nodata &
\nodata &
4000--6600 &
\nodata &
\nodata \\
\cutinhead{GRB~140623A / iPTF14cyb}
2014-06-23~08:10 &
Gemini North &
GMOS &
4000--6600 &
\ion{Mg}{2}, \ion{Fe}{2}, \ion{Al}{2}, \ion{Si}{2}, \ion{Al}{3}, \ion{C}{1}, \ion{C}{4} &
\citet{GCN16442}
%
%
\enddata
\end{deluxetable*}

\begin{figure*}
    \centering
    \includegraphics[width=\textwidth]{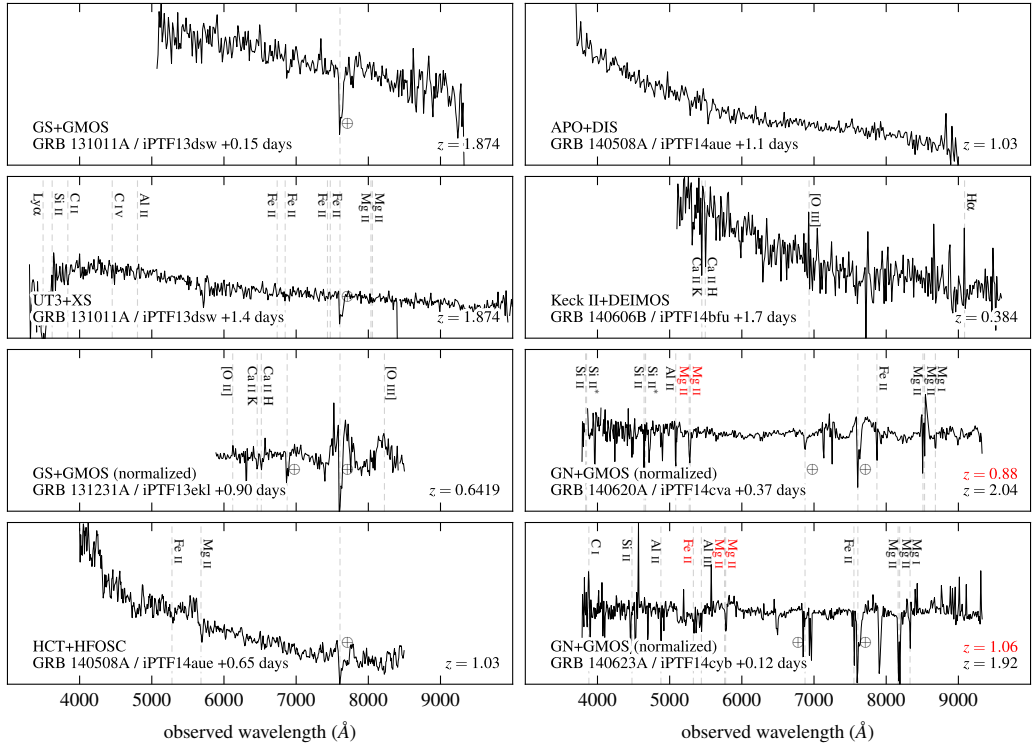}
    \caption[Afterglow spectra]{\label{fig:spectra}Afterglow spectra. The horizontal axis shows wavelength in vacuum in the observer frame, and the vertical axis shows scaled flux. Lines at the redshift of the putative host are labeled in black; lines corresponding to any intervening absorbing systems are labeled in red. Note that in cases where one or fewer lines are discernible in our spectra, the redshifts have been reported in \acp{GCN} by other groups.}
\end{figure*}

\subsection{\ac{GRB}~130702A / iPTF13bxl}

This is the first \ac{GBM} burst whose afterglow we discovered with iPTF \citep{iPTF13bxl}, indeed the first afterglow ever to be pinpointed based solely on a \emph{Fermi} \ac{GBM} localization. It is also the lowest redshift \ac{GRB} in our sample, so it has the richest and most densely sampled broadband afterglow data. It has two other major distinctions: its associated \ac{SN} (\ac{SN}~2013dx, \citealt{GCN14994,GCN14996,GCN14998,GCN15000}) was detected spectroscopically, and its prompt energetics are intermediate between \acp{llGRB} and standard cosmic bursts (see below).

Based on the \emph{Fermi} \ac{GBM} ground localization with an error radius of $4\arcdeg$, we imaged ten fields twice with the \ac{P48} at $\Delta t = t - t_\mathrm{GBM} = 4.2$~hr after the burst.\footnote{Our tiling algorithm at the time selected fields based on an empirical calibration of \emph{Fermi} \ac{GBM}'s systematic errors. We had selected bursts that were detected by both \emph{Swift} and \emph{Fermi}, and constructed a fit to a cumulative histogram of the number of bursts whose \ac{BAT} or \ac{XRT} positions were within a given number of nominal $1\sigma$ statistical radii of the center of the \emph{Fermi} error circle. Our tiling algorithm scaled this fit by the $1\sigma$ radius of the burst in question, and then constructed a 2D angular probability distribution from it. For sufficiently large error radii, this prescription produced probability distributions that had a hole in the middle. For this reason, the tiling algorithm picked out \ac{P48} fields that formed an annulus around the \ac{GBM} 1\nobreakdashes-$\sigma$ error circle (not, as we stated in \citealt{iPTF13bxl}, because of a lack of reference images).} We scheduled P60 imaging and P200 spectroscopy for three significantly varying sources. Of the three, iPTF13bxl showed the clearest evidence of fading in the \ac{P48} images. Its spectrum at $\Delta t = 1.2$~days consisted of a featureless blue continuum. We triggered \emph{Swift}, which found a bright X-ray source at the position of iPTF13bxl \citep{GCN14967,GCN14973}. Shortly after we issued our \ac{GCN} circular \citep{GCN14967}, \citet{GCN14971} announced that the burst had entered the FOV of LAT at $\Delta t = 250$~s. The LAT error circle had a radius of $0.5\arcdeg$, and its center was $0.8\arcdeg$ from iPTF13bxl. An \ac{IPN} triangulation with \emph{MESSENGER} (GRNS), \emph{INTEGRAL} (SPI\nobreakdashes--ACS), \emph{Fermi}\nobreakdashes--\ac{GBM}, and Konus\nobreakdashes--\emph{WIND} \citep{GCN14974} yielded a $0.46\arcdeg$\nobreakdashes-wide annulus that was also consistent with the OT.

The afterglow's position is $0\farcs 6$ from an $R = 23.01$~mag source that is just barely discernible in the \ac{P48} reference images. A NOT+ALFOSC spectrum \citep{GCN14983} determined a redshift of $z = 0.145$ for a galaxy $7\farcs 6$ to the south of iPTF13bxl. At $\Delta t = 2.0$~days, we obtained a Magellan+IMACS spectrum \citep{GCN14985} and found weak emission lines at the location of the afterglow that we interpreted as H$\alpha$ and [\ion{O}{3}] at the same redshift. \citet{13bxlhost} characterized the burst's host environment in detail, and concluded that it exploded in a dwarf satellite galaxy.

Joining the two \ac{P48} observations at $\Delta t < 1$~day to the late\nobreakdashes-time P60 light curve requires a break at $\Delta t = 1.17 \pm 0.09$~days, with slopes $\alpha_{\mathrm{O},1}=0.57 \pm 0.03$ and $\alpha_{\mathrm{O},2}=1.05 \pm 0.03$ before and after the break respectively. The XRT light curve begins just prior to this apparent break and seems to follow the late\nobreakdashes-time optical decay (until the \ac{SN} begins to dominate at $\Delta t = 5$~days), although the automated \emph{Swift} light curve analysis \citep{2009MNRAS.397.1177E} also suggests a possible X\nobreakdashes-ray break with about the same time and slopes. This hints at an achromatic break, normally a signature of a jet. However, the late slope and the change in slope are both unusually shallow for a jet break. The change in slope is also a little too large for cooling frequency crossing the band (for which one would expect $\Delta \alpha = 1/4$). An energy injection or a structured jet model may provide a better fit \citep{ModelsForAchromaticLightCurveBreaks}.

Late\nobreakdashes-time $\Delta t > 1$~day observations include several P60 $gri$ observations, three RATIR $r'i'ZYJH$ epochs, an extensive \emph{Swift} XRT and UVOT light curve, as well as radio observations with VLA and \ac{CARMA} (although of the VLA data, we only have access to the first observation). The optical and X-ray spectral slopes are similar, $\beta_\mathrm{O} = 0.7 \pm 0.1$ and $\beta_\mathrm{X} = 0.8 \pm 0.1$. An SED at $2 < \Delta t < 2.3$~days is well explained by the standard external shock model \citep{AfterglowSpectra} in the slow cooling regime, with $\nu_m$ lying between the VLA and \ac{CARMA} frequencies and $\nu_c$ in the optical. This fit requires a relatively flat electron spectrum, $d n_\mathrm{e} / d \gamma_\mathrm{e} \propto {\gamma_\mathrm{e}}^{-p}$ with $p \approx 1.6$, cut off at high energies. Applying the relevant closure relations (for the case of $1 < p < 2$, see \citealt{AfterglowSpectraFlatSpectrum}) to $\alpha_X$ and $\beta_X$ permits either an ISM or wind environment. However, the fact that the X-ray and optical continue to track each other in time suggests that the $\nu_c$ is decreasing with time, arguing for an ISM density profile.

This model has difficulty explaining the constant or slowly varying 93~GHz \ac{CARMA} light curve.

Our late\nobreakdashes-time spectroscopy and analysis of the \ac{SN} will be published separately (Cenko~et~al., in~preparation).

\begin{figure}
    \centering
    \includegraphics{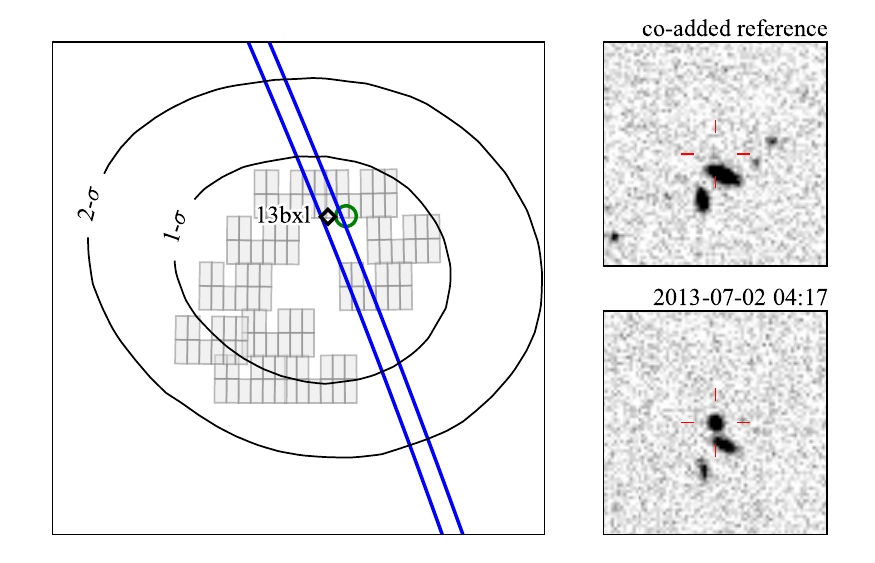}
    \caption[Discovery of \acs{GRB}~130702A~/~iPTF13bxl]{\label{fig:GRB130702A}\emph{Fermi} \ac{GBM} localization (black contours), \ac{P48} tiling (gray rectangles), 3$\sigma$ \ac{IPN} triangulation (blue), \ac{LAT} 1$\sigma$ error circle (green), and discovery images for \ac{GRB}~130702A~/~iPTF13bxl.}
\end{figure}

\subsection{\ac{GRB}~131011A / iPTF13dsw}

We started \ac{P48} observations of \emph{Fermi} trigger 403206457 \citep{GCN15331} about 11.6 hours after the burst. The optical transient iPTF13dsw \citep{GCN15324} faded from $R=19.7$~mag to $R=20.2$~mag from 11.6~to~14.3 hours. The latest pre\nobreakdashes-trigger image on 25~September~2013 had no source at this location to a limit of $R > 20.6$~mag. The optical transient continued to fade as it was monitored by several facilities \citep{GCN15325,GCN15326,GCN15327,GCN15328,GCN15341}.

At 15.1~hours after the burst, we obtained a spectrum of iPTF13dsw with the \acf{GMOS} on the Gemini\nobreakdashes--South telescope. \ac{GMOS} was configured with the R400 grating with a central wavelength of 7200~$\AA$ and the 1\arcsec slit, providing coverage over the wavelength range from 5100\nobreakdashes--9300~$\AA$ with a resolution of $\approx 3$~$\AA$.  No prominent features were detected over this bandpass, while the spectrum had a typical SNR of $\approx 3$\nobreakdashes--4 per 1.4~$\AA$ pixel.

\citet{GCN15330} observed the optical transient with the X\nobreakdashes-Shooter instrument on the ESO 8.2\nobreakdashes-m Very Large Telescope (VLT). In their spectrum extending from $\sim$3000~to~$\sim24000$\AA, they identified several weak absorption lines from which they derived a redshift of $z = 1.874$.

The source was detected by \emph{Swift} XRT \citep{GCN15329}, but with insufficient photons for spectral analysis. There are no radio observations. Largely because in our sample this is the oldest afterglow at the time discovery, there are not enough broadband data to constrain the blast wave physics.

\begin{figure}
    \centering
    \includegraphics{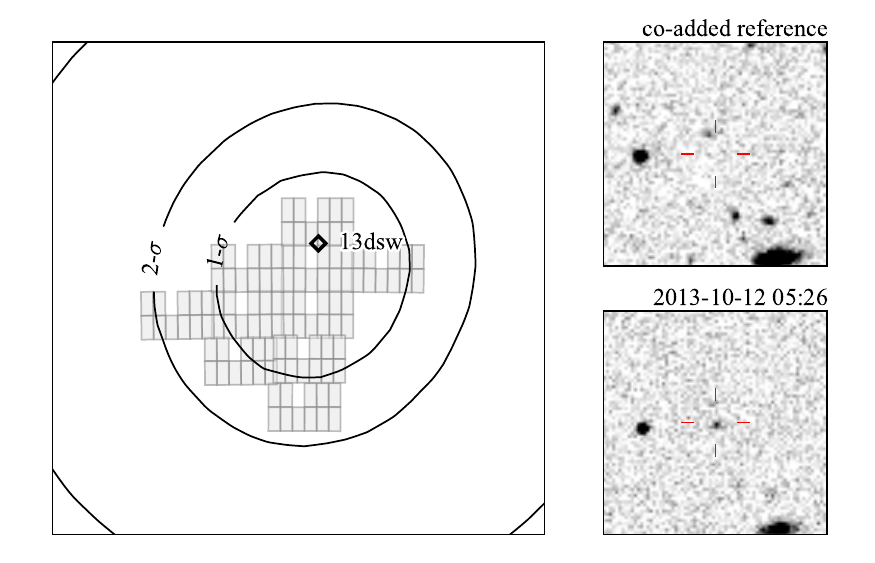}
    \caption[Discovery of \acs{GRB}~131011A~/~iPTF13dsw]{\label{fig:GRB131011A}\emph{Fermi} \ac{GBM} localization (black contours), \ac{P48} tiling (gray rectangles), and discovery images for \ac{GRB}~131011A~/~iPTF13dsw.}
\end{figure}

\subsection{\ac{GRB}~131231A / iPTF13ekl}

\ac{GRB}~131231A was detected by \emph{Fermi} \ac{LAT} \citep{GCN15640} and \ac{GBM} \citep{GCN15644}, with photons detected up to 9.7~GeV. \citet{GCN15641} observed the \ac{LAT} error circle with the 1\nobreakdashes-m telescope at Mt. Nanshan, Xinjiang, China. At 7.9~hours after the burst, they detected a single $R = \sim17.6$~mag source that was not present in SDSS images. At 17.3~hours after the burst, \citet{GCN15642} observed the afterglow candidate with the MOSaic CAmera (MOSCA) on the 2.56\nobreakdashes-m Nordic Optical Telescope (NOT). The source had faded to $R = 18.6$.

Although we had imaged ten \ac{P48} fields shortly after the \emph{Fermi} trigger \citep{GCN15643}, due to the short visibility window at Palomar we were only able to obtain one epoch. At 1.45~hours after the burst, we detected an $R = 15.7$~mag optical transient iPTF13ekl at the position of the Nanshan candidate. Though our single detection of iPTF13ekl could not by itself rule out that the source was a moving solar system object, the Nanshan detection at 6.46~hours, fitting a decay with a power law index of $\alpha = 1.03$, was strong evidence that the transient was the optical afterglow of \ac{GRB}~131231A.

On January 1.09 UT (21.5~hours after the trigger), we observed the afterglow with Gemini\nobreakdashes--South using the \ac{GMOS} camera (Hook et al. 2004) in Nod\&Shuffle mode: we obtained 32 dithered observations of 30 seconds each at an average airmass of 2. We analyzed this dataset using the dedicated GEMINI package under the IRAF environment and extracted the 1-dimensional spectrum using the APALL task. We determined the redshift of the \ac{GRB}, based on the simultaneous identification of forbidden nebular emission lines ([\ion{O}{2}], [\ion{O}{3}]) and absorption features (CaH\&K) at the same redshift of $z=0.6419$. In Figure~\ref{fig:spectra}, we show the normalized spectrum. In black we show a smoothed continuum. Emission lines are indicated as well as CaH\&K absorption features. Also atmospheric bands are marked.

The source was also detected by \emph{Swift} XRT \citep{GCN15648} and UVOT \citep{GCN15673}, as well as \ac{CARMA} \citep{GCN15680}.

With only the \ac{CARMA} observations, the SED is highly degenerate. Contributing to the degeneracy, the X\nobreakdashes-ray and optical observations appear to fall on the same power\nobreakdashes-law segment. It is consistent with either fast or slow cooling if the greater of $\nu_c$ or $\nu_m$ is near the optical, assuming a flat electron distribution with $p \sim 1.5$. It is also consistent with slow cooling if $\nu_c$ is above the X\nobreakdashes-ray band and $p \sim 2.6$.

\begin{figure}
    \centering
    \includegraphics{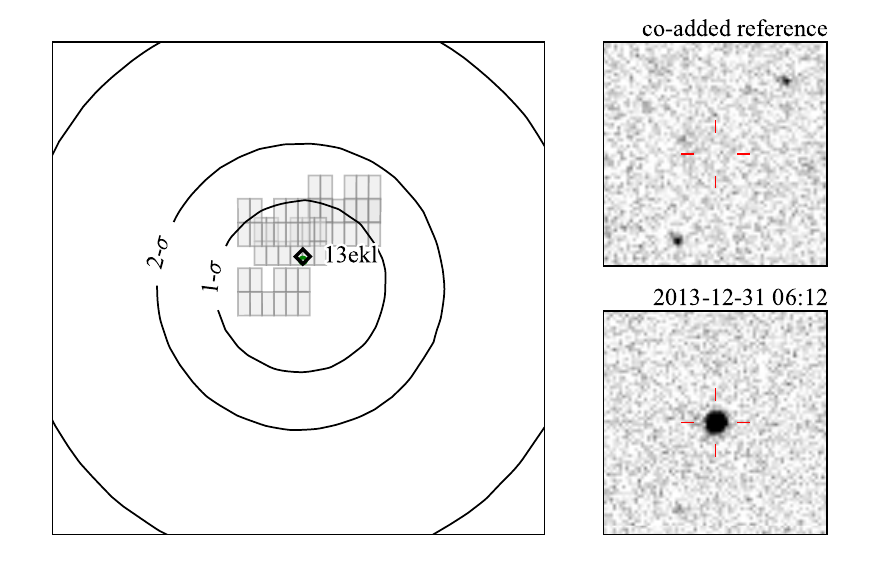}
    \caption[Discovery of \acs{GRB}~131231A~/~iPTF13ekl]{\label{fig:GRB131231A}\emph{Fermi} \ac{GBM} localization (black contours), \ac{P48} tiling (gray rectangles), \ac{LAT} 1$\sigma$ error circle (green), and discovery images for \ac{GRB}~131231A~/~iPTF13ekl.}
\end{figure}

\subsection{\ac{GRB}~140508A / iPTF14aue}

This burst was detected by \emph{Fermi} \ac{GBM} and \emph{INTEGRAL} SPI\nobreakdashes-ACS \citep{GCN16224}, as well as by Konus\nobreakdashes-\emph{WIND}, Mars Odyssey (not included in the \ac{GCN} circular), \emph{Swift} \ac{BAT} (outside the coded field of view), and \emph{MESSENGER}, yielding a $1 \fdg 5 \times 12\arcmin$ \ac{IPN} error box \citep{GCN16225}.

Due to poor weather early in the night, \ac{P48} observations started 6.7 hours after the trigger \citep{GCN16226}. We found one optical transient candidate within the \ac{IPN} triangulation, iPTF14aue, which faded from $r = 17.89 \pm 0.01$~mag with a power law fit of $\alpha = 1.12 \pm 0.1$.

We triggered a \emph{Swift} \ac{TOO}. From 0.8~to~8.1~days after the trigger, \emph{Swift} XRT detected a coincident X\nobreakdashes-ray source that faded with a power law $\alpha = 1.48\;(+0.15, -0.14)$ \citep{GCN16232,GCN16254}. The source was also detected by \emph{Swift} UVOT \citep{GCN16243}.

\citet{GCN16228} obtained a 20~min, 3800\nobreakdashes--7200~\AA{} spectrum of iPTF14aue with the 6\nobreakdashes-m BTA telescope in Zelenchukskaia. Exhibiting no absorption features, this established an upper limit of $z < 2.1$. \citet{GCN16229} used the Andalucia Faint Object Spectrograph and Camera (ALFOSC) on NOT to get an 1800~s spectrum spanning 3200\nobreakdashes--9100~\AA, and found several absorption features at redshift $z = 1.03$. Consistent redshifts were reported by \citet{GCN16231} with the ACAM instrument on the 4.2\nobreakdashes-m William Herschel Telescope and by \citet{GCN16244} with \ac{HFOSC} on the 2\nobreakdashes-m \ac{HCT}.

Due to the brightness of the optical transient, optical photometry was available from several facilities up to 4.5 days after the burst \citep{GCN16227,GCN16228,GCN16229,GCN16235,GCN16236,GCN16246,GCN16259,GCN16260}.

\citet{GCN16266} detected the source with \ac{VLA} 5.2 days after the Fermi trigger, at 6.1 GHz (C-band) and at 22 GHz (K-band). A broadband \ac{SED} constructed from P60 and \ac{XRT} data from around this time is consistent with $p \approx 2$. Because $p$ is not distinguishable from 2, we cannot discriminate between fast and slow cooling based on this one time slice. However, given the late time of this observation, the slow cooling interpretation is more likely, putting $\nu_m$ between the radio and optical bands and $\nu_c$ between the optical and X\nobreakdashes-ray. Because the \ac{VLA} light curve is decreasing with time, an \ac{ISM} circumburst density profile is favored.

\begin{figure}
    \centering
    \includegraphics{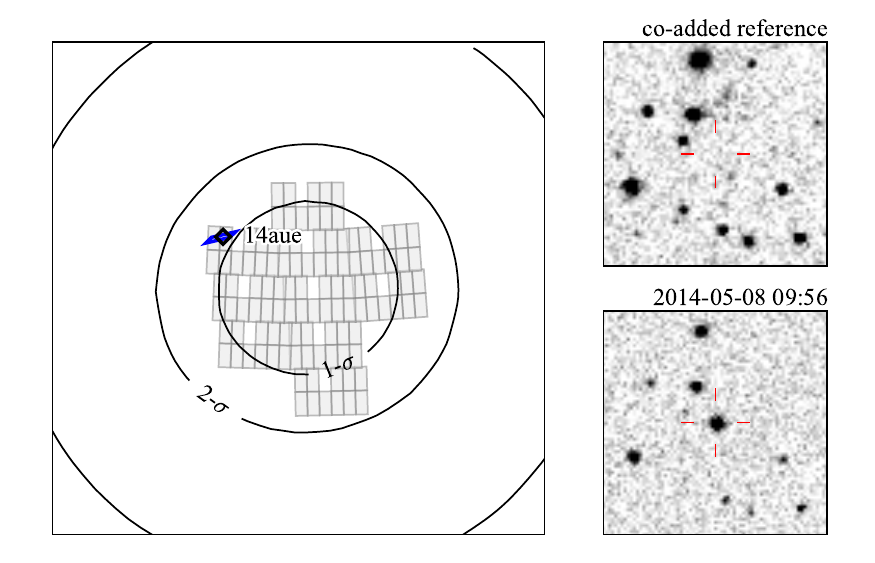}
    \caption[Discovery of \acs{GRB}~140508A~/~iPTF14aue]{\label{fig:GRB140508A}\emph{Fermi} \ac{GBM} localization (black contours), \ac{P48} tiling (gray rectangles), 3$\sigma$ \ac{IPN} triangulation (blue), and discovery images for \ac{GRB}~140508A~/~iPTF14aue.}
\end{figure}

\subsection{\ac{GRB}~140606B / iPTF14bfu}

\emph{Fermi} trigger 423717114 \citep{GCN16363} was observable from Palomar for several hours, starting about 4.3 hours after the time of the burst. Based on the final \ac{GBM} localization, we searched ten \ac{P48} fields and found several plausible optical transient candidates \citep{GCN16360}.

iPTF14bfu had no previous detections in iPTF between 23~May and 13~October~2013. Its position was outside the SDSS survey footprint, but it had no plausible host associations in VizieR \citep{VizieR}. From 4.3 to 5.5 hours after the burst, it faded from $R = 19.89 \pm 0.10$ to $20.32 \pm 0.14$ mag, fitting a power law of $\alpha = -1.6 \pm 0.7$ relative to the time of the \ac{GBM} trigger. iPTF14bfw ($R = 19.96 \pm 0.06$ mag) was coincident with an $r = 21.27$ galaxy in SDSS DR10, and displayed no statistically significant photometric variation over the course of our \ac{P48} observations. iPTF14bgc ($R = 18.44 \pm 0.02$ mag) was coincident with an $R = 21.07 \pm 0.08$ mag point source in our coadded reference image composed of exposures from July~31 through 24~September~2013. iPTF14bga ($R = 19.75 \pm 0.06$ mag) was likewise coincident with a $R = 20.42 \pm 0.17$ mag point source in our reference image composed of exposures from 29~July through 20~October~2011.

On the following night, we observed all four candidates again with \ac{P48} and P60 \citep{GCN16362}. iPTF14bfw and iPTF14bga had not faded relative to the previous night. iPTF14bgc had faded to $R = 20.68 \pm 0.21$ mag, consistent with the counterpart in our reference images but significantly fainter than the previous night. A power law fit to the decay gave a temporal index of $\alpha = -1.1 \pm 0.1$, entirely consistent with typical \ac{GRB} afterglows. iPTF14bfu was not detected in our \ac{P48} images to a limiting magnitude of $R < 21.1$, but it was detected in stacked P60 images ($r = 21.1 \pm 0.2$), consistent with a power law of $\alpha \sim -0.5$.

An \ac{IPN} triangulation from \emph{Fermi}, Konus\nobreakdashes--\emph{WIND}, and MESSENGER yielded a long, slender $14\fdg18 \times 0\fdg414$ error box that contained iPTF14bfu and iPTF14bfw \citep{GCN16369}.

We obtained two 900~s spectra with the DEIMOS spectrograph on the Keck~II 10\nobreakdashes-m telescope \citep{GCN16365}. On a blue continuum, we found \ion{O}{2}, \ion{O}{3}, and H\,$\alpha$ emission features, and \ion{Ca}{2} absorption features, at a common redshift of $z = 0.384$. A galaxy offset by $\sim 2\arcsec$ along the slit showed the same emission lines at the same redshift.

\emph{Swift} XRT observed the location of iPTF14bfu for a total of 9~ks from 2.1~to~9.3 days after the \ac{GBM} trigger, and found a source that faded with a power-law fit of $\alpha = -1.0\;(+0.7,-0.6)$ \citep{GCN16366,GCN16373,GCN16412}.

At 18.4 days after the trigger, we obtained a 1200~s spectrum of iPTF14bfu with the Low Resolution Imaging Spectrometer (LRIS) on the Keck~I 10\nobreakdashes-meter telescope \citep{GCN16454}. The spectrum had developed broad emission features. A comparison using Superfit \citep{Superfit} showed a good match to \ac{SN}~1998bw near maximum light, indicating that the source had evolved into an \ac{SNIcBL}. Our late\nobreakdashes-time photometry and spectroscopy will published separately (Cano~et~al., in~preparation).

Although there were three radio detections of \ac{GRB}~140606B, only during the first \ac{CARMA} detection does the optical emission appear to be dominated by the afterglow. We can construct an \ac{SED} around this time using nearly coeval \ac{DCT} and \ac{XRT} data. Because of the faintness of the X\nobreakdashes-ray afterglow, the spectral slopes $\beta_\mathrm{X}$ and $\beta_\mathrm{OX}$ are only weakly determined. As a result, there is a degeneracy between two plausible fits. The first has $\nu_m$ anywhere below the \ac{CARMA} band, $\nu_c$ just below the X\nobreakdashes-rays, and $p \approx 2$. The second has $\nu_m$ just above the radio and $\nu_c$ in the middle of the \ac{XRT} band, with $p \approx 2.2$.

The early \ac{P48} observations do not connect smoothly with the P60 and \ac{DCT} observations from $\Delta t = 1$~to~4~days. This may indicate a steep\nobreakdashes--shallow\nobreakdashes--steep time evolution requiring late time energy injection, or may just indicate that the afterglow is contaminated by light from the host galaxy or the \ac{SN} at relatively early times.

\begin{figure}
    \centering
    \includegraphics{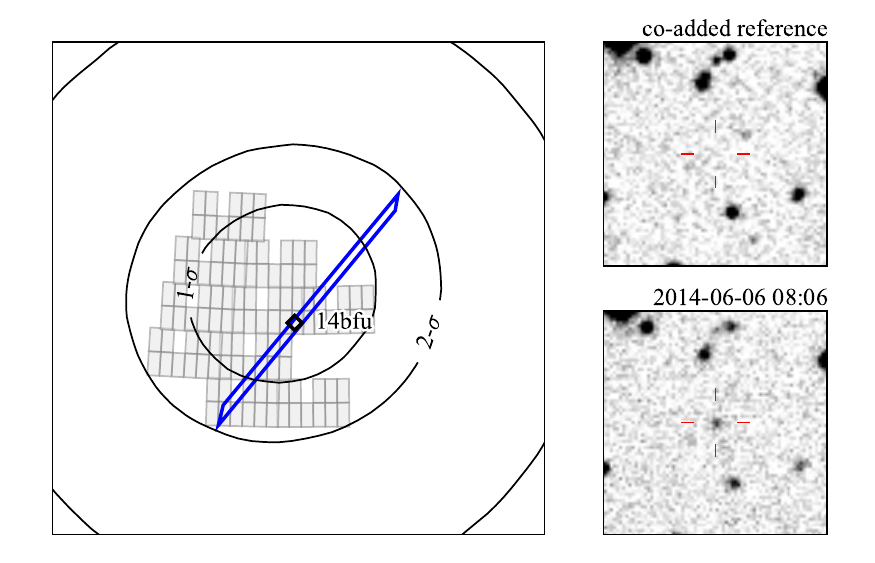}
    \caption[Discovery of \acs{GRB}~140606B~/~iPTF14bfu]{\label{fig:GRB140606B}\emph{Fermi} \ac{GBM} localization (black contours), \ac{P48} tiling (gray rectangles), 3$\sigma$ \ac{IPN} triangulation (blue), and discovery images for \ac{GRB}~140606B~/~iPTF14bfu.}
\end{figure}

\subsection{\ac{GRB}~140620A / iPTF14cva}

This burst is distinctive in our sample for two reasons. First, it is the earliest afterglow detection in the \ac{IPTF} sample at $\Delta t = 0.25$~hours. Second, its broadband \ac{SED} is not clearly explainable by the standard forward shock model.

\emph{Fermi} trigger 424934131 \citep{GCN16426} was observable from Palomar for about 6 hours from the time of the burst. Based on the ground localization, we started observing ten \ac{P48} fields about ten minutes after the trigger. Based on the final localization, we added ten more fields, for a total of twenty, about an hour after the trigger.

The candidate iPTF14cva \citep{GCN16425} was contained within one of the early ten fields. From 14.9 to 87.2 minutes after the trigger, the candidate faded from $R = 17.60 \pm 0.01$ to $18.80 \pm 0.02$~mag, consistent with a somewhat slow power law of $\alpha = 0.62 \pm 0.01$.

We observed the candidate with GMOS on the 8\nobreakdashes-m Gemini North telescope. Starting 8.8~hours after the trigger, we obtained two 900~s spectra extending from 4000 to 9300~\AA. We detected \ion{Mg}{2} and \ion{Fe}{2} absorption lines at $z = 0.88$ and many absorption features at a common redshift of $z = 2.04$. The lack of Ly-$\alpha$ absorption implied an upper limit of $z \sim 2.3$, and suggested that $z = 2.04$ was the redshift of the source.

We triggered \emph{Swift} and \ac{VLA} follow\nobreakdashes-up. In a 3~ks exposure starting 10.4~hours after the \emph{Fermi} trigger, \emph{Swift} XRT detected an X\nobreakdashes-ray source with a count rate of $1.2 \times 10^{-1}$~cts\,s$^{-1}$ \citep{GCN16428}. Over the next four days of \emph{Swift} observations, the X-ray source faded with a slope $\alpha = 1.32 \pm 0.16$ \citep{GCN16455}. A fading source was also detected by \emph{Swift} UVOT \citep{GCN16432}.

The source was detected by \ac{VLA} on June 23 at 6.1~GHz (C band) at $108 \pm 15$~$\mu$Jy and at 22~GHz (K band) at $62 \pm 15$~$\mu$Jy.  On June 30, there was a marginal detection in C band with $48 \pm 12$~$\mu$Jy and no detection in K-band with a noise level of 15~$\mu$Jy \ac{RMS}. 

The optical transient was also observed in $R$ band by the Konkoly Observatory \citep{GCN16440} and the 1\nobreakdashes-m telescope at the Tien Shan Astronomical Observatory \citep{GCN16453}.

The \ac{SED} of this afterglow cannot be explained by a standard forward shock model. If we place the peak frequency near the radio band, the optical and X\nobreakdashes-ray fluxes are drastically under\nobreakdashes-predicted, whereas if we place the peak frequency between the optical and X\nobreakdashes-ray bands, we miss the radio observations by orders of magnitude. This seems to require an additional component. One possibility is that there is a forward shock peak in the UV and a reverse shock peak at low frequencies (similar to \ac{GRB}~130427A; see \citealt{LaskarGRB130427A,PerleyGRB130427A}). Another possibility is that there is an inverse Compton peak in the UV (similar to \ac{GRB}~120326A; \citealt{UrataGRB120326A}).

\begin{figure}
    \centering
    \includegraphics{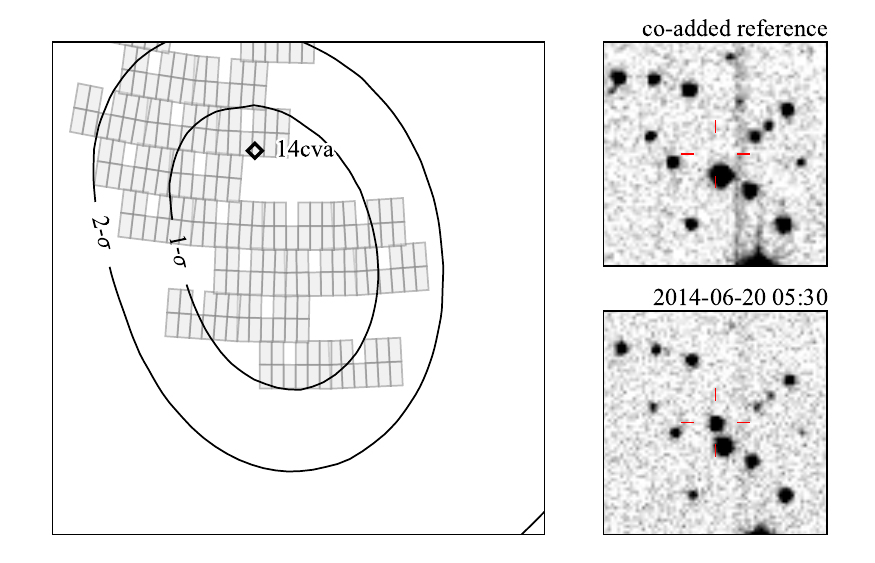}
    \caption[Discovery of \acs{GRB}~140620A~/~iPTF14cva]{\label{fig:GRB140620A}\emph{Fermi} \ac{GBM} localization (black contours), \ac{P48} tiling (gray rectangles), and discovery images for \ac{GRB}~140620A~/~iPTF14cva.}
\end{figure}

\subsection{\ac{GRB}~140623A / iPTF14cyb}

\emph{Fermi} trigger 425193729 \citep{GCN16450} was observable from Palomar for about 6 hours from the time of the burst. Based on the ground localization, we started imaging ten fields 11 minutes after the trigger. The final \emph{Fermi} localization, which was avilable 2.6 hours later, shifted by 13\fdg4. Due to the large change in the localization, we calculated only a 4\% chance that the source was contained within the \ac{P48} fields.

Candidate iPTF14cyb \citep{GCN16425}, situated at an extreme edge of the \ac{P48} tiling, was within the 1\nobreakdashes-$\sigma$ confidence region for both the ground and final localizations. From 16 to 83 minutes after the trigger, the source faded from $R = 18.04 \pm 0.01$ to $19.69 \pm 0.06$~mag, consistent with a power-law decay with an index $\alpha = 0.94 \pm 0.03$.

Starting 2.8~hours after the trigger, we obtained two 900~s GMOS spectra extending from 4000 to 9300~\AA. We detected \ion{Mg}{2} and \ion{Fe}{2} absorption lines at $z = 1.06$ and many absorption features at $z = 1.92$. The lack of Ly-$\alpha$ absorption implied that this was the redshift of the burst.

We triggered \emph{Swift}, \ac{VLA}, and \ac{CARMA} follow\nobreakdashes-up. In a 3~ks exposure starting 10.7~hours after the burst, \emph{Swift} XRT detected an uncatalogued X\nobreakdashes-ray source with a count rate of $(2.2 \pm 0.6) \times 10^{-3}$~cts\,s$^{-1}$ \citep{GCN16451}. By 79~hours after the trigger, the source was no longer detected in a 5~ks exposure \citep{GCN16464}. No radio source was detected with \ac{VLA} in C band (6.1~GHz) to an \ac{RMS} level of 17~$\mu$Jy, or in K band (22~GHz) to an \ac{RMS} level of 18~$\mu$Jy.

Because of the lack of radio detections and the extreme faintness of the X\nobreakdashes-ray afterglow, the broadband behavior of the afterglow does not constrain the shock physics.

\begin{figure}
    \centering
    \includegraphics{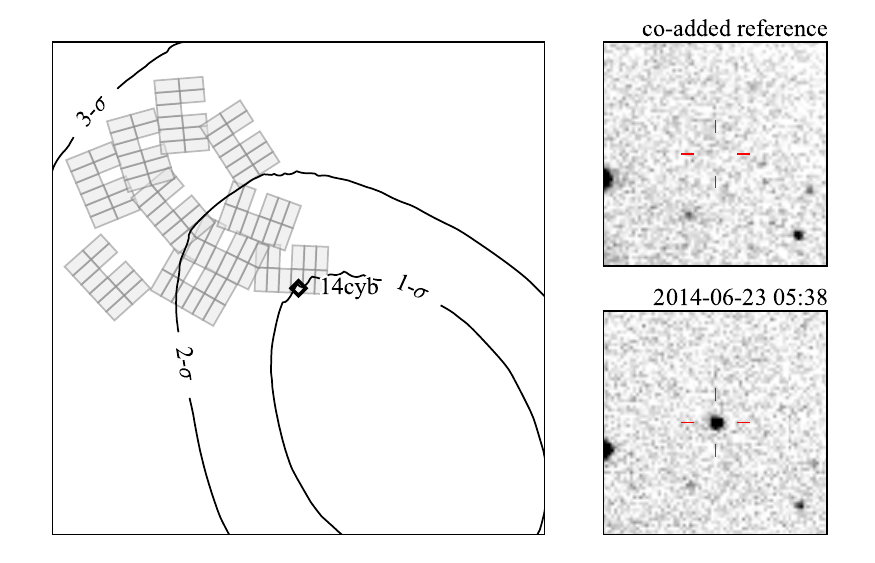}
    \caption[Discovery of \acs{GRB}~140623A~/~iPTF14cyb]{\label{fig:GRB140623A}\emph{Fermi} \ac{GBM} localization (black contours), \ac{P48} tiling (gray rectangles), and discovery images for \ac{GRB}~140623A~/~iPTF14cyb.}
\end{figure}

\subsection{\ac{GRB}~140808A / iPTF14eag}

\emph{Fermi} trigger 429152043 \citep{GCN16669} was observable from Palomar about 3 hours after the burst. We imaged 13 fields with \ac{P48} and found one promising optical transient. iPTF14eag was situated on the extreme edge of one of the \ac{P48} tiles that was just outside the \ac{GBM} 1\nobreakdashes-sigma contour. It faded from $R = 18.91 \pm 0.06$ to $19.29 \pm 0.10$~mag from 3.35~to~4.91 hours after the trigger and had no archival counterparts in SDSS or in our own reference images.

We were unable to use our \ac{TOO} programs on Keck or Gemini North because Hawaii was being struck by Hurricane Iselle. We requested photometric confirmation of the fading from \ac{HCT} \citep{GCN16684}, submitted a \emph{Swift} \ac{TOO}, and sent our \ac{GCN} circular \citep{GCN16668} to encourage others to obtain a spectrum.

\emph{Swift} observed the position of iPTF14eag from 11.6~to~14.4 hours after the burst \citep{GCN16670}. An X-ray source was detected with a count rate of $1.5\times10^{-2}$~counts\,s$^{-1}$. In a second observation starting 62.2 hours after the trigger \citep{GCN16682}, the source had faded to below $2.46 \times 10^{-3}$~counts\,s$^{-1}$. No source was detected by UVOT \citep{GCN16672}.

Meanwhile, from 20.8~to~21.9 hours after the burst, \citet{GCN16671} obtained a spectrum from 3630~to~7500~\AA\ with the OSIRIS instrument on the Gran Telescopio Canarias (GTC) and determined a redshift of $z=3.29$.

The source was detected in radio with \ac{VLA} \citep{GCN16694} and \ac{AMI} \citep{GCN16725}. The broadband \ac{SED} around the time of the VLA detection broadly fits a forward shock model, but is poorly constrained due to the lack of a contemporaneous X\nobreakdashes-ray detection. The spectral slope between the two VLA bands is somewhat steeper than the standard low\nobreakdashes-frequency value of $\beta = -1/3$, possibly indicating that the radio emission is self\nobreakdashes-absorbed. We obtained 14 \ac{AMI} observations every 2 or 3 days from 2014~August~8 until 2014~Sept~12. Observations were 2\nobreakdashes--4 hours in duration. \ac{AMI} first detected the afterglow 4.6 days post-burst. The \ac{AMI} light curve peaked $\sim$10.6~days post\nobreakdashes-burst at 15.7 GHz, which is characteristic of forward shock emission at radio wavelengths \citep{RadioSelectedGRBAfterglows}.

A peculiar feature of the optical light curve is that the P60 $r$ and $i$ band observations at $\Delta t \approx 2$~days appears to be inverted, with a rising rather than falling spectral shape, compared to the earlier P60 photometry at $\Delta t \approx 1$~day. However, this feature is within the error bars and may be merely a statistical fluctuation.

This is the highest redshift burst in our sample, and also had the weakest prompt emission in terms of the fluence observed by \ac{GBM}.

\begin{figure}
    \centering
    \includegraphics{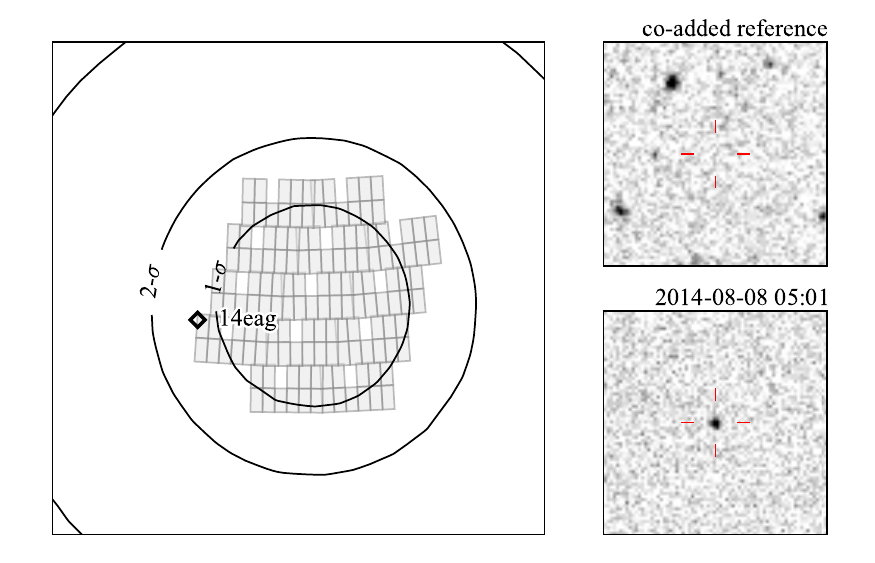}
    \caption[Discovery of \acs{GRB}~140808A~/~iPTF14eag]{\label{fig:GRB140808A}\emph{Fermi} \ac{GBM} localization (black contours), \ac{P48} tiling (gray rectangles), and discovery images for \ac{GRB}~140808A~/~iPTF14eag.}
\end{figure}

\section{The population in context}

\subsection{Selection effects}

First, we investigate the properties of the subset of \ac{GBM} bursts followed up by \ac{IPTF} compared to the \ac{GBM} bursts as a whole. It is known that, on average, GRBs with larger prompt fluences have brighter optical afterglows, though the correlation is very weak \citep{ComparisonAfterglows}. In Figure~\ref{fig:fluence-radius}, we plot the fluence in the 10\nobreakdashes--1000~keV band and 1\nobreakdashes-$\sigma$ localization radius of all \ac{GBM} bursts from the beginning of our experiment, retrieved from the \emph{Fermi} \ac{GBM} Burst Catalog at HEASARC\footnote{\url{http://heasarc.gsfc.nasa.gov/W3Browse/fermi/fermigbrst.html}}. As expected, there is a weak but clearly discernible correlation between fluence and radius, $F \propto r^{-1.3}$, with a Pearson correlation coefficient of $R = 0.64$.\footnote{In a separate sample of \ac{GBM} \acp{GRB} compiled by \citet{GBMLocalization}, the correlation between error radius and photon fluence is slightly stronger than the correlation between error radius and fluence. However, we use fluence rather than photon fluence here because the latter is not available for all bursts in the online \emph{Fermi} GBM archive.} The subset of bursts that we followed up spans a wide range in fluence, and error radii up to $\sim10\arcdeg$. The bursts for which we detected optical afterglows are preferentially brighter, with the faintest burst having a fluence as low as $3\times10^{-6}$~erg\,cm$^{-2}$. There are some bright ($> 3 \times 10^{-5}$~erg\,cm$^{-2}$) and well confined ($< 1.8\arcdeg$) events for which we did not find afterglows: those at 2013-08-28~07:19:56, 2013-11-27~14:12:14, and 2014-01-04~17:32:00 (see Table~\ref{table:nondetections}). However, these non\nobreakdashes-detections are not constraining given their ages of 20.28, 13.46, and 18.57 hours respectively. Conversely, there were two especially young bursts (followed up at $\Delta t = 0.11$~and~0.17~hours) for which we did not detect afterglows. The non\nobreakdashes-detection of the burst at 2014-07-16~07:20:13 makes sense because we searched only 28\% of the \ac{GBM} localization. The non\nobreakdashes-detection on 2014-04-04~04:06:48, for which we observed 69\% of the localization, is a little more surprising, especially given its relatively high fluence of $8\times10^{-6}$~erg\,cm$^{-2}$; this is a possible candidate for a ``dark \ac{GRB}''. On the whole, however, we can see that (1) we have followed up bursts with a large range of error radii and fluences, (2), there is a weak preference toward detecting bursts with small error radii, and (3) the detections tend toward bursts with high fluences. Naively one might expect higher fluences to translate into lower redshifts, but the interplay between the \ac{GRB} luminosity function and detector threshold greatly complicate such inferences \citep{GRBLuminosityFunction}.

\begin{figure}
    \centering
    \includegraphics{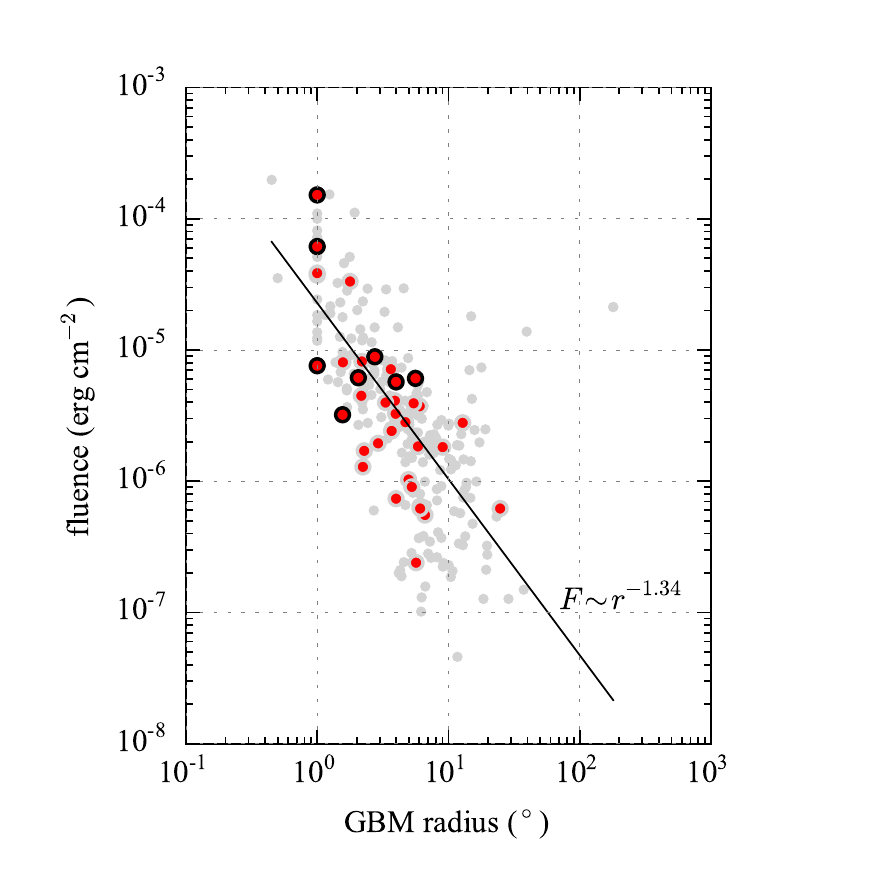}
    \caption[Fluence and error radius]{\label{fig:fluence-radius}Fluence and statistical error radius of \ac{GBM} bursts. Red dots mark bursts that were followed up with \ac{IPTF}; black circles around red dots mark bursts whose afterglows were detected by \ac{IPTF}. The black line is a power\nobreakdashes-law fit to the fluence\nobreakdashes--error~radius relation.}
\end{figure}

Second, the rich sample of all of the \ac{GRB} afterglows that we have today is undeniably the result of the success of the \emph{Swift} mission. It is therefore interesting to consider how the \ac{GBM}\nobreakdashes--\ac{IPTF} sample is similar to or different from the \emph{Swift} sample, given the differences in bandpasses and our increased reliance on the optical afterglow. In Figure~\ref{fig:redshift-distribution}, we plot the cumulative redshift distribution of our sample, alongside the distribution of redshifts of long \acp{GRB} detected by \emph{Swift}.\footnote{This sample was extracted from the \emph{Swift} GRB Table, \url{http://swift.gsfc.nasa.gov/archive/grb_table/}.} Indeed, we find that our sample is at lower redshifts; the former distribution lies almost entirely to the left of the latter, and the ratio of the median redshifts ($z = 1.5$ versus $z = 1.9$) of the two populations is about 0.75. However, with the small sample size, the difference between the two redshift distributions is not significant: a two\nobreakdashes-sample Kolmogorov\nobreakdashes--Smirnov test yields a $p$\nobreakdashes-value of 0.26, meaning that there is a 26\% chance of obtaining these two empirical samples from the same underlying distribution. More \ac{GBM}\nobreakdashes--\ac{IPTF} events are needed to determine whether the redshift distribution is significantly different.

\begin{figure}
    \centering
    \includegraphics{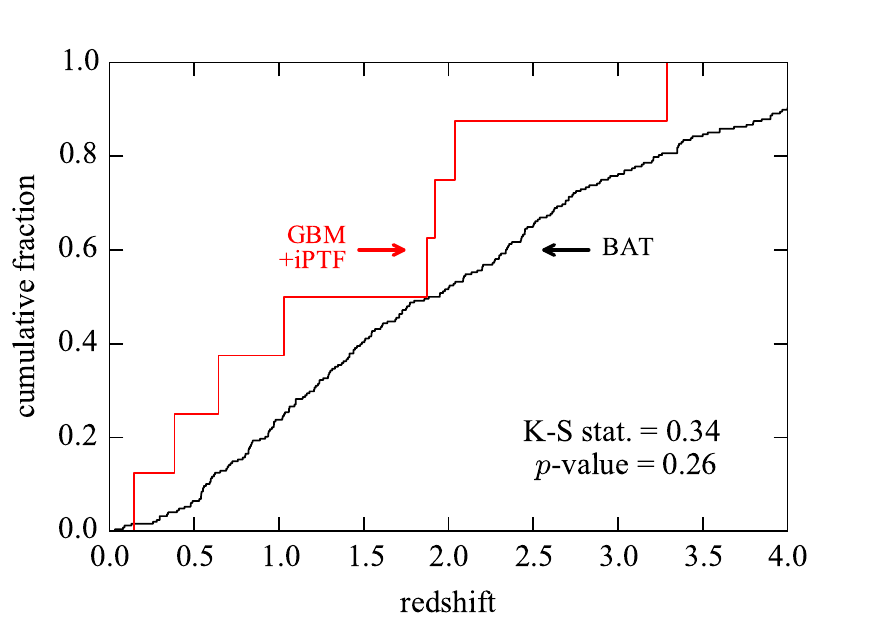}
    \caption[Cumulative redshift distribution]{\label{fig:redshift-distribution}Cumulative distribution of redshifts of long \acp{GRB} observed by \emph{Swift} \ac{BAT} (gray) and the \ac{GBM}\nobreakdashes--\ac{IPTF} experiment (red).}
\end{figure}

\subsection{\acp{GRB} as standard candles?}

\begin{figure*}
    \centering
    \ifinthesis
    \includegraphics[width=\textwidth]{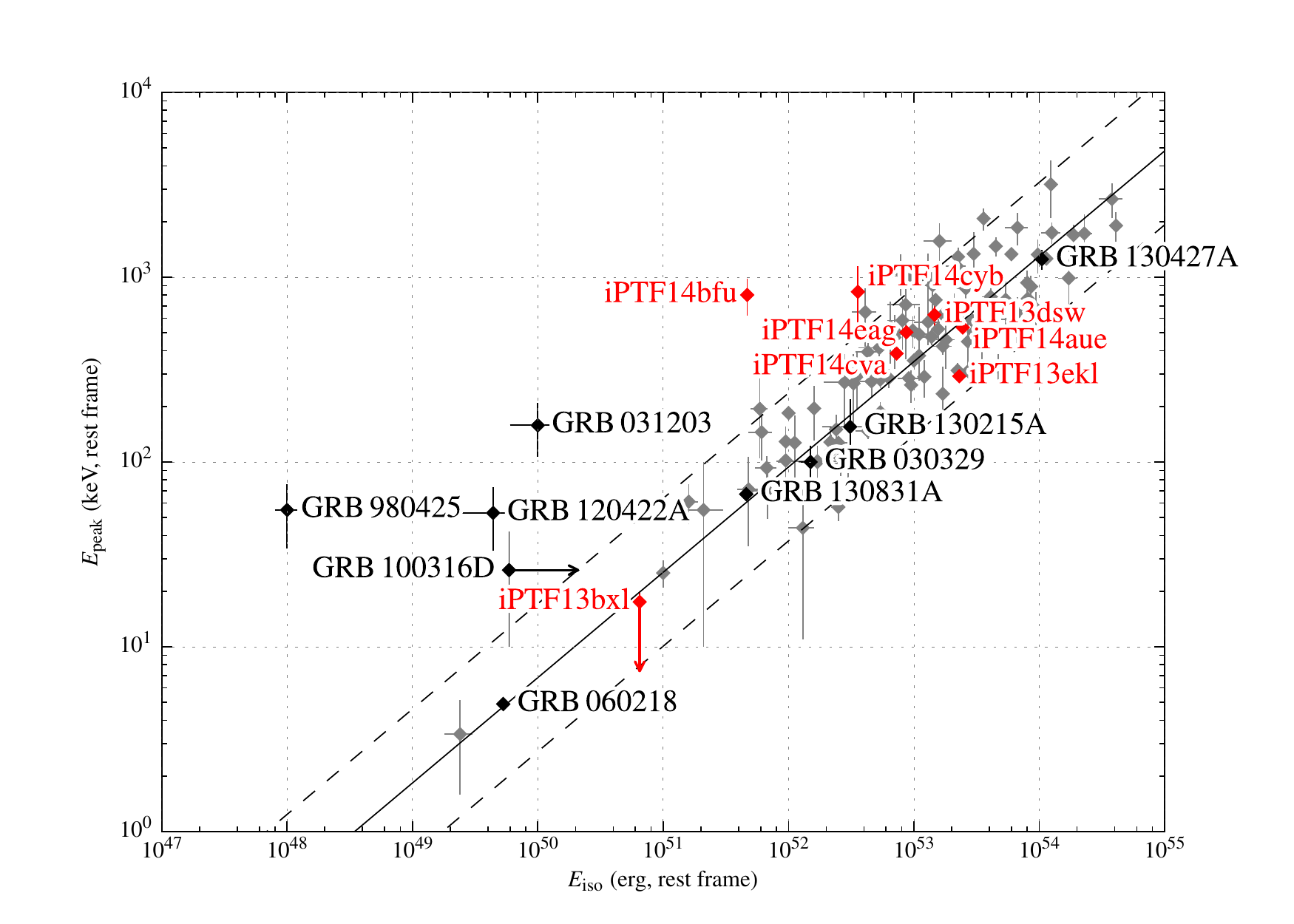}
    \fi
    \caption[Amati relation]{\label{fig:amati}Rest\nobreakdashes-frame energetics of \ac{GBM}\nobreakdashes--iPTF bursts (red) in comparison to an illustrative sample of previous \ac{GRB}\nobreakdashes--\acp{SN}\footnote{Including \ac{GRB}~060218~/~SN2006aj \citep{GRB060218-SN2006aj-1,GRB060218-SN2006aj-2,GRB060218-SN2006aj-3}, \ac{GRB}~100316D~/~SN2010bh \citep{GRB100316D-SN2010bh-1,GRB100316D-SN2010bh-2}, \ac{GRB}~120422A~/~SN2012bz \citep{GRB120422A-SN2012bz-1,GRB120422A-SN2012bz-2}, \ac{GRB}~130215A~/~SN2013ez \citep{GRB130215A-SN2013ez}, \ac{GRB}~130427A~/~SN2013cq \citep{GRB130427A-SN2013cq}, and \ac{GRB}~130831A~/~SN2013fu \citep{GCN15320}.} (black). A general long \ac{GRB} sample from \citet{2006MNRAS.372..233A,2008MNRAS.391..577A,2009A&A...508..173A} is shown in gray. The dashed lines denote $1\sigma$ confidence bands around the Amati relation \citep{AmatiRelation}.}
\end{figure*}

\citet{AmatiRelation} pointed out a striking empirical correlation in the rest\nobreakdashes-frame prompt emission spectra of BeppoSAX \acp{GRB}, with the peak energy (in the $\nu F_\nu$ sense) $E_\mathrm{peak}$ related to the bolometric, isotropic\nobreakdashes-equivalent energy release $E_\mathrm{iso}$ by $E_\mathrm{peak} \propto {E_\mathrm{iso}}^{0.52 \pm 0.06}$. It was quickly realized that such a relation, if intrinsic to the bursts, could be used to measure the redshifts of \acp{GRB} non\nobreakdashes-spectroscopically \citep{EmpiricalRedshiftIndicators}. As with the Phillips relation for Type Ia \acp{SN} \citep{PhillipsRelation}, with such a relation \acp{GRB} could serve as standardizable candles in order to measure cosmological parameters (\citealt{DaiGRBsStandardCandles,FriedmanGRBsStandardCandles,GhirlandaGRBsStandardCandles}; etc.).

However, there has been a vigorous debate about whether the Amati relation and related correlations are innate to \acp{GRB} or reflect a detector\nobreakdashes-dependent selection bias \citep{2005ApJ...627..319B,2005MNRAS.360L..73N,2005MNRAS.361L..10G,2006ApJ...636L..73S,2007ApJ...671..656B,2007MNRAS.382..342C,2007ApJ...656L..53S,2009ApJ...694...76B,2009MNRAS.393.1209F,2009ApJ...704.1405K,GRBLuminosityFunction,2011MNRAS.411.1843S,2012ApJ...747...39C,2012ApJ...747..146K}. One alternative interpretation is that bursts to the upper\nobreakdashes-left boundary of the Amati relation are selected against by photon\nobreakdashes-counting instruments because, being relatively hard, there are fewer photons. The lack of bursts to the lower\nobreakdashes-right of the Amati line may be due to a genuine lack of relativistic explosions that are much softer than, but as energetic as, standard \acp{GRB}.

It has been difficult to directly test the Amati relation in the context of \emph{Fermi} bursts because most lack known redshifts, aside from bursts that were coincidentally also observed and localized by the \emph{Swift} \ac{BAT}, which do not directly sample the selection bias of \emph{Fermi} \ac{GBM}. However, \citet{FermiAmatiDebate} showed that many \emph{Fermi} bursts without known redshifts would be inconsistent with the Amati relation at any distance. (See also \citealt{UrataEnergeticWAMLATGRBs} for outlier events detected by \emph{Fermi}~\ac{LAT} and \emph{Suzaku}~\acs{WAM}.) Here, we have a small sample of \emph{Fermi} bursts with \emph{known} redshifts. One of them, \ac{GRB}~140606B~/~iPTF14bfu at $z=0.384$, is a clear outlier, over $2\sigma$ away from the mean Amati relation. This burst is not alone: in Figure~\ref{fig:amati}, we have marked a selection previous long \acp{GRB} with spectroscopically identified \acp{SN}. Three among them are also outliers. (A possible caveat is that the prompt emission mechanism for \ac{GRB}~140606B could be different from typical cosmological bursts; we explore this in the next section.) To be sure, most of the bursts in our \ac{GBM}\nobreakdashes--\ac{IPTF} sample fall within a $1\sigma$ band of the Amati relation. This includes the nearest event to date, \ac{GRB}~130702A~/~iPTF13bxl at $z=0.145$. However, the one outlier in our admittedly small sample strengthens the case that the boundary of the Amati relation is somewhat influenced by the detector thresholds and bandpasses.

\subsection{Shock breakout}

Two \acp{GRB} in our sample, \ac{GRB}~130702A~/~iPTF13bxl and \ac{GRB}~140606B~/~iPTF14bfu, have $E_\mathrm{iso} \sim 10^{51}$~erg (rest frame), energetically intermediate between ``standard'' luminous, cosmically distant bursts and nearby \acp{llGRB}. Prototypes of the latter class include \ac{GRB}~980425~/~\ac{SN}~1998bw \citep{SN1998bw,1998bwShockBreakout}, which was also the first \ac{SN} discovered in association with a \ac{GRB}. They offer an interesting test case for competing theories to explain the wide range of prompt gamma\nobreakdashes-ray energy releases observed from \acp{GRB} (e.g., \citealt{2014A&A...566A.102S}).

It has been suggested that the two luminosity regimes correspond to different prompt emission mechanisms \citep{AreLowLuminosityBurstsGeneratedByRelativisticJets}. The \acp{llGRB} could be explained by the breakout of a mildly relativistic shock from the progenitor envelope \citep{RelativisticShockBreakoutRelation}. High\nobreakdashes-luminosity bursts, on the other hand, are thought to be produced by internal shocks within an ultra\nobreakdashes-relativistic jet \citep{ReesInternalShocks} that has successfully punched through the star. A central engine that sometimes fails to launch an ultra\nobreakdashes-relativistic jet is one way to unify the luminosity functions of standard \acp{GRB} and \acp{llGRB} \citep{LuminosityFunctionJetStructureGRBs}.

The smoking gun for the relativistic shock breakout model is a cooling, thermal component to the prompt X\nobreakdashes-ray emission, as in the case of \ac{GRB}~060218 \citep{GRB060218ShockBreakoutNature}. Unfortunately, this diagnostic is not possible for \acp{GRB}~130702A~and~140606B because we lack early\nobreakdashes-time \emph{Swift} observations.

However, \citet{RelativisticShockBreakoutRelation} propose a closure relation (their Equation 18) between the prompt energy, temperature, and timescale that is valid for shock breakout\nobreakdashes-powered \acp{GRB}. We reproduce it here:
\begin{equation}
    t^\mathrm{obs}_\mathrm{bo} \sim
        20~\text{s}
        \left(\frac{E_\mathrm{bo}}{10^{46}~\text{erg}}\right)^{\frac{1}{2}}
        \left(\frac{T_\mathrm{bo}}{50~keV}\right)^{-\frac{9+\sqrt{3}}{4}}.
\end{equation}
If we very crudely assume that all of the prompt emission is from a shock escaping from the progenitor envelope, then we can use $E_\mathrm{iso}$, $E_\mathrm{peak}$, and $T_{90}$ as proxies for these observables. This gives us a simple discriminator of which bursts are plausible shock breakout candidates, the ratio
\begin{equation}
    \xi = (1 + z) t^\mathrm{obs}_\mathrm{bo} / T_\mathrm{90},
\end{equation}
which should be close to 1. As expected, most of the energetic ($E_\mathrm{iso} > 10^{52}~\textrm{erg}$), cosmic ($z > 0.5$) \acp{GRB} in our sample are inconsistent with the closure relation. They are all much \emph{shorter} in duration, given their $\gamma$\nobreakdashes-ray spectra, than would be expected for a shock breakout. The exception is \ac{GRB}~140623A~/~iPTF14cyb, which yields $\xi = 0.5 \pm 0.5$.

Surprisingly, of the two low\nobreakdashes-luminosity, low\nobreakdashes-redshift bursts in our sample, \ac{GRB}~130702A~/~iPTF13bxl's prompt emission was also much too brief to be consistent with this shock breakout model, with $\xi = (1.6 \pm 0.7) \times 10^3$. Most likely, this means that the prompt emission of \ac{GRB}~130702A is simply a very soft, very sub\nobreakdashes-luminous version of an otherwise `ordinary' long \ac{GRB}. Any early\nobreakdashes-time shock breakout signature, if present, was unobserved either because it occurred at energies below \ac{GBM}'s bandpass, or because it was much weaker than the emission from the standard \ac{GRB} mechanism. However, \ac{GRB}~140606B~/~iPTF14bfu's prompt emission is consistent with the closure relation, with $\xi = 0.5 \pm 0.3$. Though we must interpret this with caution because we cannot disentangle a thermal component from the \ac{GBM} data, if we naively apply linear least squares to (the logarithm of) Equations (14, 16, 17) of \citeauthor{RelativisticShockBreakoutRelation},
\begin{align}
    E_\mathrm{bo} &\approx 2 \times 10^{45} R_5^2 \gamma_{f,0}^\frac{1+\sqrt{3}}{2}~\text{erg}, \\
    T_\mathrm{bo} &\sim 50 \gamma_{f,0}~\text{keV}, \\
    t_\mathrm{bo}^\mathrm{obs} &\approx 10 \frac{R_5}{\gamma_{f,0}^2}~\text{s},
\end{align}
then we find the breakout radius and Lorentz factor to be:
\begin{eqnarray*}
    R_\mathrm{bo} &=& (1.3 \pm 0.2) \times 10^3 \, R_\sun, \\
    \gamma_{f,0} &=& 14 \pm 2.
\end{eqnarray*}

The breakout radius is comparable to that which \citet{RelativisticShockBreakoutRelation} find for \acp{GRB}~060218 and 100316D, suggestive of breakout from a dense wind environment, rather than the star itself. However, the derived Lorentz factor of \ac{GRB}~140606B is a bit higher than those of the other two examples.

Another way to constrain the nature of the explosion is to look at the kinetic energy $E_{k,\mathrm{iso}}$ of the blast compared to the promptly radiated energy $E_{\gamma,\mathrm{iso}} \equiv E_\mathrm{iso}$, and the radiative efficiency $\eta = E_{\gamma,\mathrm{iso}} / (E_{k,\mathrm{iso}} + E_\mathrm{iso})$. After the end of any plateau phase, the X\nobreakdashes-ray flux is a fairly clean diagnostic of $E_{k,\mathrm{iso}}$ assuming the X\nobreakdashes-rays are above the cooling frequency \citep{EnergyOfGammaRayBursts}. During the slow\nobreakdashes-cooling phase and under the typical conditions where $p \approx 2$ and $\nu_\mathrm{c} < \nu_\mathrm{X}$, the X\nobreakdashes-ray flux is only weakly sensitive to global parameters such as the fraction of the internal energy partitioned to electrons and to the magnetic ($\epsilon_e$, $\epsilon_B$). Even the radiative losses, necessary for extrapolating from the late\nobreakdashes-time afterglow to the end of the prompt phase, are minor, amounting to order unity at $\Delta t = 1$~day \citep{KineticEnergyRadiativeEfficiencyOfGammaRayBursts}. We calculate the isotropic-equivalent rest frame X\nobreakdashes-ray luminosity from the flux at $\Delta t = 1$~day using Equation~(1) of \citet{FermiSwiftPopulationStudies}, reproduced below:
\begin{equation}\label{eq:LX}
    L_\mathrm{X}(t) = 4 \pi {D_\mathrm{L}}^2 F_\mathrm{X} (t) (1 + z)^{-\alpha_\mathrm{X} + \beta_\mathrm{X} - 1}.
\end{equation}
Then we estimate the kinetic energy at the end of the prompt emission phase using Equation~(7) of \citet{KineticEnergyRadiativeEfficiencyOfGammaRayBursts}:
\begin{multline}\label{eq:Ek}
    E_{k,\mathrm{iso}} = \left(10^{52}~\text{ergs}\right) \times R \times
        \left(\frac{L_\mathrm{X} (1~\text{day})}{10^{46}~\text{ergs}~\text{s}^{-1}}\right)^{-4/(p+2)}
        \left(\frac{1 + z}{2}\right)^{-1} \\
        \times \epsilon_{e,-1}^{4(1-p)/(2+p)}
        \epsilon_{B,-2}^{(2-p)/(2+p)} t_\mathrm{10~hr}^{(3p-2)/(p+2)}
        \nu_{18}^{2(p-2)/(p+2)}.
\end{multline}
The correction factor $R$ for radiative losses is given by Equation~(8) of \citet{KineticEnergyRadiativeEfficiencyOfGammaRayBursts}, adopted here:
\begin{equation}\label{eq:R}
    R = \left(\frac{t}{T_\mathrm{90}}\right)^{(17/16)\epsilon_e}.
\end{equation}
The numeric subscripts follow the usual convention for representing quantities in powers of 10 times the cgs unit, i.e., $\epsilon_{e,-1} = \epsilon_e / 10^{-1}$, $\epsilon_{B,-2} = \epsilon_B / 10^{-2}$, and $\nu_{18} \equiv \nu / (10^{18}~\text{Hz})$. We assume $\epsilon_e = 0.1$ and $\epsilon_B = 0.01$. For bursts that have \ac{XRT} detections around $\Delta t = 1$~day (\acp{GRB}~130702A, 131231A, 140508A, 140606B, and 140620A), we calculate $L_\mathrm{X}$ by interpolating a least squares power law fit to the X\nobreakdashes-ray light curve. Some of our bursts (\acp{GRB}~131011A, 140623A, and 140808A) were only weakly detected by \ac{XRT}; for these we extrapolate from the mean time of the \ac{XRT} detection assuming a typical temporal slope of $\alpha_\mathrm{X} = 1.43 \pm 0.35$ \citep{FermiSwiftPopulationStudies}. The kinetic and radiative energies of our eight bursts are shown in Figure~\ref{fig:radiative-efficiency}. Half of our bursts are reasonably well constrained in $E_\mathrm{k}$--$E_\gamma$ space; these are shown as red points. The other half (\acp{GRB}~131011A, 131231A, 140620A, and 140623A) have highly degenerate \acp{SED}, so their position in this plot is highly sensitive to model assumptions; these are shown as gray points. Dotted lines are lines of constant radiative efficiency.

Within our sample, there are at least three orders of magnitude of variation in both $E_{k,\mathrm{iso}}$ and $E_{\gamma,\mathrm{iso}}$. The two \ac{GRB}--\acp{SN} have radiative and kinetic energies of $\sim 10^{51}$~ergs, both two to three orders of magnitude lower than the other extreme in our sample or the average values for \emph{Swift} bursts. In our sample, they have two of the lowest inferred radiative efficiencies of $\eta \sim 0.1$\nobreakdashes--0.5, but these values are not atypical of BATSE bursts (e.g., \citealt{KineticEnergyRadiativeEfficiencyOfGammaRayBursts}) and are close to the median value for \emph{Swift} bursts. These are, therefore, truly less energetic than cosmological bursts, not merely less efficient at producing gamma\nobreakdashes-rays.

\begin{figure}
    \centering
    \includegraphics{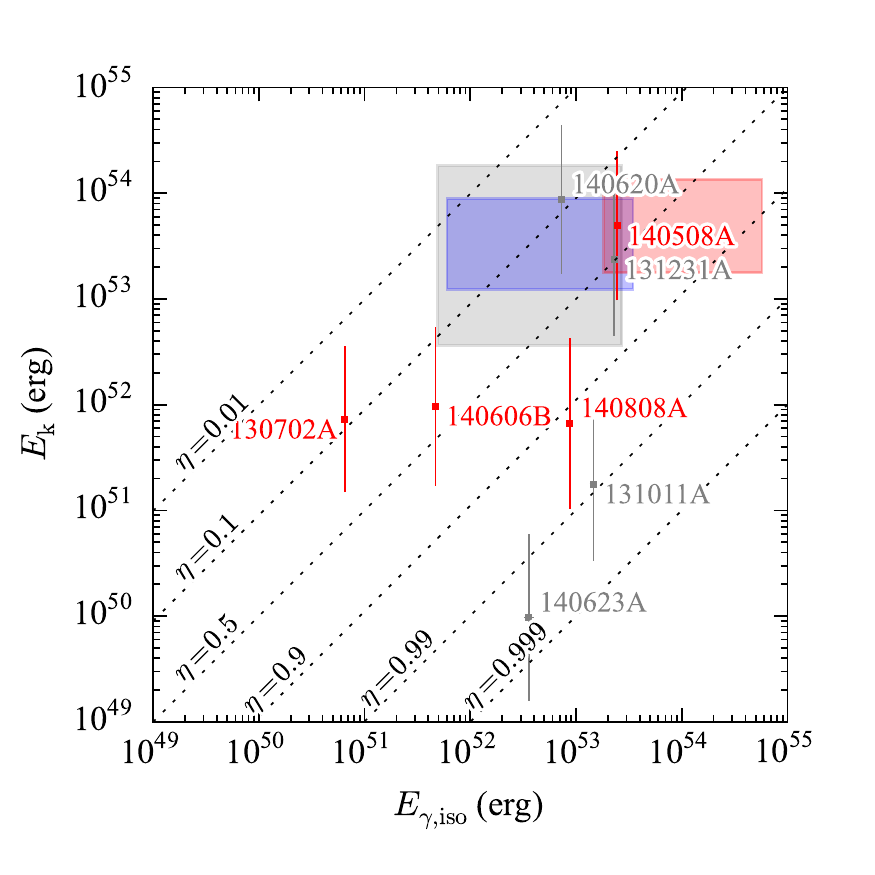}
    \caption[Radiative efficiency of prompt phase]{\label{fig:radiative-efficiency}Fireball kinetic energy $E_{k,\mathrm{iso}}$ at $t = T_{90}$ as estimated from X\nobreakdashes-ray flux versus rest\nobreakdashes-frame isotropic\nobreakdashes-equivalent gamma\nobreakdashes-ray energy $E_{\gamma,\mathrm{iso}}$. Red points denote bursts for which $E_{k,\mathrm{iso}}$ can be reliably estimated from the \emph{Swift} \ac{XRT} data; gray points denote bursts for which the calculation of $E_{k,\mathrm{iso}}$ may have extreme model dependence. Dashed lines are lines of constant radiative efficiency $\eta = E_{\gamma,\mathrm{iso}} / (E_{k,\mathrm{iso}} + E_{\gamma,\mathrm{iso}})$. The gray, blue, and red rectangles show the 1$\sigma$ parameter ranges of \emph{Swift} \ac{BAT}, \ac{BAT}+\ac{GBM}, and \ac{BAT}+\ac{LAT} long \acp{GRB} from \citet{FermiSwiftPopulationStudies}.}
\end{figure}

\section{Looking forward}

In this experiment, we have followed up 35 \emph{Fermi} \ac{GBM} bursts, scanning areas from 30~to~147~deg$^2$. To date, we have detected eight afterglows with apparent optical magnitudes as bright as $R \approx 16$ and as faint as $R \approx 20$. We have found redshifts as nearby as $z = 0.145$ and as distant as $z = 3.29$. A continuation of the project should reveal more low\nobreakdashes-redshift events, more GRB\nobreakdashes--\acp{SN}, and more relatively hard \acp{GRB}.

We aim to uncover the much fainter afterglows of short, hard bursts by using stacked \ac{P48} exposures and integrating a co-addition stage into the real-time pipeline, and by honing our follow\nobreakdashes-up to sift through the increased number of candidates. The greatest factor limiting discoveries is, of course, that \emph{Fermi} detects bursts all over the sky, only a fraction of which are visible from Palomar. Given our success so far, we enthusiastically suggest that other wide\nobreakdashes-field surveys implement a similar program. Furthermore, automatically sharing lists of candidates between longitudinally separated instruments would facilitate rapid identification and follow\nobreakdashes-up of the fastest fading events.

It is uncertain what directions future gamma\nobreakdashes-ray space missions will take. Some may be like \emph{Swift}, able to rapidly train multiple on\nobreakdashes-board follow\nobreakdashes-up instruments on new targets. Even if they lack these capabilities, we should be able to routinely locate \ac{GRB} afterglows and find their redshifts using targeted, ground\nobreakdashes-based optical transient searches similar to the one that we have described.

Looking beyond \acp{GRB}, the current \ac{IPTF} \ac{TOO} pipeline serves as a prototype for searching for optical counterparts of \ac{GW} transients. Near the end of 2015, Advanced \ac{LIGO} will begin taking data, with Advanced Virgo soon following suit. The first detections are anticipated by 2016 or later \citep{LIGOObservingScenarios}. On a similar timescale, \ac{IPTF} will become the \acl{ZTF}, featuring a new 47~deg$^2$ survey camera that can reach $R = 20.4$~mag~in~30~s. The prime \ac{GW} sources, \ac{BNS} mergers, may also produce a variety of optical transients: on- or off-axis afterglows \citep{SyntheticSGRBAfterglows}, kilonovae \citep{Kilonova,KilonovaHighOpacities}, and neutron-powered precursors \citep{KilonovaPrecursor}; see Figure~\ref{fig:sgrb-kcorrected} for some examples.

There will be two key challenges. First, \ac{GW} localizations can be even coarser than \emph{Fermi} \ac{GBM} error circles. Starting around $\sim$600~deg$^2$ in the initial (2015) two\nobreakdashes-detector configuration \citep{KasliwalTwoDetectors,FirstTwoYears}, the areas will shrink to $\sim$200~deg$^2$ with the addition of Virgo in 2016. They should reach $\sim 10$~deg$^2$ toward the end of the decade as the three detectors approach final design sensitivity and can approach $\sim 1$~deg$^2$ as additional planned \ac{GW} facilities come online (LIGO\nobreakdashes--India and KAGRA; see \citealt{ShutzThreeFiguresOfMerit,Veitch:2012,FairhurstLIGOIndia,NissankeKasliwalEMCounterparts,LIGOObservingScenarios}). Since the detection efficiency of our \ac{GBM}--\ac{IPTF} afterglow search is consistent with the areas that we searched, we expect that even the earliest Advanced \ac{LIGO} localizations will present no undue difficulties for \ac{ZTF} when we consider its 15\nobreakdashes-fold increase in areal survey rate as compared to \ac{IPTF}.

However, there is a second challenge that these optical signatures are predicted to be fainter than perhaps $22$~mag (with the exception of on-axis afterglows, which should be rare but bright due to beaming). To confront this, we must perform deep (10~min--1~hour) integrations with the \ac{P48}. To work through the larger number of contaminating sources, we propose the following strategies:
\begin{enumerate}
    \item We may adopt a mix of two or more exposure depths in order to cover the area both rapidly and deeply, to be sensitive to both bright on-axis events and the fainter isotropic signatures. For instance, we may cover the entire \ac{GW} localization once at a single\nobreakdashes-exposure depth, then repeat with deeper 10~min--1~hour exposures as the visibility allows. Note that because the \ac{LIGO} antenna pattern is preferentially sensitive above and directly opposite of North America, we are optimistic that most early Advanced \ac{LIGO} events should be promptly accessible from Palomar with long observability windows \citep{KasliwalTwoDetectors}.
    \item We will increasingly automate the selection of targets for photometric follow-up. With more advanced machine learning algorithms and better leveraging of light curve history across multiple surveys, we plan to robotically trigger follow-up by multiple telescopes.
    \item With longer \ac{TOO} time blocks, we will have to begin follow-up of the most promising candidates while the \ac{P48} observations are ongoing, to capitalize on accessibility from Palomar.
    \item To further prioritize candidates for follow-up and severely reduce false positives, we can use spatial proximity to nearby galaxies \citep{NissankeKasliwalEMCounterparts}.
    \item Our first experiences with detections and non-detections will guide decisions about the optimal filter. At the moment, kilonova models prefer redder filters (suggesting $i$-band), and precursor models prefer bluer (suggesting $g$-band).
\end{enumerate}

The combination of \ac{LIGO}, Virgo, \emph{Fermi}, \emph{Swift}, and \ac{IPTF}/\ac{ZTF} is poised to make major discoveries over the next few years, of which we have provided a small taste in this work. We offer both lessons learned and a way forward in this multimessenger effort. The ultimate reward will be joint observations of a compact binary merger in gamma, X\nobreakdashes-rays, optical, and GWs, giving us an exceptionally complete record of a complex astrophysical process: it will be almost as good as being there.

\section*{Acknowledgements}

L.P.S. thanks generous support from the \ac{NSF} in the form of a Graduate Research Fellowship.

The National Radio Astronomy Observatory is a facility of the \ac{NSF} operated under cooperative agreement by Associated Universities, Inc.

This paper is based on observations obtained with the \acl{P48} and the \acl{P60} at the Palomar Observatory as part of the Intermediate Palomar Transient Factory project, a scientific collaboration among the California Institute of Technology, Los Alamos National Laboratory, the University of Wisconsin, Milwaukee, the Oskar Klein Center, the Weizmann Institute of Science, the TANGO Program of the University System of Taiwan, and the Kavli Institute for the Physics and Mathematics of the Universe. The present work is partly funded by \emph{Swift} Guest Investigator Program Cycle 9 award 10522 (NASA grant NNX14AC24G) and Cycle 10 award 10553 (NASA grant NNX14AI99G).

Some of the data presented herein were obtained at the W. M. Keck Observatory, which is operated as a scientific partnership among the California Institute of Technology, the University of California, and NASA; the Observatory was made possible by the generous financial support of the W.~M.~Keck Foundation.

We thank Thomas Kr\"{u}hler for reducing the X-shooter spectrum of GRB~131011A~/~iPTF13dsw.

We thank the staff of the Mullard Radio Astronomy Observatory for their invaluable assistance in the operation of \ac{AMI}. G.E.A., R.P.F., and T.D.S. acknowledge the support of the European Research Council Advanced Grant 267697, ``4 Pi Sky: Extreme Astrophysics with Revolutionary Radio Telescopes.''

Support for \ac{CARMA} construction was derived from the Gordon and Betty Moore Foundation, the Kenneth T. and Eileen L. Norris Foundation, the James S. McDonnell Foundation, the Associates of the California Institute of Technology, the University of Chicago, the states of California, Illinois, and Maryland, and the \ac{NSF}. Ongoing \ac{CARMA} development and operations are supported by the \ac{NSF} under a cooperative agreement, and by the \ac{CARMA} partner universities.

These results made use of Lowell Observatory's \ac{DCT}. Lowell operates the \ac{DCT} in partnership with Boston University, Northern Arizona University, the University of Maryland, and the University of Toledo. Partial support of the \ac{DCT} was provided by Discovery Communications. \ac{LMI} was built by Lowell Observatory using funds from the \ac{NSF} (AST-1005313).

A portion of this work was carried out at the Jet Propulsion Laboratory under a Research and Technology Development Grant, under contract with NASA. Copyright 2014 California Institute of Technology. All Rights Reserved. US Government Support Acknowledged.

K.H. acknowledges support for the \ac{IPN} under the following NASA grants: NNX07AR71G, NNX13AP09G, NNX11AP96G, and NNX13AI54G.

IRAF is distributed by the National Optical Astronomy Observatory, which is operated by the Association of Universities for Research in Astronomy (AURA) under cooperative agreement with the \ac{NSF}.

This research has made use of data, software and/or web tools obtained from \ac{HEASARC}, a service of the Astrophysics Science Division at NASA/GSFC and of the Smithsonian Astrophysical Observatory's High Energy Astrophysics Division.

This research has made use of \ac{NED}, which is operated by the Jet Propulsion Laboratory, California Institute of Technology, under contract with NASA.

This work made use of data supplied by the UK Swift Science Data Centre at the University of Leicester including the \emph{Swift} \ac{XRT} \ac{GRB} catalog and light curve repository \citep{2007A&A...476.1401G,SwiftXRTRepository,2009MNRAS.397.1177E}.

This research made use of Astropy\footnote{\url{http://www.astropy.org}} \citep{astropy}, a community-developed core Python package for Astronomy. Some of the results in this paper have been derived using HEALPix \citep{HEALPix}.

\begin{figure*}
    \centering
    \includegraphics[width=\textwidth]{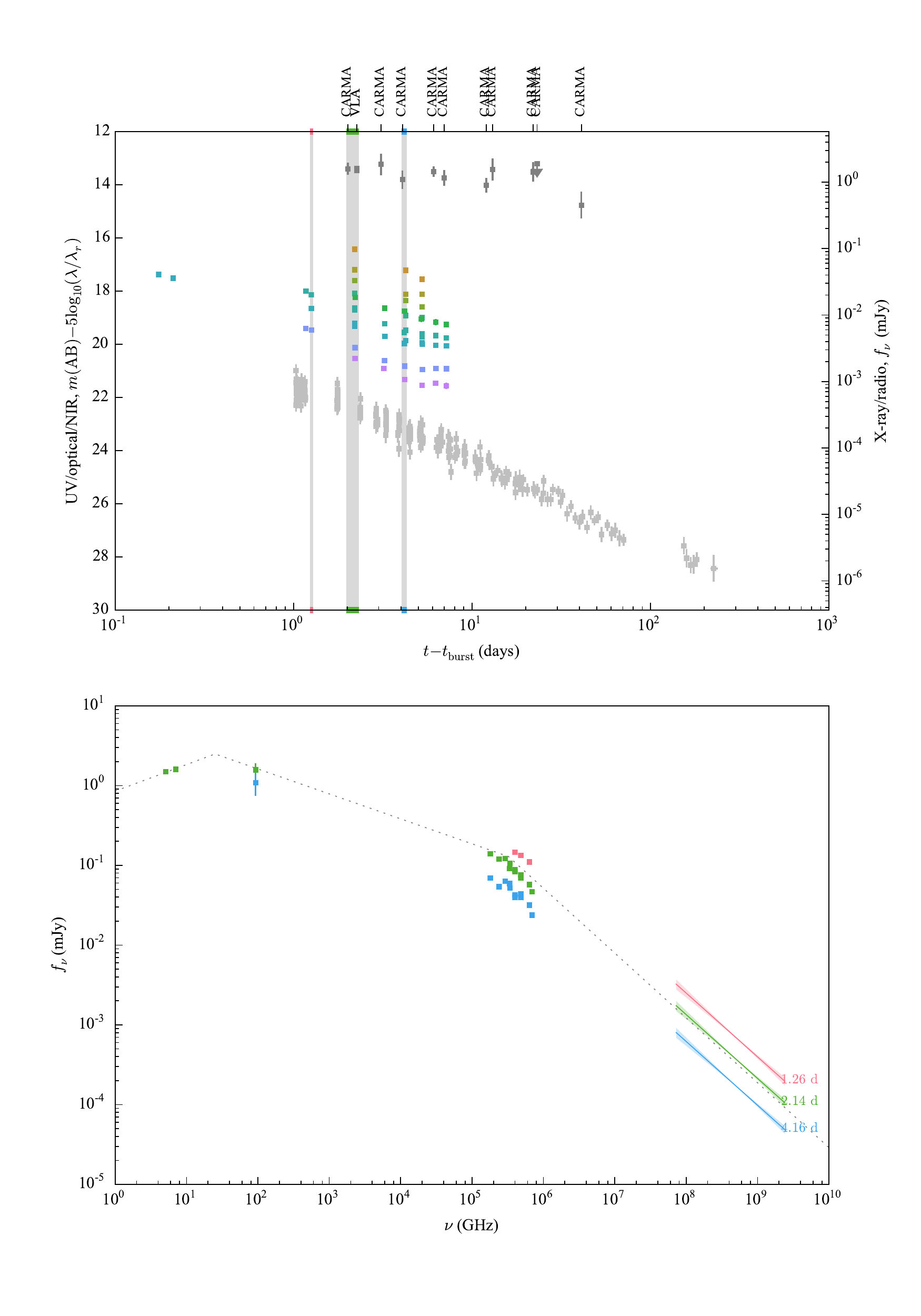}
    \caption{\label{fig:grb-130702A-sed}Light curve and \acs{SED} of \ac{GRB}~130702A~/~iPTF13bxl.}
\end{figure*}

\begin{figure*}
    \centering
    \includegraphics[width=\textwidth]{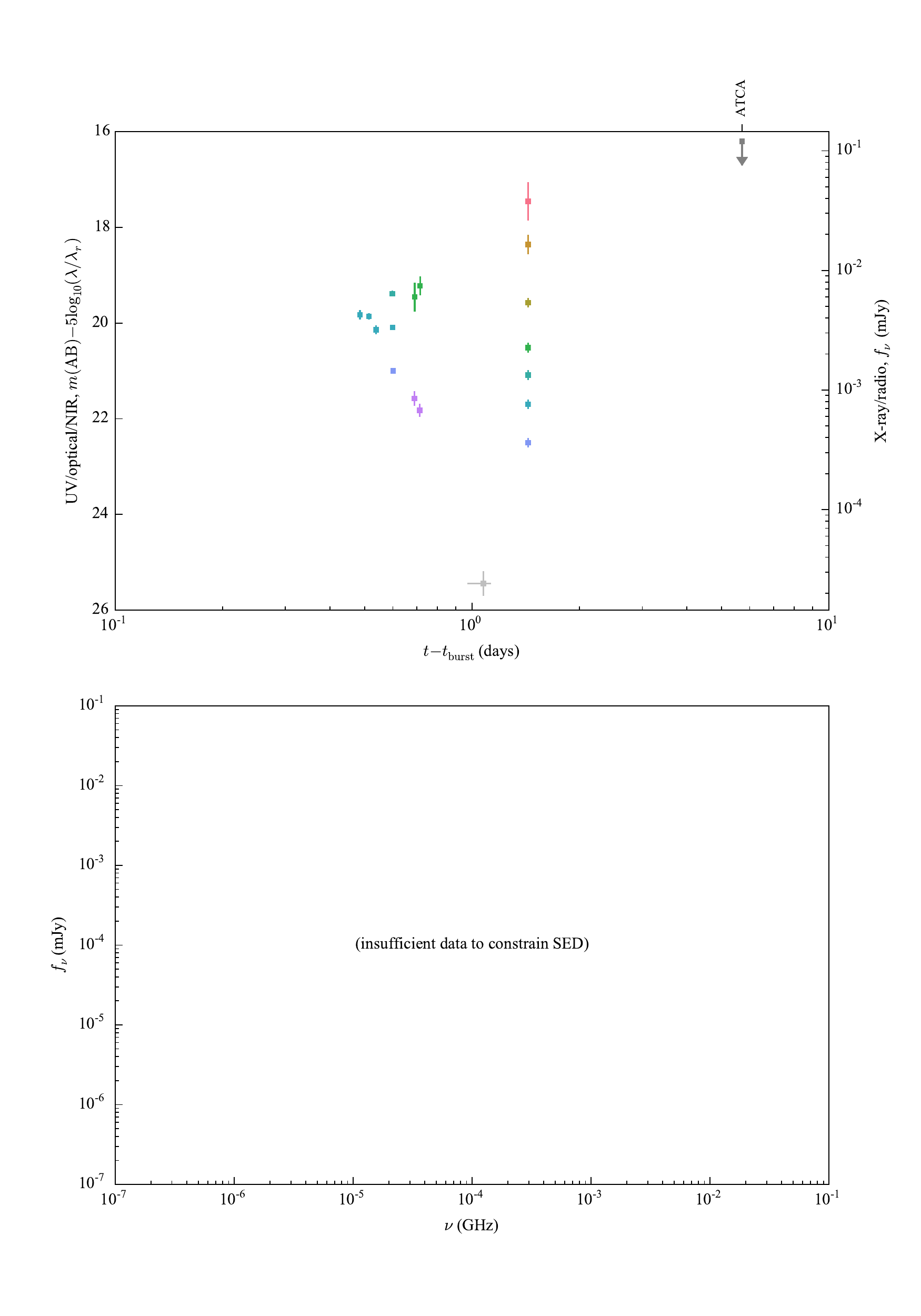}
    \caption{\label{fig:grb-131011A-sed}Light curve of \acs{GRB}~131011A~/~iPTF13dsw.}
\end{figure*}

\begin{figure*}
    \centering
    \includegraphics[width=\textwidth]{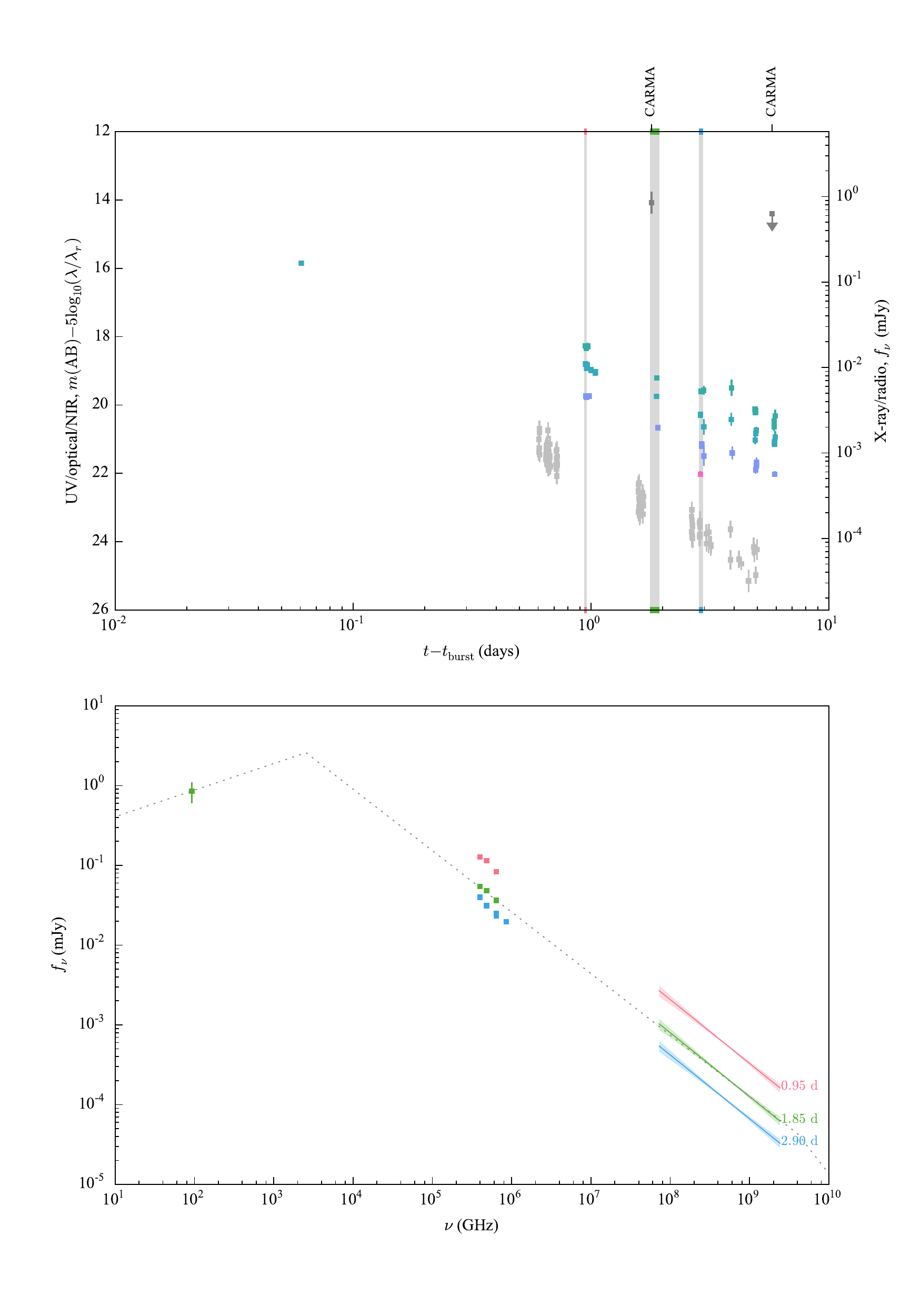}
    \caption{\label{fig:grb-131231A-sed}Light curve and \acs{SED} of \ac{GRB}~131231A~/~iPTF13ekl.}
\end{figure*}

\begin{figure*}
    \centering
    \includegraphics[width=\textwidth]{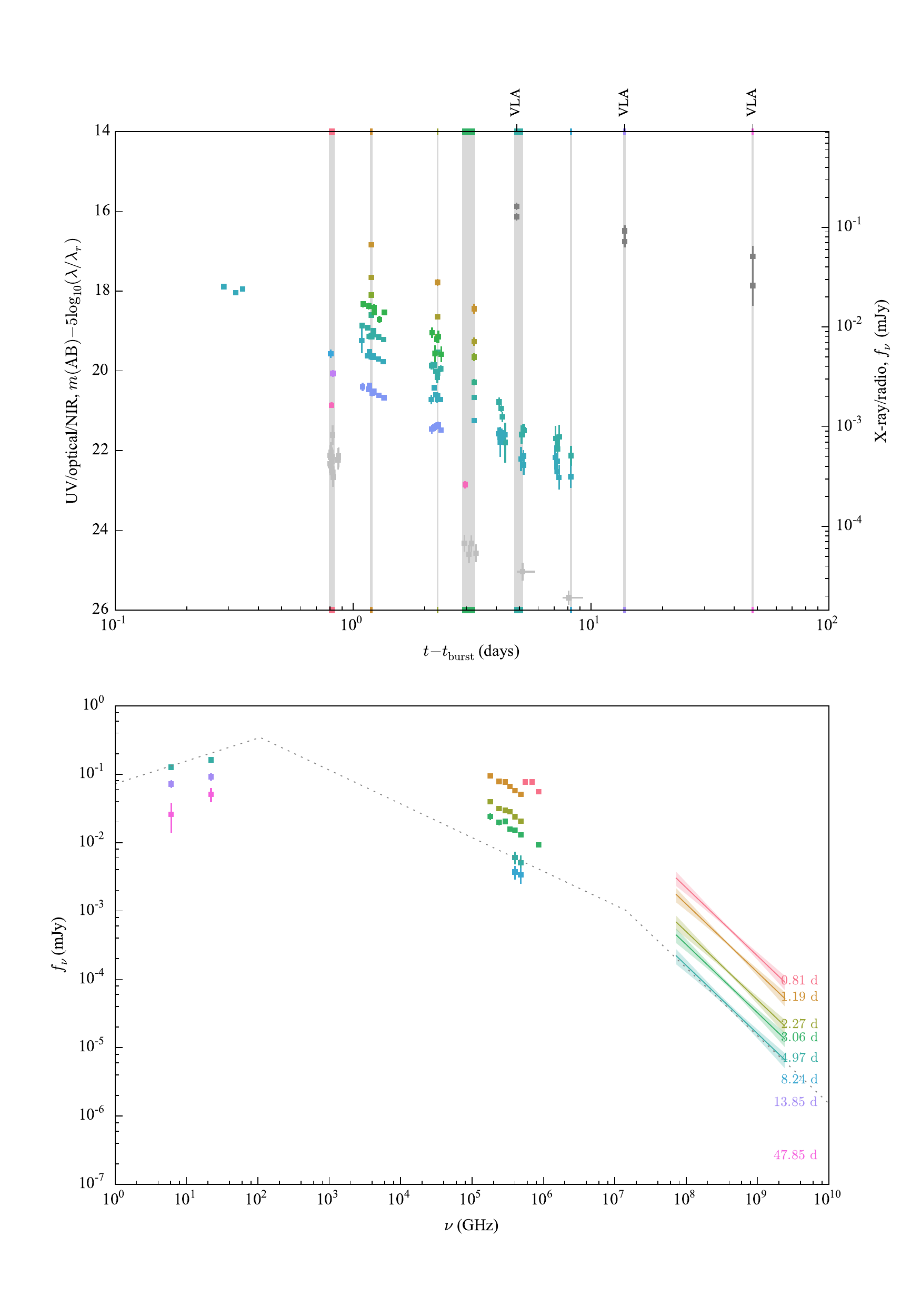}
    \caption{\label{fig:grb-GRB140508A-sed}Light curve and \acs{SED} of \ac{GRB}~140508A~/~iPTF14aue.}
\end{figure*}

\begin{figure*}
    \centering
    \includegraphics[width=\textwidth]{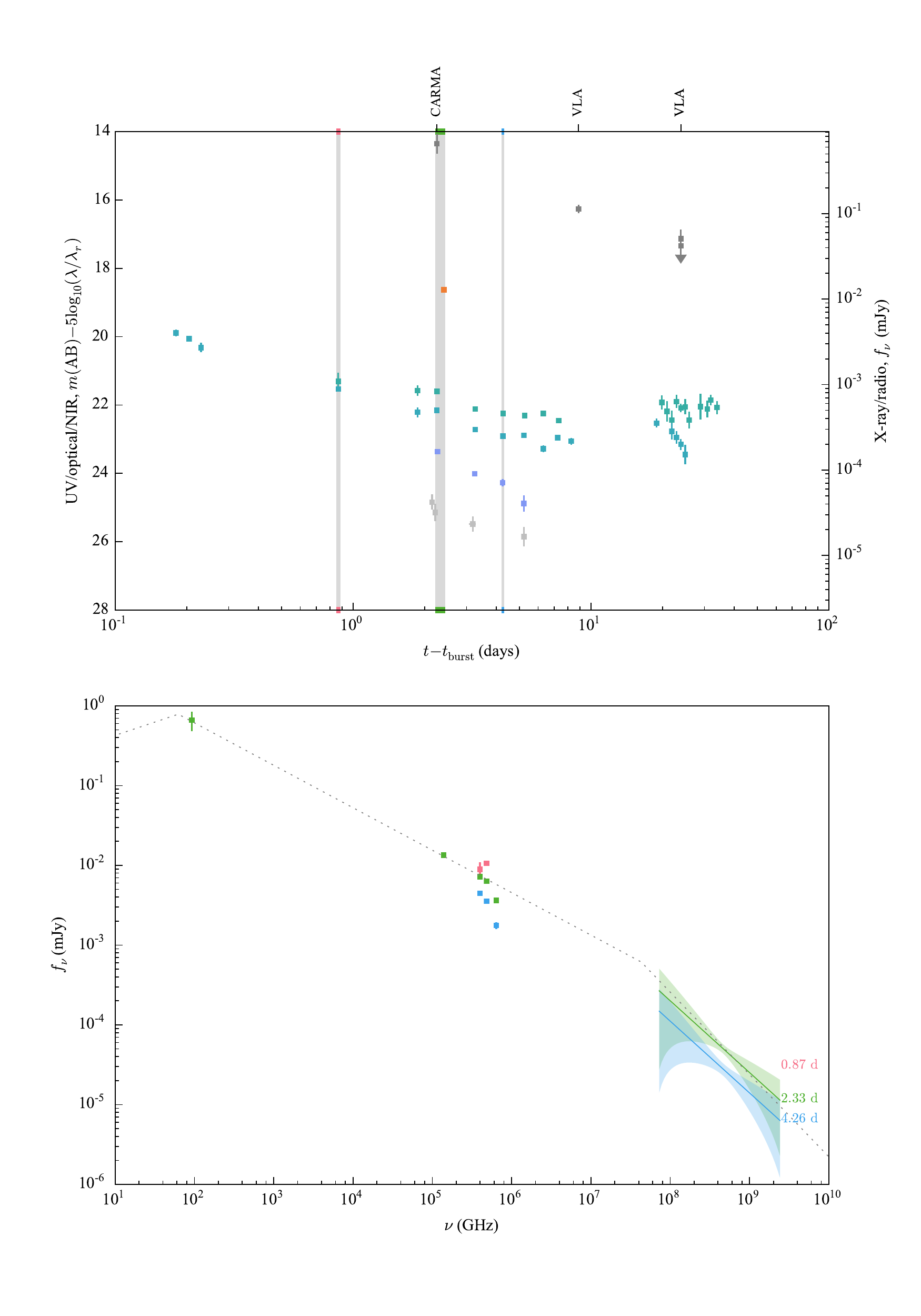}
    \caption{\label{fig:grb-140606B-sed}Light curve and \acs{SED} of \ac{GRB}~140606B~/~iPTF14bfu.}
\end{figure*}

\begin{figure*}
    \centering
    \includegraphics[width=\textwidth]{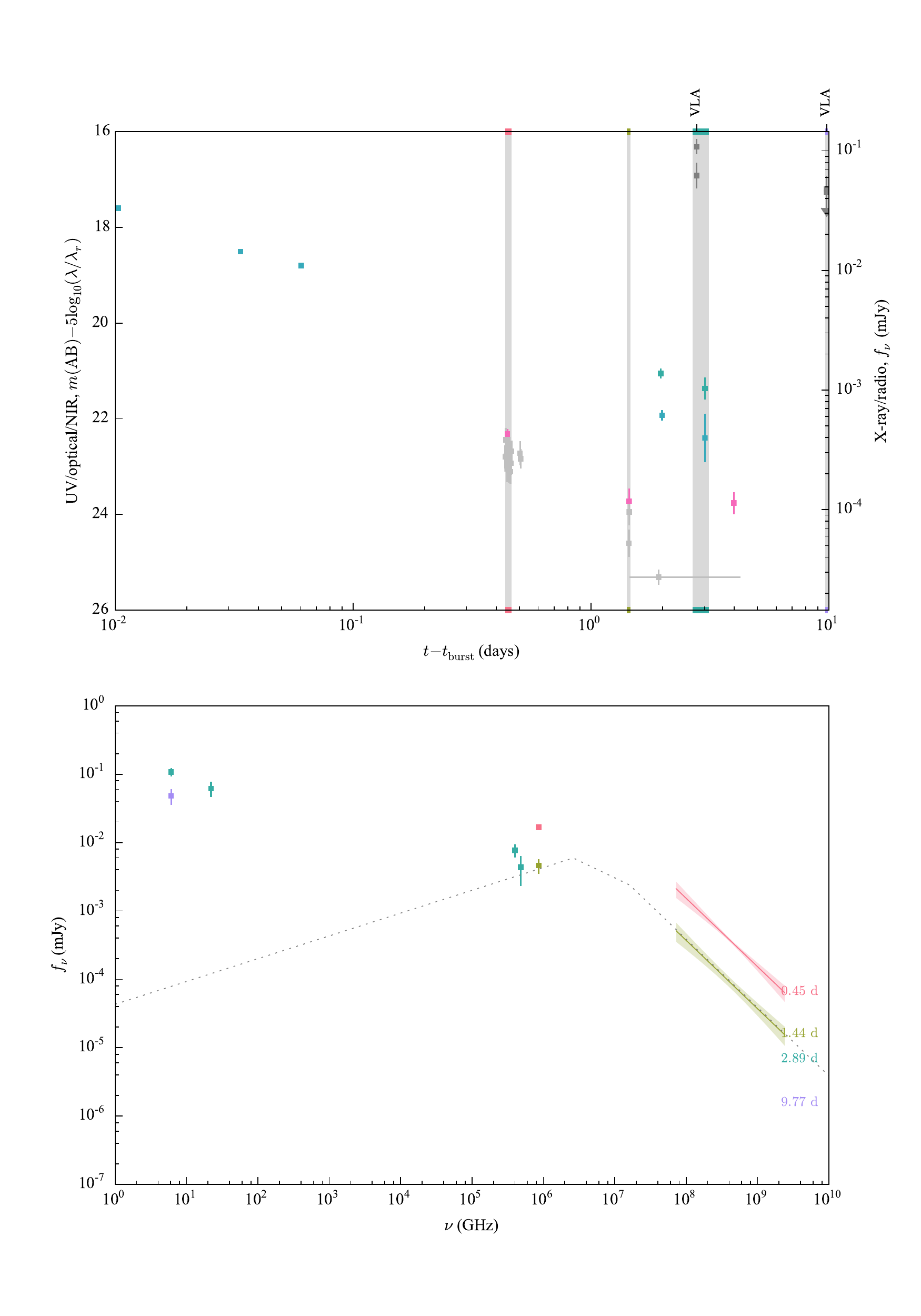}
    \caption{\label{fig:grb-140620A-sed}Light curve and \acs{SED} of \ac{GRB}~140620A~/~iPTF14cva.}
\end{figure*}

\begin{figure*}
    \centering
    \includegraphics[width=\textwidth]{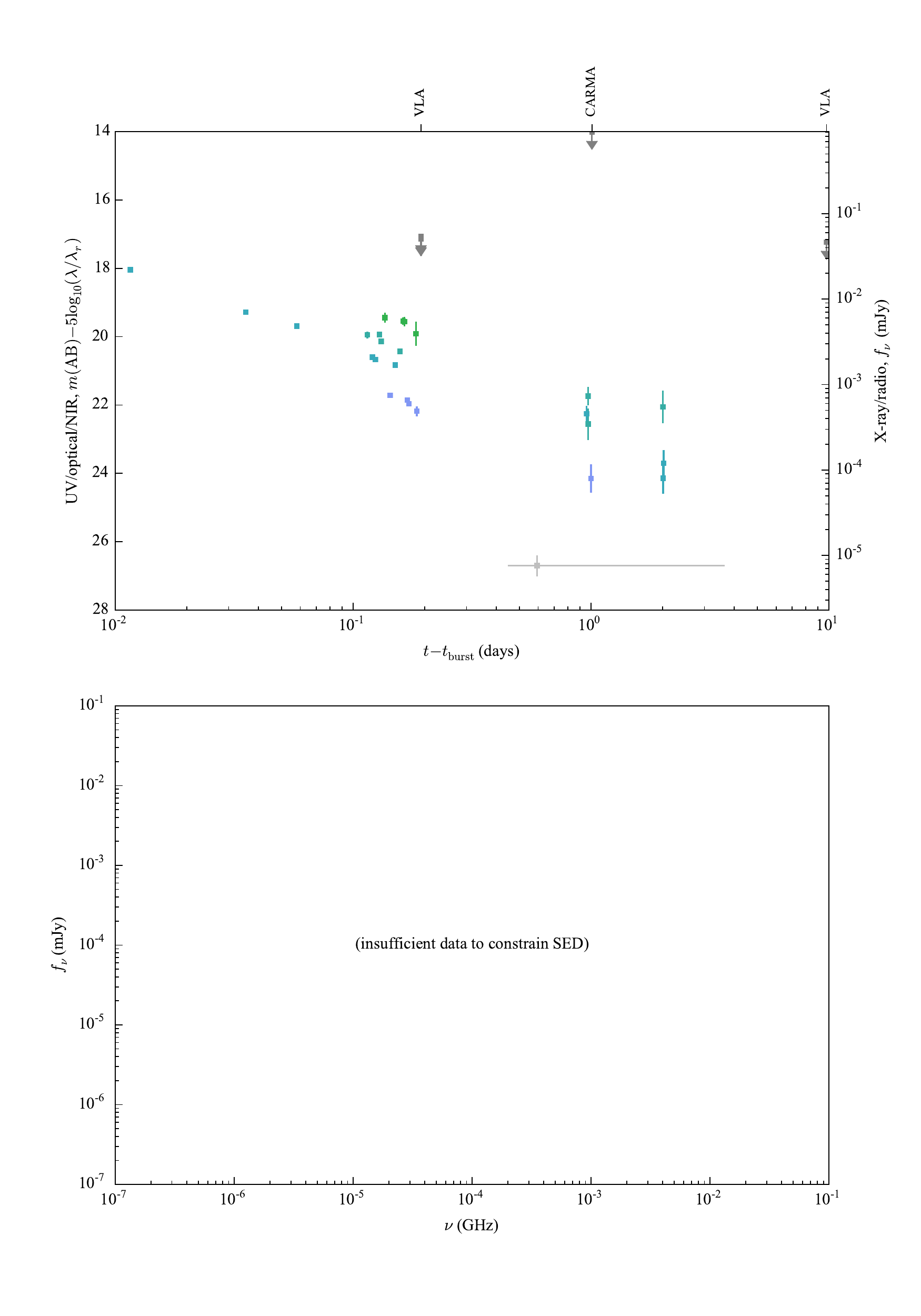}
    \caption{\label{fig:grb-140623A-sed}Light curve of \acs{GRB}~140623A~/~iPTF14cyb.}
\end{figure*}

\begin{figure*}
    \centering
    \includegraphics[width=\textwidth]{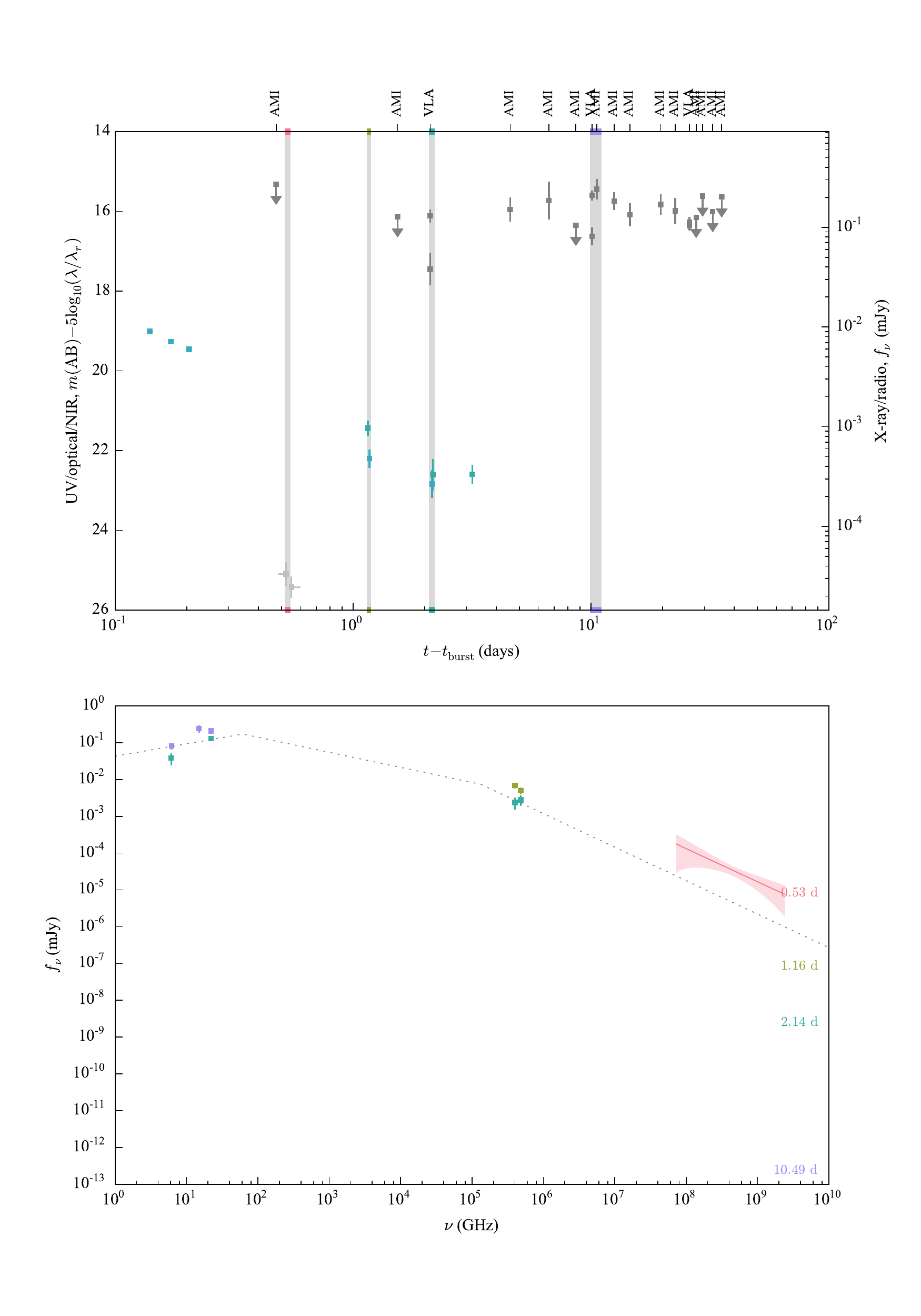}
    \caption{\label{fig:grb-140808A-sed}Light curve and \acs{SED} of \ac{GRB}~140808A~/~iPTF14eag.}
\end{figure*}

\chapter{Conclusion}

In this thesis, I have described two of the main components of the Advanced \ac{LIGO} real\nobreakdashes-time search for \acp{GW} from \ac{BNS} mergers. First, I described a novel filter design and detection pipeline that can handle long\nobreakdashes-duration inspiral waveforms while remaining computationally tractable and without sacrificing latency. Curiously, the unique multiband structure lends itself to producing detection candidates a few seconds before merger for most events, and tens to a hundred seconds before merger for exceptional events.

Second, I developed a rapid Bayesian sky localization algorithm that is a thousand times faster than the full \ac{MCMC} parameter estimation, while producing sky maps that are just as accurate. As a result of my contributions, the total response time of the whole online Advanced \ac{LIGO} analysis is now low enough that it will be possible to search for all of the predicted optical counterparts of \ac{BNS} mergers, including kilonovae, neutron-powered precursors, and afterglows.

I described in detail the detection and sky localization capabilities of the early two- and three-detector configurations of the global \ac{GW} detector network. With the computational advances described above, we were able to simulate several months of Advanced \ac{LIGO} observations and detect and localize thousands of \ac{BNS} mergers. We show how the interplay between priors and polarization break parameter degeneracies. This overturns an old misconception that sky localization requires three or more detectors. In the case of two detectors, we elucidate one surviving degeneracy between the island of probability containing the true location of the source, and its polar opposite location. We quantify the improvement in position accuracy that is gained when Virgo comes online with reduced sensitivity. The large sample size, the astrophysically motivated source population, the detector configurations based on the commissioning and observing schedule, and the use of Bayesian parameter estimation instead of the Fisher matrix or other approximations, make this work a realistic description of the position reconstruction capabilities of advanced \ac{GW} detectors. It is also constitutes a demonstration of a complete pipeline for rapid detection and localization of \ac{BNS} mergers in Advanced \ac{LIGO}.

Switching from \acp{GW} to photons, within \ac{IPTF} we have developed a unique wide-field \ac{TOO} capability with which we can now routinely pinpoint out of areas of $\sim$100~deg$^2$ the afterglows of \acp{GRB} detected by \emph{Fermi} \ac{GBM}. So far, this has been a relatively rich source of low redshift, low\nobreakdashes-luminosity \acp{GRB}, with spectroscopically detectable \acp{SN}. Two of these bursts fit more cleanly in the picture of mildly relativistic shock breakout, rather than the standard picture of an internal shock within an ultra\nobreakdashes-relativistic jet. On a technical level, this serves as a demonstration and a prototype for finding optical counterparts of \ac{GW} events with \ac{ZTF}.

\section{Next steps}

Activities over the next few years are clearly critical to maximizing the early scientific returns from Advanced \ac{LIGO} and \ac{ZTF}. I will list a few clear next steps for the Advanced \ac{LIGO} real\nobreakdashes-time \ac{BNS} search and the \ac{IPTF} \ac{TOO} program.

\subsection{Spin and \acs{NSBH} mergers}

For the reasons discussed in Section~\ref{sec:first2years-source-sensitivity}, I have ignored the effect of spin on \ac{BNS} merger waveforms. \acp{NS} in field binaries formed through isolated binary stellar evolution should have small spins (the most rapidly spinning pulsar known in a \ac{BNS} system having $\chi = |\boldsymbol{\chi}| = |\mathbf{S}| / m^2 \lesssim 0.05$; \citealt{2003Natur.426..531B}) that can be neglected for the purposes of detection and sky localization with Advanced \ac{LIGO} \citep{BerryLocalization}. However, in the dense environment of a globular cluster, a \ac{BNS} system with appreciable component spins could be formed through dynamical capture of a young or accretion-recycled \ac{NS} (up to the \ac{NS} breakup limit of $\chi \sim 0.7$; see references in \citealt{DetectingBNSSystemsWithSpin}). Furthermore, in \ac{NSBH} binaries, the spin of the \ac{BH} is expected to be an important effect because stellar\nobreakdashes-mass \acp{BH} are known to have large or nearly extremal spins ($\chi \sim 0.98$; see \citealt{MassAndSpinNSAndBHReview} and references therein). Effort is underway to incorporate spin effects into offline Advanced \ac{LIGO} \ac{CBC} searches using either non\nobreakdashes-precessing templates for spins that are aligned with the orbital angular momentum \citep{SBank,DetectingBNSSystemsWithSpin} or precessing templates for systems where one spin dominates \citep{PhysicalTemplateFamily,harry-single-spin}. It is known that failing to account for spin can result in a loss of \ac{SNR} \citep{DetectingBNSSystemsWithSpin} or can introduce biases in parameter estimation \citep{Raymond:2009}. Therefore, incorporating spin at some level into the online Advanced \ac{LIGO} detection pipeline may be essential to providing accurate rapid sky localizations for \ac{NSBH} merger events.

\subsection{Sub-threshold signals in rapid localization}

As we have noted, the condition in which the rapid localization is \emph{not} as accurate as the full parameter estimation is when one or more of the detectors is not represented by a trigger. This is especially an issue in the 2016 configuration, when Virgo is present but with a third of the sensitivity of the \ac{LIGO} detectors. In this situation, the signal will often be below threshold in Virgo and represented by triggers in only the Hanford and Livingston detectors. In these cases, using the extra information from the weak signal in Virgo can reduce the area by a factor of one third. One remedy would be to reduce or eliminate the single\nobreakdashes-detector threshold, such that times, amplitudes, and phases on arrival are available for every detector and every event. This would probably require redesigning the coincidence stage of the detection pipeline. A simpler and more useful approach might be to have the detection pipeline send to \ac{BAYESTAR} an excerpt from the \ac{SNR} time series of the best matching template, extending a small fraction of a second before and after the time of the event. Within \ac{BAYESTAR}, these small portions of the \ac{SNR} time series would be used in the likelihood instead of the template autocorrelation function. Mathematically, this trivial change would make the \ac{BAYESTAR} likelihood mathematically equivalent to what is used by the \ac{MCMC} analysis. This simple change ought to be implemented before 2016 when Virgo comes online.

\subsection{Distance-resolved rapid localizations}

Initial \ac{LIGO} \ac{EM} counterparts relied upon imaging nearby galaxies that were contained with the \ac{GW} localizations. Within the expanded detection volume of Advanced \ac{LIGO}, there are enough galaxies that this approach will be of little help. However, \citet{NissankeKasliwalEMCounterparts} have proposed to reduce the number of optical false positives by combining a galaxy catalog with the \emph{joint} posterior probability distribution of sky location \emph{and distance}. This could be an especially important technique in the early years of Advanced \ac{LIGO}, due to the initially coarse localizations of hundreds of deg$^2$. Currently, \ac{BAYESTAR} supplies flat, two\nobreakdashes-dimensional posterior distributions on the sky. This is simply because distance is treated as a marginal variable, and integrated away. There is no difficulty in calculating the distance posterior. For example, even without modifying the integration scheme one could simply run several instances of \ac{BAYESTAR} in parallel with different distance limits spanning a sequence of shells. The challenge in supplying the distance information to astronomers is simply one of data management: the sky maps already take up several hundred kilobytes per event, and encoding the distance information naively would make the data products cumbersome. Some small effort must go into picking an appropriate representation of the full three\nobreakdashes-dimensional probability distributions, and determining whether some two-plus-one dimensional distribution (for example, sky posterior plus the mode of the marginal distance posterior at every pixel) would be sufficient.

\subsection{\acp{GRB} beyond the \emph{Fermi} bandpass}

Still within the context of \acp{GRB}, there are several other possible applications of the \ac{IPTF} \ac{TOO} capability. At even higher energy scales than are probed by gamma\nobreakdashes-ray satellites, the \acf{HAWC} recently began normal observations, at any instant monitoring about 15\% of the sky for air showers induced by gamma\nobreakdashes-rays and cosmic rays with energies of $\sim$30~GeV\nobreakdashes--100~TeV \citet{HAWCFirstResults}. \ac{HAWC} is likely to detect 50--500~GeV emission for one or two \acp{GRB} per year \citep{HAWC-GRB-1,HAWC-GRB-2}, with a typical localization uncertainty radius of $\lesssim 1\arcdeg$ (Ukwatta, private communication). Although this is too large for conventional optical follow\nobreakdashes-up, it is well within the reach of \ac{IPTF}, the \acf{RAPTOR} network, or even tiled \emph{Swift} \ac{XRT} observations. We have submitted a \emph{Swift} Guest Investigator proposal for X\nobreakdashes-ray follow\nobreakdashes-up of \ac{HAWC} \acp{GRB} (P.I.: Ukwatta, LANL).

\subsection{\emph{Fermi} \acs{GBM} and \acs{IPTF} as a short \acs{GRB} factory}

As we noted in Chapter~\ref{chap:iptf-gbm}, \emph{Fermi} \ac{GBM} detects a large number of short bursts, about with perhaps 40 out of a total of 240 \acp{GRB}~year$^{-1}$ likely to be associated with \ac{CBC} events \citep{GBMLocalization}. The sample of short \ac{GRB} optical afterglows is still so small that even a few of these \emph{Fermi} bursts per year would be a significant contribution. Over the past year, with \ac{IPTF} we have followed up a handful of \emph{Fermi} short \acp{GRB}, but did not detect their afterglows. In these few cases, the optical limits were not constraining because the \ac{GBM} localizations were only observable from Palomar only after a significant fraction of a day from the trigger. This is just due to chance, but we do expect it to be more difficult to find afterglows of short bursts than long bursts for two reasons. First, short \acp{GRB} are typically less well localized, because they have lower photon fluences. Second, their afterglows are about 6~mag fainter in an absolute sense, and the brightest short \acp{GRB} afterglows have apparent magnitudes that are comparable with the faintest long \acp{GRB} \citep{KannTypeITypeIIOpticalAfterglows}.

To address the first issue, we have increased the standard \ac{P48} \ac{TOO} tiling from 10 to 20 fields. The larger \ac{FOV} of the \ac{ZTF} camera (see Figure~\ref{fig:ztf-camera}) will also be a help here. For the second issue, we need deeper \ac{P48} observations. Since the \ac{P48} does not have a guide camera, the longest useful exposure time is about 60~s; we need to stack images. The real\nobreakdashes-time transient pipeline now has an automatic co\nobreakdashes-add stage prior to the image subtraction, but we need to modify the standard \ac{TOO} program to take advantage of this. Due to the relatively long $\sim$46~s overhead between exposures, for short \acp{GRB} we should probably tile the \ac{GBM} localization several times without pauses between epochs. \ac{ZTF}, on the other hand, will be able to reach deeper limits faster; current plans call for the \ac{ZTF} transient survey to adopt a standard exposure duration of 30~s \citep{ZTFBellm}, but variable exposure times of 10~min or more have been considered in the context of following up \ac{LIGO} events \citep{KasliwalTwoDetectors}. Furthermore, with faster readout electronics and a reduced overhead of $\sim$15~s between exposures, \ac{ZTF} will be able to tile the localizations faster or reach deeper limits with stacked images.

In the context of \ac{IPTF} co\nobreakdashes-adding consecutive two or three consecutive \ac{P48} images will modestly increase the number of candidates that need to be sorted. We will need to automate some of the vetting procedures that are currently done by humans. We will also need to speed up the human vetting by streamlining how candidates are presented. These modest improvements will be an important step in the evolution of the \ac{TOO} program toward a depth that is relevant for Advanced \ac{LIGO} and achievable with \ac{ZTF}.

\section{Future directions}

In this section, I list some future directions for related research. I will first discuss two applications related to early-warning detection and sky localization. Then, I will end with some thoughts on \acp{TOO} with \ac{ZTF}.

\subsection{Early warning and dynamically tuned squeezing}

The \ac{LIGO} detectors are designed to be sensitive across a broad range of frequencies, but there exist several techniques that can improve the sensitivity across a narrow range of frequencies while sacrificing sensitivity at other frequencies. Since \ac{CBC} signals are chirps, if we coupled a time-domain filtering scheme like \ac{LLOID} to the control of the detector, then we could dynamically tune the narrowband sensitivity to track the frequency evolution of the signal. \citet{PhysRevD.47.2184} proposed dynamically detuning the signal recycling cavity, and \citet{PhysRevD.90.102003} showed with a detailed time-domain analysis that this scheme could result in \ac{SNR} gains of a factor of 17. A disadvantage of this proposal is that the detector has a nontrivial transient response to changes in length of the signal recycling cavity.

An approach without this difficulty could use squeezed vacuum states of light. Squeezed states of light have less amplitude variation but more phase variation, or vice-versa, than a minimum-uncertainty coherent state. By injecting squeezed vacuum states into the dark port of the interferometer, the noise can be reduced below the standard quantum limit in a narrow frequency band that is determined by the squeezing angle. This has recently been demonstrated on the GEO\,600 \citep{GEOSqueezing} and \ac{LIGO} Hanford detectors \citep{LIGOSqueezing}. With an optimally frequency-dependent squeezing angle, it would be possible to beat the standard quantum limit across a broad range of frequencies. Unfortunately, frequency-dependent squeezing has an added cost in terms of infrastructure because current proposals involve external 100\nobreakdashes--1000~m filter cavities \citep{FDSqueezing}. An alternative is to use the already\nobreakdashes-demonstrated frequency\nobreakdashes-independent squeezing, but dynamically tune the squeezing angle so that the narrowband sensitivity tracks the frequency evolution of a \ac{GW} chirp. This would be simpler in some ways than tuning the signal recycling cavity because the detector would have no transient response to the change in squeezing angle.

\begin{figure*}[t]
    \centering
    \includegraphics{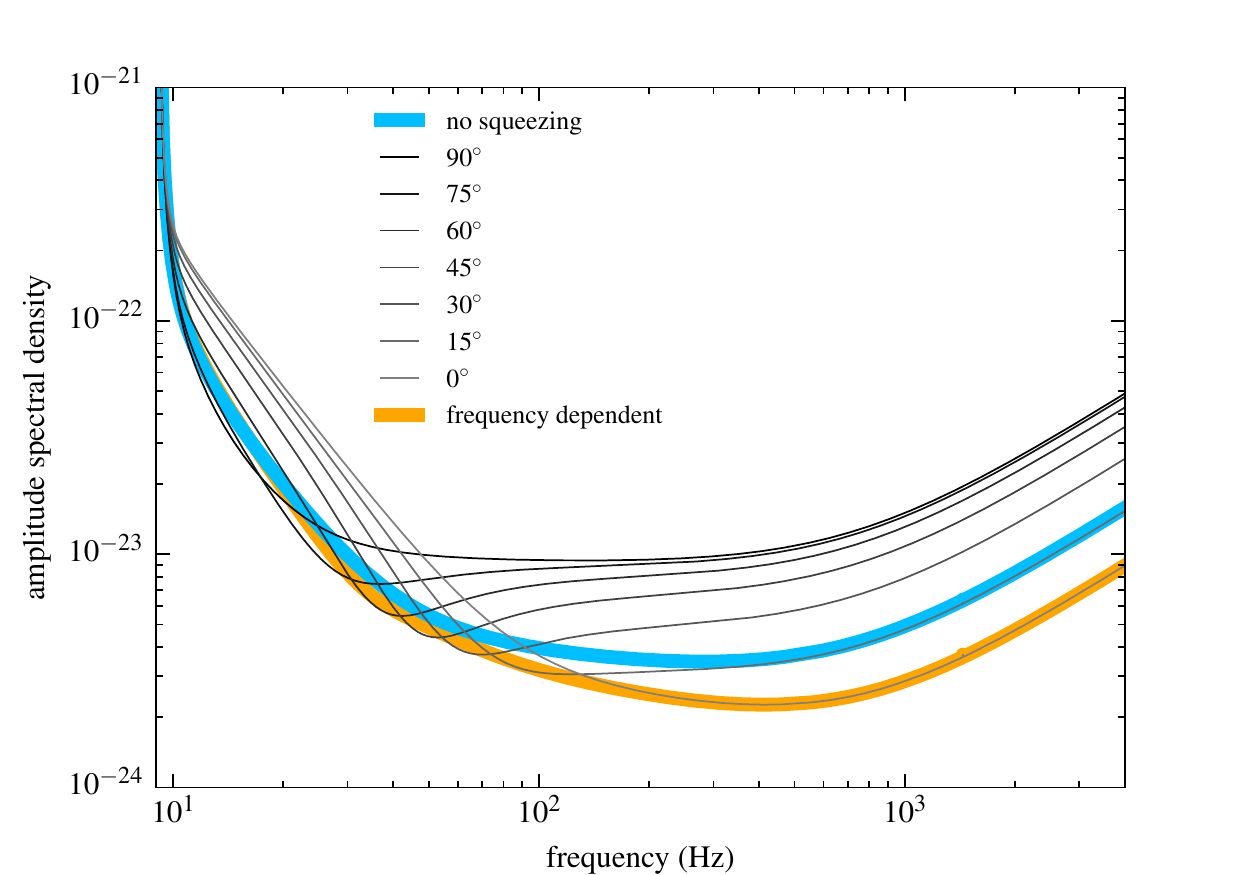}
    \caption[Sensitivity curves for various squeezing configurations]{\label{fig:squeezeasd}Sensitivity curves for various squeezing configurations. The blue line represents the baseline Advanced \ac{LIGO} design with no squeezing. The black and gray lines show configurations with 11~dB of squeezing injected with 15\% losses, with squeezing angles varying from 90\arcdeg to 0\arcdeg in 15\arcdeg increments. The orange line shows a frequency-dependent squeezing configuration obtained by sending the squeezed light through a filter cavity (length: 100~m, input mirror transmission: $1.85\times10^{-4}$, end mirror transmission: $3\times10^{-6}$, round trip losses: $1.5\times10^{-5}$).}
\end{figure*}

Noise \acp{PSD} for Advanced \ac{LIGO} with no squeezing, frequency\nobreakdashes-independent squeezing, and frequency\nobreakdashes-dependent squeezing are shown in Figure~\ref{fig:squeezeasd}. The filter cavity design with the fiducial parameters in the caption would increase the range for \ac{BNS} mergers by 24\%, almost doubling the detection rate. A dynamically tuned squeezing scheme would increase the \ac{BNS} range by 32\%, slightly better because the filter cavity itself has optical losses. The improvement in sensitivity would also enhance the estimation of all of the \ac{GW} parameters \citep{SqueezingParameterEstimation}. In both scenarios, much of the improved sensitivity relative to no squeezing is at high frequencies, with the amplitude spectral density reduced by a factor of $\approx 0.56$ at $f \gtrsim 500$~Hz. The \ac{GW} signal from a \ac{BNS} merger starts to be affected by tidal effects, deviating from the inspiral of point particles, at $f \sim$500\nobreakdashes--1000~Hz. Therefore, the injection of squeezed vacuum states could make it possible to detect tidal effects and constrain the \ac{NS} equation of state with Advanced \ac{LIGO}.

For dynamically tuned squeezing, it remains to be shown is that the binary parameters can be estimated with sufficient accuracy from the early inspiral to track the frequency evolution through the late inspiral. One must consider in what configuration to hold the detector while waiting for the start of a chirp, and at what threshold to activate squeezing or to start varying the squeezing angle. A first step would be to study the costs and benefits of dynamically tuned squeezing using Fisher matrix considerations similar to Chapter~\ref{chap:fisher} and Section~\ref{sec:prospects-detection}. For a practical implementation, one would have to think about whether the squeezing is activated based on an a coincidence of early-warning triggers from all of the sites, and whether feedback on the squeezing angle arises from the data from a single detector, from the network, or some intermediate combination. One would also need take into account that there will be some delay, due to technical sources of latency listed in Chapter~\ref{chap:detection}, between the detection pipeline and actuation on the squeezer.

\subsection{\acs{LIGO} as a short \acs{GRB} early-warning system}

In Section~\ref{sec:prospects-detection}, we briefly described using an early\nobreakdashes-warning \ac{GW} detection to position \emph{Swift} \ac{BAT} in advance of the merger. This would require a radically new, fully autonomous \ac{TOO} mode, and the rate of suitable early\nobreakdashes-warning detections would be constrained to about 1\% of \ac{BNS} events due to \emph{Swift}'s $\approx80$~s slew time\footnote{\url{http://swift.gsfc.nasa.gov/proposals/tech_appd/swiftta_v10/node15.html}}. Extremely low latency triggers might be more suitable for an instrument such as ISS-Lobster \citep{ISSLobster}, a proposed X\nobreakdashes-ray imager on the International Space Station. ISS-Lobster would have a sensitivity of $1.3 \times 10^{-11}$~erg\,cm$^{-2}$\,$s^{-1}$ over a band similar to \emph{Swift} \ac{XRT}, 0.3\nobreakdashes--5~keV. This would be sufficient to detect a short \ac{GRB} X\nobreakdashes-ray afterglow at Advanced \ac{LIGO} distances. It is envisioned as having an instantaneous \ac{FOV} of 820~deg$^2$ and a slew time of $\approx 25$~s (for a $60\arcdeg$ slew, assuming a slew rate of 4\arcdeg~s$^{-1}$ and acceleration and settling times of 5~s; Camp, personal communication). Although still only a few percent of \ac{BNS} events will be detectable 25~s before merger, for about 20\% of events it could be possible to have Lobster on settled on the \ac{GW} localization within a 15 seconds after merger\footnote{Subject, of course, to the position of the ISS and any structural constraints on pointing.}, especially if the initial pre-merger position estimate is updated with a refined localization every few seconds. Even using ordinary post\nobreakdashes-merger detections, with modest improvements to \ac{GW} data handling one could begin soft X\nobreakdashes-ray observations within 25~s after merger, earlier than is usually possible with \emph{Swift} \ac{XRT}.

\subsection{Optical counterpart search with \acs{ZTF}}

The prime sources for Advanced \ac{LIGO} and Virgo are \ac{BNS} mergers, which are also thought to be the progenitors of short \acp{GRB} \citep{1986ApJ...308L..43P,1989Natur.340..126E,1992ApJ...395L..83N,2011ApJ...732L...6R}. As we have noted, short bursts and their afterglows are much fainter than long bursts. However, the redshifts of short \acp{GRB} that we might find in connection with \ac{GW} detections would be limited to Advanced \ac{LIGO}'s range. In Figure~\ref{fig:sgrb-kcorrected}, we show Kann's complete sample of short \ac{GRB} afterglows with known redshifts and \citet{SyntheticSGRBAfterglows}'s synthetic afterglow models, both shifted to Advanced \ac{LIGO}'s range at final design sensitivity of 200~Mpc or $z = 0.045$. By comparison with Figure~\ref{fig:lightcurve-zoo}, we see that short \ac{GRB} afterglows at this distance will be just as bright as a typical long \ac{GRB} afterglow at cosmological distances.

\begin{figure}
    \centering
    \includegraphics[width=\textwidth]{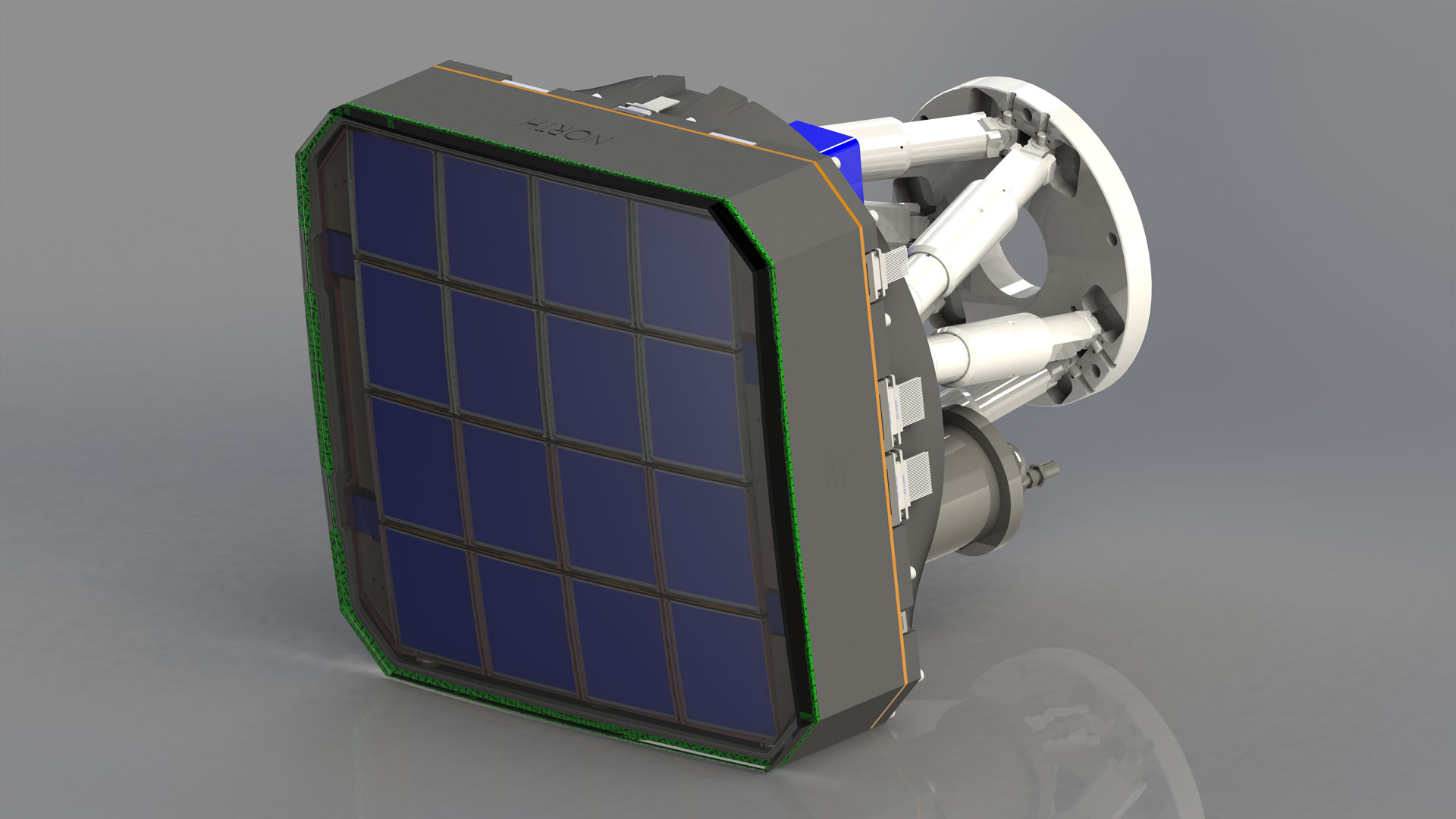}
    \caption[The \acs{ZTF} camera]{\label{fig:ztf-camera}The \ac{ZTF} camera. Rendering reproduced from E.~Bellm (private communication).}
\end{figure}

An added challenge is that \ac{GW} localizations of \ac{BNS} mergers can be more uncertain than \emph{Fermi} \ac{GBM} bursts. In the early (2015) two\nobreakdashes-detector network configuration, we expect probability maps that consist of multimodal arcs spanning up to 600~deg$^2$ \citep{KasliwalTwoDetectors,FirstTwoYears}, though with the addition of Virgo in 2016 these shrink to 200~deg$^2$ \citep{LIGOObservingScenarios}. The typical areas should reach $\lesssim 10$~deg$^2$ toward the end of the decade as the detectors approach final design sensitivity and as additional planned \ac{GW} facilities come online (LIGO\nobreakdashes--India and KAGRA; see \citealt{ShutzThreeFiguresOfMerit,Veitch:2012,FairhurstLIGOIndia,NissankeKasliwalEMCounterparts,LIGOObservingScenarios}). Since we have shown that our \ac{IPTF} afterglow search has exactly the detection efficiency that is predicted by the areas that we search and the afterglow luminosity distribution, we expect that even the earliest Advanced \ac{LIGO} localizations would present no undue difficulties for \ac{ZTF} when we consider its 15\nobreakdashes-fold increase in areal survey rate as compared to \ac{IPTF}.

\begin{figure}
    \includegraphics[width=\textwidth]{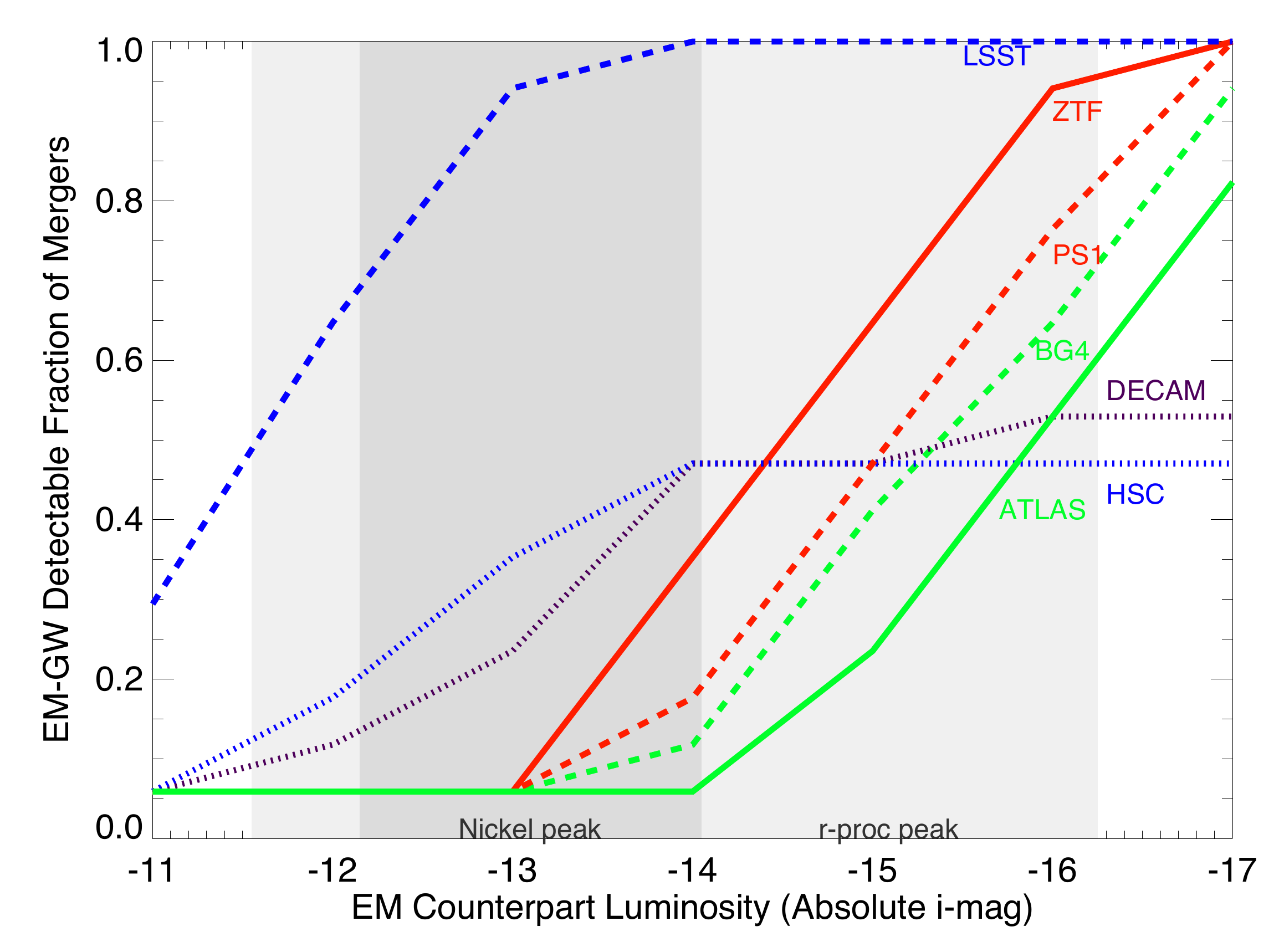}
    \caption[Kilonova phase space accessible with various optical instruments]{\label{fig:kilonova-ztf-reach}Range of kilonova absolute magnitudes detectable with various optical instruments within two\nobreakdashes-detector, HL type, \ac{GW} localization areas of hundreds of deg$^2$. The instruments shown include \ac{LSST}, \ac{ZTF}, Pan\nobreakdashes-STARRS (PS1), BlackGEM (BG4), \ac{DECAM}, \ac{HSC}, and the \acf{ATLAS}. Reproduced with permission from \citet{KasliwalTwoDetectors}.}
\end{figure}

Though \emph{Swift} may detect a \ac{GRB} in coincidence with a \ac{LIGO} trigger, a \emph{Fermi} \ac{GBM} or \ac{IPN} detection is much more likely on the basis of sky coverage. In these cases, efforts such as ours may still be necessary to pinpoint the event and initiate deeper follow\nobreakdashes-up.

Although afterglows are the brightest and most distinctive expected optical counterparts of \ac{BNS} merger, they will probably also be quite rare considering that, like the \acp{GRB} themselves, they are tightly collimated within a jet with half\nobreakdashes-opening angle $\theta_j$. If short \acp{GRB} have typical values of $\theta_j=3$\nobreakdashes--8\arcdeg\ as is suggested by recent jet break observations \citep{JetBreakShortGRB111020A}, then only a few percent of \ac{BNS} events could be accompanied by \acp{GRB}.

\citet{MostPromisingEMCounterpart} point out that ``kilonovae'', radioactively powered transients sourced in the hot $r$\nobreakdashes-process ejecta, may be a more promising counterpart because they are not expected to be beamed. When one considers realistic opacities, these signatures can be faint and very red, rising to only $R = 21$\nobreakdashes--25~mag but peaking in the infrared \citep{KilonovaHighOpacities}. These would be challenging but sometimes possible to capture with \ac{ZTF} using stacked exposures of 600~s or longer \citep{KasliwalTwoDetectors}. The ``kilonova precursor'' powered by free neutrons in the fast\nobreakdashes-moving outer ejecta \citep{KilonovaPrecursor} may be a better prospect for \ac{ZTF}: this signature could be much bluer and might rise as high as $R \approx 22$~mag in a matter of hours after the merger. A fast, bright, blue peak could also arise if the remnant persists as a hypermassive \ac{NS} for $\gtrsim 100$~ms, supplying neutrinos to the ejecta and keeping the electron fraction too high to form the high\nobreakdashes-opacity Lanthanide elements \citep{KilonovaRedOrBlue}. We have plotted some representative kilonova precursor models in Figure~\ref{fig:sgrb-kcorrected}.

Contaminants for a kilonova search will be numerous due to the required survey depth, and may be different in nature from those which we discussed in Section~\ref{sec:afterglow-search-method}. To deal with them, we will need to further automate the candidate vetting process and develop new target selection criteria. On the other hand, a search designed for kilonova precursors may be similar in many ways to our present afterglow effort, aside from using much longer and deeper exposures.

As scheduled, the first two years of Advanced \ac{LIGO} will roughly coincide with the last two years of \ac{IPTF}. The first detections could occur as early as 2016, but will have (median 90\% confidence) positions that are uncertain by 200--600~deg$^2$, depending on when Advanced Virgo comes online and with what sensitivity. These localizations could take an hour or more to tile with \ac{IPTF}, leaving little time for deep co\nobreakdashes-added observations. Given this constraint, except in the case of an exceptionally nearby and well\nobreakdashes-localized \ac{BNS} merger, \ac{IPTF} will be sensitive to on\nobreakdashes-axis afterglows but probably not kilonovae.

However, \ac{ZTF}'s fast readout and much expanded \ac{FOV} will enable much more rapid tiling; its guide camera will enable deep exposures. \ac{ZTF} is expected to see first light in early 2017 and be fully commissioned by mid 2017 to early 2018. Considering also the expected improvements in the \ac{GW} sensitivity, \ac{ZTF} should be able to tile the \ac{GW} localizations in a handful of pointings. Given present theoretical predictions, our \ac{ZTF} follow\nobreakdashes-up strategy should be optimized for searching for kilonovae, without sacrificing early\nobreakdashes-time observations that could capture an optical afterglow. A mixed depth approach involving rapidly tiling with one or two epochs of 30~s exposures and then revisiting with one\nobreakdashes-hour exposures should be sufficient to capture almost any coincident on\nobreakdashes-axis afterglow, the fiducial kilonova precursor model, and a significant fraction of the predicted kilonova phase space (see Figure~\ref{fig:kilonova-ztf-reach}).

\section{Conclusion}

The next few years should be an exciting time, seeing both the dawn of \ac{GW} transient astronomy and the maturing of synoptic optical transient surveys. Advanced \ac{LIGO} should begin observing by late 2015, with Advanced Virgo following soon thereafter. The first direct detections of \ac{GW} transients seem likely over the next few years. As a result of this thesis, we will be able to deliver \ac{GW} detection candidates and accurate sky localizations promptly enough to look for all of the predicted optical counterparts of \ac{BNS} mergers, including afterglows and kilonovae. Given current predictions for the \ac{EM} signatures, some \ac{LIGO} sources should be detectable by present and planned synoptic optical surveys, including \ac{PTF} and \ac{ZTF}, operating in \ac{TOO} mode. The \ac{GBM}\nobreakdashes--\ac{IPTF} afterglow search serves as a prototype that must now rapidly evolve to fulfill its role in the search for optical signatures of \ac{BNS} mergers.

\appendix
\chapter{Computer codes}

\section{Astropy}
\label{sec:code-astropy}

All chapters of this thesis made use of Astropy (\citealt{astropy}; \url{http://www.astropy.org}), a community-developed core Python package for Astronomy. The author's contributions to Astropy included a numerical code for inverting \ac{WCS} transformations \citep{WCS1, WCS2, WCS3}, and an implementation of the image distortion corrections used in \ac{PTF} images.

\section{GSTLAL}
\label{sec:code-gstlal}

Source code for \textls{GSTLAL} is at \url{http://www.lsc-group.phys.uwm.edu/cgit/gstlal/tree}.

\section{GStreamer}
\label{sec:code-gstreamer}

GStreamer is used in Chapter~\ref{chap:detection}. Its project page is at \url{http://gstreamer.freedesktop.org}. It is widely installed on nearly all Linux and Unix desktop configurations. The author's contributions to GStreamer include enhancements to the upsampling/downsampling element \texttt{audioresample} to support rapidly skipping over gaps.

\section{HEALPix}
\label{sec:code-healpix}

Some of the results in this paper have been derived using \ac{HEALPix} (\citealt{HEALPix}; \url{http://healpix.sourceforge.net}). The author's contributions to \ac{HEALPix} include build infrastructure enhancements to support deployment on \ac{LIGO} computing clusters and improve the availability of OpenMP\nobreakdashes-accelerated \ac{HEALPix} routines in Python on the Macintosh platform.

\section{LALSuite}
\label{sec:code-lalsuite}

\ac{BAYESTAR} is part of the \textls{LALINFERENCE} parameter estimation library, which is in turn part of the \ac{LIGO} Algorithm Library Suite (\url{http://www.lsc-group.phys.uwm.edu/cgit/lalsuite/tree}). \ac{BAYESTAR}'s C language source code is in \texttt{lalinference/src/bayestar\_sky\_map.c}. Python bindings and high\nobreakdashes-level driver codes are in the directories \texttt{lalinference/python} and \\ \texttt{lalinference/python/lalinference/bayestar}.

\section{Code listing: sky resolution from Fisher matrices}
\label{sec:fisher-matrix-code-listing}

The following Python listing calculates the coherent Fisher matrix described in Section~\ref{sec:fisher-matrix-area-outline}.

\begin{lstlisting}[language=Python,firstline=2,lastline=156]
"""
Fisher matrix calculation of detector resolution
Copyright (C) 2014  Leo Singer <leo.singer@ligo.org>
"""
from __future__ import division
import collections
import numpy as np
import scipy.optimize
import scipy.stats
import scipy.linalg


SkyLocalizationAccuracyBase = collections.namedtuple(
    'SkyLocalizationAccuracyBase',
    ('coherent_fisher_matrix', 'timing_fisher_matrix', 'snrs'))


def exp_i(phi):
    return np.cos(phi) + np.sin(phi) * 1j


class SkyLocalizationAccuracy(SkyLocalizationAccuracyBase):

    @staticmethod
    def _h(psi, u, r):
        """Polarization for phic=0 (arbitrary?)"""
        phic = 0
        cos2psi = np.cos(2*psi)
        sin2psi = np.sin(2*psi)
        f = 0.5 * (1 + np.square(u))
        pre = exp_i(2*phic) / r
        hp = pre * (f * cos2psi + 1j * u * sin2psi)
        hc = pre * (f * sin2psi - 1j * u * cos2psi)
        return hp, hc

    @staticmethod
    def _rot(phi, theta):
        """Active transformation from radiation to Earth frame."""
        c1 = np.cos(phi)
        c2 = np.cos(theta)
        s1 = np.sin(phi)
        s2 = np.sin(theta)
        return np.asarray(
            [[c1*c2, -s1, c1*s2],
             [c2*s1, c1, s1*s2],
             [-s2, 0, c2]])

    @staticmethod
    def _rescale(r, locations, r1s, w1s, w2s):
        """Rescale distance and time units."""
        wbar = np.sqrt(np.mean(w2s))
        rbar = scipy.stats.gmean(np.concatenate(([r], r1s)))
        return (
            r / rbar,
            locations * wbar,
            r1s / rbar,
            w1s / wbar,
            w2s / np.square(wbar))

    @staticmethod
    def _term(hp, hc, R, response, location, r1, w1, w2):
        # Transformed detector response tensor and position
        D = r1 * np.dot(np.dot(R.T, response), R)
        d = np.dot(R.T, location)

        # Amplitude
        z = hp*(D[0, 0] - D[1, 1]) + 2 * hc*D[0, 1]
        a = np.real(z)
        b = np.imag(z)
        snr2 = np.square(a) + np.square(b)

        # Jacobian matrix
        dz_dth = -2 * hp * D[0, 2] - 2 * hc * D[1, 2]
        dz_dph = -2 * hc * D[0, 2] + 2 * hp * D[1, 2]
        dz_drehp = D[0, 0] - D[1, 1]
        dz_drehc = 2 * D[0, 1]
        J = np.asarray([
             [np.real(dz_dth), np.real(dz_dph), dz_drehp, 0, dz_drehc, 0, 0],
             [np.imag(dz_dth), np.imag(dz_dph), 0, dz_drehp, 0, dz_drehc, 0],
             [-d[0], -d[1], 0, 0, 0, 0, 1]])

        I = np.asarray([
             [1, 0, w1*b],
             [0, 1, -w1*a],
             [w1*b, -w1*a, snr2*w2]])
        coherent_term = np.dot(np.dot(J.T, I), J)

        J = np.asarray([[-d[0], -d[1], 1]])
        I = snr2 * (w2 - np.square(w1))
        timing_term = np.dot(J.T, J) * I

        return coherent_term, timing_term, np.sqrt(snr2)

    @staticmethod
    def marginal_information(I, n):
        """Given an information matrix I, find the information matrix for the
        marginal distribution of just the first n parameters by using
        partitioned matrix inversion (the Schur complement)."""
        A = I[:n, :n]
        B = I[:n, n:]
        C = I[n:, n:]
        return A - np.dot(B, scipy.linalg.solve(C, B.T, sym_pos=True))

    def __new__(
            cls, phi, theta, psi, u, r, responses, locations, r1s, w1s, w2s):

        r, locations, r1s, w1s, w2s = cls._rescale(r, locations, r1s, w1s, w2s)
        hp, hc = cls._h(psi, u, r)
        R = cls._rot(phi, theta)

        # Loop over detectors
        coherent_terms, timing_terms, snrs = zip(*[
            cls._term(hp, hc, R, response, location, r1, w1, w2)
            for response, location, r1, w1, w2
            in zip(responses, locations, r1s, w1s, w2s)])

        I_coherent = cls.marginal_information(
            np.sum(coherent_terms, axis=0), 2)
        I_timing = cls.marginal_information(
            np.sum(timing_terms, axis=0), 2)
        snrs = tuple(snrs)

        # Done!
        return super(SkyLocalizationAccuracy, cls).__new__(
            cls, I_coherent, I_timing, snrs)

    @staticmethod
    def _area(I, quantile):
        return -2*np.log(1-quantile)*180*180/np.pi / np.sqrt(np.linalg.det(I))

    @staticmethod
    def _width(I, quantile):
        ppf = scipy.stats.halfnorm.ppf(quantile)
        return 2 * ppf * 180 / np.pi / np.sqrt(np.max(np.linalg.eigvalsh(I)))

    def coherent_area(self, quantile=0.9):
        if len(self.snrs) <= 2:
            raise ValueError(
                'Networks of <= 2 detectors are always degenerate. ' +
                'Use coherent_width instead to get the annulus width.')
        return self._area(self.coherent_fisher_matrix, quantile)

    def timing_area(self, quantile=0.9):
        if len(self.snrs) <= 2:
            raise ValueError(
                'Networks of <= 2 detectors are always degenerate. ' +
                'Use timing_width instead to get the annulus width.')
        return self._area(self.timing_fisher_matrix, quantile)

    def coherent_width(self, quantile=0.9):
        return self._width(self.coherent_fisher_matrix, quantile)

    def timing_width(self, quantile=0.9):
        return self._width(self.timing_fisher_matrix, quantile)
\end{lstlisting}

\chapter{\label{sec:low-frequency-cutoff}Low frequency cutoff for inspiral searches}

\attribution{This Appendix is reproduced in part from \citet{Cannon:2011vi}, \textcopyright{}~2012 The American Astronomical Society.}

Ground-based \ac{GW} detectors are unavoidably affected at low frequencies by seismic and anthropogenic ground motion. The \ac{LIGO} test masses are suspended from multiple-stage pendula, which attenuate ground motion down to the pole frequency. In the detector configuration in place during \ac{S6}, seismic noise dominated the instrumental background below about 40~Hz. Considerable effort is being invested in improving seismic attenuation in Advanced \ac{LIGO} using active and passive isolation~\citep{0264-9381-27-8-084006}, so that suspension thermal noise will dominate around 10--15~Hz. Inspiral waveforms are chirps of infinite duration, but since an interferometric detector's noise diverges at this so-called ``seismic wall,'' templates for matched filter searches are truncated at a low-frequency cutoff $f_\mathrm{low}$ in order to save computational overhead with negligible loss of \ac{SNR}.

The expected matched-filter \ac{SNR}, integrated from $f_\mathrm{low}$ to $f_\mathrm{high}$, is given by Equation~(\ref{eq:expected-snr}). The high-frequency cutoff for the inspiral is frequently taken to be the \ac{GW} frequency at the \ac{LSO}; for non-spinning systems, $f_\mathrm{LSO} = 4400 (\Msun / M)$~Hz, where $M$ is the total mass of the binary \citep[Section 3.4.1 of][]{livrev12}. The choice of $f_\mathrm{low}$ is based on the fraction of the total SNR that is accumulated at frequencies above $f_\mathrm{low}$. To illustrate the relative contributions to the \ac{SNR} at different frequencies for a (1.4,~1.4)~$\Msun$ binary, we normalized and plotted the integrand of Equation~(\ref{eq:expected-snr}), the noise-weighted power spectral density of the inspiral waveform, in Figure~\ref{fig:low-frequency-cutoff}(b). This is the quantity
$$
	\frac{1}{\rho^2}\frac{\mathrm{d}\rho^2}{\mathrm{d}f} = \frac{f^{-7/3}}{S(f)} \left( \int_0^{f_\mathrm{LSO}} \frac{{f'}^{-7/3}}{S(f')} \, \mathrm{d}f' \right)^{-1},
$$
which is normalized by the total \ac{SNR} squared in order to put detectors with different absolute sensitivities on the same footing.  We used several different noise power spectra: all of the envisioned Advanced \ac{LIGO} configurations from~\citet{ALIGONoise}; the best-achieved sensitivity at \ac{LHO} in \ac{S5}, measured by \citet{S5InspiralRange}; and the best-achieved sensitivity at \ac{LHO} during \ac{S6}, measured by \citet{S6InspiralRange}.  (The noise spectra themselves are shown in Figure~\ref{fig:low-frequency-cutoff}(a).)  It is plain that during \ac{S5} and \ac{S6} the greatest contribution to the \ac{SNR} was between 100 and 150~Hz, but for all of the proposed \ac{LIGO} configurations the bulk of the \ac{SNR} is accumulated below 60~Hz.  This information is presented in a complementary way in Figure~\ref{fig:low-frequency-cutoff}(c), as the square root of the cumulative integral from $f_\mathrm{low}$ to $f_\mathrm{LSO}$, interpreted as a fraction of the total ``available'' \ac{SNR},
$$
	\rho_\mathrm{frac}(f_\mathrm{low}) = \sqrt{\left( \int_{f_\mathrm{low}}^{f_\mathrm{LSO}} \frac{{f}^{-7/3}}{S(f)} \, \mathrm{d}f \right) \left( \int_0^{f_\mathrm{LSO}} \frac{{f}^{-7/3}}{S(f)} \, \mathrm{d}f \right)^{-1}}.
$$
Table~\ref{table:accum_snr} shows the fractional accumulated \ac{SNR} for four selected low-frequency cutoffs, 40~Hz, 30~Hz, 20~Hz, and 10~Hz.  In \ac{S5} and \ac{S6}, all of the \ac{SNR} is accumulated above 40~Hz.  For the `high frequency' Advanced \ac{LIGO} configuration, scarcely \emph{half} of the \ac{SNR} is accumulated above 40~Hz.  For the preferred final configuration, `zero detuning, high power,' 86.1\% of the \ac{SNR} is above 40~Hz, 93.2\% is above 30~Hz, and 98.1\% is above 20~Hz.  (Since \ac{SNR} accumulates in quadrature, this means, on the other hand, that under the `high frequency' configuration a template encompassing \emph{just the early inspiral} from 10 to 40~Hz would accumulate $\sqrt{1 - 0.533^2} \approx 84.6\%$ of the total \ac{SNR}!  In the `zero detuning, high power,' configuration, integration from 10 to 40~Hz alone would yield 50.9\% of the total \ac{SNR}, from 10 to 30~Hz, 36.2\%, and from 10 to 20~Hz, 19.4\%.)

\begin{deluxetable}{rcccc}
\tablewidth{0pt}
\tablecaption{\label{table:accum_snr}Fractional accumulated \acs{SNR} for selected low frequency cutoffs}
\tablehead{
\colhead{Noise model} & \colhead{40 Hz} & \colhead{30 Hz} & \colhead{20 Hz} & \colhead{10 Hz}
}
\startdata
\ac{LHO} (best S5) & 100.0 & 100.0 & 100.0 & 100.0 \\
\ac{LHO} (best S6) & 100.0 & 100.0 & 100.0 & 100.0 \\
High frequency & 53.3 & 80.1 & 97.6 & 100.0 \\
No SRM & 87.8 & 95.1 & 98.7 & 100.0 \\
BHBH 20$^\circ$ & 71.1 & 84.2 & 96.2 & 100.0 \\
NSNS optimized & 91.5 & 96.3 & 99.0 & 100.0 \\
Zero detuning, low power & 67.9 & 80.0 & 93.5 & 100.0 \\
Zero detuning, high power & 86.1 & 93.2 & 98.1 & 100.0 \\
\enddata
\end{deluxetable}

Since the \ac{GW} amplitude is inversely proportional to the luminosity distance of the source, and the sensitive volume is proportional to distance cubed, the rate of detectable coalescences depends on the choice of low-frequency cutoff.  An inspiral search that is designed with a low-frequency cutoff at the seismic wall would gain an increase in detection rate of $\rho_\mathrm{frac}^{-3}(f_\mathrm{low})$ relative to a search with a low-frequency cutoff of $f_\mathrm{low}$.  This would represent almost a twofold increase in the rate of detection over a search with a fractional accumulated \ac{SNR} of 80\%, and still a 37\% increase over a search with $\rho_\mathrm{frac} = 90\%$.  Existing coalescing binary detection pipelines strive to sacrifice no more than 3\% of the available \ac{SNR}; this forfeits less than a 10\% gain in detection rate.  In order to satisfy this constraint, the low-frequency cutoff would have to be placed below 30~Hz for all of the conceived Advanced LIGO configurations.

The instantaneous \ac{GW} frequency, given by Equation~(\ref{eq:fgw}), is a power law function of time, so the amount of time for the GW frequency to evolve from $f_\mathrm{low}$ to $f_\mathrm{LSO}$ depends strongly on $f_\mathrm{low}$.  The duration of a (1.4,~1.4)~$\Msun$ inspiral is shown in Figure~\ref{fig:low-frequency-cutoff}(d).  The inspiral takes only 25~s to evolve from 40~Hz to $f_\mathrm{LSO}$, but takes 54~s to evolve from 30~Hz to $f_\mathrm{LSO}$, 158~s from 20~Hz, and 1002~s from 10~Hz.

\begin{figure*}[b]
	\begin{center}
	\includegraphics[width=0.5\textwidth]{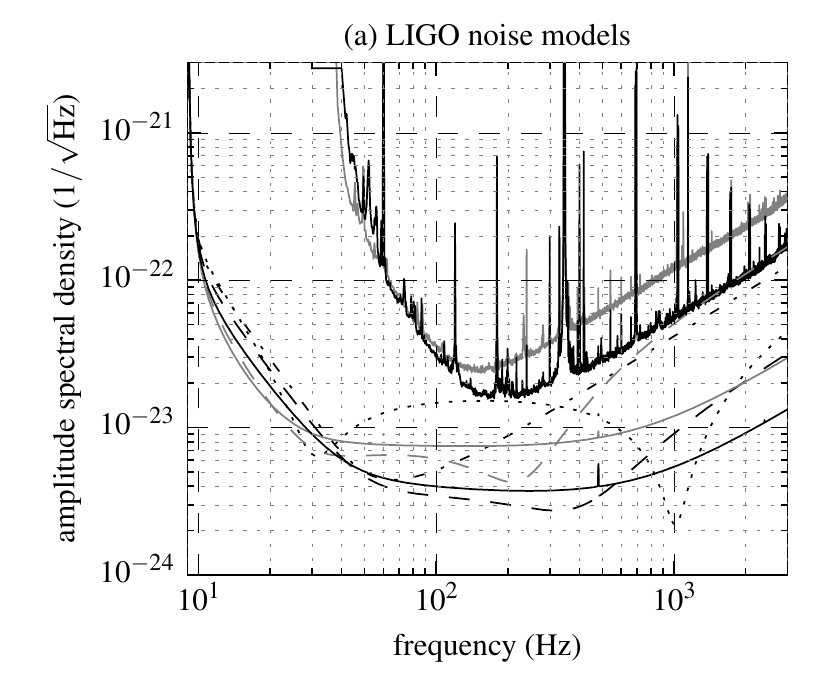}%
	\includegraphics[width=0.5\textwidth]{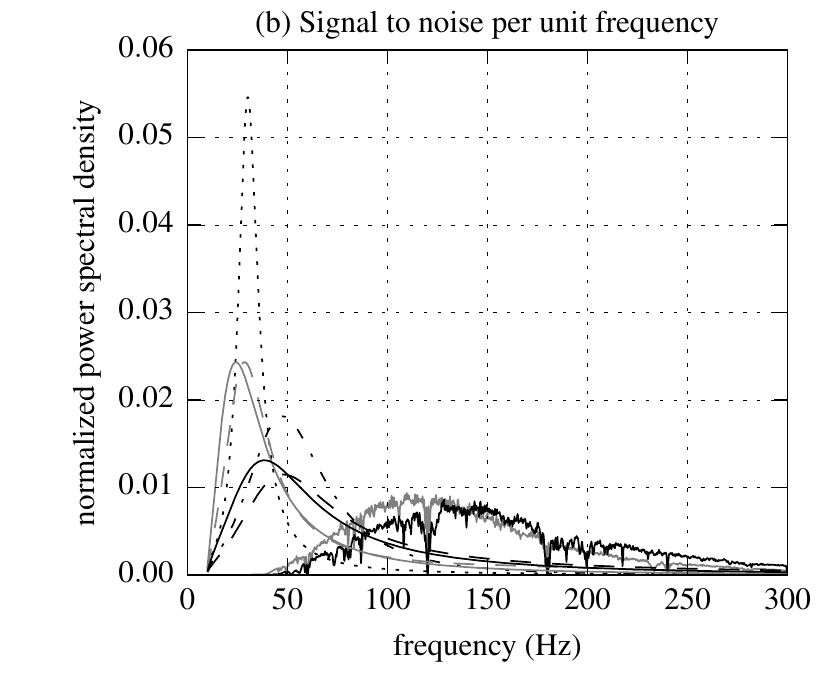}\\[1em]
                    
	\includegraphics[width=0.5\textwidth]{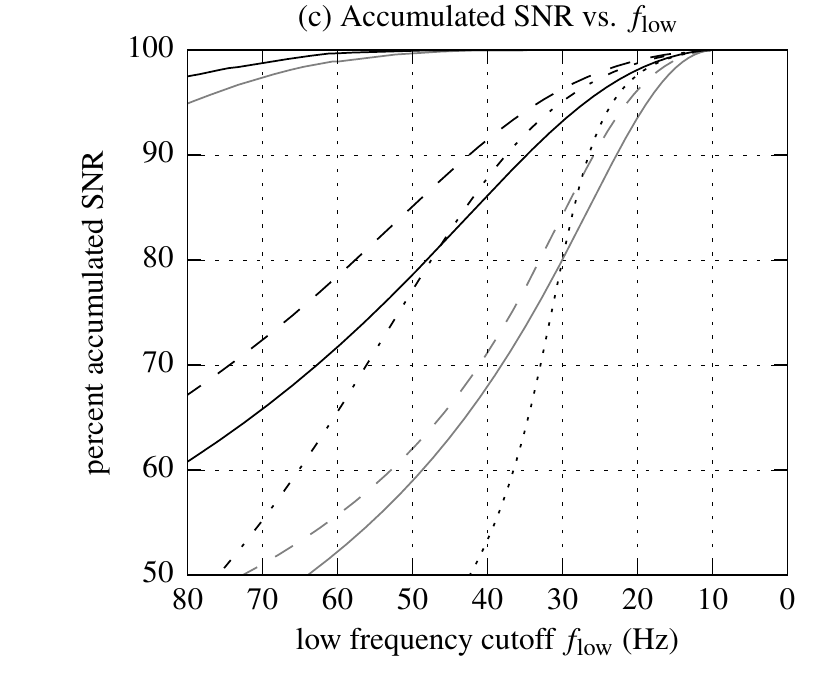}%
	\includegraphics[width=0.5\textwidth]{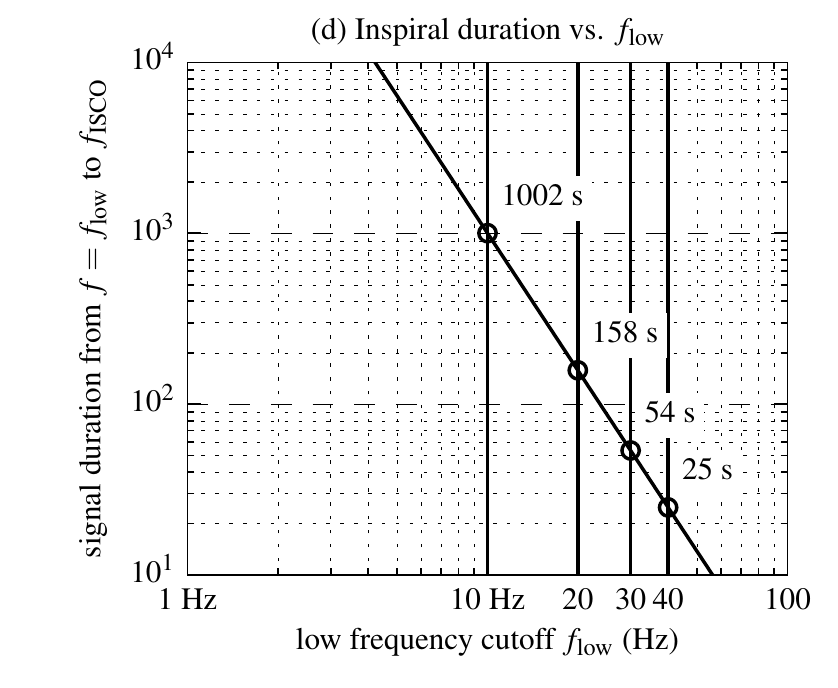}\\[1em]
                    
	\includegraphics[width=\textwidth]{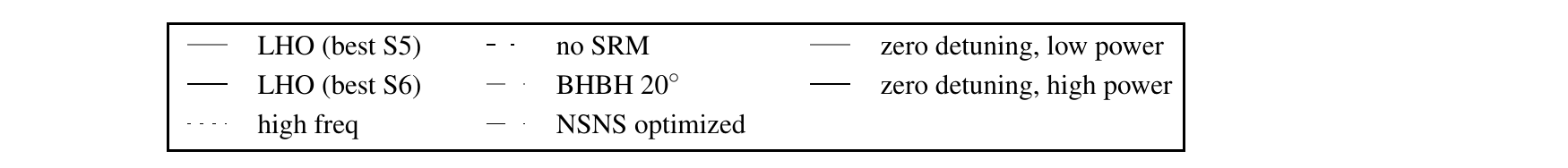}
	\end{center}

	\caption[Accumulated \acs{SNR} as a function of low frequency cutoff]{\label{fig:low-frequency-cutoff}From top left:  (a)~noise amplitude spectral density for a variety of Advanced \ac{LIGO} noise models, \ac{S5}, and \ac{S6}.  (b)~Normalized signal-to-noise per unit frequency, $(\mathrm{d}\rho^2/\mathrm{d}f)/\rho^2$, for a (1.4,~1.4)~$\Msun$ inspiral.  (c)~Percentage of \ac{SNR} that is accumulated from $f_\mathrm{low}$ to $f_\mathrm{LSO}$, relative to \ac{SNR} accumulated from $f_\mathrm{low} = 0~\mathrm{Hz}$ to $f_\mathrm{LSO}$.    (d)~Amount of time for a \ac{NS}--\ac{NS} inspiral signal to evolve from frequency $f_\mathrm{low}$ to $f_\mathrm{LSO}$, as a function of $f_\mathrm{low}$.  For (a)--(c), the line style indicates which noise model was used.}
\end{figure*}

\chapter{Sky map file format}
\label{chap:sky-map-format}

\attribution{This Appendix is reproduced from the technical document LIGO\nobreakdashes-T1300512\nobreakdashes-v7.}

We are proposing to adopt \ac{HEALPix} in \ac{FITS} files as the official format for \ac{LIGO}/Virgo \ac{GW} sky maps, in order to facilitate exchange with astronomers. We suggest that we adopt the following conventions.

\emph{Note:} See \url{http://archive.stsci.edu/fits/fits_standard/} for a reference to the FITS standard and \url{http://heasarc.gsfc.nasa.gov/docs/fcg/standard_dict.html} for a dictionary of standard \ac{FITS} header keywords.

\section{Mandatory}

\begin{enumerate}
\item Sky may use either the \texttt{RING} or \texttt{NESTED} indexing scheme.
\item Sky maps must use \texttt{IMPLICIT} indexing (full HEALPix array; not sparse indices).
\item Sky maps must use equatorial coordinates (\texttt{COORDSYS = C} in the \ac{FITS} header).
\item Pixel data should be placed in a binary table with dimensions reshaped to rows of 1024 elements for backwards compatibility with IDL (pass \texttt{use\_IDL=True} when calling \texttt{healpy.write\_map}).
\item Pixel values should denote the probability per square degree.
\item The name of the column containing the HEALPix pixel data should be \texttt{PROB}.
\item The units of the column containing the HEALPix pixel data should be given in the header as \texttt{pix-1}, to indicate that the sky map is given in units of probability per pixel. This goes in the \texttt{TUNIT}$n$ keyword in the header.
\item Skymaps should be viewable and reasonable in the Aladin\footnote{\url{http://aladin.u-strasbg.fr/}} sky browser.
\end{enumerate}

\section{Optional}

\begin{enumerate}
\item Sky maps may use any \ac{HEALPix} resolution, at the creator's discretion.
\item Pixels that contribute negligibly to the total probability may be set to 0.
\begin{enumerate}
   \item If all pixels are nonzero, then the sum of all of the pixel values should
      be equal to 1.
   \item If some pixels are zero, then the sum of all pixel values should be less
      than or equal to 1.
\end{enumerate}
\item The following optional entries may be added to the \ac{FITS} header:
\begin{enumerate}
   \item \texttt{DATE}, UTC file creation date in ISO 8601.
   \item \texttt{DATE-OBS}, UTC trigger time in ISO 8601.
   \item \texttt{INSTRUME}, a comma-separated string of the style \texttt{H1,L1,V1} denoting
      which detectors contributed data to the event.
   \item \texttt{MJD-OBJS}, trigger time in modified Julian days.
   \item \texttt{OBJECT}, the GraceDB ID of the event, or any other unique identifier.
   \item \texttt{CREATOR}, the name of the program that created the \ac{FITS} file.
   \item \texttt{REFERENC}, the URL of the GraceDB entry, or any other URL that provides
      further information about the event.
   \item \texttt{ORIGIN}, the organization responsible for the data (in most cases,
      probably \texttt{LIGO/Virgo}).
   \item \texttt{RUNTIME} (nonstandard), the elapsed time in seconds that the program
      took to create the sky map.
\end{enumerate}
\item Sky maps may be gzip\nobreakdashes-compressed, signaled by the file extension \texttt{.fits.gz}.
\end{enumerate}

\section{Example code in Python}

Example code to read and write \ac{HEALPix}/\ac{FITS} sky maps with all optional metadata is in the Python module \texttt{lalinference.bayestar.fits}\footnote{\url{http://www.lsc-group.phys.uwm.edu/cgit/lalsuite/tree/lalinference/python/bayestar/fits.py}}.

\section{Specimen}

Below is an example of what the \ac{FITS} headers should look like.

\begin{verbatim}
$ funhead -a skymap.fits.gz
SIMPLE  =                    T / conforms to FITS standard
BITPIX  =                    8 / array data type
NAXIS   =                    0 / number of array dimensions
EXTEND  =                    T
END
        Extension: xtension

XTENSION= 'BINTABLE'           / binary table extension
BITPIX  =                    8 / array data type
NAXIS   =                    2 / number of array dimensions
NAXIS1  =                 4096 / length of dimension 1
NAXIS2  =                  192 / length of dimension 2
PCOUNT  =                    0 / number of group parameters
GCOUNT  =                    1 / number of groups
TFIELDS =                    1 / number of table fields
TTYPE1  = 'PROB    '
TFORM1  = '1024E   '
TUNIT1  = 'pix-1   '
PIXTYPE = 'HEALPIX '           / HEALPIX pixelisation
ORDERING= 'RING    '           / Pixel ordering scheme, either RING or NESTED
COORDSYS= 'C       '           / Ecliptic, Galactic or Celestial (equatorial)
EXTNAME = 'xtension'           / name of this binary table extension
NSIDE   =                  128 / Resolution parameter of HEALPIX
FIRSTPIX=                    0 / First pixel # (0 based)
LASTPIX =               196607 / Last pixel # (0 based)
INDXSCHM= 'IMPLICIT'           / Indexing: IMPLICIT or EXPLICIT
OBJECT  = 'T73435  '           / Unique identifier for this event
REFERENC= 'https://gracedb.ligo.org/events/T73435' / URL of this event
INSTRUME= 'H1,L1,V1'           / Instruments that triggered this event
DATE-OBS= '2013-02-16T03:19:17.438259' / UTC date of the observation
MJD-OBS =    56339.13839627615 / modified Julian date of the observation
DATE    = '2013-06-28T18:32:57' / UTC date of file creation
CREATOR = 'bayestar_localize_lvalert' / Program that created this file
ORIGIN  = 'LIGO/Virgo'         / Organization responsible for this FITS file
RUNTIME =   0.6783878803253174 / Runtime in seconds of the CREATOR program
END
\end{verbatim}

\chapter{``First Two Years'' data release}
\label{chap:first2years-data}

\attribution{This Appendix is reproduced from \citet{FirstTwoYears}, copyright~\textcopyright{}~2014 The American Astronomical Society.}

We describe a catalog of all simulated events, detections, and sky maps that were generated for Chapter~\ref{chap:first2years}.

For the 2015 scenario, parameters of simulated signals are given in Table~\ref{table:2015-sim}. In the same order, parameters of the detection including the operating detector network, false alarm rate, $\rho_\mathrm{net}$, \ac{SNR} in each detector, recovered masses, and sky localization areas are given in Table~\ref{table:2015-coinc}. For the 2016 scenario, the simulated signals are recorded in Table~\ref{table:2016-sim} and the detections in Table~\ref{table:2016-coinc}. In the print journal, parameters are given for just the four sample events that appear earlier in the text (see Figures~\ref{fig:typical},~\ref{fig:typical-unimodal},~\ref{fig:typical-hlv},~and~\ref{fig:typical-hv}). In the machine readable tables in the online journal, parameters are given for all detected signals.

The tables give two integer IDs. The ``event ID'' column corresponds to a field in that scenario's full \textls{gstlal} output that identifies one coincident detection candidate. The ``simulation ID'' likewise identifies one simulated signal. In the full \textls{gstlal} output, there may be zero or many event candidates that match any given simulated signal. However, in our catalog there is one\nobreakdashes-to\nobreakdashes-one correspondence between simulation and event IDs because we have retained only simulated signals that are detected above threshold, and only the highest \ac{SNR} detection candidate for each signal.

Note that the arbitrary dates of the simulated signals range from August~21 through 2010~October~19. This reflects the two\nobreakdashes-month duration of the simulated data stream, not the dates or durations of the anticipated Advanced LIGO/Virgo observing runs.

For convenience, we also provide a browsable sky map catalog\footnote{\url{http://www.ligo.org/scientists/first2years}}. This Web page provides a searchable version of Tables~\ref{table:2015-sim}\nobreakdashes--\ref{table:2016-coinc}, with posterior sky map images from both the rapid parameter estimation and the stochastic samplers.

The Web page also provides, for each localization, a FITS file representing the posterior in the HEALPix projection~\citep{HEALPix} using the NESTED indexing scheme. For reading these files, the authors recommend the Python package Healpy\footnote{\url{http://healpy.readthedocs.org}} or the HEALPix C/C++/IDL/Java/Fortran library\footnote{\url{http://healpix.sourceforge.net}}. They can also be displayed by many standard imaging programs such as DS9\footnote{\url{http://ds9.si.edu}} and Aladin\footnote{\url{http://aladin.u-strasbg.fr}}.

Synthetic GW time series data and posterior sample chains are available upon request.

\begin{landscape}

\begin{deluxetable*}{rr|rrrrrrr|rr|rrr|rrr}
\tabletypesize{\footnotesize}
\tablecaption{\label{table:2015-sim}Simulated BNS signals for 2015 scenario}
\tablehead{
    \colhead{Event} &
    \colhead{sim} &
    \colhead{\,} &
    \multicolumn{5}{c}{Orientation\tablenotemark{d}} &
    \colhead{$d$} &
    \multicolumn{2}{c}{Masses ($M_\odot$)} &
    \multicolumn{3}{c}{Spin 1} &
    \multicolumn{3}{c}{Spin 2} \\
    \colhead{ID\tablenotemark{a}} &
    \colhead{ID\tablenotemark{b}} &
    \colhead{MJD\tablenotemark{c}} &
    \colhead{$\alpha$} &
    \colhead{$\delta$} &
    \colhead{$\iota$} &
    \colhead{$\psi$} &
    \colhead{$\phi_c$} &
    \colhead{(Mpc)} &
    \colhead{$m_1$} &
    \colhead{$m_2$} &
    \colhead{$S_1^x$} &
    \colhead{$S_1^y$} &
    \colhead{$S_1^z$} &
    \colhead{$S_2^x$} &
    \colhead{$S_2^y$} &
    \colhead{$S_2^z$}
}
\startdata
18951&10807&55442.25864&137.8&-39.9&139&43&42&75&1.40&1.51&-0.01&-0.01&-0.04&-0.05&-0.01&+0.01\\
20342&21002&55454.76654&19.8&-23.7&145&197&145&75&1.34&1.48&-0.03&+0.01&-0.03&-0.01&+0.02&-0.01\\

\dots & \dots & \dots & \dots & \dots & \dots & \dots & \dots & \dots & \dots & \dots & \dots & \dots & \dots & \dots & \dots & \dots
\enddata
\tablenotetext{a}{Identifier for detection candidate. This is the same value as the \texttt{coinc\_event\_id} column in the \textls{gstlal} output database and the \texttt{OBJECT} cards in sky map FITS headers, with the \texttt{coinc\_event:coinc\_event\_id:} prefix stripped.}
\tablenotetext{b}{Identifier for simulated signal. This is the same value as the \texttt{simulation\_id} column in the \textls{gstlal} output database, with the \texttt{sim\_inspiral:simulation\_id:} prefix stripped.}
\tablenotetext{c}{Time of arrival at geocenter of GWs from last stable orbit.}
\tablenotetext{d}{$\alpha$: RA, $\delta$: Dec (J2000), $\iota$: binary orbital inclination angle, $\psi$: polarization angle \citep[Appendix B]{ExcessPower}, $\phi_c$: orbital phase at coalescence.}
\end{deluxetable*}

\begin{deluxetable*}{rr|r|rrr|rr|rrr|rrr}
\tabletypesize{\footnotesize}
\tablecaption{\label{table:2015-coinc}Detections and sky localization areas for 2015 scenario}
\tablehead{
    \colhead{Event} &
    \colhead{sim} &
    \colhead{\,} &
    \multicolumn{3}{c}{SNR} &
    \multicolumn{2}{c}{Masses\tablenotemark{b}} &
    \multicolumn{3}{c}{\ac{BAYESTAR}} &
    \multicolumn{3}{c}{\textls{LALINFERENCE\_NEST}} \\
    \colhead{ID} &
    \colhead{ID} &
    \colhead{Network} &
    \colhead{Net\tablenotemark{a}} &
    \colhead{H} &
    \colhead{L} &
    \colhead{$m_1$} &
    \colhead{$m_2$} &
    \colhead{50\%} &
    \colhead{90\%} &
    \colhead{\footnotesize Searched} &
    \colhead{50\%} &
    \colhead{90\%} &
    \colhead{\footnotesize Searched}
}
\startdata
18951&10807&HL&15.0&10.3&10.9&1.67&1.27&159&630&127&158&683&81.2\\
20342&21002&HL&12.7&7.3&10.3&1.59&1.25&126&526&16.9&168&618&12.3\\

\dots & \dots & \dots & \dots & \dots & \dots & \dots & \dots & \dots & \dots & \dots & \dots & \dots & \dots
\enddata
\tablenotetext{a}{Network SNR, or root-sum-squared SNR over all detectors.}
\tablenotetext{b}{Maximum likelihood estimate of masses as reported by \textls{gstlal}.}
\end{deluxetable*}

\begin{deluxetable*}{rr|rrrrrrr|rr|rrr|rrr}
\tabletypesize{\footnotesize}
\tablecaption{\label{table:2016-sim}Simulated BNS signals for 2016 scenario}
\tablehead{
    \colhead{Event} &
    \colhead{sim} &
    \colhead{\,} &
    \multicolumn{5}{c}{Orientation} &
    \colhead{$d$} &
    \multicolumn{2}{c}{Masses ($M_\odot$)} &
    \multicolumn{3}{c}{Spin 1} &
    \multicolumn{3}{c}{Spin 2} \\
    \colhead{ID} &
    \colhead{ID} &
    \colhead{MJD} &
    \colhead{$\alpha$} &
    \colhead{$\delta$} &
    \colhead{$\iota$} &
    \colhead{$\psi$} &
    \colhead{$\phi_c$} &
    \colhead{(Mpc)} &
    \colhead{$m_1$} &
    \colhead{$m_2$} &
    \colhead{$S_1^x$} &
    \colhead{$S_1^y$} &
    \colhead{$S_1^z$} &
    \colhead{$S_2^x$} &
    \colhead{$S_2^y$} &
    \colhead{$S_2^z$}
}
\startdata
655803&45345&55484.63177&79.2&+5.0&121&321&69&66&1.60&1.29&+0.00&+0.00&-0.00&+0.00&+0.00&-0.00\\
821759&8914&55439.93634&18.3&-15.1&158&257&230&187&1.60&1.45&-0.00&+0.02&-0.01&+0.04&+0.03&-0.02\\

\dots & \dots & \dots & \dots & \dots & \dots & \dots & \dots & \dots & \dots & \dots & \dots & \dots & \dots & \dots & \dots & \dots
\enddata
\end{deluxetable*}

\begin{deluxetable*}{rr|r|rrrr|rr|rrr|rrr}
\tabletypesize{\footnotesize}
\tablecaption{\label{table:2016-coinc}Detections and sky localization areas for 2016 scenario}
\tablehead{
    \colhead{Event} &
    \colhead{sim} &
    \colhead{\,} &
    \multicolumn{4}{c}{SNR} &
    \multicolumn{2}{c}{Masses} &
    \multicolumn{3}{c}{\ac{BAYESTAR}} &
    \multicolumn{3}{c}{\textls{LALINFERENCE\_MCMC}} \\
    \colhead{ID} &
    \colhead{ID} &
    \colhead{Network} &
    \colhead{Net} &
    \colhead{H\tablenotemark{a}} &
    \colhead{L\tablenotemark{a}} &
    \colhead{V\tablenotemark{a}} &
    \colhead{$m_1$} &
    \colhead{$m_2$} &
    \colhead{50\%} &
    \colhead{90\%} &
    \colhead{\footnotesize Searched} &
    \colhead{50\%} &
    \colhead{90\%} &
    \colhead{\footnotesize Searched}
}
\startdata
655803&45345&HV&12.2&11.5&&4.2&1.52&1.35&478&4570&65.5&304&3960&20.6\\
821759&8914&HLV&13.4&8.5&10.4&&1.57&1.47&336&1070&473&91.0&515&93.8\\

\dots & \dots & \dots & \dots & \dots & \dots & \dots & \dots & \dots & \dots & \dots & \dots & \dots & \dots & \dots
\enddata
\tablenotetext{a}{Blank if $\mathrm{SNR} < 4$ or detector is not online.}
\end{deluxetable*}

\end{landscape}

\chapter{Observations of iPTF/GBM afterglows}
\label{chap:iptf-gbm-observations}

\attribution{This Appendix contains the optical and radio observations from Chapter~\ref{chap:iptf-gbm}, which is in preparation as an article for the \textnormal{The Astrophysical Journal}.}

\newgeometry{twocolumn}
\tabletypesize{\footnotesize}
\tablewidth{0pt}
\begin{deluxetable}{lrrr@{\enspace}r@{\enspace}l@{$\,\pm\,$}l}
\tablecolumns{7}
\tablecaption{\label{tab:photometry}Optical observations of \acs{GBM}--\acs{IPTF} afterglows}
\tablehead{
    \colhead{Date (mid)} & \colhead{Inst.\tablenotemark{a}} & \colhead{$\Delta t$\tablenotemark{b}} & \multicolumn{4}{c}{mag\tablenotemark{c}}
}
\startdata
        \cutinhead{GRB\,130702A / iPTF13bxl} \\
        2013-07-02 04:18 &
        P48 &
        0.18 &
        $R$ &
        $=$ &17.38 &
        0.04 \\
        
        2013-07-02 05:10 &
        P48 &
        0.21 &
        $R$ &
        $=$ &17.52 &
        0.04 \\
        
        2013-07-03 04:13 &
        P60 &
        1.17 &
        $g$ &
        $=$ &18.80 &
        0.04 \\
        
        2013-07-03 04:15 &
        P60 &
        1.17 &
        $i$ &
        $=$ &18.42 &
        0.04 \\
        
        2013-07-03 06:16 &
        P60 &
        1.26 &
        $i$ &
        $=$ &18.56 &
        0.06 \\
        
        2013-07-03 06:17 &
        P60 &
        1.26 &
        $r$ &
        $=$ &18.66 &
        0.05 \\
        
        2013-07-03 06:20 &
        P60 &
        1.26 &
        $g$ &
        $=$ &18.86 &
        0.04 \\
\enddata
\tablenotetext{a}{RATIR data is from \citet{GCN14980,GCN14993,GCN16236}. GROND data is from \citet{GCN15328}.}
\tablenotetext{b}{Time in days relative to \ac{GBM} trigger.}
\tablenotetext{c}{Magnitudes are in the AB system \citep{ABMags}.}
\end{deluxetable}

\tabletypesize{\footnotesize}
\tablewidth{0pt}
\begin{deluxetable}{llrr@{\enspace}r@{\enspace}r@{}c@{}l}
\tablecolumns{8}
\tablecaption{\label{tab:radio}Radio observations of \acs{GBM}--\acs{IPTF} afterglows}
\tablehead{
    \colhead{Date (start)} & Inst.\tablenotemark{a} & \colhead{$\Delta t$\tablenotemark{b}} & \multicolumn{5}{c}{flux density\tablenotemark{c}}
}
\startdata
        \cutinhead{GRB\,130702A / iPTF13bxl} \\
        2013-07-04 &
        CARMA &
        2 &
        $f_\nu(93)$ &
        
        $=$ &
        1580 &
        $\pm$ &
        330
         \\
        
        2013-07-04 &
        VLA &
        2.3 &
        $f_\nu(5.1)$ &
        
        $=$ &
        1490 &
        $\pm$ &
        75
         \\
        
        2013-07-04 &
        VLA &
        2.3 &
        $f_\nu(7.1)$ &
        
        $=$ &
        1600 &
        $\pm$ &
        81
         \\
        
        2013-07-05 &
        CARMA &
        3.1 &
        $f_\nu(93)$ &
        
        $=$ &
        1850 &
        $\pm$ &
        690
         \\
        
        2013-07-06 &
        CARMA &
        4.1 &
        $f_\nu(93)$ &
        
        $=$ &
        1090 &
        $\pm$ &
        350
         \\
        
        2013-07-08 &
        CARMA &
        6.1 &
        $f_\nu(93)$ &
        
        $=$ &
        1440 &
        $\pm$ &
        260
         \\
        
        2013-07-08 &
        CARMA &
        7 &
        $f_\nu(93)$ &
        
        $=$ &
        1160 &
        $\pm$ &
        320
         \\
\enddata
\tablenotetext{a}{The \ac{ATCA} observation is from \citet{GCN15395}.}
\tablenotetext{b}{Time in days relative to \ac{GBM} trigger.}
\tablenotetext{c}{Flux density in $\mu$Jy as a function of frequency in GHz. For detections, the confidence intervals are $1\sigma$ statistical uncertainties added in quadrature with an estimated 5\% systematic error. For non\nobreakdashes-detections, we show $3\sigma$ upper limits.}
\end{deluxetable}

\restoregeometry

\extrachapter{Acronyms}
\providecommand{\acrolowercase}[1]{\lowercase{#1}}

\begin{acronym}
\acro{2MASS}[2MASS]{2 Micron All\nobreakdashes-Sky Survey}
\acro{AdVirgo}[AdVirgo]{Advanced Virgo}
\acro{AMI}[AMI]{Arcminute Microkelvin Imager}
\acro{AGN}[AGN]{active galactic nucleus}
\acroplural{AGN}[AGN\acrolowercase{s}]{active galactic nuclei}
\acro{aLIGO}[aLIGO]{Advanced \acs{LIGO}}
\acro{ATCA}[ATCA]{Australia Telescope Compact Array}
\acro{ATLAS}[ATLAS]{Asteroid Terrestrial-impact Last Alert System}
\acro{BAT}[BAT]{Burst Alert Telescope\acroextra{ (instrument on \emph{Swift})}}
\acro{BATSE}[BATSE]{Burst and Transient Source Experiment\acroextra{ (instrument on \acs{CGRO})}}
\acro{BAYESTAR}[BAYESTAR]{BAYESian TriAngulation and Rapid localization}
\acro{BBH}[BBH]{binary black hole}
\acro{BHBH}[BHBH]{\acl{BH}\nobreakdashes--\acl{BH}}
\acro{BH}[BH]{black hole}
\acro{BNS}[BNS]{binary neutron star}
\acro{CARMA}[CARMA]{Combined Array for Research in Millimeter\nobreakdashes-wave Astronomy}
\acro{CASA}[CASA]{Common Astronomy Software Applications}
\acro{CFH12k}[CFH12k]{Canada--France--Hawaii $12\,288 \times 8\,192$ pixel CCD mosaic\acroextra{ (instrument formerly on the Canada--France--Hawaii Telescope, now on the \ac{P48})}}
\acro{CRTS}[CRTS]{Catalina Real-Time Sky Survey}
\acro{CTIO}[CTIO]{Cerro Tololo Inter-American Observatory}
\acro{CBC}[CBC]{compact binary coalescence}
\acro{CCD}[CCD]{charge coupled device}
\acro{CDF}[CDF]{cumulative distribution function}
\acro{CGRO}[CGRO]{Compton Gamma Ray Observatory}
\acro{CMB}[CMB]{cosmic microwave background}
\acro{CRLB}[CRLB]{Cram\'{e}r\nobreakdashes--Rao lower bound}
\acro{cWB}[\acrolowercase{c}WB]{Coherent WaveBurst}
\acro{DASWG}[DASWG]{Data Analysis Software Working Group}
\acro{DBSP}[DBSP]{Double Spectrograph\acroextra{ (instrument on \acs{P200})}}
\acro{DCT}[DCT]{Discovery Channel Telescope}
\acro{DECAM}[DECam]{Dark Energy Camera\acroextra{ (instrument on the Blanco 4\nobreakdashes-m telescope at \acs{CTIO})}}
\acro{DFT}[DFT]{discrete Fourier transform}
\acro{EM}[EM]{electromagnetic}
\acro{FD}[FD]{frequency domain}
\acro{FAR}[FAR]{false alarm rate}
\acro{FFT}[FFT]{fast Fourier transform}
\acro{FIR}[FIR]{finite impulse response}
\acro{FITS}[FITS]{Flexible Image Transport System}
\acro{FLOPS}[FLOPS]{floating point operations per second}
\acro{FOV}[FOV]{field of view}
\acroplural{FOV}[FOV\acrolowercase{s}]{fields of view}
\acro{FTN}[FTN]{Faulkes Telescope North}
\acro{GBM}[GBM]{Gamma-ray Burst Monitor\acroextra{ (instrument on \emph{Fermi})}}
\acro{GCN}[GCN]{Gamma-ray Coordinates Network}
\acro{GMOS}[GMOS]{Gemini Multi-Object Spectrograph\acroextra{ (instrument on the Gemini telescopes)}}
\acro{GRB}[GRB]{gamma-ray burst}
\acro{GSL}[GSL]{GNU Scientific Library}
\acro{GW}[GW]{gravitational wave}
\acro{HAWC}[HAWC]{High\nobreakdashes-Altitude Water \v{C}erenkov Gamma\nobreakdashes-Ray Observatory}
\acro{HCT}[HCT]{Himalayan Chandra Telescope}
\acro{HEALPix}[HEALP\acrolowercase{ix}]{Hierarchical Equal Area isoLatitude Pixelization}
\acro{HEASARC}[HEASARC]{High Energy Astrophysics Science Archive Research Center}
\acro{HETE}[HETE]{High Energy Transient Explorer}
\acro{HFOSC}[HFOSC]{Himalaya Faint Object Spectrograph and Camera\acroextra{ (instrument on \acs{HCT})}}
\acro{HMXB}[HMXB]{high\nobreakdashes-mass X\nobreakdashes-ray binary}
\acroplural{HMXB}[HMXB\acrolowercase{s}]{high\nobreakdashes-mass X\nobreakdashes-ray binaries}
\acro{HSC}[HSC]{Hyper Suprime\nobreakdashes-Cam\acroextra{ (instrument on the 8.2\nobreakdashes-m Subaru telescope)}}
\acro{IACT}[IACT]{imaging atmospheric \v{C}erenkov telescope}
\acro{IIR}[IIR]{infinite impulse response}
\acro{IMACS}[IMACS]{Inamori-Magellan Areal Camera \& Spectrograph\acroextra{ (instrument on the Magellan Baade telescope)}}
\acro{IPAC}[IPAC]{Infrared Processing and Analysis Center}
\acro{IPN}[IPN]{InterPlanetary Network}
\acro{IPTF}[\acrolowercase{i}PTF]{Intermediate \acl{PTF}}
\acro{ISM}[ISM]{interstellar medium}
\acro{KAGRA}[KAGRA]{KAmioka GRAvitational\nobreakdashes-wave observatory}
\acro{LAT}[LAT]{Large Area Telescope}
\acro{LHO}[LHO]{\ac{LIGO} Hanford Observatory}
\acro{LIGO}[LIGO]{Laser Interferometer \acs{GW} Observatory}
\acro{llGRB}[\acrolowercase{ll}GRB]{low\nobreakdashes-luminosity \ac{GRB}}
\acro{LLOID}[LLOID]{Low Latency Online Inspiral Detection}
\acro{LLO}[LLO]{\ac{LIGO} Livingston Observatory}
\acro{LMI}[LMI]{Large Monolithic Imager\acroextra{ (instrument on \ac{DCT})}}
\acro{LOFAR}[LOFAR]{Low Frequency Array}
\acro{LSB}[LSB]{long, soft burst}
\acro{LSC}[LSC]{\acs{LIGO} Scientific Collaboration}
\acro{LSO}[LSO]{last stable orbit}
\acro{LSST}[LSST]{Large Synoptic Survey Telescope}
\acro{LTI}[LTI]{linear time invariant}
\acro{MAP}[MAP]{maximum a posteriori}
\acro{MBTA}[MBTA]{Multi-Band Template Analysis}
\acro{MCMC}[MCMC]{Markov chain Monte Carlo}
\acro{MLE}[MLE]{\ac{ML} estimator}
\acro{ML}[ML]{maximum likelihood}
\acro{NED}[NED]{NASA/IPAC Extragalactic Database}
\acro{NSBH}[NSBH]{neutron star\nobreakdashes--black hole}
\acro{NSBH}[NSBH]{\acl{NS}\nobreakdashes--\acl{BH}}
\acro{NSF}[NSF]{National Science Foundation}
\acro{NSNS}[NSNS]{\acl{NS}\nobreakdashes--\acl{NS}}
\acro{NS}[NS]{neutron star}
\acro{OT}[OT]{optical transient}
\acro{P48}[P48]{Palomar 48\nobreakdashes-inch Oschin telescope}
\acro{P60}[P60]{robotic Palomar 60\nobreakdashes-inch telescope}
\acro{P200}[P200]{Palomar 200\nobreakdashes-inch Hale telescope}
\acro{PC}[PC]{photon counting}
\acro{PSD}[PSD]{power spectral density}
\acro{PSF}[PSF]{point spread function}
\acro{PTF}[PTF]{Palomar Transient Factory}
\acro{QUEST}[QUEST]{Quasar Equatorial Survey Team}
\acro{RAPTOR}[RAPTOR]{Rapid Telescopes for Optical Response}
\acro{REU}[REU]{Research Experiences for Undergraduates}
\acro{RMS}[RMS]{root mean square}
\acro{ROTSE}[ROTSE]{Robotic Optical Transient Search}
\acro{S5}[S5]{\ac{LIGO}'s fifth science run}
\acro{S6}[S6]{\ac{LIGO}'s sixth science run}
\acro{SHB}[SHB]{short, hard burst}
\acro{SHGRB}[SHGRB]{short, hard \acl{GRB}}
\acro{SMT}[SMT]{Slewing Mirror Telescope\acroextra{ (instrument on \acs{UFFO} Pathfinder)}}
\acro{SNR}[SNR]{signal\nobreakdashes-to\nobreakdashes-noise ratio}
\acro{SSC}[SSC]{synchrotron self\nobreakdashes-Compton}
\acro{SDSS}[SDSS]{Sloan Digital Sky Survey}
\acro{SED}[SED]{spectral energy distribution}
\acro{SN}[SN]{supernova}
\acroplural{SN}[SN\acrolowercase{e}]{supernova}
\acro{SNIcBL}[\acs{SN}\,I\acrolowercase{c}\nobreakdashes-BL]{broad\nobreakdashes-line type~Ic \ac{SN}}
\acroplural{SNIcBL}[\acsp{SN}\,I\acrolowercase{c}\nobreakdashes-BL]{broad\nobreakdashes-line type~Ic \acp{SN}}
\acro{SVD}[SVD]{singular value decomposition}
\acro{TAROT}[TAROT]{T\'{e}lescopes \`{a} Action Rapide pour les Objets Transitoires}
\acro{TDOA}[TDOA]{time delay on arrival}
\acroplural{TDOA}[TDOA\acrolowercase{s}]{time delays on arrival}
\acro{TD}[TD]{time domain}
\acro{TOA}[TOA]{time of arrival}
\acroplural{TOA}[TOA\acrolowercase{s}]{times of arrival}
\acro{TOO}[TOO]{target of opportunity}
\acroplural{TOO}[TOO\acrolowercase{s}]{targets of opportunity}
\acro{UFFO}[UFFO]{Ultra Fast Flash Observatory}
\acro{UHE}[UHE]{ultra high energy}
\acro{UVOT}[UVOT]{UV/Optical Telescope\acroextra{ (instrument on \emph{Swift})}}
\acro{VHE}[VHE]{very high energy}
\acro{VLA}[VLA]{Karl G. Jansky Very Large Array}
\acro{WAM}[WAM]{Wide\nobreakdashes-band All\nobreakdashes-sky Monitor\acroextra{ (instrument on \emph{Suzaku})}}
\acro{WCS}[WCS]{World Coordinate System}
\acro{WSS}[w.s.s.]{wide\nobreakdashes-sense stationary}
\acro{XRF}[XRF]{X\nobreakdashes-ray flash}
\acroplural{XRF}[XRF\acrolowercase{s}]{X\nobreakdashes-ray flashes}
\acro{XRT}[XRT]{X\nobreakdashes-ray Telescope\acroextra{ (instrument on \emph{Swift})}}
\acro{ZTF}[ZTF]{Zwicky Transient Facility}
\end{acronym}

\bibliographystyle{apj}
\bibliography{ms,iptf-gbm-paper/gcn}

\begin{thebibliography}{374}
\expandafter\ifx\csname natexlab\endcsname\relax\def\natexlab#1{#1}\fi

\bibitem[{Aasi {et~al.}(2013{\natexlab{a}})Aasi, {Abadie}, {Abbott}, {Abbott},
  {Abbott}, {Abernathy}, {Adams}, {Adams}, {Addesso}, {Adhikari}, \&
  et~al.}]{LIGOSqueezing}
Aasi, J., {et~al.} 2013{\natexlab{a}}, Nature Photonics, 7, 613

\bibitem[{Aasi {et~al.}(2013{\natexlab{b}})}]{S6PE}
---. 2013{\natexlab{b}}, \prd, 88, 062001

\bibitem[{Aasi {et~al.}(2013{\natexlab{c}})}]{LIGOObservingScenarios}
---. 2013{\natexlab{c}}, arXiv:1304.0670

\bibitem[{Aasi {et~al.}(2014)Aasi, {Abadie}, {Abbott}, {Abbott}, {Abbott},
  {Abernathy}, {Accadia}, {Acernese}, {Adams}, {Adams}, \&
  et~al.}]{FirstSearchesOpticalCounterparts}
---. 2014, \apjs, 211, 7

\bibitem[{Abadie {et~al.}(2010{\natexlab{a}})}]{S5InspiralRange}
Abadie, J., {et~al.} 2010{\natexlab{a}}, LIGO-T0900499-v19

\bibitem[{Abadie {et~al.}(2010{\natexlab{b}})}]{LIGORates}
---. 2010{\natexlab{b}}, Class. Quantum Grav., 27, 173001

\bibitem[{Abadie {et~al.}(2012{\natexlab{a}})}]{CBCLowLatency}
---. 2012{\natexlab{a}}, \aap, 541, A155

\bibitem[{Abadie
  {et~al.}(2012{\natexlab{b}})}]{FirstPromptSearchGWTransientsEMCounterparts}
---. 2012{\natexlab{b}}, \aap, 539, A124

\bibitem[{Abadie {et~al.}(2012{\natexlab{c}})}]{LowMassSearchS6VSR23}
---. 2012{\natexlab{c}}, \prd, 85, 082002

\bibitem[{Abadie {et~al.}(2012{\natexlab{d}})}]{S6InspiralRange}
---. 2012{\natexlab{d}}, arXiv:1203.2674

\bibitem[{{Abeysekara} {et~al.}(2012){Abeysekara}, {Aguilar}, {Aguilar},
  {Alfaro}, {Almaraz}, {{\'A}lvarez}, {{\'A}lvarez-Romero}, {{\'A}lvarez},
  {Arceo}, {Arteaga-Vel{\'a}zquez}, {Badillo}, {Barber}, {Baughman},
  {Bautista-Elivar}, {Belmont}, {Ben{\'{\i}}tez}, {BenZvi}, {Berley}, {Bernal},
  {Bonamente}, {Braun}, {Caballero-Lopez}, {Cabrera}, {Carrami{\~n}ana},
  {Carrasco}, {Castillo}, {Chambers}, {Conde}, {Condreay}, {Cotti}, {Cotzomi},
  {D'Olivo}, {de la Fuente}, {De Le{\'o}n}, {Delay}, {Delepine}, {DeYoung},
  {Diaz}, {Diaz-Cruz}, {Dingus}, {Duvernois}, {Edmunds}, {Ellsworth}, {Fick},
  {Fiorino}, {Flandes}, {Fraija}, {Galindo}, {Garc{\'{\i}}a-Luna},
  {Garc{\'{\i}}a-Torales}, {Garfias}, {Gonz{\'a}lez}, {Gonz{\'a}lez},
  {Goodman}, {Grabski}, {Gussert}, {Guzm{\'a}n-Ceron}, {Hampel-Arias},
  {Harris}, {Hays}, {Hernandez-Cervantes}, {H{\"u}ntemeyer}, {Imran},
  {Iriarte}, {Jimenez}, {Karn}, {Kelley-Hoskins}, {Kieda}, {Langarica}, {Lara},
  {Lauer}, {Lee}, {Linares}, {Linnemann}, {Longo}, {Luna-Garc{\'{\i}}a},
  {Mart{\'{\i}}nez}, {Mart{\'{\i}}nez}, {Mart{\'{\i}}nez}, {Mart{\'{\i}}nez},
  {Mart{\'{\i}}nez-Castro}, {Martos}, {Matthews}, {McEnery}, {Medina-Tanco},
  {Mendoza-Torres}, {Miranda-Romagnoli}, {Montaruli}, {Moreno}, {Mostafa},
  {Napsuciale}, {Nava}, {Nellen}, {Newbold}, {Noriega-Papaqui},
  {Oceguera-Becerra}, {Olmos Tapia}, {Orozco}, {P{\'e}rez},
  {P{\'e}rez-P{\'e}rez}, {Perkins}, {Pretz}, {Ramirez}, {Ram{\'{\i}}rez},
  {Rebello}, {Renter{\'{\i}}a}, {Reyes}, {Rosa-Gonz{\'a}lez}, {Rosado}, {Ryan},
  {Sacahui}, {Salazar}, {Salesa}, {Sandoval}, {Santos}, {Schneider}, {Shoup},
  {Silich}, {Sinnis}, {Smith}, {Sparks}, {Springer}, {Su{\'a}rez}, {Suarez},
  {Taboada}, {Tellez}, {Tenorio-Tagle}, {Tepe}, {Toale}, {Tollefson}, {Torres},
  {Ukwatta}, {Valdes-Galicia}, {Vanegas}, {Vasileiou}, {V{\'a}zquez},
  {V{\'a}zquez}, {Villase{\~n}or}, {Wall}, {Walters}, {Warner}, {Westerhoff},
  {Wisher}, {Wood}, {Yodh}, {Zaborov}, \& {Zepeda}}]{HAWC-GRB-1}
{Abeysekara}, A.~U., {et~al.} 2012, Astroparticle Physics, 35, 641

\bibitem[{{Abeysekara} {et~al.}(2014){Abeysekara}, {Alfaro}, {Alvarez},
  {{\'A}lvarez}, {Arceo}, {Arteaga-Vel{\'a}zquez}, {Ayala Solares}, {Barber},
  {Baughman}, {Bautista-Elivar}, {Belmont}, {BenZvi}, {Berley}, {Bonilla
  Rosales}, {Braun}, {Caballero-Mora}, {Carrami{\~n}ana}, {Castillo}, {Cotti},
  {Cotzomi}, {de la Fuente}, {De Le{\'o}n}, {DeYoung}, {Diaz Hernandez},
  {D{\'{\i}}az-V{\'e}lez}, {Dingus}, {DuVernois}, {Ellsworth}, {Fiorino},
  {Fraija}, {Galindo}, {Garfias}, {Gonz{\'a}lez}, {Goodman}, {Gussert},
  {Hampel-Arias}, {Harding}, {H{\"u}ntemeyer}, {Hui}, {Imran}, {Iriarte},
  {Karn}, {Kieda}, {Kunde}, {Lara}, {Lauer}, {Lee}, {Lennarz}, {Le{\'o}n
  Vargas}, {Linnemann}, {Longo}, {Luna-Garc{\'{\i}}a}, {Malone}, {Marinelli},
  {Marinelli}, {Martinez}, {Martinez}, {Mart{\'{\i}}nez-Castro}, {Matthews},
  {McEnery}, {Mendoza Torres}, {Miranda-Romagnoli}, {Moreno}, {Mostaf{\'a}},
  {Nellen}, {Newbold}, {Noriega-Papaqui}, {Oceguera-Becerra}, {Patricelli},
  {Pelayo}, {P{\'e}rez-P{\'e}rez}, {Pretz}, {Rivi{\`e}re}, {Rosa-Gonz{\'a}lez},
  {Ruiz-Velasco}, {Ryan}, {Salazar}, {Salesa Greus}, {Sandoval}, {Schneider},
  {Sinnis}, {Smith}, {Sparks Woodle}, {Springer}, {Taboada}, {Toale},
  {Tollefson}, {Torres}, {Ukwatta}, {Villase{\~n}or}, {Weisgarber},
  {Westerhoff}, {Wisher}, {Wood}, {Yodh}, {Younk}, {Zaborov}, {Zepeda}, {Zhou},
  \& {The HAWC Collaboration}}]{HAWCFirstResults}
---. 2014, \apj, 796, 108

\bibitem[{Acernese {et~al.}(2013)Acernese, Alshourbagy, Antonucci,
  {et~al.}}]{aVirgo}
Acernese, F., {et~al.} 2013, Tech. Rep. VIR-0027A-09

\bibitem[{{Adhikari}(2014)}]{GWDetectionLaserInterferometry}
{Adhikari}, R.~X. 2014, Rev. Mod. Phys., 86, 121

\bibitem[{{Ajith} \& {Bose}(2009)}]{2009PhRvD..79h4032A}
{Ajith}, P., \& {Bose}, S. 2009, \prd, 79, 084032

\bibitem[{{Ajith} {et~al.}(2014){Ajith}, {Fotopoulos}, {Privitera}, {Neunzert},
  {Mazumder}, \& {Weinstein}}]{SBank}
{Ajith}, P., {et~al.} 2014, \prd, 89, 084041

\bibitem[{Allen(2005)}]{AllenChiSq}
Allen, B. 2005, \prd, 71, 062001

\bibitem[{{Allen} {et~al.}(2012){Allen}, {Anderson}, {Brady}, {Brown}, \&
  {Creighton}}]{FINDCHIRP}
{Allen}, B., {et~al.} 2012, \prd, 85, 122006

\bibitem[{{Amaral-Rogers}(2014{\natexlab{a}})}]{GCN16232}
{Amaral-Rogers}, A. 2014{\natexlab{a}}, GCN, 16232, 1

\bibitem[{{Amaral-Rogers}(2014{\natexlab{b}})}]{GCN16254}
---. 2014{\natexlab{b}}, GCN, 16254, 1

\bibitem[{Amati(2006)}]{2006MNRAS.372..233A}
Amati, L. 2006, \mnras, 372, 233

\bibitem[{{Amati} {et~al.}(2013){Amati}, {Dichiara}, {Frontera}, {Guidorzi},
  {Izzo}, \& {Della Valle}}]{GCN15025}
{Amati}, L., {et~al.} 2013, GCN, 15025, 1

\bibitem[{Amati {et~al.}(2009)Amati, {Frontera}, \&
  {Guidorzi}}]{2009A&A...508..173A}
Amati, L., {Frontera}, F., \& {Guidorzi}, C. 2009, \aap, 508, 173

\bibitem[{Amati {et~al.}(2008)Amati, {Guidorzi}, {Frontera}, {Della Valle},
  {Finelli}, {Landi}, \& {Montanari}}]{2008MNRAS.391..577A}
Amati, L., {et~al.} 2008, \mnras, 391, 577

\bibitem[{Amati {et~al.}(2002)Amati, {Frontera}, {Tavani}, {in't Zand},
  {Antonelli}, {Costa}, {Feroci}, {Guidorzi}, {Heise}, {Masetti}, {Montanari},
  {Nicastro}, {Palazzi}, {Pian}, {Piro}, \& {Soffitta}}]{AmatiRelation}
---. 2002, \aap, 390, 81

\bibitem[{{Anderson} {et~al.}(2014{\natexlab{a}}){Anderson}, {Staley},
  {Fender}, {Kasliwal}, \& {Horesh}}]{GCN16725}
{Anderson}, G.~E., {et~al.} 2014{\natexlab{a}}, GCN, 16725, 1

\bibitem[{{Anderson} {et~al.}(2014{\natexlab{b}}){Anderson}, {van der Horst},
  {Staley}, {Fender}, {Wijers}, {Scaife}, {Rumsey}, {Titterington},
  {Rowlinson}, \& {Saunders}}]{GRB130427A-AMI}
---. 2014{\natexlab{b}}, \mnras, 440, 2059

\bibitem[{{Anderson} {et~al.}(2001){Anderson}, {Brady}, {Creighton}, \&
  {Flanagan}}]{ExcessPower}
{Anderson}, W.~G., {et~al.} 2001, \prd, 63, 042003

\bibitem[{{Atteia}(2003)}]{EmpiricalRedshiftIndicators}
{Atteia}, J.-L. 2003, \aap, 407, L1

\bibitem[{{Atwood} {et~al.}(2009){Atwood}, {Abdo}, {Ackermann}, {Althouse},
  {Anderson}, {Axelsson}, {Baldini}, {Ballet}, {Band}, {Barbiellini}, \&
  et~al.}]{LAT}
{Atwood}, W.~B., {et~al.} 2009, \apj, 697, 1071

\bibitem[{Babak {et~al.}(2013)Babak, Biswas, Brady, Brown, Cannon, Capano,
  Clayton, Cokelaer, Creighton, Dent, Dietz, Fairhurst, Fotopoulos, Gonz\'alez,
  Hanna, Harry, Jones, Keppel, McKechan, Pekowsky, Privitera, Robinson,
  Rodriguez, Sathyaprakash, Sengupta, Vallisneri, Vaulin, \& Weinstein}]{ihope}
Babak, S., {et~al.} 2013, Phys. Rev. D, 87, 024033

\bibitem[{{Balasubramanian} {et~al.}(1996){Balasubramanian}, {Sathyaprakash},
  \& {Dhurandhar}}]{1996PhRvD..53.3033B}
{Balasubramanian}, R., {Sathyaprakash}, B.~S., \& {Dhurandhar}, S.~V. 1996,
  \prd, 53, 3033

\bibitem[{{Band} \& {Preece}(2005)}]{2005ApJ...627..319B}
{Band}, D.~L., \& {Preece}, R.~D. 2005, \apj, 627, 319

\bibitem[{Barankin(1949)}]{barankin1949locally}
Barankin, E. 1949, Annals Math. Stat., 477

\bibitem[{{Barnes} \& {Kasen}(2013{\natexlab{a}})}]{KilonovaHighOpacities}
{Barnes}, J., \& {Kasen}, D. 2013{\natexlab{a}}, \apj, 775, 18

\bibitem[{{Barnes} \&
  {Kasen}(2013{\natexlab{b}})}]{BarnesKasenKilonovaOpacities}
---. 2013{\natexlab{b}}, \apj, 775, 18

\bibitem[{Barsotti \& Fritschel(2012)}]{EarlyAdvancedLIGONoiseCurves}
Barsotti, L., \& Fritschel, P. 2012, LIGO-T1200307-v4

\bibitem[{{Barthelmy} {et~al.}(2005){Barthelmy}, {Barbier}, {Cummings},
  {Fenimore}, {Gehrels}, {Hullinger}, {Krimm}, {Markwardt}, {Palmer},
  {Parsons}, {Sato}, {Suzuki}, {Takahashi}, {Tashiro}, \& {Tueller}}]{BAT}
{Barthelmy}, S.~D., {et~al.} 2005, \ssr, 120, 143

\bibitem[{{Belczynski} {et~al.}(2011){Belczynski}, {Bulik}, \&
  {Bailyn}}]{CygX1NSBHBBH}
{Belczynski}, K., {Bulik}, T., \& {Bailyn}, C. 2011, \apjl, 742, L2

\bibitem[{{Belczynski} {et~al.}(2013){Belczynski}, {Bulik}, {Mandel},
  {Sathyaprakash}, {Zdziarski}, \& {Miko{\l}ajewska}}]{CygX3NSBHBBH}
{Belczynski}, K., {et~al.} 2013, \apj, 764, 96

\bibitem[{Belczynski {et~al.}(2002)Belczynski, Kalogera, \&
  Bulik}]{0004-637X-572-1-407}
Belczynski, K., Kalogera, V., \& Bulik, T. 2002, \apj, 572, 407

\bibitem[{{Bellm}(2014)}]{ZTFBellm}
{Bellm}, E.~C. 2014, arXiv:1410.8185

\bibitem[{{Berry} {et~al.}(2014){Berry}, {Mandel}, {Middleton}, {Singer},
  {Urban}, {Vecchio}, {Vitale}, {Cannon}, {Farr}, {Farr}, {Graff}, {Hanna},
  {Haster}, {Mohapatra}, {Pankow}, {Price}, {Sidery}, \&
  {Veitch}}]{BerryLocalization}
{Berry}, C.~P.~L., {et~al.} 2014, arXiv:1411.6934

\bibitem[{{Bertin}(2006)}]{SCAMP}
{Bertin}, E. 2006, in ASP Conf. Series, Vol. 351, Astronomical Data Analysis
  Software and Systems XV, ed. C.~{Gabriel}, C.~{Arviset}, D.~{Ponz}, \&
  S.~{Enrique}, 112

\bibitem[{{Bertin} {et~al.}(2002){Bertin}, {Mellier}, {Radovich}, {Missonnier},
  {Didelon}, \& {Morin}}]{SWarp}
{Bertin}, E., {et~al.} 2002, in ASP Conf. Series, Vol. 281, Astronomical Data
  Analysis Software and Systems XI, ed. D.~A. {Bohlender}, D.~{Durand}, \&
  T.~H. {Handley}, 228

\bibitem[{{Bhalerao} \& {Sahu}(2014)}]{GCN16244}
{Bhalerao}, V., \& {Sahu}, D.~K. 2014, GCN, 16244, 1

\bibitem[{{Bhalerao} {et~al.}(2014){Bhalerao}, {Singer}, {Kasliwal}, {Cenko},
  {Horesh}, \& {Perley}}]{GCN16442}
{Bhalerao}, V., {et~al.} 2014, GCN, 16442, 1

\bibitem[{{Blackman} {et~al.}(2014){Blackman}, {Szilagyi}, {Galley}, \&
  {Tiglio}}]{Blackman:2014maa}
{Blackman}, J., {et~al.} 2014, \prl, 113, 021101

\bibitem[{{Bloom} {et~al.}(2001){Bloom}, {Frail}, \& {Sari}}]{KCorrectionGRBs}
{Bloom}, J.~S., {Frail}, D.~A., \& {Sari}, R. 2001, \aj, 121, 2879

\bibitem[{{Bloom} {et~al.}(2012){Bloom}, {Richards}, {Nugent}, {Quimby},
  {Kasliwal}, {Starr}, {Poznanski}, {Ofek}, {Cenko}, {Butler}, {Kulkarni},
  {Gal-Yam}, \& {Law}}]{BloomMachineLearning}
{Bloom}, J.~S., {et~al.} 2012, \pasp, 124, 1175

\bibitem[{{Bock} {et~al.}(2006){Bock}, {Bolatto}, {Hawkins}, {Kemball}, {Lamb},
  {Plambeck}, {Pound}, {Scott}, {Woody}, \& {Wright}}]{CARMA1}
{Bock}, D.~C.-J., {et~al.} 2006, in SPIE Conf. Series, Vol. 6267, SPIE Conf.
  Series, 13

\bibitem[{Bork {et~al.}(2001)Bork, Abbott, Barker, \& Heefner}]{Bork2001}
Bork, R., {et~al.} 2001, in Proc. 8th Int. Conf. on Accelerator and Large
  Experimental Physics Control Systems, ed. H.~Shoaee, Vol. C011127 (Menlo
  Park, CA: Stanford Linear Accelerator Center Technical Publications), 19--23

\bibitem[{{Brink} {et~al.}(2013){Brink}, {Richards}, {Poznanski}, {Bloom},
  {Rice}, {Negahban}, \& {Wainwright}}]{RB2}
{Brink}, H., {et~al.} 2013, \mnras, 435, 1047

\bibitem[{{Bromberg} {et~al.}(2011){Bromberg}, {Nakar}, \&
  {Piran}}]{AreLowLuminosityBurstsGeneratedByRelativisticJets}
{Bromberg}, O., {Nakar}, E., \& {Piran}, T. 2011, \apjl, 739, L55

\bibitem[{{Brown} {et~al.}(2012){Brown}, {Harry}, {Lundgren}, \&
  {Nitz}}]{DetectingBNSSystemsWithSpin}
{Brown}, D.~A., {et~al.} 2012, \prd, 86, 084017

\bibitem[{Bue {et~al.}(2014)Bue, Wagstaff, Rebbapragada, Thompson, \&
  Tang}]{RB4}
Bue, B.~D., {et~al.} 2014, in Proceedings of the 2014 conference on Big Data
  from Space (BiDS'14)

\bibitem[{{Bufano} {et~al.}(2011){Bufano}, {Benetti}, {Sollerman}, {Pian}, \&
  {Cupani}}]{GRB100316D-SN2010bh-2}
{Bufano}, F., {et~al.} 2011, Astronomische Nachrichten, 332, 262

\bibitem[{{Buonanno} \& {Chen}(2001)}]{QuantumNoiseSecondGeneration}
{Buonanno}, A., \& {Chen}, Y. 2001, \prd, 64, 042006

\bibitem[{{Buonanno} {et~al.}(2003){Buonanno}, {Chen}, \&
  {Vallisneri}}]{SpinTaylorT4}
{Buonanno}, A., {Chen}, Y., \& {Vallisneri}, M. 2003, \prd, 67, 104025

\bibitem[{{Buonanno} {et~al.}(2006){Buonanno}, {Chen}, \&
  {Vallisneri}}]{SpinTaylorT4Erratum}
---. 2006, \prd, 74, 029904

\bibitem[{Buonanno {et~al.}(2009)Buonanno, Iyer, Ochsner, Pan, \&
  Sathyaprakash}]{TaylorF2}
Buonanno, A., {et~al.} 2009, \prd, 80, 084043

\bibitem[{{Burgay} {et~al.}(2003){Burgay}, {D'Amico}, {Possenti}, {Manchester},
  {Lyne}, {Joshi}, {McLaughlin}, {Kramer}, {Sarkissian}, {Camilo}, {Kalogera},
  {Kim}, \& {Lorimer}}]{2003Natur.426..531B}
{Burgay}, M., {et~al.} 2003, \nat, 426, 531

\bibitem[{{Burns}(2014)}]{GCN16363}
{Burns}, E. 2014, GCN, 16363, 1

\bibitem[{{Burrows} {et~al.}(2005){Burrows}, {Hill}, {Nousek}, {Kennea},
  {Wells}, {Osborne}, {Abbey}, {Beardmore}, {Mukerjee}, {Short}, {Chincarini},
  {Campana}, {Citterio}, {Moretti}, {Pagani}, {Tagliaferri}, {Giommi},
  {Capalbi}, {Tamburelli}, {Angelini}, {Cusumano}, {Br{\"a}uninger}, {Burkert},
  \& {Hartner}}]{XRT}
{Burrows}, D.~N., {et~al.} 2005, \ssr, 120, 165

\bibitem[{Buskulic {et~al.}(2010)Buskulic, the LIGO Scientific~Collaboration,
  \& the Virgo~Collaboration}]{Buskulic2010}
Buskulic, D., the LIGO Scientific~Collaboration, \& the Virgo~Collaboration.
  2010, Class. Quantum Grav., 27, 194013

\bibitem[{{Butler} {et~al.}(2013{\natexlab{a}}){Butler}, {Watson}, {Kutyrev},
  {Lee}, {Richer}, {Klein}, {Fox}, {Prochaska}, {Bloom}, {Cucchiara}, {Troja},
  {Littlejohns}, {Ramirez-Ruiz}, {de Diego}, {Georgiev}, {Gonzalez},
  {Roman-Zuniga}, {Gehrels}, \& {Moseley}}]{GCN14993}
{Butler}, N., {et~al.} 2013{\natexlab{a}}, GCN, 14993, 1

\bibitem[{{Butler} {et~al.}(2013{\natexlab{b}}){Butler}, {Watson}, {Kutyrev},
  {Lee}, {Richer}, {Klein}, {Fox}, {Prochaska}, {Bloom}, {Cucchiara}, {Troja},
  {Littlejohns}, {Ramirez-Ruiz}, {de Diego}, {Georgiev}, {Gonzalez},
  {Roman-Zuniga}, {Gehrels}, \& {Moseley}}]{GCN14980}
---. 2013{\natexlab{b}}, GCN, 14980, 1

\bibitem[{{Butler} {et~al.}(2014{\natexlab{a}}){Butler}, {Watson}, {Kutyrev},
  {Lee}, {Richer}, {Klein}, {Fox}, {Prochaska}, {Bloom}, {Cucchiara}, {Troja},
  {Littlejohns}, {Ramirez-Ruiz}, {de Diego}, {Georgiev}, {Gonzalez},
  {Roman-Zuniga}, {Gehrels}, \& {Moseley}}]{GCN16246}
---. 2014{\natexlab{a}}, GCN, 16246, 1

\bibitem[{{Butler} {et~al.}(2014{\natexlab{b}}){Butler}, {Watson}, {Kutyrev},
  {Lee}, {Richer}, {Klein}, {Fox}, {Prochaska}, {Bloom}, {Cucchiara}, {Troja},
  {Littlejohns}, {Ramirez-Ruiz}, {de Diego}, {Georgiev}, {Gonzalez},
  {Roman-Zuniga}, {Gehrels}, \& {Moseley}}]{GCN16236}
---. 2014{\natexlab{b}}, GCN, 16236, 1

\bibitem[{{Butler} {et~al.}(2010){Butler}, {Bloom}, \&
  {Poznanski}}]{GRBLuminosityFunction}
{Butler}, N.~R., {Bloom}, J.~S., \& {Poznanski}, D. 2010, \apj, 711, 495

\bibitem[{{Butler} {et~al.}(2009){Butler}, {Kocevski}, \&
  {Bloom}}]{2009ApJ...694...76B}
{Butler}, N.~R., {Kocevski}, D., \& {Bloom}, J.~S. 2009, \apj, 694, 76

\bibitem[{{Butler} {et~al.}(2007){Butler}, {Kocevski}, {Bloom}, \&
  {Curtis}}]{2007ApJ...671..656B}
{Butler}, N.~R., {et~al.} 2007, \apj, 671, 656

\bibitem[{{Cabrera} {et~al.}(2007){Cabrera}, {Firmani}, {Avila-Reese},
  {Ghirlanda}, {Ghisellini}, \& {Nava}}]{2007MNRAS.382..342C}
{Cabrera}, J.~I., {et~al.} 2007, \mnras, 382, 342

\bibitem[{{Calabretta} \& {Greisen}(2002)}]{WCS2}
{Calabretta}, M.~R., \& {Greisen}, E.~W. 2002, \aap, 395, 1077

\bibitem[{{Cameron}(2011)}]{BinomialConfidenceIntervalsAstronomy}
{Cameron}, E. 2011, \pasa, 28, 128

\bibitem[{{Camp} {et~al.}(2013){Camp}, {Barthelmy}, {Blackburn}, {Carpenter},
  {Gehrels}, {Kanner}, {Marshall}, {Racusin}, \& {Sakamoto}}]{ISSLobster}
{Camp}, J., {et~al.} 2013, Experimental Astronomy, 36, 505

\bibitem[{{Campana} {et~al.}(2006){Campana}, {Mangano}, {Blustin}, {Brown},
  {Burrows}, {Chincarini}, {Cummings}, {Cusumano}, {Della Valle}, {Malesani},
  {M{\'e}sz{\'a}ros}, {Nousek}, {Page}, {Sakamoto}, {Waxman}, {Zhang}, {Dai},
  {Gehrels}, {Immler}, {Marshall}, {Mason}, {Moretti}, {O'Brien}, {Osborne},
  {Page}, {Romano}, {Roming}, {Tagliaferri}, {Cominsky}, {Giommi}, {Godet},
  {Kennea}, {Krimm}, {Angelini}, {Barthelmy}, {Boyd}, {Palmer}, {Wells}, \&
  {White}}]{GRB060218ShockBreakoutNature}
{Campana}, S., {et~al.} 2006, \nat, 442, 1008

\bibitem[{{Canizares} {et~al.}(2013){Canizares}, {Field}, {Gair}, \&
  {Tiglio}}]{roq-pe}
{Canizares}, P., {et~al.} 2013, \prd, 87, 124005

\bibitem[{Cannon {et~al.}(2012)Cannon, Cariou, Chapman, Crispin-Ortuzar,
  Fotopoulos, {et~al.}}]{Cannon:2011vi}
Cannon, K., {et~al.} 2012, \apj, 748, 136

\bibitem[{Cannon {et~al.}(2010)Cannon, Chapman, Hanna, Keppel, Searle, \&
  Weinstein}]{Cannon:2010p10398}
---. 2010, \prd, 82, 44025

\bibitem[{Cannon {et~al.}(2013)Cannon, Hanna, \& Keppel}]{Cannon:2012zt}
Cannon, K., Hanna, C., \& Keppel, D. 2013, \prd, 88, 024025

\bibitem[{Cannon {et~al.}(2011)Cannon, Hanna, Keppel, \&
  Searle}]{svd-compdetstat}
Cannon, K., {et~al.} 2011, \prd, 83, 084053

\bibitem[{{Cano} {et~al.}(2014){Cano}, {de Ugarte Postigo}, {Pozanenko},
  {Butler}, {Th{\"o}ne}, {Guidorzi}, {Kr{\"u}hler}, {Gorosabel}, {Jakobsson},
  {Leloudas}, {Malesani}, {Hjorth}, {Melandri}, {Mundell}, {Wiersema},
  {D'Avanzo}, {Schulze}, {Gomboc}, {Johansson}, {Zheng}, {Kann}, {Knust},
  {Varela}, {Akerlof}, {Bloom}, {Burkhonov}, {Cooke}, {de Diego}, {Dhungana},
  {Farina}, {Ferrante}, {Flewelling}, {Fox}, {Fynbo}, {Gehrels}, {Georgiev},
  {Gonz{\'a}lez}, {Greiner}, {G{\"u}ver}, {Hartoog}, {Hatch}, {Jelinek},
  {Kehoe}, {Klose}, {Klunko}, {Kopa{\v c}}, {Kutyrev}, {Krugly}, {Lee},
  {Levan}, {Linkov}, {Matkin}, {Minikulov}, {Molotov}, {Prochaska}, {Richer},
  {Rom{\'a}n-Z{\'u}{\~n}iga}, {Rumyantsev}, {S{\'a}nchez-Ram{\'{\i}}rez},
  {Steele}, {Tanvir}, {Volnova}, {Watson}, {Xu}, \&
  {Yuan}}]{GRB130215A-SN2013ez}
{Cano}, Z., {et~al.} 2014, \aap, 568, A19

\bibitem[{{Cenko} {et~al.}(2013{\natexlab{a}}){Cenko}, {Gal-Yam}, {Kasliwal},
  {Stern}, {Markey}, {Alduena}, {Alduena}, \& {Kuo}}]{GCN14998}
{Cenko}, S.~B., {et~al.} 2013{\natexlab{a}}, GCN, 14998, 1

\bibitem[{Cenko {et~al.}(2014)Cenko, Kasliwal, Perley, Jewitt,
  {et~al.}}]{iPTF14yb}
Cenko, S.~B., {et~al.} 2014, GCN, 15883, 1

\bibitem[{{Cenko} {et~al.}(2012){Cenko}, {Ofek}, \& {Nugent}}]{GCN13489}
{Cenko}, S.~B., {Ofek}, E.~O., \& {Nugent}, P.~E. 2012, GCN, 13489, 1

\bibitem[{{Cenko} {et~al.}(2006){Cenko}, {Fox}, {Moon}, {Harrison}, {Kulkarni},
  {Henning}, {Guzman}, {Bonati}, {Smith}, {Thicksten}, {Doyle}, {Petrie},
  {Gal-Yam}, {Soderberg}, {Anagnostou}, \& {Laity}}]{P60Automation}
{Cenko}, S.~B., {et~al.} 2006, \pasp, 118, 1396

\bibitem[{{Cenko} {et~al.}(2009){Cenko}, {Kelemen}, {Harrison}, {Fox},
  {Kulkarni}, {Kasliwal}, {Ofek}, {Rau}, {Gal-Yam}, {Frail}, \&
  {Moon}}]{DarkBurstsSwiftEra}
---. 2009, \apj, 693, 1484

\bibitem[{{Cenko} {et~al.}(2013{\natexlab{b}}){Cenko}, {Kulkarni}, {Horesh},
  {Corsi}, {Fox}, {Carpenter}, {Frail}, {Nugent}, {Perley}, {Gruber},
  {Gal-Yam}, {Groot}, {Hallinan}, {Ofek}, {Rau}, {MacLeod}, {Miller}, {Bloom},
  {Filippenko}, {Kasliwal}, {Law}, {Morgan}, {Polishook}, {Poznanski},
  {Quimby}, {Sesar}, {Shen}, {Silverman}, \& {Sternberg}}]{PTF11agg}
---. 2013{\natexlab{b}}, \apj, 769, 130

\bibitem[{{Cenko} {et~al.}(2014){Cenko}, {Kasliwal}, {Perley}, {Jewitt},
  {Filippenko}, {Horesh}, {De Cia}, {Rubin}, {Yaron}, {Arcavi}, {Cao}, \&
  {Nugent}}]{GCN15883}
---. 2014, GCN, 15883, 1

\bibitem[{{Chandra} \& {Frail}(2012)}]{RadioSelectedGRBAfterglows}
{Chandra}, P., \& {Frail}, D.~A. 2012, \apj, 746, 156

\bibitem[{{Cheung} {et~al.}(2013){Cheung}, {Vianello}, {Zhu}, {Racusin},
  {Connaughton}, \& {Carpenter}}]{GCN14971}
{Cheung}, T., {et~al.} 2013, GCN, 14971, 1

\bibitem[{{Chevalier} \& {Li}(1999)}]{AfterglowSpectraWind}
{Chevalier}, R.~A., \& {Li}, Z.-Y. 1999, \apjl, 520, L29

\bibitem[{{Chornock} {et~al.}(2010){Chornock}, {Berger}, {Levesque},
  {Soderberg}, {Foley}, {Fox}, {Frebel}, {Simon}, {Bochanski}, {Challis},
  {Kirshner}, {Podsiadlowski}, {Roth}, {Rutledge}, {Schmidt}, {Sheppard}, \&
  {Simcoe}}]{GRB100316D-SN2010bh-1}
{Chornock}, R., {et~al.} 2010, arXiv:1004.2262

\bibitem[{Cokelaer(2007)}]{PhysRevD.76.102004}
Cokelaer, T. 2007, \prd, 76, 102004

\bibitem[{{Collazzi} \& {Connaughton}(2013)}]{GCN14972}
{Collazzi}, A.~C., \& {Connaughton}, V. 2013, GCN, 14972, 1

\bibitem[{{Collazzi} {et~al.}(2012){Collazzi}, {Schaefer}, {Goldstein}, \&
  {Preece}}]{2012ApJ...747...39C}
{Collazzi}, A.~C., {et~al.} 2012, \apj, 747, 39

\bibitem[{{Connaughton} {et~al.}(2014){Connaughton}, {Briggs}, {Goldstein},
  {Meegan}, {Paciesas}, {Preece}, {Wilson-Hodge}, {Gibby}, {Greiner}, {Gruber},
  {Jenke}, {Kippen}, {Pelassa}, {Xiong}, {Yu}, {Bhat}, {Burgess}, {Byrne},
  {Fitzpatrick}, {Foley}, {Giles}, {Guiriec}, {van der Horst}, {von Kienlin},
  {McBreen}, {McGlynn}, {Tierney}, \& {Zhang}}]{GBMLocalization}
{Connaughton}, V., {et~al.} 2014, arXiv:1411.2685

\bibitem[{{Corder} {et~al.}(2010){Corder}, {Wright}, \& {Carpenter}}]{CARMA2}
{Corder}, S.~A., {Wright}, M.~C.~H., \& {Carpenter}, J.~M. 2010, in SPIE Conf.
  Series, Vol. 7733, SPIE Conf. Series, 3

\bibitem[{Corsi \& Horesh(2014)}]{GCN16694}
Corsi, A., \& Horesh, A. 2014, GCN, 16694, 1

\bibitem[{{Corsi} {et~al.}(2012){Corsi}, {Ofek}, {Gal-Yam}, {Frail},
  {Poznanski}, {Mazzali}, {Kulkarni}, {Kasliwal}, {Arcavi}, {Ben-Ami}, {Cenko},
  {Filippenko}, {Fox}, {Horesh}, {Howell}, {Kleiser}, {Nakar}, {Rabinak},
  {Sari}, {Silverman}, {Xu}, {Bloom}, {Law}, {Nugent}, \& {Quimby}}]{PTF10vgv}
{Corsi}, A., {et~al.} 2012, \apjl, 747, L5

\bibitem[{{Costa} {et~al.}(1997){Costa}, {Frontera}, {Heise}, {Feroci}, {in't
  Zand}, {Fiore}, {Cinti}, {Dal Fiume}, {Nicastro}, {Orlandini}, {Palazzi},
  {Rapisarda\#}, {Zavattini}, {Jager}, {Parmar}, {Owens}, {Molendi},
  {Cusumano}, {Maccarone}, {Giarrusso}, {Coletta}, {Antonelli}, {Giommi},
  {Muller}, {Piro}, \& {Butler}}]{GRBsHaveXrayAfterglows}
{Costa}, E., {et~al.} 1997, \nat, 387, 783

\bibitem[{{Cucchiara}(2014)}]{GCN15652}
{Cucchiara}, A. 2014, GCN, 15652, 1

\bibitem[{{Cutri} \& {et al.}(2014)}]{ALLWISE}
{Cutri}, R.~M., \& {et al.} 2014, VizieR Online Data Catalog, 2328, 0

\bibitem[{{Dai} \& {Cheng}(2001)}]{AfterglowSpectraFlatSpectrum}
{Dai}, Z.~G., \& {Cheng}, K.~S. 2001, \apjl, 558, L109

\bibitem[{{Dai} {et~al.}(2004){Dai}, {Liang}, \& {Xu}}]{DaiGRBsStandardCandles}
{Dai}, Z.~G., {Liang}, E.~W., \& {Xu}, D. 2004, \apjl, 612, L101

\bibitem[{{Dalal} {et~al.}(2006){Dalal}, {Holz}, {Hughes}, \&
  {Jain}}]{DalalStandardSirens}
{Dalal}, N., {et~al.} 2006, \prd, 74, 063006

\bibitem[{{D'Avanzo} {et~al.}(2013){D'Avanzo}, {Porterfield}, {Burrows},
  {Siegel}, {Melandri}, \& {Evans}}]{GCN14973}
{D'Avanzo}, P., {et~al.} 2013, GCN, 14973, 1

\bibitem[{{De Pasquale}(2014{\natexlab{a}})}]{GCN16455}
{De Pasquale}, M. 2014{\natexlab{a}}, GCN, 16455, 1

\bibitem[{{De Pasquale}(2014{\natexlab{b}})}]{GCN16428}
---. 2014{\natexlab{b}}, GCN, 16428, 1

\bibitem[{{Del Pozzo} {et~al.}(2013){Del Pozzo}, {Li}, {Agathos}, {Van Den
  Broeck}, \& {Vitale}}]{MeasuringNSEquationOfStateCanBeDone}
{Del Pozzo}, W., {et~al.} 2013, \prl, 111, 071101

\bibitem[{{D'Elia} \& {Izzo}(2014)}]{GCN16464}
{D'Elia}, V., \& {Izzo}, L. 2014, GCN, 16464, 1

\bibitem[{{D'Elia} {et~al.}(2014){D'Elia}, {Marshall}, \&
  {Malesani}}]{GCN16451}
{D'Elia}, V., {Marshall}, F.~E., \& {Malesani}, D. 2014, GCN, 16451, 1

\bibitem[{{D'Elia} {et~al.}(2013){D'Elia}, {D'Avanzo}, {Melandri}, {Della
  Valle}, {Tagliaferri}, {Malesani}, {Pian}, {Antonelli}, {Piranomonte},
  {Harutyunyan}, \& {Carosati}}]{GCN15000}
{D'Elia}, V., {et~al.} 2013, GCN, 15000, 1

\bibitem[{{\relax DLMF}(2014)}]{NIST:DLMF}
{\relax DLMF}. 2014, {NIST Digital Library of Mathematical Functions},
  http://dlmf.nist.gov/, Release 1.0.8 of 2014-04-25, online companion to
  \cite{Olver:2010:NHMF}

\bibitem[{{Drake} {et~al.}(2009){Drake}, {Djorgovski}, {Mahabal}, {Beshore},
  {Larson}, {Graham}, {Williams}, {Christensen}, {Catelan}, {Boattini},
  {Gibbs}, {Hill}, \& {Kowalski}}]{CRTS}
{Drake}, A.~J., {et~al.} 2009, \apj, 696, 870

\bibitem[{{Dressler} {et~al.}(2011){Dressler}, {Bigelow}, {Hare}, {Sutin},
  {Thompson}, {Burley}, {Epps}, {Oemler}, {Bagish}, {Birk}, {Clardy},
  {Gunnels}, {Kelson}, {Shectman}, \& {Osip}}]{IMACS}
{Dressler}, A., {et~al.} 2011, \pasp, 123, 288

\bibitem[{East {et~al.}(2013)East, McWilliams, Levin, \&
  Pretorius}]{PhysRevD.87.043004}
East, W.~E., {et~al.} 2013, \prd, 87, 043004

\bibitem[{{Eichler} {et~al.}(1989){Eichler}, {Livio}, {Piran}, \&
  {Schramm}}]{1989Natur.340..126E}
{Eichler}, D., {et~al.} 1989, \nat, 340, 126

\bibitem[{{Evans} {et~al.}(2007){Evans}, {Beardmore}, {Page}, {Tyler},
  {Osborne}, {Goad}, {O'Brien}, {Vetere}, {Racusin}, {Morris}, {Burrows},
  {Capalbi}, {Perri}, {Gehrels}, \& {Romano}}]{SwiftXRTRepository}
{Evans}, P.~A., {et~al.} 2007, \aap, 469, 379

\bibitem[{{Evans} {et~al.}(2009){Evans}, {Beardmore}, {Page}, {Osborne},
  {O'Brien}, {Willingale}, {Starling}, {Burrows}, {Godet}, {Vetere}, {Racusin},
  {Goad}, {Wiersema}, {Angelini}, {Capalbi}, {Chincarini}, {Gehrels}, {Kennea},
  {Margutti}, {Morris}, {Mountford}, {Pagani}, {Perri}, {Romano}, \&
  {Tanvir}}]{2009MNRAS.397.1177E}
---. 2009, \mnras, 397, 1177

\bibitem[{{Evans} {et~al.}(2012){Evans}, {Fridriksson}, {Gehrels}, {Homan},
  {Osborne}, {Siegel}, {Beardmore}, {Handbauer}, {Gelbord}, {Kennea}, \&
  et~al.}]{SwiftFollowupGWTransients}
---. 2012, \apjs, 203, 28

\bibitem[{Fairhurst(2009)}]{FairhurstTriangulation}
Fairhurst, S. 2009, New J. Phys., 11, 123006

\bibitem[{Fairhurst(2011)}]{FairhurstLocalizationAdvancedLIGO}
---. 2011, Class. Quantum Grav., 28, 105021

\bibitem[{Fairhurst(2014)}]{FairhurstLIGOIndia}
---. 2014, J. Phys. Conf. Series, 484, 012007

\bibitem[{{Farr} \& {Kalogera}(2013)}]{HierarchicalParameterEstimation}
{Farr}, B., \& {Kalogera}, V. 2013, in APS Meeting Abstracts, 10003

\bibitem[{{Farr} {et~al.}(2014){Farr}, {Kalogera}, \&
  {Luijten}}]{KDEJumpProposal}
{Farr}, B., {Kalogera}, V., \& {Luijten}, E. 2014, \prd, 90, 024014

\bibitem[{Field {et~al.}(2011)Field, Galley, Herrmann, Hesthaven, Ochsner, \&
  Tiglio}]{rbf}
Field, S.~E., {et~al.} 2011, \prl, 106, 221102

\bibitem[{{Firmani} {et~al.}(2009){Firmani}, {Cabrera}, {Avila-Reese},
  {Ghisellini}, {Ghirlanda}, {Nava}, \& {Bosnjak}}]{2009MNRAS.393.1209F}
{Firmani}, C., {et~al.} 2009, \mnras, 393, 1209

\bibitem[{{Fitzpatrick} \& {Connaughton}(2014)}]{GCN16426}
{Fitzpatrick}, G., \& {Connaughton}, V. 2014, GCN, 16426, 1

\bibitem[{{Flesch}(2010)}]{ARXA}
{Flesch}, E. 2010, \pasa, 27, 283

\bibitem[{{Fong} {et~al.}(2012){Fong}, {Berger}, {Margutti}, {Zauderer},
  {Troja}, {Czekala}, {Chornock}, {Gehrels}, {Sakamoto}, {Fox}, \&
  {Podsiadlowski}}]{JetBreakShortGRB111020A}
{Fong}, W., {et~al.} 2012, \apj, 756, 189

\bibitem[{{Frail} {et~al.}(1997){Frail}, {Kulkarni}, {Nicastro}, {Feroci}, \&
  {Taylor}}]{GRBsHaveRadioAfterglows}
{Frail}, D.~A., {et~al.} 1997, \nat, 389, 261

\bibitem[{{Frater} {et~al.}(1992){Frater}, {Brooks}, \& {Whiteoak}}]{ATCA}
{Frater}, R.~H., {Brooks}, J.~W., \& {Whiteoak}, J.~B. 1992, Journal of
  Electrical and Electronics Engineering Australia, 12, 103

\bibitem[{{Freedman} \& {Waxman}(2001)}]{EnergyOfGammaRayBursts}
{Freedman}, D.~L., \& {Waxman}, E. 2001, \apj, 547, 922

\bibitem[{{Friedman} \& {Bloom}(2005)}]{FriedmanGRBsStandardCandles}
{Friedman}, A.~S., \& {Bloom}, J.~S. 2005, \apj, 627, 1

\bibitem[{{Fujiwara} {et~al.}(2014){Fujiwara}, {Yoshii}, {Saito}, {Tachibana},
  {Ohuchi}, {Kurita}, {Ono}, {Yatsu}, \& {Kawai}}]{GCN16259}
{Fujiwara}, T., {et~al.} 2014, GCN, 16259, 1

\bibitem[{Gabriel \& Zamir(1979)}]{WeightedSVD}
Gabriel, K.~R., \& Zamir, S. 1979, Technometrics, 21, 489

\bibitem[{{Gal-Yam} {et~al.}(2011){Gal-Yam}, {Kasliwal}, {Arcavi}, {Green},
  {Yaron}, {Ben-Ami}, {Xu}, {Sternberg}, {Quimby}, {Kulkarni}, {Ofek},
  {Walters}, {Nugent}, {Poznanski}, {Bloom}, {Cenko}, {Filippenko}, {Li},
  {Silverman}, {Walker}, {Sullivan}, {Maguire}, {Howell}, {Mazzali}, {Frail},
  {Bersier}, {James}, {Akerlof}, {Yuan}, {Law}, {Fox}, \&
  {Gehrels}}]{2011ApJ...736..159G}
{Gal-Yam}, A., {et~al.} 2011, \apj, 736, 159

\bibitem[{{Gal-Yam} {et~al.}(2014){Gal-Yam}, {Arcavi}, {Ofek}, {Ben-Ami},
  {Cenko}, {Kasliwal}, {Cao}, {Yaron}, {Tal}, {Silverman}, {Horesh}, {De Cia},
  {Taddia}, {Sollerman}, {Perley}, {Vreeswijk}, {Kulkarni}, {Nugent},
  {Filippenko}, \& {Wheeler}}]{iPTF13ast}
---. 2014, \nat, 509, 471

\bibitem[{{Galama} {et~al.}(1998){Galama}, {Vreeswijk}, {van Paradijs},
  {Kouveliotou}, {Augusteijn}, {B{\"o}hnhardt}, {Brewer}, {Doublier},
  {Gonzalez}, {Leibundgut}, {Lidman}, {Hainaut}, {Patat}, {Heise}, {in't Zand},
  {Hurley}, {Groot}, {Strom}, {Mazzali}, {Iwamoto}, {Nomoto}, {Umeda},
  {Nakamura}, {Young}, {Suzuki}, {Shigeyama}, {Koshut}, {Kippen}, {Robinson},
  {de Wildt}, {Wijers}, {Tanvir}, {Greiner}, {Pian}, {Palazzi}, {Frontera},
  {Masetti}, {Nicastro}, {Feroci}, {Costa}, {Piro}, {Peterson}, {Tinney},
  {Boyle}, {Cannon}, {Stathakis}, {Sadler}, {Begam}, \& {Ianna}}]{SN1998bw}
{Galama}, T.~J., {et~al.} 1998, \nat, 395, 670

\bibitem[{Gardner(1995)}]{gardner1995efficient}
Gardner, W.~G. 1995, J. Audio Eng. Soc, 43, 127

\bibitem[{{Gehrels} {et~al.}(2004){Gehrels}, {Chincarini}, {Giommi}, {Mason},
  {Nousek}, {Wells}, {White}, {Barthelmy}, {Burrows}, {Cominsky}, {Hurley},
  {Marshall}, {M{\'e}sz{\'a}ros}, {Roming}, {Angelini}, {Barbier}, {Belloni},
  {Campana}, {Caraveo}, {Chester}, {Citterio}, {Cline}, {Cropper}, {Cummings},
  {Dean}, {Feigelson}, {Fenimore}, {Frail}, {Fruchter}, {Garmire}, {Gendreau},
  {Ghisellini}, {Greiner}, {Hill}, {Hunsberger}, {Krimm}, {Kulkarni}, {Kumar},
  {Lebrun}, {Lloyd-Ronning}, {Markwardt}, {Mattson}, {Mushotzky}, {Norris},
  {Osborne}, {Paczynski}, {Palmer}, {Park}, {Parsons}, {Paul}, {Rees},
  {Reynolds}, {Rhoads}, {Sasseen}, {Schaefer}, {Short}, {Smale}, {Smith},
  {Stella}, {Tagliaferri}, {Takahashi}, {Tashiro}, {Townsley}, {Tueller},
  {Turner}, {Vietri}, {Voges}, {Ward}, {Willingale}, {Zerbi}, \&
  {Zhang}}]{Swift}
{Gehrels}, N., {et~al.} 2004, \apj, 611, 1005

\bibitem[{{Ghirlanda} {et~al.}(2005){Ghirlanda}, {Ghisellini}, \&
  {Firmani}}]{2005MNRAS.361L..10G}
{Ghirlanda}, G., {Ghisellini}, G., \& {Firmani}, C. 2005, \mnras, 361, L10

\bibitem[{{Ghirlanda} {et~al.}(2006){Ghirlanda}, {Ghisellini}, \&
  {Firmani}}]{GhirlandaGRBsStandardCandles}
---. 2006, New J. Phys., 8, 123

\bibitem[{{Goad} {et~al.}(2007){Goad}, {Tyler}, {Beardmore}, {Evans}, {Rosen},
  {Osborne}, {Starling}, {Marshall}, {Yershov}, {Burrows}, {Gehrels}, {Roming},
  {Moretti}, {Capalbi}, {Hill}, {Kennea}, {Koch}, \& {vanden
  Berk}}]{2007A&A...476.1401G}
{Goad}, M.~R., {et~al.} 2007, \aap, 476, 1401

\bibitem[{{Goldstein} {et~al.}(2012){Goldstein}, {Burgess}, {Preece}, {Briggs},
  {Guiriec}, {van der Horst}, {Connaughton}, {Wilson-Hodge}, {Paciesas},
  {Meegan}, {von Kienlin}, {Bhat}, {Bissaldi}, {Chaplin}, {Diehl}, {Fishman},
  {Fitzpatrick}, {Foley}, {Gibby}, {Giles}, {Greiner}, {Gruber}, {Kippen},
  {Kouveliotou}, {McBreen}, {McGlynn}, {Rau}, \&
  {Tierney}}]{GBMSpectralCatalog}
{Goldstein}, A., {et~al.} 2012, \apjs, 199, 19

\bibitem[{Gorosabel {et~al.}(2014{\natexlab{a}})Gorosabel, de~Ugarte~Postigo,
  Thoene, Perley, \& Rodriguez}]{GCN16671}
Gorosabel, J., {et~al.} 2014{\natexlab{a}}, GCN, 16671, 1

\bibitem[{Gorosabel {et~al.}(2014{\natexlab{b}})Gorosabel, {Sanchez-Lavega},
  {Perez-Hoyos}, {Hueso}, {Ugarte}, {Ordonez}, {Arandia}, \& {Perez de
  Nanclares}}]{GCN16227}
---. 2014{\natexlab{b}}, GCN, 16227, 1

\bibitem[{{G{\'o}rski} {et~al.}(2005){G{\'o}rski}, {Hivon}, {Banday},
  {Wandelt}, {Hansen}, {Reinecke}, \& {Bartelmann}}]{HEALPix}
{G{\'o}rski}, K.~M., {et~al.} 2005, \apj, 622, 759

\bibitem[{{Graff} {et~al.}(2012){Graff}, {Feroz}, {Hobson}, \&
  {Lasenby}}]{BAMBI}
{Graff}, P., {et~al.} 2012, \mnras, 421, 169

\bibitem[{{Graff} {et~al.}(2014){Graff}, {Feroz}, {Hobson}, \&
  {Lasenby}}]{SKYNET}
---. 2014, \mnras, 441, 1741

\bibitem[{{Greisen} \& {Calabretta}(2002)}]{WCS1}
{Greisen}, E.~W., \& {Calabretta}, M.~R. 2002, \aap, 395, 1061

\bibitem[{{Greisen} {et~al.}(2006){Greisen}, {Calabretta}, {Valdes}, \&
  {Allen}}]{WCS3}
{Greisen}, E.~W., {et~al.} 2006, \aap, 446, 747

\bibitem[{{Grover} {et~al.}(2014){Grover}, {Fairhurst}, {Farr}, {Mandel},
  {Rodriguez}, {Sidery}, \& {Vecchio}}]{Grover:2013}
{Grover}, K., {et~al.} 2014, \prd, 89, 042004

\bibitem[{{Hancock} {et~al.}(2013){Hancock}, {Murphy}, {Gaensler}, {Bell}, \&
  {Burlon}}]{GCN15395}
{Hancock}, P., {et~al.} 2013, GCN, 15395, 1

\bibitem[{Hanna(2008)}]{HannaThesis}
Hanna, C. 2008, PhD thesis, Louisiana State University

\bibitem[{{Hannam} {et~al.}(2014){Hannam}, {Schmidt}, {Boh{\'e}}, {Haegel},
  {Husa}, {Ohme}, {Pratten}, \& {P{\"u}rrer}}]{Hannam:2013waveform}
{Hannam}, M., {et~al.} 2014, \prl, 113, 151101

\bibitem[{Harry(2010)}]{aLIGO}
Harry, G.~M. 2010, Class. Quantum Grav., 27, 084006

\bibitem[{Harry \& the LIGO
  Scientific~Collaboration(2010)}]{0264-9381-27-8-084006}
Harry, G.~M., \& the LIGO Scientific~Collaboration. 2010, Class. Quantum Grav.,
  27, 084006

\bibitem[{{Harry} {et~al.}(2009){Harry}, {Allen}, \&
  {Sathyaprakash}}]{2009PhRvD..80j4014H}
{Harry}, I.~W., {Allen}, B., \& {Sathyaprakash}, B.~S. 2009, \prd, 80, 104014

\bibitem[{Harry \& Fairhurst(2011{\natexlab{a}})}]{harry-single-spin}
Harry, I.~W., \& Fairhurst, S. 2011{\natexlab{a}}, Class. Quantum Grav., 28,
  134008

\bibitem[{Harry \& Fairhurst(2011{\natexlab{b}})}]{PhysRevD.83.084002}
---. 2011{\natexlab{b}}, \prd, 83, 084002

\bibitem[{{Harry} {et~al.}(2014){Harry}, {Nitz}, {Brown}, {Lundgren},
  {Ochsner}, \& {Keppel}}]{Harry:2013tca}
{Harry}, I.~W., {et~al.} 2014, \prd, 89, 024010

\bibitem[{{Hessels} {et~al.}(2006){Hessels}, {Ransom}, {Stairs}, {Freire},
  {Kaspi}, \& {Camilo}}]{FastestSpinningMillisecondPulsar}
{Hessels}, J.~W.~T., {et~al.} 2006, Science, 311, 1901

\bibitem[{{Heussaff} {et~al.}(2013){Heussaff}, {Atteia}, \&
  {Zolnierowski}}]{FermiAmatiDebate}
{Heussaff}, V., {Atteia}, J.-L., \& {Zolnierowski}, Y. 2013, \aap, 557, A100

\bibitem[{Hinderer {et~al.}(2010)Hinderer, Lackey, Lang, \&
  Read}]{PhysRevD.81.123016}
Hinderer, T., {et~al.} 2010, \prd, 81, 123016

\bibitem[{{Hinshaw} {et~al.}(2013){Hinshaw}, {Larson}, {Komatsu}, {Spergel},
  {Bennett}, {Dunkley}, {Nolta}, {Halpern}, {Hill}, {Odegard}, {Page}, {Smith},
  {Weiland}, {Gold}, {Jarosik}, {Kogut}, {Limon}, {Meyer}, {Tucker}, {Wollack},
  \& {Wright}}]{WMAP9}
{Hinshaw}, G., {et~al.} 2013, \apjs, 208, 19

\bibitem[{{Hjorth} \& {Bloom}(2012)}]{HjorthGRBSNConnection}
{Hjorth}, J., \& {Bloom}, J.~S. 2012, in Cambridge Astrophysics Series,
  Vol.~51, Gamma-Ray Bursts, ed. C.~Kouveliotou, R.~A.~M.~J. Wijers, \&
  S.~Woosley (Cambridge: Cambridge University Press), 169--190

\bibitem[{{Holland} \& {Mangano}(2014)}]{GCN15673}
{Holland}, S.~T., \& {Mangano}, V. 2014, GCN, 15673, 1

\bibitem[{{Holz} \& {Hughes}(2005)}]{HolzStandardSirens}
{Holz}, D.~E., \& {Hughes}, S.~A. 2005, \apj, 629, 15

\bibitem[{Hooper {et~al.}(2010)Hooper, Wen, Blair, Chung, Chen, \&
  Luan}]{shaunIIR}
Hooper, S., {et~al.} 2010, in AIP Conf. Proc., Vol. 1246, Frontiers of
  Fundamental and Computational Physics: 10th Intl. Symp., ed. J.~G. Hartnett
  \& P.~C. Abbott (Melville, NY: AIP), 211--214

\bibitem[{{Horesh} {et~al.}(2014){Horesh}, {Singer}, {Cenko}, {Kasliwal}, \&
  {Perley}}]{GCN16266}
{Horesh}, A., {et~al.} 2014, GCN, 16266, 1

\bibitem[{{Howell} {et~al.}(2005){Howell}, {Sullivan}, {Perrett}, {Bronder},
  {Hook}, {Astier}, {Aubourg}, {Balam}, {Basa}, {Carlberg}, {Fabbro},
  {Fouchez}, {Guy}, {Lafoux}, {Neill}, {Pain}, {Palanque-Delabrouille},
  {Pritchet}, {Regnault}, {Rich}, {Taillet}, {Knop}, {McMahon}, {Perlmutter},
  \& {Walton}}]{Superfit}
{Howell}, D.~A., {et~al.} 2005, \apj, 634, 1190

\bibitem[{Hughey(2011)}]{HugheyGWPAW2011}
Hughey, B. 2011, in Gravitational Wave Physics and Astronomy Workshop (GWPAW)

\bibitem[{{Hulse} \& {Taylor}(1975)}]{1975ApJ...195L..51H}
{Hulse}, R.~A., \& {Taylor}, J.~H. 1975, \apjl, 195, L51

\bibitem[{{Hurley} {et~al.}(2013){Hurley}, {Goldsten}, {Connaughton}, {Briggs},
  {Meegan}, {Pelassa}, {Golenetskii}, {Aptekar}, {Mazets}, {Pal'Shin},
  {Frederiks}, {Svinkin}, {Cline}, {von Kienlin}, {Zhang}, {Rau}, {Savchenko},
  {Bozzo}, \& {Ferrigno}}]{GCN14974}
{Hurley}, K., {et~al.} 2013, GCN, 14974, 1

\bibitem[{{Hurley} {et~al.}(2014{\natexlab{a}}){Hurley}, {Golenetskii},
  {Aptekar}, {Pal'Shin}, {Frederiks}, {Svinkin}, {Cline}, {Mitrofanov},
  {Golovin}, {Litvak}, {Sanin}, {Boynton}, {Fellows}, {Harshman}, {Enos},
  {Starr}, {von Kienlin}, {Zhang}, {Rau}, {Savchenko}, {Bozzo}, {Ferrigno},
  {Barthelmy}, {Cummings}, {Gehrels}, {Krimm}, {Palmer}, {Connaughton},
  {Briggs}, {Meegan}, \& {Pelassa}}]{GCN15888}
---. 2014{\natexlab{a}}, GCN, 15888, 1

\bibitem[{{Hurley} {et~al.}(2014{\natexlab{b}}){Hurley}, {Goldsten},
  {Golenetskii}, {Aptekar}, {Pal'Shin}, {Frederiks}, {Svinkin}, {Cline}, {von
  Kienlin}, {Zhang}, {Rau}, {Savchenko}, {Bozzo}, {Ferrigno}, {Connaughton},
  {Briggs}, {Meegan}, {Pelassa}, {Barthelmy}, {Cummings}, {Gehrels}, {Krimm},
  \& {Palmer}}]{GCN16225}
---. 2014{\natexlab{b}}, GCN, 16225, 1

\bibitem[{{Hurley} {et~al.}(2014{\natexlab{c}}){Hurley}, {Goldsten},
  {Golenetskii}, {Aptekar}, {Pal'Shin}, {Frederiks}, {Svinkin}, {Cline},
  {Connaughton}, {Briggs}, {Meegan}, {Pelassa}, \& {Goldstein}}]{GCN16369}
---. 2014{\natexlab{c}}, GCN, 16369, 1

\bibitem[{{Ivezic} {et~al.}(2008){Ivezic}, {Tyson}, {Acosta}, {Allsman},
  {Anderson}, {Andrew}, {Angel}, {Axelrod}, {Barr}, {Becker}, {Becla},
  {Beldica}, {Blandford}, {Bloom}, {Borne}, {Brandt}, {Brown}, {Bullock},
  {Burke}, {Chandrasekharan}, {Chesley}, {Claver}, {Connolly}, {Cook},
  {Cooray}, {Covey}, {Cribbs}, {Cutri}, {Daues}, {Delgado}, {Ferguson},
  {Gawiser}, {Geary}, {Gee}, {Geha}, {Gibson}, {Gilmore}, {Gressler}, {Hogan},
  {Huffer}, {Jacoby}, {Jain}, {Jernigan}, {Jones}, {Juric}, {Kahn}, {Kalirai},
  {Kantor}, {Kessler}, {Kirkby}, {Knox}, {Krabbendam}, {Krughoff}, {Kulkarni},
  {Lambert}, {Levine}, {Liang}, {Lim}, {Lupton}, {Marshall}, {Marshall}, {May},
  {Miller}, {Mills}, {Monet}, {Neill}, {Nordby}, {O'Connor}, {Oliver},
  {Olivier}, {Olsen}, {Owen}, {Peterson}, {Petry}, {Pierfederici},
  {Pietrowicz}, {Pike}, {Pinto}, {Plante}, {Radeka}, {Rasmussen}, {Ridgway},
  {Rosing}, {Saha}, {Schalk}, {Schindler}, {Schneider}, {Schumacher}, {Sebag},
  {Seppala}, {Shipsey}, {Silvestri}, {Smith}, {Smith}, {Strauss}, {Stubbs},
  {Sweeney}, {Szalay}, {Thaler}, {Vanden Berk}, {Walkowicz}, {Warner},
  {Willman}, {Wittman}, {Wolff}, {Wood-Vasey}, {Yoachim}, {Zhan}, \& {for the
  LSST Collaboration}}]{LSST}
{Ivezic}, Z., {et~al.} 2008, arXiv:0805.2366

\bibitem[{Jackson(2003)}]{jackson2003user}
Jackson, J. 2003, A User's Guide to Principal Components, Wiley Series in
  Probability and Mathematical Statistics (New York: Wiley-Interscience)

\bibitem[{{Jenke}(2013)}]{GCN15331}
{Jenke}, P. 2013, GCN, 15331, 1

\bibitem[{{Jenke} \& {Xiong}(2014)}]{GCN15644}
{Jenke}, P., \& {Xiong}, S. 2014, GCN, 15644, 1

\bibitem[{Johnson \& Frigo(2007)}]{Johnson:2007p9654}
Johnson, S., \& Frigo, M. 2007, IEEE Trans. Signal Process., 55, 111

\bibitem[{Jovanovic-Dolecek(2002)}]{jovanovic2002multirate}
Jovanovic-Dolecek, G. 2002, Multirate Systems: Design and Applications
  (Hershey, PA: Idea Group Publ.)

\bibitem[{{Kaiser} {et~al.}(2010){Kaiser}, {Burgett}, {Chambers}, {Denneau},
  {Heasley}, {Jedicke}, {Magnier}, {Morgan}, {Onaka}, \& {Tonry}}]{PanSTARRS}
{Kaiser}, N., {et~al.} 2010, in SPIE Conf. Series, Vol. 7733, SPIE Conf.
  Series, 0

\bibitem[{Kalogera {et~al.}(2004)Kalogera, Kim, Lorimer, Burgay, D’Amico,
  Possenti, Manchester, Lyne, Joshi, McLaughlin, Kramer, Sarkissian, \&
  Camilo}]{KalogeraRates}
Kalogera, V., {et~al.} 2004, \apjl, 601, L179, erratum in 614, L137

\bibitem[{{Kann} {et~al.}(2010){Kann}, {Klose}, {Zhang}, {Malesani}, {Nakar},
  {Pozanenko}, {Wilson}, {Butler}, {Jakobsson}, {Schulze}, {Andreev},
  {Antonelli}, {Bikmaev}, {Biryukov}, {B{\"o}ttcher}, {Burenin}, {Castro
  Cer{\'o}n}, {Castro-Tirado}, {Chincarini}, {Cobb}, {Covino}, {D'Avanzo},
  {D'Elia}, {Della Valle}, {de Ugarte Postigo}, {Efimov}, {Ferrero}, {Fugazza},
  {Fynbo}, {G{\aa}lfalk}, {Grundahl}, {Gorosabel}, {Gupta}, {Guziy}, {Hafizov},
  {Hjorth}, {Holhjem}, {Ibrahimov}, {Im}, {Israel}, {Je{\'l}inek}, {Jensen},
  {Karimov}, {Khamitov}, {Kizilo{\v g}lu}, {Klunko}, {Kub{\'a}nek}, {Kutyrev},
  {Laursen}, {Levan}, {Mannucci}, {Martin}, {Mescheryakov}, {Mirabal},
  {Norris}, {Ovaldsen}, {Paraficz}, {Pavlenko}, {Piranomonte}, {Rossi},
  {Rumyantsev}, {Salinas}, {Sergeev}, {Sharapov}, {Sollerman}, {Stecklum},
  {Stella}, {Tagliaferri}, {Tanvir}, {Telting}, {Testa}, {Updike}, {Volnova},
  {Watson}, {Wiersema}, \& {Xu}}]{KanSwiftAfterglowsI}
{Kann}, D.~A., {et~al.} 2010, \apj, 720, 1513

\bibitem[{{Kann} {et~al.}(2011){Kann}, {Klose}, {Zhang}, {Covino}, {Butler},
  {Malesani}, {Nakar}, {Wilson}, {Antonelli}, {Chincarini}, {Cobb}, {D'Avanzo},
  {D'Elia}, {Della Valle}, {Ferrero}, {Fugazza}, {Gorosabel}, {Israel},
  {Mannucci}, {Piranomonte}, {Schulze}, {Stella}, {Tagliaferri}, \&
  {Wiersema}}]{KannTypeITypeIIOpticalAfterglows}
---. 2011, \apj, 734, 96

\bibitem[{Kanner {et~al.}(2008)Kanner, Huard, M\'{a}rka, Murphy, Piscionere,
  Reed, \& Shawhan}]{kanner2008}
Kanner, J., {et~al.} 2008, Class. Quantum Grav., 25, 184034

\bibitem[{{Kasliwal}(2011)}]{KasliwalThesis}
{Kasliwal}, M.~M. 2011, PhD thesis, California Institute of Technology

\bibitem[{{Kasliwal} {et~al.}(2014){Kasliwal}, {Cenko}, \& {Singer}}]{GCN16425}
{Kasliwal}, M.~M., {Cenko}, S.~B., \& {Singer}, L.~P. 2014, GCN, 16425, 1

\bibitem[{{Kasliwal} \& {Nissanke}(2014)}]{KasliwalTwoDetectors}
{Kasliwal}, M.~M., \& {Nissanke}, S. 2014, \apjl, 789, L5

\bibitem[{{Kasliwal} {et~al.}(2013){Kasliwal}, {Singer}, \& {Cenko}}]{GCN15324}
{Kasliwal}, M.~M., {Singer}, L.~P., \& {Cenko}, S.~B. 2013, GCN, 15324, 1

\bibitem[{{Kasliwal} {et~al.}(2012){Kasliwal}, {Kulkarni}, {Gal-Yam}, {Nugent},
  {Sullivan}, {Bildsten}, {Yaron}, {Perets}, {Arcavi}, {Ben-Ami}, {Bhalerao},
  {Bloom}, {Cenko}, {Filippenko}, {Frail}, {Ganeshalingam}, {Horesh}, {Howell},
  {Law}, {Leonard}, {Li}, {Ofek}, {Polishook}, {Poznanski}, {Quimby},
  {Silverman}, {Sternberg}, \& {Xu}}]{CaRichGapTransients}
{Kasliwal}, M.~M., {et~al.} 2012, \apj, 755, 161

\bibitem[{{Kelemen}(2014)}]{GCN16440}
{Kelemen}, J. 2014, GCN, 16440, 1

\bibitem[{{Kelly} {et~al.}(2013){Kelly}, {Filippenko}, {Fox}, {Zheng}, \&
  {Clubb}}]{13bxlhost}
{Kelly}, P.~L., {et~al.} 2013, \apjl, 775, L5

\bibitem[{{Keppel}(2012)}]{2012PhRvD..86l3010K}
{Keppel}, D. 2012, \prd, 86, 123010

\bibitem[{{Keppel}(2013)}]{2013arXiv1307.4158K}
---. 2013, arXiv:1307.4158

\bibitem[{{Kimble} {et~al.}(2002){Kimble}, {Levin}, {Matsko}, {Thorne}, \&
  {Vyatchanin}}]{FDSqueezing}
{Kimble}, H.~J., {et~al.} 2002, \prd, 65, 022002

\bibitem[{Klimenko {et~al.}(2011)Klimenko, Vedovato, Drago, Mazzolo,
  Mitselmakher, Pankow, Prodi, Re, Salemi, \& Yakushin}]{CWBLocalization}
Klimenko, S., {et~al.} 2011, Phys. Rev. D, 83, 102001

\bibitem[{{Klose} {et~al.}(2013){Klose}, {Nicuesa Guelbenzu}, {Kruehler},
  {Greiner}, {Kann}, {Rau}, {Olivares}, \& {Schulze}}]{GCN15320}
{Klose}, S., {et~al.} 2013, GCN, 15320, 1

\bibitem[{{Kocevski}(2012)}]{2012ApJ...747..146K}
{Kocevski}, D. 2012, \apj, 747, 146

\bibitem[{{Kopparapu} {et~al.}(2008){Kopparapu}, {Hanna}, {Kalogera},
  {O'Shaughnessy}, {Gonz{\'a}lez}, {Brady}, \&
  {Fairhurst}}]{GWHostGalaxyCatalog}
{Kopparapu}, R.~K., {et~al.} 2008, \apj, 675, 1459

\bibitem[{{Krimm} {et~al.}(2009){Krimm}, {Yamaoka}, {Sugita}, {Ohno},
  {Sakamoto}, {Barthelmy}, {Gehrels}, {Hara}, {Norris}, {Ohmori}, {Onda},
  {Sato}, {Tanaka}, {Tashiro}, \& {Yamauchi}}]{2009ApJ...704.1405K}
{Krimm}, H.~A., {et~al.} 2009, \apj, 704, 1405

\bibitem[{{Kulkarni}(2012)}]{ZTF}
{Kulkarni}, S.~R. 2012, in IAU Symposium, Vol. 285, IAU Symposium, ed.
  E.~{Griffin}, R.~{Hanisch}, \& R.~{Seaman}, 55--61

\bibitem[{{Kulkarni}(2013)}]{iPTF}
{Kulkarni}, S.~R. 2013, ATEL, 4807, 1

\bibitem[{{Kulkarni} {et~al.}(1998){Kulkarni}, {Frail}, {Wieringa}, {Ekers},
  {Sadler}, {Wark}, {Higdon}, {Phinney}, \& {Bloom}}]{1998bwShockBreakout}
{Kulkarni}, S.~R., {et~al.} 1998, \nat, 395, 663

\bibitem[{{Laskar} {et~al.}(2013){Laskar}, {Berger}, {Zauderer}, {Margutti},
  {Soderberg}, {Chakraborti}, {Lunnan}, {Chornock}, {Chandra}, \&
  {Ray}}]{LaskarGRB130427A}
{Laskar}, T., {et~al.} 2013, \apj, 776, 119

\bibitem[{{Law} {et~al.}(2009){Law}, {Kulkarni}, {Dekany}, {Ofek}, {Quimby},
  {Nugent}, {Surace}, {Grillmair}, {Bloom}, {Kasliwal}, {Bildsten}, {Brown},
  {Cenko}, {Ciardi}, {Croner}, {Djorgovski}, {van Eyken}, {Filippenko}, {Fox},
  {Gal-Yam}, {Hale}, {Hamam}, {Helou}, {Henning}, {Howell}, {Jacobsen},
  {Laher}, {Mattingly}, {McKenna}, {Pickles}, {Poznanski}, {Rahmer}, {Rau},
  {Rosing}, {Shara}, {Smith}, {Starr}, {Sullivan}, {Velur}, {Walters}, \&
  {Zolkower}}]{PTFLaw}
{Law}, N.~M., {et~al.} 2009, \pasp, 121, 1395

\bibitem[{{Leloudas} {et~al.}(2013){Leloudas}, {Fynbo}, {Schulze}, {Xu},
  {Malesani}, {Geier}, {Cano}, \& {Jakobsson}}]{GCN14983}
{Leloudas}, G., {et~al.} 2013, GCN, 14983, 1

\bibitem[{{Li} \& {Paczy{\'n}ski}(1998)}]{Kilonova}
{Li}, L.-X., \& {Paczy{\'n}ski}, B. 1998, \apjl, 507, L59

\bibitem[{{Li} {et~al.}(2011){Li}, {Leaman}, {Chornock}, {Filippenko},
  {Poznanski}, {Ganeshalingam}, {Wang}, {Modjaz}, {Jha}, {Foley}, \&
  {Smith}}]{LickSupernovaLuminosityFunction}
{Li}, W., {et~al.} 2011, \mnras, 412, 1441

\bibitem[{{Li} \& {Hjorth}(2014)}]{LightCurvesGRBSNe}
{Li}, X., \& {Hjorth}, J. 2014, arXiv:1407.3506

\bibitem[{{Ligo Scientific Collaboration} {et~al.}(2011){Ligo Scientific
  Collaboration}, {Abadie}, {Abbott}, {Abbott}, {Abbott}, {Abernathy}, {Adams},
  {Adhikari}, {Affeldt}, {Allen}, \& et~al.}]{GEOSqueezing}
{Ligo Scientific Collaboration} {et~al.} 2011, Nature Physics, 7, 962

\bibitem[{{Lloyd-Ronning} \&
  {Zhang}(2004)}]{KineticEnergyRadiativeEfficiencyOfGammaRayBursts}
{Lloyd-Ronning}, N.~M., \& {Zhang}, B. 2004, \apj, 613, 477

\bibitem[{{Luan} {et~al.}(2012){Luan}, {Hooper}, {Wen}, \& {Chen}}]{linqingIIR}
{Luan}, J., {et~al.} 2012, \prd, 85, 102002

\bibitem[{{Lynch} {et~al.}(2014){Lynch}, {Vitale}, {Barsotti}, {Evans}, \&
  {Dwyer}}]{SqueezingParameterEstimation}
{Lynch}, R., {et~al.} 2014, arXiv:1410.8503

\bibitem[{Maggiore(2008)}]{maggiore2008gravitational}
Maggiore, M. 2008, Gravitational Waves: Volume 1: Theory and Experiments,
  Gravitational Waves (OUP Oxford)

\bibitem[{{Malesani} {et~al.}(2014){Malesani}, {Xu}, {D'Avanzo}, {Palazzi}, \&
  {Perna}}]{GCN16229}
{Malesani}, D., {et~al.} 2014, GCN, 16229, 1

\bibitem[{{Malesani} {et~al.}(2013){Malesani}, {Xu}, {Losada}, \&
  {Duval}}]{GCN15642}
---. 2013, GCN, 15642, 1

\bibitem[{{Manca} \& {Vallisneri}(2010)}]{2010PhRvD..81b4004M}
{Manca}, G.~M., \& {Vallisneri}, M. 2010, \prd, 81, 024004

\bibitem[{{Mangano}(2014)}]{GCN16412}
{Mangano}, V. 2014, GCN, 16412, 1

\bibitem[{{Mangano} \& {Burrows}(2014)}]{GCN16373}
{Mangano}, V., \& {Burrows}, D.~N. 2014, GCN, 16373, 1

\bibitem[{{Mangano} {et~al.}(2014{\natexlab{a}}){Mangano}, {Evans}, \&
  {Goad}}]{GCN16366}
{Mangano}, V., {Evans}, P.~A., \& {Goad}, M.~R. 2014{\natexlab{a}}, GCN, 16366,
  1

\bibitem[{{Mangano} {et~al.}(2014{\natexlab{b}}){Mangano}, {Page}, \&
  {Malesani}}]{GCN15648}
{Mangano}, V., {Page}, K., \& {Malesani}, D. 2014{\natexlab{b}}, GCN, 15648, 1

\bibitem[{Marion \& the Virgo~Collaboration(2003)}]{Marion2004}
Marion, F., \& the Virgo~Collaboration. 2003, in Proc. Rencontres de Moriond
  2003, ed. J.~Dumarchez, Les Arcs

\bibitem[{{Marshall} \& {Amarel-Rogers}(2014)}]{GCN16243}
{Marshall}, F.~E., \& {Amarel-Rogers}, A. 2014, GCN, 16243, 1

\bibitem[{{Masi}(2014)}]{GCN16235}
{Masi}, G. 2014, GCN, 16235, 1

\bibitem[{McAulay \& Hofstetter(1971)}]{mcaulay1971barankin}
McAulay, R., \& Hofstetter, E. 1971, IEEE Trans. Inf. Theory, 17, 669

\bibitem[{McAulay \& Seidman(1969)}]{mcaulay1969barankin}
McAulay, R., \& Seidman, L. 1969, IEEE Trans. Inf. Theory, 15, 273

\bibitem[{{Meegan} {et~al.}(2009{\natexlab{a}}){Meegan}, {Lichti}, {Bhat},
  {Bissaldi}, {Briggs}, {Connaughton}, {Diehl}, {Fishman}, {Greiner}, {Hoover},
  {van der Horst}, {von Kienlin}, {Kippen}, {Kouveliotou}, {McBreen},
  {Paciesas}, {Preece}, {Steinle}, {Wallace}, {Wilson}, \&
  {Wilson-Hodge}}]{GBM}
{Meegan}, C., {et~al.} 2009{\natexlab{a}}, \apj, 702, 791

\bibitem[{{Meegan} {et~al.}(2009{\natexlab{b}}){Meegan}, {Lichti}, {Bhat},
  {Bissaldi}, {Briggs}, {Connaughton}, {Diehl}, {Fishman}, {Greiner}, {Hoover},
  {van der Horst}, {von Kienlin}, {Kippen}, {Kouveliotou}, {McBreen},
  {Paciesas}, {Preece}, {Steinle}, {Wallace}, {Wilson}, \&
  {Wilson-Hodge}}]{FermiGBM}
---. 2009{\natexlab{b}}, \apj, 702, 791

\bibitem[{{Meegan} {et~al.}(1992){Meegan}, {Fishman}, {Wilson}, {Horack},
  {Brock}, {Paciesas}, {Pendleton}, \& {Kouveliotou}}]{GRBsAreExtragalactic}
{Meegan}, C.~A., {et~al.} 1992, \nat, 355, 143

\bibitem[{Meers {et~al.}(1993)Meers, Krolak, \& Lobo}]{PhysRevD.47.2184}
Meers, B.~J., Krolak, A., \& Lobo, J.~A. 1993, \prd, 47, 2184

\bibitem[{{Melandri} {et~al.}(2012){Melandri}, {Pian}, {Ferrero}, {D'Elia},
  {Walker}, {Ghirlanda}, {Covino}, {Amati}, {D'Avanzo}, {Mazzali}, {Della
  Valle}, {Guidorzi}, {Antonelli}, {Bernardini}, {Bersier}, {Bufano},
  {Campana}, {Castro-Tirado}, {Chincarini}, {Deng}, {Filippenko}, {Fugazza},
  {Ghisellini}, {Kouveliotou}, {Maeda}, {Marconi}, {Masetti}, {Nomoto},
  {Palazzi}, {Patat}, {Piranomonte}, {Salvaterra}, {Saviane}, {Starling},
  {Tagliaferri}, {Tanaka}, \& {Vergani}}]{GRB120422A-SN2012bz-2}
{Melandri}, A., {et~al.} 2012, \aap, 547, A82

\bibitem[{{Messenger} \& {Read}(2012)}]{MessengerStandardSirens}
{Messenger}, C., \& {Read}, J. 2012, Physical Review Letters, 108, 091101

\bibitem[{{Metzger} {et~al.}(2015){Metzger}, {Bauswein}, {Goriely}, \&
  {Kasen}}]{KilonovaPrecursor}
{Metzger}, B.~D., {et~al.} 2015, \mnras, 446, 1115

\bibitem[{{Metzger} \& {Berger}(2012)}]{MostPromisingEMCounterpart}
{Metzger}, B.~D., \& {Berger}, E. 2012, \apj, 746, 48

\bibitem[{{Metzger} \& {Fern{\'a}ndez}(2014)}]{KilonovaRedOrBlue}
{Metzger}, B.~D., \& {Fern{\'a}ndez}, R. 2014, \mnras, 441, 3444

\bibitem[{{Miller} \& {Miller}(2014)}]{MassAndSpinNSAndBHReview}
{Miller}, M.~C., \& {Miller}, J.~M. 2014, arXiv:1408.4145

\bibitem[{{Mirabal} {et~al.}(2007){Mirabal}, {Halpern}, \&
  {O'Brien}}]{2007ApJ...661L.127M}
{Mirabal}, N., {Halpern}, J.~P., \& {O'Brien}, P.~T. 2007, \apjl, 661, L127

\bibitem[{{Modjaz} {et~al.}(2006){Modjaz}, {Stanek}, {Garnavich}, {Berlind},
  {Blondin}, {Brown}, {Calkins}, {Challis}, {Diamond-Stanic}, {Hao}, {Hicken},
  {Kirshner}, \& {Prieto}}]{GRB060218-SN2006aj-2}
{Modjaz}, M., {et~al.} 2006, \apjl, 645, L21

\bibitem[{{Moskvitin} {et~al.}(2014){Moskvitin}, {Makarov}, {Valeev},
  {Sokolov}, {Castro-Tirado}, {Gorosabel}, {Gorbovskoy}, {Denisenko}, \&
  {Lipunov}}]{GCN16228}
{Moskvitin}, A.~S., {et~al.} 2014, GCN, 16228, 1

\bibitem[{{Mulchaey} {et~al.}(2013){Mulchaey}, {Kasliwal}, {Arcavi}, {Bellm},
  \& {Kelson}}]{GCN14985}
{Mulchaey}, J., {et~al.} 2013, GCN, 14985, 1

\bibitem[{{Nakar} \& {Piran}(2005)}]{2005MNRAS.360L..73N}
{Nakar}, E., \& {Piran}, T. 2005, \mnras, 360, L73

\bibitem[{{Nakar} \& {Piran}(2011)}]{NakarPiranRadioFlares}
---. 2011, \nat, 478, 82

\bibitem[{{Nakar} \& {Sari}(2012)}]{RelativisticShockBreakoutRelation}
{Nakar}, E., \& {Sari}, R. 2012, \apj, 747, 88

\bibitem[{{Narayan} {et~al.}(1992){Narayan}, {Paczynski}, \&
  {Piran}}]{1992ApJ...395L..83N}
{Narayan}, R., {Paczynski}, B., \& {Piran}, T. 1992, \apjl, 395, L83

\bibitem[{{Nissanke} {et~al.}(2010){Nissanke}, {Holz}, {Hughes}, {Dalal}, \&
  {Sievers}}]{StandardSirens}
{Nissanke}, S., {et~al.} 2010, \apj, 725, 496

\bibitem[{Nissanke {et~al.}(2013)Nissanke, Kasliwal, \&
  Georgieva}]{NissankeKasliwalEMCounterparts}
Nissanke, S., Kasliwal, M., \& Georgieva, A. 2013, \apj, 767, 124

\bibitem[{{Nissanke} {et~al.}(2011){Nissanke}, {Sievers}, {Dalal}, \&
  {Holz}}]{NissankeLocalization}
{Nissanke}, S., {et~al.} 2011, \apj, 739, 99

\bibitem[{{Nitz} {et~al.}(2013){Nitz}, {Lundgren}, {Brown}, {Ochsner},
  {Keppel}, \& {Harry}}]{Nitz:2013mxa}
{Nitz}, A.~H., {et~al.} 2013, \prd, 88, 124039

\bibitem[{{Nugent} {et~al.}(2011){Nugent}, {Sullivan}, {Cenko}, {Thomas},
  {Kasen}, {Howell}, {Bersier}, {Bloom}, {Kulkarni}, {Kandrashoff},
  {Filippenko}, {Silverman}, {Marcy}, {Howard}, {Isaacson}, {Maguire},
  {Suzuki}, {Tarlton}, {Pan}, {Bildsten}, {Fulton}, {Parrent}, {Sand},
  {Podsiadlowski}, {Bianco}, {Dilday}, {Graham}, {Lyman}, {James}, {Kasliwal},
  {Law}, {Quimby}, {Hook}, {Walker}, {Mazzali}, {Pian}, {Ofek}, {Gal-Yam}, \&
  {Poznanski}}]{PTF11fe}
{Nugent}, P.~E., {et~al.} 2011, \nat, 480, 344

\bibitem[{Nuttall \& Sutton(2010)}]{galaxy-catalog}
Nuttall, L.~K., \& Sutton, P.~J. 2010, \prd, 82, 102002

\bibitem[{{Nysewander} {et~al.}(2009){Nysewander}, {Fruchter}, \&
  {Pe'er}}]{ComparisonAfterglows}
{Nysewander}, M., {Fruchter}, A.~S., \& {Pe'er}, A. 2009, \apj, 701, 824

\bibitem[{Oates \& Cenko(2014)}]{GCN16672}
Oates, S.~R., \& Cenko, S.~B. 2014, GCN, 16672, 1

\bibitem[{{Ochsenbein} {et~al.}(2000){Ochsenbein}, {Bauer}, \&
  {Marcout}}]{VizieR}
{Ochsenbein}, F., {Bauer}, P., \& {Marcout}, J. 2000, \aaps, 143, 23

\bibitem[{{Ofek} {et~al.}(2012){Ofek}, {Laher}, {Law}, {Surace}, {Levitan},
  {Sesar}, {Horesh}, {Poznanski}, {van Eyken}, {Kulkarni}, {Nugent},
  {Zolkower}, {Walters}, {Sullivan}, {Ag{\"u}eros}, {Bildsten}, {Bloom},
  {Cenko}, {Gal-Yam}, {Grillmair}, {Helou}, {Kasliwal}, \&
  {Quimby}}]{PTFPhotometricCalibration}
{Ofek}, E.~O., {et~al.} 2012, \pasp, 124, 62

\bibitem[{{Ofek} {et~al.}(2013){Ofek}, {Sullivan}, {Cenko}, {Kasliwal},
  {Gal-Yam}, {Kulkarni}, {Arcavi}, {Bildsten}, {Bloom}, {Horesh}, {Howell},
  {Filippenko}, {Laher}, {Murray}, {Nakar}, {Nugent}, {Silverman}, {Shaviv},
  {Surace}, \& {Yaron}}]{PTF10tel}
---. 2013, \nat, 494, 65

\bibitem[{{Oke} \& {Gunn}(1983)}]{ABMags}
{Oke}, J.~B., \& {Gunn}, J.~E. 1983, \apj, 266, 713

\bibitem[{Olver {et~al.}(2010)Olver, Lozier, Boisvert, \&
  Clark}]{Olver:2010:NHMF}
Olver, F.~W.~J., {et~al.}, eds. 2010, {NIST Handbook of Mathematical Functions}
  (New York, NY: Cambridge University Press), print companion to
  \cite{NIST:DLMF}

\bibitem[{Oppenheim {et~al.}(1997)Oppenheim, Willsky, \&
  Nawab}]{oppenheim1997signals}
Oppenheim, A., Willsky, A., \& Nawab, S. 1997, Signals and systems, Prentice
  Hall Signal Processing Series (Upper Saddle River, NJ: Prentice Hall)

\bibitem[{Owen \& Sathyaprakash(1999)}]{Owen:1998dk}
Owen, B.~J., \& Sathyaprakash, B.~S. 1999, \prd, 60, 022002

\bibitem[{{{\"O}zel} {et~al.}(2012){{\"O}zel}, {Psaltis}, {Narayan}, \& {Santos
  Villarreal}}]{NeutronStarMass2}
{{\"O}zel}, F., {et~al.} 2012, \apj, 757, 55

\bibitem[{{Paciesas} {et~al.}(2012{\natexlab{a}}){Paciesas}, {Meegan}, {von
  Kienlin}, {Bhat}, {Bissaldi}, {Briggs}, {Burgess}, {Chaplin}, {Connaughton},
  {Diehl}, {Fishman}, {Fitzpatrick}, {Foley}, {Gibby}, {Giles}, {Goldstein},
  {Greiner}, {Gruber}, {Guiriec}, {van der Horst}, {Kippen}, {Kouveliotou},
  {Lichti}, {Lin}, {McBreen}, {Preece}, {Rau}, {Tierney}, \&
  {Wilson-Hodge}}]{FermiGBMCatalog}
{Paciesas}, W.~S., {et~al.} 2012{\natexlab{a}}, \apjs, 199, 18

\bibitem[{{Paciesas} {et~al.}(2012{\natexlab{b}}){Paciesas}, {Meegan}, {von
  Kienlin}, {Bhat}, {Bissaldi}, {Briggs}, {Burgess}, {Chaplin}, {Connaughton},
  {Diehl}, {Fishman}, {Fitzpatrick}, {Foley}, {Gibby}, {Giles}, {Goldstein},
  {Greiner}, {Gruber}, {Guiriec}, {van der Horst}, {Kippen}, {Kouveliotou},
  {Lichti}, {Lin}, {McBreen}, {Preece}, {Rau}, {Tierney}, \&
  {Wilson-Hodge}}]{FermiGBMFirstTwoYears}
---. 2012{\natexlab{b}}, \apjs, 199, 18

\bibitem[{{Paczynski}(1986)}]{1986ApJ...308L..43P}
{Paczynski}, B. 1986, \apjl, 308, L43

\bibitem[{Page(2013)}]{GCN15329}
Page, K.~L. 2013, GCN, 15329, 1

\bibitem[{Page \& Cenko(2014)}]{GCN16682}
Page, K.~L., \& Cenko, S.~B. 2014, GCN, 16682, 1

\bibitem[{Page {et~al.}(2014)Page, Evans, \& Cenko}]{GCN16670}
Page, K.~L., Evans, P.~A., \& Cenko, S.~B. 2014, GCN, 16670, 1

\bibitem[{{Pan} {et~al.}(2004){Pan}, {Buonanno}, {Chen}, \&
  {Vallisneri}}]{PhysicalTemplateFamily}
{Pan}, Y., {et~al.} 2004, \prd, 69, 104017

\bibitem[{{Panaitescu}(2005)}]{ModelsForAchromaticLightCurveBreaks}
{Panaitescu}, A. 2005, \mnras, 362, 921

\bibitem[{{Pejcha} {et~al.}(2012){Pejcha}, {Thompson}, \&
  {Kochanek}}]{NeutronStarMass1}
{Pejcha}, O., {Thompson}, T.~A., \& {Kochanek}, C.~S. 2012, \mnras, 424, 1570

\bibitem[{Penn {et~al.}(2007)Penn, Kipperman, Newman, Stephens, Harry, Saulson,
  \& Gretarsson}]{ELIGOSusp}
Penn, S., {et~al.} 2007, LIGO-G070553-00-Z

\bibitem[{{Perley}(2014)}]{GCN15680}
{Perley}, D.~A. 2014, GCN, 15680, 1

\bibitem[{{Perley} {et~al.}(2014{\natexlab{a}}){Perley}, {Cao}, {Kasliwal}, \&
  {Kirby}}]{GCN16365}
{Perley}, D.~A., {et~al.} 2014{\natexlab{a}}, GCN, 16365, 1

\bibitem[{{Perley} {et~al.}(2013){Perley}, {Cenko}, \& {Kasliwal}}]{GCN15327}
{Perley}, D.~A., {Cenko}, S.~B., \& {Kasliwal}, M.~M. 2013, GCN, 15327, 1

\bibitem[{{Perley} {et~al.}(2014{\natexlab{b}}){Perley}, {Graham},
  {Filippenko}, \& {Cenko}}]{GCN16454}
{Perley}, D.~A., {et~al.} 2014{\natexlab{b}}, GCN, 16454, 1

\bibitem[{{Perley} \& {Singer}(2014)}]{GCN16362}
{Perley}, D.~A., \& {Singer}, L. 2014, GCN, 16362, 1

\bibitem[{{Perley} {et~al.}(2014{\natexlab{c}}){Perley}, {Cenko}, {Corsi},
  {Tanvir}, {Levan}, {Kann}, {Sonbas}, {Wiersema}, {Zheng}, {Zhao}, {Bai},
  {Bremer}, {Castro-Tirado}, {Chang}, {Clubb}, {Frail}, {Fruchter}, {G{\"o}{\u
  g}{\"u}{\c s}}, {Greiner}, {G{\"u}ver}, {Horesh}, {Filippenko}, {Klose},
  {Mao}, {Morgan}, {Pozanenko}, {Schmidl}, {Stecklum}, {Tanga}, {Volnova},
  {Volvach}, {Wang}, {Winters}, \& {Xin}}]{PerleyGRB130427A}
{Perley}, D.~A., {et~al.} 2014{\natexlab{c}}, \apj, 781, 37

\bibitem[{{Pescalli} {et~al.}(2014){Pescalli}, {Ghirlanda}, {Salafia},
  {Ghisellini}, {Nappo}, \& {Salvaterra}}]{LuminosityFunctionJetStructureGRBs}
{Pescalli}, A., {et~al.} 2014, ArXiv e-prints

\bibitem[{Peters(1964)}]{PhysRev.136.B1224}
Peters, P.~C. 1964, Phys. Rev., 136, B1224

\bibitem[{{Phillips}(1993)}]{PhillipsRelation}
{Phillips}, M.~M. 1993, \apjl, 413, L105

\bibitem[{{Pian} {et~al.}(2000){Pian}, {Amati}, {Antonelli}, {Butler}, {Costa},
  {Cusumano}, {Danziger}, {Feroci}, {Fiore}, {Frontera}, {Giommi}, {Masetti},
  {Muller}, {Nicastro}, {Oosterbroek}, {Orlandini}, {Owens}, {Palazzi},
  {Parmar}, {Piro}, {in't Zand}, {Castro-Tirado}, {Coletta}, {Dal Fiume}, {Del
  Sordo}, {Heise}, {Soffitta}, \& {Torroni}}]{paa+00}
{Pian}, E., {et~al.} 2000, \apj, 536, 778

\bibitem[{{Pian} {et~al.}(2006){Pian}, {Mazzali}, {Masetti}, {Ferrero},
  {Klose}, {Palazzi}, {Ramirez-Ruiz}, {Woosley}, {Kouveliotou}, {Deng},
  {Filippenko}, {Foley}, {Fynbo}, {Kann}, {Li}, {Hjorth}, {Nomoto}, {Patat},
  {Sauer}, {Sollerman}, {Vreeswijk}, {Guenther}, {Levan}, {O'Brien}, {Tanvir},
  {Wijers}, {Dumas}, {Hainaut}, {Wong}, {Baade}, {Wang}, {Amati}, {Cappellaro},
  {Castro-Tirado}, {Ellison}, {Frontera}, {Fruchter}, {Greiner}, {Kawabata},
  {Ledoux}, {Maeda}, {M{\o}ller}, {Nicastro}, {Rol}, \&
  {Starling}}]{GRB060218-SN2006aj-1}
---. 2006, \nat, 442, 1011

\bibitem[{{Piran} {et~al.}(2013){Piran}, {Nakar}, \&
  {Rosswog}}]{PiranNakarRosswogEMSignals}
{Piran}, T., {Nakar}, E., \& {Rosswog}, S. 2013, \mnras, 430, 2121

\bibitem[{{Pozanenko} {et~al.}(2013){Pozanenko}, {Volnova}, {Burhonov}, \&
  {Molotov}}]{GCN14996}
{Pozanenko}, A., {et~al.} 2013, GCN, 14996, 1

\bibitem[{Press {et~al.}(2007{\natexlab{a}})Press, Teukolsky, Vetterling, \&
  Flannery}]{numerical-recipes-inversion-partition}
Press, W.~H., {et~al.} 2007{\natexlab{a}}, Numerical Recipes, 3rd edn.
  (Cambridge Univ. Press), 81--82

\bibitem[{Press {et~al.}(2007{\natexlab{b}})Press, Teukolsky, Vetterling, \&
  Flannery}]{numerical-recipes-chapter-13}
---. 2007{\natexlab{b}}, Numerical Recipes, 3rd edn. (Cambridge: Cambridge
  Univ. Press)

\bibitem[{{Privitera} {et~al.}(2014){Privitera}, {Mohapatra}, {Ajith},
  {Cannon}, {Fotopoulos}, {Frei}, {Hanna}, {Weinstein}, \&
  {Whelan}}]{2014PhRvD..89b4003P}
{Privitera}, S., {et~al.} 2014, \prd, 89, 024003

\bibitem[{{Quimby} {et~al.}(2011){Quimby}, {Kulkarni}, {Kasliwal}, {Gal-Yam},
  {Arcavi}, {Sullivan}, {Nugent}, {Thomas}, {Howell}, {Nakar}, {Bildsten},
  {Theissen}, {Law}, {Dekany}, {Rahmer}, {Hale}, {Smith}, {Ofek}, {Zolkower},
  {Velur}, {Walters}, {Henning}, {Bui}, {McKenna}, {Poznanski}, {Cenko}, \&
  {Levitan}}]{SLSNe}
{Quimby}, R.~M., {et~al.} 2011, \nat, 474, 487

\bibitem[{{Racusin} {et~al.}(2011){Racusin}, {Oates}, {Schady}, {Burrows}, {de
  Pasquale}, {Donato}, {Gehrels}, {Koch}, {McEnery}, {Piran}, {Roming},
  {Sakamoto}, {Swenson}, {Troja}, {Vasileiou}, {Virgili}, {Wanderman}, \&
  {Zhang}}]{FermiSwiftPopulationStudies}
{Racusin}, J.~L., {et~al.} 2011, \apj, 738, 138

\bibitem[{{Rahmer} {et~al.}(2008){Rahmer}, {Smith}, {Velur}, {Hale}, {Law},
  {Bui}, {Petrie}, \& {Dekany}}]{P48PTF}
{Rahmer}, G., {et~al.} 2008, in SPIE Conf. Series, Vol. 7014, SPIE Conf.
  Series, 4

\bibitem[{{Rau} {et~al.}(2013){Rau}, {Kruehler}, \& {Greiner}}]{GCN15330}
{Rau}, A., {Kruehler}, T., \& {Greiner}, J. 2013, GCN, 15330, 1

\bibitem[{{Rau} {et~al.}(2009){Rau}, {Kulkarni}, {Law}, {Bloom}, {Ciardi},
  {Djorgovski}, {Fox}, {Gal-Yam}, {Grillmair}, {Kasliwal}, {Nugent}, {Ofek},
  {Quimby}, {Reach}, {Shara}, {Bildsten}, {Cenko}, {Drake}, {Filippenko},
  {Helfand}, {Helou}, {Howell}, {Poznanski}, \& {Sullivan}}]{PTFRau}
{Rau}, A., {et~al.} 2009, \pasp, 121, 1334

\bibitem[{{Raymond} {et~al.}(2009){Raymond}, {van der Sluys}, {Mandel},
  {Kalogera}, {R{\"o}ver}, \& {Christensen}}]{Raymond:2009}
{Raymond}, V., {et~al.} 2009, Class. Quantum Grav., 26, 114007

\bibitem[{{Read} {et~al.}(2009){Read}, {Markakis}, {Shibata}, {Ury{\= u}},
  {Creighton}, \& {Friedman}}]{MeasuringNSEquationOfState}
{Read}, J.~S., {et~al.} 2009, \prd, 79, 124033

\bibitem[{{Rees} \& {Meszaros}(1994)}]{ReesInternalShocks}
{Rees}, M.~J., \& {Meszaros}, P. 1994, \apjl, 430, L93

\bibitem[{{Rezzolla} {et~al.}(2011){Rezzolla}, {Giacomazzo}, {Baiotti},
  {Granot}, {Kouveliotou}, \& {Aloy}}]{2011ApJ...732L...6R}
{Rezzolla}, L., {et~al.} 2011, \apjl, 732, L6

\bibitem[{{Rhoads}(1997)}]{offaxis}
{Rhoads}, J.~E. 1997, \apjl, 487, L1

\bibitem[{{Richardson} {et~al.}(2002){Richardson}, {Branch}, {Casebeer},
  {Millard}, {Thomas}, \& {Baron}}]{RichardsonComparativeSupernovae}
{Richardson}, D., {et~al.} 2002, \aj, 123, 745

\bibitem[{{Robitaille} {et~al.}(2013){Robitaille}, {Tollerud}, {Greenfield},
  {Droettboom}, {Bray}, {Aldcroft}, {Davis}, {Ginsburg}, {Price-Whelan},
  {Kerzendorf}, {Conley}, {Crighton}, {Barbary}, {Muna}, {Ferguson},
  {Grollier}, {Parikh}, {Nair}, {Unther}, {Deil}, {Woillez}, {Conseil},
  {Kramer}, {Turner}, {Singer}, {Fox}, {Weaver}, {Zabalza}, {Edwards}, {Azalee
  Bostroem}, {Burke}, {Casey}, {Crawford}, {Dencheva}, {Ely}, {Jenness},
  {Labrie}, {Lian Lim}, {Pierfederici}, {Pontzen}, {Ptak}, {Refsdal},
  {Servillat}, \& {Streicher}}]{astropy}
{Robitaille}, T.~P., {et~al.} 2013, \aap, 558, A33

\bibitem[{{Rodriguez} {et~al.}(2014){Rodriguez}, {Farr}, {Raymond}, {Farr},
  {Littenberg}, {Fazi}, \& {Kalogera}}]{RodriguezBasicParameterEstimation}
{Rodriguez}, C.~L., {et~al.} 2014, \apj, 784, 119

\bibitem[{{Roming} {et~al.}(2005){Roming}, {Kennedy}, {Mason}, {Nousek}, {Ahr},
  {Bingham}, {Broos}, {Carter}, {Hancock}, {Huckle}, {Hunsberger}, {Kawakami},
  {Killough}, {Koch}, {McLelland}, {Smith}, {Smith}, {Soto}, {Boyd},
  {Breeveld}, {Holland}, {Ivanushkina}, {Pryzby}, {Still}, \& {Stock}}]{UVOT}
{Roming}, P.~W.~A., {et~al.} 2005, \ssr, 120, 95

\bibitem[{Sahu {et~al.}(2014)Sahu, Bhalerao, \& Anupama}]{GCN16684}
Sahu, D.~K., Bhalerao, V., \& Anupama, G.~C. 2014, GCN, 16684, 1

\bibitem[{{Sakamoto} {et~al.}(2006){Sakamoto}, {Barbier}, {Barthelmy},
  {Cummings}, {Fenimore}, {Gehrels}, {Hullinger}, {Krimm}, {Markwardt},
  {Palmer}, {Parsons}, {Sato}, \& {Tueller}}]{2006ApJ...636L..73S}
{Sakamoto}, T., {et~al.} 2006, \apjl, 636, L73

\bibitem[{{Sari} {et~al.}(1998){Sari}, {Piran}, \&
  {Narayan}}]{AfterglowSpectra}
{Sari}, R., {Piran}, T., \& {Narayan}, R. 1998, \apjl, 497, L17

\bibitem[{Sathyaprakash \& Schutz(2009)}]{livrev12}
Sathyaprakash, B., \& Schutz, B. 2009, Living Rev. Relativity 12,
  arXiv:0903.0338v1

\bibitem[{{Savaglio} {et~al.}(2009){Savaglio}, {Glazebrook}, \& {Le
  Borgne}}]{sgl09}
{Savaglio}, S., {Glazebrook}, K., \& {Le Borgne}, D. 2009, \apj, 691, 182

\bibitem[{{Schaefer} \& {Collazzi}(2007)}]{2007ApJ...656L..53S}
{Schaefer}, B.~E., \& {Collazzi}, A.~C. 2007, \apjl, 656, L53

\bibitem[{Schlafly \& Finkbeiner(2011)}]{SchlaflyExtinction}
Schlafly, E.~F., \& Finkbeiner, D.~P. 2011, \apj, 737, 103

\bibitem[{{Schulze} {et~al.}(2013){Schulze}, {Leloudas}, {Xu}, {Fynbo},
  {Geier}, \& {Jakobsson}}]{GCN14994}
{Schulze}, S., {et~al.} 2013, GCN, 14994, 1

\bibitem[{{Schulze} {et~al.}(2014{\natexlab{a}}){Schulze}, {Malesani},
  {Cucchiara}, {Tanvir}, {Kr{\"u}hler}, {de Ugarte Postigo}, {Leloudas},
  {Lyman}, {Bersier}, {Wiersema}, {Perley}, {Schady}, {Gorosabel}, {Anderson},
  {Castro-Tirado}, {Cenko}, {De Cia}, {Ellerbroek}, {Fynbo}, {Greiner},
  {Hjorth}, {Kann}, {Kaper}, {Klose}, {Levan}, {Mart{\'{\i}}n}, {O'Brien},
  {Page}, {Pignata}, {Rapaport}, {S{\'a}nchez-Ram{\'{\i}}rez}, {Sollerman},
  {Smith}, {Sparre}, {Th{\"o}ne}, {Watson}, {Xu}, {Bauer}, {Bayliss},
  {Bj{\"o}rnsson}, {Bremer}, {Cano}, {Covino}, {D'Elia}, {Frail}, {Geier},
  {Goldoni}, {Hartoog}, {Jakobsson}, {Korhonen}, {Lee}, {Milvang-Jensen},
  {Nardini}, {Nicuesa Guelbenzu}, {Oguri}, {Pandey}, {Petitpas}, {Rossi},
  {Sandberg}, {Schmidl}, {Tagliaferri}, {Tilanus}, {Winters}, {Wright}, \&
  {Wuyts}}]{GRB120422A-SN2012bz-1}
---. 2014{\natexlab{a}}, \aap, 566, A102

\bibitem[{{Schulze} {et~al.}(2014{\natexlab{b}}){Schulze}, {Malesani},
  {Cucchiara}, {Tanvir}, {Kr{\"u}hler}, {de Ugarte Postigo}, {Leloudas},
  {Lyman}, {Bersier}, {Wiersema}, {Perley}, {Schady}, {Gorosabel}, {Anderson},
  {Castro-Tirado}, {Cenko}, {De Cia}, {Ellerbroek}, {Fynbo}, {Greiner},
  {Hjorth}, {Kann}, {Kaper}, {Klose}, {Levan}, {Mart{\'{\i}}n}, {O'Brien},
  {Page}, {Pignata}, {Rapaport}, {S{\'a}nchez-Ram{\'{\i}}rez}, {Sollerman},
  {Smith}, {Sparre}, {Th{\"o}ne}, {Watson}, {Xu}, {Bauer}, {Bayliss},
  {Bj{\"o}rnsson}, {Bremer}, {Cano}, {Covino}, {D'Elia}, {Frail}, {Geier},
  {Goldoni}, {Hartoog}, {Jakobsson}, {Korhonen}, {Lee}, {Milvang-Jensen},
  {Nardini}, {Nicuesa Guelbenzu}, {Oguri}, {Pandey}, {Petitpas}, {Rossi},
  {Sandberg}, {Schmidl}, {Tagliaferri}, {Tilanus}, {Winters}, {Wright}, \&
  {Wuyts}}]{2014A&A...566A.102S}
---. 2014{\natexlab{b}}, \aap, 566, A102

\bibitem[{{Schutz}(1986)}]{SchutzStandardSirens}
{Schutz}, B.~F. 1986, \nat, 323, 310

\bibitem[{{Schutz}(2011)}]{ShutzThreeFiguresOfMerit}
---. 2011, Class. Quantum Grav., 28, 125023

\bibitem[{{Shahmoradi} \& {Nemiroff}(2011)}]{2011MNRAS.411.1843S}
{Shahmoradi}, A., \& {Nemiroff}, R.~J. 2011, \mnras, 411, 1843

\bibitem[{Shoemaker(2010)}]{ALIGONoise}
Shoemaker, D. 2010, LIGO-T0900288-v3

\bibitem[{{Sidery} {et~al.}(2014){Sidery}, {Aylott}, {Christensen}, {Farr},
  {Farr}, {Feroz}, {Gair}, {Grover}, {Graff}, {Hanna}, {Kalogera}, {Mandel},
  {O'Shaughnessy}, {Pitkin}, {Price}, {Raymond}, {R{\"o}ver}, {Singer}, {van
  der Sluys}, {Smith}, {Vecchio}, {Veitch}, \&
  {Vitale}}]{SiderySkyLocalizationComparison}
{Sidery}, T., {et~al.} 2014, \prd, 89, 084060

\bibitem[{{Siegel} \& {De Pasquale}(2014)}]{GCN16432}
{Siegel}, M.~H., \& {De Pasquale}, M. 2014, GCN, 16432, 1

\bibitem[{Simakov(2014)}]{PhysRevD.90.102003}
Simakov, D.~A. 2014, \prd, 90, 102003

\bibitem[{{Singer} {et~al.}(2013){Singer}, {Cenko}, \& {Kasliwal}}]{GCN15643}
{Singer}, L.~P., {Cenko}, S.~B., \& {Kasliwal}, M.~M. 2013, GCN, 15643, 1

\bibitem[{Singer {et~al.}(2014{\natexlab{a}})Singer, {Cenko}, {Kasliwal},
  {Fremling}, \& {Dzigan}}]{GCN16226}
Singer, L.~P., {et~al.} 2014{\natexlab{a}}, GCN, 16226, 1

\bibitem[{Singer {et~al.}(2014{\natexlab{b}})Singer, Kasliwal, Bhalerao, Cenko,
  Perley, \& Johansson}]{GCN16668}
---. 2014{\natexlab{b}}, GCN, 16668, 1

\bibitem[{Singer {et~al.}(2014{\natexlab{c}})Singer, {Kasliwal}, \&
  {Cenko}}]{GCN16360}
Singer, L.~P., {Kasliwal}, M.~M., \& {Cenko}, S.~B. 2014{\natexlab{c}}, GCN,
  16360, 1

\bibitem[{Singer {et~al.}(2013)Singer, {Cenko}, {Kasliwal}, {Perley}, {Ofek},
  {Brown}, {Nugent}, {Kulkarni}, {Corsi}, {Frail}, {Bellm}, {Mulchaey},
  {Arcavi}, {Barlow}, {Bloom}, {Cao}, {Gehrels}, {Horesh}, {Masci}, {McEnery},
  {Rau}, {Surace}, \& {Yaron}}]{iPTF13bxl}
Singer, L.~P., {et~al.} 2013, \apjl, 776, L34

\bibitem[{{Singer} {et~al.}(2013){Singer}, {Cenko}, {Kasliwal}, {Brown},
  {Yaron}, {Bellm}, {Caudill}, {Tinyanont}, {Khatami}, \&
  {Weinstein}}]{GCN14967}
{Singer}, L.~P., {et~al.} 2013, GCN, 14967, 1

\bibitem[{{Singer} {et~al.}(2014){Singer}, {Price}, {Farr}, {Urban}, {Pankow},
  {Vitale}, {Veitch}, {Farr}, {Hanna}, {Cannon}, {Downes}, {Graff}, {Haster},
  {Mandel}, {Sidery}, \& {Vecchio}}]{FirstTwoYears}
---. 2014, \apj, 795, 105

\bibitem[{{Skrutskie} {et~al.}(2006){Skrutskie}, {Cutri}, {Stiening},
  {Weinberg}, {Schneider}, {Carpenter}, {Beichman}, {Capps}, {Chester},
  {Elias}, {Huchra}, {Liebert}, {Lonsdale}, {Monet}, {Price}, {Seitzer},
  {Jarrett}, {Kirkpatrick}, {Gizis}, {Howard}, {Evans}, {Fowler}, {Fullmer},
  {Hurt}, {Light}, {Kopan}, {Marsh}, {McCallon}, {Tam}, {Van Dyk}, \&
  {Wheelock}}]{2MASS}
{Skrutskie}, M.~F., {et~al.} 2006, \aj, 131, 1163

\bibitem[{{Smith} {et~al.}(2014{\natexlab{a}}){Smith}, {Hanna}, {Mandel}, \&
  {Vecchio}}]{interpolation-pe}
{Smith}, R.~J.~E., {et~al.} 2014{\natexlab{a}}, \prd, 90, 044074

\bibitem[{{Smith} {et~al.}(2014{\natexlab{b}}){Smith}, {Dekany}, {Bebek},
  {Bellm}, {Bui}, {Cromer}, {Gardner}, {Hoff}, {Kaye}, {Kulkarni}, {Lambert},
  {Levi}, \& {Reiley}}]{ZTFSmith}
{Smith}, R.~M., {et~al.} 2014{\natexlab{b}}, in SPIE Conf. Series, Vol. 9147,
  SPIE Conf. Series, 79

\bibitem[{{Soderberg} {et~al.}(2004){Soderberg}, {Kulkarni}, {Berger}, {Fox},
  {Sako}, {Frail}, {Gal-Yam}, {Moon}, {Cenko}, {Yost}, {Phillips}, {Persson},
  {Freedman}, {Wyatt}, {Jayawardhana}, \& {Paulson}}]{2004Natur.430..648S}
{Soderberg}, A.~M., {et~al.} 2004, \nat, 430, 648

\bibitem[{{Soderberg} {et~al.}(2010){Soderberg}, {Chakraborti}, {Pignata},
  {Chevalier}, {Chandra}, {Ray}, {Wieringa}, {Copete}, {Chaplin},
  {Connaughton}, {Barthelmy}, {Bietenholz}, {Chugai}, {Stritzinger}, {Hamuy},
  {Fransson}, {Fox}, {Levesque}, {Grindlay}, {Challis}, {Foley}, {Kirshner},
  {Milne}, \& {Torres}}]{scp+10}
---. 2010, \nat, 463, 513

\bibitem[{{Sollerman} {et~al.}(2006){Sollerman}, {Jaunsen}, {Fynbo}, {Hjorth},
  {Jakobsson}, {Stritzinger}, {F{\'e}ron}, {Laursen}, {Ovaldsen}, {Selj},
  {Th{\"o}ne}, {Xu}, {Davis}, {Gorosabel}, {Watson}, {Duro}, {Ilyin}, {Jensen},
  {Lysfjord}, {Marquart}, {Nielsen}, {N{\"a}r{\"a}nen}, {Schwarz}, {Walch},
  {Wold}, \& {{\"O}stlin}}]{GRB060218-SN2006aj-3}
{Sollerman}, J., {et~al.} 2006, \aap, 454, 503

\bibitem[{{Sonbas} {et~al.}(2013){Sonbas}, {Racusin}, {Kocevski}, \&
  {McEnery}}]{GCN15640}
{Sonbas}, E., {et~al.} 2013, GCN, 15640, 1

\bibitem[{Strain \& Cagnoli(2006)}]{ALIGOSusp}
Strain, K., \& Cagnoli, G. 2006, LIGO-T050267-01-K

\bibitem[{{Sudilovsky} {et~al.}(2013){Sudilovsky}, {Tanga}, \&
  {Greiner}}]{GCN15328}
{Sudilovsky}, V., {Tanga}, M., \& {Greiner}, J. 2013, GCN, 15328, 1

\bibitem[{{Svensson} {et~al.}(2010){Svensson}, {Levan}, {Tanvir}, {Fruchter},
  \& {Strolger}}]{GRBSNHostGalaxies}
{Svensson}, K.~M., {et~al.} 2010, \mnras, 405, 57

\bibitem[{{Taboada} \& {Gilmore}(2014)}]{HAWC-GRB-2}
{Taboada}, I., \& {Gilmore}, R.~C. 2014, Nuclear Instruments and Methods in
  Physics Research A, 742, 276

\bibitem[{{Taracchini} {et~al.}(2014){Taracchini}, {Buonanno}, {Pan},
  {Hinderer}, {Boyle}, {Hemberger}, {Kidder}, {Lovelace}, {Mrou{\'e}},
  {Pfeiffer}, {Scheel}, {Szil{\'a}gyi}, {Taylor}, \&
  {Zenginoglu}}]{Taracchini:2013}
{Taracchini}, A., {et~al.} 2014, \prd, 89, 061502

\bibitem[{{Taylor} \& {Weisberg}(1982)}]{1982ApJ...253..908T}
{Taylor}, J.~H., \& {Weisberg}, J.~M. 1982, \apj, 253, 908

\bibitem[{Turin(1960)}]{matched-filter}
Turin, G. 1960, IRE Trans. Inf. Theory, 6, 311

\bibitem[{{Urata} {et~al.}(2012){Urata}, {Huang}, {Yamaoka}, {Tsai}, \&
  {Tashiro}}]{UrataEnergeticWAMLATGRBs}
{Urata}, Y., {et~al.} 2012, \apjl, 748, L4

\bibitem[{{Urata} {et~al.}(2014){Urata}, {Huang}, {Takahashi}, {Im}, {Yamaoka},
  {Tashiro}, {Kim}, {Jang}, \& {Pak}}]{UrataGRB120326A}
---. 2014, \apj, 789, 146

\bibitem[{{van den Broeck} {et~al.}(2009){van den Broeck}, {Brown}, {Cokelaer},
  {Harry}, {Jones}, {Sathyaprakash}, {Tagoshi}, \&
  {Takahashi}}]{2009PhRvD..80b4009V}
{van den Broeck}, C., {et~al.} 2009, \prd, 80, 024009

\bibitem[{{van der Sluys} {et~al.}(2008{\natexlab{a}}){van der Sluys},
  {Raymond}, {Mandel}, {R{\"o}ver}, {Christensen}, {Kalogera}, {Meyer}, \&
  {Vecchio}}]{2008CQGra..25r4011V}
{van der Sluys}, M., {et~al.} 2008{\natexlab{a}}, Class. Quantum Grav., 25,
  184011

\bibitem[{{van der Sluys} {et~al.}(2008{\natexlab{b}}){van der Sluys},
  {R{\"o}ver}, {Stroeer}, {Raymond}, {Mandel}, {Christensen}, {Kalogera},
  {Meyer}, \& {Vecchio}}]{2008ApJ...688L..61V}
{van der Sluys}, M.~V., {et~al.} 2008{\natexlab{b}}, \apjl, 688, L61

\bibitem[{{van Eerten} \& {MacFadyen}(2011)}]{SyntheticSGRBAfterglows}
{van Eerten}, H.~J., \& {MacFadyen}, A.~I. 2011, \apjl, 733, L37

\bibitem[{{van Paradijs} {et~al.}(1997){van Paradijs}, {Groot}, {Galama},
  {Kouveliotou}, {Strom}, {Telting}, {Rutten}, {Fishman}, {Meegan}, {Pettini},
  {Tanvir}, {Bloom}, {Pedersen}, {N{\o}rdgaard-Nielsen}, {Linden-V{\o}rnle},
  {Melnick}, {van der Steene}, {Bremer}, {Naber}, {Heise}, {in't Zand},
  {Costa}, {Feroci}, {Piro}, {Frontera}, {Zavattini}, {Nicastro}, {Palazzi},
  {Bennett}, {Hanlon}, \& {Parmar}}]{GRBsHaveOpticalAfterglows}
{van Paradijs}, J., {et~al.} 1997, \nat, 386, 686

\bibitem[{{Veitch} \& {Vecchio}(2010)}]{LALINFERENCE_NEST}
{Veitch}, J., \& {Vecchio}, A. 2010, \prd, 81, 062003

\bibitem[{{Veitch} {et~al.}(2012){Veitch}, {Mandel}, {Aylott}, {Farr},
  {Raymond}, {Rodriguez}, {van der Sluys}, {Kalogera}, \&
  {Vecchio}}]{Veitch:2012}
{Veitch}, J., {et~al.} 2012, \prd, 85, 104045

\bibitem[{{V{\'e}ron-Cetty} \& {V{\'e}ron}(2010)}]{QuasarAtlas}
{V{\'e}ron-Cetty}, M.-P., \& {V{\'e}ron}, P. 2010, \aap, 518, A10

\bibitem[{{Vitale} \& {Zanolin}(2011)}]{2011PhRvD..84j4020V}
{Vitale}, S., \& {Zanolin}, M. 2011, \prd, 84, 104020

\bibitem[{Vogt(1989)}]{LIGOProposal}
Vogt, R.~E. 1989, A Laser Interferometer Gravitational-Wave Observatory, Tech.
  Rep. PHY-8803557, Proposal to the National Science Foundation

\bibitem[{{Volnova} {et~al.}(2013){Volnova}, {Inasaridze}, {Inasaridze},
  {Zhuzhunadze}, {Krugly}, {Khnu}, {Molotov}, \& {Pozanenko}}]{GCN15341}
{Volnova}, A., {et~al.} 2013, GCN, 15341, 1

\bibitem[{{Volnova} {et~al.}(2014{\natexlab{a}}){Volnova}, {Klunko},
  {Eselevich}, {Korobtsev}, \& {Pozanenko}}]{GCN16260}
---. 2014{\natexlab{a}}, GCN, 16260, 1

\bibitem[{{Volnova} {et~al.}(2014{\natexlab{b}}){Volnova}, {Mundrzyjewski},
  {Kusakin}, \& {Pozanenko}}]{GCN16453}
---. 2014{\natexlab{b}}, GCN, 16453, 1

\bibitem[{{von Kienlin}(2014)}]{GCN16450}
{von Kienlin}, A. 2014, GCN, 16450, 1

\bibitem[{{Wen} \& {Chen}(2010)}]{WenLocalizationAdvancedLIGO}
{Wen}, L., \& {Chen}, Y. 2010, \prd, 81, 082001

\bibitem[{{White} {et~al.}(2011){White}, {Daw}, \& {Dhillon}}]{GWGC}
{White}, D.~J., {Daw}, E.~J., \& {Dhillon}, V.~S. 2011, Class. Quantum Grav.,
  28, 085016

\bibitem[{{Wiersema} {et~al.}(2014){Wiersema}, {Tanvir}, {Levan}, \&
  {Karjalainen}}]{GCN16231}
{Wiersema}, K., {et~al.} 2014, GCN, 16231, 1

\bibitem[{Will(2006)}]{lrr-2006-3}
Will, C.~M. 2006, Living Rev. Rel., 9

\bibitem[{{Wright} {et~al.}(2010){Wright}, {Eisenhardt}, {Mainzer}, {Ressler},
  {Cutri}, {Jarrett}, {Kirkpatrick}, {Padgett}, {McMillan}, {Skrutskie},
  {Stanford}, {Cohen}, {Walker}, {Mather}, {Leisawitz}, {Gautier}, {McLean},
  {Benford}, {Lonsdale}, {Blain}, {Mendez}, {Irace}, {Duval}, {Liu}, {Royer},
  {Heinrichsen}, {Howard}, {Shannon}, {Kendall}, {Walsh}, {Larsen}, {Cardon},
  {Schick}, {Schwalm}, {Abid}, {Fabinsky}, {Naes}, \& {Tsai}}]{WISEOnOrbit}
{Wright}, E.~L., {et~al.} 2010, \aj, 140, 1868

\bibitem[{Xu {et~al.}(2013)Xu, {Malesani}, {Kruehler}, {Fynbo}, {Jakobsson}, \&
  {Telting}}]{GCN15325}
Xu, D., {et~al.} 2013, GCN, 15325, 1

\bibitem[{{Xu} {et~al.}(2013){Xu}, {Niu}, {Zhang}, \& {Esamdin}}]{GCN15641}
{Xu}, D., {et~al.} 2013, GCN, 15641, 1

\bibitem[{Xu {et~al.}(2013{\natexlab{a}})Xu, {Zhang}, \& {Cao}}]{GCN15326}
Xu, D., {Zhang}, C.-M., \& {Cao}, C. 2013{\natexlab{a}}, GCN, 15326, 1

\bibitem[{Xu {et~al.}(2013{\natexlab{b}})Xu, {de Ugarte Postigo}, {Leloudas},
  {Kr{\"u}hler}, {Cano}, {Hjorth}, {Malesani}, {Fynbo}, {Th{\"o}ne},
  {S{\'a}nchez-Ram{\'{\i}}rez}, {Schulze}, {Jakobsson}, {Kaper}, {Sollerman},
  {Watson}, {Cabrera-Lavers}, {Cao}, {Covino}, {Flores}, {Geier}, {Gorosabel},
  {Hu}, {Milvang-Jensen}, {Sparre}, {Xin}, {Zhang}, {Zheng}, \&
  {Zou}}]{GRB130427A-SN2013cq}
Xu, D., {et~al.} 2013{\natexlab{b}}, \apj, 776, 98

\bibitem[{{Yaron} \& {Gal-Yam}(2012)}]{yg12}
{Yaron}, O., \& {Gal-Yam}, A. 2012, \pasp, 124, 668

\bibitem[{{Yu} \& {Goldstein}(2014)}]{GCN16224}
{Yu}, H.-F., \& {Goldstein}, A. 2014, GCN, 16224, 1

\bibitem[{Zhang(2014)}]{GCN16669}
Zhang, B. 2014, GCN, 16669, 1

\bibitem[{{Zhang} {et~al.}(2012){Zhang}, {Fan}, {Shen}, {Xu}, {Zhang}, {Wei},
  {Burrows}, {Zhang}, \& {Gehrels}}]{zfs+12}
{Zhang}, B.-B., {et~al.} 2012, \apj, 756, 190

\bibitem[{{Zwart} {et~al.}(2008){Zwart}, {Barker}, {Biddulph}, {Bly}, {Boysen},
  {Brown}, {Clementson}, {Crofts}, {Culverhouse}, {Czeres}, {Dace}, {Davies},
  {D'Alessandro}, {Doherty}, {Duggan}, {Ely}, {Felvus}, {Feroz}, {Flynn},
  {Franzen}, {Geisb{\"u}sch}, {G{\'e}nova-Santos}, {Grainge}, {Grainger},
  {Hammett}, {Hills}, {Hobson}, {Holler}, {Hurley-Walker}, {Jilley}, {Jones},
  {Kaneko}, {Kneissl}, {Lancaster}, {Lasenby}, {Marshall}, {Newton}, {Norris},
  {Northrop}, {Odell}, {Petencin}, {Pober}, {Pooley}, {Pospieszalski}, {Quy},
  {Rodr{\'{\i}}guez-Gonz{\'a}lvez}, {Saunders}, {Scaife}, {Schofield}, {Scott},
  {Shaw}, {Shimwell}, {Smith}, {Taylor}, {Titterington}, {Veli{\'c}},
  {Waldram}, {West}, {Wood}, {Yassin}, \& {AMI Consortium}}]{AMI}
{Zwart}, J.~T.~L., {et~al.} 2008, \mnras, 391, 1545

\end{thebibliography}

\end{document}